\documentclass{supporting-files/stylesheet}
\usepackage{subfiles}
\usepackage{fancyhdr} %for header

\newcommand{\lp}{\left(}
\newcommand{\rp}{\right)}

\renewcommand{\>}{\rangle}
\newcommand{\<}{\langle}
\newcommand{\stabcode}[3]{\llbracket #1, #2, #3 \rrbracket}
\newcommand{\subsyscode}[4]{\llbracket #1, #2, #3, #4 \rrbracket}

\newcommand{\CC}{\mathbb{C}}
\newcommand{\FF}{\mathbb{F}}
\newcommand{\NN}{\mathbb{N}}
\newcommand{\RR}{\mathbb{R}}
\newcommand{\ZZ}{\mathbb{Z}}

\newcommand{\spn}{{\text{span}}}
\newcommand{\defsum}[3]{\sum_{#1}^{#2}{#3}}

\newcommand{\XOR}{\textsc{xor}}

\newcommand{\code}{\mathcal{C}}
\newcommand{\Pauli}{\mathcal{P}}

\newcommand{\barX}{\overline{X}}
\newcommand{\barZ}{\overline{Z}}
\newcommand{\barH}{\overline{H}}
\newcommand{\barS}{\overline{S}}

\newcommand{\bigOh}[1]{O\!\lp #1 \rp}
\newcommand{\bigOmega}[1]{\Omega\!\lp #1 \rp}
\newcommand{\bigTheta}[1]{\Theta\!\lp #1 \rp}

\newcommand{\cnot}{\textsc{cnot}\xspace}

\newcommand{\numth}{\textsuperscript{th}}

\renewcommand{\ketbra}[1]{|{#1}\>\mkern-4mu\<{#1}|}
\renewcommand{\tr}{\textup{Tr}}

\renewcommand{\>}{\rangle}
\renewcommand{\<}{\langle}
\newcommand{\N}{{\mathbb{N}}}
\renewcommand{\C}{{\mathbb{C}}}
\renewcommand{\R}{{\mathbb{R}}}

\newcommand{\Z}{{\mathbb{Z}}}

\newcommand{\GL}{{\operatorname{GL}}}

\renewcommand{\spn}{{\textrm{span}\,}}
\newcommand{\im}{{\textrm{im}\,}}
\renewcommand{\poly}{{\textrm{poly}\,}}

\newcommand{\Hg}{{\mathtt{H}}}
\newcommand{\Tg}{{\mathtt{T}}}
\newcommand{\cnotg}{{\mathrm{CNOT}}}
\newcommand{\Sg}{{\mathtt{S}}}
\newcommand{\swapg}{{\mathtt{SWAP}}}
\newcommand{\czg}{{\mathrm{CZ}}}
\newcommand{\Id}{{\mathtt{Id}}}

\newcommand{\loc}{{\texttt{loc}}}
\newcommand{\supp}{{\mathrm{supp}}}
\newcommand{\mdc}{{\mathrm{MDC}}}
\newcommand{\wt}{{\mathrm{wt}}}
\newcommand{\mrk}{{\mathrm{mrk}}}

\newcommand{\qnoise}[1]{\gate[1,style={dashed,rounded corners,fill=red!10}]{#1}}

\newcommand{\zmeas}[2]{ \gate[#1,style={fill=blue!10}]{#2} }

\newcommand{\invgate}[1]{ \gate[1,style={blue!0,fill=blue!0,fill opacity=0,opacity=0}]{#1} }
\newcommand{\bshade}[3]{{ \gategroup[#1,steps=#2,style={blue!10,rounded corners,fill=blue!10, inner xsep=1pt},background,label style={label position=above,anchor=north,yshift=0.3cm}]{{#3}} }}
\newcommand{\bshadeth}[3]{{ \gategroup[#1,steps=#2,style={blue!10,rounded corners,fill=blue!10, inner xsep=1pt,inner ysep=2.5pt},background,label style={label position=above,anchor=north,yshift=0.4cm}]{{#3}} }}

\newcommand{\gshade}[3]{{ \gategroup[#1,steps=#2,style={green!10,fill=green!10, inner xsep=1pt,inner ysep=1pt},background,label style={label position=below,anchor=north,yshift=-0.2cm}]{{#3}} }}

\newcommand{\rgate}[1]{\gate[1,style={red!20,fill=red!10}]{#1}}

\providetoggle{@accepted}
\providetoggle{@unpublished}
\togglefalse{@accepted}
\togglefalse{@unpublished}

\definecolor{dgray}{RGB}{64,64,64}
\renewcommand{\subsectionmark}[1]{}
\renewcommand{\headrulewidth}{0.4pt}
\renewcommand{\headrule}{\hbox to\headwidth{\color{black}\leaders\hrule height \headrulewidth\hfill}}
\fancypagestyle{plain}{
    \fancyhead{}
    \renewcommand{\headrulewidth}{0pt}
    \iftoggle{@accepted}
    {\fancyfoot[R]{\iftoggle{@unpublished}{\thepage}{\thepage}}}
    {\fancyfoot[C]{\iftoggle{@unpublished}{\thepage}{\thepage}}}
}
\fancypagestyle{openingstyle}{
    \fancyhead{}
    \renewcommand{\headrulewidth}{0pt}
    \iftoggle{@accepted}
    {\fancyfoot[R]{\iftoggle{@unpublished}{\thepage}{\thepage}}}
    {\fancyfoot[C]{\iftoggle{@unpublished}{\thepage}{\thepage}}}
}
\pretocmd{\section}{\clearpage\thispagestyle{plain}}{}{}
\bibliography{supporting-files/references.bib}

\begin{document}
\title{Quantum error correction and fault tolerance: A comprehensive tutorial}

\author[1,2,3]{Daniel J. Spencer\thanks{These authors contributed equally}}
\author[1,2]{Shubham P. Jain\protect\footnotemark[1]}
\author[4,5]{Andrew Tanggara\protect\footnotemark[1]}
\author[4]{Zeen Sun\protect\footnotemark[1]}
\author[6]{Tobias Haug}
\author[7]{Derek Khu}
\author[1,8]{Kishor Bharti\thanks{Corresponding author: \href{mailto:kishor.bharti1@gmail.com}{\texttt{kishor.bharti1@gmail.com}}}}

\affil[1]{Joint Center for Quantum Information and Computer Science, NIST/University of Maryland, College Park, MD 20742, USA}
\affil[2]{Department of Physics, University of Maryland, College Park, MD 20742, USA}
\affil[3]{Joint Quantum Institute, NIST/University of Maryland, College Park, MD 20742, USA}
\affil[4]{Centre for Quantum Technologies, National University of Singapore, 3 Science Drive 2, Singapore 117543}
\affil[5]{Nanyang Quantum Hub, School of Physical and Mathematical Sciences, Nanyang Technological University, Singapore 639673}
\affil[6]{Quantum Research Center, Technology Innovation Institute, Abu Dhabi, UAE}
\affil[7]{Institute for Infocomm Research (I\textsuperscript{2}R), Agency for Science, Technology and Research (A*STAR),
1 Fusionopolis Way, \#21-01, Connexis South Tower, Singapore 138632, Republic of Singapore}
\affil[8]{Department of Computer Science and Institute for Advanced Computer Studies, University of Maryland, College Park, Maryland 20742, USA}
\date{}
\maketitle

\begin{abstract}
    Noise is one of the central obstacles to building useful quantum computers, and quantum error correction (QEC) provides the framework for protecting quantum information against it. Unlike classical error correction, QEC must preserve fragile quantum states without copying them, measuring them directly, or destroying the information they encode. Driven by rapid progress in both theory and experiment, this challenge has grown into one of the most active areas of quantum information science. This tutorial gives a guided introduction to modern QEC, developing the core concepts of codes, syndromes, stabilizers, decoding, and fault tolerance before connecting them to major code families and current research directions. We cover both established constructions and newer developments, including topological and subsystem codes, bosonic and qudit codes, dynamical codes, and quantum low-density parity-check (qLDPC) codes. The emphasis is on building operational understanding: explaining not only what the main objects are, but how they are used in code design, error diagnosis, decoding, and fault-tolerant computation. The tutorial is intended for newcomers seeking a first path through QEC, as well as researchers looking for a coherent reference for the concepts, code families, and tools that arise in current work.
\end{abstract}

\thispagestyle{openingstyle} %for header

\newpage
\thispagestyle{openingstyle} %for header
\begin{center}
    \vspace*{\fill}
    \emph{The following tutorial is derived in part from an online lecture series held between August 20, 2025 and December 17, 2025. These lectures are now archived on YouTube and can be found~\href{https://www.youtube.com/playlist?list=PLKl95OY6CIkJWDk1y4xgN0ftb7i0slc2a}{here}.} 
    \vspace*{\fill}
\end{center}
\newpage

\pdfbookmark[1]{Contents}{Contents}
\tableofcontents
\thispagestyle{plain} %for header

\clearpage
\section{Motivation}\label{sec:motivation}

Quantum computing is one of the boldest ideas in modern science. Quantum computers can perform calculations beyond what is possible with classical computers by harnessing hitherto abstract concepts of quantum mechanics, such as interference, entanglement, and superposition. This promises tremendous breakthroughs in a wide range of fields, including  chemistry~\cite{lanyon2010towards,cao2019quantum,bauer2020quantum,mcardle2020quantum,motta2021emerging}, many-body physics~\cite{smith2019simulating,ayral2023quantum,fauseweh2024quantum}, cryptography~\cite{shor1994algorithms,gisin2002quantum,portmann2022security,xu2020secure}, sensing~\cite{degen2017quantum}, optimization~\cite{symons2023practitioner,abbas2024challenges}, and simulation~\cite{feynman1982simulating,lloyd1996universal,aspuru-guzik2005simulated,ma2020quantum,daley2022practical}. Initially introduced as a curious theoretical idea, quantum computing has since exploded into a fast-moving industry, with real quantum devices becoming widely available across various physical platforms. Progress appears almost daily: larger processors, higher gate fidelities, longer qubit coherence times, and more sophisticated algorithmic demonstrations.

However, it must contend with a fundamental issue. The very same quantum effects that give quantum computing its power also make it fragile and susceptible to noise. This noise is unavoidable, so the quantum computer has to constantly deal with errors trying to corrupt its internal state. Examples of such errors are bit flips, phase flips, decoherence, crosstalk, and even external events like cosmic-ray impacts. These errors accumulate over the course of a computation, at every gate, while a qubit idles waiting for the next gate, and at the point of measurement to extract the relevant information from the computation. Without some way to fix this problem, the accumulation of errors destroys the information in the computation, rendering the output useless.

Initially, this appeared to be a devastating obstacle to overcome. While classical computation allows for error correction via the redundant encoding of information by copying it and checking it, the very rules of quantum mechanics forbid this~\cite{wootters1982single}. In particular, unknown quantum states cannot be cloned and the act of measuring a quantum state disturbs the state being protected. Despite this, \emph{quantum error correction} (QEC) resolves this seemingly paradoxical problem. Quantum error-correcting codes encode quantum information distributed across a larger system, which remarkably allows one to detect and correct errors afflicting the state, all without learning or destroying the encoded state itself. Thus, while advancements in materials, control, qubit architecture, cryogenics, and algorithms help in accelerating the field forward, progress towards large-scale, useful quantum computation ultimately depends on quantum error correction.

Today's devices, despite exciting progress over the past few decades, still exhibit physical qubit error rates on the order of $10^{-4}$ to $10^{-3}$~\cite{ransford2025helios}. While a ``one-in-a-thousand'' or even ``one-in-ten-thousand'' error rate may sound acceptable, useful large-scale quantum computations may require logical failure rates of $10^{-9}$ or even less than that, depending on the architecture and algorithm being deployed. This is \emph{five to six orders of magnitude} lower than what has been achieved so far. Barring a major alternative way of achieving large-scale quantum computation, the most promising way forward is through quantum error correction.

We can quantify the physical error rate needed to achieve \emph{fault-tolerant quantum computation} (FTQC) through the \emph{threshold theorem}~\cite{aharonov2008fault-tolerant,knill1996threshold,kitaev1997fault-tolerant}. Informally, this cornerstone theorem guarantees the existence of a physical error rate below which any arbitrarily long quantum computation can be performed reliably by scaling up the number of qubits of the quantum computer. Thus, noise does not need to be eliminated entirely, it only needs to be suppressed to a sufficiently low level. The specific value depends on the quantum error-correcting code being used. For example, the surface code has a threshold value on the order of 0.1\% to 1\% and can require on the order of 1,000--10,000 \emph{physical qubits per logical qubit} to implement~\cite{fowler2012surface}. While this makes the surface code highly attractive for experimental demonstrations of early fault tolerance, quantum computations requiring hundreds or thousands of \emph{logical} qubits would require hundreds of thousands or \emph{millions} of physical qubits, which is out of reach for current devices. Recent progress in promising code constructions like \emph{quantum low-density parity-check} (qLDPC) codes has lowered these numbers significantly~\cite{webster2026pinnacle,cain2026shors}, but there remains a lot of work to be done.

The tension between the theoretical scalability and the practical, experimental overhead required to realize large-scale quantum computation is one of the primary reasons that quantum error correction remains a popular and active field of research. Over the past several years, there has been vast progress made in both theory, with the introduction of novel code constructions, improvements in code parameters, and decoding algorithms, and experiment, with advancements in the quality of qubits, coherence times, and gate fidelities, and even some early demonstrations of logical error suppression below threshold and small fault-tolerant primitives~\cite{sundaresan2023demonstrating,googlequantumai2023suppressing,bluvstein2024logical,hong2024entangling,rodriguez2024experimental,paetznik2024demonstration,lacroix2025scaling,eickbusch2025demonstration,googlequantumai2025quantum,butt2026demonstration}. However, the pace of development can make entry into the field difficult for newcomers and even for seasoned experts to keep track of.

This tutorial is written with this challenge in mind. It aims to bring a fresh, modern perspective on quantum error correction, focusing not only on the basics, but also diving deeper into recent breakthroughs and modern techniques. We take the reader on a journey through the basic formalism of quantum error correction, several promising code constructions, decoding techniques, the limitations of quantum error-correcting codes, and fault tolerance, the pinnacle of quantum error correction. This tutorial is designed to be a comprehensive entry point for those new to quantum error correction and a pedagogical reference for researchers actively working in quantum information more broadly.

\newpage

\section{Brief introduction to quantum information}\label{sec:quantum-intro}

While we assume the reader is familiar with the basics of quantum information and quantum computing, we nonetheless provide a brief overview as a helpful reference if needed and to declare the notation that we use throughout the rest of the tutorial. For a more comprehensive introduction to quantum information and quantum computing, we refer the reader to the many excellent textbooks~\cite{nielsen2010quantum,kitaev2002classical} and lecture notes~\cite{aaronson2018introduction,dewolf2019quantum,jozsa2019,watrous2025understanding,preskill2025course,childs2025lecture}.

\subsection{Quantum states, operators, and measurements}\label{subsec:quantum-intro-states-operators-measurements}
A quantum system is described mathematically by a state vector, which is an element of a complex-valued Hilbert space. Informally, a Hilbert space is a vector space with an inner product, together with a completeness property.\footnote{A vector space is complete if every Cauchy sequence in the space converges to a limit that also lies in the space. In particular, every finite-dimensional normed vector space is automatically complete.} For a single qubit, the relevant Hilbert space is two-dimensional. We will use \emph{Dirac notation} for state vectors: a state is written as a \emph{ket}, $\ket{\psi}$, and its dual vector is written as a \emph{bra}, $\bra{\psi}$. The bra is the Hermitian conjugate of the ket, meaning its complex conjugate transpose, that is, $\bra{\psi} = \ket{\psi}^{\dagger}$.

Any single-qubit state can be written as a \emph{superposition} of \emph{computational basis} vectors, which are usually labeled as $\ket{0}$ and $\ket{1}$. Formally, a superposition of states is just a linear combination of the state vectors. A general state $\ket{\psi}$ can thus be written as
\begin{equation}
    \ket{\psi} = \alpha\ket{0} + \beta\ket{1}\,,
\end{equation}
where $\alpha, \beta \in \C$. The state is \emph{physical} if it is normalized, meaning its $\ell_2$-norm is equal to one. In that case, the coefficients $\alpha$ and $\beta$ are \emph{probability amplitudes}, and their squared magnitudes give the probabilities of obtaining the corresponding outcomes when measuring in the computational basis. Thus, the probability of obtaining outcome $0$ is $\abs{\alpha}^2$ and the probability of obtaining outcome $1$ is $\abs{\beta}^2$. Since these two outcomes are exhaustive and probabilities must sum to $1$, any physical quantum state satisfies
\begin{equation}
    \abs{\alpha}^2 + \abs{\beta}^2 = 1\,.
\end{equation}
The \emph{dual} of the state $\ket{\psi}$ is given by
\begin{equation}
    \bra{\psi} = \alpha^\ast\bra{0} + \beta^\ast\bra{1}\,,
\end{equation}
where, recall, $\alpha$ and $\beta$ are complex-valued, so we must take their complex conjugate, which we denote with the asterisk. Dirac notation is remarkably convenient for the \emph{inner product} between two vectors $\ket{\psi}$ and $\ket{\phi}$, written as the complex-valued scalar $\braket{\phi}{\psi} = \braket{\psi}{\phi}^\ast$. The norm of a vector is obtained from its inner product with itself:
\begin{equation}
    \norm{\ket{\psi}}_2 = \sqrt{\braket{\psi}{\psi}}\,.
\end{equation}
If two vectors $\ket{\psi}$ and $\ket{\phi}$ are normalized, that is,
\begin{equation}
    \braket{\psi}{\psi} = \braket{\phi}{\phi} = 1\,,
\end{equation}
then the magnitude of their inner product $\abs{\braket{\phi}{\psi}}$ quantifies the degree to which these two vectors \emph{overlap}. The normalization enforces that the squared magnitude of the probability amplitudes adds up to 1. If the inner product between two vectors is $0$, then we say that the vectors are \emph{orthogonal}. The computational basis vectors are both orthogonal and normalized, so they form an \emph{orthonormal set}:
\begin{align}
    \braket{0}{1} &= \braket{1}{0} = 0\,,\\
    \braket{0}{0} &= \braket{1}{1} = 1\,.
\end{align}
If we have more than one qubit, then we naturally have a larger Hilbert space of dimension greater than $2$. In particular, if we have $n$ qubits, each with its own two-dimensional Hilbert space $\mathcal{H}_i$, then their joint Hilbert space is the \emph{tensor product} of the $\mathcal{H}_i$:
\begin{equation}
    \mathcal{H} = \mathcal{H}_1 \otimes \cdots \otimes \mathcal{H}_n
\end{equation}
of dimension $2^n$. As $\mathcal{H}$ is a vector space, any vector $\ket{\psi} \in \mathcal{H}$ can be written as a linear combination of basis elements of $\mathcal{H}$. The \emph{joint computational basis} for $\mathcal{H}$ is given by the tensor products of the single-qubit computational basis vectors of the form $\ket{i_1}\otimes\cdots\otimes\ket{i_n}$, where $i_j \in \{0,1\}$ for all $j$. In particular, any $n$-qubit state can be expanded in the joint computational basis as
\begin{equation}
    \ket{\psi}=\sum_{i_1,\ldots,i_n \in \{0,1\}}c_{i_1\cdots i_n}\ket{i_1}\otimes\cdots\otimes\ket{i_n}\,.
\end{equation}

\begin{example}[label={ex:two-qubit-joint-state}]{Two-qubit joint state}
    A general two-qubit state can be written as a linear combination of the four joint computational-basis states $\ket{00},\ket{01},\ket{10},\ket{11}$. As a special case, suppose the state is a \emph{product state} $\ket{\psi}=\ket{\psi_1}\otimes\ket{\psi_2}$, where $\ket{\psi_1} = \alpha'\ket{0} + \beta'\ket{1}$ and $\ket{\psi_2} = \gamma'\ket{0} + \delta'\ket{1}$. Then,
    \begin{align}
        \ket{\psi} &= \lp \alpha'\ket{0} + \beta'\ket{1} \rp \otimes \lp \gamma'\ket{0} + \delta'\ket{1} \rp\\
        &= \alpha'\gamma'\ket{00} + \alpha'\delta'\ket{01} + \beta'\gamma'\ket{10} + \beta'\delta'\ket{11}\\
        &= \alpha\ket{00} + \beta\ket{01} + \gamma\ket{10} + \delta\ket{11}\,,
    \end{align}
     where we have introduced the shorthand notation $\ket{a} \otimes \ket{b} \equiv \ket{ab}$, which we make use of throughout this tutorial.
\end{example}

\begin{exerbox}[label={exer:2-qubit-probability}]{2-qubit measurement}
    Given a 2-qubit state of the form
    \begin{equation}
        \ket{\psi} = \alpha\ket{00} + \beta\ket{01} + \gamma\ket{10} + \delta\ket{11}\,,
    \end{equation}
    what is the probability of measuring $\ket{\psi}$ in \emph{either} the state $\ket{00}$ or $\ket{10}$?
\end{exerbox}

For the state vector description used in this section, if a multipartite state can be expressed as $|\psi\rangle=|\psi_1\rangle\otimes\cdots\otimes|\psi_n\rangle$, then it is called a \emph{product state}. Pure states that cannot be written in this way are called \emph{entangled states}.\footnote{Pure states are states that are fully described by a single state vector. The more general quantum states are known as \emph{mixed states} and will be introduced in the next chapter.} Among the most familiar examples of entangled states are the Bell states, introduced in~\cref{ex:two-qubit-entangled-state} below. Entanglement is one of the key features that distinguishes quantum mechanics from classical mechanics and, as will become clear throughout this tutorial, it is a central ingredient in quantum error correction.

\begin{example}[label={ex:two-qubit-entangled-state}]{Two-qubit entangled state}
    Taking the two-qubit state from~\cref{ex:two-qubit-joint-state}, if we set $\beta = \gamma = 0$, then we have the state
    \begin{align}
        \ket{\psi} = \alpha\ket{00} + \delta\ket{11}\,,
    \end{align}
    which cannot be written as a tensor product when both $\alpha$ and $\delta$ are nonzero. This is an example of an entangled state. A particularly important special case is the Bell state $\ket{\Phi^+}=\frac{1}{\sqrt{2}}(\ket{00}+\ket{11})$.
\end{example}

So far, we have discussed static quantum states. To describe dynamics, and in particular quantum computations, we need transformations of those states. These transformations are described by \emph{operators}, which we denote by capital letters such as $A$.\footnote{Many references also put a ``hat'' on top of operators, as in $\hat{A}$, but because we will be writing many operators, we omit the hats for readability.} Mathematically, operators are matrices acting on states by matrix multiplication. For an $n$-qubit state, the relevant operators are $2^n\times 2^n$ complex-valued matrices acting on the $2^n$-dimensional Hilbert space. Operators act on kets from the left and on bras from the right: $A\ket{\psi}$ and $\bra{\psi}A$, respectively. We summarize several useful properties of operators in \cref{box:operator-properties}.

\begin{explbox}[label={box:operator-properties}]{Properties of operators}
    Let $A$, $B$, and $C$ be operators. Then, they satisfy the following properties:
    \begin{itemize}
        \item \textbf{Associativity}: $A+(B+C) = (A+B)+C$ and $A(BC)=(AB)C$
        \item \textbf{Additive commutativity}: $A+B=B+A$
        \item \textbf{Non-commutativity (in general)}: $AB \neq BA$
    \end{itemize}
    The last property above has to do with the \emph{commutator} of two operators, which is given as
    \begin{equation}
        [A,B] := AB - BA\,.
    \end{equation}
    If $[A,B] = 0$, that is, $AB = BA$, then we say that $A$ and $B$ \emph{commute}. If $AB = -BA$, then we say that $A$ and $B$ \emph{anticommute}. The Hermitian conjugate applies to operators in the following ways:
    \begin{itemize}
        \item \textbf{Antidistributivity}: $(AB)^\dagger = B^\dagger A^\dagger$
        \item \textbf{Antilinearity}: $(cA)^\dagger = c^\ast A^\dagger$
        \item \textbf{Involution}: $(A^\dagger)^\dagger = A$
    \end{itemize}
    In particular, if $A = A^\dagger$, then we say that $A$ is \emph{Hermitian}. Hermitian observables have real eigenvalues and, as such, all physical observables in quantum mechanics are Hermitian.
\end{explbox}

In quantum computation, we often care about \emph{unitary} operators, where an operator is unitary if $A^{-1} = A^\dagger$. Equivalently, $AA^\dagger = A^\dagger A = I$, where $I$ is the identity operator. As we mentioned above, operators are matrices and so can be written in a basis using the \emph{outer product}. For example, we can decompose the operator $A$ as
\begin{equation}
    A = \defsum{m,n}{}{\ket{m}\mel{m}{A}{n}\bra{n}} = \defsum{m,n}{}{A_{m,n}\ket{m}\bra{n}}\,,
\end{equation}
where the matrix elements are $A_{mn}=\mel{m}{A}{n}$ and the outer products $\ket{m}\bra{n}$ form a basis for the space of operators. Just as we did for state vectors, so too can we tensor together operators to act on individual Hilbert spaces. For example, we can tensor together several single-qubit gates that act individually on a set of qubits. Operators that cannot be decomposed into a tensor product of single-subsystem operators can generate entanglement. One very important example is the CNOT operator, which can be used to entangle two qubits.

Given an operator $A$, an important concept is that of its \emph{eigenstates} and \emph{eigenvalues} (e.g., for the stabilizer codes that we will meet in \cref{sec:stabilizer}). An eigenstate $\ket{a_i}$ of the operator $A$ is a state such that $A\ket{a_i} = a_i\ket{a_i}$, where $a_i \in \C$ is called the eigenvalue. If $A$ is Hermitian (as mentioned in \cref{box:operator-properties}), then $A$ has real eigenvalues: $a_i \in \R\ \forall i$. Furthermore, the set of eigenstates of any Hermitian operator constitute its \emph{eigenbasis}, which is also a basis for the space of all states. When written in this basis, the operator becomes diagonal with the matrix elements being its eigenvalues, that is, $A=\sum_i a_i\ket{a_i}\bra{a_i}$. The set of all eigenstates sharing the same eigenvalue forms a subspace called an \emph{eigenspace}. If multiple orthogonal eigenstates share the same eigenvalue, then the eigenvalue is termed \emph{degenerate}, meaning the choice of an orthonormal basis for this eigenspace is no longer unique. One useful thing to do with the eigenbasis is to take its expectation value with respect to a specific state $\ket{\psi}$. If we expand $\ket{\psi}$ in the eigenbasis of $A$ as $\ket{\psi} = \defsum{i}{}{c_i\ket{a_i}}$, then the expectation value of $A$ with respect to $\ket{\psi}$ is
\begin{equation}
    \mel{\psi}{A}{\psi} = \defsum{i}{}{a_i\abs{c_i}^2}\,,
\end{equation}
so the expectation value is the average of the eigenvalues $a_i$ with respect to the probability distribution $\{|c_i|^2\}_i$ defined by the state in the eigenbasis of $A$, where $\defsum{i}{}{\abs{c_i}^2} = 1$ is the normalization condition.

\subsection{Postulates of quantum mechanics}\label{subsec:quantum-intro-postulates}
We now formalize the above notions and condense several themes into a series of postulates, giving us the \emph{postulates of quantum mechanics}. This section follows the presentation in \emph{Quantum Computation and Quantum Information} by Nielsen and Chuang~\cite{nielsen2010quantum}. The first postulate has to do with the state space that we work with in quantum mechanics:

\begin{postulate}[label={post:qm-state-space}]{State space}
    An isolated quantum system can be fully described by a \emph{state vector}, which is a unit vector in a complex-valued vector space called a \emph{Hilbert space}. A Hilbert space is a complete vector space equipped with an inner product.
\end{postulate}

As we saw in the previous section, the state space is where the kets and bras live. Mathematically, we wrote $\ket{\psi} \in \mathcal{H}$, where $\ket{\psi}$ is the state vector and $\mathcal{H}$ is the Hilbert space. The specific state space for a given quantum system is determined by the physics of the system. For example, a qubit has a natural state space of a two-dimensional Hilbert space. Other physical systems, such as photons or atoms, come with different spaces. Of course, quantum states do not just exist statically without changing; to have interesting dynamics like a quantum computation, the state needs to \emph{evolve}. Evolution of quantum states is the subject of the next postulate: 

\begin{postulate}[label={post:qm-evolution}]{Evolution}
    Given a closed quantum system described by the state vector $\ket{\psi}$, its evolution from time $t_0$ to time $t_1$ is described by a unitary operator $U$, which depends only on $t_0$ and $t_1$:
    \begin{equation}
        \ket{\psi(t_1)} = U(t_0; t_1)\ket{\psi(t_0)}\,.
    \end{equation}
\end{postulate}

For a single qubit, this evolution is described by a unitary operator, which in the language of quantum computing is called a \emph{quantum gate}. We discuss quantum gates and quantum circuits in more detail in the next section. Mathematically, the evolution of the state $\ket{\psi}$ of a \emph{closed} quantum system, meaning one idealized as isolated from its environment, is described by the \emph{Schr\"odinger equation}:
\begin{equation}
    i\hbar\dv{\ket{\psi}}{t} = H\ket{\psi}\,,
\end{equation}
where $H$ is called the system's \emph{Hamiltonian} and $\hbar$ is the reduced Planck's constant. Thus, given an initial state and the Hamiltonian, we can find the state at any given later time. This description applies to \emph{closed} systems. By contrast, an \emph{open} quantum system can interact with external degrees of freedom, so the system of interest cannot in general be described by unitary evolution alone. A widely used model for many such situations is \emph{Lindbladian} dynamics. In this tutorial, we will not need to work directly with these equations very often and will mostly use only unitary gates for evolution of quantum states. At the end of the evolution of our quantum system, we have some final state that encodes information we would like to extract. The only way to access this information is by \emph{measuring} the state, which leads to the next postulate:

\begin{postulate}[label={post-qm-measurement}]{Measurement}
    A quantum state $\ket{\psi}$ can be measured, where the set of measurements are described by a set $\{M_m\}$ of measurement operators, one for each possible measurement outcome $m$. The probability that the specific measurement outcome $m$ occurs is given by $p(m) = \mel{\psi}{M_m^\dagger M_m}{\psi}$ and the state of the system right after the measurement is given by
    \begin{equation}
        \ket{\psi'} = \frac{M_m\ket{\psi}}{\sqrt{\mel{\psi}{M_m^\dagger M_m}{\psi}}}\,.
    \end{equation}
    The set of measurements $\{M_m\}$ is \emph{complete}:
    \begin{equation}
        \defsum{m}{}{M_m^\dagger M_m} = I\,,
    \end{equation}
    which means that one of the possible measurement outcomes described by $\{M_m\}$ must occur.
\end{postulate}

This post-measurement update is often described informally as the \emph{collapse of the wavefunction}: after a measurement outcome is obtained, the state is updated to the corresponding post-measurement state. There are several types of quantum measurements. Three notable types are projective measurements, positive operator-valued measures (POVMs), and more general measurement channels. It often suffices to consider projective measurements, and we will introduce more general formalisms only when needed. A projective measurement is described by \emph{projection operators} $P$, which satisfy $P^2 = P$. These projectors are the outer products of the observable's eigenstates mapping a superposition onto one of the allowed outcomes. We demonstrate these ideas through the following example:

\begin{example}[label={ex:single-qubit-measurement}]{Single-qubit measurement}
    Take a qubit in a superposition state
    \begin{equation}
        \ket{\psi} = \alpha\ket{0} + \beta\ket{1}\,,
    \end{equation}
    where $\abs{\alpha}^2 + \abs{\beta}^2 = 1$. The probability of measuring the system in the state $\ket{0}$ is
    \begin{equation}
        p(0) = \abs{\braket{0}{\psi}}^2 = \abs{\alpha}^2
    \end{equation}
    and the probability of measuring the system in the state $\ket{1}$ is
    \begin{equation}
        p(1) = \abs{\braket{1}{\psi}}^2 = \abs{\beta}^2\,.
    \end{equation}
    If, for example, we measured the state in $\ket{1}$, then the post-measurement state is 
    \begin{equation}
        \ket{\psi'} = \ket{1}\,.
    \end{equation}
\end{example}

The final postulate tells us how to combine multiple quantum systems into a \emph{composite system}:

\begin{postulate}[label={post:qm-composite-systems}]{Composite systems}
    Independent quantum systems can be combined into a \emph{composite system}, whose state space is described by the tensor product of the state spaces of the component systems.
\end{postulate}

We saw composite systems in the previous section by way of taking the \emph{tensor product} of individual quantum states. That is, given $n$ \emph{independent} quantum systems $\ket{\psi_i}$, the composite system is described by the following state:
\begin{equation}
    \ket{\psi} = \ket{\psi_1} \otimes \ldots \otimes \ket{\psi_n}\,.
\end{equation}
We often shorten the notation by omitting the tensor-product symbols, for example $\ket{0}\otimes\ket{1}\equiv\ket{01}$. As we saw in the previous section, some states cannot be written as tensor products of subsystem states. Such states are called \emph{entangled}, and entanglement is one of the key features that gives quantum computation its power.

\subsection{Quantum circuit model}\label{subsec:quantum-intro-circuit-model}
Fundamentally, a quantum computation is a process that evolves an initial state to a final state from which we hope to extract useful information. We can describe this process using a tool analogous to the Boolean logic circuits of classical computing: the \emph{quantum circuit model}. There are three basic steps in a quantum computation: initialization of the starting state, evolution of the state (i.e., the \emph{quantum algorithm}), and measurement of the final state.

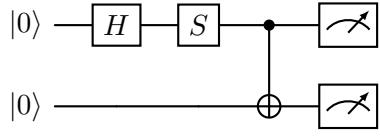
\begin{figure}
    \centering
    \begin{quantikz}
        \lstick{$\ket{0}$} & \gate{H} & \gate{S} & \ctrl{1} & \meter{}\\
        \lstick{$\ket{0}$} & & & \targ{} & \meter{}
    \end{quantikz}
    \caption{Example of a quantum circuit. The figure illustrates the basic structure of a circuit diagram: wires denote qubits, gates represent unitary operations, and meter symbols indicate measurements.}

    \label{fig:quantum-circuit}
\end{figure}

We represent a quantum circuit using a circuit diagram like the one seen in~\cref{fig:quantum-circuit}. We read the circuit from left to right, where each qubit is assigned to one wire. The quantum algorithm is a sequence of quantum \emph{gates}, which transform the initial state of the qubits to a final state. We then measure the qubits, represented by the meter symbols in the figure. Gates can act on one, two, or more qubits, though single-qubit and two-qubit gates are the most common and multi-qubit gates can be decomposed into a sequence of simpler gates.

Quantum gates are unitary operators, which we saw above in~\cref{subsec:quantum-intro-states-operators-measurements}, and any unitary operator qualifies as a valid quantum gate. However, there are several specific gates that we encounter repeatedly in quantum computing. A very important set of gates, especially for quantum error correction, are the \emph{Pauli gates}. We will formally meet the Pauli group and its constituent matrices in~\cref{subsec:stabilizer-group-theory-primer}, but we introduce the single-qubit Pauli gates now, of which there are four: the identity matrix $I$ and the Pauli-$X$, $Y$, and $Z$ matrices. In matrix form, these are
\begin{align}
    I &=
    \begin{pmatrix}
        1 & 0\\
        0 & 1
    \end{pmatrix}\,,
    \quad
    X =
    \begin{pmatrix}
        0 & 1\\
        1 & 0
    \end{pmatrix}\,,\\
    Y &= 
    \begin{pmatrix}
        0 & -i\\
        i & 0
    \end{pmatrix}\,,
    \quad
    Z = 
    \begin{pmatrix}
        1 & 0\\
        0 & -1
    \end{pmatrix}\,.
\end{align}
The identity matrix does nothing to a qubit, leaving it in whatever state it started in. The Pauli-$X$ gate acts as a bit flip operator in classical computation, taking the state $\ket{0}$ to $\ket{1}$ and \emph{vice versa}. The Pauli-$Z$ gate is a uniquely quantum mechanical operator and is also known as a \emph{phase flip}, where it adds a phase of $-1$ to the state $\ket{1}$, bringing it to $-\ket{1}$, while it does nothing to the state $\ket{0}$. The Pauli-$Y$ gate in a sense acts as both a bit flip and phase flip, where it brings the state $\ket{0}$ to $-i\ket{1}$ and the state $\ket{1}$ to $i\ket{0}$; in fact, one can calculate that $Y = iXZ = -iZX$. Observe that $I$ commutes with each of $X, Y, Z$, but any two of $X, Y, Z$ anticommute.

Another important quantum gate is the \emph{Hadamard gate}, which is often used to create superposition states. In the computational basis, the matrix representation is
\begin{equation}
    H = \frac{1}{\sqrt{2}}
    \begin{pmatrix}
        1 & 1\\
        1 & -1
    \end{pmatrix}\label{eqn:hadamard-matrix}
\end{equation}
For example, applying the Hadamard gate to the state $\ket{0}$ brings it to the superposition state $\frac{1}{\sqrt{2}}(\ket{0} + \ket{1})$. This is the ``plus state,'' denoted $\ket{+}$, so the Hadamard gate can be viewed as a basis transformation between the computational ($Z$) basis and the $X$ basis.

\begin{exerbox}[label={exer:hadamard-multiplication}]{}
    Show by direct matrix-vector calculation that the Hadamard matrix applied to the computational basis state
    \begin{equation}
        \ket{0} = 
        \begin{pmatrix}
            1\\
            0
        \end{pmatrix}
    \end{equation}
    produces the state $\ket{+} = \frac{1}{\sqrt{2}}(\ket{0} + \ket{1})$.
\end{exerbox}

\begin{figure}[tb]
    \centering
    \begin{quantikz}
        \lstick{$\ket{0}$} & \gate{H} & \ctrl{1} & \rstick[2]{$\ket{\Phi^+}$}\\
        \lstick{$\ket{0}$} & & \targ{} &
    \end{quantikz}
    \caption{Quantum circuit that prepares the two-qubit Bell state $\ket{\Phi^+} = \frac{1}{\sqrt{2}}(\ket{00} + \ket{11})$.}
    \label{fig:bell-state-circuit}
\end{figure}
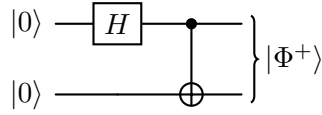

Other important gates include the phase or $S$ gate and the $\pi/8$ or $T$ gate, which are given by
\begin{equation}
    S = 
    \begin{pmatrix}
        1 & 0\\
        0 & i
    \end{pmatrix}\,,
    \quad
    T = 
    \begin{pmatrix}
        1 & 0\\
        0 & e^{i\pi/4}
    \end{pmatrix}\,.
\end{equation}
There are also important two-qubit gates, in particular the \emph{controlled} gates. In a controlled gate, there are two qubits, one called the \emph{control} and the other called the \emph{target}, where the gate is applied to the target qubit only if the control qubit is in a particular state. Common examples include the CNOT and CZ gates, which apply the $X$ gate or $Z$ gate, respectively, only if the control qubit is in a particular state, usually $\ket{1}$. In matrix form, these gates are
\begin{equation}
    \mathrm{CNOT} = 
    \begin{pmatrix}
        1 & 0 & 0 & 0\\
        0 & 1 & 0 & 0\\
        0 & 0 & 0 & 1\\
        0 & 0 & 1 & 0
    \end{pmatrix}\,,
    \quad
    \mathrm{CZ} = 
    \begin{pmatrix}
        1 & 0 & 0 & 0\\
        0 & 1 & 0 & 0\\
        0 & 0 & 1 & 0\\
        0 & 0 & 0 & -1
    \end{pmatrix}\,.
\end{equation}
Such gates can be used to create \emph{entangled} states and are thus very important for quantum computation. For example, in~\cref{fig:bell-state-circuit}, we use a CNOT gate to create a Bell state, a useful two-qubit entangled state.

\begin{exerbox}[label={exer:ghz-circuit}]{A simple entangling circuit}
    Determine the state prepared by the circuit in \cref{fig:quantum-circuit} immediately before measurement. What probabilities would you expect if the two qubits are then measured in the computational basis?
\end{exerbox}

The concepts in this chapter will appear repeatedly throughout the rest of the tutorial. Quantum codes are subspaces of larger Hilbert spaces, encoding and recovery are implemented by circuits and operators, and error diagnosis relies on carefully chosen measurements. In particular, the idea of extracting information about an error without learning the encoded quantum state is one of the central themes of quantum error correction.

\newpage

\section{Fundamentals of classical and quantum error correction}\label{sec:fundamentals}

We now shift our attention to error correction. As we saw in the previous chapter, quantum information is powerful because it can be stored in superpositions, take advantage of entanglement, and transformed by coherent quantum dynamics, but this is also precisely what makes it fragile. While we would like our quantum states to remain isolated from the external environment, this is impossible in practice. Realistic quantum systems experience undesired interactions with the environment, imperfect gates and measurements, and errors in state preparation such that long quantum computations become challenging to perform reliably.

As such, we need a way to protect quantum states and information. This is exactly where error correction comes in. The basic idea of QEC is to encode quantum information into a larger quantum system such that physical errors can be detected and appropriately corrected. By addressing these physical errors, we can thus avoid incurring \emph{logical} errors, which would corrupt our computation and final results. Before getting to the rich field of QEC, though, we turn to classical error correction, from which many ideas like redundancy, codespaces, distances, and logical operations, can be extended to the quantum setting. In particular, we will see how some of these ideas can be directly extended in the quantum regime, but also how many cannot, where unique quantum mechanical effects like the impossibility of copying quantum information, the disturbing of quantum states via measurement, and other errors like phase-flip errors greatly enrich the QEC program.

\subsection{Classical error correction}\label{subsec:classical-error-correction}
In classical computation~\cite{sipser2012introduction}, we encode information using \emph{bits}. A bit is an abstract concept that takes one of two values (or \emph{states}), where two is the minimal number of possible states needed for information processing. Bits can be represented by the two sides of a coin, the on/off settings of a light switch, the low/high voltages of a transistor, or the elements of the binary field $\mathbb{F}_2$ (i.e., the numbers 0 and 1), to name a few. If we (the \emph{sender}) wish to transmit a message to someone (the \emph{receiver}) over a communication channel, we can write that message in \emph{binary}, that is, as a series of 0's and 1's according to some mapping in which letters, numbers, punctuation, and other symbols each have a unique binary representation. Once the receiver gets the binary message, they can use the mapping to recover the original symbols and read the message. However, no communication channel is perfect, and there is some (usually small) probability that one or more transmitted bits will flip. While this may be acceptable for a few bits representing letters, if we are trying to send numerical information, even one bit-flip can dramatically change the value we are trying to send. We will assume that we have already reduced the bit-flip rate in the communication channel as much as possible and now seek mechanisms to \emph{detect} and \emph{correct} the remaining errors. Those mechanisms are, of course, what error correction provides.

The field of error correction~\cite{macwilliams1977theory,huffman2003fundamentals,lin2021fundamentals} was initiated by Richard Hamming, who introduced~\cite{hamming1950error} the 7-bit \eczoo[Hamming code]{hamming743} in 1947.\footnote{Throughout this tutorial, when referring to specific codes, we link to the corresponding entry in the Error Correction Zoo with an ``EC Zoo'' symbol \eczooIcon[height=1.2em]} Shortly after, Claude Shannon incorporated error-correcting codes into his mathematical theory of communication in 1948~\cite{shannon1948mathematical}. The Hamming code, the \eczoo[repetition code]{repetition}, and many other classical error-correcting codes belong to the class of \eczoo[\emph{linear codes}]{binary_linear}. Other important examples include \eczoo[Reed-Solomon codes]{reed_solomon}, \eczoo[low-density parity-check]{ldpc} (LDPC) codes, and \eczoo[turbo codes]{turbo}. Linear codes have many useful properties, and we will see their quantum analogue in the form of \emph{stabilizer codes} below. \eczoo[Stabilizer codes]{stabilizer} are one of the most important classes of quantum error-correcting codes.

Today, classical error-correcting codes are used across many important technologies, several of which we encounter in our everyday lives. These include cellular networks, fiber-optic communication, QR codes, Wi-Fi, flash memory, hard drives, random-access memory (RAM), and even deep space telecommunication systems like those on the Voyager 2 spacecraft. Without error-correcting codes, many of these technologies would be very difficult or even impossible to realize at a practical scale.

\subsubsection{Classical repetition code}\label{subsubsec:classical-repetition-code}
The central idea behind the Hamming code and many others is \emph{redundancy}, which is especially easy to see in \emph{repetition codes}. This is perhaps the most basic family of classical error-correcting codes. At this stage, we only need to consider a single kind of error: the flipping of a bit from ``0'' to ``1'' or \emph{vice versa}. This is called a \emph{bit-flip error}. The repetition code works exactly as one would expect, where we just repeat the message that we want to send over a noisy channel. We do this already, for example, when we are talking to someone in a noisy room, where if the person does not hear us, we simply repeat whatever we are trying to tell them. With enough repetitions (hopefully not more than 2!), the person should have heard enough of the noisy messages to piece together the true message we are trying to send them.

To formalize these ideas a bit, consider a message $m$ consisting of several bits. In the three-bit repetition code, each logical bit is encoded using three physical bits: 0 is encoded as $\overline{0} = 000$ and 1 as $\overline{1} = 111$. Here, the overline denotes the \emph{logical} bit, that is, the underlying ``true'' information we wish to protect.

\begin{example}[label={ex:3-bit-repetition-code}]{Three-bit repetition code}
    Imagine we wish to send the number 24 as the message. This message in binary is $m = 11000$. To encode it in a three-bit repetition code, we encode each bit in $m$ using three bits, so that the actual message that we send over the noisy channel is
    \begin{equation}
        m \to \underbrace{111}_1\ \underbrace{111}_1\ \underbrace{000}_0\ \underbrace{000}_0\ \underbrace{000}_0\,.
    \end{equation}
\end{example}

We call the three-bit blocks encoding each logical bit a \emph{codeword}. Thus, the three-bit repetition code has two codewords: 000 and 111. A bit-flip error results in the flip of one or more of the bits. For example, a single bit-flip error can occur in either the first, second, or third bit, resulting in:
\begin{align}
    \overline{0} &= 000 \to
    \begin{cases}
        100 &\\
        010 &\\
        001
    \end{cases}\\
    \overline{1} &= 111 \to
    \begin{cases}
        011 &\\
        101 &\\
        110
    \end{cases}\,.
\end{align}
The first step in error correction is \emph{error detection}, that is, determining whether an error has occurred. The next step is \emph{decoding}: inferring the most likely error from the received word. For the three-bit repetition code, if any received codeword is not all 0's or all 1's, then we detect that an error has occurred. Decoding can be done by taking a majority vote, which tells us which logical bit was most likely sent. For example, if we receive the codeword 011, then because there are more 1's than 0's, we infer that the first bit-flipped. We then correct the error by flipping that bit back, recovering 111, which encodes the logical bit 1.

When discussing error-correcting codes, an important concept is the \emph{code distance}, which is the smallest number of bit-flips required to transform one codeword into another. Sticking with the three-bit repetition code, this means the smallest number of bit-flips required to transform 000 into 111, or \emph{vice versa}, which is three. Thus, we say that the three-bit repetition code has distance $d=3$. In general, an $n$-bit repetition code has distance $n$. We can formalize this idea using the \emph{Hamming distance}, which will be useful throughout this tutorial. The Hamming distance $d_H(a,b)$ between two $n$-bit strings $a,b \in \{0,1\}^n$ is the minimum number of bit-flips required to transform $a$ into $b$. The code distance of a classical code is the minimum Hamming distance between any two distinct codewords. Thus, the code distance for the three-bit repetition code is the Hamming distance between its two codewords: $d = d_H(000,111) = 3$. 

The code distance determines two important quantities. The first one is the maximum number of errors that can be detected, which is $d-1$ for a code with distance $d$. The second is the maximum number of errors that can be corrected, which is $\lfloor\frac{d-1}{2}\rfloor$, where $\lfloor\cdot\rfloor$ is the floor operator. Naturally, we want a high code distance! We often summarize a classical code that encodes $k$ logical bits into $n$ physical bits and has distance $d$ by its \emph{code parameters} $[n,k,d]$. For example, the three-bit repetition code is a $[3,1,3]$ code.

We need to actually do something with our encoded data in order to perform a computation, and we can do so using a set of \emph{logical operators}, which transform the logical bits. For example, the logical \textsc{not} operator flips a logical bit by flipping each of the physical bits:
\begin{align}
    \textsc{not}(\overline{0}) &= \textsc{not}(000) = 111 \equiv \overline{1}\\
    \textsc{not}(\overline{1}) &= \textsc{not}(111) = 000 \equiv \overline{0}\,.
\end{align}
Logical operations (or logical \emph{gates}) are an important part of developing a quantum error-correcting code.

\subsubsection{Classical linear codes}\label{subsubsec:classical-linear-codes}
The three-bit repetition code is the simplest example of a broader and very important class of classical codes: \emph{linear codes}. We focus on \emph{binary} linear codes. A binary linear code is a subspace $\mathcal{C} \subseteq \FF_2^n$, where $\FF_2^n$ is the $n$-dimensional vector space over the field $\FF_2$ (equivalently, $\{0,1\}^n$ with bitwise addition modulo 2). Thus, if $u, v \in \mathcal{C}$, then $u+v \in \mathcal{C}$ as well. The elements of $\mathcal{C}$ are called \emph{codewords}. In particular, taking $u=v$ shows that $u+u=0^n$ must belong to every binary linear code.

There are three common ways to describe a linear code. The most direct is simply to list all of its codewords, although this quickly becomes impractical for larger codes. The second and more compact description is to specify a set of generators. We say that $u_1, \ldots, u_m \in \FF_2^n$ \emph{generate} the code $\mathcal{C}$ if
\begin{equation}
    \mathcal{C} = \{\alpha_1 u_1 + \cdots + \alpha_m u_m \mid \alpha_1, \ldots, \alpha_m \in \FF_2\}\,.
\end{equation}
If the generators are linearly independent, then they form a minimal generating set, and it is often convenient to arrange them as the rows of a matrix $G$, called the \emph{generator matrix}. The generator description is also useful for understanding the code distance. Since a linear code contains the all-zero codeword, the distance of a binary linear code is equal to the minimum \emph{Hamming weight} of any nonzero codeword, where the Hamming weight of a bit string is the number of 1's it contains. Indeed, if $a$ and $b$ are codewords, then $a+b$ is also a codeword, and its Hamming weight is exactly the Hamming distance $d_H(a,b)$ that we introduced above. Thus, to compute the distance, it suffices to examine the nonzero words generated by $G$. For the three-bit repetition code, the generator matrix is simply $G=(111)$, so the only nonzero codeword is $111$, which has Hamming weight 3, and hence the code distance is $d=3$.

The third kind of description is in terms of \emph{parity checks}. A set of parity checks is a collection of vectors $v_1, \ldots, v_r \in \FF_2^n$ such that
\begin{equation}
    \mathcal{C} = \{u \in \FF_2^n \mid u \cdot v_1 = \cdots = u \cdot v_r = 0 \}\,,
\end{equation}
where the dot product is taken over $\FF_2$. In other words, a string belongs to the code exactly when it satisfies all parity checks. Arranging these parity checks as the rows of a matrix $H$ gives the \emph{parity-check matrix}. The linear span of these rows form the \emph{dual code}, denoted $\mathcal{C}^\perp$, which consists of \emph{all} vectors orthogonal to $\mathcal{C}$. Thus, $\mathcal{C}^\perp$ is exactly the dual space of the codespace, and $GH^\intercal = 0$. If a code over $\FF_2^n$ has $m$ independent generators and $r$ independent parity checks, then it follows from the rank-nullity theorem~\cite{strang2022introduction} that $n=m+r$. Thus, the code encodes $m$ bits and has $2^m=2^{n-r}$ codewords.

The parity-check matrix also gives a compact way to detect and diagnose errors. Given a received word $w\in \FF_2^n$, we define its \emph{syndrome} by
\begin{equation}
    s = Hw^\intercal \in \FF_2^r\,.
\end{equation}
If $w$ is a valid codeword, then $s=0$ because it satisfies all parity checks. A nonzero syndrome indicates that an error has occurred, and different syndromes can be used to infer different likely error patterns. For the three-bit repetition code, a convenient parity-check matrix is
\begin{equation}
    H =
    \begin{pmatrix}
        1 & 1 & 0\\
        1 & 0 & 1
    \end{pmatrix}\,.
\end{equation}
Its two rows simply check whether the first bit agrees with the second and third bits, respectively. Accordingly, both codewords satisfy $H(000)^\intercal=0$ and $H(111)^\intercal=0$. If instead we receive, for example, the corrupted word $011$, then
\begin{equation}
    H(011)^\intercal =
    \begin{pmatrix}
        1\\
        1
    \end{pmatrix}\,,
\end{equation}
which tells us that the first bit disagrees with both of the others. The most likely error is therefore a flip of the first bit, and we decode by flipping it back to recover $111$. In this way, syndrome decoding gives a compact algebraic version of the majority-vote rule. We will return to generator and parity-check descriptions in~\cref{subsec:stabilizer-css-codes}, where they provide the classical input data for quantum CSS codes.

\subsection{Quantum repetition code}\label{subsec:fundamentals-quantum-repetition-code}
We can naturally extend the classical repetition code to qubits and get the \eczoo[\emph{quantum repetition code}]{quantum_repetition}. However, there is an immediate obstacle in the quantum setting. Classically, repetition works by copying the information several times, but for an unknown quantum state, this is impossible in general because of the \emph{no-cloning theorem}: there is no physical operation that takes an arbitrary state $\ket{\psi}$ to $\ket{\psi}\ket{\psi}$. Instead, a quantum repetition code stores redundancy in correlations among several qubits rather than in independent copies of the original state. With that distinction in mind, we now introduce the three-qubit repetition code, where the extension to more than three qubits is straightforward. We start by defining the codewords, which are the same codewords as in the classical repetition code but written as kets:
\begin{equation}
    \ket{\overline{0}} = \ket{000} \text{ and } \ket{\overline{1}} = \ket{111}\,.
\end{equation}
Here, $\ket{\overline{0}}$ and $\ket{\overline{1}}$ are the \emph{logical computational basis states} of the encoded \emph{logical qubit}, and the error-free encoded state lives in the \emph{codespace} that they span. That is, we can write a general single-qubit quantum state $\ket{\psi} = \alpha\ket{0} + \beta\ket{1}$ encoded into the three-qubit repetition code as
\begin{align}
    \ket{\overline{\psi}} &= \alpha\ket{\overline{0}} + \beta\ket{\overline{1}}\\
    &= \alpha\ket{000} + \beta\ket{111}\,.
\end{align}
In the classical repetition code, the only error we needed to consider was a bit-flip. For quantum states, however, we must also contend with \emph{phase-flip errors}, which have no classical analogue. We represent bit-flip and phase-flip errors using the Pauli operators $X$ and $Z$, respectively, introduced in \cref{subsec:quantum-intro-circuit-model}. Their actions on the computational basis are
\begin{align}
    X\ket{0} &= \ket{1},\quad  X\ket{1} = \ket{0}\,,\\
    Z\ket{0} &= \ket{0},\quad Z\ket{1} =-\ket{1}\,.
\end{align}
Correcting these two errors turns out to be sufficient for correcting more general errors. We will formalize this statement later in this tutorial.

Just as in the classical repetition code, the first step in QEC is error \emph{detection}. We again consider bit-flip errors in the three-qubit repetition code. However, for encoded quantum states we cannot simply inspect the individual qubits to determine which one was flipped. Doing so would reveal information about the encoded state and collapse the superposition (remember~\cref{post-qm-measurement}), destroying the quantum information we are trying to protect.

To get around this, we perform carefully chosen measurements that reveal information about the error while leaving the encoded information intact. A basic example is a \emph{parity measurement} on a pair of qubits. There are two possible outcomes: even parity (i.e., $00$ or $11$) and odd parity (i.e., $01$ or $10$). This measurement reveals only whether two qubits are the same or different, rather than the value of either qubit individually. The state is therefore projected onto either the even-parity subspace spanned by $\ket{00}$ and $\ket{11}$ or the odd-parity subspace spanned by $\ket{01}$ and $\ket{10}$. If the encoded state already lies entirely within one of these parity sectors, then the measurement does not disturb the logical information.

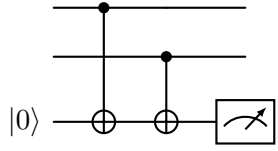
\begin{figure}[tb]
    \centering
    \begin{quantikz}
        & \ctrl{2} & & \\
        & & \ctrl{1} & \\
        \lstick{$\ket{0}$} & \targ{} & \targ{} & \meter{}
    \end{quantikz}
    \caption{Quantum circuit for measuring the joint parity of two qubits. Ancilla measurement of $1$ implies odd parity, while $0$ implies even parity.}
    \label{fig:parity-measurement-circuit}
\end{figure}

In practice, such measurements are often implemented using \emph{ancilla qubits}, which is one reason QEC requires more overhead than its classical counterpart. A quantum circuit to measure the parity of the first two qubits of the encoded state using the three-qubit repetition code is given in~\cref{fig:parity-measurement-circuit}. Note that the parity measurement circuit is equivalent to measuring the operator $Z \otimes Z \otimes I$ on the first two qubits:
\begin{align}
    Z \otimes Z \otimes I\ket{000} &= \ket{000}\\
    Z \otimes Z \otimes I\ket{010} &= -\ket{010}\\
    Z \otimes Z \otimes I\ket{100} &= -\ket{100}\\
    Z \otimes Z \otimes I\ket{110} &= \ket{110}\,.
\end{align}
Measurement of a $+1$ eigenvalue implies even parity, while a $-1$ eigenvalue implies odd parity. We can similarly probe the other pairs of qubits by measuring $Z \otimes I \otimes Z$ and $I \otimes Z \otimes Z$. These are examples of \emph{syndrome measurements}: their outcomes help us decode which error occurred. We will often use the shorthand $A \otimes B \otimes C \equiv ABC$, which is \emph{not} matrix multiplication of $A$, $B$, and $C$. Suppose an error $E$ occurs on the encoded state; for bit-flip errors, the possibilities are $XII$, $IXI$, and $IIX$. If $E$ commutes with a syndrome operator, the corresponding measurement yields $+1$, whereas if $E$ anticommutes with it, the outcome is $-1$; we characterize all of these in~\cref{tab:three-qubit-repetition-code-syndromes}. Once the error has been identified, we correct it by applying its inverse, which for Pauli operators is the same operator again.

\begin{table}[tb]
    \centering
    \begin{tblr}{colspec={Q[c,m]Q[c,m]Q[c,m]Q[c,m]}, vline{2}={1pt}, hline{2,3}={1pt}}
    & \SetCell[c=3]{c} \textbf{Syndrome} & & \\
    \textbf{Error} & $Z \otimes Z \otimes I$ & $Z \otimes I \otimes Z$ & $I \otimes Z \otimes Z$\\
    $X \otimes I \otimes I$ & $-1$ & $-1$ & $+1$\\
    $I \otimes X \otimes I$ & $-1$ & $+1$ & $-1$\\
    $I \otimes I \otimes X$ & $+1$ & $-1$ & $-1$
\end{tblr}
    \caption{Syndrome measurements for single-qubit bit-flip error.}
    \label{tab:three-qubit-repetition-code-syndromes}
\end{table}

Encoded quantum information is useful only if we can also manipulate it. This is done using \emph{logical operators}, which act on encoded states in the same way that ordinary gates act on unencoded qubits. For example, for the three-qubit repetition code, we need to be able to implement logical $X$ and logical $Z$ operations, meaning we need $\overline{X}$ and $\overline{Z}$ with the action
\begin{align}
    \overline{X}: \ket{\overline{0}} &\to \ket{\overline{1}} \text{ and } \ket{\overline{1}} \to \ket{\overline{0}}\,,\\
    \overline{Z}: \ket{\overline{0}} &\to \ket{\overline{0}} \text{ and } \ket{\overline{1}} \to -\ket{\overline{1}}\,.
\end{align}

Let us start with logical $X$ operations, written as $\overline{X}$. Recall that the logical states are encoded as $\ket{\overline{0}} = \ket{000}$ and $\ket{\overline{1}} = \ket{111}$. It is easy to see that $\overline{X} = X \otimes X \otimes X \equiv XXX$ takes $\ket{\overline{0}}$ to $\ket{\overline{1}}$ and \emph{vice versa}. Thus, it has the desired logical action. For the logical $Z$ operator, written $\overline{Z}$, one can choose $\overline{Z} = Z \otimes I \otimes I \equiv ZII$. In the next chapter we will see that logical operators are not unique: multiplying a logical operator by a stabilizer produces another operator with the same logical action. For example, $ZZZ$, $IIZ$, and $IZI$ all have the same logical action as $ZII$ on this code.

The final notion we consider in this section is the code distance, which we already encountered for the classical repetition code. For a quantum code, we first define the \emph{Pauli weight} of an operator as the number of qubits on which it acts nontrivially, that is, with a non-identity operator. For example, the weight of $XII$ is 1, while the weight of $ZIZ$ is 2. The distance of a quantum code is the minimum Pauli weight of any nontrivial logical operator. For the three-qubit repetition code, it is useful to distinguish the $X$-type and $Z$-type distances, denoted $d_X$ and $d_Z$, respectively. The logical $X$ operator is $XXX$, so $d_X=3$, which means the code can detect up to $d_X-1=2$ bit-flip errors and correct up to $\lfloor (d_X-1)/2 \rfloor = 1$ bit-flip error. By contrast, the logical $Z$ operator can be implemented by any of $ZII$, $IZI$, $IIZ$, or $ZZZ$, where the minimum weight is 1 and hence $d_Z=1$, meaning that the code cannot detect or correct any phase-flip errors. The overall code distance is therefore
\begin{equation}
    d=\min\{d_X,d_Z\}=1.
\end{equation}
Thus, although the three-qubit repetition code protects against bit-flips, it does not protect against phase-flips. To handle both error types, we need more sophisticated QEC codes.

% The final notion we consider in this section is the code distance, which we already encountered for the classical repetition code. To define it for a quantum code, we first introduce the \emph{Pauli weight} of an operator: the number of qubits on which the operator acts nontrivially. For example, the weight of $XII$ is 1, while the weight of $ZIZ$ is 2. The distance of a quantum code is the minimum weight of any nontrivial logical operator. For the three-qubit repetition code, it is useful to distinguish the $X$-type and $Z$-type distances, denoted $d_X$ and $d_Z$, respectively. For the three-qubit repetition code, the unique logical $X$ operator $XXX$, so $d_X = 3$. The set of logical $Z$ operators $\{ZII,IZI,IIZ,ZZZ\}$ has minimum weight 1, so $d_Z = 1$. Recall that a code of distance $d$ can detect up to $d-1$ errors and correct up to $\lfloor\frac{d-1}{2}\rfloor$ errors, we conclude that the three-qubit repetition code can detect 2 and correct 1 $X$ error, but it cannot detect or correct any $Z$ errors. The code distance $d$ is defined as 
% \begin{equation}
%     d:=\min{d_X,d_Z}.
% \end{equation}
% Thus, the distance of the three-qubit repetition code is 1, implying that it cannot correct even single qubit errors. This is beacuse it can correct $X$ but not $Z$ errors even on a single qubit. To handle both error types, we need more sophisticated QEC codes.

Before presenting such a code, we note that we can use a modified repetition code to detect phase-flip errors at the cost of not being able to detect bit-flip errors. The change is quite simple: we take the encoding circuit for the three-qubit repetition code for detecting bit-flip errors and we simply apply a Hadamard gate to all three qubits after the CNOT gates. This is just a basis change, and so the logical qubit that we are encoding into our three-qubit repetition code is now
\begin{equation}
    \ket{\overline{\psi}} = \alpha\ket{+++} + \beta\ket{---}\,,
\end{equation}
with the basis code states $HHH\ket{000}=\ket{+++}$ and $HHH\ket{111}=\ket{---}$. 
Then, a phase-flip error on a single qubit takes the state from $\ket{+}$ to $\ket{-}$, or \emph{vice versa}. The error detection and correction is then the same as it is for the bit-flip error three-qubit repetition code, just with the syndrome operators conjugated by Hadamard gates on each qubit.

\subsection{Nine-qubit Shor code}\label{subsec:fundamentals-nine-qubit-shor-code}

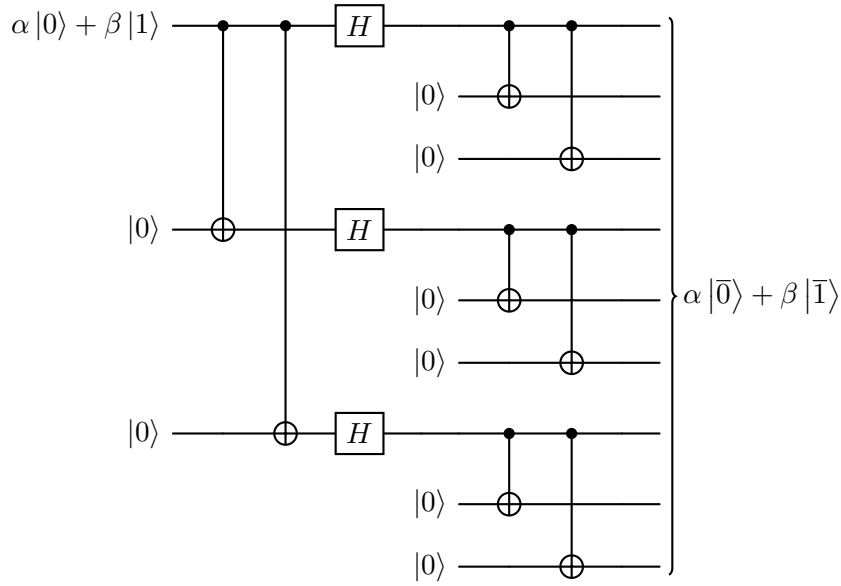
\begin{figure}[tb]
    \centering
    \begin{quantikz}
        \lstick{$\alpha\ket{0} + \beta\ket{1}$} & \ctrl{3} & \ctrl{6} & \gate{H} & & & \ctrl{1} & \ctrl{2} & & \rstick[9]{$\alpha\ket{\overline{0}} + \beta\ket{\overline{1}}$} \\
        \setwiretype{n} & & & & & \lstick{$\ket{0}$} & \targ{}\setwiretype{q} & & & \\
         \setwiretype{n} & & & & & \lstick{$\ket{0}$} & \setwiretype{q} & \targ{} & & \\
         \lstick{$\ket{0}$} & \targ{} & & \gate{H} & & & \ctrl{1} & \ctrl{2} & & \\
        \setwiretype{n} & & & & & \lstick{$\ket{0}$} & \targ{}\setwiretype{q} & & & \\
         \setwiretype{n} & & & & & \lstick{$\ket{0}$} & \setwiretype{q} & \targ{} & & \\
         \lstick{$\ket{0}$} & & \targ{} & \gate{H} & & & \ctrl{1} & \ctrl{2} & & \\
        \setwiretype{n} & & & & & \lstick{$\ket{0}$} & \targ{}\setwiretype{q} & & & \\
         \setwiretype{n} & & & & & \lstick{$\ket{0}$} & \setwiretype{q} & \targ{} & &
    \end{quantikz}
    \caption{Quantum circuit for encoding a single qubit into the nine-qubit Shor code.}
    \label{fig:shor-code-encoding}
\end{figure}

We saw in the previous subsection that the three-qubit repetition code can only correct one type of error at a time, so we need a quantum error-correcting code that can correct both bit-flip and phase-flip errors. The first code proposed to do just this was the nine-qubit \eczoo[Shor code]{shor_nine}, which was introduced by Peter Shor in 1995~\cite{shor1995scheme}. The Shor code is a $\stabcode{9}{1}{3}$ code, meaning it encodes 1 logical qubit into 9 physical qubits and has a distance of 3, such that it can detect any two-qubit error and correct any single-qubit error. The logical codewords for the Shor code are
\begin{equation}
    \ket{\overline{0}} = \frac{1}{2\sqrt{2}}\lp \ket{000} + \ket{111} \rp^{\otimes 3} \text{ and } \ket{\overline{1}} = \frac{1}{2\sqrt{2}}\lp \ket{000} - \ket{111} \rp^{\otimes 3}\,,
\end{equation}
which is obtained by combining two simpler codes: the three-qubit repetition code for bit-flip errors and the modified version of that code for phase-flip errors. The process of combining existing codes in this way is called \emph{concatenation}, and we will see several examples of it throughout this tutorial. At the first level of the code, which we call the \emph{outer code}, we encode one qubit into three qubits. At the second level, which we call the \emph{inner code}, we encode each of those three qubits into three more qubits, giving a total of nine qubits. Each layer corrects a different type of error, either bit-flips or phase-flips, and this can be done in either order. To be concrete, we take the outer code to be the modified three-qubit repetition code for phase-flip errors and the inner code to be the original three-qubit repetition code for bit-flip errors. The circuit diagram for this encoding is given in~\cref{fig:shor-code-encoding}.

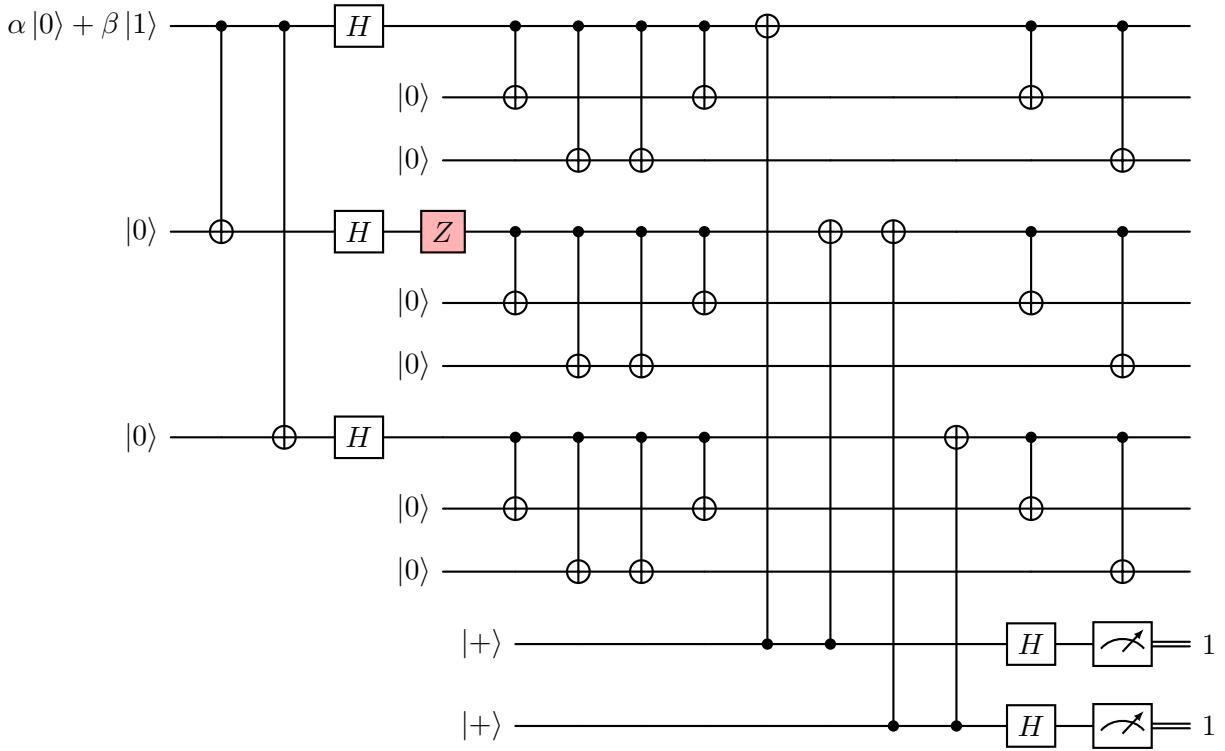
\begin{figure}[tb]
    \centering
    \resizebox{\linewidth}{!}{
    \begin{quantikz}
        \lstick{$\alpha\ket{0} + \beta\ket{1}$} & \ctrl{3} & \ctrl{6} & \gate{H} & & \ctrl{1} & \ctrl{2} & \ctrl{2} & \ctrl{1} & \targ{} & & & & \ctrl{1} & \ctrl{2} & \\
        \setwiretype{n} & & & & \lstick{$\ket{0}$} & \targ{}\setwiretype{q} & & & \targ{} & & & & & \targ{} & & \\
        \setwiretype{n} & & & & \lstick{$\ket{0}$} & \setwiretype{q} & \targ{} & \targ{} & & & & & & & \targ{} & \\
        \lstick{$\ket{0}$} & \targ{} &  & \gate{H} & \gate[style={fill=red!30}]{Z} & \ctrl{1} & \ctrl{2} & \ctrl{2} & \ctrl{1} & & \targ{} & \targ{} & & \ctrl{1} & \ctrl{2} & \\
        \setwiretype{n} & & & & \lstick{$\ket{0}$} & \targ{}\setwiretype{q} & & & \targ{} & & & & & \targ{} & & \\
        \setwiretype{n} & & & & \lstick{$\ket{0}$} & \setwiretype{q} & \targ{} & \targ{} & & & & & & & \targ{} & \\
        \lstick{$\ket{0}$} & & \targ{} & \gate{H} & & \ctrl{1} & \ctrl{2} & \ctrl{2} & \ctrl{1} & & & & \targ{} & \ctrl{1} & \ctrl{2} & \\
        \setwiretype{n} & & & & \lstick{$\ket{0}$} & \targ{}\setwiretype{q} & & & \targ{} & & & & & \targ{} & & \\
        \setwiretype{n} & & & & \lstick{$\ket{0}$} & \setwiretype{q} & \targ{} & \targ{} & & & & & & & \targ{} & \\
        \setwiretype{n} & & & & & \lstick{$\ket{+}$} & \setwiretype{q} & & & \ctrl{-9} & \ctrl{-6} & & & \gate{H} & \meter{} & \setwiretype{c}\rstick{1} \\
        \setwiretype{n} & & & & & \lstick{$\ket{+}$} & \setwiretype{q} & & & & & \ctrl{-7} & \ctrl{-4} & \gate{H} & \meter{} & \setwiretype{c}\rstick{1}
    \end{quantikz}
    }
    \caption{Quantum circuit illustrating a single $Z$ error on the middle block of the nine-qubit Shor code, followed by a syndrome measurement that identifies which block the error occurred in. The syndrome outcome \texttt{11} indicates that the $Z$ error occurred on the middle block of three qubits. The full syndrome-extraction circuit is larger and measures all relevant $X$- and $Z$-type syndromes; here we show only the part relevant to the illustrated $Z$ error.}
    \label{fig:shor-code-phase-flip-error}
\end{figure}

Using this encoding, we can write an encoded logical qubit as
\begin{align}
    \ket{\overline{\psi}} = \alpha\ket{\overline{0}} + \beta\ket{\overline{1}}\,,
\end{align}
where $\ket{\overline{0}}$ and $\ket{\overline{1}}$ are given above. The encoding produces three blocks of qubits, as indicated in~\cref{fig:shor-code-encoding}, where each block is a three-qubit repetition encoding used to protect against bit-flips. Since the single-qubit Pauli operators $X$, $Y$, and $Z$ form a basis for single-qubit errors, and since $Y$ simply combines both a bit-flip and a phase-flip, it suffices to understand how the code handles bit-flip and phase-flip errors. Here, a bit-flip corresponds to $X$, a phase-flip to $Z$, and a combined bit-and-phase-flip to $Y$.

Let's now see how we can detect and correct a single $X$ error. First, note that each block encodes a single qubit into three qubits using the usual three-qubit repetition code. Thus, we can treat each of these blocks independently and, within each block, perform the syndrome measurement and $X$ error correction as described above in~\cref{subsec:fundamentals-quantum-repetition-code}. This allows us to detect and correct at most one bit-flip error.

For $Z$ errors, things are slightly more complicated because of the inner encoding used to handle $X$ errors. In particular, a single $Z$ error acting on any one qubit within a given block has the same logical effect as a $Z$ error on either of the other two qubits in that block. One way to detect and correct such an error is to first decode the inner code, leaving the three qubits of the outer modified repetition code for $Z$ errors. We then perform the appropriate syndrome measurement and correction, and finally re-encode using the inner code to recover the original nine-qubit encoding. We illustrate a single $Z$ error and the corresponding syndrome measurement in~\cref{fig:shor-code-phase-flip-error}.

\subsection{Noise channels}\label{subsec:fundamentals-noise-channels}
So far we have been somewhat vague about what exactly we mean by an ``error,'' so we now make that notion precise. In particular, we describe how errors act on quantum states and how to model the resulting noisy states. Up to this point, we have described quantum states using state vectors $\ket{\psi}$, which are sufficient for fully-known, isolated states, called \emph{pure states}. After noise acts, however, we may only know that the system is in one of several possible pure states with certain classical probabilities. This motivates the \emph{density matrix}, which provides a unified description of both pure and mixed states.

\begin{defbox}[label={def:density-matrix}]{Density matrix}
    A \emph{density matrix} (or \emph{density operator}), typically denoted by $\rho$, is a positive semidefinite matrix (i.e., $\rho \succeq 0$) acting on the state space of a quantum system, such that its trace is equal to one ($\Tr[\rho] = 1$). For a statistical ensemble in which the quantum system is in the pure state $\ket{\psi_i}$ with classical probability $p_i$, the density matrix is
    \begin{equation}
       \rho = \sum_i p_i \ketbra{\psi_i}\,,
    \end{equation}
    where the probabilities satisfy $p_i \geq 0$ and $\sum_i p_i = 1$. Here, the $p_i$ quantify classical ignorance about which pure state was prepared; they are not the same as Born-rule probabilities extracted from amplitudes inside any one of the states $\ket{\psi_i}$.
\end{defbox}

The two defining properties of a density matrix have clear physical meanings. Positive semidefiniteness guarantees that all measurement probabilities computed from $\rho$ are nonnegative, while the trace-one condition guarantees that these probabilities sum to one. Thus, density matrices are precisely the operators that can represent physical quantum states. Pure states are the special case $\rho = \ketbra{\psi}$, while more general density matrices describe \emph{mixed states}, which encode classical uncertainty or correlations with an environment.

Density matrices also let us describe composite systems beyond the pure-state setting. A bipartite mixed state $\rho_{AB}$ is called \emph{separable} if it can be written as
\begin{equation}
    \rho_{AB}=\sum_i p_i\, \rho_A^{(i)} \otimes \rho_B^{(i)},
\end{equation}
with $p_i \geq 0$ and $\sum_i p_i=1$. If no such decomposition exists, the state is \emph{entangled}. This distinction will matter below when we explain why quantum channels must remain physical even when the system of interest is entangled with another one.

With this mathematical tool, we can describe noisy dynamics in a unified way. Errors may transform pure states into mixed states, and the resulting evolution of the density matrix $\rho$ is captured by the formalism of \emph{quantum channels}~\cite{wilde2019from}.

\begin{defbox}[label={def:quant-chan}]{Quantum channel}
   Let $\mathcal{E}$ be a quantum channel, $\rho$ (and any $\rho$ with a subscript) be a quantum state, and let $c_1$ and $c_2$ be complex numbers. A quantum channel is a map
    \begin{equation}
        \mathcal{E} : \mathcal{L}(\mathcal{H}_{S}) \to \mathcal{L}(\mathcal{H}_{S'})\,,
    \end{equation}
    where the input Hilbert space $\mathcal{H}_{S}$ and output Hilbert space $\mathcal{H}_{S'}$ satisfy the following properties: 
    \begin{enumerate}
        \item \textbf{Linearity}: $\mathcal{E}(c_1\rho_1 + c_2\rho_2) = c_1\mathcal{E}(\rho_1) + c_2\mathcal{E}(\rho_2)$,
        \item \textbf{Trace-preservation}: $\Tr[\mathcal{E}(\rho)] = \Tr[\rho]$,
        \item \textbf{Complete positivity}: $(\mathcal{E} \otimes \mathcal{I}_R)(\rho_{SR}) \succeq 0$ for every auxiliary system $R$ and every joint state $\rho_{SR} \succeq 0$ on $\mathcal{H}_{S} \otimes \mathcal{H}_{R}$.
    \end{enumerate}
    $\rho \succeq 0$ means that $\rho$ is positive semidefinite.
\end{defbox}

Thus, a quantum channel is a completely positive trace-preserving (CPTP) map. Each part of this definition is needed for physical consistency. Linearity reflects the fact that the evolution of a statistical mixture is the corresponding mixture of the evolved states. Trace preservation guarantees that total probability remains 1. Positivity ensures that valid density matrices are mapped to valid density matrices. Finally, complete positivity is stronger than ordinary positivity: it guarantees that the map remains physical even when the system is part of a larger entangled state. Without complete positivity, a map could appear valid on isolated systems but produce an unphysical output, such as a matrix with negative eigenvalues, when applied to one half of an entangled pair.

While the above definition is mathematically clean, it is not especially convenient for calculations. For that purpose, we use the \emph{Kraus representation}:

\begin{defbox}[label={def:kraus-rep}]{Kraus Representation}
    Any CPTP map can be expressed in the \emph{Kraus representation} (or operator-sum representation) as:
    \begin{equation}\label{eq:kraus_rep}
        \mathcal{E}(\rho) = \defsum{i}{}{E_i \rho E_i^\dagger}\,,
    \end{equation}
    where $\rho$ is a quantum state in a Hilbert space $\mathcal{H}_S$ and the operators $\{E_i\}$ are linear operators mapping $\mathcal{H}_S$ to $\mathcal{H}_{S'}$, known as \emph{Kraus operators}. For the map to be trace-preserving, the Kraus operators must satisfy the completeness relation:
    \begin{equation}
        \defsum{i}{}{E_i^\dagger E_i} = I_S\,,\label{eqn:kraus-completeness}
    \end{equation}
    where $I_S$ is the identity operator on the input Hilbert space $\mathcal{H}_S$.
\end{defbox}

We practice using this definition in the following exercise:

\begin{exerbox}[label={exer:trace-preserving-kraus}]{Trace preservation in the Kraus representation}
    Prove that the map in~\cref{eq:kraus_rep} is trace-preserving, meaning that $\Tr[\mathcal{E}(\rho)] = \Tr[\rho]$ for all density matrices $\rho$, if and only if the Kraus operators satisfy the completeness relation in~\cref{eqn:kraus-completeness}.

    [Hint: use the cyclic property of the trace.]
\end{exerbox}

Using this Kraus representation of quantum channels, we introduce our first quantum noise channel, the \emph{dephasing channel}, in the following example.

\begin{example}[label={ex:dephasing}]{Dephasing channel}
    The dephasing channel models the decay of the coherence encoded in a density matrix. The Kraus operators for a single-qubit dephasing channel are
    \begin{equation}\label{eq:depha_kraus}
        E_0 = \sqrt{1-p}I \text{ and } E_1 = \sqrt{p} Z\,,
    \end{equation}
    where $p$ is the probability of a phase-flip occurring. We can check that the completeness relation is satisfied:
    \begin{equation}
        E_0^\dagger E_0 + E_1^\dagger E_1 = (1-p)I + pZ^2 = I\,.
    \end{equation}
    The action of the channel on a density matrix
    \begin{equation}
        \rho =
        \begin{pmatrix} 
            \rho_{00} & \rho_{01}\\ 
            \rho_{10} & \rho_{11}
        \end{pmatrix}
    \end{equation}
    is
    \begin{align}\label{eq:depha_channel}
        \mathcal{E}(\rho) &= E_0\rho E_0^\dagger + E_1\rho E_1^\dagger\\
        &= (1-p)\rho + pZ\rho Z\\
        &=
        \begin{pmatrix} 
            \rho_{00} & (1-2p)\rho_{01} \\ 
            (1-2p)\rho_{10} & \rho_{11}
        \end{pmatrix}\,.
    \end{align}
    When $p = \frac{1}{2}$, this becomes a completely dephasing channel because the off-diagonal elements vanish, meaning that coherence is entirely lost. In experimental settings, dephasing is often used as a simple model for $T_2$ relaxation, which characterizes the timescale over which phase information between the states $\ket{0}$ and $\ket{1}$ decays.
\end{example}

Another common quantum noise channel is the \emph{depolarizing channel}, which we explore in the next example:

\begin{example}[label={ex:depolarizing}]{Depolarizing channel}
    The \emph{depolarizing channel} is a noise channel in which a qubit state is replaced by the maximally-mixed state $\frac{1}{2}I$ with probability $p$. The original state is then preserved with probability $1-p$. The Kraus operators are
    \begin{equation}\label{eq:depola_kraus}
       E_0 = \sqrt{1-\frac{3p}{4}}I, \quad E_i = \frac{\sqrt{p}}{2}\sigma_i \quad \text{for } i \in \{1, 2, 3\}
    \end{equation}
    The action of the channel on an arbitrary single-qubit state is then
    \begin{align}\label{eq:depola_channel}
        \mathcal{E}(\rho) &= (1-p)\rho + p\frac{I}{2}=
        \begin{pmatrix}
            (1-p/2)\rho_{00} & (1-p)\rho_{01}\\
            (1-p)\rho_{10} & (1-p/2)\rho_{11}
        \end{pmatrix}\,.
    \end{align}
    The depolarizing channel models a symmetric loss of information: the state remains unchanged with probability $1-\frac{3p}{4}$, while each of the errors $X$, $Y$, and $Z$ occurs with probability $\frac{p}{4}$. When $p=1$, the channel is \emph{completely depolarizing} because every input state is replaced by the maximally mixed state.
\end{example}

The final channel we consider is the \emph{amplitude damping channel}, which also has some energy dissipation in addition to the loss of quantum information:

\begin{example}[label={ex:amp-damp}]{Amplitude damping channel}
    Given a single qubit, the \emph{amplitude damping channel} models the process of spontaneous emission, where the excited state $\ket{1}$ decays to the ground state $\ket{0}$. The Kraus operators for this channel are
    \begin{equation}\label{eq:kraus_amp_damp}
        E_0 =
        \begin{pmatrix}
            1 & 0\\
            0 & \sqrt{1-p}
        \end{pmatrix}
        \text{ and }
        E_1 =
        \begin{pmatrix} 
            0 & \sqrt{p}\\
            0 & 0
        \end{pmatrix}\,,
    \end{equation}
    where $p$ is the probability that the excited state $\ket{1}$ decays to the ground state $\ket{0}$. We can check that the completeness relation is satisfied:
    \begin{align}
        E_0^\dagger E_0 + E_1^\dagger E_1 = I\,.
    \end{align}
    The action on an arbitrary quantum state $\rho$ is given by
    \begin{align}\label{eq:amp_damp_channel}
        \mathcal{E}(\rho) &= E_0\rho E_0^\dagger + E_1\rho E_1^\dagger\\
        &=
        \begin{pmatrix}
            \rho_{00} + p\rho_{11} & \sqrt{1-p}\rho_{01}\\
            \sqrt{1-p}\rho_{10} & (1-p)\rho_{11}
        \end{pmatrix}\,.
    \end{align}
    The amplitude damping channel models the $T_1$ relaxation, also called the longitudinal relaxation. $T_1$ is thus the timescale of energy loss that occurs when the excited state $\ket{1}$ decays to the ground state $\ket{0}$. In real-world quantum computing platforms, both $T_1$ and $T_2$ represent critical performance metrics that directly determine the feasibility and scalability of quantum information processing.
\end{example}

The previous examples all acted on a single qubit, but the same formalism extends naturally to many-qubit systems where one often models the multi-qubit channel by applying the relevant single-qubit channels independently to each qubit. One important family is the \emph{Pauli channel}, defined as follows:

\begin{defbox}[label={def:pauli-channel}]{Pauli channel}
    An $n$-qubit \emph{Pauli channel} is given by
    \begin{align}
        \mathcal{E}(\rho)=\defsum{i}{}{p_i P_i \rho P_i^\dagger}\,,
    \end{align}
    where $P_i$ are tensor products of $I$, $X$, $Y$ and $Z$, also called Pauli strings: $P_i \in \{I,X,Y,Z\}^{\otimes n}$, and probabilities $p_i$ with $\sum_i p_i=1$
\end{defbox}

When $n=1$, this family includes familiar examples such as dephasing and depolarizing noise.

\begin{exerbox}[label={exer:single-qubit-pauli-distributions}]{Single-qubit Pauli channels}
    Consider the single-qubit version of the Pauli channel in~\cref{def:pauli-channel}, where the Pauli operators are $\{I,X,Y,Z\}$ with probabilities $\{p_I,p_X,p_Y,p_Z\}$. Determine these probabilities for the channel to become:
    \begin{enumerate}
        \item The dephasing channel in~\cref{ex:dephasing},
        \item The depolarizing channel in~\cref{ex:depolarizing}.
    \end{enumerate}
\end{exerbox}

We conclude this section with the \emph{Knill-Laflamme condition}, which formalizes what errors a quantum code can correct:

\begin{thmbox}[label={thm:knill-laflamme-condition}]{Knill--Laflamme condition}
    Let $P$ be the projector onto the codespace $\mathcal{C}$, and let $\{E_a\}$ be a set of errors. Then, $\mathcal{C}$ corrects these errors if and only if
    \begin{equation}
        P E_a^\dagger E_b P = C_{ab} P\,,
    \end{equation}
    where $C_{ab} \in \CC$.
\end{thmbox}

We interpret this to mean that the combined action of any pair of correctable errors is indistinguishable from a scalar multiple of the identity. That is, the error process must not reveal any information about the encoded quantum state.

\subsection{Brief history of QEC}\label{subsec:qec-history}
We conclude this chapter with a brief history of QEC, for those interested in its historical development. QEC was introduced shortly after Peter Shor kick-started the field of quantum computing with his factoring algorithm in 1994 (though finally published in 1997)~\cite{shor1997polynomial}. While the three-qubit bit-flip code had been introduced in 1985 by Asher Peres~\cite{peres1985reversible}, ten years later, Shor~\cite{shor1995scheme} introduced the first QEC code that could correct both bit-flip and phase-flip errors with his nine-qubit code. Shortly after, in 1996, Andrew Steane also pointed out that decoherence is a fundamental aspect of quantum mechanics and so any computation based on quantum mechanical systems must inevitably deal with this problem, which he proposed can be done with his seven-qubit code~\cite{steane1996error}. The years 1996 and 1997 were then quite busy for QEC, where after Steane proposed his seven-qubit code, several other advancements were made. The five-qubit code---the smallest possible QEC code that can protect a single logical qubit against single-qubit errors---was independently proposed by Laflamme, Miquel, Paz, and Zurek~\cite{laflamme1996perfect} and Bennett, DiVincenzo, Smolin, and Wootters~\cite{bennett1996mixed-state} in 1996. A more general framework for these codes was then introduced by Calderbank and Shor~\cite{calderbank1996good} and Steane~\cite{steane1996multiple-particle} in the form of Calderbank-Shor-Steane (CSS) codes. Further generalizations were then made independently by Daniel Gottesman in his seminal doctoral thesis~\cite{gottesman1997stabilizer} and by Calderbank, Rains, Shor, and Sloane~\cite{calderbank1996good}.

Fault-tolerant quantum memory was then proposed by Alexei Kitaev~\cite{kitaev1997fault-tolerant}, where he also proposed the toric code, still a highly relevant code construction today. Later, Shor again made a major contribution, showing how to do universal logical FTQC without using concatenation~\cite{shor1996fault-tolerant}. Further details of FTQC were then developed by Kitaev~\cite{kitaev1997quantumcomputations}; Cirac, Pellizzari and Zoller~\cite{cirac1996enforcing}; Zurek and Laflamme~\cite{zurek1996quantum}; DiVincenzo and Shor; and Gottesman~\cite{gottesman1998theory}. All of this work was then put together independently by Knill, Laflamme, and Zurek~\cite{knill1996threshold}; Aharonov and Ben-Or~\cite{aharonov1996fault}; and Kitaev~\cite{kitaev1997fault-tolerant} in 1996 and 1997 in the form of the fault-tolerant threshold theorem, one of the most important results of QEC. The proof for the threshold theorem was subsequently made more rigorous and improved upon by Aliferis, Gottesman, and Preskill in 2005~\cite{aliferis2005quantum}.

The above advancements laid much of the groundwork for QEC upon which further advancements were made in the late 1990s. Beyond the stabilizer formalism, Rains, Hardin, Shor, and Sloane~\cite{rains1997nonadditive} introduced a non-additive (i.e., non-stabilizer) construction in 1997. Also in 1997, Leung, Nielsen, Chuang, and Yamamoto~\cite{leung1997approximate} introduced approximate error correction, expanding the somewhat narrow consideration of single-qubit Pauli error channels. All of the above advancements were made for qubits, but of course, quantum systems can have more than two states, so Gottesman~\cite{gottesman1998fault-tolerant} and Rains~\cite{rains1997nonbinary} independently introduced formalism to consider higher dimensional QEC codes. Furthermore, in the late 1990s, there was abundant work on quantum analogues of error correction bounds; see, for example, references~\cite{ekert1996quantum,gottesman1996class,shor1997quantum,ashikhmin1997remarks,ashikhmin1999upper,rains1998quantum,rains1999quantum,rains1999monotonicity}.

Over the next few years, in the early 2000s, several other improvements and advancements were made by many researchers. In 2002, Dennis, Kitaev, Landahl, and Preskill~\cite{dennis2002topological} studied many details of the toric code, which was introduced by Kitaev a few years earlier. Then, in 2004, Bravyi and Kitaev~\cite{bravyi2005universal} introduced magic state distillation, still a hot topic of research today. In 2006, Bombin and Delgado~\cite{bombin2006topological,bombin2008statistical} proposed a new type of code, the color code. While the surface code was introduced by Kitaev years before, it was popularized in 2012 by Fowler, Mariantoni, Martinis, and Cleland~\cite{fowler2012surface}.

We stop our early history of QEC there, though that is not to say that this is where the advancements in QEC research stopped. QEC remains one of the fastest growing subfields of quantum information science and quantum computing, and is finally even seeing exciting experimental demonstrations of high quality error correction and early fault tolerance~\cite{sundaresan2023demonstrating,googlequantumai2023suppressing,bluvstein2024logical,hong2024entangling,rodriguez2024experimental,paetznik2024demonstration,lacroix2025scaling,eickbusch2025demonstration,googlequantumai2025quantum,butt2026demonstration}.

Finally, we highlight several introductory references specific to QEC. The books by Lidar and Brun~\cite{lidar2013quantum} and Gaitan~\cite{gaitan2013quantum} are well-known references, though much progress has been made in QEC since their publication in 2013. Though the textbook is not yet available in print (at the time of writing), Daniel Gottesman presents a more recent view of QEC in his \emph{Surviving as a Quantum Computer in a Classical World}~\cite{gottesman2024surviving}. There are also several freely available online references, such as lecture notes by Michael Kastoryano~\cite{kastoryano2019quantum} and the introductory guide by Joschka Roffe~\cite{roffe2019quantum}. Finally, the Error Correction Zoo~\cite{eczoo} is an invaluable resource for researchers and students alike.

\newpage

\section{Stabilizer codes}\label{sec:stabilizer}

We now turn to one of the most important and successful classes of quantum error-correcting codes: the \emph{stabilizer codes}, first introduced by Daniel Gottesman (see~\cref{subsec:qec-history} for more history). The stabilizer formalism is one of the most useful tools in quantum error correction and several of the codes that we have already encountered can be described using it. We make use of the notions in this chapter, including group theory, the stabilizer formalism, and CSS codes, throughout this tutorial.

\subsection{Primer on group theory}\label{subsec:stabilizer-group-theory-primer}
Before introducing stabilizer codes, we need to understand some basic \emph{group theory}~\cite{dummit2003abstract}. Group theory is a field of mathematics that studies highly-structured collections of objects equipped with a binary operation satisfying certain properties. Since we primarily focus on discrete errors in this tutorial, this primer deals exclusively with finite groups. We begin with the concept of an \emph{equivalence relation}:

\begin{defbox}[label={def:equivalence-relation}]{Equivalence relation}
    A relation $\sim$ on a set $G$ is an \emph{equivalence relation} if it satisfies the following conditions:
    \begin{enumerate}
        \item \textbf{Reflexivity}: $g \sim g$ for all $g \in G$.
        \item \textbf{Symmetry}: $g \sim g'$ if and only if $g' \sim g$.
        \item \textbf{Transitivity}: if $g \sim g'$ and $g' \sim g''$, then $g \sim g''$.
    \end{enumerate}
    For any element $g \in G$, its equivalence class, denoted $[g]$, is defined as $[g] = \{g'\in G : g' \sim g\}$. 
\end{defbox}

Two equivalence classes are either \emph{disjoint} or \emph{equal}. Consequently, any equivalence relation partitions a set into disjoint equivalence classes. This construction is closely connected to cosets in group theory and will play a crucial role in helping us organize errors and logical operators. With this in mind, we now formally define a group:

\begin{defbox}[label={def:group}]{Group}
    For a set $G$ and a binary operation $\ast: G \times G \to G$, the tuple $(G,\ast)$ is a \emph{group} if all of the following hold:
    \begin{enumerate}
        \item \textbf{Associativity}: $(a \ast b) \ast c = a \ast (b \ast c)$ for any $a,b,c\in G$.
        \item \textbf{Identity}: There exists a unique \emph{identity} element $e \in G$ such that $e \ast a = a \ast e = a$ for any $a \in G$.
        \item \textbf{Inverse}: For each $a \in G$ there exists a unique \emph{inverse} element $b \in G$ such that $a \ast b = b \ast a = e$. This unique element $b$ is denoted by $a^{-1}$.
    \end{enumerate}
    In addition, we say that $a, b\in G$ \emph{commute} if $a \ast b = b \ast a$. A group $(G,\ast)$ in which all $a, b\in G$ commute is called an \emph{abelian} group; otherwise, it is \emph{non-abelian}.
\end{defbox}

When it is clear from the context, we simply refer to the group $(G, \ast)$ as $G$. Also, where appropriate, we suppress the ``$\ast$'' operation and write $a \ast b$ simply as $ab$. For example, when $\ast$ is the multiplication operator, $\times$, we write $ab$ instead of $a \times b$ for group elements $a, b \in G$. Let's look at a few examples of groups, starting with the integers under addition:

\begin{example}[label={ex:integer-addition}]{Integer addition}
    The set of integers $\Z = \{\dots,-2,-1,0,1,2,\dots\}$ under the addition operation $+$ is an abelian group $(\Z,+)$ with identity element $0$. The inverse element of $a \in \Z$ is $-a$ such that $a + (-a) = (-a) + a = 0$.
\end{example}

By restricting the integers to a finite set and using modular arithmetic, we can construct an important class of finite abelian groups known as the cyclic groups.

\begin{example}[label={ex:integers-modulo-n}]{Integers modulo $n$}
    The group $(\mathbb{Z}_n, +)$ (also written as $\mathbb{Z}/n\mathbb{Z}$) represents the integers modulo a positive integer $n$ under modular addition. For example, $\mathbb{Z}_2 = \{0, 1\}$ contains only two elements, where the group operation is defined by $0+0=0$, $0+1=1$, and $1+1=0$.
\end{example}

As a more sophisticated example, we now look at matrix multiplication over real matrices:

\begin{example}[label={ex:invertible-real-matrix-mult}]{Invertible real matrix multiplication}
    The \emph{general linear group} $\GL_n(\R)$ is the set of $n\times n$ invertible real matrices under matrix multiplication. Its identity element is the $n\times n$ identity matrix $I_n$, and the inverse of $A\in\GL_n(\R)$ is its matrix inverse $A^{-1}$.
    The group $\GL_n(\R)$ is non-abelian for $n\geq2$.
    For example, matrices
    \begin{equation}
        A=\left[\begin{array}{cc}
            1 & 0 \\
            1 & 1
        \end{array}\right]
        \quad\text{and}\quad
        B=\left[\begin{array}{cc}
            1 & 1 \\
            0 & 1
        \end{array}\right]
    \end{equation}
    are invertible, so both belong to $\GL_2(\R)$.
    However,
    \begin{equation}
        AB=\left[\begin{array}{cc}
            1 & 1 \\
            1 & 2
        \end{array}\right]
        \neq
        BA=\left[\begin{array}{cc}
            2 & 1 \\
            1 & 1
        \end{array}\right]\,,
    \end{equation}
    which means that $\GL_2(\RR)$ is a non-abelian group.
\end{example}

Fundamental to the stabilizer formalism and the study of quantum error-correcting codes in general is the concept of a \emph{generator}, which define as:

\begin{defbox}[label={def:generator}]{Generator}
    Given a group $G$, a subset $S \subseteq G$ is a \emph{generating set} if every element of $G$ can be written as a finite product of elements of $S$ and their inverses. Equivalently,
    \begin{equation}
        G =\langle S \rangle = \{s_1^{e_1}s_2^{e_2} \cdots s_k^{e_k} : k \geq 0, s_i \in S, e_i \in \ZZ\}\,.
    \end{equation}
\end{defbox}

As an example of how to produce a group from a generating set, consider the Pauli group, which we have already seen as an important group in our study of quantum error correction:

\begin{example}[label={ex:n-qubit-pauli-group}]{Pauli group}
    The $n$-qubit Pauli group $\mathcal{P}_n$ consists of all $n$-fold tensor products of $I$, $X$, $Y$, and $Z$, together with overall phases $\pm1$ and $\pm i$, that is,
    \begin{equation}
        \mathcal{P}_n=\{I,X,Y,Z\}^{\otimes n}\times\{\pm1,\pm i\}\,.
    \end{equation} 
    A generating set for $\mathcal{P}_n$ is
    \begin{equation}
        \langle i, X_1, Z_1, X_2, Z_2, \ldots, X_n, Z_n \rangle\,,
    \end{equation}
    where $X_i$ and $Z_i$ are the $n$-qubit operators that act as $X$ or $Z$ on the $i\numth$ qubit and as the identity on all other qubits. Up to an overall phase, Pauli operators are self-inverse: $P^2 = \pm I$ for any $P \in \mathcal{P}_n$. Taking $n=2$ as an example, we can check that $X_1X_2$ and $Z_1X_2$ do not commute, that is, $(X_1X_2) (Z_1X_2) = - (Z_1X_2) (X_1X_2)$, so the Pauli group is overall a non-abelian group.
\end{example}

Another important structure is the \emph{subgroup}. Since stabilizers of a code constitute a particular subgroup of the Pauli group, understanding subgroup properties is essential for stabilizer code theory. We define a subgroup as follows:

\begin{defbox}[label={def:subgroup}]{Subgroup}
    Given a group $(G,\ast)$, a subset $H \subseteq G$ is a \emph{subgroup} if $H$ is itself a group under the binary operation induced by $G$.
\end{defbox}

Related is the concept of \emph{cosets}:

\begin{defbox}[label={def:coset-quotient}]{Coset}
    Let $G$ be a group and $H$ a subgroup of $G$. For any element $g \in G$:
    \begin{enumerate}
        \item The \emph{left coset} of $H$ containing $g$ is defined as $gH = \{gh : h \in H\}$.
        \item The \emph{right coset} of $H$ containing $g$ is defined as $Hg = \{hg : h \in H\}$.
    \end{enumerate}
 \end{defbox}

We can now define normal subgroups and quotient groups:

\begin{defbox}[label={def:normal-subgroup}]{Normal subgroup and quotient group}
    A subgroup $N$ of a group $G$ is called \emph{normal} if $gN = Ng$ for all $g \in G$. Equivalently, $gNg^{-1} = N$ for all $g \in G$. If $H$ is a normal subgroup of $G$, then the set of all left cosets
    \begin{equation}\label{eq:quotient}
       G/H = \{gH : g \in G\} 
    \end{equation}
    forms a group under the operation $(g_1H) \cdot (g_2H) = (g_1g_2)H$, called the \emph{quotient group} of $G$ by $H$.
\end{defbox}

If we define an equivalence relation on $G$ by declaring that $g\sim g'$ if and only if $g^{-1}g' \in H$, then the corresponding equivalence class is
\begin{equation}
    [g] = \{x \in G \mid g^{-1}x \in H \}.
\end{equation}
Writing $g^{-1}x=h$ for some $h\in H$ gives $x=gh$, so
\begin{equation}
    [g]=\{gh\mid h\in H\}=gH.
\end{equation}
Thus, the cosets of a subgroup are exactly the equivalence classes induced by that subgroup, and they partition the group into disjoint classes of equal size. In the next section, this viewpoint will help us classify logical operators. Before that, we introduce two related concepts: the normalizer and the centralizer.
\begin{defbox}[label={def:normalizer-centralizer}]{Normalizer and centralizer}
    Let $G$ be a group and $H$ a subgroup of $G$. The \emph{centralizer} of $H$ in $G$ is
    \begin{equation}\label{eqn:centralizer}
        C_G(H)=\{g\in G:gh=hg \text{ }\forall h\in H\}.
    \end{equation}
    It contains all elements of $G$ that commute with every element of $H$. The \emph{normalizer} of $H$ in $G$ is
    \begin{equation}
        N_G(H)=\{g\in G:gHg^{-1}=H\}.
    \end{equation}
\end{defbox}
In general, the centralizer is contained in the normalizer: commuting with every element of $H$ is a stronger condition than merely preserving $H$ under conjugation. Besides studying the internal structure of a group, we will also need maps between groups that preserve the group operation. These are called \emph{homomorphisms}:

\begin{defbox}[label={def:group-homo-isomorphsim}]{Group homomorphism and isomorphism}
    Let $(G, *)$ and $(H, \cdot)$ be groups. A map $\phi: G \to H$ is a \emph{group homomorphism} if for all $g_1, g_2 \in G$, it satisfies the property:
    \begin{equation}
        \phi(g_1* g_2) = \phi(g_1) \cdot \phi(g_2) 
    \end{equation}
    A homomorphism $\phi$ that is also bijective (both one-to-one and onto) is called a \emph{group isomorphism}. If such a map exists between two groups $G$ and $H$, they are said to be isomorphic, denoted $G \cong H$.
\end{defbox}

We will also find the concept of a \emph{kernel} useful:

\begin{defbox}[label={def:kernel}]{Kernel}
    Let $\phi: G \to H$ be a group homomorphism. The \emph{kernel} of $\phi$, defined as $\ker(\phi) = \{g \in G \mid \phi(g) = e_H\}$, where $e_H$ is the identity element in $H$, is a normal subgroup of $G$.
\end{defbox}

In words, the kernel consists of those elements of $G$ that vanish under the map $\phi$, in the sense that they are sent to the identity in $H$. Since homomorphisms respect the group operation, conjugating such an element yields a pair of inverse elements multiplied together, which still yields the identity. So the kernel is closed under conjugation and is therefore normal.

\begin{thmbox}[label={thm:isomorphism-thm}]{First isomorphism theorem}
    Let $\phi: G \to H$ be a group homomorphism. Then the quotient group $G/\ker(\phi)$ is isomorphic to the image of $\phi$, which is a subgroup of $H$. This is expressed as:
    \begin{equation}
        G/\ker(\phi) \cong \operatorname{im}(\phi)
    \end{equation}
\end{thmbox}

For complete proofs of~\cref{def:kernel} and~\cref{thm:isomorphism-thm}, see~\cite{pinter_book_1982}. Here we give a simple example motivated by QEC. Suppose we measure the two-qubit parity operator $ZZ$. On computational-basis states, this measurement does not tell us the state of the qubits; it only tells us whether the parity is even or odd. In particular, any superposition of $\ket{00}$ and $\ket{11}$ produce the same measurement outcome, while that of $\ket{01}$ and $\ket{10}$ produce the other outcome. We can model this using a group homomorphism. Let
\begin{equation}
    G = \{00,01,10,11\}
\end{equation}
with group operation given by bitwise \XOR, and let
\begin{equation}
    H = \{0,1\}
\end{equation}
with addition modulo 2. Define the map $\phi:G\to H$ by
\begin{equation}
    \phi(00)=\phi(11)=0,\qquad \phi(01)=\phi(10)=1\,.
\end{equation}
This map records exactly the outcome of the $ZZ$ measurement: 0 for even parity and 1 for odd parity. The kernel is
\begin{equation}
    \ker(\phi)=\{00,11\}\,,
\end{equation}
which is the set of strings with even parity. The other coset is
\begin{equation}
    01+\ker(\phi)=\{01,10\}\,,
\end{equation}
which is the set of strings with odd parity. Therefore,
\begin{equation}
    G/\ker(\phi) = \left\{\{00,11\},\{01,10\}\right\} = \{[00], [01]\}\,,
\end{equation}
so the quotient group is exactly the set of parity classes.

The induced map $\overline{\phi}: G/\ker(\phi)\to \operatorname{im}(\phi)$ defined by $\overline{\phi}([g])=\phi(g)$ is bijective, illustrating the first isomorphism theorem in this setting. In the language of quantum error correction, the theorem formalizes the idea that a syndrome measurement groups physical configurations into equivalence classes that produce the same measurement outcome. The syndrome does not identify the exact error, only its class, which is why decoding generally involves choosing a representative correction. This is closely related to the degeneracy of quantum codes. To save resources, we sometimes choose not to decode exactly but choose the most likely logical class and apply a representative correction. More details on decoding will be covered in~\cref{sec:decoders}.

\subsection{Stabilizer codes}\label{subsec:stabilizer-stabilizer-codes}
We are now ready to introduce stabilizer codes and the stabilizer formalism~\cite{gottesman1997stabilizer,bradshaw2025introduction}. Many of the most important QEC codes can be described this way, including the Shor code, the Steane code, the five-qubit code, CSS codes, and surface code-type constructions. At its heart, the stabilizer formalism describes a code by the Pauli operators that leave every code state unchanged. We formalize this as follows:

\begin{defbox}[label={def:stabilizer-code}]{Stabilizer code}
    An $[n,k]$ \emph{stabilizer code} is a quantum error-correcting code $C \subseteq \mathcal{H}_{2^n}$ that encodes $k$ logical qubits into $n$ physical qubits, where the codespace is of dimension $2^k$ and is defined by the +1 eigenspace of the \emph{stabilizer group} $\mathcal{S}$, which is an abelian subgroup of the $n$-qubit Pauli group $\mathcal{P}_n$ (excluding $-I^{\otimes n}$). That is,
    \begin{equation}
        \mathcal{S}(C) = \{P \in \mathcal{P}_n : P\ket{\psi} = \ket{\psi}, \forall\ket{\psi} \in C\}\,.
    \end{equation}
\end{defbox}

Thus, the stabilizer group is the set of Pauli operators that leave every code state invariant. We refer to these Pauli operators as \emph{stabilizer operators}, \emph{stabilizer elements}, or simply \emph{stabilizers}. Since the codespace is a simultaneous eigenspace of all stabilizers, the stabilizer group must be abelian. As with any group, we can describe it in terms of generators. The following fact is especially useful:

\begin{factbox}[label={fact:abelian-group-generators}]{Abelian group generators}
    Let $G$ be an abelian group of self-inverse operators generated by $m$ generators $g_1,\ldots,g_m$. Then any element $g \in G$ can be written as
    \begin{equation}
        g = \prod_{i=1}^m g_i^{\alpha_i}\,,
    \end{equation}
    where $\alpha_i \in \{0,1\}$.
\end{factbox}

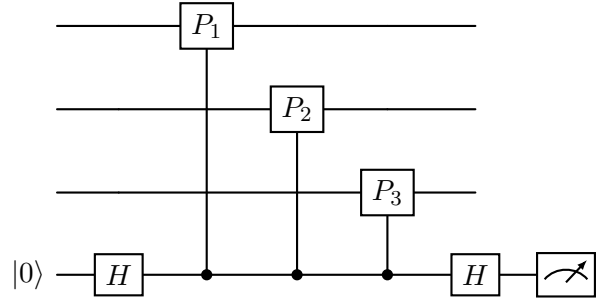
\begin{figure}
    \centering
    \begin{quantikz}
        & & \gate{P_1} & & & \\
        & & & \gate{P_2} & & \\
        & & & & \gate{P_3} & \\
        \lstick{$\ket{0}$} & \gate{H} & \ctrl{-3} & \ctrl{-2} & \ctrl{-1} & \gate{H} & \meter{}
    \end{quantikz}
    \caption{Example circuit for measuring the stabilizer $P_1 \otimes P_2 \otimes P_3$.}
    \label{fig:syndrome-measurement}
\end{figure}

This fact applies directly to stabilizer codes because stabilizer generators are Pauli operators, which are self-inverse up to a phase. Since stabilizer groups exclude $-I^{\otimes n}$, every stabilizer can be written as a product of generators with exponents in $\{0,1\}$, so $\abs{\mathcal{S}} = 2^m$. Moreover, if a stabilizer code on $n$ physical qubits has $m$ independent generators, then each generator halves the Hilbert-space dimension, so
\begin{equation}
    \frac{2^n}{2^m} = 2^k \implies k=n-m\,,
\end{equation}
directly giving us the number of logical qubits from the knowledge of the size of the stabilizer group. Let us now apply the stabilizer formalism to the three-qubit repetition code. Recall, the codewords are
\begin{equation}
    \ket{\overline{0}} = \ket{000} \text{ and } \ket{\overline{1}} = \ket{111}\,.
\end{equation}
This is a $\stabcode{3}{1}{1}$ code, so $n=3$, $k=1$, and hence $m=n-k=2$. Therefore, the stabilizer group has $\abs{\mathcal{S}}=2^2=4$ elements, namely
\begin{equation}
    \mathcal{S}=\{III,ZIZ,IZZ,ZZI\}.
\end{equation}
Once we know the stabilizers, we can measure them to detect errors and extract a syndrome. An example circuit for measuring the stabilizer $P_1 \otimes P_2 \otimes P_3$ is shown in~\cref{fig:syndrome-measurement}.

The stabilizer formalism also organizes the logical operators. These are the operators that act on the encoded states in the same way that ordinary Pauli operators act on an unencoded qubit. For the three-qubit repetition code, a logical bit flip and a logical phase flip act as
\begin{align}
    \overline{X}\ket{\overline{0}} &= \ket{\overline{1}}, \quad \overline{X}\ket{\overline{1}} = \ket{\overline{0}}\\
    \overline{Z}\ket{\overline{0}} &= \ket{\overline{0}}, \quad \overline{Z}\ket{\overline{1}} = -\ket{\overline{1}}\,.
\end{align}
One choice of logical operators is $\{XXX,ZII\}$. However, logical operators are not unique: multiplying a logical operator by any stabilizer gives another operator with the same action on the codespace. For example, multiplying $ZII$ by the stabilizer $ZIZ$ gives $IIZ$, which is also a valid logical $Z$. More generally, if $\ket{\psi}$ lies in the codespace and $S_j\in\mathcal{S}$, then $S_j\ket{\psi}=\ket{\psi}$, so
\begin{equation}
    LS_j\ket{\psi}=L\ket{\psi}.
\end{equation}
Thus, every logical operator $L$ comes with an entire equivalence class $\{LS_j : S_j\in\mathcal{S}\}$ of physically equivalent representatives.

Finally, we can describe logical operators and code distance in group-theoretic terms. The logical Pauli operators are the Pauli operators that preserve the codespace but are not themselves stabilizers, so they form the set $N(\mathcal{S}) \setminus \mathcal{S}$, where $A \setminus B$ denotes the set of elements of $A$ that are not in $B$. For Pauli stabilizer groups, this is the same as $C(\mathcal{S}) \setminus \mathcal{S}$. The reason is that Pauli conjugation can only introduce a sign, and a stabilizer group never contains $-I$; thus a Pauli operator cannot normalize $\mathcal{S}$ without actually commuting with every stabilizer. As a result, the centralizer and the normalizer of any stabilizer group coincide. The distance of a stabilizer code is therefore the minimum weight of any element of $N(\mathcal{S}) \setminus \mathcal{S}$. For the three-qubit repetition code, this minimum is 1, since $ZII$, $IZI$, and $IIZ$ are all weight-one logical operators.

\subsection{Five-qubit perfect code}\label{subsec:stabilizer-five-qubit-perfect-code} 
To see another example of the stabilizer formalism, we now consider the \eczoo[five-qubit code]{stab_5_1_3}, the smallest code that can correct an arbitrary single-qubit error. It does so using four fewer qubits than the nine-qubit Shor code from~\cref{subsec:fundamentals-nine-qubit-shor-code}. This code, denoted as a $\stabcode{5}{1}{3}$ code~\cite{laflamme1996perfect}, is especially important because it is a \emph{perfect code}, a notion we discuss at the end of this section.

The five-qubit code is elegantly described in the stabilizer language. Since it encodes one logical qubit into five physical qubits, the stabilizers reduce the physical Hilbert-space dimension from $2^5=32$ to a two-dimensional codespace. Using the relation from the previous section, we need $5-1=4$ independent stabilizer generators, which can be written as
\begin{align}\label{eqn:five-qubit-stabilizers}
    g_1 &= XZZXI, \\
    g_2 &= IXZZX, \\
    g_3 &= XIXZZ, \\
    g_4 &= ZXIXZ\,.
\end{align}
Any valid logical state $\ket{\overline{\psi}}$ in the codespace must satisfy $g_i\ket{\overline{\psi}} = \ket{\overline{\psi}}\ \forall i \in \{1, 2, 3, 4\}$. It can be easily verified that all these generators commute with each other, which is also a necessary condition.

\begin{exerbox}[label={exer:five-qubit-mutual-commutation}]{Mutual commutation of five-qubit code stabilizers}
    Verify that $g_1$, $g_2$, $g_3$, and $g_4$ all mutually commute.
\end{exerbox}

To manipulate the encoded qubit, we need logical Pauli operators $\overline{X}$ and $\overline{Z}$. These operators must commute with all stabilizer generators but must not themselves be stabilizers. Additionally, $\overline{X}$ and $\overline{Z}$ must anticommute with each other, just like normal Pauli operators. In group-theoretic language, logical operators are represented by cosets of $\mathcal{S}$ inside its normalizer. For the five-qubit code, one convenient choice is
\begin{align}\label{eq:5qubit_logical_ops}
    \overline{Z} &= ZZZZZ, \\
    \overline{X} &= XXXXX.
\end{align}

\begin{exerbox}[label={exer:five-qubit-logical-commutation}]{$\stabcode{5}{1}{3}$ logical Pauli operators}
    Verify that $\overline{X}$ and $\overline{Z}$ of the five-qubit code commute with all stabilizer generators $\{g_1,g_2,g_3,g_4\}$ and anticommute with each other.
\end{exerbox}
The power of the stabilizer formalism lies in the detection of errors. Suppose a single-qubit error $E$ occurs on the state $\ket{\overline{\psi}}$, transforming it to $E\ket{\overline{\psi}}$. To detect this error, we measure the eigenvalues of the four stabilizer generators. If a generator $g_i$ commutes with the error $E$, then
\begin{equation}\label{eq:stabilizer_commutes}
    g_i E\ket{\overline{\psi}} = E g_i\ket{\overline{\psi}} = E\ket{\overline{\psi}}\,.
\end{equation}
If $g_i$ anticommutes with $E$, then
\begin{equation}\label{eq:stabilizer_anticommutes}
    g_i E\ket{\overline{\psi}} = -E g_i\ket{\overline{\psi}} = -E\ket{\overline{\psi}}\,.
\end{equation}
Thus, by measuring the eigenvalues of all four generators, we can determine the error syndrome. Each unique single-qubit error ($X_i$, $Y_i$, and $Z_i$ for $i = 1, \ldots, 5$) produces a unique, non-trivial syndrome. \cref{tab:five-qubit-syndrome} lists the syndromes for all 15 possible single-qubit Pauli errors, where $+1$ indicates commutation and $-1$ indicates anticommutation. Once a syndrome is measured, we simply look it up in the table to identify the error that occurred. For example, if we measure the syndrome $(+1, -1, +1, -1)$, we know that a $Z_2$ error has occurred. To correct it, we simply apply the same operator $Z_2$ again to correct the error.

\begin{table}[tb]
    \centering
    \begin{tblr}{colspec={Q[c,m]Q[c,m]Q[c,m]Q[c,m]Q[c,m]},hline{2}={1pt},vlines,row{2,3,4,8,9,10,14,15,16}={bg=\colortheme!30},row{1}={font=\bfseries}}
        \hline
        Error & $g_1$ & $g_2$ & $g_3$ & $g_4$ \\
        $X_1$ & $+1$ & $+1$ & $+1$ & $-1$ \\
        $Z_1$ & $-1$ & $+1$ & $-1$ & $+1$ \\
        $Y_1$ & $-1$ & $+1$ & $-1$ & $-1$ \\
        \hline
        $X_2$ & $-1$ & $+1$ & $+1$ & $+1$ \\
        $Z_2$ & $+1$ & $-1$ & $+1$ & $-1$ \\
        $Y_2$ & $-1$ & $-1$ & $+1$ & $-1$ \\
        \hline
        $X_3$ & $-1$ & $-1$ & $+1$ & $+1$ \\
        $Z_3$ & $+1$ & $+1$ & $-1$ & $+1$ \\
        $Y_3$ & $-1$ & $-1$ & $-1$ & $+1$ \\
        \hline
        $X_4$ & $+1$ & $-1$ & $-1$ & $+1$ \\
        $Z_4$ & $-1$ & $+1$ & $+1$ & $-1$ \\
        $Y_4$ & $-1$ & $-1$ & $-1$ & $-1$ \\
        \hline
        $X_5$ & $+1$ & $+1$ & $-1$ & $-1$ \\
        $Z_5$ & $+1$ & $-1$ & $-1$ & $+1$ \\
        $Y_5$ & $+1$ & $-1$ & $-1$ & $-1$ \\
        \hline
    \end{tblr}
    \caption{The 15 error syndromes for the $\stabcode{5}{1}{3}$ code. Each row shows the resulting eigenvalues from measuring the four stabilizer generators against a specific single-qubit Pauli error.}
    \label{tab:five-qubit-syndrome}
\end{table}

Finally, we return to the statement that the five-qubit code is \emph{perfect}. As \cref{tab:five-qubit-syndrome} shows, each of the 15 nontrivial single-qubit Pauli errors gives a unique syndrome. There is therefore a one-to-one correspondence between correctable errors and nontrivial syndromes: no syndrome is wasted, and every correctable error is uniquely identified. By contrast, the nine-qubit Shor code has eight stabilizer generators and therefore $2^8-1=255$ nontrivial syndromes available, far more than the 27 nontrivial single-qubit Pauli errors it is designed to correct. Thus the five-qubit code uses its syndrome space optimally. This efficient use of syndrome space is rare and occurs only for certain codes. Equivalently, the five-qubit code saturates the \emph{quantum Hamming bound}, which we discuss later in the tutorial (see~\cref{subsec:bounds-quantum-hamming-bound}).

\subsection{CSS codes}\label{subsec:stabilizer-css-codes}
We conclude this chapter with an introduction to an important family of stabilizer codes known as \emph{CSS codes}, named after their discoverers, Robert Calderbank, Peter Shor, and Andrew Steane. Common examples include the Steane code and the Shor code, and the toric code, which we will introduce in the next chapter.

A CSS code is a quantum stabilizer code built from a pair of classical linear codes. One classical code supplies the $Z$-type stabilizer generators, which are used to detect bit-flip errors, while the other supplies the $X$-type stabilizer generators, which are used to detect phase-flip errors. This separation is what makes CSS codes especially transparent and useful. More concretely, suppose we start with two binary linear codes of length $n$, with parity-check matrices $H_Z$ and $H_X$. Each row of $H_Z$ is interpreted as the support of a $Z$-type stabilizer generator: wherever the row has a 1, we place a $Z$, and wherever it has a 0, we place an $I$. Similarly, each row of $H_X$ defines an $X$-type stabilizer generator. For example, the row $(1,0,1,1)$ of $H_Z$ would define the stabilizer $ZIZZ$, while the same row in $H_X$ would define $XIXX$.

These generators define a valid stabilizer code only if every $X$-type generator commutes with every $Z$-type generator. Since single-qubit Pauli operators $X$ and $Z$ anticommute on the same qubit and commute on different qubits, this commutation condition is equivalent to requiring that each row of $H_X$ have even overlap with each row of $H_Z$. In matrix form, this is the condition
\begin{equation}
    H_X H_Z^\intercal = 0
\end{equation}
over $\FF_2$. When this condition holds, the $X$-type and $Z$-type parity checks together define a CSS stabilizer group.

The parameters of the resulting CSS code can be read off from the underlying classical codes. If the two classical codes have parameters $[n,k_X,d_X]$ and $[n,k_Z,d_Z]$, then the number of encoded qubits is
\begin{equation}\label{eq:css-logical-dimension}
    k = k_X + k_Z - n.
\end{equation}

\begin{exerbox}[label={exer:css-parameters}]{Parameters of a CSS code}
    Prove the relation in~\cref{eq:css-logical-dimension}.

    [Hint: think of the number of $X$- and $Z$-type stabilizer generators.]
\end{exerbox}

The distances of the underlying classical codes also influence the error-correcting properties of the CSS code, but the relation is subtler than the formula for $k$. Roughly speaking, one classical code governs protection against bit-flip ($X$-type) errors, while the other governs protection against phase-flip ($Z$-type) errors, so a CSS code can have different $X$- and $Z$-type logical distances. However, the true quantum distances are not obtained simply by taking the classical distances of the seed codes. Classical codewords of any one kind correspond only to stabilizers and therefore do not provide information about nontrivial logical operators.

More precisely, if the two classical codes are $C_X$ and $C_Z$, then the relevant logical distances are
\begin{align}
    d_X^{(\mathrm{CSS})} &= \min\{\wt(c) : c \in C_Z \setminus C_X^\perp\},\\
    d_Z^{(\mathrm{CSS})} &= \min\{\wt(c) : c \in C_X \setminus C_Z^\perp\},
\end{align}
where $\wt(c)$ denotes the Hamming weight of $c$. The overall CSS code distance is then
\begin{equation}
    d=\min\{d_X^{(\mathrm{CSS})},d_Z^{(\mathrm{CSS})}\}\geq \min\{d_X,d_Z\} .
\end{equation}
Thus, the classical distances $d_X$ and $d_Z$ provide useful intuition and a lower bound, but the actual quantum distance is determined by the minimum weight of a nontrivial logical operator rather than by the minimum weight of a nonzero classical codeword.

\begin{example}[label={ex:css-steane-code}]{Seven-qubit Steane code}
    An example of a CSS code is the seven-qubit \eczoo[Steane code]{steane}, which is constructed by taking two copies of the $[7,4,3]$ Hamming code, whose parity-check matrix is
    \begin{equation}
        H =
        \begin{pmatrix}
            0 & 0 & 0 & 1 & 1 & 1 & 1\\
            0 & 1 & 1 & 0 & 0 & 1 & 1\\
            1 & 0 & 1 & 0 & 1 & 0 & 1
        \end{pmatrix}\,.
    \end{equation}
    Thus, we take $H_X = H_Z = H$. Because the Hamming code is self-orthogonal in the required sense, the condition $H_XH_Z^\intercal=0$ is satisfied, so the corresponding $X$- and $Z$-type stabilizers commute. The resulting CSS code has parameters $\stabcode{7}{1}{3}$, where the number of physical qubits is inherited from the block length of the Hamming code. The number of encoded logical qubits is
    \begin{equation}
        k = k_X + k_Z - n = 4 + 4 - 7 = 1\,.
    \end{equation}
    In this example, the smallest nontrivial $X$- and $Z$-type logical operators both have weight 3, so
    \begin{equation}
        d=\min\{d_X^{(\mathrm{CSS})},d_Z^{(\mathrm{CSS})}\}=3\,.
    \end{equation}
    Thus, the Steane code is a $\stabcode{7}{1}{3}$ code.
\end{example}

One of the main advantages of CSS codes is therefore both conceptual and practical: they reduce much of the quantum construction to familiar classical coding data while still producing genuinely quantum codes. This makes them one of the most useful bridges between classical coding theory and quantum error correction, and it is one reason they appear so often in both theory and practice.

\newpage

\section{Topological codes}\label{sec:topological}

In the first four chapters of this tutorial, we gave a general overview of quantum error correction. We now dive deeper into specific code constructions, starting with \emph{topological codes}. Important examples include the toric and planar/surface codes~\cite{dennis2002topological,bravyi2005universal,fowler2012surface}, color codes~\cite{bombin2006topological,landahl2011fault}, topological subsystem codes~\cite{poulin2005stabilizer,bombin2010topological,bombin_gauge_2015}, and hyperbolic surface codes~\cite{Breuckmann_2016,higgott2024constructions}.

\subsection{Primer on topology}\label{subsec:topological-primer-topology}
We begin with an introduction to some concepts from topology and homology~\cite{munkres2000topology} that we will need in this chapter and several other chapters throughout the tutorial.

\subsubsection{Introduction to topology}\label{subsubsec:topological-overview-topology}
Topology is a branch of mathematics dealing with the preserved properties of an object undergoing continuous transformations, such as stretching, bending, or twisting the material itself. For example, a solid square can be deformed into a solid rectangle by using such transformations. Thus, we say that these shapes are \emph{topologically equivalent}. Other examples of topologically equivalent objects include a circle and an ellipse, as well as a sphere and an ellipsoid. However, a sphere cannot be continuously deformed into a torus; such a transition would require puncturing and gluing the surface to create a new hole, which is not allowed. As such, we say that a sphere and a torus are \emph{topologically inequivalent}. Some other questions that arise in topology include asking how many holes an object has and what the boundary of the object is, which are answered by \emph{homology}, which we explore in~\cref{subsubsec:topological-homology}.

\begin{figure}[b]
    \centering
    \includegraphics[width=0.5\textwidth]{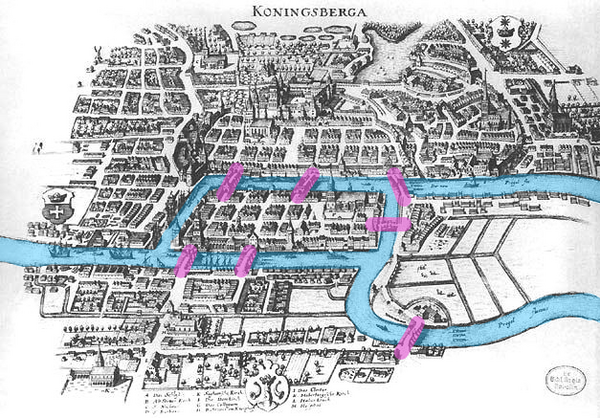}
    \caption{Map of the seven bridges of K\"{o}nigsberg. Photo credit: \href{https://en.wikipedia.org/wiki/Seven_Bridges_of_Königsberg}{Wikipedia}.}
    \label{fig:seven-bridges}
\end{figure}

A famous early problem in topology, which also led to the branch of mathematics called \emph{graph theory}, is the ``Seven Bridges of K\"{o}nigsberg,'' which was proposed by Euler. Given a map of the then-Prussian city of K\"{o}nigsberg (now Kaliningrad in Russia) in~\cref{fig:seven-bridges}, we can see that there are seven bridges connecting the various land masses comprising the city. Euler asked an interesting question: is it possible to traverse a path that crosses each bridge once and only once? Try it for yourself and you should find that there is no such path! Euler abstracted the problem and modeled it as a graph and determined that a path that crosses each bridge exactly once is only possible if there is a maximum of two vertices (the land masses) of the graph that are of odd degree. However, for the Seven Bridges of K\"{o}nigsberg problem, all four of the vertices are of odd degree, and there are four vertices, so such a path is not possible.

From this example, we can see that abstracting the problem and dealing only with the vertices and edges of a graph was enough to arrive at a useful result. Other details like the lengths of the bridges, the total length of the path, the distance between each bridge, etc. were not needed to answer the original question posed of whether it is possible to take a path that crosses each bridge exactly once. All we needed was the set of four vertices and seven edges, demonstrating an application of topology.

Another early result (also from Euler) in topology has to do with the relation of components of different topological dimension in a geometric object. Given the two-dimensional surface of a geometric shape like a polyhedron (treating it as a hollow shell with flat surfaces), we get several components of varying \emph{dimension}. Thus, a 0-dimensional object is a vertex, a 1-dimensional object is an edge, a 2-dimensional object is a face or plaquette. Euler found that the quantity
\begin{equation}\label{eqn:euler_torus}
    \chi = V - E + F \,,
\end{equation}
which is now called the \emph{Euler characteristic}, is invariant for a given topological surface. Here, $V$ is the number of vertices in the object, $E$ is the number of edges, and $F$ is the number of faces. For example, for convex polyhedra (tetrahedra, cubes, icosahedra, dodecahedra, etc.), $\chi = 2$. Since this quantity does not change (up to a topological equivalence which we will introduce shortly), we call it a \emph{topological invariant}.

The canonical example of topology is, of course, the idea that a coffee mug and a donut are topologically equivalent. That is, one can be deformed into the other without making any cuts or tears. We mention this example to introduce the concept of a \emph{homeomorphism}, which is exactly what the transformation from donut to coffee mug or \emph{vice versa} is. Informally, we can think of a homeomorphism as an intrinsic deformation of one object or space into another without ``breaking'' or ``gluing'' anything. A homeomorphism is the standard function used to define this \emph{topological equivalence}. More formally, a homeomorphism is a function between two objects or topological spaces that
\begin{enumerate}
    \item is continuous,
    \item is a bijection (i.e., it is one-to-one and onto and therefore invertible), and
    \item has a continuous inverse function.
\end{enumerate}
Thus, when we say that two objects are topologically equivalent, we mean that they are equivalent \emph{up to a homeomorphism}. Related are the topological invariants that we mentioned above, where we saw that the Euler characteristic was one such example. Now that we understand what a homeomorphism is, we can more precisely define a topological invariant as a property or numerical quantity that remains unchanged under homeomorphism. For convex polyhedra, we can stretch and deform them into other convex polyhedra (e.g., a tetrahedron to a cube), but the Euler characteristic remains 2 because it is invariant under the underlying homeomorphism.

\begin{figure}[tb]
    \centering
    \includegraphics[width=0.3\textwidth]{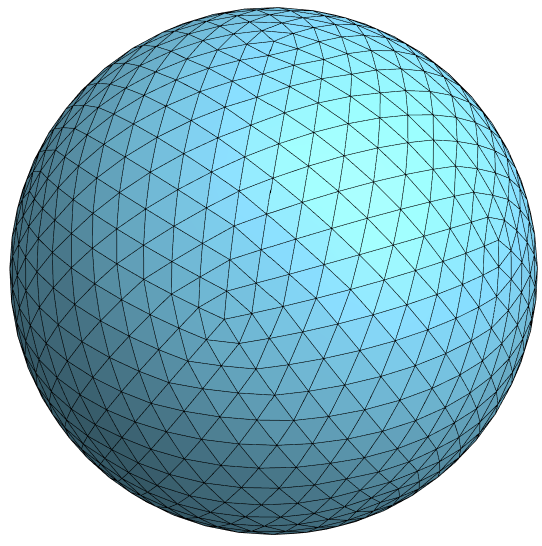}
    \caption{Triangulated sphere.}
    \label{fig:sphere-triangulation}
\end{figure}

We can generalize the Euler characteristic to distinguish general spaces via a process called \emph{cellulation}, which partitions a space into a collection of constituent cells. For a two-dimensional surface, this amounts to decomposing the space into polygonal cells. One example is the triangulation of the surface of a sphere, as in~\cref{fig:sphere-triangulation}. If we count the number of vertices, edges, and faces of our triangulated sphere, we should find that the Euler characteristic is $\chi_{\text{sphere}} = 2$. On the other hand, if we consider a torus using a square $L \times L$ cellulation and periodic boundary conditions, as we will see when we study the toric code in~\cref{subsec:toric-code}, we will find that the number of vertices is $L^2$, the number of edges is $2L^2$, and the number of faces is $L^2$, where $L$ is a parameter we can choose. This gives us an Euler characteristic of $\chi_{\text{torus}} = L^2 - 2L^2 + L^2 = 0$, which is not equal to $\chi_{\text{sphere}}$. Thus, we can conclude that a sphere and a torus are \emph{topologically inequivalent}.

Finally, we highlight an additional property that differentiates topological objects: the number of \emph{holes} an object has, which we characterize by its \emph{genus}. In the sphere and torus example from above, we can see that a sphere has 0 holes and a torus has 1 hole. Thus, we say that a sphere is a genus-0 object while a torus is a genus-1 object. One important result that we will find helpful when studying topological quantum error-correcting codes is that the Euler characteristic $\chi$ of any \emph{closed, orientable} surface of genus $g$ is $\chi=2-2g$. We will explore this idea further when we use it in our development of topological codes.

\subsubsection{Introduction to homology}\label{subsubsec:topological-homology}
We now move onto a branch of topology called \emph{homology}, which abstracts and generalizes the notion of a \emph{boundary}. In this section, we will focus on concrete aspects of homology that are relevant to qubit topological codes, which means we will restrict ourselves to $\ZZ_2$ homology and cohomology over square lattice cellulations. Homology is a rich field with extensions to other groups, lattice layouts, etc., but for our purposes we will only need to consider these regimes.

The most basic object in homology is an \emph{$n$-cell}, which is an $n$-dimensional object in a cellulation. For example, a 0-cell is a vertex, a 1-cell is an edge, and a 2-cell is a face (or, more frequently in QEC, a \emph{plaquette}). Given a lattice or graph with a set $K_n$ of $n$-cells, we define an \emph{$n$-chain} to be a formal $\ZZ_2$-linear combination of these $n$-cells:
\begin{equation}
    c = \sum_{\sigma \in K_n} c_\sigma \sigma\,.
\end{equation}
Since we work throughout with $\ZZ_2$ homology, each coefficient $c_\sigma$ is either 0 or 1. 
Equivalently, an $n$-chain $c$ can be viewed as an assignment of 0 or 1 to each $n$-cell: a 0-chain assigns 0 or 1 to each vertex, a 1-chain assigns 0 or 1 to each edge, and so on. Intuitively, an $n$-chain $c$ simply specifies a subset of $n$-cells: $\{\sigma : c_\sigma\neq0\}$.

Choosing $\ZZ_2$ coefficients introduces a massive geometric simplification: cell orientations can be completely ignored. In general homology theories (such as over $\ZZ$ or $\ZZ_3$), each cell must be assigned a choice of direction or orientation, and reversing this orientation corresponds to an algebraic sign change of $-1$. However, because $-1 \equiv 1 \pmod 2$, the distinction between opposite orientations vanishes entirely, allowing us to compute boundaries and sum chains without tracking directed arrows on the lattice cells.

Given two $n$-chains, we add them by adding their coefficients cell by cell modulo 2. For example, if a square plaquette has boundary edges $e_1,e_2,e_3,e_4$ (1-cells), then the 1-chain $c''$ consisting of those four edges is
\begin{equation}
c'' =
\vcenter{\hbox{\begin{tikzpicture}
    % Square
    \draw[thick, dashed] (0,0) -- (1,0) -- (1,1) -- (0,1) -- cycle;
    % Red left side
    \draw[red, very thick] (0,0) -- (0,1);
    \draw[red, very thick] (0,0) -- (1,0);
    \draw[red, very thick] (1,0) -- (1,1);
    \draw[red, very thick] (0,1) -- (1,1);
    % Filled circles at corners
    \fill (0,0) circle (3pt);
    \fill (1,0) circle (3pt);
    \fill (1,1) circle (3pt);
    \fill (0,1) circle (3pt);
\end{tikzpicture}}} =
\vcenter{\hbox{\begin{tikzpicture}
    % Square
    \draw[thick, dashed] (0,0) -- (1,0) -- (1,1) -- (0,1) -- cycle;
    % Red left side
    \draw[red, very thick] (0,0) -- (0,1);
    % Label for the red edge
    \node[right, red] at (0,0.5) {$e_1$};
    % Filled circles at corners
    \fill (0,0) circle (3pt);
    \fill (1,0) circle (3pt);
    \fill (1,1) circle (3pt);
    \fill (0,1) circle (3pt);
\end{tikzpicture}}} +
\vcenter{\hbox{\begin{tikzpicture}
    % Square
    \draw[thick, dashed] (0,0) -- (1,0) -- (1,1) -- (0,1) -- cycle;
    % Red left side
    \draw[red, very thick] (0,0) -- (1,0);
    % Label for the red edge
    \node[above, red] at (0.5,0) {$e_2$};
    % Filled circles at corners
    \fill (0,0) circle (3pt);
    \fill (1,0) circle (3pt);
    \fill (1,1) circle (3pt);
    \fill (0,1) circle (3pt);
\end{tikzpicture}}} +
\vcenter{\hbox{\begin{tikzpicture}
    % Square
    \draw[thick, dashed] (0,0) -- (1,0) -- (1,1) -- (0,1) -- cycle;
    % Red left side
    \draw[red, very thick] (1,0) -- (1,1);
    % Label for the red edge
    \node[left, red] at (1,0.5) {$e_3$};
    % Filled circles at corners
    \fill (0,0) circle (3pt);
    \fill (1,0) circle (3pt);
    \fill (1,1) circle (3pt);
    \fill (0,1) circle (3pt);
\end{tikzpicture}}} +
\vcenter{\hbox{\begin{tikzpicture}
    % Square
    \draw[thick, dashed] (0,0) -- (1,0) -- (1,1) -- (0,1) -- cycle;
    % Red left side
    \draw[red, very thick] (0,1) -- (1,1);
    % Label for the red edge
    \node[below, red] at (0.5,1) {$e_4$};
    % Filled circles at corners
    \fill (0,0) circle (3pt);
    \fill (1,0) circle (3pt);
    \fill (1,1) circle (3pt);
    \fill (0,1) circle (3pt);
\end{tikzpicture}}} 
=
e_1 + e_2 + e_3 + e_4.
\end{equation}
As another example, for two 1-chains $c$ and $c'$ given by
\begin{equation}
c =
\vcenter{\hbox{\begin{tikzpicture}
    % Square
    \draw[thick, dashed] (0,0) -- (1,0) -- (1,1) -- (0,1) -- cycle;
    % Red left side
    \draw[red, very thick] (0,0) -- (0,1);
    \draw[red, very thick] (0,0) -- (1,0);
    % Filled circles at corners
    \fill (0,0) circle (3pt);
    \fill (1,0) circle (3pt);
    \fill (1,1) circle (3pt);
    \fill (0,1) circle (3pt);
\end{tikzpicture}}} =
\vcenter{\hbox{\begin{tikzpicture}
    % Square
    \draw[thick, dashed] (0,0) -- (1,0) -- (1,1) -- (0,1) -- cycle;
    % Red left side
    \draw[red, very thick] (0,0) -- (0,1);
    % Label for the red edge
    \node[right, red] at (0,0.5) {$e_1$};
    % Filled circles at corners
    \fill (0,0) circle (3pt);
    \fill (1,0) circle (3pt);
    \fill (1,1) circle (3pt);
    \fill (0,1) circle (3pt);
\end{tikzpicture}}} +
\vcenter{\hbox{\begin{tikzpicture}
    % Square
    \draw[thick, dashed] (0,0) -- (1,0) -- (1,1) -- (0,1) -- cycle;
    % Red left side
    \draw[red, very thick] (0,0) -- (1,0);
    % Label for the red edge
    \node[above, red] at (0.5,0) {$e_2$};
    % Filled circles at corners
    \fill (0,0) circle (3pt);
    \fill (1,0) circle (3pt);
    \fill (1,1) circle (3pt);
    \fill (0,1) circle (3pt);
\end{tikzpicture}}}
= e_1 + e_2
, \qquad c' =
\vcenter{\hbox{\begin{tikzpicture}
    % Square
    \draw[thick, dashed] (0,0) -- (1,0) -- (1,1) -- (0,1) -- cycle;
    % Red left side
    \draw[red, very thick] (1,0) -- (1,1);
    \draw[red, very thick] (0,0) -- (1,0);
    % Filled circles at corners
    \fill (0,0) circle (3pt);
    \fill (1,0) circle (3pt);
    \fill (1,1) circle (3pt);
    \fill (0,1) circle (3pt);
\end{tikzpicture}}} =
\vcenter{\hbox{\begin{tikzpicture}
    % Square
    \draw[thick, dashed] (0,0) -- (1,0) -- (1,1) -- (0,1) -- cycle;
    % Red left side
    \draw[red, very thick] (0,0) -- (1,0);
    % Label for the red edge
    \node[above, red] at (0.5,0) {$e_2$};
    % Filled circles at corners
    \fill (0,0) circle (3pt);
    \fill (1,0) circle (3pt);
    \fill (1,1) circle (3pt);
    \fill (0,1) circle (3pt);
\end{tikzpicture}}} +
\vcenter{\hbox{\begin{tikzpicture}
    % Square
    \draw[thick, dashed] (0,0) -- (1,0) -- (1,1) -- (0,1) -- cycle;
    % Red left side
    \draw[red, very thick] (1,0) -- (1,1);
    % Label for the red edge
    \node[left, red] at (1,0.5) {$e_3$};
    % Filled circles at corners
    \fill (0,0) circle (3pt);
    \fill (1,0) circle (3pt);
    \fill (1,1) circle (3pt);
    \fill (0,1) circle (3pt);
\end{tikzpicture}}}
= e_2 + e_3,
\end{equation}
respectively, their sum is a 1-chain
\begin{equation}
c + c' =
\vcenter{\hbox{\begin{tikzpicture}
    % Square
    \draw[thick, dashed] (0,0) -- (1,0) -- (1,1) -- (0,1) -- cycle;
    % Red left side
    \draw[red, very thick] (0,0) -- (0,1);
    \draw[red, very thick] (0,0) -- (1,0);
    % Filled circles at corners
    \fill (0,0) circle (3pt);
    \fill (1,0) circle (3pt);
    \fill (1,1) circle (3pt);
    \fill (0,1) circle (3pt);
\end{tikzpicture}}} +
\vcenter{\hbox{\begin{tikzpicture}
    % Square
    \draw[thick, dashed] (0,0) -- (1,0) -- (1,1) -- (0,1) -- cycle;
    % Red left side
    \draw[red, very thick] (1,0) -- (1,1);
    \draw[red, very thick] (0,0) -- (1,0);
    % Filled circles at corners
    \fill (0,0) circle (3pt);
    \fill (1,0) circle (3pt);
    \fill (1,1) circle (3pt);
    \fill (0,1) circle (3pt);
\end{tikzpicture}}} =
\vcenter{\hbox{\begin{tikzpicture}
    % Square
    \draw[thick, dashed] (0,0) -- (1,0) -- (1,1) -- (0,1) -- cycle;
    % Red left side
    \draw[red, very thick] (0,0) -- (0,1);
    \draw[red, very thick] (1,0) -- (1,1);
    % Filled circles at corners
    \fill (0,0) circle (3pt);
    \fill (1,0) circle (3pt);
    \fill (1,1) circle (3pt);
    \fill (0,1) circle (3pt);
\end{tikzpicture}}} =
\vcenter{\hbox{\begin{tikzpicture}
    % Square
    \draw[thick, dashed] (0,0) -- (1,0) -- (1,1) -- (0,1) -- cycle;
    % Red left side
    \draw[red, very thick] (0,0) -- (0,1);
    % Label for the red edge
    \node[right, red] at (0,0.5) {$e_1$};
    % Filled circles at corners
    \fill (0,0) circle (3pt);
    \fill (1,0) circle (3pt);
    \fill (1,1) circle (3pt);
    \fill (0,1) circle (3pt);
\end{tikzpicture}}} +
\vcenter{\hbox{\begin{tikzpicture}
    % Square
    \draw[thick, dashed] (0,0) -- (1,0) -- (1,1) -- (0,1) -- cycle;
    % Red left side
    \draw[red, very thick] (1,0) -- (1,1);
    % Label for the red edge
    \node[left, red] at (1,0.5) {$e_3$};
    % Filled circles at corners
    \fill (0,0) circle (3pt);
    \fill (1,0) circle (3pt);
    \fill (1,1) circle (3pt);
    \fill (0,1) circle (3pt);
\end{tikzpicture}}}
= e_1 + e_3,
\end{equation}
where $e_2 + e_2 = 0$ because we are doing arithmetic over $\ZZ_2$. 
In fact, the set of $n$-chains forms an abelian group, with the following properties:
\begin{enumerate}
    \item \textbf{Group composition}: cell-wise addition modulo 2
    \item \textbf{Associativity}: follows from associativity of addition
    \item \textbf{Identity element}: the all-zero $n$-chain, denoted by $0_n$
    \item \textbf{Inverse}: every $n$-chain is self-inverse (for $\ZZ_2$)
\end{enumerate}
We denote the group of $n$-chains as $C_n$ and we use lowercase Latin letters to label specific chains in the group: $c \in C_n$. With $n$-cells and $n$-chains defined, we can now consider \emph{boundary maps}, which are the fundamental operators that we consider in homology. The $n$-boundary map $\partial_n$ is a group-structure-preserving map (i.e., a group homomorphism) from the set of $n$-chains to the set of $(n-1)$-chains. That is, $\partial_n : C_n \to C_{n-1}$. We can think of a boundary map as breaking up a higher dimensional object. For example, for a triangular face (i.e., a 2-chain consisting of a single 2-cell) $f \in C_2$, the boundary map $\partial_2$ maps $f$ to its constituent edges (1-chains) $e_1$, $e_2$, and $e_3$. That is,
\begin{equation}\label{eqn:triangle_boundary}
    f =
    \vcenter{\hbox{
    \begin{tikzpicture}
        % Vertices of an equilateral triangle with side length 1
        \coordinate (A) at (0,0);
        \coordinate (B) at (1,0);
        \coordinate (C) at ({1/2},{sqrt(3)/2});
        % Light blue fill
        \fill[blue!15] (A) -- (B) -- (C) -- cycle;
        \node[black] at (0.5,0.3) {$f$};
        % Edges
        \draw[black, thick, dashed] (A) -- (C);
        \draw[black, thick, dashed] (A) -- (B);
        \draw[black, thick, dashed] (B) -- (C);
        % Filled circles at corners
        \fill (A) circle (3pt);
        \fill (B) circle (3pt);
        \fill (C) circle (3pt);
    \end{tikzpicture}
    }} \mapsto
    \partial_2 f =
    \vcenter{\hbox{
    \begin{tikzpicture}
        % Vertices of an equilateral triangle with side length 1
        \coordinate (A) at (0,0);
        \coordinate (B) at (1,0);
        \coordinate (C) at ({1/2},{sqrt(3)/2});
        % Edges
        \draw[red, very thick] (A) -- (C);
        \draw[red, very thick] (A) -- (B);
        \draw[red, very thick] (B) -- (C);
        % Filled circles at corners
        \fill (A) circle (3pt);
        \fill (B) circle (3pt);
        \fill (C) circle (3pt);
    \end{tikzpicture}
    }} =
    \vcenter{\hbox{
    \begin{tikzpicture}
        % Vertices of an equilateral triangle with side length 1
        \coordinate (A) at (0,0);
        \coordinate (B) at (1,0);
        \coordinate (C) at ({1/2},{sqrt(3)/2});
        % Edges
        \draw[red, very thick] (A) -- (C);
        \draw[black, thick, dashed] (A) -- (B);
        \draw[black, thick, dashed] (B) -- (C);
        % Label for the red edge
        \node[left, red] at ($(A)!0.5!(C)$) {$e_1$};
        % Filled circles at corners
        \fill (A) circle (3pt);
        \fill (B) circle (3pt);
        \fill (C) circle (3pt);
    \end{tikzpicture}
    }} +
    \vcenter{\hbox{
    \begin{tikzpicture}
        % Vertices of an equilateral triangle with side length 1
        \coordinate (A) at (0,0);
        \coordinate (B) at (1,0);
        \coordinate (C) at ({1/2},{sqrt(3)/2});
        % Edges
        \draw[black, thick, dashed] (A) -- (C);
        \draw[red, very thick] (A) -- (B);
        \draw[black, thick, dashed] (B) -- (C);
        % Label for the red edge
        \node[above, red] at ($(B)!0.5!(A)$) {$e_2$};
        % Filled circles at corners
        \fill (A) circle (3pt);
        \fill (B) circle (3pt);
        \fill (C) circle (3pt);
    \end{tikzpicture}
    }} +
    \vcenter{\hbox{
    \begin{tikzpicture}
        % Vertices of an equilateral triangle with side length 1
        \coordinate (A) at (0,0);
        \coordinate (B) at (1,0);
        \coordinate (C) at ({1/2},{sqrt(3)/2});
        % Edges
        \draw[black, thick, dashed] (A) -- (C);
        \draw[black, thick, dashed] (A) -- (B);
        \draw[red, very thick] (B) -- (C);
        % Label for the red edge
        \node[right, red] at ($(B)!0.5!(C)$) {$e_3$};
        % Filled circles at corners
        \fill (A) circle (3pt);
        \fill (B) circle (3pt);
        \fill (C) circle (3pt);
    \end{tikzpicture}
    }}
    = e_1 + e_2 + e_3\,,
\end{equation}
where $e_i \in C_1$ for $i = 1,2,3$.

As the reader may note, a boundary map $\partial_n$ quite literally maps an $n$-chain $c\in C_n$ to its boundaries in $C_{n-1}$.
For an $n$-chain $c$, we call $\partial_nc$ the \emph{boundary} of $c$.
The set of all $(n-1)$-chains that arise as boundaries of $n$-chains is $B_{n-1} = \im{\partial_n} \subseteq C_{n-1}$ which we call the $(n-1)$-boundaries.
Some chains have a \emph{null boundary} (i.e., they have no boundary), that is, $\partial_nc = 0_{n-1}$.
For example, the 1-chain $e_1+e_2+e_3$ in~\cref{eqn:triangle_boundary} is a null boundary since
\begin{equation}
\begin{aligned}
    \partial_1(e_1 + e_2 + e_3) &= 
    \vcenter{\hbox{
    \begin{tikzpicture}
        % Vertices of an equilateral triangle with side length 1
        \coordinate (A) at (0,0);
        \coordinate (B) at (1,0);
        \coordinate (C) at ({1/2},{sqrt(3)/2});
        % Edges
        \draw[black, thick, dashed] (A) -- (C);
        \draw[black, thick, dashed] (A) -- (B);
        \draw[black, thick, dashed] (B) -- (C);
        % Label for the red edge
        \node[left, red] at ({1/2},{sqrt(3)/2}) {$v_1$};
        \node[below, red] at (0,0) {$v_2$};
        % Filled circles at corners
        \fill[red] (A) circle (3pt);
        \fill (B) circle (3pt);
        \fill[red] (C) circle (3pt);
    \end{tikzpicture}
    }} +
    \vcenter{\hbox{
    \begin{tikzpicture}
        % Vertices of an equilateral triangle with side length 1
        \coordinate (A) at (0,0);
        \coordinate (B) at (1,0);
        \coordinate (C) at ({1/2},{sqrt(3)/2});
        % Edges
        \draw[black, thick, dashed] (A) -- (C);
        \draw[black, thick, dashed] (A) -- (B);
        \draw[black, thick, dashed] (B) -- (C);
        % Label for the red edge
        \node[below, red] at (1,0) {$v_3$};
        \node[below, red] at (0,0) {$v_2$};
        % Filled circles at corners
        \fill[red] (A) circle (3pt);
        \fill[red] (B) circle (3pt);
        \fill (C) circle (3pt);
    \end{tikzpicture}
    }} +
    \vcenter{\hbox{
    \begin{tikzpicture}
        % Vertices of an equilateral triangle with side length 1
        \coordinate (A) at (0,0);
        \coordinate (B) at (1,0);
        \coordinate (C) at ({1/2},{sqrt(3)/2});
        % Edges
        \draw[black, thick, dashed] (A) -- (C);
        \draw[black, thick, dashed] (A) -- (B);
        \draw[black, thick, dashed] (B) -- (C);
        % Label for the red edge
        \node[left, red] at ({1/2},{sqrt(3)/2}) {$v_1$};
        \node[below, red] at (1,0) {$v_3$};
        % Filled circles at corners
        \fill (A) circle (3pt);
        \fill[red] (B) circle (3pt);
        \fill[red] (C) circle (3pt);
    \end{tikzpicture}
    }}\\
    &= (v_1+v_2) + (v_2+v_3) + (v_3+v_1) = 0_0
\end{aligned}
\end{equation}
because $\partial_1 e_j = v_j + v_{j+1\pmod{3}}$.
Chains with no boundaries are called \emph{cycles}. 
In particular, an \emph{$n$-cycle} is an $n$-chain $c$ satisfying $\partial_n c = 0_{n-1}$.
The set of all $n$-cycles is denoted as $Z_n=\ker\partial_n$.
So the 1-chain $e_1+e_2+e_3$ above is an example of a 1-cycle.
Note that sometimes we will omit the subscript $n$ for boundary map on $n$-cycles when it is clear from the context, e.g. $\partial c$ means $\partial_nc$ for $c \in C_n$.

We now connect all of these groups together. Given a boundary map $\partial_n$, its domain is $C_n$, its image is $B_{n-1}$, and its kernel is $Z_n$. This leads to the \emph{fundamental lemma of homology}: the boundary of a boundary is a null chain, or 
\begin{equation}\label{eqn:homology_boundary_map_null}
    \partial_{n}\partial_{n+1}c = 0_{n-1} 
\end{equation}
for all $c_n \in C_{n+1}$, which indicates that the image of $\partial_{n+1}$ is contained in the kernel of $\partial_{n}$, that is, $B_n\subseteq Z_n$.
Thus $B_n$ is a subgroup of $Z_n$. $B_n$ is then a \emph{normal} subgroup of $Z_n$ since $Z_n$ is an abelian group, so we can take the quotient group $H_n = Z_n/B_n$ which is called the \emph{$n\numth$ homology group}.

Elements of $H_n$ are equivalence classes of elements of $Z_n$ where any pair of $n$-cycles $c,c'\in Z_n$ is equivalent, denoted as $c\sim c'$ if they belong to the same coset of $B_n$ in $Z_n$, that is, $c,c'\in c''+B_n = \{c''+c^*:c^*\in B_n\}$ for some $c''\in Z_n$.
Note that if $c\in c''+B_n$, then there is some $c^*\in B_n$ such that $c=c''+c^*$ which implies that $c''=c+c^*$ (since we are working over $\Z_2$), so $c''\in c+B_n$.
This holds in general for any pair of elements in a coset of $B_n$, so we have $c'+B_n=c+B_n$ if $c\sim c'$ and thus all $n$-cycles in a coset is equivalent.
Therefore we have
\begin{equation}
    c\sim c' \Leftrightarrow c+B_n = c'+B_n \,.
\end{equation}
Writing the equivalence class of an $n$-cycle $c$ as $[c]=c+B_n$, addition between any two elements of $H_n$ is therefore given by $[c]+[c']=[c+c']=(c+c')+B_n$.
For any given coset, one can take any of its element as a representative for this group addition.
Thus, any coset containing the identity element $0_n$ is the identity element of $H_n$, denoted as $[0_n]=0_n+B_n=B_n$.
The homology group $H_n$ therefore consists of all equivalence classes $\{[c]=c+B_n: c\in Z_n\}$ that are closed under addition, where all distinct cosets $c+B_n \in H_n$ partition $Z_n$, that is, $\bigcup_{g\in H_n} g = Z_n$ and $g\cap g'=\emptyset$ if $g\neq g'$.
Thus, the homology group $H_n$ consists of $|H_n| = |Z_n|/|B_n|$ elements.

Taking the simple triangle example in~\cref{eqn:triangle_boundary}, the only $1$-boundaries are the boundary of the triangle and the null-boundary: $B_1=\{c,0_1\}$ where $c=e_1+e_2+e_3$, whereas the only $1$-cycles are also the boundary of the triangle and the null-boundary: $Z_1=\{c,0_1\}$.
Its first homology group therefore only consists of one (identity) element since
\begin{equation}
    c+B_1=\{c+c,c+0_1\} = \{0_1,c\} = \{0_1+c,0_1+0_1\} = 0_1+B_1 = B_1\,.
\end{equation}
We can consider a more interesting example: a $3\times3$ square cellulation of a torus, which when laid flat on a 2-D surface can be illustrated as a grid with periodic boundary condition
\begin{equation}\label{eqn:33_lattice_torus}
    \vcenter{\hbox{
    \begin{tikzpicture}
        % 3x3 square grid (so vertices run from 0 to 3)
        \draw[] (0,0) grid (3,3);
        % Red left boundary, matching the previous square's style
        \draw[red, very thick] (0,0) -- (0,3);
        \draw[red, very thick] (0,0) -- (3,0);
        \draw[red, very thick] (3,0) -- (3,3);
        \draw[red, very thick] (0,3) -- (3,3);
        % Optional label for the red side
        \node[left, red] at (0,0.5) {$e_1$};
        \node[left, red] at (0,1.5) {$e_2$};
        \node[left, red] at (0,2.5) {$e_3$};
        \node[below, red] at (0.5,0) {$e_1'$};
        \node[below, red] at (1.5,0) {$e_2'$};
        \node[below, red] at (2.5,0) {$e_3'$};
        \node[right, red] at (3,0.5) {$e_1$};
        \node[right, red] at (3,1.5) {$e_2$};
        \node[right, red] at (3,2.5) {$e_3$};
        \node[above, red] at (0.5,3) {$e_1'$};
        \node[above, red] at (1.5,3) {$e_2'$};
        \node[above, red] at (2.5,3) {$e_3'$};
        % Filled circles at all vertices
        \foreach \x in {0,1,2,3} {
            \foreach \y in {0,1,2,3} {
                \fill (\x,\y) circle (2.5pt);
            }
        }
    \end{tikzpicture}
    }}\,,
\end{equation}
where the top and bottom edges are the same and the left and right edges are the same, similarly for the top-bottom and right-left vertices.
Its $1$-boundary group $B_1$ is generated by the boundaries of the square plaquettes:
\begin{equation}\label{eqn:33_torus_1boundaries}
    \partial_2(p_1) =
    \vcenter{\hbox{
    \begin{tikzpicture}[scale=0.7]
        % 3x3 square grid
        \draw[] (0,0) grid (3,3);
        \node[] at (0.5,0.5) {$p_1$};
        % Red edges: bottom left square
        \draw[red, very thick] (0,0) -- (1,0) -- (1,1) -- (0,1) -- cycle;
        % Label
        \node at (0.5,0.5) {$p_1$};
        % Filled circles at all vertices
        \foreach \x in {0,1,2,3} {
            \foreach \y in {0,1,2,3} {
                \fill (\x,\y) circle (2.5pt);
            }
        }
    \end{tikzpicture}
    }} ,\;
    \partial_2(p_2) =
    \vcenter{\hbox{
    \begin{tikzpicture}[scale=0.7]
        % 3x3 square grid
        \draw[] (0,0) grid (3,3);
        % Red edges: bottom middle square
        \draw[red, very thick] (1,0) -- (2,0) -- (2,1) -- (1,1) -- cycle;
        % Label
        \node at (1.5,0.5) {$p_2$};
        % Filled circles at all vertices
        \foreach \x in {0,1,2,3} {
            \foreach \y in {0,1,2,3} {
                \fill (\x,\y) circle (2.5pt);
            }
        }
    \end{tikzpicture}
    }} ,\;
    \dots \,,\;
    \partial_2(p_9) =
    \vcenter{\hbox{
    \begin{tikzpicture}[scale=0.7]
        % 3x3 square grid
        \draw[] (0,0) grid (3,3);
        % Red edges: top right square
        \draw[red, very thick] (2,2) -- (3,2) -- (3,3) -- (2,3) -- cycle;
        % Label
        \node at (2.5,2.5) {$p_9$};
        % Filled circles at all vertices
        \foreach \x in {0,1,2,3} {
            \foreach \y in {0,1,2,3} {
                \fill (\x,\y) circle (2.5pt);
            }
        }
    \end{tikzpicture}
    }}\,,
\end{equation}
since the boundary of two plaquettes (a 2-chain) $p_1+p_2$ is the sum of the boundaries of $p_1$ and $p_2$:
\begin{equation}
    \partial_2(p_1+p_2)=
    \partial_2 
    \vcenter{\hbox{
    \begin{tikzpicture}[scale=0.7]
        % 3x3 square grid
        \draw[] (0,0) grid (3,3);
        \node[] at (0.5,0.5) {$p_1$};
        % Label
        \node at (0.5,0.5) {$p_1$};
        \node at (1.5,0.5) {$p_2$};
        % Filled circles at all vertices
        \foreach \x in {0,1,2,3} {
            \foreach \y in {0,1,2,3} {
                \fill (\x,\y) circle (2.5pt);
            }
        }
    \end{tikzpicture}
    }} =
    \vcenter{\hbox{
    \begin{tikzpicture}[scale=0.7]
        % 3x3 square grid
        \draw[] (0,0) grid (3,3);
        \node[] at (0.5,0.5) {$p_1$};
        % Red edges: bottom left square
        \draw[red, very thick] (0,0) -- (2,0) -- (2,1) -- (0,1) -- cycle;
        % Label
        \node at (0.5,0.5) {$p_1$};
        \node at (1.5,0.5) {$p_2$};
        % Filled circles at all vertices
        \foreach \x in {0,1,2,3} {
            \foreach \y in {0,1,2,3} {
                \fill (\x,\y) circle (2.5pt);
            }
        }
    \end{tikzpicture}
    }}
    =
    \vcenter{\hbox{
    \begin{tikzpicture}[scale=0.7]
        % 3x3 square grid
        \draw[] (0,0) grid (3,3);
        \node[] at (0.5,0.5) {$p_1$};
        % Red edges: bottom left square
        \draw[red, very thick] (0,0) -- (1,0) -- (1,1) -- (0,1) -- cycle;
        % Label
        \node at (0.5,0.5) {$p_1$};
        % Filled circles at all vertices
        \foreach \x in {0,1,2,3} {
            \foreach \y in {0,1,2,3} {
                \fill (\x,\y) circle (2.5pt);
            }
        }
    \end{tikzpicture}
    }} +
    \vcenter{\hbox{
    \begin{tikzpicture}[scale=0.7]
        % 3x3 square grid
        \draw[] (0,0) grid (3,3);
        % Red edges: bottom middle square
        \draw[red, very thick] (1,0) -- (2,0) -- (2,1) -- (1,1) -- cycle;
        % Label
        \node at (1.5,0.5) {$p_2$};
        % Filled circles at all vertices
        \foreach \x in {0,1,2,3} {
            \foreach \y in {0,1,2,3} {
                \fill (\x,\y) circle (2.5pt);
            }
        }
    \end{tikzpicture}
    }}\,.
\end{equation}
On the other hand, the 1-cycle group $Z_1$ is generated by the boundaries of the square plaquettes in~\cref{eqn:33_torus_1boundaries}, as well as six \emph{non-contractible} loops, namely loops that wrap around the torus in vertical and horizontal directions:
\begin{equation}
    \vcenter{\hbox{
    \begin{tikzpicture}[scale=0.7]
        % 3x3 square grid (so vertices run from 0 to 3)
        \draw[] (0,0) grid (3,3);
        % Red left boundary, matching the previous square's style
        \draw[red, very thick] (0,0) -- (0,3);
        % Optional label for the red side
        \node[left, red] at (0,1.5) {$\ell_1$};
        % Filled circles at all vertices
        \foreach \x in {0,1,2,3} {
            \foreach \y in {0,1,2,3} {
                \fill (\x,\y) circle (2.5pt);
            }
        }
    \end{tikzpicture}
    }} ,\dots,
    \vcenter{\hbox{
    \begin{tikzpicture}[scale=0.7]
        % 3x3 square grid (so vertices run from 0 to 3)
        \draw[] (0,0) grid (3,3);
        % Red left boundary, matching the previous square's style
        \draw[red, very thick] (2,0) -- (2,3);
        % Optional label for the red side
        \node[left, red] at (2,1.5) {$\ell_3$};
        % Filled circles at all vertices
        \foreach \x in {0,1,2,3} {
            \foreach \y in {0,1,2,3} {
                \fill (\x,\y) circle (2.5pt);
            }
        }
    \end{tikzpicture}
    }} ,
    \vcenter{\hbox{
    \begin{tikzpicture}[scale=0.7]
        % 3x3 square grid (so vertices run from 0 to 3)
        \draw[] (0,0) grid (3,3);
        % Red left boundary, matching the previous square's style
        \draw[red, very thick] (0,0) -- (3,0);
        % Optional label for the red side
        \node[above, red] at (1.5,0) {$\ell_1'$};
        % Filled circles at all vertices
        \foreach \x in {0,1,2,3} {
            \foreach \y in {0,1,2,3} {
                \fill (\x,\y) circle (2.5pt);
            }
        }
    \end{tikzpicture}
    }} ,\dots,
    \vcenter{\hbox{
    \begin{tikzpicture}[scale=0.7]
        % 3x3 square grid (so vertices run from 0 to 3)
        \draw[] (0,0) grid (3,3);
        % Red left boundary, matching the previous square's style
        \draw[red, very thick] (0,2) -- (3,2);
        % Optional label for the red side
        \node[above, red] at (1.5,2) {$\ell_3'$};
        % Filled circles at all vertices
        \foreach \x in {0,1,2,3} {
            \foreach \y in {0,1,2,3} {
                \fill (\x,\y) circle (2.5pt);
            }
        }
    \end{tikzpicture}
    }}\,.
\end{equation}
where $\ell_1=e_1+e_2+e_3$, $\ell_1'=e_1'+e_2'+e_3'$, and so on.

Now let us look into the homology group $H_1=Z_1/B_1$.
Since $B_1$ is generated by the plaquette boundaries $\partial_2(p_1),\dots,\partial_2(p_9)$, we have the identity element $[0_1]=B_1$ being the equivalence class of all plaquette boundaries, that is, $\partial_2(p_i)\sim \partial_2(p_j)$ for all $i,j\in\{1,\dots,9\}$.
However, it does not contain any non-contractible loops since they cannot be generated from the plaquette boundaries.
Now note that the coset $\ell_j+B_1\in H_1$ containing the vertical non-contractible loop $\ell_j$ contains all other vertical non-contractible loops, as well their sum with any 1-boundaries.
For example, $[\ell_1]=\ell_1+B_1$ contains:
\begin{equation}\label{eqn:homology_loop_plus_boundary}
\begin{gathered}
    \vcenter{\hbox{
    \begin{tikzpicture}[scale=0.7]
        % 3x3 square grid (so vertices run from 0 to 3)
        \draw[] (0,0) grid (3,3);
        % Red left boundary, matching the previous square's style
        \draw[red, very thick] (0,0) -- (0,3);
        % Optional label for the red side
        \node[left, red] at (0,1.5) {$\ell_1$};
        % Filled circles at all vertices
        \foreach \x in {0,1,2,3} {
            \foreach \y in {0,1,2,3} {
                \fill (\x,\y) circle (2.5pt);
            }
        }
    \end{tikzpicture}
    }} +
    \vcenter{\hbox{
    \begin{tikzpicture}[scale=0.7]
        % 3x3 square grid
        \draw[] (0,0) grid (3,3);
        % Red edges: bottom left square
        \draw[red, very thick] (0,0) -- (1,0) -- (1,3) -- (0,3) -- cycle;
        % Filled circles at all vertices
        \foreach \x in {0,1,2,3} {
            \foreach \y in {0,1,2,3} {
                \fill (\x,\y) circle (2.5pt);
            }
        }
    \end{tikzpicture}
    }} =
    \vcenter{\hbox{
    \begin{tikzpicture}[scale=0.7]
        % 3x3 square grid (so vertices run from 0 to 3)
        \draw[] (0,0) grid (3,3);
        % Red left boundary, matching the previous square's style
        \draw[red, very thick] (1,0) -- (1,3);
        % Optional label for the red side
        \node[left, red] at (1,1.5) {$\ell_2$};
        % Filled circles at all vertices
        \foreach \x in {0,1,2,3} {
            \foreach \y in {0,1,2,3} {
                \fill (\x,\y) circle (2.5pt);
            }
        }
    \end{tikzpicture}
    }} 
    \\\text{and}\\
    \vcenter{\hbox{
    \begin{tikzpicture}[scale=0.7]
        % 3x3 square grid (so vertices run from 0 to 3)
        \draw[] (0,0) grid (3,3);
        % Red left boundary, matching the previous square's style
        \draw[red, very thick] (2,0) -- (2,3);
        % Optional label for the red side
        \node[left, red] at (2,1.5) {$\ell_3$};
        % Filled circles at all vertices
        \foreach \x in {0,1,2,3} {
            \foreach \y in {0,1,2,3} {
                \fill (\x,\y) circle (2.5pt);
            }
        }
    \end{tikzpicture}
    }} +
    \vcenter{\hbox{
    \begin{tikzpicture}[scale=0.7]
        % 3x3 square grid
        \draw[] (0,0) grid (3,3);
        % Red edges: bottom middle square
        \draw[red, very thick] (1,1) -- (2,1) -- (2,2) -- (1,2) -- cycle;
        % Filled circles at all vertices
        \foreach \x in {0,1,2,3} {
            \foreach \y in {0,1,2,3} {
                \fill (\x,\y) circle (2.5pt);
            }
        }
    \end{tikzpicture}
    }} =
    \vcenter{\hbox{
    \begin{tikzpicture}[scale=0.7]
        % 3x3 square grid
        \draw[] (0,0) grid (3,3);
        \draw[red, very thick] (2,0) -- (2,1);
        \draw[red, very thick] (2,2) -- (2,3);
        % Red edges: bottom middle square
        \draw[red, very thick] (2,1) -- (1,1) -- (1,2) -- (2,2) ;
        % Filled circles at all vertices
        \foreach \x in {0,1,2,3} {
            \foreach \y in {0,1,2,3} {
                \fill (\x,\y) circle (2.5pt);
            }
        }
    \end{tikzpicture}
    }}\,.
\end{gathered}
\end{equation}
By a similar argument, we also have $\ell_i\sim \ell_j$ which correspond to coset $[\ell_1]=\ell_1+B_1$.
Thus, the homology group $H_1$ is generated by $[\ell_1]$ and $[v_1]$.

For the $n\numth$ homology group, we define the $n\numth$ \emph{Betti number} as $\beta_n = \dim{H_n} = \dim{Z_n} - \dim{B_n}$, where the dimension $\dim{G}$ of a group $G$ is the minimum number of elements needed to generate the group.
The Betti number is a topologically invariant quantity that can informally be thought of as characterizing the number of holes in an object and is an important quantity for studying topological codes. The calculation of Betti numbers has also been given much attention recently in the quantum algorithms, where Lloyd, Garnerone, and Zanardi~\cite{lloyd2016quantum} introduced a quantum algorithm for calculating Betti numbers, giving rise to the field of quantum topological data analysis.
The Betti numbers then allow us to calculate the Euler characteristic as an alternating sum of the Betti numbers. That is,

\begin{thmbox}[label={thm:euler-characteristic-betti-number}]{Euler characteristic and Betti numbers}
    Given a \emph{finite, two-dimensional} cell complex (such as a finite lattice composed of vertices, edges, and plaquettes), the Euler characteristic is $\chi = \beta_2 - \beta_1 + \beta_0$.
\end{thmbox}

\begin{proof}
    This identity stems from an application of the rank-nullity theorem; see, for example, Ref.~\cite[Sec. 2.2]{hatcher2001algebraic}. We prove it here nonetheless. Starting with the rank-nullity theorem applied to the boundary operator, we have
    \begin{equation}
        \dim{C_n} = \dim{\ker{\partial_n}} + \dim{\im{\partial_n}}
    \end{equation}
    Recall from above that for a boundary map $\partial_n$, we took the domain to be $C_n$, the kernel to be $Z_n$, and the range (or image) to be $B_{n-1}$. Thus, we have
    \begin{equation}
        \dim{C_n} = \dim{Z_n} + \dim{B_{n-1}}\,.
    \end{equation}
    Relating this to the Betti number, we have
    \begin{equation}
        \beta_n = \dim{H_n} = \dim{Z_n} - \dim{B_n}\,.
    \end{equation}
    Then, recalling the Euler characteristic from~\cref{eqn:euler_torus} before, which relates the number of vertices, edges, and faces, we have
    \begin{equation}
        \chi = F - E + V\,,
    \end{equation}
    which we can write as
    \begin{equation}
        \chi = a_2 - a_1 + a_0\,,
    \end{equation}
    where $a_n$ is the number of $n$-cells. Each of these is exactly equal to the rank of the chain group: $a_n = \dim{C_n}$, which we can write as
    \begin{equation}
        a_n = \dim{C_n} = \dim{Z_n} + \dim{B_{n-1}}\,.
    \end{equation}
    Since we have no 3-cells, we have $\dim{B_2} = 0$, so $\beta_2 = \dim{Z_2}$. Similarly, $Z_0 = C_0$ and so $\dim{Z_0} = a_0$. Plugging in for $n = 0, 1, 2$, we have
    \begin{align}
        a_2 &= \beta_2 + \dim{B_1},\\
        a_1 &= \beta_1 + \dim{B_1} + \dim{B_0},\\
        a_0 &= \beta_0 + \dim{B_0}\,.
    \end{align}
    Plugging in everything, we have
    \begin{align}
        \chi &= a_2 - a_1 + a_0\\
        &= \beta_2 + \dim{B_1} - \beta_1 - \dim{B_1} - \dim{B_0} + \beta_0 + \dim{B_0}\\
        &= \beta_2 - \beta_1 + \beta_0\,,
    \end{align}
    as claimed.
\end{proof}

This result can be generalized to any dimension to give the following alternating sum
\begin{equation}
    \chi = \sum_{n\geq0}{(-1)^n\beta_n}\,.
\end{equation}

The dual theory of homology is called \emph{cohomology}.
Since we work over $\mathbb{Z}_2$, the chain groups $C_n$ can be viewed as vector spaces over the finite field $\mathbb{F}_2$ (which is simply $\ZZ_2$ along with the usual multiplication operation). The corresponding \emph{$n$-cochain group} is given by
\begin{equation}
C^n := \mathrm{Hom}(C_n,\mathbb{Z}_2),
\end{equation}
where $\mathrm{Hom}(C_n,\mathbb{Z}_2)$ is the set of all linear functionals that map an $n$-chain to an element of $\Z_2$ with identity element $0^n$. 
Thus if $\alpha \in C^n$ and $c\in C_n$, the action of $\alpha$ on $c$ is given by $\langle \alpha,c\rangle \in \mathbb{Z}_2$ so that an $n$-cochain assigns a binary value to each $n$-chain. 
It is useful to introduce a ``dual basis'' as follows. 
For each $n$-chain $c_\sigma$ consisting of a single $n$-cell $\sigma$, let $c_\sigma^\ast\in C^n$ be the cochain that evaluates to 1 on $c_\sigma$ and to 0 on all other $n$-chains $c_\tau$ consisting of a single $n$-cell $\tau\neq\sigma$:
\begin{equation}
    \langle c_\sigma^\ast,c_\tau\rangle =
    \begin{cases}
    1, & \tau=\sigma,\\
    0, & \tau\neq \sigma.
    \end{cases}
\end{equation}
Therefore, just as an $n$-chain can be viewed as a subset of $n$-cells, an $n$-cochain can also be viewed as a subset of $n$-cells, defined by its functional relationship with the $n$-chains. 

The dual of the boundary map is called the \emph{coboundary map} $\delta^n : C^n \to C^{n+1}$ which is defined by the adjoint relation
\begin{equation}
    \langle \delta^n \alpha, c\rangle
    =
    \langle \alpha, \partial_{n+1} c\rangle,
    \qquad
    \alpha\in C^n,\quad c\in C_{n+1}.
\end{equation}
In words, $\delta^n \alpha$ evaluates an $(n+1)$-chain by first taking its boundary and then applying $\alpha$.
For example, let $v$ be a vertex and let $v^\ast\in C^0$ be the corresponding dual 0-cochain. If an edge $e$ has endpoint vertices $u$ and $w$, then over $\mathbb{Z}_2$, $\partial_1 e = u+w$. Hence
\begin{equation}
    \langle \delta^0 v^\ast,e\rangle
    =
    \langle v^\ast,\partial_1 e\rangle
    =
    \langle v^\ast,u+w\rangle.
\end{equation}
This is equal to 1 exactly when $v$ is one of the endpoints of $e$. Therefore, $\delta^0 v^\ast$ is the 1-cochain supported on all edges incident on $v$. 
The coboundary maps satisfy the relation $\delta^{n+1}\delta^n=0^{n+2}$, which follows directly from the boundary map identity $\partial_n\partial_{n+1}=0_{n-1}$ for any $n\in\N$ in~\cref{eqn:homology_boundary_map_null}. Indeed, for any $\alpha\in C^n$ and any $c\in C_{n+2}$,
\begin{equation}
    \langle \delta^{n+1}\delta^n\alpha,c\rangle
    =
    \langle \alpha,\partial_{n+1}\partial_{n+2}c\rangle
    =
    0.
\end{equation}
We can therefore define \emph{cocycles}, \emph{coboundaries}, and \emph{cohomology groups} by
\begin{equation}
    Z^n := \ker \delta^n,
    \qquad
    B^n := \operatorname{im}\delta^{n-1},
    \qquad
    H^n := Z^n/B^n \,,
\end{equation}
respectively.

For the square-cellulated torus example, we can identify the $0$-cochains, $1$-cochains, and $2$-cochains with the vertices, edges, and squares, respectively, similar to the chains.
However, the coboundary map now maps vertices to the edges around it and maps edges to the squares adjacent to it.
For example, a coboundary map on a vertex $v$ surrounded by edges $e_1,e_2,e_3,e_4$ is given by
\begin{equation}
\begin{aligned}
    v =
    \vcenter{\hbox{
    \begin{tikzpicture}[scale=0.80]
        % Cross edges (each of length 1)
        \draw[thick, dashed] (0,0) -- (1,0);
        \draw[thick, dashed] (0,0) -- (0,1);
        \draw[thick, dashed] (0,0) -- (0,-1);
        \draw[thick, dashed] (0,0) -- (-1,0);
        % Optional label for the red edge
        \node[above, red] at (0.2,0) {$v$};
        % Filled circles at vertices
        \fill[red] (0,0) circle (3pt);
        \fill (1,0) circle (3pt);
        \fill (-1,0) circle (3pt);
        \fill (0,1) circle (3pt);
        \fill (0,-1) circle (3pt);
    \end{tikzpicture}
    }} 
    \mapsto
    \delta^0 v &=
    \vcenter{\hbox{
    \begin{tikzpicture}[scale=0.80]
        % Cross edges (each of length 1)
        \draw[] (0,0) -- (1,0);
        \draw[] (0,0) -- (0,1);
        \draw[] (0,0) -- (0,-1);
        \draw[red, very thick] (0,0) -- (-1,0);
        % Optional label for the red edge
        \node[above, red] at (-0.5,0) {$e_1$};
        % Filled circles at vertices
        \fill (0,0) circle (3pt);
        \fill (1,0) circle (3pt);
        \fill (-1,0) circle (3pt);
        \fill (0,1) circle (3pt);
        \fill (0,-1) circle (3pt);
    \end{tikzpicture}
    }} +
    \vcenter{\hbox{
    \begin{tikzpicture}[scale=0.80]
        % Cross edges (each of length 1)
        \draw[] (0,0) -- (1,0);
        \draw[] (0,0) -- (0,1);
        \draw[red, very thick] (0,0) -- (0,-1);
        \draw[] (0,0) -- (-1,0);
        % Optional label for the red edge
        \node[left, red] at (0,-0.5) {$e_2$};
        % Filled circles at vertices
        \fill (0,0) circle (3pt);
        \fill (1,0) circle (3pt);
        \fill (-1,0) circle (3pt);
        \fill (0,1) circle (3pt);
        \fill (0,-1) circle (3pt);
    \end{tikzpicture}
    }} +
    \vcenter{\hbox{
    \begin{tikzpicture}[scale=0.80]
        % Cross edges (each of length 1)
        \draw[red, very thick] (0,0) -- (1,0);
        \draw[] (0,0) -- (0,1);
        \draw[] (0,0) -- (0,-1);
        \draw[] (0,0) -- (-1,0);
        % Optional label for the red edge
        \node[above, red] at (0.5,0) {$e_3$};
        % Filled circles at vertices
        \fill (0,0) circle (3pt);
        \fill (1,0) circle (3pt);
        \fill (-1,0) circle (3pt);
        \fill (0,1) circle (3pt);
        \fill (0,-1) circle (3pt);
    \end{tikzpicture}
    }} +
    \vcenter{\hbox{
    \begin{tikzpicture}[scale=0.80]
        % Cross edges (each of length 1)
        \draw[] (0,0) -- (1,0);
        \draw[red, very thick] (0,0) -- (0,1);
        \draw[] (0,0) -- (0,-1);
        \draw[] (0,0) -- (-1,0);
        % Optional label for the red edge
        \node[left, red] at (0,0.5) {$e_4$};
        % Filled circles at vertices
        \fill (0,0) circle (3pt);
        \fill (1,0) circle (3pt);
        \fill (-1,0) circle (3pt);
        \fill (0,1) circle (3pt);
        \fill (0,-1) circle (3pt);
    \end{tikzpicture}
    }} \\
    &= e_1+e_2+e_3+e_4 \,.
\end{aligned}
\end{equation}
Taking the example of a $3 \times 3$ square cellulation of a torus that we discussed for homology (c.f.~\cref{eqn:33_lattice_torus}), the coboundaries of the nine vertices (recall that the top vertices are the same vertices as the ones at the bottom, so are the right and left vertices) generate the 1-coboundary group $B^1$:
\begin{equation}\label{eqn:33_torus_1coboundaries}
    \delta^0(v_1) =
    \vcenter{\hbox{
    \begin{tikzpicture}[scale=0.7]
        % 3x3 square grid
        \draw[] (0,0) grid (3,3);
        % Red edges: bottom left square
        \draw[red, very thick] (0,0) -- (1,0);
        \draw[red, very thick] (0,0) -- (0,1);
        \draw[red, very thick] (3,0) -- (2,0);
        \draw[red, very thick] (0,3) -- (0,2);
        % Label
        \node at (0.3,0.25) {$v_1$};
        % Filled circles at all vertices
        \foreach \x in {0,1,2,3} {
            \foreach \y in {0,1,2,3} {
                \fill (\x,\y) circle (2.5pt);
            }
        }
    \end{tikzpicture}
    }} ,\;
    \delta^0(v_2) =
    \vcenter{\hbox{
    \begin{tikzpicture}[scale=0.7]
        % 3x3 square grid
        \draw[] (0,0) grid (3,3);
        % Red edges: bottom middle square
        \draw[red, very thick] (1,0) -- (2,0);
        \draw[red, very thick] (1,0) -- (1,1);
        \draw[red, very thick] (1,0) -- (0,0);
        \draw[red, very thick] (1,3) -- (1,2);
        % Label
        \node at (1.3,0.25) {$v_2$};
        % Filled circles at all vertices
        \foreach \x in {0,1,2,3} {
            \foreach \y in {0,1,2,3} {
                \fill (\x,\y) circle (2.5pt);
            }
        }
    \end{tikzpicture}
    }} ,\;
    \dots \,,\;
    \delta^0(v_9) =
    \vcenter{\hbox{
    \begin{tikzpicture}[scale=0.7]
        % 3x3 square grid
        \draw[] (0,0) grid (3,3);
        % Red edges: top right square
        \draw[red, very thick] (2,2) -- (3,2);
        \draw[red, very thick] (2,2) -- (1,2);
        \draw[red, very thick] (2,2) -- (2,3);
        \draw[red, very thick] (2,2) -- (2,1);
        % Label
        \node at (2.3,2.25) {$v_9$};
        % Filled circles at all vertices
        \foreach \x in {0,1,2,3} {
            \foreach \y in {0,1,2,3} {
                \fill (\x,\y) circle (2.5pt);
            }
        }
    \end{tikzpicture}
    }}\,.
\end{equation}
On the other hand, the 1-cocycle group $Z^1$ is generated by the coboundaries above, as well as six 1-cocycle loops around the torus in the vertical ($\lambda_1,\lambda_2,\lambda_3$) and horizontal ($\lambda_1',\lambda_2',\lambda_3'$) directions:
\begin{equation}
    \vcenter{\hbox{
    \begin{tikzpicture}[scale=0.7]
        % 3x3 square grid (so vertices run from 0 to 3)
        \draw[] (0,0) grid (3,3);
        \node[red] at (-0.35,1) {$\lambda_1$};
        % Red left boundary, matching the previous square's style
        \foreach \y in {0,1,2} {
            \draw[red, very thick] (0,\y) -- (1,\y);
        }
        % Filled circles at all vertices
        \foreach \x in {0,1,2,3} {
            \foreach \y in {0,1,2,3} {
                \fill (\x,\y) circle (2.5pt);
            }
        }
    \end{tikzpicture}
    }} ,\dots,
    \vcenter{\hbox{
    \begin{tikzpicture}[scale=0.7]
        % 3x3 square grid (so vertices run from 0 to 3)
        \draw[] (0,0) grid (3,3);
        \node[red] at (1.65,1.3) {$\lambda_3$};
        % Red left boundary, matching the previous square's style
        \foreach \y in {0,1,2} {
            \draw[red, very thick] (2,\y) -- (3,\y);
        }
        % Filled circles at all vertices
        \foreach \x in {0,1,2,3} {
            \foreach \y in {0,1,2,3} {
                \fill (\x,\y) circle (2.5pt);
            }
        }
    \end{tikzpicture}
    }} ,
    \vcenter{\hbox{
    \begin{tikzpicture}[scale=0.7]
        % 3x3 square grid (so vertices run from 0 to 3)
        \draw[] (0,0) grid (3,3);
        \node[red] at (1.3,1.3) {$\lambda_1'$};
        % Red left boundary, matching the previous square's style
        \foreach \x in {0,1,2} {
            \draw[red, very thick] (\x,0) -- (\x,1);
        }
        % Filled circles at all vertices
        \foreach \x in {0,1,2,3} {
            \foreach \y in {0,1,2,3} {
                \fill (\x,\y) circle (2.5pt);
            }
        }
    \end{tikzpicture}
    }} ,\dots,
    \vcenter{\hbox{
    \begin{tikzpicture}[scale=0.7]
        % 3x3 square grid (so vertices run from 0 to 3)
        \draw[] (0,0) grid (3,3);
        \node[red] at (1.3,1.7) {$\lambda_3'$};
        % Red left boundary, matching the previous square's style
        \foreach \x in {0,1,2} {
            \draw[red, very thick] (\x,2) -- (\x,3);
        }
        % Filled circles at all vertices
        \foreach \x in {0,1,2,3} {
            \foreach \y in {0,1,2,3} {
                \fill (\x,\y) circle (2.5pt);
            }
        }
    \end{tikzpicture}
    }}\,.
\end{equation}
The cohomology group $H^1=Z^1/B^1$ here can be obtained by a similar argument to the analysis of the homology group $H_1$, where now we are dealing with cosets of $B^1$ in $Z^1$.
Elements of $H^1$ are equivalence classes of elements of $Z^1$, denoted as $[c^*] = c^*+B^1$ for $c^*\in Z^1$, where $c^*,\Tilde{c}^*$ are equivalent, that is, $c^*\sim\Tilde{c}^*$, whenever they belong to the same coset.
Also, the identity element of $H^1$ is given by $[0^1]=0^1+B^1=B^1$ where $0^1$ is the identity element of group $C^1$.
So, any 1-cochain that is the sum of vertex coboundaries is equivalent to $0^1$.
Again we note that coset $\lambda_i+B^1$ contains all 1-cochain vertical loops, as well as the sum of loops with vertex coboundaries.
For example, $[\lambda_1]$ contains
\begin{equation}\label{eqn:cohomology_loop_plus_boundary}
\begin{gathered}
    \vcenter{\hbox{
    \begin{tikzpicture}[scale=0.7]
        % 3x3 square grid (so vertices run from 0 to 3)
        \draw[] (0,0) grid (3,3);
        \node[red] at (-0.35,1) {$\lambda_1$};
        % Red left boundary, matching the previous square's style
        \foreach \y in {0,1,2} {
            \draw[red, very thick] (0,\y) -- (1,\y);
        }
        % Filled circles at all vertices
        \foreach \x in {0,1,2,3} {
            \foreach \y in {0,1,2,3} {
                \fill (\x,\y) circle (2.5pt);
            }
        }
    \end{tikzpicture}
    }} +
    \vcenter{\hbox{
    \begin{tikzpicture}[scale=0.7]
        % 3x3 square grid (so vertices run from 0 to 3)
        \draw[] (0,0) grid (3,3);
        % Red left boundary, matching the previous square's style
        \foreach \y in {0,1,2} {
            \draw[red, very thick] (0,\y) -- (1,\y);
        }
        \foreach \y in {0,1,2} {
            \draw[red, very thick] (1,\y) -- (2,\y);
        }
        % Filled circles at all vertices
        \foreach \x in {0,1,2,3} {
            \foreach \y in {0,1,2,3} {
                \fill (\x,\y) circle (2.5pt);
            }
        }
    \end{tikzpicture}
    }}
    =
    \vcenter{\hbox{
    \begin{tikzpicture}[scale=0.7]
        % 3x3 square grid (so vertices run from 0 to 3)
        \draw[] (0,0) grid (3,3);
        \node[red] at (0.65,1.3) {$\lambda_2$};
        % Red left boundary, matching the previous square's style
        \foreach \y in {0,1,2} {
            \draw[red, very thick] (1,\y) -- (2,\y);
        }
        % Filled circles at all vertices
        \foreach \x in {0,1,2,3} {
            \foreach \y in {0,1,2,3} {
                \fill (\x,\y) circle (2.5pt);
            }
        }
    \end{tikzpicture}
    }} \\
    \text{and}\\
    \vcenter{\hbox{
    \begin{tikzpicture}[scale=0.7]
        % 3x3 square grid (so vertices run from 0 to 3)
        \draw[] (0,0) grid (3,3);
        \node[red] at (0.65,1.3) {$\lambda_2$};
        % Red left boundary, matching the previous square's style
        \foreach \y in {0,1,2} {
            \draw[red, very thick] (1,\y) -- (2,\y);
        }
        % Filled circles at all vertices
        \foreach \x in {0,1,2,3} {
            \foreach \y in {0,1,2,3} {
                \fill (\x,\y) circle (2.5pt);
            }
        }
    \end{tikzpicture}
    }} +
    \vcenter{\hbox{
    \begin{tikzpicture}[scale=0.7]
        % 3x3 square grid (so vertices run from 0 to 3)
        \draw[] (0,0) grid (3,3);
        \node at (1.65,1.3) {$v_6$};
        % Red left boundary, matching the previous square's style
        \draw[red, very thick] (1,1) -- (2,1);
        \draw[red, very thick] (3,1) -- (2,1);
        \draw[red, very thick] (2,2) -- (2,1);
        \draw[red, very thick] (2,0) -- (2,1);
        % Filled circles at all vertices
        \foreach \x in {0,1,2,3} {
            \foreach \y in {0,1,2,3} {
                \fill (\x,\y) circle (2.5pt);
            }
        }
    \end{tikzpicture}
    }} =
    \vcenter{\hbox{
    \begin{tikzpicture}[scale=0.7]
        % 3x3 square grid (so vertices run from 0 to 3)
        \draw[] (0,0) grid (3,3);
        % Red left boundary, matching the previous square's style
        \draw[red, very thick] (1,2) -- (2,2);
        \draw[red, very thick] (1,0) -- (2,0);
        \draw[red, very thick] (3,1) -- (2,1);
        \draw[red, very thick] (2,2) -- (2,1);
        \draw[red, very thick] (2,0) -- (2,1);
        % Filled circles at all vertices
        \foreach \x in {0,1,2,3} {
            \foreach \y in {0,1,2,3} {
                \fill (\x,\y) circle (2.5pt);
            }
        }
    \end{tikzpicture}
    }}
\end{gathered}
\end{equation}
Similarly, all horizontal loops and its sum with any coboundaries is in $[\lambda_1']$.
Thus the first cohomology group $H^1$ is generated by $[\lambda_1],[\lambda_1']$.

Alternatively, we can define a \emph{dual lattice} of the original square lattice, which we call the \emph{primal lattice}, on the surface of the torus where each vertex of the dual lattice is centered on a square plaquette of the primal lattice, whereas a square plaquette of the dual lattice has a vertex of the original lattice at its center (see~\cref{fig:toric-code-lattice}).
The square lattice is a \emph{self-dual} lattice since its dual lattice is a square lattice.
However, this does not hold for other types of lattices, such as a hexagonal lattice, whose dual lattice is triangular.

\subsection{Toric code}\label{subsec:toric-code}
\begin{figure}
    \centering
    \subfloat[]{\includegraphics[width=0.4\textwidth]{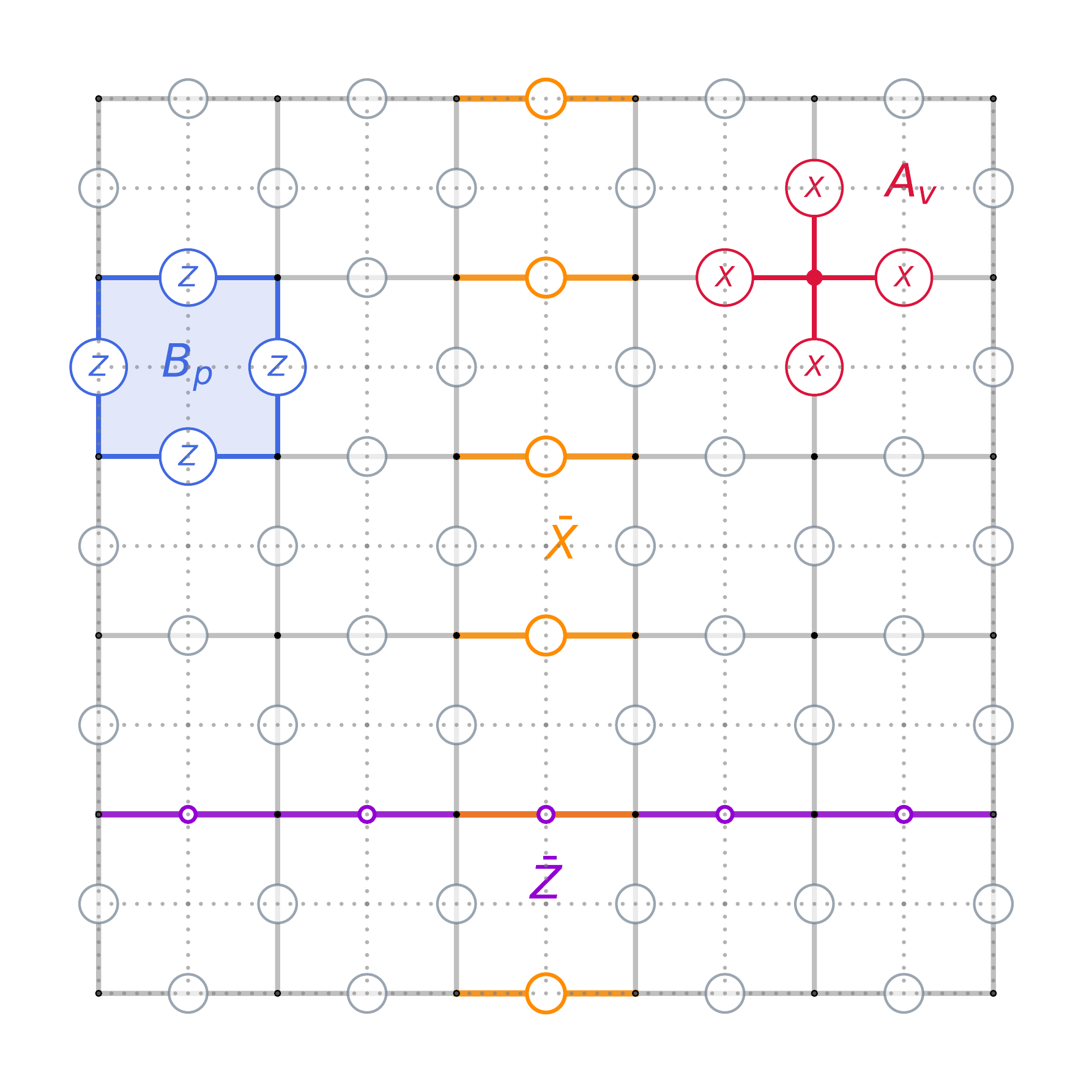}\label{fig:toric-code-lattice}}
    \hspace{8mm}
    \subfloat[]{\includegraphics[width=0.45\textwidth]{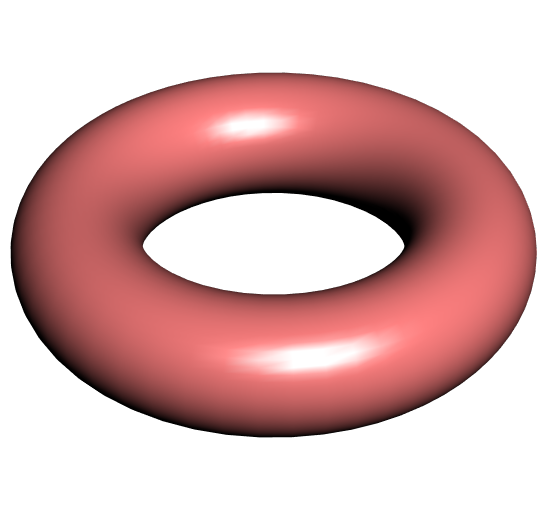}\label{fig:torus}}
    \caption{
    (a) $5 \times 5$ lattice with periodic boundary conditions for the toric code where qubits (hollow circles) are placed on the edges and $Z_p$ and $X_v$ are stabilizer generators comprised of Pauli-$Z$ (blue) and Pauli-$X$ (red) operators, respectively. These operators surround either a vertex or a plaquette. The dashed lines indicate the dual lattice. (b) A torus, shown as a geometric visualization of the periodic boundary conditions in (a). Connecting the left and right edges and the top and bottom edges of the square lattice produces the closed surface with one hole shown in the figure.}
    \label{fig:toric-code-stuff}
\end{figure}

We now introduce our first topological code, the \eczoo[\emph{toric code}]{toric}, which is one of the simplest topological codes.
The toric code is defined on an $L \times L$ square cellulation on a torus which gives a regular square lattice with  periodic boundary conditions as illustrated in~\cref{fig:toric-code-stuff}. 
As we have discussed, the lattice has 0-cells (vertices), 1-cells (edges), and 2-cells (faces or plaquettes). 
With an $L \times L$ lattice, we have $L^2$ vertices, $L^2$ plaquettes, and $L^2 + L^2 = 2L^2$ edges. 
We place physical qubits on the edges of the lattice (the circles in~\cref{fig:toric-code-lattice}) so that the code consists of $n = 2L^2$ physical qubits.

\subsubsection{Stabilizers}\label{subsubsec:toric-code-stabilizers}
The toric code is a CSS code, which means that there are two types of stabilizer generators: weight-4 $Z$-type stabilizers supported on the edges around a square plaquette and weight-4 $X$-type stabilizers supported on the edges around a vertex, which we will often refer to as \emph{plaquette stabilizers} and \emph{vertex stabilizers}, respectively.
Since there are $L^2$ plaquettes and $L^2$ vertices, there are $L^2$ stabilizer generators of each type. 
A plaquette stabilizer $Z_p$ on plaquette $p$ can be defined using the boundary map over the 2-cells (square plaquettes) and a vertex stabilizer $X_v$ can be defined using the coboundary map over the 0-cells (vertices) as
\begin{equation}
\begin{gathered}
    Z_p = \prod_{e\in\partial p} Z(e)
    \quad\text{and}\quad
    X_v = \prod_{e\in\delta v} X(e) \,,
\end{gathered}
\end{equation}
respectively\footnote{Note that here we make a slight abuse of notation in $e \in \partial p$ and $e \in \delta v$ where the boundary $\partial p$ and coboundary $\delta v$ are considered sets of 1-cells (edges) with non-zero coefficient.}, where $Z(e)$ (resp., $X(e)$) is an $n$-qubit Pauli operator with Pauli $Z$ (resp., Pauli $X$) on qubit on edge $e$ and identity everywhere else.
Note that commutation between the plaquette and vertex stabilizers is guaranteed by the action of 1-cochain $\delta^0 v$ on 1-chain $\partial_2 p$ for a vertex $v$ on the corner of a plaquette $p$: 
\begin{equation}
    \<\delta^0 v,\partial_2 p\> = \<v,\partial_1\partial_2 p\> = 0
\end{equation}
since $\partial_1\partial_2=0$.

In our $3\times3$ square lattice example, such operators on plaquette $p$ and vertex $v$ are
\begin{equation}
\begin{gathered}
    Z_p
    = \vcenter{\hbox{
    \begin{tikzpicture}[scale=0.9]
        % 3x3 square grid
        \draw[] (0,0) grid (3,3);
        % Red edges: bottom middle square
        \draw[red, very thick] (1,1) -- (2,1) -- (2,2) -- (1,2) -- cycle;
        % Label
        \node at (1.5,1.5) {$p$};
        \node[circle, draw, fill=white, minimum size=3mm, inner sep=0pt, font=\tiny] at (1.5,1) {$Z$};
        \node[circle, draw, fill=white, minimum size=3mm, inner sep=0pt, font=\tiny] at (1,1.5) {$Z$};
        \node[circle, draw, fill=white, minimum size=3mm, inner sep=0pt, font=\tiny] at (1.5,2) {$Z$};
        \node[circle, draw, fill=white, minimum size=3mm, inner sep=0pt, font=\tiny] at (2,1.5) {$Z$};
        % Filled circles at all vertices
        \foreach \x in {0,1,2,3} {
            \foreach \y in {0,1,2,3} {
                \fill (\x,\y) circle (2.5pt);
            }
        }
    \end{tikzpicture}
    }}
    \quad\text{and}\quad
    X_v
    = \vcenter{\hbox{
    \begin{tikzpicture}[scale=0.9]
        % 3x3 square grid
        \draw[] (0,0) grid (3,3);
        % Red edges: top right square
        \draw[red, very thick] (2,2) -- (3,2);
        \draw[red, very thick] (2,2) -- (1,2);
        \draw[red, very thick] (2,2) -- (2,3);
        \draw[red, very thick] (2,2) -- (2,1);
        % Label
        \node at (2.3,2.25) {$v$};
        \node[circle, draw, fill=white, minimum size=3mm, inner sep=0pt, font=\tiny] at (1.5,2) {$X$};
        \node[circle, draw, fill=white, minimum size=3mm, inner sep=0pt, font=\tiny] at (2,1.5) {$X$};
        \node[circle, draw, fill=white, minimum size=3mm, inner sep=0pt, font=\tiny] at (2.5,2) {$X$};
        \node[circle, draw, fill=white, minimum size=3mm, inner sep=0pt, font=\tiny] at (2,2.5) {$X$};
        % Filled circles at all vertices
        \foreach \x in {0,1,2,3} {
            \foreach \y in {0,1,2,3} {
                \fill (\x,\y) circle (2.5pt);
            }
        }
    \end{tikzpicture}
    }}\,.
\end{gathered}
\end{equation}
Thus, the toric code is a stabilizer code with stabilizer group $\mathcal{S}_\mathrm{tc}$ generated by the plaquette stabilizers $Z_p$ and vertex stabilizers $X_v$:
\begin{equation}
    \mathcal{S}_\mathrm{tc} = \< Z_p,X_v : \text{plaquettes }p\,,\;\text{vertices }v\> \,.
\end{equation}
Since the plaquette and vertex stabilizers are defined by the boundaries and coboundaries of $C_2$ and $C^0$, respectively, we can write each $Z$-type and $X$-type stabilizer in terms of elements of the boundary group $B_1\subseteq C_1$ (1-chains) and coboundary group $B^1\subseteq C^1$ (1-cochains), respectively, as follows
\begin{equation}
    Z_c = \prod_{e\in c} Z(e)
    \quad\text{and}\quad
    X_{c^*} = \prod_{e\in c^*} X(e)
\end{equation}
for $c\in B_1$ and $c^*\in B^1$.
Therefore we can write the entire stabilizer group $\mathcal{S}_\mathrm{tc}$ in terms of the boundary group $B_1$ and coboundary group $B^1$:

\begin{lembox}[label={lemma:stabilizer_boundaries}]{Stabilizer-boundary relation}
    The stabilizer group $\mathcal{S}_\mathrm{tc}$ of a toric code can be expressed as
    \begin{equation}
        \mathcal{S}_\mathrm{tc} = \{Z_cX_{c^*} : c\in B_1 \,,\; c^*\in B^1 \} \,.
    \end{equation}
\end{lembox}

\subsubsection{Encoded logical qubits}\label{subsubsec:toric-code-encoded-logicals}
From its stabilizer group $\mathcal{S}_\mathrm{tc}$, we can then determine the logical operators and the codespace dimension (number of encoded logical qubits) of the toric code through its normalizer in the Pauli group: $\mathcal{N}(\mathcal{S}_\mathrm{tc})$.
We show below that we can write the normalizer in terms of cycle and cocycle groups $Z_1,Z^1$.

\begin{lembox}[label={lemma:normalizer_cycles}]{Normalizer-cycle relation}
    For a toric code stabilizer group $\mathcal{S}_\mathrm{tc}$, its normalizer in the Pauli group is given by
    \begin{equation}
        \mathcal{N}(\mathcal{S}_\mathrm{tc}) = \< \{Z_c,X_{c^*} : c\in Z_1 \,,\; c^*\in Z^1\} \cup \{\pm i\} \> \,.
    \end{equation}
\end{lembox}
\begin{proof}
    Note that for any $c^*\in Z^1$ and $c\in Z_1$ we have
    \begin{equation}\label{eqn:chain_cochain_commute_boundary_coboundary}
        \<c^*,\partial_2 p\> = \<\delta^1 c^*,p\> = 0
        \quad\text{and}\quad
        \<\delta^0 v, c\> = \<v, \partial_1 c\> = 0
    \end{equation}
    for any plaquette $p$ and vertex $v$, since $c^* \in Z^1 = \ker{\delta^1}$ and $c \in Z_1 = \ker{\partial_1}$ by definition.
    So for any $c\in Z_1$ and $c^*\in Z^1$, the Pauli operators $Z_c$ and $X_{c^*}$ commute with any stabilizer in $\mathcal{S}_\mathrm{tc}$ since the boundary group $B_1$ and coboundary group $B^1$ are generated by $\{\partial_2 p\}_p$ and $\{\delta^0 v\}_v$, respectively.
    
    Moreover, the only 1-chains (resp. 1-cochains) satisfying the relations in~\cref{eqn:chain_cochain_commute_boundary_coboundary} are those that belong in $Z_1$ (resp. in $Z^1$).
    This is because a 1-chain that is not in $Z_1$, that is, $c\notin Z_1$ must have vertices in its boundaries, that is, $\partial_1c \neq 0$, whereas a non-cocycle 1-cochain $c^*\notin Z^1$ must have plaquettes in its coboundaries, that is, $\delta^1c^* \neq 0$
    For example,
    \begin{equation}
        c = 
        \vcenter{\hbox{
        \begin{tikzpicture}[scale=0.9]
            % 3x3 square grid
            \draw[] (0,0) grid (3,3);
            % Red edges: top right square
            \draw[red, very thick] (2,2) -- (1,2);
            \draw[red, very thick] (2,2) -- (2,1);
            % Label
            \node[red] at (2.3,1.25) {$v_6$};
            \node[red] at (0.7,2.25) {$v_8$};
            \foreach \x in {0,1,2,3} {
                \foreach \y in {0,1,2,3} {
                    \fill (\x,\y) circle (2.5pt);
                }
            }
        \end{tikzpicture}
        }}
        \quad\text{and}\quad
        c^* =
        \vcenter{\hbox{
        \begin{tikzpicture}[scale=0.9]
            % 3x3 square grid
            \draw[] (0,0) grid (3,3);
            % Red edges: bottom middle square
            \draw[red, very thick] (2,2) -- (1,2) -- (1,1) -- (0,1);
            % Label
            \node[red] at (1.5,2.5) {$p_8$};
            \node[red] at (0.5,0.5) {$p_1$};
            % Filled circles at all vertices
            \foreach \x in {0,1,2,3} {
                \foreach \y in {0,1,2,3} {
                    \fill (\x,\y) circle (2.5pt);
                }
            }
        \end{tikzpicture}
        }} \,.
    \end{equation}
    So, for the 1-chain $c$ we have
    \begin{equation}
    \begin{gathered}
        \<\delta^0 v_6,c\> = \<v_6,\partial_1c\> = \<v_6,v_6+v_8\> \neq0
        \quad\text{and}\quad
        \<\delta^0 v_8,c\> = \<v_8,\partial_1c\> = \<v_8,v_6+v_8\> \neq0 \,,
    \end{gathered}
    \end{equation}
    whereas for 1-cochain $c^*$ we have
    \begin{equation}
        \<c^*,\partial_2 p_1\> = \<\delta^1c^*, p_1\> = \<p_1+p_8 , p_1\> \neq0
        \quad\text{and}\quad
        \<c^*,\partial_2 p_8\> = \<\delta^1c^*, p_8\> = \<p_1+p_8 , p_8\> \neq0 \,.
    \end{equation}
    In a similar fashion, for any non-cycle 1-chain $c$ and any non-cocycle 1-cochain $c^*$, there exists some plaquette $p$ and some vertex $v$ such that $\<c^*,\partial_2 p\>\neq0$ and $\<\delta^0 v, c\>\neq0$.

    Finally, note that we can write the cycle and cocycle groups as
    \begin{equation}
    \begin{gathered}
        Z_1 = \{c\in C_1 : \<\delta^0v,c\>=0 \,, \text{ for all vertices }v\}
        \\\text{and}\\
        Z^1 = \{c^*\in C^1 : \<c^*,\partial_2 p\>=0 \,, \text{ for all plaquettes }p\} \,.
    \end{gathered}
    \end{equation}
    Therefore for any $c\in C_1$ and any $c^*\in C^1$, it holds that
    \begin{equation}
    \begin{gathered}
        Z_cX_v=X_vZ_c\,,\text{ for all vertices }v \Leftrightarrow c\in Z_1
        \\\text{and}\\
        X_{c^*}Z_p=Z_pX_{c^*}\,,\text{ for all plaquettes }p \Leftrightarrow c^*\in Z^1 \,.
    \end{gathered}
    \end{equation}
    Since $\mathcal{S}_\mathrm{tc}$ is generated by the $Z_p$'s and $X_v$'s, and by considering Paulis with phases, the claim of the Lemma follows.
\end{proof}

Now let us recall that the logical operators of a stabilizer code with stabilizer group $\mathcal{S}$ are elements of the quotient group of $\mathcal{N}(\mathcal{S})/\mathcal{S}$.
In the toric code, the quotient group $\mathcal{N}(\mathcal{S}_\mathrm{tc})/\mathcal{S}_\mathrm{tc}$ consists of cosets of $\mathcal{S}_\mathrm{tc}$ in $\mathcal{N}(\mathcal{S}_\mathrm{tc})$ of the form\footnote{Note that this form of elements of the normalizer holds for CSS stabilizer codes, but not necessarily for general stabilizer codes. 
This is implicit in the stabilizer-boundary relation in~\cref{lemma:stabilizer_boundaries} and normalizer-cycle relation in~\cref{lemma:normalizer_cycles} that the $X$- and $Z$- type Paulis are ``independently'' generated.
The toric code is but a special case of this more general property of the CSS codes.
}
\begin{equation}
    \begin{aligned}
        P \mathcal{S}_\mathrm{tc} &= \{Ps : s\in\mathcal{S}_\mathrm{tc} \} 
        = \{PZ_bX_{b^*} : b\in B_1 \,,\; b^*\in B^1 \}
    \end{aligned}
\end{equation}
for $P\in\mathcal{N}(\mathcal{S}_\mathrm{tc})$, by~\cref{lemma:stabilizer_boundaries} linking the stabilizers to the boundaries $B_1$ and coboundaries $B^1$.
We can then write the elements of normalizer $\mathcal{N}(\mathcal{S}_\mathrm{tc})$ in terms of cycles $Z_1$ and cocycles $Z^1$, so that these cosets can be written as
\begin{equation}
\begin{gathered}
    \{ i^a Z_{c+b}X_{c^*+b^*} : b\in B_1 \,,\; b^*\in B^1 \} = 
    i^a\cdot [Z_c] \cdot [X_{c^*}] \\
    \text{for}\\
    [Z_c] = \{Z_{c+b} : b\in B_1\}
    \quad\text{and}\quad
    [X_{c^*}] = \{X_{c^*+b^*} : b^*\in B^1\}
\end{gathered}
\end{equation}
for some $a\in\{0,\dots,3\}$, $c\in Z_1$, and $c^*\in Z^1$ (here we used the fact that the pair $Z_b,X_{c^*}$ commute).

Now consider the direct sum between the homology and cohomology group:
\begin{equation}
    H_1 \oplus H^1 = \{g \oplus g^* : g\in H_1 \,,\; g^*\in H^1\}
\end{equation}
with group operation ``$+$'' given by $(g\oplus g') + (h\oplus h') = (g+h)\oplus(g'+h')$ for $g,h\in H_1$ and $g',h'\in H^1$.
Now we show that the group $H_1 \oplus H^1$ is isomorphic to the group $\mathcal{L}_\mathrm{tc}$, which is the quotient group $\mathcal{N}(\mathcal{S}_\mathrm{tc})/\mathcal{S}_\mathrm{tc}$ up to phases, that consists of elements of the form $[Z_c] \cdot [X_{c^*}]$ with group operation ``$\ast$'' given by
\begin{equation}
    ([Z_c] \cdot [X_{c^*}]) \ast ([Z_a] \cdot [X_{a^*}]) = [Z_{c+a}] \cdot [X_{c^*+a^*}] \,.
\end{equation}

\begin{lembox}[label={lemma:isomorphism_logical_homology_oplus_cohomology}]{$\mathcal{L}_\mathrm{tc}$ - $H_1 \oplus H^1$ isomorphism}
    The group $\mathcal{L}_\mathrm{tc}$ (the quotient group $\mathcal{N}(\mathcal{S}_\mathrm{tc})/\mathcal{S}_\mathrm{tc}$, up to phases) is isomorphic to $H_1 \oplus H^1$.
\end{lembox}

\begin{proof}
    Consider a map $\varphi$ between $\mathcal{L}_\mathrm{tc}$ and group $H_1 \oplus H^1$ defined by
    \begin{equation}
        \varphi([Z_c] \cdot [X_{c^*}]) = [c]\oplus[c^*] \,.
    \end{equation}
    $\varphi$ is well-defined because if $\varphi([Z_c] \cdot [X_{c^*}]) = [c]\oplus[c^*]$ and $\varphi([Z_c] \cdot [X_{c^*}]) = [a]\oplus[a^*]$, then
    \begin{equation}
        [a]\oplus[a^*] = \varphi([Z_c] \cdot [X_{c^*}]) = [c]\oplus[c^*]
    \end{equation}
    which means that $[a]=[c]$ and $[a^*]=[c^*]$.
    Now we will show that $\varphi$ is a one-to-one function (a bijection) and that it is a group homomorphism.

    To show that $\varphi$ is bijective, we show that it is both injective and surjective.
    First, we show that $\varphi$ is injective, namely that, given any distinct $g,g'\in\mathcal{L}_\mathrm{tc}$, we have $\varphi(g)\neq\varphi(g')$.
    Note that for any pair of distinct elements $[Z_a]\cdot[X_{a^*}],[Z_c]\cdot[X_{c^*}]$ of $\mathcal{L}_\mathrm{tc}$ it holds that $[Z_a]\neq[Z_c]$ or $[X_{a^*}]\neq[X_{c^*}]$.
    So we have 
    \begin{equation}
    \begin{aligned}
        \varphi([Z_c] \cdot [X_{c^*}]) = [c]\oplus[c^*] \neq [a]\oplus[a^*] = \varphi([Z_a] \cdot [X_{a^*}])
    \end{aligned}
    \end{equation}
    since $[Z_a]\neq[Z_c]$ implies $[a]\neq[c]$ and $[X_{a^*}]\neq[X_{c^*}]$ implies $[a^*]\neq[c^*]$.
    Now, to show that $\varphi$ is surjective, we need to show that for any $[c]\oplus[c^*]\in H_1\oplus H^1$, there exists some $[Z_c]\cdot[X_{c^*}]\in\mathcal{L}_\mathrm{tc}$ such that $\varphi([Z_c] \cdot [X_{c^*}]) = [c]\oplus[c^*]$.
    This follows directly from the definition of $\varphi$ that for any $[c]\oplus[c^*]\in H_1\oplus H^1$ we have
    \begin{equation}
        [c]\oplus[c^*] = \varphi([Z_c] \cdot [X_{c^*}]) \,.
    \end{equation}
    Lastly, to show that $\varphi$ is a group homomorphism, consider again distinct $Z_a,Z_c$ and distinct $X_{a^*},X_{c^*}$ so that
    \begin{equation}
    \begin{aligned}
        \varphi\big( ([Z_c] \cdot [X_{c^*}]) \ast ([Z_a] \cdot [X_{a^*}]) \big) &= \varphi( [Z_{c+a}] \cdot [X_{c^*+a^*}] ) \\
        &= [c+a]\oplus[c^*+a^*] \\
        &= ([c]\oplus[c^*]) + ([a]\oplus[a^*])
    \end{aligned}
    \end{equation}
    since $[Z_a]\neq[Z_c]$ implies $[a]\neq[c]$ and $[X_{a^*}]\neq[X_{c^*}]$ implies $[a^*]\neq[c^*]$ (i.e. $a\notin[c]$ and $a^*\notin[c^*]$).
    Hence we have shown that $\varphi$ is an isomorphism between $H_1 \oplus H^1$ and $\mathcal{L}_\mathrm{tc}$, up to phases.
\end{proof}

Recall our discussion at the end of~\cref{subsubsec:topological-homology} that $H_1$ is generated by cosets of 1-cycle loops $[\ell_1],[\ell_1']$ and $H^1$ is generated by cosets of 1-cocycle loops $[\lambda_1],[\lambda_1']$.
By the isomorphism in~\cref{lemma:isomorphism_logical_homology_oplus_cohomology}, we can write $\mathcal{L}_\mathrm{tc}$ as
\begin{equation}
    \mathcal{L}_\mathrm{tc} = \< [Z_{\ell_1}],[Z_{\ell_1'}],[X_{\lambda_1}],[X_{\lambda_1'}] \> \,,
\end{equation}
that is, it is generated by $4=2\times2$ independent elements of the form $[Z_\ell]\cdot[X_\lambda]$, and therefore the toric code encodes $k=2$ logical qubits. Alternatively, we can determine the number of logical qubits by counting the number of independent stabilizers in the toric code.
Note that the number of independent vertex (resp., plaquette) stabilizers in the toric code is $L^2 - 1$ since the product of all vertex (resp., plaquette) stabilizers is the identity.
So, we can directly calculate the number of logical qubits $k$ as
\begin{equation}
    k = n- 2(L^2-1) = 2L^2-2L^2+2 = 2 \,.
\end{equation}
The number of logical qubits encoded in the toric code can also be verified using the topological invariant in~\cref{subsec:topological-primer-topology}, namely the Euler characteristic $\chi$.
Recall that the number of logical qubits is given by $k = n - m$, where $n$ is the number of physical qubits and $m$ is the number of independent stabilizers. Note that $n = E$ by construction and $m = (V-1) + (F-1)$, where the first term is the number of independent vertex stabilizers and the second term is the number of independent plaquette stabilizers, where $V = F = L^2$. 
Then, the number of encoded logical qubits is
\begin{equation}
    k = E - V - F + 2\,.
\end{equation}
By rearranging and plugging in the Euler characteristic formula given in~\cref{eqn:euler_torus}, we obtain
\begin{equation}
    k = 2 - \chi\,.
\end{equation}
If we take $\chi = 2 - 2g$ for a genus-$g$ surface, then the number of encoded qubits is again $k = 2g = 2$, since the torus is a genus $g=1$ surface.

By identifying the independent generators of $\mathcal{L}_\mathrm{tc}$, we can also identify the $X$- and $Z$- logical operators and find the ones with minimum weight to determine the code distance.
Note that for a 1-chain $c$, the weight of $Z_c$ is simply the number of edges (1-cells) in $c$.
For example, for a 1-chain $c = e_1+e_2+e_3$, the weight of $Z_c = Z(e_1)Z(e_2)Z(e_3)$ is three.
Similar statements can be made for 1-cochain $c^*$ and $X_{c^*}$. 
Now note that for a toric code, the coset $[\ell_1]$ consists of a non-contractible 1-chain loop of the form $\ell_1+b$ for a boundary $b\in B_1$, whereas $[\lambda_1]$ consists of a non-contractible 1-cochain loop of the form $\lambda_1+b^*$ for a coboundary $b^*\in B^1$; see, for example,~\cref{eqn:homology_loop_plus_boundary,eqn:cohomology_loop_plus_boundary}.
Note that non-contractible loops in $\{\ell_i\}_{i=1}^3\subseteq[\ell_1]$ and $\{\lambda_i\}_{i=1}^3\subseteq[\lambda_1]$ are the ones with the least number of 1-cells (edges) in it.
On an $L\times L$ torus, each of these non-contractible 1-chains 1-cochains has exactly $L$ 1-cells (edges) in it.
Thus, for an $L \times L$ toric code, its code distance must be $d=L$.
Therefore, the toric code has parameters $\llbracket 2L^2, 2, L\rrbracket$.

\subsubsection{Detecting and correcting errors}\label{subsubsec:toric-code-detecting-errors}
The final thing about the toric code that we discuss is how errors are detected and corrected.
Since the toric code is a CSS code, we can treat bit-flip and phase-flip errors separately. The $X$-type errors are detected by plaquette measurements, whereas $Z$-type errors are detected by vertex measurements. To see this, consider first a single $Z$ error on a qubit (edge).
A $Z$ operator commutes with every plaquette operator, but it anticommutes with the two vertex operators adjacent to that edge, since each of those contains a Pauli-$X$ acting on the same qubit. 
Therefore, a single $Z$ error flips the measurement outcome of exactly two vertex stabilizers from $+1$ to $-1$. 
In the $3\times3$ toric code example, this can be illustrated as:
\begin{equation}
    \vcenter{\hbox{
    \begin{tikzpicture}[scale=0.9]
        % 3x3 square grid
        \draw[] (0,0) grid (3,3);
        % Red edges: top right square
        % \draw[red, very thick] (2,2) -- (3,2);
        \draw[red, very thick] (2,2) -- (1,2);
        % \draw[red, very thick] (2,2) -- (2,3);
        % \draw[red, very thick] (2,2) -- (2,1);
        % Label
        \node[red] at (2.3,2.25) {$v_9$};
        \node[red] at (0.7,2.25) {$v_8$};
        \node[circle, red, draw, fill=white, minimum size=3mm, inner sep=0pt, font=\tiny] at (1.5,2) {$Z$};
        % Filled circles at all vertices
        \foreach \x in {0,1,2,3} {
            \foreach \y in {0,1,2,3} {
                \fill (\x,\y) circle (2.5pt);
            }
        }
    \end{tikzpicture}
    }}\,,
\end{equation}
where the $Z$ error flips vertex stabilizers $X_{v_8}$ and $X_{v_9}$.
Now consider two adjacent $Z$ errors, that is, the edges where the $Z$ errors happened share a common vertex:
\begin{equation}
    \vcenter{\hbox{
    \begin{tikzpicture}[scale=0.9]
        % 3x3 square grid
        \draw[] (0,0) grid (3,3);
        % Red edges: top right square
        % \draw[red, very thick] (2,2) -- (3,2);
        \draw[red, very thick] (2,2) -- (1,2);
        % \draw[red, very thick] (2,2) -- (2,3);
        \draw[red, very thick] (2,2) -- (2,1);
        % Label
        \node at (2.3,2.25) {$v_9$};
        \node[red] at (2.3,1.25) {$v_6$};
        \node[red] at (0.7,2.25) {$v_8$};
        \node[circle, red, draw, fill=white, minimum size=3mm, inner sep=0pt, font=\tiny] at (1.5,2) {$Z$};
        \node[circle, red, draw, fill=white, minimum size=3mm, inner sep=0pt, font=\tiny] at (2,1.5) {$Z$};
        % Filled circles at all vertices
        \foreach \x in {0,1,2,3} {
            \foreach \y in {0,1,2,3} {
                \fill (\x,\y) circle (2.5pt);
            }
        }
    \end{tikzpicture}
    }}\,.
\end{equation}
In this case, the vertex stabilizer $X_{v_9}$ is no longer flipped since the two $Z$ errors commute with it, however there are still two flipped vertex stabilizers: $X_{v_8}$ and $X_{v_6}$.
If there are $Z$ errors on two non-adjacent edges, we have
\begin{equation}
    \vcenter{\hbox{
    \begin{tikzpicture}[scale=0.9]
        % 3x3 square grid
        \draw[] (0,0) grid (3,3);
        % Red edges: top right square
        % \draw[red, very thick] (2,2) -- (3,2);
        \draw[red, very thick] (2,2) -- (1,2);
        % \draw[red, very thick] (2,2) -- (2,3);
        \draw[red, very thick] (2,0) -- (2,1);
        % Label
        \node[red] at (2.3,2.25) {$v_9$};
        \node[red] at (2.3,1.25) {$v_6$};
        \node[red] at (2.3,0.25) {$v_3$};
        \node[red] at (0.7,2.25) {$v_8$};
        \node[circle, red, draw, fill=white, minimum size=3mm, inner sep=0pt, font=\tiny] at (1.5,2) {$Z$};
        \node[circle, red, draw, fill=white, minimum size=3mm, inner sep=0pt, font=\tiny] at (2,0.5) {$Z$};
        % Filled circles at all vertices
        \foreach \x in {0,1,2,3} {
            \foreach \y in {0,1,2,3} {
                \fill (\x,\y) circle (2.5pt);
            }
        }
    \end{tikzpicture}
    }}\,,
\end{equation}
where the flipped vertex stabilizers are the pairs at the endpoints of the two strings of $Z$ errors.
In general, any combination of $Z$ errors flips an even number of vertex stabilizers.

Similarly, a single $X$ error anticommutes with the two plaquette operators adjacent to the corresponding edge, and so it creates two violated plaquette stabilizers. 
In the $3\times3$ toric code, examples of $X$ errors can be illustrated as:
\begin{equation}
    \vcenter{\hbox{
    \begin{tikzpicture}[scale=0.9]
        % 3x3 square grid
        \draw[] (0,0) grid (3,3);
        % Red edges: bottom middle square
        \draw[red, very thick] (2,2) -- (1,2);
        % Label
        \node[red] at (1.5,1.5) {$p_5$};
        \node[red] at (1.5,2.5) {$p_8$};
        \node[circle, red, draw, fill=white, minimum size=3mm, inner sep=0pt, font=\tiny] at (1.5,2) {$X$};
        % Filled circles at all vertices
        \foreach \x in {0,1,2,3} {
            \foreach \y in {0,1,2,3} {
                \fill (\x,\y) circle (2.5pt);
            }
        }
    \end{tikzpicture}
    }}
    \quad\text{and}\quad
    \vcenter{\hbox{
    \begin{tikzpicture}[scale=0.9]
        % 3x3 square grid
        \draw[] (0,0) grid (3,3);
        % Red edges: bottom middle square
        \draw[red, very thick] (2,2) -- (1,2) -- (1,1) -- (0,1);
        % Label
        \node at (1.5,1.5) {$p_5$};
        \node[red] at (1.5,2.5) {$p_8$};
        \node at (0.5,1.5) {$p_4$};
        \node[red] at (0.5,0.5) {$p_1$};
        \node[circle, red, draw, fill=white, minimum size=3mm, inner sep=0pt, font=\tiny] at (1.5,2) {$X$};
        \node[circle, red, draw, fill=white, minimum size=3mm, inner sep=0pt, font=\tiny] at (1,1.5) {$X$};
        \node[circle, red, draw, fill=white, minimum size=3mm, inner sep=0pt, font=\tiny] at (0.5,1) {$X$};
        % Filled circles at all vertices
        \foreach \x in {0,1,2,3} {
            \foreach \y in {0,1,2,3} {
                \fill (\x,\y) circle (2.5pt);
            }
        }
    \end{tikzpicture}
    }}\,,
\end{equation}
where the single $X$ error flips plaquette stabilizers $Z_{p_5}$ and $Z_{p_8}$, whereas the string of three $X$ errors flips plaquette stabilizers $Z_{p_1}$ and $Z_{p_8}$.
Finally, a $Y$ error is simply the product of an $X$ and a $Z$ error (up to a phase), and so it flips both types of stabilizers.
For example,
\begin{equation}
    \vcenter{\hbox{
    \begin{tikzpicture}[scale=0.9]
        % 3x3 square grid
        \draw[] (0,0) grid (3,3);
        % Red edges: bottom middle square
        \draw[red, very thick] (2,2) -- (1,2);
        % Label
        \node[red] at (1.5,1.5) {$p_5$};
        \node[red] at (1.5,2.5) {$p_8$};
        \node[red] at (2.3,2.25) {$v_9$};
        \node[red] at (0.7,2.25) {$v_8$};
        \node[circle, red, draw, fill=white, minimum size=3mm, inner sep=0pt, font=\tiny] at (1.5,2) {$Y$};
        % Filled circles at all vertices
        \foreach \x in {0,1,2,3} {
            \foreach \y in {0,1,2,3} {
                \fill (\x,\y) circle (2.5pt);
            }
        }
    \end{tikzpicture}
    }}
    \quad\text{and}\quad
    \vcenter{\hbox{
    \begin{tikzpicture}[scale=0.9]
        % 3x3 square grid
        \draw[] (0,0) grid (3,3);
        % Red edges: bottom middle square
        \draw[red, very thick] (2,1) -- (2,2) -- (1,2);
        % Label
        \node at (1.5,1.5) {$p_5$};
        \node[red] at (1.5,2.5) {$p_8$};
        \node at (2.3,2.25) {$v_9$};
        \node[red] at (0.7,2.25) {$v_8$};
        \node[red] at (2.5,1.5) {$p_6$};
        \node[red] at (2.3,0.75) {$v_6$};
        \node[circle, red, draw, fill=white, minimum size=3mm, inner sep=0pt, font=\tiny] at (1.5,2) {$Y$};
        \node[circle, red, draw, fill=white, minimum size=3mm, inner sep=0pt, font=\tiny] at (2,1.5) {$Y$};
        % Filled circles at all vertices
        \foreach \x in {0,1,2,3} {
            \foreach \y in {0,1,2,3} {
                \fill (\x,\y) circle (2.5pt);
            }
        }
    \end{tikzpicture}
    }}\,.
\end{equation}
Naturally we can generalize this to any chain of $X$ (resp., $Z$) errors where it flips an even number of plaquette (resp., vertex) stabilizers.
In the primal-dual lattice picture, $Z$ errors form strings on the edges of the primal lattice, while $X$ errors form strings on the edges of the dual lattice.

Error correction proceeds by first measuring all stabilizer generators to obtain error syndromes used by the decoder to infer errors.
An error (which may happen on multiple qubits/edges) therefore causes syndrome flips on vertices and plaquettes.
As we have established in the discussion above, the number of syndromes from errors is even.
A common decoding strategy pairs syndromes globally and chooses a likely set of recovery paths by assigning error probabilities on each edge as its weight.
For example, if the probability of $X$ and $Z$ error on any edge is equal the inferred corrections are
\begin{equation}
\begin{gathered}
    \vcenter{\hbox{
    \begin{tikzpicture}[scale=0.9]
        % 3x3 square grid
        \draw[] (0,0) grid (3,3);
        % Red edges: top right square
        \draw[red, very thick] (2,2) -- (1,2);
        \draw[red, very thick] (2,2) -- (2,1);
        % Label
        \node[red] at (2.3,1.25) {$v_6$};
        \node[red] at (0.7,2.25) {$v_8$};
        \node[circle, red, draw, fill=white, minimum size=3mm, inner sep=0pt, font=\tiny] at (1.5,2) {$Z$};
        \node[circle, red, draw, fill=white, minimum size=3mm, inner sep=0pt, font=\tiny] at (2,1.5) {$Z$};
        % Filled circles at all vertices
        \foreach \x in {0,1,2,3} {
            \foreach \y in {0,1,2,3} {
                \fill (\x,\y) circle (2.5pt);
            }
        }
    \end{tikzpicture}
    }} +
    \vcenter{\hbox{
    \begin{tikzpicture}[scale=0.9]
        % 3x3 square grid
        \draw[] (0,0) grid (3,3);
        % Red edges: top right square
        \draw[blue, very thick] (2,1) -- (1,1) -- (1,2);
        % Label
        \node[red] at (2.3,1.25) {$v_6$};
        \node[red] at (0.7,2.25) {$v_8$};
        \node[circle, blue, draw, fill=white, minimum size=3mm, inner sep=0pt, font=\tiny] at (1.5,1) {$Z$};
        \node[circle, blue, draw, fill=white, minimum size=3mm, inner sep=0pt, font=\tiny] at (1,1.5) {$Z$};
        % Filled circles at all vertices
        \foreach \x in {0,1,2,3} {
            \foreach \y in {0,1,2,3} {
                \fill (\x,\y) circle (2.5pt);
            }
        }
    \end{tikzpicture}
    }}
    =
    \vcenter{\hbox{
    \begin{tikzpicture}[scale=0.9]
        % 3x3 square grid
        \draw[] (0,0) grid (3,3);
        % Red edges: top right square
        \draw[red, very thick] (2,2) -- (1,2);
        \draw[red, very thick] (2,2) -- (2,1);
        \draw[blue, very thick] (2,1) -- (1,1) -- (1,2);
        % Label
        \node at (2.3,1.25) {$v_6$};
        \node at (0.7,2.25) {$v_8$};
        \node at (1.5,1.5) {$p_5$};
        \node[circle, red, draw, fill=white, minimum size=3mm, inner sep=0pt, font=\tiny] at (1.5,2) {$Z$};
        \node[circle, red, draw, fill=white, minimum size=3mm, inner sep=0pt, font=\tiny] at (2,1.5) {$Z$};
        \node[circle, blue, draw, fill=white, minimum size=3mm, inner sep=0pt, font=\tiny] at (1.5,1) {$Z$};
        \node[circle, blue, draw, fill=white, minimum size=3mm, inner sep=0pt, font=\tiny] at (1,1.5) {$Z$};
        % Filled circles at all vertices
        \foreach \x in {0,1,2,3} {
            \foreach \y in {0,1,2,3} {
                \fill (\x,\y) circle (2.5pt);
            }
        }
    \end{tikzpicture}
    }}
    \\\text{and}\\
    \vcenter{\hbox{
    \begin{tikzpicture}[scale=0.9]
        % 3x3 square grid
        \draw[] (0,0) grid (3,3);
        % Red edges: bottom middle square
        \draw[red, very thick] (2,2) -- (1,2);
        % Label
        \node[red] at (1.5,1.5) {$p_5$};
        \node[red] at (1.5,2.5) {$p_8$};
        \node[circle, red, draw, fill=white, minimum size=3mm, inner sep=0pt, font=\tiny] at (1.5,2) {$X$};
        % Filled circles at all vertices
        \foreach \x in {0,1,2,3} {
            \foreach \y in {0,1,2,3} {
                \fill (\x,\y) circle (2.5pt);
            }
        }
    \end{tikzpicture}
    }} +
    \vcenter{\hbox{
    \begin{tikzpicture}[scale=0.9]
        % 3x3 square grid
        \draw[] (0,0) grid (3,3);
        % Red edges: bottom middle square
        \draw[blue, very thick] (2,2) -- (1,2);
        % Label
        \node[red] at (1.5,1.5) {$p_5$};
        \node[red] at (1.5,2.5) {$p_8$};
        \node[circle, blue, draw, fill=white, minimum size=3mm, inner sep=0pt, font=\tiny] at (1.5,2) {$X$};
        % Filled circles at all vertices
        \foreach \x in {0,1,2,3} {
            \foreach \y in {0,1,2,3} {
                \fill (\x,\y) circle (2.5pt);
            }
        }
    \end{tikzpicture}
    }}
    =
    \vcenter{\hbox{
    \begin{tikzpicture}[scale=0.9]
        % 3x3 square grid
        \draw[] (0,0) grid (3,3);
        % Red edges: bottom middle square
        \draw[ very thick] (2,2) -- (1,2);
        % Label
        \node at (1.5,1.5) {$p_5$};
        \node at (1.5,2.5) {$p_8$};
        \node[circle, draw, fill=white, minimum size=3mm, inner sep=0pt, font=\tiny] at (1.5,2) {$I$};
        % Filled circles at all vertices
        \foreach \x in {0,1,2,3} {
            \foreach \y in {0,1,2,3} {
                \fill (\x,\y) circle (2.5pt);
            }
        }
    \end{tikzpicture}
    }}\,,
\end{gathered}
\end{equation}
where the product between the correction on the blue edges for the 2-qubit $Z$ error on the red edges results in a 4-qubit $Z$ Pauli operator that is equal to the plaquette stabilizer $Z_{p_5}$, thus it acts trivially on any codestate. 
On the other hand, the correction on the single-qubit $X$ error, directly cancels it out to an identity on the faulty qubit.

Note that distinct errors may flip the same stabilizers.
For example, the following distinct two-qubit $Z$ error flips stabilizers $X_{v_8}$ and $X_{v_6}$:
\begin{equation}
    \vcenter{\hbox{
    \begin{tikzpicture}[scale=0.9]
        % 3x3 square grid
        \draw[] (0,0) grid (3,3);
        % Red edges: top right square
        \draw[red, very thick] (2,2) -- (1,2);
        \draw[red, very thick] (2,2) -- (2,1);
        % Label
        \node[red] at (2.3,1.25) {$v_6$};
        \node[red] at (0.7,2.25) {$v_8$};
        \node[circle, red, draw, fill=white, minimum size=3mm, inner sep=0pt, font=\tiny] at (1.5,2) {$Z$};
        \node[circle, red, draw, fill=white, minimum size=3mm, inner sep=0pt, font=\tiny] at (2,1.5) {$Z$};
        % Filled circles at all vertices
        \foreach \x in {0,1,2,3} {
            \foreach \y in {0,1,2,3} {
                \fill (\x,\y) circle (2.5pt);
            }
        }
    \end{tikzpicture}
    }}
    \quad\text{and}\quad
    \vcenter{\hbox{
    \begin{tikzpicture}[scale=0.9]
        % 3x3 square grid
        \draw[] (0,0) grid (3,3);
        % Red edges: top right square
        \draw[red, very thick] (2,1) -- (1,1) -- (1,2);
        % Label
        \node[red] at (2.3,1.25) {$v_6$};
        \node[red] at (0.7,2.25) {$v_8$};
        \node[circle, red, draw, fill=white, minimum size=3mm, inner sep=0pt, font=\tiny] at (1.5,1) {$Z$};
        \node[circle, red, draw, fill=white, minimum size=3mm, inner sep=0pt, font=\tiny] at (1,1.5) {$Z$};
        % Filled circles at all vertices
        \foreach \x in {0,1,2,3} {
            \foreach \y in {0,1,2,3} {
                \fill (\x,\y) circle (2.5pt);
            }
        }
    \end{tikzpicture}
    }} \,.
\end{equation}
However, this is not an issue in decoding when the error probability on any edge is equal, since the two possible shortest strings between $v_6$ and $v_8$ are
\begin{equation}
    \vcenter{\hbox{
    \begin{tikzpicture}[scale=0.9]
        % 3x3 square grid
        \draw[] (0,0) grid (3,3);
        % Red edges: top right square
        \draw[blue, very thick] (2,1) -- (2,2) -- (1,2);
        % Label
        \node[red] at (2.3,1.25) {$v_6$};
        \node[red] at (0.7,2.25) {$v_8$};
        \node[circle, blue, draw, fill=white, minimum size=3mm, inner sep=0pt, font=\tiny] at (1.5,2) {$Z$};
        \node[circle, blue, draw, fill=white, minimum size=3mm, inner sep=0pt, font=\tiny] at (2,1.5) {$Z$};
        % Filled circles at all vertices
        \foreach \x in {0,1,2,3} {
            \foreach \y in {0,1,2,3} {
                \fill (\x,\y) circle (2.5pt);
            }
        }
    \end{tikzpicture}
    }}
    \quad\text{and}\quad
    \vcenter{\hbox{
    \begin{tikzpicture}[scale=0.9]
        % 3x3 square grid
        \draw[] (0,0) grid (3,3);
        % Red edges: top right square
        \draw[blue, very thick] (2,1) -- (1,1) -- (1,2);
        % Label
        \node[red] at (2.3,1.25) {$v_6$};
        \node[red] at (0.7,2.25) {$v_8$};
        \node[circle, blue, draw, fill=white, minimum size=3mm, inner sep=0pt, font=\tiny] at (1.5,1) {$Z$};
        \node[circle, blue, draw, fill=white, minimum size=3mm, inner sep=0pt, font=\tiny] at (1,1.5) {$Z$};
        % Filled circles at all vertices
        \foreach \x in {0,1,2,3} {
            \foreach \y in {0,1,2,3} {
                \fill (\x,\y) circle (2.5pt);
            }
        }
    \end{tikzpicture}
    }} \,.
\end{equation}
Whichever two-qubit $Z$ error occurred, any of the two-qubit $Z$ corrections either cancel out the error or turn it into the plaquette stabilizer $Z_{p_5}$, that is, a contractible boundary:
\begin{equation}
\begin{gathered}
    \vcenter{\hbox{
    \begin{tikzpicture}[scale=0.9]
        % 3x3 square grid
        \draw[] (0,0) grid (3,3);
        % Red edges: top right square
        \draw[red, very thick] (2,2) -- (1,2);
        \draw[red, very thick] (2,2) -- (2,1);
        % Label
        \node[red] at (2.3,1.25) {$v_6$};
        \node[red] at (0.7,2.25) {$v_8$};
        \node[circle, red, draw, fill=white, minimum size=3mm, inner sep=0pt, font=\tiny] at (1.5,2) {$Z$};
        \node[circle, red, draw, fill=white, minimum size=3mm, inner sep=0pt, font=\tiny] at (2,1.5) {$Z$};
        % Filled circles at all vertices
        \foreach \x in {0,1,2,3} {
            \foreach \y in {0,1,2,3} {
                \fill (\x,\y) circle (2.5pt);
            }
        }
    \end{tikzpicture}
    }} +
    \vcenter{\hbox{
    \begin{tikzpicture}[scale=0.9]
        % 3x3 square grid
        \draw[] (0,0) grid (3,3);
        % Red edges: top right square
        \draw[blue, very thick] (2,1) -- (2,2) -- (1,2);
        % Label
        \node[red] at (2.3,1.25) {$v_6$};
        \node[red] at (0.7,2.25) {$v_8$};
        \node[circle, blue, draw, fill=white, minimum size=3mm, inner sep=0pt, font=\tiny] at (1.5,2) {$Z$};
        \node[circle, blue, draw, fill=white, minimum size=3mm, inner sep=0pt, font=\tiny] at (2,1.5) {$Z$};
        % Filled circles at all vertices
        \foreach \x in {0,1,2,3} {
            \foreach \y in {0,1,2,3} {
                \fill (\x,\y) circle (2.5pt);
            }
        }
    \end{tikzpicture}
    }} =
    \vcenter{\hbox{
    \begin{tikzpicture}[scale=0.9]
        % 3x3 square grid
        \draw[] (0,0) grid (3,3);
        % Red edges: top right square
        \draw[very thick] (2,1) -- (2,2) -- (1,2);
        % Label
        \node at (2.3,1.25) {$v_6$};
        \node at (0.7,2.25) {$v_8$};
        \node[circle, draw, fill=white, minimum size=3mm, inner sep=0pt, font=\tiny] at (1.5,2) {$I$};
        \node[circle, draw, fill=white, minimum size=3mm, inner sep=0pt, font=\tiny] at (2,1.5) {$I$};
        % Filled circles at all vertices
        \foreach \x in {0,1,2,3} {
            \foreach \y in {0,1,2,3} {
                \fill (\x,\y) circle (2.5pt);
            }
        }
    \end{tikzpicture}
    }}
    \\ \text{and}\\
    \vcenter{\hbox{
    \begin{tikzpicture}[scale=0.9]
        % 3x3 square grid
        \draw[] (0,0) grid (3,3);
        % Red edges: top right square
        \draw[red, very thick] (2,2) -- (1,2);
        \draw[red, very thick] (2,2) -- (2,1);
        % Label
        \node[red] at (2.3,1.25) {$v_6$};
        \node[red] at (0.7,2.25) {$v_8$};
        \node[circle, red, draw, fill=white, minimum size=3mm, inner sep=0pt, font=\tiny] at (1.5,2) {$Z$};
        \node[circle, red, draw, fill=white, minimum size=3mm, inner sep=0pt, font=\tiny] at (2,1.5) {$Z$};
        % Filled circles at all vertices
        \foreach \x in {0,1,2,3} {
            \foreach \y in {0,1,2,3} {
                \fill (\x,\y) circle (2.5pt);
            }
        }
    \end{tikzpicture}
    }} +
    \vcenter{\hbox{
    \begin{tikzpicture}[scale=0.9]
        % 3x3 square grid
        \draw[] (0,0) grid (3,3);
        % Red edges: top right square
        \draw[blue, very thick] (2,1) -- (1,1) -- (1,2);
        % Label
        \node[red] at (2.3,1.25) {$v_6$};
        \node[red] at (0.7,2.25) {$v_8$};
        \node[circle, blue, draw, fill=white, minimum size=3mm, inner sep=0pt, font=\tiny] at (1.5,1) {$Z$};
        \node[circle, blue, draw, fill=white, minimum size=3mm, inner sep=0pt, font=\tiny] at (1,1.5) {$Z$};
        % Filled circles at all vertices
        \foreach \x in {0,1,2,3} {
            \foreach \y in {0,1,2,3} {
                \fill (\x,\y) circle (2.5pt);
            }
        }
    \end{tikzpicture}
    }} =
    \vcenter{\hbox{
    \begin{tikzpicture}[scale=0.9]
        % 3x3 square grid
        \draw[] (0,0) grid (3,3);
        % Red edges: top right square
        \draw[very thick] (2,1) -- (1,1) -- (1,2);
        \draw[very thick] (2,1) -- (2,2) -- (1,2);
        % Label
        \node at (2.3,1.25) {$v_6$};
        \node at (0.7,2.25) {$v_8$};
        \node[circle, draw, fill=white, minimum size=3mm, inner sep=0pt, font=\tiny] at (1.5,1) {$Z$};
        \node[circle, draw, fill=white, minimum size=3mm, inner sep=0pt, font=\tiny] at (1,1.5) {$Z$};
        % Filled circles at all vertices
        \foreach \x in {0,1,2,3} {
            \foreach \y in {0,1,2,3} {
                \fill (\x,\y) circle (2.5pt);
            }
        }
        \node[circle, draw, fill=white, minimum size=3mm, inner sep=0pt, font=\tiny] at (1.5,2) {$Z$};
        \node[circle, draw, fill=white, minimum size=3mm, inner sep=0pt, font=\tiny] at (2,1.5) {$Z$};
    \end{tikzpicture}
    }} \,.
\end{gathered}
\end{equation}

There are other errors that flip the same stabilizers that cause a logical error when using this decoding strategy.
For example, 
\begin{equation}
\begin{gathered}
    \vcenter{\hbox{
    \begin{tikzpicture}[scale=0.9]
        % 3x3 square grid
        \draw[] (0,0) grid (3,3);
        % Red edges: top right square
        \draw[red, very thick] (2,2) -- (1,2);
        \draw[red, very thick] (2,2) -- (2,1);
        % Label
        \node[red] at (2.3,1.25) {$v_6$};
        \node[red] at (0.7,2.25) {$v_8$};
        \node[circle, red, draw, fill=white, minimum size=3mm, inner sep=0pt, font=\tiny] at (1.5,2) {$Z$};
        \node[circle, red, draw, fill=white, minimum size=3mm, inner sep=0pt, font=\tiny] at (2,1.5) {$Z$};
        % Filled circles at all vertices
        \foreach \x in {0,1,2,3} {
            \foreach \y in {0,1,2,3} {
                \fill (\x,\y) circle (2.5pt);
            }
        }
    \end{tikzpicture}
    }}
    \quad\text{and}\quad
    \vcenter{\hbox{
    \begin{tikzpicture}[scale=0.9]
        % 3x3 square grid
        \draw[] (0,0) grid (3,3);
        % Red edges: top right square
        \draw[red, very thick] (2,3) -- (1,3) -- (1,2);
        \draw[red, very thick] (2,0) -- (2,1);
        % Label
        \node[red] at (2.3,1.25) {$v_6$};
        \node[red] at (0.7,2.25) {$v_8$};
        \node[circle, red, draw, fill=white, minimum size=3mm, inner sep=0pt, font=\tiny] at (1.5,3) {$Z$};
        \node[circle, red, draw, fill=white, minimum size=3mm, inner sep=0pt, font=\tiny] at (1,2.5) {$Z$};
        \node[circle, red, draw, fill=white, minimum size=3mm, inner sep=0pt, font=\tiny] at (2,0.5) {$Z$};
        % Filled circles at all vertices
        \foreach \x in {0,1,2,3} {
            \foreach \y in {0,1,2,3} {
                \fill (\x,\y) circle (2.5pt);
            }
        }
    \end{tikzpicture}
    }} \,.
\end{gathered}
\end{equation}
If the probability of a $Z$ error on any edge is equal, then the decoder will infer a correction for the 3-qubit $Z$ error similarly to how it will infer a correction for the 2-qubit $Z$ error (since it only sees that $X_{v_8}$ and $X_{v_6}$ are flipped):
\begin{equation}
\begin{gathered}
    \vcenter{\hbox{
    \begin{tikzpicture}[scale=0.9]
        % 3x3 square grid
        \draw[] (0,0) grid (3,3);
        % Red edges: top right square
        \draw[red, very thick] (2,3) -- (1,3) -- (1,2);
        \draw[red, very thick] (2,0) -- (2,1);
        % Label
        \node[red] at (2.3,1.25) {$v_6$};
        \node[red] at (0.7,2.25) {$v_8$};
        \node[circle, red, draw, fill=white, minimum size=3mm, inner sep=0pt, font=\tiny] at (1.5,3) {$Z$};
        \node[circle, red, draw, fill=white, minimum size=3mm, inner sep=0pt, font=\tiny] at (1,2.5) {$Z$};
        \node[circle, red, draw, fill=white, minimum size=3mm, inner sep=0pt, font=\tiny] at (2,0.5) {$Z$};
        % Filled circles at all vertices
        \foreach \x in {0,1,2,3} {
            \foreach \y in {0,1,2,3} {
                \fill (\x,\y) circle (2.5pt);
            }
        }
    \end{tikzpicture}
    }} +
    \vcenter{\hbox{
    \begin{tikzpicture}[scale=0.9]
        % 3x3 square grid
        \draw[] (0,0) grid (3,3);
        % Red edges: top right square
        \draw[blue, very thick] (2,1) -- (1,1) -- (1,2);
        % Label
        \node[red] at (2.3,1.25) {$v_6$};
        \node[red] at (0.7,2.25) {$v_8$};
        \node[circle, blue, draw, fill=white, minimum size=3mm, inner sep=0pt, font=\tiny] at (1.5,1) {$Z$};
        \node[circle, blue, draw, fill=white, minimum size=3mm, inner sep=0pt, font=\tiny] at (1,1.5) {$Z$};
        % Filled circles at all vertices
        \foreach \x in {0,1,2,3} {
            \foreach \y in {0,1,2,3} {
                \fill (\x,\y) circle (2.5pt);
            }
        }
    \end{tikzpicture}
    }}
    = 
    \vcenter{\hbox{
    \begin{tikzpicture}[scale=0.9]
        % 3x3 square grid
        \draw[] (0,0) grid (3,3);
        % Red edges: top right square
        \draw[red, very thick] (2,3) -- (1,3) -- (1,2);
        \draw[red, very thick] (2,0) -- (2,1);
        \draw[blue, very thick] (2,1) -- (1,1) -- (1,2);
        % Label
        % \node[red] at (2.3,1.25) {$v_6$};
        % \node[red] at (0.7,2.25) {$v_8$};
        \node[circle, red, draw, fill=white, minimum size=3mm, inner sep=0pt, font=\tiny] at (1.5,3) {$Z$};
        \node[circle, red, draw, fill=white, minimum size=3mm, inner sep=0pt, font=\tiny] at (1,2.5) {$Z$};
        \node[circle, red, draw, fill=white, minimum size=3mm, inner sep=0pt, font=\tiny] at (2,0.5) {$Z$};\node[circle, blue, draw, fill=white, minimum size=3mm, inner sep=0pt, font=\tiny] at (1.5,1) {$Z$};
        \node[circle, blue, draw, fill=white, minimum size=3mm, inner sep=0pt, font=\tiny] at (1,1.5) {$Z$};
        % Filled circles at all vertices
        \foreach \x in {0,1,2,3} {
            \foreach \y in {0,1,2,3} {
                \fill (\x,\y) circle (2.5pt);
            }
        }
    \end{tikzpicture}
    }} \,.
\end{gathered}
\end{equation}
Note that the product between the 3-qubit error and the 2-qubit correction forms a non-contractible loop, namely a logical $Z$ operator.
This then causes a logical error, since we are effectively applying a logical $Z$ unitary on the code through a bad correction. For a larger code and more errors, decoding gets more complicated as the number of flipped vertices and plaquettes increases, which can be addressed using a decoder such as \emph{minimum-weight perfect matching} (MWPM).
A more detailed discussion on MWPM and other decoding strategies will be covered in~\cref{sec:decoders}.

\begin{figure}
    \centering
    \subfloat[]{\includegraphics[width=0.4\textwidth]{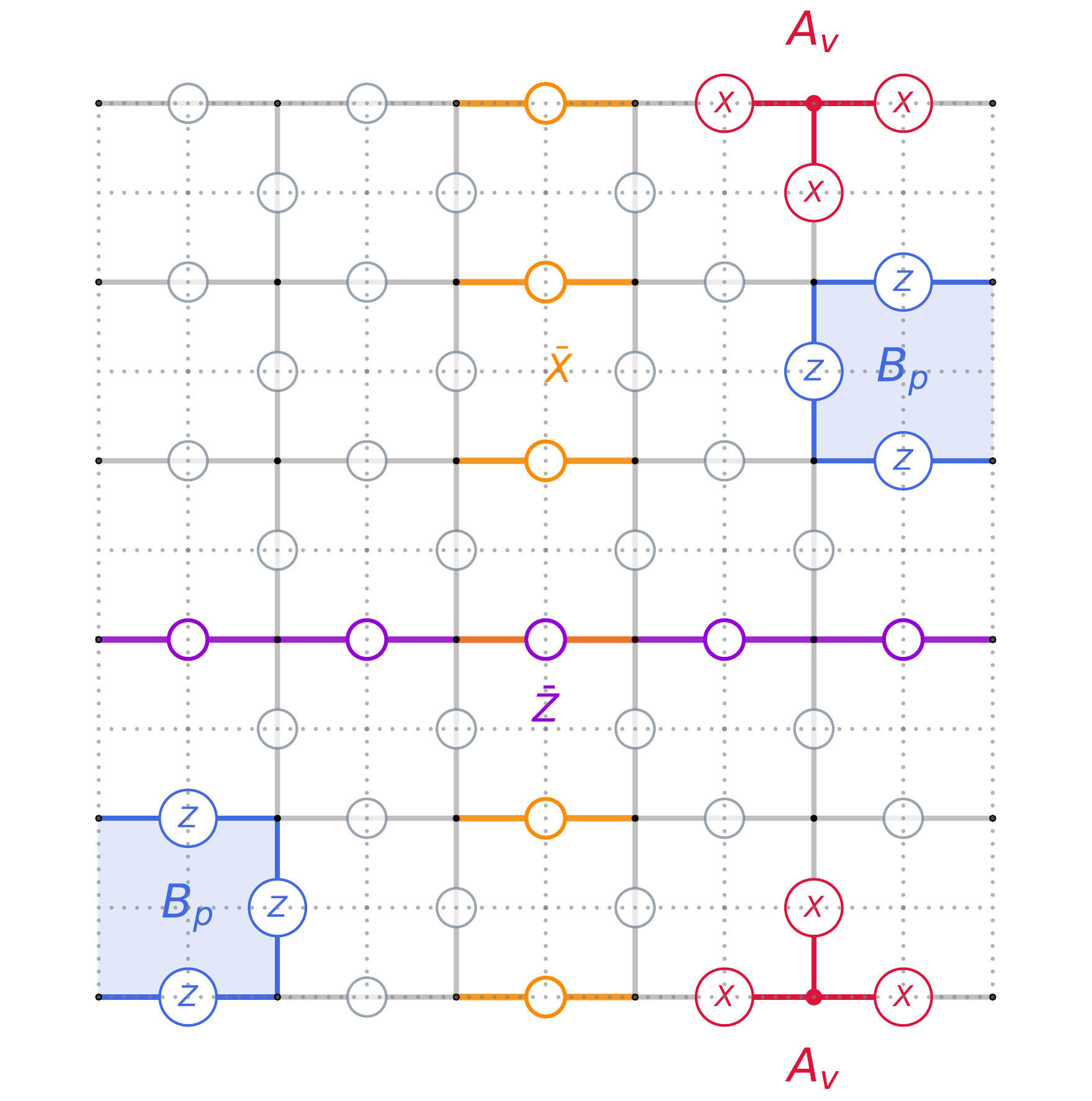}\label{fig:planar-code-lattice}}
    \hspace{8mm}
    \subfloat[]{\includegraphics[width=0.4\textwidth]{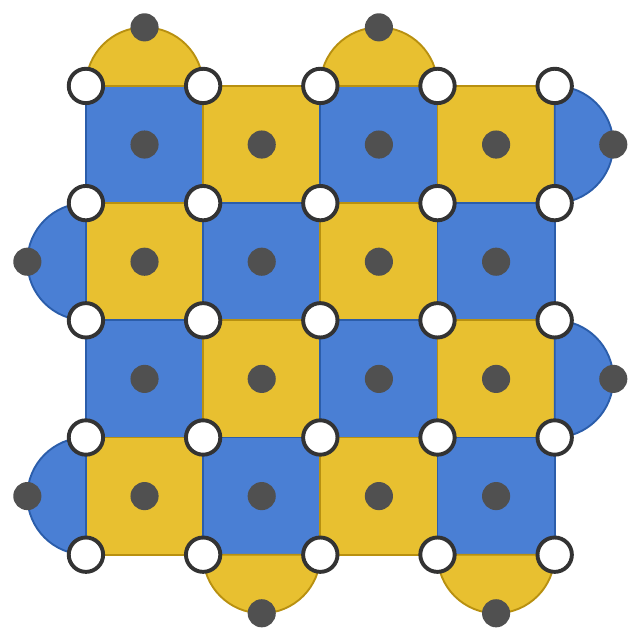}\label{fig:rotated-surface-code}}
    \caption{Distance-5 unrotated and rotated surface codes. (a) A $5 \times 5$ unrotated planar patch with open boundary conditions. Qubits are placed on edges. Interior plaquette and vertex checks have weight 4, while boundary checks have weight 3. The left and right boundaries are rough and the top and bottom boundaries are smooth. (b) A rotated surface code of distance $5$. Hollow circles and gray circles denote data and syndrome qubits, respectively. Boundary syndrome measurements involve weight-2 checks. Blue and yellow plaquettes represent $Z$- and $X$-type stabilizer measurements.}
    \label{fig:surface-code}
\end{figure}

\subsection{Surface code}\label{subsec:surface-code}
Practically, implementing a toric code on a physical device with connectivity or architectural constraints is challenging.
One solution is to cut the torus open, laying it flat to build the code on a planar graph. 
This leads to a planar variant of the \eczoo[surface code]{eczoo_surface} on a square lattice with \emph{open} boundaries, which is commonly referred to as the \emph{unrotated surface code}. 
These open boundary conditions become the defining feature that results in a topology and code structure that are different from that of the toric code. 
As we will see, two important distinctions of the unrotated surface code from the toric code are that its stabilizers are of two different types: \emph{interior} and \emph{boundary} stabilizers, and that it encodes only one logical qubit instead of two.

To define the stabilizers, we first need to look into how boundaries and coboundaries are defined.
Consider the following $4\times4$ planar lattice with 12 vertices, 12 faces\footnote{
We are counting the three spaces between the edges sticking out to the left and three other spaces between the edges sticking out to the right as a special type of faces, which we will discuss in more detail shortly.
}, and $16+9 = 25$ edges:
\begin{equation}
    \vcenter{\hbox{
    \begin{tikzpicture}[scale=0.9]
        % 3x3 square grid
        \draw[] (0,0) grid (2,3);
        \foreach \y in {0,1,2,3} {
            \draw (0,\y) -- (-0.8,\y) ;
            \draw (2,\y) -- (2.8,\y) ;
        }
        % Filled circles at all vertices
        \foreach \x in {0,1,2} {
            \foreach \y in {0,1,2,3} {
                \fill (\x,\y) circle (2.5pt);
            }
        }
    \end{tikzpicture}
    }} \,.
\end{equation}
As opposed to the lattices on the surface of the torus with periodic boundary conditions, here we have a lattice with open boundaries.
In fact, as shown above, there are two types of boundaries, the top and bottom boundaries of the lattice are \emph{smooth} boundaries, whereas the right and left boundaries of the lattice are \emph{rough} boundaries.
Also, note that on each rough boundary we have three ``half-plaquettes'' or ``half-faces,'' which are plaquettes with only three edges around them.
In total, there are six such plaquettes, which we label as $\overline{p}_1,\dots,\overline{p}_6$ besides the ordinary plaquettes $p_1,\dots,p_6$ in the interior:
\begin{equation}
\begin{gathered}
    \vcenter{\hbox{
    \begin{tikzpicture}[scale=0.7]
        % 3x3 square grid
        \draw[] (0,0) grid (2,3);
        \foreach \y in {0,1,2,3} {
            \draw (0,\y) -- (-0.8,\y) ;
            \draw (2,\y) -- (2.8,\y) ;
        }
        % Label
        \node at (-0.5,0.5) {$\overline{p}_1$};
        % Filled circles at all vertices
        \foreach \x in {0,1,2} {
            \foreach \y in {0,1,2,3} {
                \fill (\x,\y) circle (2.5pt);
            }
        }
    \end{tikzpicture}
    }} \,,
    \vcenter{\hbox{
    \begin{tikzpicture}[scale=0.7]% 3x3 square grid
        \draw[] (0,0) grid (2,3);
        \foreach \y in {0,1,2,3} {
            \draw (0,\y) -- (-0.8,\y) ;
            \draw (2,\y) -- (2.8,\y) ;
        }
        % Label
        \node at (2.5,0.5) {$\overline{p}_2$};
        % Filled circles at all vertices
        \foreach \x in {0,1,2} {
            \foreach \y in {0,1,2,3} {
                \fill (\x,\y) circle (2.5pt);
            }
        }
    \end{tikzpicture}
    }} ,\dots,\;
    \vcenter{\hbox{
    \begin{tikzpicture}[scale=0.7]
        % 3x3 square grid
        \draw[] (0,0) grid (2,3);
        \foreach \y in {0,1,2,3} {
            \draw (0,\y) -- (-0.8,\y) ;
            \draw (2,\y) -- (2.8,\y) ;
        }
        % Label
        \node at (2.5,2.5) {$\overline{p}_6$};
        % Filled circles at all vertices
        \foreach \x in {0,1,2} {
            \foreach \y in {0,1,2,3} {
                \fill (\x,\y) circle (2.5pt);
            }
        }
    \end{tikzpicture}
    }} 
    \\\text{and}\\
    \vcenter{\hbox{
    \begin{tikzpicture}[scale=0.7]
        % 3x3 square grid
        \draw[] (0,0) grid (2,3);
        \foreach \y in {0,1,2,3} {
            \draw (0,\y) -- (-0.8,\y) ;
            \draw (2,\y) -- (2.8,\y) ;
        }
        % Label
        \node at (0.5,0.5) {$p_1$};
        % Filled circles at all vertices
        \foreach \x in {0,1,2} {
            \foreach \y in {0,1,2,3} {
                \fill (\x,\y) circle (2.5pt);
            }
        }
    \end{tikzpicture}
    }} \,,
    \vcenter{\hbox{
    \begin{tikzpicture}[scale=0.7]% 3x3 square grid
        \draw[] (0,0) grid (2,3);
        \foreach \y in {0,1,2,3} {
            \draw (0,\y) -- (-0.8,\y) ;
            \draw (2,\y) -- (2.8,\y) ;
        }
        % Label
        \node at (1.5,0.5) {$p_2$};
        % Filled circles at all vertices
        \foreach \x in {0,1,2} {
            \foreach \y in {0,1,2,3} {
                \fill (\x,\y) circle (2.5pt);
            }
        }
    \end{tikzpicture}
    }} ,\dots,\;
    \vcenter{\hbox{
    \begin{tikzpicture}[scale=0.7]
        % 3x3 square grid
        \draw[] (0,0) grid (2,3);
        \foreach \y in {0,1,2,3} {
            \draw (0,\y) -- (-0.8,\y) ;
            \draw (2,\y) -- (2.8,\y) ;
        }
        % Label
        \node at (1.5,2.5) {$p_6$};
        % Filled circles at all vertices
        \foreach \x in {0,1,2} {
            \foreach \y in {0,1,2,3} {
                \fill (\x,\y) circle (2.5pt);
            }
        }
    \end{tikzpicture}
    }}\,.
\end{gathered}
\end{equation}
These are all of the 2-cells.
Similarly, on each smooth boundary, we have three ``half-vertices,'' which are vertices with only three edges around them.
In total, there are six such vertices which we label as $\overline{v}_1,\dots,\overline{v}_6$ besides the ordinary vertices $v_1,\dots,v_6$ in the interior:
\begin{equation}
\begin{gathered}
    \vcenter{\hbox{
    \begin{tikzpicture}[scale=0.7]
        % 3x3 square grid
        \draw[] (0,0) grid (2,3);
        \foreach \y in {0,1,2,3} {
            \draw (0,\y) -- (-0.8,\y) ;
            \draw (2,\y) -- (2.8,\y) ;
        }
        % Label
        \node at (0.3,0.25) {$\overline{v}_1$};
        % Filled circles at all vertices
        \foreach \x in {0,1,2} {
            \foreach \y in {0,1,2,3} {
                \fill (\x,\y) circle (2.5pt);
            }
        }
    \end{tikzpicture}
    }} ,\;
    \vcenter{\hbox{
    \begin{tikzpicture}[scale=0.7]
        % 3x3 square grid
        \draw[] (0,0) grid (2,3);
        \foreach \y in {0,1,2,3} {
            \draw (0,\y) -- (-0.8,\y) ;
            \draw (2,\y) -- (2.8,\y) ;
        }
        % Label
        \node at (1.3,0.25) {$\overline{v}_2$};
        % Filled circles at all vertices
        \foreach \x in {0,1,2} {
            \foreach \y in {0,1,2,3} {
                \fill (\x,\y) circle (2.5pt);
            }
        }
    \end{tikzpicture}
    }} ,
    \dots ,\;
    \vcenter{\hbox{
    \begin{tikzpicture}[scale=0.7]
        % 3x3 square grid
        \draw[] (0,0) grid (2,3);
        \foreach \y in {0,1,2,3} {
            \draw (0,\y) -- (-0.8,\y) ;
            \draw (2,\y) -- (2.8,\y) ;
        }
        % Label
        \node at (2.3,2.65) {$\overline{v}_6$};
        % Filled circles at all vertices
        \foreach \x in {0,1,2} {
            \foreach \y in {0,1,2,3} {
                \fill (\x,\y) circle (2.5pt);
            }
        }
    \end{tikzpicture}
    }} 
    \\\text{and}\\
    \vcenter{\hbox{
    \begin{tikzpicture}[scale=0.7]
        % 3x3 square grid
        \draw[] (0,0) grid (2,3);
        \foreach \y in {0,1,2,3} {
            \draw (0,\y) -- (-0.8,\y) ;
            \draw (2,\y) -- (2.8,\y) ;
        }
        % Label
        \node at (0.3,1.25) {$v_1$};
        % Filled circles at all vertices
        \foreach \x in {0,1,2} {
            \foreach \y in {0,1,2,3} {
                \fill (\x,\y) circle (2.5pt);
            }
        }
    \end{tikzpicture}
    }} ,\;
    \vcenter{\hbox{
    \begin{tikzpicture}[scale=0.7]
        % 3x3 square grid
        \draw[] (0,0) grid (2,3);
        \foreach \y in {0,1,2,3} {
            \draw (0,\y) -- (-0.8,\y) ;
            \draw (2,\y) -- (2.8,\y) ;
        }
        % Label
        \node at (1.3,1.25) {$v_2$};
        % Filled circles at all vertices
        \foreach \x in {0,1,2} {
            \foreach \y in {0,1,2,3} {
                \fill (\x,\y) circle (2.5pt);
            }
        }
    \end{tikzpicture}
    }} ,
    \dots ,\;
    \vcenter{\hbox{
    \begin{tikzpicture}[scale=0.7]
        % 3x3 square grid
        \draw[] (0,0) grid (2,3);
        \foreach \y in {0,1,2,3} {
            \draw (0,\y) -- (-0.8,\y) ;
            \draw (2,\y) -- (2.8,\y) ;
        }
        % Label
        \node at (2.3,1.65) {$v_6$};
        % Filled circles at all vertices
        \foreach \x in {0,1,2} {
            \foreach \y in {0,1,2,3} {
                \fill (\x,\y) circle (2.5pt);
            }
        }
    \end{tikzpicture}
    }} \,.
\end{gathered}
\end{equation}
These are all of the 0-cells.
Since the half-plaquettes and half-vertices have only three edges around them, their boundaries are given by
\begin{equation}
\begin{gathered}
    \partial_2(\overline{p}_1) =
    \vcenter{\hbox{
    \begin{tikzpicture}[scale=0.7]
        % 3x3 square grid
        \draw[] (0,0) grid (2,3);
        \foreach \y in {0,1,2,3} {
            \draw (0,\y) -- (-0.8,\y) ;
            \draw (2,\y) -- (2.8,\y) ;
        }
        % Red edges: bottom left square
        \draw[red, very thick] (-0.8,0) -- (0,0) -- (0,1) -- (-0.8,1) ;
        % Label
        \node at (-0.5,0.5) {$\overline{p}_1$};
        % Filled circles at all vertices
        \foreach \x in {0,1,2} {
            \foreach \y in {0,1,2,3} {
                \fill (\x,\y) circle (2.5pt);
            }
        }
    \end{tikzpicture}
    }} ,\;
    \partial_2(\overline{p}_2) =
    \vcenter{\hbox{
    \begin{tikzpicture}[scale=0.7]% 3x3 square grid
        \draw[] (0,0) grid (2,3);
        \foreach \y in {0,1,2,3} {
            \draw (0,\y) -- (-0.8,\y) ;
            \draw (2,\y) -- (2.8,\y) ;
        }
        % Red edges: bottom left square
        \draw[red, very thick] (2.8,0) -- (2,0) -- (2,1) -- (2.8,1) ;
        % Label
        \node at (2.5,0.5) {$\overline{p}_2$};
        % Filled circles at all vertices
        \foreach \x in {0,1,2} {
            \foreach \y in {0,1,2,3} {
                \fill (\x,\y) circle (2.5pt);
            }
        }
    \end{tikzpicture}
    }} ,\dots,\;
    \partial_2(\overline{p}_6) =
    \vcenter{\hbox{
    \begin{tikzpicture}[scale=0.7]
        % 3x3 square grid
        \draw[] (0,0) grid (2,3);
        \foreach \y in {0,1,2,3} {
            \draw (0,\y) -- (-0.8,\y) ;
            \draw (2,\y) -- (2.8,\y) ;
        }
        % Red edges: bottom left square
        \draw[red, very thick] (2.8,2) -- (2,2) -- (2,3) -- (2.8,3) ;
        % Label
        \node at (2.5,2.5) {$\overline{p}_6$};
        % Filled circles at all vertices
        \foreach \x in {0,1,2} {
            \foreach \y in {0,1,2,3} {
                \fill (\x,\y) circle (2.5pt);
            }
        }
    \end{tikzpicture}
    }}
    \\\text{and}\\\delta^0(\overline{v}_1) =
    \vcenter{\hbox{
    \begin{tikzpicture}[scale=0.7]
        % 3x3 square grid
        \draw[] (0,0) grid (2,3);
        \foreach \y in {0,1,2,3} {
            \draw (0,\y) -- (-0.8,\y) ;
            \draw (2,\y) -- (2.8,\y) ;
        }
        % Red edges: bottom left square
        \draw[red, very thick] (0,0) -- (1,0);
        \draw[red, very thick] (0,0) -- (0,1);
        \draw[red, very thick] (0,0) -- (-0.8,0);
        % Label
        \node at (0.3,0.25) {$\overline{v}_1$};
        % Filled circles at all vertices
        \foreach \x in {0,1,2} {
            \foreach \y in {0,1,2,3} {
                \fill (\x,\y) circle (2.5pt);
            }
        }
    \end{tikzpicture}
    }} ,\;
    \delta^0(\overline{v}_2) =
    \vcenter{\hbox{
    \begin{tikzpicture}[scale=0.7]
        % 3x3 square grid
        \draw[] (0,0) grid (2,3);
        \foreach \y in {0,1,2,3} {
            \draw (0,\y) -- (-0.8,\y) ;
            \draw (2,\y) -- (2.8,\y) ;
        }
        % Red edges: bottom middle square
        \draw[red, very thick] (1,0) -- (2,0);
        \draw[red, very thick] (1,0) -- (1,1);
        \draw[red, very thick] (1,0) -- (0,0);
        % Label
        \node at (1.3,0.25) {$\overline{v}_2$};
        % Filled circles at all vertices
        \foreach \x in {0,1,2} {
            \foreach \y in {0,1,2,3} {
                \fill (\x,\y) circle (2.5pt);
            }
        }
    \end{tikzpicture}
    }} ,
    \dots ,\;
    \delta^0(\overline{v}_6) =
    \vcenter{\hbox{
    \begin{tikzpicture}[scale=0.7]
        % 3x3 square grid
        \draw[] (0,0) grid (2,3);
        \foreach \y in {0,1,2,3} {
            \draw (0,\y) -- (-0.8,\y) ;
            \draw (2,\y) -- (2.8,\y) ;
        }
        % Red edges: bottom middle square
        \draw[red, very thick] (2,3) -- (2.8,3);
        \draw[red, very thick] (2,3) -- (2,2);
        \draw[red, very thick] (2,3) -- (1,3);
        % Label
        \node at (2.3,2.65) {$\overline{v}_6$};
        % Filled circles at all vertices
        \foreach \x in {0,1,2} {
            \foreach \y in {0,1,2,3} {
                \fill (\x,\y) circle (2.5pt);
            }
        }
    \end{tikzpicture}
    }} \,.
\end{gathered}
\end{equation}
Then, the boundary and coboundary groups are defined as
\begin{equation}
\begin{gathered}
    B_1 = \<\partial_2p : \text{plaquette }p\>
    \quad\text{and}\quad
    B^1 = \<\delta^0v : \text{vertex }v\> \,,
\end{gathered}
\end{equation}
respectively.
By again placing a qubit on the edges, the vertex and plaquette stabilizers can then be defined similarly to the toric code on the (half-) vertices and (half-) plaquettes (see also~\cref{fig:planar-code-lattice}):
\begin{equation}
    X_v = \prod_{e\in\delta^0v} X(e)
    \quad\text{and}\quad
    Z_p = \prod_{e\in\partial_2p} Z(e) \,,
\end{equation}
respectively.
So, the stabilizer group of the unrotated surface code is given by
\begin{equation}
    \mathcal{S}_\mathrm{sur} = \<X_{c^*},Z_c : c\in B_1\,,\; c^*\in B^1\> \,.
\end{equation}
The underlying planar region is topologically a disk. If we were to compute its standard first homology group (resp., cohomology group) as we did with the torus, then the result would be trivial because every closed loop on a disk is the boundary of a patch of faces (and dually, every dual loop is a coboundary). Furthermore, standard homology and cohomology require cycles to be strictly closed loops, whereas the logical operators of the planar surface code are open strings spanning opposite sides of the primary or dual lattices. 

To accommodate these open strings rigorously, we use \emph{relative homology} and \emph{relative cohomology}. Under this framework, any $1$-chain whose boundary endpoints terminate entirely within the rough boundaries is treated as having a ``null'' boundary, allowing it to function mathematically as a non-trivial relative cycle. Symmetrically, $1$-cochains terminating on the smooth boundaries function as relative cocycles.
From this point on, $C_n,H_1,Z_1,B_1$ (resp., $C^n,H^1,Z^1,B^1$) denote the relative $n$-chains, homology, cycle, and boundary groups modulo the rough boundaries (resp., relative $n$-cochains, cohomology, cocycle, and coboundary groups modulo the smooth boundaries) on the planar surface.
These non-contractible 1-chains (connecting rough boundaries) and 1-cochains (connecting smooth boundaries) are
\begin{equation}
\begin{gathered}
    \vcenter{\hbox{
    \begin{tikzpicture}[scale=0.7]
        % 3x3 square grid
        \draw[] (0,0) grid (2,3);
        \foreach \y in {0,1,2,3} {
            \draw (0,\y) -- (-0.8,\y) ;
            \draw (2,\y) -- (2.8,\y) ;
        }
        \draw[red, very thick] (-0.8,0) -- (2.8,0) ;
        % Label
        \node[red] at (1.4,0.4) {$\ell_1$};
        % Filled circles at all vertices
        \foreach \x in {0,1,2} {
            \foreach \y in {0,1,2,3} {
                \fill (\x,\y) circle (2.5pt);
            }
        }
    \end{tikzpicture}
    }} ,
    \dots ,\;
    \vcenter{\hbox{
    \begin{tikzpicture}[scale=0.7]
        % 3x3 square grid
        \draw[] (0,0) grid (2,3);
        \foreach \y in {0,1,2,3} {
            \draw (0,\y) -- (-0.8,\y) ;
            \draw (2,\y) -- (2.8,\y) ;
        }
        \draw[red, very thick] (-0.8,3) -- (2.8,3) ;
        % Label
        \node[red] at (1.4,2.6) {$\ell_4$};
        % Filled circles at all vertices
        \foreach \x in {0,1,2} {
            \foreach \y in {0,1,2,3} {
                \fill (\x,\y) circle (2.5pt);
            }
        }
    \end{tikzpicture}
    }} 
    \\\text{and}\\
    \vcenter{\hbox{
    \begin{tikzpicture}[scale=0.7]
        % 3x3 square grid
        \draw[] (0,0) grid (2,3);
        \foreach \y in {0,1,2,3} {
            \draw (0,\y) -- (-0.8,\y) ;
            \draw (2,\y) -- (2.8,\y) ;
        }
        \foreach \y in {0,1,2,3} {
            \draw[red, very thick] (-0.8,\y) -- (0,\y) ;
        }
        % Label
        \node[red] at (-0.5,1.5) {$\lambda_1$};
        % Filled circles at all vertices
        \foreach \x in {0,1,2} {
            \foreach \y in {0,1,2,3} {
                \fill (\x,\y) circle (2.5pt);
            }
        }
    \end{tikzpicture}
    }} ,
    \dots ,\;
    \vcenter{\hbox{
    \begin{tikzpicture}[scale=0.7]
        % 3x3 square grid
        \draw[] (0,0) grid (2,3);
        \foreach \y in {0,1,2,3} {
            \draw (0,\y) -- (-0.8,\y) ;
            \draw (2,\y) -- (2.8,\y) ;
        }
        \foreach \y in {0,1,2,3} {
            \draw[red, very thick] (2.8,\y) -- (2,\y) ;
        }
        % Label
        \node[red] at (2.5,1.5) {$\lambda_4$};
        % Filled circles at all vertices
        \foreach \x in {0,1,2} {
            \foreach \y in {0,1,2,3} {
                \fill (\x,\y) circle (2.5pt);
            }
        }
    \end{tikzpicture}
    }} \,,
\end{gathered}
\end{equation}
as well as their sum with boundaries and coboundaries, respectively, (e.g., $\ell_1 + \partial_2p$ and $\lambda_2+\delta^0v$).
These non-contractible 1-chains define the cycle and cocycle groups:
\begin{equation}
\begin{gathered}
    Z_1 = \<\{\ell_1,\dots,\ell_4\}\cup B_1\>
    \quad\text{and}\quad
    Z^1 = \<\{\lambda_1,\dots,\lambda_4\}\cup B^1\> \,.
\end{gathered}
\end{equation}
Similar to the toric code, the normalizer is given by
\begin{equation}
    \mathcal{N}(\mathcal{S}_\mathrm{sur}) = \<\{X_{c^*},Z_c : c\in Z_1\,,\; c^*\in Z^1 \} \cup\{\pm i\}\> \,.
\end{equation}

Similar to our analysis for the toric code in~\cref{subsec:toric-code}, we can identify the logical operators, the number of encoded logical qubits $k$, and the distance $d$ of the unrotated surface code from the homology group $H_1=Z_1/B_1$ and cohomology group $H^1=Z^1/B^1$.
By the same analysis as we have done for the torus homology we can find that the cosets in $H_1$ are of the form 
\begin{equation}
    [\ell_1] = \{\ell_1 + b : b\in B_1\}
\end{equation}
and since the non-contractible 1-chains are equivalent up to a boundary, the homology group $H_1$ is generated by $[\ell_1]$.
Similarly, the cosets in $H^1$ are of the form 
\begin{equation}
    [\lambda_1] = \{\lambda_1 + b^* : b^*\in B^1\}
\end{equation}
and since the non-contractible 1-cochains are equivalent up to a coboundary, the cohomology group $H^1$ is generated by $[\lambda_1]$.
Similar to the toric code, it can also be shown that the quotient group $\mathcal{N}(\mathcal{S}_\mathrm{sur})/\mathcal{S}_\mathrm{sur}$, up to phases, denoted by $\mathcal{L}_\mathrm{sur}$, is isomorphic to $H_1\oplus H^1$.
So, representatives of the logical $X$ and $Z$ operators of the unrotated surface code are 
\begin{equation}
    X_{\lambda_1} =
     \vcenter{\hbox{
    \begin{tikzpicture}[scale=0.7]
        % 3x3 square grid
        \draw[] (0,0) grid (2,3);
        \foreach \y in {0,1,2,3} {
            \draw (0,\y) -- (-0.8,\y) ;
            \draw (2,\y) -- (2.8,\y) ;
        }
        \foreach \y in {0,1,2,3} {
            \draw[red, very thick] (-0.8,\y) -- (0,\y) ;
        }
        % Label
        \node[red] at (-0.5,1.5) {$\lambda_1$};
        % Filled circles at all vertices
        \foreach \x in {0,1,2} {
            \foreach \y in {0,1,2,3} {
                \fill (\x,\y) circle (2.5pt);
            }
        }
    \end{tikzpicture}
    }}
    \quad\text{and}\quad 
    Z_{\ell_1} = 
    \vcenter{\hbox{
    \begin{tikzpicture}[scale=0.7]
        % 3x3 square grid
        \draw[] (0,0) grid (2,3);
        \foreach \y in {0,1,2,3} {
            \draw (0,\y) -- (-0.8,\y) ;
            \draw (2,\y) -- (2.8,\y) ;
        }
        \draw[red, very thick] (-0.8,0) -- (2.8,0) ;
        % Label
        \node[red] at (1.4,0.4) {$\ell_1$};
        % Filled circles at all vertices
        \foreach \x in {0,1,2} {
            \foreach \y in {0,1,2,3} {
                \fill (\x,\y) circle (2.5pt);
            }
        }
    \end{tikzpicture}
    }} \,.
\end{equation}
See also~\cref{fig:planar-code-lattice} for a primal-dual lattice illustration.

Now, we can determine the number of encoded logical qubits $k$ and the code distance $d$ from $H_1$ and $H^1$.
Since there are $2k=2$ independent generators of $H_1\oplus H^1$, there is $k=1$ encoded logical qubit.
On the other hand, the minimum weight of non-contractible 1-chains and 1-cochains in the cosets $[\ell_1]$ and $[\lambda_1]$ is $4$.
A pair of 1-chain and 1-cochain with weight 4 are $\ell_1$ and $\lambda_1$, so that the logical operators $Z_{\ell_1}$ and $X_{\lambda_1}$ are the minimum weight logical operators.
Thus, we conclude that the code distance is $d=4$.
Finally, since there are $16$ horizontal edges and $9$ vertical edges, there are $n=16+9=25$ physical qubits.
So, the $4\times4$ unrotated surface code is a $\llbracket25,1,4\rrbracket$ code.

One can easily generalize this analysis to an $L \times L$ unrotated surface code.
We will find that for any $L$, the homology and cohomology groups are each generated by one coset.
Thus, the code always encodes $k=1$ logical qubit for any $L$.
The minimum-weight non-contractible 1-chains/1-cochains are always $L$ since any non-contractible 1-chains/1-cochains terminating at the same type of boundaries must have a minimum weight of $L$.
So the code distance must be $d=L$.
Since there are $L^2$ horizontal edges and $(L-1)^2$ vertical edges in such lattice, an $L\times L$ unrotated surface code is an $\stabcode{L^2 + (L-1)^2}{1}{L}$ code.

We note that a variant of the unrotated surface code that has been rotated $45^\circ$ is a popular choice in experimental demonstrations of QEC~\cite{fowler2012surface,googlequantumai2025quantum,bluvstein_fault-tolerant_2026,bluvstein2024logical}.  
Compared to its unrotated counterpart, the rotated surface code requires fewer physical qubits to encode the same number of logical qubits $k$ with the same distance $d$.
For example, in~\cref{fig:rotated-surface-code}, the rotated surface code with distance $5$ uses $25$ physical qubits while the unrotated surface code with distance $5$ uses $41$ physical qubits. 

Finally, we briefly discuss the errors on the lattice. Similar to the toric code, errors on the lattice are represented as 1-chains of Pauli operators. 
However, there are errors that only flip one stabilizer, namely those on the qubits on boundaries.
For example,
\begin{equation}
\begin{gathered}
    \vcenter{\hbox{
    \begin{tikzpicture}[scale=0.7]
        % 3x3 square grid
        \draw[] (0,0) grid (2,3);
        \foreach \y in {0,1,2,3} {
            \draw (0,\y) -- (-0.8,\y) ;
            \draw (2,\y) -- (2.8,\y) ;
        }
        % Red edges: bottom left square
        % \draw[red, very thick] (0,0) -- (1,0);
        % \draw[red, very thick] (0,0) -- (0,1);
        \draw[red, very thick] (0,0) -- (-0.8,0);
        % Label
        \node[red] at (0.3,0.25) {$\overline{v}_1$};
        % Filled circles at all vertices
        \foreach \x in {0,1,2} {
            \foreach \y in {0,1,2,3} {
                \fill (\x,\y) circle (2.5pt);
            }
        }
        \fill[red] (0,0) circle (2.5pt);
        % errors
        \node[circle, red, draw, fill=white, minimum size=3mm, inner sep=0pt, font=\tiny] at (-0.5,0) {$Z$};
    \end{tikzpicture}
    }}
    \quad\text{and}\quad
    \vcenter{\hbox{
    \begin{tikzpicture}[scale=0.7]
        % 3x3 square grid
        \draw[] (0,0) grid (2,3);
        \foreach \y in {0,1,2,3} {
            \draw (0,\y) -- (-0.8,\y) ;
            \draw (2,\y) -- (2.8,\y) ;
        }
        % Red edges: bottom left square
        \draw[red, very thick] (0,3) -- (1,3) ;
        % Label
        \node[red] at (0.5,2.5) {$p_5$};
        % Filled circles at all vertices
        \foreach \x in {0,1,2} {
            \foreach \y in {0,1,2,3} {
                \fill (\x,\y) circle (2.5pt);
            }
        }
        % errors
        \node[circle, red, draw, fill=white, minimum size=3mm, inner sep=0pt, font=\tiny] at (0.5,3) {$X$};
    \end{tikzpicture}
    }} \;.
\end{gathered}
\end{equation}
On the other hand, errors in the interior are detected similarly to the toric code as they flip two stabilizers.
For example,
\begin{equation}
\begin{gathered}
    \vcenter{\hbox{
    \begin{tikzpicture}[scale=0.7]
        % 3x3 square grid
        \draw[] (0,0) grid (2,3);
        \foreach \y in {0,1,2,3} {
            \draw (0,\y) -- (-0.8,\y) ;
            \draw (2,\y) -- (2.8,\y) ;
        }
        % Red edges: bottom left square
        % \draw[red, very thick] (0,0) -- (1,0);
        % \draw[red, very thick] (0,0) -- (0,1);
        \draw[red, very thick] (0,1) -- (1,1);
        % Label
        \node[red] at (-0.4,1.25) {$v_1$};
        \node[red] at (1.3,1.25) {$v_2$};
        % Filled circles at all vertices
        \foreach \x in {0,1,2} {
            \foreach \y in {0,1,2,3} {
                \fill (\x,\y) circle (2.5pt);
            }
        }
        \fill[red] (0,1) circle (2.5pt);
        \fill[red] (1,1) circle (2.5pt);
        % errors
        \node[circle, red, draw, fill=white, minimum size=3mm, inner sep=0pt, font=\tiny] at (0.5,1) {$Z$};
    \end{tikzpicture}
    }}
    \quad\text{and}\quad
    \vcenter{\hbox{
    \begin{tikzpicture}[scale=0.7]
        % 3x3 square grid
        \draw[] (0,0) grid (2,3);
        \foreach \y in {0,1,2,3} {
            \draw (0,\y) -- (-0.8,\y) ;
            \draw (2,\y) -- (2.8,\y) ;
        }
        % Red edges: bottom left square
        \draw[red, very thick] (0,2) -- (1,2) ;
        \draw[red, very thick] (0,1) -- (1,1) ;
        % Label
        \node[red] at (0.5,2.5) {$p_5$};
        % \node[red] at (0.5,1.5) {$p_3$};
        \node[red] at (0.5,0.5) {$p_1$};
        % Filled circles at all vertices
        \foreach \x in {0,1,2} {
            \foreach \y in {0,1,2,3} {
                \fill (\x,\y) circle (2.5pt);
            }
        }
        % errors
        \node[circle, red, draw, fill=white, minimum size=3mm, inner sep=0pt, font=\tiny] at (0.5,2) {$X$};
        \node[circle, red, draw, fill=white, minimum size=3mm, inner sep=0pt, font=\tiny] at (0.5,1) {$X$};
    \end{tikzpicture}
    }} \;.
\end{gathered}
\end{equation}
So, a multi-qubit error that forms a 1-chain that terminates on a boundary flips a stabilizer:
\begin{equation}\label{eqn:surface_code_boundary_error_chains}
\begin{gathered}
    \vcenter{\hbox{
    \begin{tikzpicture}[scale=0.7]
        % 3x3 square grid
        \draw[] (0,0) grid (2,3);
        \foreach \y in {0,1,2,3} {
            \draw (0,\y) -- (-0.8,\y) ;
            \draw (2,\y) -- (2.8,\y) ;
        }
        % Red edges: bottom left square
        % \draw[red, very thick] (0,0) -- (1,0);
        % \draw[red, very thick] (0,0) -- (0,1);
        \draw[red, very thick] (0,1) -- (1,1) -- (2,1) -- (2.8,1);
        % Label
        \node[red] at (-0.4,1.25) {$v_1$};
        % \node[red] at (1.3,1.25) {$v_2$};
        % Filled circles at all vertices
        \foreach \x in {0,1,2} {
            \foreach \y in {0,1,2,3} {
                \fill (\x,\y) circle (2.5pt);
            }
        }
        \fill[red] (0,1) circle (2.5pt);
        % errors
        \node[circle, red, draw, fill=white, minimum size=3mm, inner sep=0pt, font=\tiny] at (0.5,1) {$Z$};
        \node[circle, red, draw, fill=white, minimum size=3mm, inner sep=0pt, font=\tiny] at (1.5,1) {$Z$};
        \node[circle, red, draw, fill=white, minimum size=3mm, inner sep=0pt, font=\tiny] at (2.5,1) {$Z$};
    \end{tikzpicture}
    }}
    \quad\text{and}\quad
    \vcenter{\hbox{
    \begin{tikzpicture}[scale=0.7]
        % 3x3 square grid
        \draw[] (0,0) grid (2,3);
        \foreach \y in {0,1,2,3} {
            \draw (0,\y) -- (-0.8,\y) ;
            \draw (2,\y) -- (2.8,\y) ;
        }
        % Red edges: bottom left square
        \draw[red, very thick] (0,2) -- (1,2) ;
        \draw[red, very thick] (0,3) -- (1,3) ;
        % Label
        % \node[red] at (0.5,2.5) {$p_5$};
        \node[red] at (0.5,1.5) {$p_3$};
        % \node[red] at (0.5,0.5) {$p_1$};
        % Filled circles at all vertices
        \foreach \x in {0,1,2} {
            \foreach \y in {0,1,2,3} {
                \fill (\x,\y) circle (2.5pt);
            }
        }
        % errors
        \node[circle, red, draw, fill=white, minimum size=3mm, inner sep=0pt, font=\tiny] at (0.5,2) {$X$};
        \node[circle, red, draw, fill=white, minimum size=3mm, inner sep=0pt, font=\tiny] at (0.5,3) {$X$};
    \end{tikzpicture}
    }} \;.
\end{gathered}
\end{equation}
% A single $Z$ error on an edge anti-commutes with the two $X_v$ vertex operators sharing that edge, resulting in a $-1$ eigenvalue syndrome. 
% More generally, an error $Z_c$ consisting of a chain of $Z$ errors on edges in $c\in C_1$ only triggers the vertex stabilizers at its endpoints (thus, an even number of them), similar to the toric code. 
% Similarly, an $X$-type error triggers an even number of plaquette stabilizers. 
Similar to the toric code, we need to choose a 1-chain of Pauli-$Z$ (resp., 1-cochain of Pauli-$X$) corrections that connect each pair of flipped vertex stabilizers (resp., plaquette stabilizers) caused by $Z$ errors (resp., $X$ errors) such that the product between the corrections and the errors is a stabilizer.
However, a chain of $Z$ errors (resp., a cochain of $X$ errors) with exactly one endpoint on the rough (resp., smooth) boundary only flips one vertex stabilizer (resp., plaquette stabilizer), such as the ones in~\cref{eqn:surface_code_boundary_error_chains}.
Thus, a similar decoding strategy to the one we had for the toric code can be used.
Similarly, we assign weights to edges corresponding to the probability of a certain error.
However, now we need to find an error chain (resp., cochain) with an endpoint on a rough (resp., smooth) boundary, so the decoder must find a path from the flipped vertex stabilizer (resp., plaquette stabilizer) to a rough boundary (resp., smooth boundary) with the lowest weight.
More details on matching decoders can be found in~\cref{sec:decoders}.
% Following the measurement of these syndromes, a recovery Pauli operator $E_{c'}$ is applied to reconnect the detected endpoints. 
% As we have pointed out for error-correction in toric codes, for any given pair of endpoints, there exist multiple 1-chains $c'$ that connect them. 
% For $Z$ errors, the choice of correction $Z_{c'}$ may not match the original error $Z_c$, the decoding is successful if the combined 1-cycle $c{c'}$ is a boundary (i.e., $c+c' \in B_1$), so that $Z_cZ_{c'}$ is in the stabilizer group $\mathcal{S}$.
% Similarly, for $X$ error $X_{c^*}$, the decoding is successful if it infers $X_{\Tilde{c}^*}$ such that $c^*+\Tilde{c}^*\in B^1$.
% On the other hand, the decoding fails if it chooses a correction $E_{c'}$ for error $E_c$ that causes a logical error.
% This happens when $c + {c'}$ forms a non-trivial cycle, that is, $c+c'$ is in $H_1$ or $H^1$.

\begin{explbox}[label={box:color-code}]{Color code}
    The surface code is not the only important two-dimensional topological CSS code. A closely related family is the \eczoo[color code]{color}, which is defined on a trivalent lattice whose faces can be colored with three colors, usually red, green, and blue, such that adjacent faces have different colors. Unlike the toric and surface codes discussed above, where physical qubits live on edges, a 2D color code places physical qubits on vertices. In the honeycomb example below, each hexagonal face supports two stabilizer generators with the same geometric support:
    \begin{equation}
        X_f=\bigotimes_{v\in f}X_v,\qquad
        Z_f=\bigotimes_{v\in f}Z_v .
    \end{equation}
    Because the lattice is trivalent and face-3-colorable, any two faces overlap on either zero or two vertices. This even-overlap property ensures that all \(X_f\) and \(Z_{f'}\) stabilizers commute, so the construction gives a CSS stabilizer code.

    \begin{center}
        \captionsetup{type=figure}
        \subfloat[]{\includegraphics[width=0.45\textwidth]{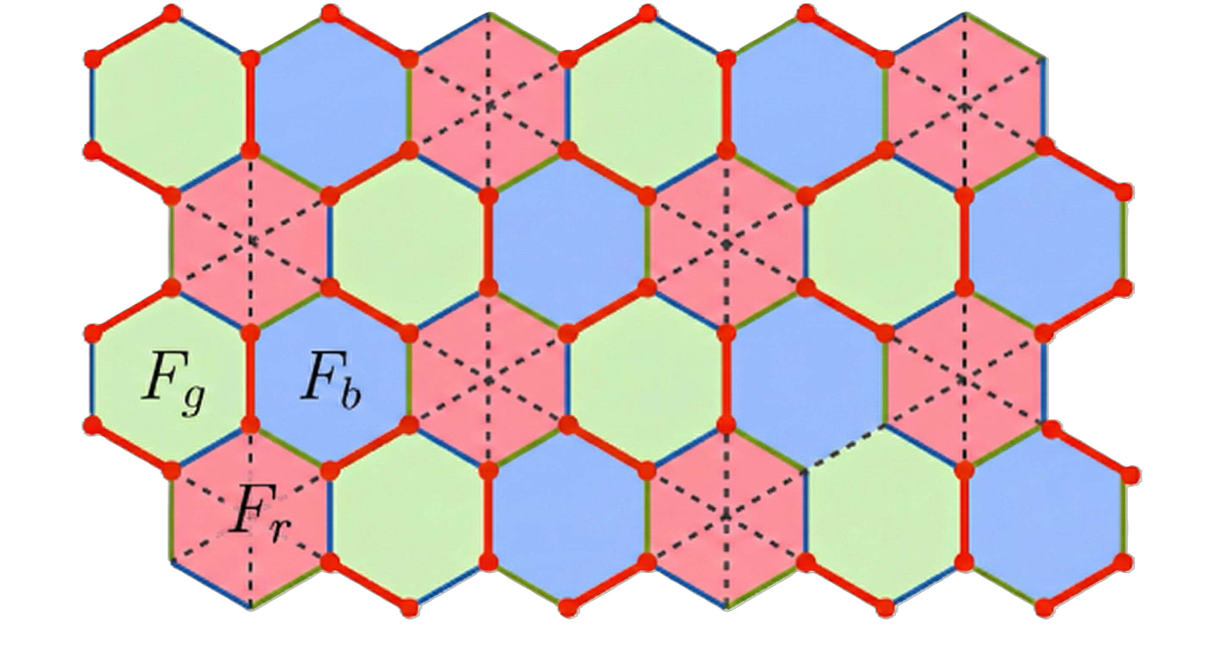}}
        \hfill
        \subfloat[]{\includegraphics[width=0.45\textwidth]{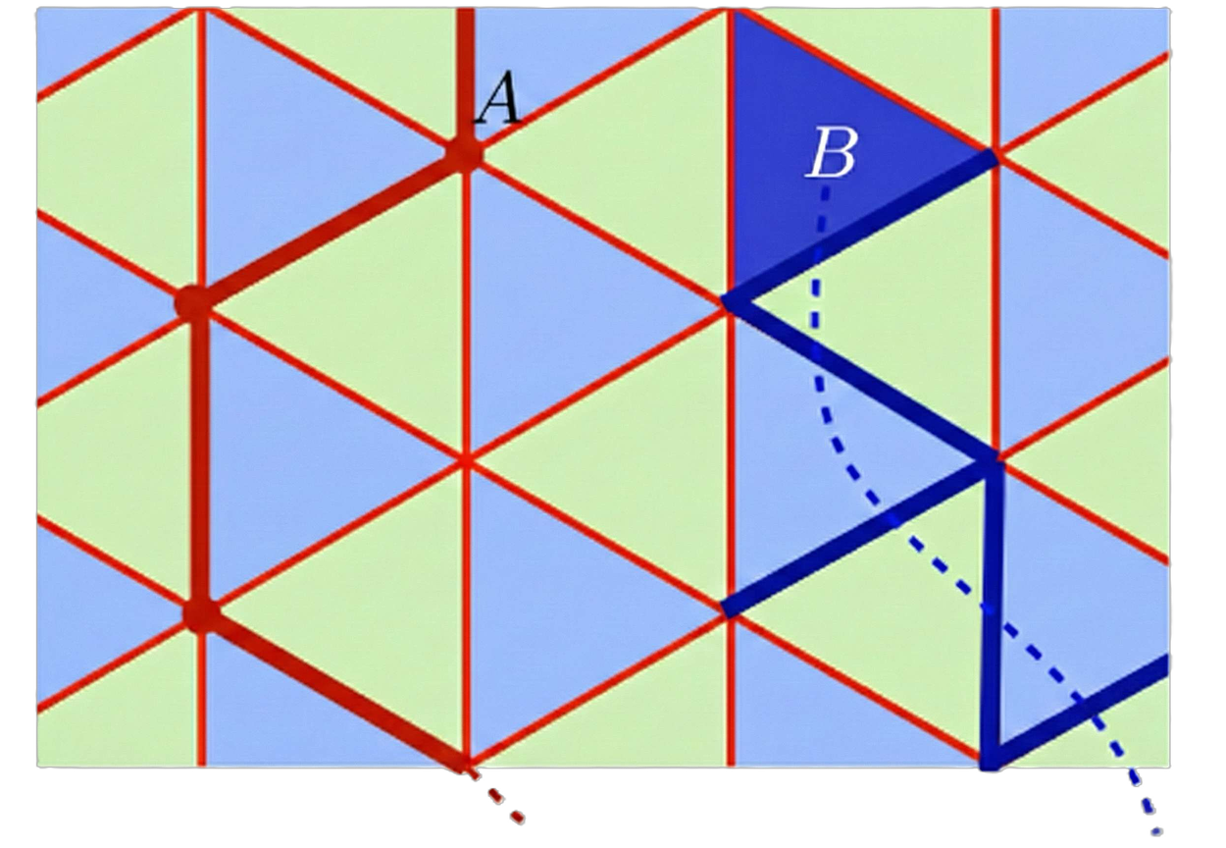}}
        \caption{(a) A 3-colorable honeycomb lattice for a 2D color code, with \(F_r\), \(F_g\), and \(F_b\) denoting the sets of red, green, and blue faces, respectively. (b) String operators drawn on the red shrunk lattice. Figures are adapted from Ref.~\cite[Figs. 1 and 3]{davydova2023floquet}}
        \label{fig:color-code-lattices}
    \end{center}

    The logical operators can also be understood topologically. Instead of using strings on the original lattice directly, it is often convenient to use a \emph{shrunk lattice}: choose one face color, shrink every face of that color to a vertex, and connect neighboring faces through the qubits between them. Closed strings on the shrunk lattice define Pauli string operators on the original physical qubits. As in the surface code, contractible closed strings are products of stabilizers, while noncontractible closed strings represent logical operators. The key difference is that each homology class has colored representatives, and only two of the three color choices are independent. This extra color degree of freedom is why, on a closed surface with Euler characteristic \(\chi\), the 2D color code encodes
    \begin{equation}
        k_{\mathrm{color}}=4-2\chi
    \end{equation}
    logical qubits, twice the number \(k_{\mathrm{surface}}=2-\chi\) for the corresponding surface-code construction on the same closed surface. For example, on a torus where \(\chi=0\), the toric/surface-code construction encodes two logical qubits, while the 2D color code encodes four.

    Thus, color codes share the same broad topological principle as surface codes: local stabilizer violations mark endpoints of string-like errors, and nontrivial homology classes support logical operators. Their distinctive feature is the colored, self-dual lattice structure, with qubits on vertices and both \(X\)- and \(Z\)-type checks attached to each face. Color codes also have distinctive logical-gate properties, but those depend on lattice and boundary details.
\end{explbox}

\newpage

\section{Subsystem codes}\label{sec:subsystem}

The central idea of a subsystem code is to encode quantum information not in a subspace of the physical Hilbert space itself, but in a tensor factor of a subspace. This generalization of the standard stabilizer formalism offers significant advantages, most notably the ability to construct codes where the check operators (measurements) have much lower weight than the stabilizers required to define the same codespace. In this chapter, we motivate subsystem codes using the standard Shor code, before rigorously defining the formalism and introducing the canonical example of a subsystem code: the Bacon--Shor code. We also discuss the Knill--Laflamme conditions in the context of subsystem codes (c.f.~\cref{thm:knill-laflamme-condition}) and other examples of subsystem codes.

\subsection{Motivation: the 9-qubit Shor code}\label{subsec:subsystem-motivation}
To understand the utility of subsystem codes, let us revisit the standard 9-qubit Shor code, which we introduced in~\cref{subsec:fundamentals-nine-qubit-shor-code}. Recall, the Shor code protects against arbitrary single-qubit errors by concatenating a 3-qubit phase-flip repetition code with a 3-qubit bit-flip repetition code. The logical states are:
\begin{equation}
    \ket{\overline{0}} = \frac{1}{\sqrt{8}}(\ket{000} + \ket{111})^{\otimes 3}\,, \quad \ket{\overline{1}} = \frac{1}{\sqrt{8}}(\ket{000} - \ket{111})^{\otimes 3}\,.
\end{equation}
We may arrange the 9 physical qubits as three blocks of three qubits each, or more precisely, in a $3 \times 3$ grid. The stabilizer group $\mathcal{S}$ for the Shor code is generated by local weight-2 $Z$-checks (detecting bit-flips within blocks) and nonlocal weight-6 $X$-checks (detecting phase-flips between blocks). Specifically, the generators are:
\begin{itemize}
    \item $X$-checks: $\begin{smallmatrix} X & X & X \\  X & X & X \\ I & I & I \end{smallmatrix} \, , \, \begin{smallmatrix} I & I & I \\  X & X & X \\  X & X & X \end{smallmatrix}$
    % $X_1X_2X_3X_4X_5X_6$ and $X_4X_5X_6X_7X_8X_9$
    \item $Z$-checks: $\begin{smallmatrix} Z & Z & I \\ I & I & I \\ I & I & I \end{smallmatrix} \, , \, \begin{smallmatrix} I & Z & Z \\ I & I & I \\ I & I & I \end{smallmatrix} \, , \, \begin{smallmatrix} I & I & I \\ Z & Z & I \\ I & I & I \end{smallmatrix} \, , \, \begin{smallmatrix} I & I & I \\ I & Z & Z \\ I & I & I \end{smallmatrix} \, , \, \begin{smallmatrix} I & I & I \\ I & I & I \\ Z & Z & I \end{smallmatrix} \, , \, \begin{smallmatrix} I & I & I \\ I & I & I \\ I & Z & Z \end{smallmatrix}$
    % $Z_1Z_2, Z_2Z_3, Z_4Z_5, Z_5Z_6, Z_7Z_8, Z_8Z_9$
\end{itemize}
While the $Z$-checks are local, the $X$-checks are high-weight. High-weight stabilizers are experimentally challenging, since they require deeper quantum circuits for syndrome measurement, increasing the probability of errors occurring during the measurement process itself.

% \subsubsection{Reducing stabilizer weight}

Can we reduce the weight of the $X$-stabilizers? Suppose we attempt to measure operators with lower weight that multiply to form the original stabilizers. For the 9-qubit Shor code, we could replace the weight-6 $X$-checks with weight-2 checks acting on neighbors.

Let us define a modified Shor code (or the \emph{$3 \times 3$ Bacon--Shor code}). As before, we arrange the 9 qubits in a $3 \times 3$ grid. Instead of measuring weight-6 operators, we measure the following weight-2 operators:
\begin{itemize}
    \item Gauge $X$-checks: $\begin{smallmatrix} X & I & I \\ X & I & I \\ I & I & I \end{smallmatrix} \, , \, \begin{smallmatrix} I & X & I \\  I & X & I \\ I & I & I \end{smallmatrix} \, , \, \begin{smallmatrix} I & I & X \\ I & I & X \\ I & I & I \end{smallmatrix} \, , \, \begin{smallmatrix} I & I & I \\ X & I & I \\ X & I & I \end{smallmatrix} \, , \, \begin{smallmatrix} I & I & I \\  I & X & I \\  I & X & I \end{smallmatrix} \, , \, \begin{smallmatrix} I & I & I \\ I & I & X \\ I & I & X \end{smallmatrix}$ (vertical neighbors)
    \item Gauge $Z$-checks: $\begin{smallmatrix} Z & Z & I \\ I & I & I \\ I & I & I \end{smallmatrix} \, , \, \begin{smallmatrix} I & Z & Z \\ I & I & I \\ I & I & I \end{smallmatrix} \, , \, \begin{smallmatrix} I & I & I \\ Z & Z & I \\ I & I & I \end{smallmatrix} \, , \, \begin{smallmatrix} I & I & I \\ I & Z & Z \\ I & I & I \end{smallmatrix} \, , \, \begin{smallmatrix} I & I & I \\ I & I & I \\ Z & Z & I \end{smallmatrix} \, , \, \begin{smallmatrix} I & I & I \\ I & I & I \\ I & Z & Z \end{smallmatrix}$ (horizontal neighbors)
\end{itemize}
These lower-weight operators do \emph{not} all commute. For example, a vertical $XX$ pair might anticommute with a horizontal $ZZ$ pair if they share a qubit. Because they do not commute, they cannot all be elements of a stabilizer group (which must be abelian).

However, notice that the original weight-6 $X$-stabilizers of the Shor code can be reconstructed from these lower-weight operators. For example, the product of the three vertical $XX$ checks along the first two rows generates the weight-6 $X$-stabilizer on those two rows. This introduces the core concept of subsystem codes: we measure non-commuting ``gauge'' operators. The actual stabilizers of the code are the specific products of these gauge operators that commute with all other gauge operators. The measurement values of the individual non-commuting gauge operators do not carry any logical information by themselves.

\subsection{The mathematical framework of subsystem codes}\label{subsec:subsystem-framework}
We now formalize the structure of subsystem codes~\cite{poulin2005stabilizer,kribs2005unified,kribs2005operator}.

\subsubsection{Hilbert space decomposition}\label{subsubsec:hilbert-space-decomposition}
Let $\mathcal{H} = (\CC^2)^{\otimes n}$ be the Hilbert space of $n$ qubits. Rather than protecting a subspace $\mathcal{C} \subseteq \mathcal{H}$, we decompose $\mathcal{C}$ as $\mathcal{C} \cong \mathcal{H}_L \otimes \mathcal{H}_G$, where $\mathcal{H}_L$ carries the logical information we wish to protect and $\mathcal{H}_G$ is a gauge subsystem whose state can be changed without affecting the encoded information. Formally, we define a subsystem code as follows.

\begin{defbox}[label={def:subsystem-code}]{Subsystem code}
    An $\subsyscode{n}{k}{g}{d}$ subsystem code is defined by a decomposition of the Hilbert space $\mathcal{H} = (\CC^2)^{\otimes n}$ of $n$ qubits as:
    \begin{equation}
        \mathcal{H} = \mathcal{C} \oplus \mathcal{C}^\perp = (\mathcal{H}_L \otimes \mathcal{H}_G) \oplus \mathcal{C}^\perp
    \end{equation}
    where:
    \begin{itemize}
        \item $\mathcal{C}$ is the code subspace,
        \item $\mathcal{H}_L$ is the logical subsystem with $\dim \mathcal{H}_L = 2^k$, and
        \item $\mathcal{H}_G$ is the gauge subsystem with $\dim \mathcal{H}_G = 2^g$.
    \end{itemize}
    The parameters $n,k,g,d$ are defined as follows:
    \begin{itemize}
        \item $n$: Number of physical qubits, i.e., $\log_2 \dim \mathcal{H}$.
        \item $k$: Number of logical qubits, i.e., $\log_2 \dim \mathcal{H}_L$.
        \item $g$: Number of gauge qubits, i.e., $\log_2 \dim \mathcal{H}_G$.
        \item $d$: Distance of the code, i.e., minimum weight of a nontrivial logical operator on the code.
    \end{itemize}
\end{defbox}

\subsubsection{Operator description}\label{subsubsec:operator-description}
As with stabilizer codes, it is more common to define a subsystem code in terms of Pauli operators. However, beyond having to define the stabilizer group, we also need to define the \emph{gauge group} for the code. Both of these are subgroups of the $n$-qubit Pauli group $\mathcal{P}_n$ and are defined as follows:
\begin{itemize}
    \item Stabilizer group $\mathcal{S}$: An abelian subgroup of $\mathcal{P}_n$ \emph{not} containing $-I$. The common $(+1)$-eigenspace of the operators in $\mathcal{S}$ is the codespace $\mathcal{C}$ of the subsystem code. Thus, elements of $\mathcal{S}$ act on the codespace $\mathcal{C} = \mathcal{H}_L \otimes \mathcal{H}_G$ as the identity $I_{\mathcal{H}_L} \otimes I_{\mathcal{H}_G}$.
    \item Gauge group $\mathcal{G}$: A (not necessarily abelian) subgroup of $\mathcal{P}_n$ whose center $Z(\mathcal{G})$ is equal to $\langle \mathcal{S}, iI \rangle$. Elements of $\mathcal{G}$ act on the codespace $\mathcal{C} = \mathcal{H}_L \otimes \mathcal{H}_G$ as $I_{\mathcal{H}_L} \otimes g_{\mathcal{H}_G}$, where $g_{\mathcal{H}_G}$ is an operator on $\mathcal{H}_G$. These are the operators we are allowed to measure.
\end{itemize}

A standard stabilizer code is simply the special case where $\dim(\mathcal{H}_G) = 1$ (the gauge subsystem is trivial), and consequently $\mathcal{G} = \langle \mathcal{S}, iI \rangle$.

\subsubsection{Logical operators}\label{subsubsec:logical-operators}
Logical operators are operations that act nontrivially on subsystem $\mathcal{H}_L$ while preserving the codespace $\mathcal{C} = \mathcal{H}_L \otimes \mathcal{H}_G$. They are elements of the centralizer of the stabilizer group, denoted $C(\mathcal{S})$, that are not in $\mathcal{G}$ (since elements in $\mathcal{G}$ act trivially on $\mathcal{H}_L$). We distinguish between two types:
\begin{itemize}
    \item Bare logicals: Operators $L \in C(\mathcal{G})$ that are not in $\mathcal{G}$. These commute with all gauge operators and act as $L_{\mathcal{H}_L} \otimes I_{\mathcal{H}_G}$, where $L_{\mathcal{H}_L} \ne I_{\mathcal{H}_L}$.
    \item Dressed logicals: Operators $L \in C(\mathcal{S})$ that are not in $\mathcal{G}$. These commute with the stabilizers but may anticommute with some gauge operators. They act as $L_{\mathcal{H}_L} \otimes L_{\mathcal{H}_G}$, where $L_{\mathcal{H}_L} \ne I_{\mathcal{H}_L}$. (We adopt the convention that bare logicals are also dressed logicals, but it should be noted that some authors exclude bare logicals from the definition of dressed logicals.)
\end{itemize}
In practice, dressed logicals are important because a bare logical multiplied by any gauge operator is a dressed logical that performs the same logical operation on the logical subsystem $\mathcal{H}_L$.

\subsubsection{Description in terms of virtual Pauli operators}\label{subsubsec:virtual-pauli-operators}
A convenient way to understand subsystem codes is to introduce the notion of virtual Pauli operators~\cite{poulin2005stabilizer}. Given an $\subsyscode{n}{k}{g}{d}$ subsystem code, one can construct $n$ pairs of ``virtual'' Pauli $X$ and $Z$ operators $\{X_1', Z_1', X_2', Z_2', \cdots, X_n', Z_n'\}$ from $\mathcal{P}_n$ satisfying the usual commutation and anticommutation relations:
\begin{enumerate}
    \item $[X_j', X_\ell'] = [Z_j', Z_\ell'] = [X_j', Z_\ell'] = 0$ for all distinct $j, \ell \in \{1, \cdots, n\}$,
    \item $X_j' Z_j' = - Z_j' X_j'$ for all $j \in \{1, \cdots, n\}$, and
    \item $(X_j')^2 = (Z_j')^2 = I$ for all $j \in \{1, \cdots, n\}$,
\end{enumerate}
such that the stabilizer group is given by $\mathcal{S} = \langle Z_1', \cdots, Z_s' \rangle$ for $s = n - k - g$, and the gauge group is given by $\mathcal{G} = \langle iI, \mathcal{S}, X_{s+1}', Z_{s+1}', \cdots, X_{s+g}', Z_{s+g}'\rangle$. A set of check measurements for the code is thus $\{Z_1', \cdots, Z_s', X_{s+1}', Z_{s+1}', \cdots, X_{s+g}', Z_{s+g}'\}$. Representatives for the logical group are given by $\{X_{s+g+1}', Z_{s+g+1}', \cdots, X_{n}', Z_{n}'\}$.

\subsection{The Bacon--Shor code}\label{subsec:bacon-shor-code}
The \eczoo[Bacon--Shor code]{bacon_shor}~\cite{bacon2006operator} is the canonical example of a subsystem code, defined on an $M \times N$ rectangular lattice of qubits.

\subsubsection{Construction}\label{subsec:construction_bacon_shor}
\begin{figure}[tb]
    \centering
    \begin{tikzpicture}[scale=0.8]
        \draw[step=1cm,black,thin] (0,0) grid (5,4);
        \node[left] at (0,0) {(0,0)};
        % \foreach \x in {1,...,6} {
        %     \node[below] at (\x,0) {(\x,0)};
        % }
        % \foreach \y in {1,...,5} {
        %     \node[left] at (0,\y) {(0,\y)};
        % }
        \node[right] at (5,0) {(5,0)};
        \node[left] at (0,4) {(0,4)};
        
        \foreach \i in {0,...,4} {
            \foreach \j in {0,...,4} {
                \draw[very thick, red] (\i,\j) -- (\i+1,\j);
            }
        }        
        \foreach \i in {0,...,5} {
            \foreach \j in {0,...,3} {
                \draw[very thick, blue] (\i,\j) -- (\i,\j+1);
            }
        }
        \foreach \i in {0,...,5} {
            \foreach \j in {0,...,4} {
                \fill[black] (\i,\j) circle (2pt);
            }
        }
    \end{tikzpicture}
    \caption{The lattice for a $6 \times 5$ Bacon--Shor code. Red links represent $XX$ gauge generators, and blue links represent $ZZ$ gauge generators. Qubits are located at vertices.}
    \label{fig:bacon_shor}
\end{figure}
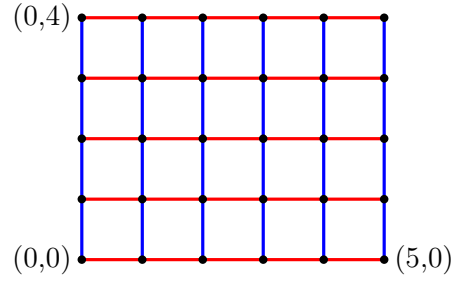

Let qubits be placed at coordinates $(i,j)$ for $0 \le i \le M-1$ and $0 \le j \le N-1$ (with the qubit at the lower left corner labeled $(0,0)$). The code is defined by the gauge group $\mathcal{G}$ generated by weight-2 Pauli operators acting on adjacent qubits, namely:
\begin{itemize}
    \item Horizontal checks: $X_{i,j} X_{i+1, j}$ for $0 \le i \le M-2$ and $0 \le j \le N-1$.
    \item Vertical checks: $Z_{i,j} Z_{i, j+1}$ for $0 \le i \le M-1$ and $0 \le j \le N-2$.
\end{itemize}
See~\cref{fig:bacon_shor} for the Bacon--Shor code lattice when $M = 6, N = 5$.

\subsubsection{Stabilizers and logicals}\label{subsubsec:stabilizers-and-logicals}
The stabilizers of the Bacon--Shor code are formed by products of gauge operators that commute with all operators in $\mathcal{G}$.
\begin{itemize}
    \item An $X$-stabilizer is formed by taking the product of horizontal $XX$ checks across two adjacent columns: $\prod_{j=0}^{N-1} X_{i,j} X_{i+1, j}$ for $0 \le i \le M-2$. Notice this commutes with all the vertical $ZZ$ checks.
    \item A $Z$-stabilizer is formed by taking the product of vertical $ZZ$ gauges across two adjacent rows: $\prod_{i=0}^{M-1} Z_{i,j} Z_{i, j+1}$ for $0 \le j \le N-2$. This commutes with all the horizontal $XX$ checks.
\end{itemize}
These $M+N-2$ stabilizers form a minimal generating set for the stabilizer group $\mathcal{S}$ of the Bacon--Shor code. The $(M-1)N + M(N-1) = 2MN - M - N$ gauge check operators listed in~\cref{subsec:construction_bacon_shor} form a minimal generating set for the gauge group up to phase. Subtracting the $M+N-2$ stabilizers gives $2MN - 2M - 2N + 2$, so the number of gauge qubits is $\frac{1}{2}(2MN - 2M - 2N + 2) = MN - M - N + 1 = (M-1)(N-1)$. In fact, it is possible to explicitly define the $(M-1)(N-1)$ pairs of virtual gauge operators in such a way that there is a clear one-to-one correspondence between the gauge qubits and the square plaquettes of the lattice; see~\cite[Sec. II]{alam2024dynamical} for a careful exposition.

There is exactly one logical qubit for the Bacon--Shor code. The logical operators are $\overline{X} = \prod_{j=0}^{N-1} X_{0,j}$ and $\overline{Z} = \prod_{i=0}^{M-1} Z_{i,0}$, which respectively correspond to a chain of $X$ operators spanning the lattice vertically and a chain of $Z$ operators spanning the lattice horizontally.

\begin{exerbox}[label={exer:bacon_shor_distance}]{Distance of Bacon--Shor code}
    Show that the distance of the $M \times N$ Bacon--Shor code is $\min(M, N)$. 
    
    [Hint: recall that the distance of the code is the minimum weight of a dressed logical of the code.]
\end{exerbox}

The parameters of the Bacon--Shor code are thus $\subsyscode{MN}{1}{(M-1)(N-1)}{\min(M, N)}$.

\subsection{Knill--Laflamme conditions for subsystem codes}\label{subsec:kl-conditions-subsystem}
How do we determine if a subsystem code can correct a specific set of errors $\mathcal{E}$? For this, we have to generalize the standard Knill--Laflamme conditions.

\begin{thmbox}[label={thm:subsystem-kl-conditions}]{Knill--Laflamme conditions for subsystem codes}
    Let $\mathcal{H} = (\mathcal{H}_L \otimes \mathcal{H}_G) \oplus \mathcal{C}^\perp$ be a subsystem code with projector $P$ onto $\mathcal{C} = \mathcal{H}_L \otimes \mathcal{H}_G$. The code $\mathcal{C}$ can correct a set of errors $\mathcal{E}$ if and only if for any pair of errors $E_j, E_\ell \in \mathcal{E}$, there exists an operator $g_{j\ell}$ acting on the gauge subsystem $\mathcal{H}_G$ such that
    \begin{equation}
        P E_j^\dagger E_\ell P = I_L \otimes g_{j\ell}\,.
    \end{equation}
\end{thmbox}

In the standard stabilizer case ($\dim \mathcal{H}_G = 1$), $g_{j\ell}$ is simply a scalar $\alpha_{j\ell}$, recovering the Knill--Laflamme condition $P E_j^\dagger E_\ell P = \alpha_{j\ell} P$ for subspace codes. In the subsystem case, the error product $E_j^\dagger E_\ell$ is allowed to disturb the gauge subsystem $\mathcal{H}_G$ arbitrarily, provided it acts as the identity (or a scalar) on the logical subsystem $\mathcal{H}_L$.

\begin{exerbox}[label={exer:knill_laflamme_subsyscode}]{Knill--Laflamme condition for Pauli errors}
    If the errors are Pauli operators, prove that the Knill--Laflamme condition simplifies to the following statement: A set of Pauli errors $\mathcal{E}$ is correctable if and only if for all $E_j, E_\ell \in \mathcal{E}$, the product $E_j E_\ell$ is not a dressed logical operator, that is, $E_j E_\ell \notin C(\mathcal{S}) \setminus \mathcal{G}$.
\end{exerbox}

\subsection{Topological subsystem codes}\label{subsec:topological-subsystem}
Subsystem codes naturally extend to topological settings, where we desire spatially local check operators of low weight. In this section, we introduce subsystem versions of the toric and surface codes introduced in~\cref{subsec:surface-code}.

\subsubsection{Subsystem toric code}\label{subsubsec:subsystem-toric}
\begin{figure}[!b]
    \centering
    \begin{tikzpicture}[scale=1.5]
        % Loop for a 3x3 grid
        \foreach \x in {0,1,2} {
            \foreach \y in {0,1,2} {
                % Draw internal triangles for each cell
                
                % Top-left triangle (Blue)
                \fill[fill=blue] (\x,\y+1) -- (\x+0.5,\y+1) -- (\x,\y+0.5) -- cycle;
                \draw (\x,\y+1) -- (\x+0.5,\y+1) -- (\x,\y+0.5) -- cycle;
                
                % Top-right triangle (Red)
                \filldraw[fill=red] (\x+0.5,\y+1) -- (\x+1,\y+1) -- (\x+1,\y+0.5) -- cycle;
                
                % Bottom-left triangle (Red)
                \filldraw[fill=red] (\x,\y+0.5) -- (\x,\y) -- (\x+0.5,\y) -- cycle;
                
                % Bottom-right triangle (Blue)
                \fill[fill=blue] (\x+0.5,\y) -- (\x+1,\y) -- (\x+1,\y+0.5) -- cycle;
                \draw (\x+0.5,\y) -- (\x+1,\y) -- (\x+1,\y+0.5) -- cycle;
                
                % Draw the cell border
                \draw[thick] (\x,\y) rectangle (\x+1,\y+1);
            }
        }

        % --- Logical Operator Annotations ---

        \begin{scope}[line width=3pt, color=cyan]
            % Vertical logical line
            \draw (1, 0) -- (1, 3);
            \node[below, font=\large\bfseries] at (1, 0) {$\Gamma_v$};

            % Horizontal logical line
            \draw (0, 1) -- (3, 1);
            \node[left, font=\large\bfseries] at (0, 1) {$\Gamma_h$};
        \end{scope}
        
    \end{tikzpicture}
    \caption{The lattice for the subsystem toric code with $M = 3$, with two noncontractible loops $\Gamma_h$ and $\Gamma_v$ labeled. Note that the left and right boundaries of the lattice are identified with each other, and so are the top and bottom boundaries. Qubits are placed at each vertex and the center of each edge of the lattice. Red triangles represent $XXX$ check operators, with an $X$ check at each vertex of the triangle. Blue triangles represent $ZZZ$ check operators, with a $Z$ check at each vertex of the triangle. Each bold cyan line represents a noncontractible loop of the torus made up of $2M$ qubits and is used in the definition of the logical operators of the code.}
    \label{fig:subsystem_toric_code}
\end{figure}
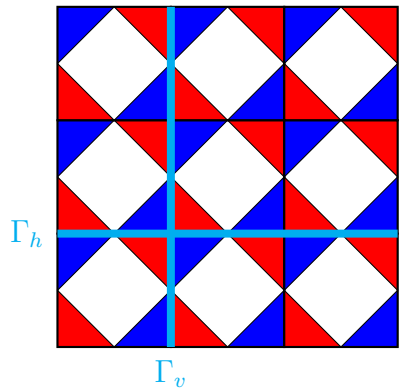

Defined on a lattice of size\footnote{This refers to the number of \emph{distinct} qubits in each row and column of the lattice. This means that lattice points which are identified (e.g., those on the top and bottom boundaries) are only counted once.} $M \times M$ with periodic boundary conditions, the subsystem toric code~\cite{bravyi2012subsystem} uses a layout involving triangular checks of weight $3$. See~\cref{fig:subsystem_toric_code} for an example where $M = 3$. Qubits are placed on both the vertices and the centers of the edges of the lattice, for a total of $n = 3M^2$ physical qubits ($M^2$ vertices and $2M^2$ edges). For every triangle in the lattice, we have a $3$-qubit check operator (which could be either $X$ or $Z$) involving the vertex qubit and the adjacent edge qubits. These are sometimes called $X$ or $Z$ ``triangle operators'', and form a generating set for the gauge group (up to phase). There are in total $2M^2$ $X$ and $Z$ triangle operators each. For the purposes of calculating the number of gauge qubits, it should be noted that these $4M^2$ triangle operators do not form a \emph{minimal} generating set for the gauge group (up to phase); any $X$ (resp.,~$Z$) triangle operator is the product of all the other $2M^2 - 1$ $X$ (resp.,~$Z$) triangle operators. To get a minimal generating set, one can remove any one $X$ and any one $Z$ triangle operator from the set of $4M^2$ check operators.

\begin{exerbox}[label={exer:stab_subsys_toric_code}]{Stabilizers of the subsystem toric code}
    \begin{enumerate}[label=(\alph*)]
        \item Each plaquette of the subsystem toric code consists of a pair of $X$ triangle operators and a pair of $Z$ triangle operators. Show that the weight-6 operator defined by the product of the two weight-3 $X$ (or $Z$) triangle operators on a single plaquette commutes with all of the triangle operators, and thus are stabilizers of the code.
        \item Show also that the set of such operators generates the stabilizer group of the subsystem toric code, that is, any stabilizer is a product of such weight-6 operators.
        
        [Hint: Each stabilizer is a product of triangle operators. Prove that if a particular triangle operator appears in the product, then the other triangle operator of the same type ($X$ or $Z$) in the same plaquette must also appear in order for the stabilizer to commute with either triangle operator of the opposite type in the same plaquette.]
        \item There are $M^2$ such weight-6 $X$ stabilizers and $M^2$ such weight-6 $Z$ stabilizers. Show that any $M^2 - 1$ of the $X$ stabilizers and $M^2 - 1$ of the $Z$ stabilizers form a minimal generating set of the stabilizer group, so $s = 2(M^2 - 1)$.
        
        [Hint: The product of all $M^2$ $X$ or $Z$ stabilizers is the identity.]
    \end{enumerate}
\end{exerbox}

Since the minimal generating set for the gauge group (up to phase) has size $2(2M^2-1)$ and that for the stabilizer group has size $2(M^2 - 1)$, we have
\begin{equation}
    g = \frac{1}{2} \left(2(2M^2 - 1) - 2(M^2 - 1)\right) = M^2\,.
\end{equation}
Therefore, this code encodes
\begin{equation}
    k = n - g - s = 3M^2 - M^2 - 2(M^2 - 1) = 2
\end{equation}
logical qubits.

\begin{figure}[tb]
    \centering
    \begin{tikzpicture}[scale=1.5]
        % --- 1. Grid and Triangles ---
        \foreach \x in {0,1,2} {
            \foreach \y in {0,1,2} {
                % Top-left triangle (Blue)
                \filldraw[fill=blue] (\x,\y+1) -- (\x+0.5,\y+1) -- (\x,\y+0.5) -- cycle;
                % Top-right triangle (Red)
                \filldraw[fill=red] (\x+0.5,\y+1) -- (\x+1,\y+1) -- (\x+1,\y+0.5) -- cycle;
                % Bottom-left triangle (Red)
                \filldraw[fill=red] (\x,\y+0.5) -- (\x,\y) -- (\x+0.5,\y) -- cycle;
                % Bottom-right triangle (Blue)
                \filldraw[fill=blue] (\x+0.5,\y) -- (\x+1,\y) -- (\x+1,\y+0.5) -- cycle;
                
                % Draw the cell border
                \draw[thick] (\x,\y) rectangle (\x+1,\y+1);
            }
        }

        % --- 2. Logical Operator Annotations (Green Lines) ---
        \begin{scope}[line width=3pt, color=cyan]
            \foreach \x in {0,1,2} {
                \draw (\x, 0) -- (\x, 3);
                \node[below, font=\large\bfseries] at (\x, 0) {$\Gamma_{v,\x}$};
            }
            \draw (3, 0) -- (3, 3);
            \node[below, font=\large\bfseries] at (3, 0) {$\Gamma_{v,0}$};
        \end{scope}

        % --- 3. Error Illustration (Distance 3) ---
        \begin{scope}[red]
            % Draw circles at (0,1.5), (1,1.5),(2,1.5), (3,1.5)
            \foreach \x in {0,...,3} {
                \fill (\x, 1.5) circle (3pt);
            }
            
            % Draw a dashed line across the three points to represent the error string
            \draw[dashed, ultra thick] (-0.3,1.5) -- (3.3,1.5) node[right, font=\Large\bfseries] {$E$};
        \end{scope}

    \end{tikzpicture}
    \caption{Distance calculation for the subsystem toric code with $M = 3$, showing a weight-3 $X$-error $E$ along the vertical edges. Note that the left and right boundaries of the lattice are identified with each other.}
    \label{fig:dist_subsystem_toric_code}
\end{figure}
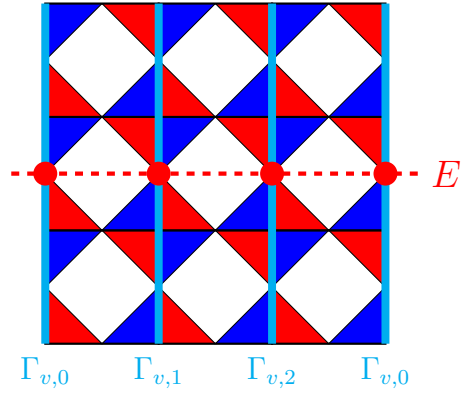

Let $\Gamma_h$ (resp.,~$\Gamma_v$) denote the set of $2M$ qubits that lie along a fixed horizontal (resp.,~vertical) line of the lattice. This represents a noncontractible loop of the torus. The two pairs of (bare) logical operators of the code can be represented by:
\begin{itemize}
    \item $\overline{X}_1 = \prod_{j \in \Gamma_h} X_j$ and $\overline{Z}_1 = \prod_{j \in \Gamma_v} Z_j$,
    \item $\overline{X}_2 = \prod_{j \in \Gamma_v} X_j$ and $\overline{Z}_2 = \prod_{j \in \Gamma_h} Z_j$.
\end{itemize}
To determine the distance of the code, we need to calculate the minimum distance of any nontrivial logical operator. Observe that such a Pauli operator $E$ must commute with all the stabilizers but must anticommute with at least one other logical operator among $\overline{X}_1, \overline{X}_2, \overline{Z}_1, \overline{Z}_2$. Let us assume without loss of generality that $E$ anticommutes with $\overline{Z}_1 = \prod_{j \in \Gamma_{v,0}} Z_j$, where $\Gamma_{v,0}$ is a vertical line of qubits as labeled in~\cref{fig:dist_subsystem_toric_code}. (The vertical lines of qubits are labeled $\Gamma_{v,0}, \ldots, \Gamma_{v,M-1}$.)

Note then that $E$ also anticommutes with $ \prod_{j \in \Gamma_{v,i}} Z_j$ for any $i \in \{1, \ldots, M-1\}$. To see this, observe that it suffices to show $E$ commutes with the product $\left(\prod_{j \in \Gamma_{v,0}} Z_j\right) \left(\prod_{j \in \Gamma_{v,i}} Z_j\right)$, and this latter statement is true because that product is simply the product of all the weight-$6$ $Z$ plaquette stabilizers lying between the lines $\Gamma_{v,0}$ and $\Gamma_{v,i}$.

Since $E$ anticommutes with each $\Gamma_{v,i}$, it must act nontrivially on at least one qubit of each $\Gamma_{v,i}$ (specifically, via an $X$ or $Y$). Thus, $E$ must act nontrivially on at least $M$ qubits, showing that the distance of the subsystem toric code is at least $M$. If we take $E$ to be the operator acting as $X$ on all $M$ qubits intersecting a fixed horizontal line at half-integer coordinates of the lattice (see~\cref{fig:dist_subsystem_toric_code} for an example), then it can be checked that $E$ is indeed a nontrivial logical operator (as it commutes with all the stabilizers and anticommutes with $\overline{Z}_1$). In fact, it is not hard to directly check that $E$ and $\overline{X}_1$ differ by a product of gauge operators, so $E$ is a dressed logical that performs the same logical operation as $\overline{X}_1$. In light of the discussion above, we see that the subsystem toric code on a lattice of size $M \times M$ has parameters $\subsyscode{3M^2}{2}{M^2}{M}$.

\subsubsection{Subsystem surface code}\label{subsubsec:subsystem-surface-code}
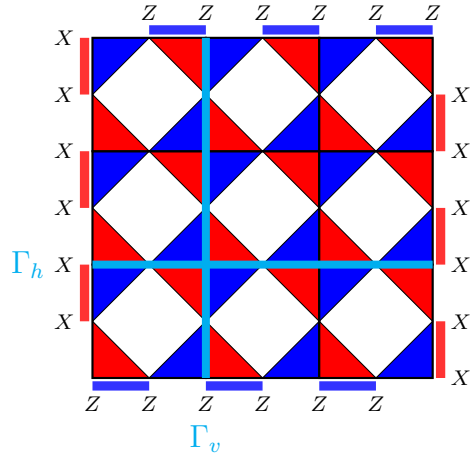
\begin{figure}[tb]
    \centering
    \begin{tikzpicture}[scale=1.5]
        % --- 1. Main Lattice (3x3 Grid) ---
        \foreach \x in {0,1,2} {
            \foreach \y in {0,1,2} {
                % Top-left triangle (Blue/Z)
                \filldraw[fill=blue] (\x,\y+1) -- (\x+0.5,\y+1) -- (\x,\y+0.5) -- cycle;
                % Top-right triangle (Red/X)
                \filldraw[fill=red] (\x+0.5,\y+1) -- (\x+1,\y+1) -- (\x+1,\y+0.5) -- cycle;
                % Bottom-left triangle (Red/X)
                \filldraw[fill=red] (\x,\y+0.5) -- (\x,\y) -- (\x+0.5,\y) -- cycle;
                % Bottom-right triangle (Blue/Z)
                \filldraw[fill=blue] (\x+0.5,\y) -- (\x+1,\y) -- (\x+1,\y+0.5) -- cycle;
                
                % Draw the cell border
                \draw[thick] (\x,\y) rectangle (\x+1,\y+1);
            }
        }

        % --- 2. Boundary Gauge Operators (Weight-2 Bars) ---
        % Using a smaller offset of 0.07 to bring them closer to the edge.
        \pgfmathsetmacro{\off}{0.07}
        
        % Vertical Red Bars (X) on Left/Right
        \foreach \y in {0,1,2} {
            % Left boundary bars
            \draw[line width=3.5pt, red!80] (-\off, \y+0.5) -- (-\off, \y+1);
            \node[left, font=\scriptsize] at (-\off, \y+0.5) {$X$};
            \node[left, font=\scriptsize] at (-\off, \y+1) {$X$};
            
            % Right boundary bars
            \draw[line width=3.5pt, red!80] (3+\off, \y) -- (3+\off, \y+0.5);
            \node[right, font=\scriptsize] at (3+\off, \y) {$X$};
            \node[right, font=\scriptsize] at (3+\off, \y+0.5) {$X$};
        }

        % Horizontal Blue Bars (Z) on Top/Bottom
        \foreach \x in {0,1,2} {
            % Top boundary bars
            \draw[line width=3.5pt, blue!80] (\x+0.5, 3+\off) -- (\x+1, 3+\off);
            \node[above, font=\scriptsize] at (\x+0.5, 3+\off) {$Z$};
            \node[above, font=\scriptsize] at (\x+1, 3+\off) {$Z$};
            
            % Bottom boundary bars
            \draw[line width=3.5pt, blue!80] (\x, -\off) -- (\x+0.5, -\off);
            \node[below, font=\scriptsize] at (\x, -\off) {$Z$};
            \node[below, font=\scriptsize] at (\x+0.5, -\off) {$Z$};
        }

        % --- Logical Operator Annotations ---

        \begin{scope}[line width=3pt, color=cyan]
            % Vertical logical line
            \draw (1, 0) -- (1, 3);
            \node[below, font=\large\bfseries] at (1, -0.3) {$\Gamma_v$};

            % Horizontal logical line
            \draw (0, 1) -- (3, 1);
            \node[left, font=\large\bfseries] at (-0.3, 1) {$\Gamma_h$};
        \end{scope}

    \end{tikzpicture}
    \caption{The lattice for the subsystem surface code with $M = 4$. Like the subsystem toric code, qubits are placed at each vertex and the center of each edge of the lattice. Red triangles represent $XXX$ check operators, with an $X$ check at each vertex of the triangle. Blue triangles represent $ZZZ$ check operators, with a $Z$ check at each vertex of the triangle. Boundary operators of weight-$2$ are indicated by thick bars with $X$ or $Z$ checks at each end. As before, each bolded cyan line is made up of $2M-1$ qubits and is used in the definition of the logical operators of the code.}
    \label{fig:subsystem_surface_code}
\end{figure}

The subsystem surface code~\cite{bravyi2012subsystem} on an $M \times M$ lattice\footnote{Specifically, the $M \times M$ lattice for the subsystem surface code comprises $M$ vertices and $M-1$ plaquettes along each row and column, as the lattice boundaries are not identified.} is defined similarly to the subsystem toric code, but with open boundary conditions. See~\cref{fig:subsystem_surface_code} for an example where $M = 4$. As before, qubits are placed on both the vertices and the centers of the edges of the lattice, but since the boundaries of the lattice are not identified, there are now a total of $n = 3M^2 - 2M$ physical qubits ($M^2$ vertices and $2M(M-1)$ edges). Besides the $4(M-1)^2$ weight-3 $X$ and $Z$ triangle operators, there are additional $4(M-1)$ weight-2 operators on the boundaries of the lattice, with $M-1$ such operators of type $X$ on each of the left and right boundaries and $M-1$ such operators of type $Z$ on each of the top and bottom boundaries. These form all the check operators for the subsystem surface code, and in fact form a minimal generating set (of size $4(M-1)^2 + 4(M-1) = 4M^2 - 4M$) for the gauge group (up to a phase).

\begin{exerbox}[label={exer:stab_subsys_surface_code}]{Stabilizers of the subsystem surface code}
    \begin{enumerate}[label=(\alph*)]
        \item Check that the $(M-1)^2$ plaquette operators of type $X$ and the $(M-1)^2$ plaquette operators of type $Z$ are stabilizers for the subsystem surface code. Verify that this also holds for the $4(M-1)$ boundary operators of weight 2.
        \item Show that these form a minimal generating set for the stabilizer group, and thus $s = 2M^2 - 2$.
    \end{enumerate}
    \end{exerbox}

By~\cref{exer:stab_subsys_surface_code}, we have
\begin{equation}
    g = \frac{1}{2} \left((4M^2 - 4M) - (2M^2 - 2)\right) = M^2 - 2M + 1 = (M-1)^2\,,
\end{equation}
and so the subsystem surface code encodes exactly
\begin{equation}
    k = n - g - s = (3M^2 - 2M) - (M^2 - 2M + 1 ) - (2M^2 - 2) = 1
\end{equation}
logical qubit.

Define $\Gamma_h$ and $\Gamma_v$ as before. Then, a pair of (bare) logical operators can be defined by $\overline{X} = \prod_{j \in \Gamma_h} X_j$ and $\overline{Z} = \prod_{j \in \Gamma_v} Z_j$. It is straightforward to verify that these do indeed commute with all check operators but not with each other. By a similar argument as in the subsystem toric code, the distance of the subsystem surface code is $M$. As such, the subsystem surface code on an $M \times M$ lattice has parameters $\subsyscode{3M^2 - 2M}{1}{(M-1)^2}{M}$.

\subsection{Why subsystem codes?}\label{subsec:why-subsystem-codes}
We have seen that subsystem codes mathematically generalize standard stabilizer codes by partitioning the codespace into a logical subsystem and a gauge subsystem. But why introduce this extra complexity? In this section, we outline the primary practical and theoretical advantages of subsystem codes.

\vspace{1em}
\noindent \textbf{Reduced weight of check operators.}
Historically, the first major motivation for subsystem codes was to simplify error syndrome measurements. As demonstrated in our transition from the 9-qubit Shor code to the Bacon--Shor code, introducing gauge degrees of freedom allows us to replace high-weight stabilizers (e.g., weight-6 $X$-checks) with low-weight gauge operators (weight-2 $XX$ and $ZZ$ checks). Because measuring high-weight operators requires deeper, more error-prone quantum circuits, subsystem codes often yield more robust fault-tolerant threshold behavior in near-term hardware.

\vspace{1em}
\noindent \textbf{Gauge fixing.} 
The defining feature of a subsystem code is the presence of the gauge group $\mathcal{G}$. Because the encoded logical information is invariant under the action of $\mathcal{G}$, we are free to measure any commuting subset of gauge operators. When we do this, we project the system into an eigenstate of those operators, a process called \emph{gauge fixing}. Effectively, the measured gauge operators are temporarily promoted to the stabilizer group $\mathcal{S}$. This ability to dynamically alter the code's stabilizer group on the fly offers several additional advantages, which we detail below.

\vspace{1em}
\noindent \textbf{Flexibility in handling biased noise.}
Gauge fixing allows a subsystem code to adapt to the dominant noise in the physical hardware. Consider the rectangular Bacon--Shor code described earlier in this chapter. If we choose to measure only the vertical $XX$-checks, the code collapses into a state stabilized by those $XX$-checks and the logical $Z$-stabilizers. Conversely, measuring only the horizontal $ZZ$-checks stabilizes the system against $X$ errors.

If a quantum processor suffers from highly biased noise, for example, if $Z$-errors occur much more frequently than $X$-errors, we can adjust our measurement schedule to perform the $XX$-checks much more frequently. This flexibility is impossible in standard stabilizer codes where the check operators are strictly fixed.

\vspace{1em}
\noindent \textbf{Dynamical logical qubits.}
By carefully scheduling which gauge operators are measured at which times, we can create \emph{dynamical logical qubits}. As demonstrated by Alam and Rieffel~\cite{alam2024dynamical}, one can apply a specific period-4 measurement schedule to the Bacon--Shor lattice. Instead of fixing all gauges at once, the schedule alternates between measuring different sets of $XX$ and $ZZ$ gauges.

At each round, only $(M-1)(N-1)-1$ gauge degrees of freedom are fixed (as opposed to $(M-1)(N-1)$, for the traditional Bacon--Shor construction). This leaves an extra degree of freedom that can store an additional logical qubit. Because the set of measured gauge operators changes at each time step, the logical operators of this extra qubit are ``dynamical'': they change at each step of the measurement round. This concept is the foundational mechanism behind the recent explosion of interest in \emph{Floquet codes}. See~\cref{sec:dynamical} for more details.

\vspace{1em}
\noindent \textbf{Enhanced decoding via schedule-induced gauge fixing.}
In the standard decoding of subsystem codes, the error syndrome is extracted by multiplying the outcomes of several gauge operator measurements to reconstruct a composite stabilizer. For instance, in the subsystem toric code, a weight-6 $X$-stabilizer is formed by taking the parity of two adjacent weight-3 $X$-triangle operators. By only passing this final product to the decoder, we discard the individual measurement outcomes of the triangles, which contain finer-grained information about an error's exact spatial and temporal location. Higgott and Breuckmann demonstrated~\cite{higgott2021subsystem} that this lost information can be recovered using \emph{schedule-induced gauge fixing}. By adjusting the measurement schedule to repeat a specific gauge measurement before any anti-commuting operators are introduced, the gauge measurement outcome becomes deterministic relative to its previous value.

Geometrically, this changes the structure of the decoding matching graph. In a minimum weight perfect matching (MWPM) decoder, the subsystem toric code without gauge information produces a dense triangular matching graph, where each weight-6 stabilizer is a vertex of degree 6. When the weight-3 triangle gauge operators are ``fixed'' by the schedule, each weight-6 vertex effectively ``splits'' into two distinct vertices of degree 3. This transforms the matching graph into a sparse hexagonal lattice. With a lower connectivity, there are fewer intersecting paths for the decoder to confuse, which significantly boosts the decoding accuracy and increases the circuit-level depolarizing threshold.

\vspace{1em}
\noindent \textbf{Code switching for fault-tolerant logical gates.}
To achieve universal FTQC, we need a universal set of logical gates (e.g., Clifford + $T$). However, the Eastin--Knill theorem dictates that no single quantum error-correcting code can have a universal set of transversal gates. Certain subsystem codes (e.g., gauge color codes) offer a workaround via \emph{code-switching}. For example, a 3D color code defined on a tetrahedral $3$-complex~\cite{bombin2007exact} allows for a transversal $T$ gate but lacks a transversal Hadamard ($H$) gate. Conversely, a 2D color code~\cite{bombin2006topological} allows for a transversal $H$ gate but lacks a transversal $T$ gate. As studied by Butt \emph{et al.}~\cite{butt2024fault}, by treating the 2D color code as a gauge-fixed version of a 3D subsystem color code, one can fault-tolerantly switch the encoded information between the 3D and 2D geometries. This allows the computer to execute a transversal $T$ gate in the 3D geometry, project down to the 2D geometry via gauge fixing to execute a transversal $H$ gate, and switch back. Similar code-switching principles apply to other families, such as the doubled color codes~\cite{bravyi2015doubled} and quadratic residue codes~\cite{sullivan2024codea,jain2025transversal}. Code-switching thus offers a pathway to universality that significantly reduces the reliance on incredibly costly magic state distillation protocols. We will return to code switching briefly in~\cref{subsec:code_switching}.

\vspace{1em}
In all, subsystem codes represent a powerful paradigm shift in quantum error correction. By relaxing the strict requirements of the stabilizer formalism and allowing certain degrees of freedom to remain unprotected (the gauge subsystem), we gain a substantial amount of flexibility. This flexibility translates into lower-weight check operators, dynamic responses to biased noise, enhanced decoding graphs, and novel pathways to fault-tolerant universal gate sets via code-switching. As hardware continues to mature, leveraging these gauge degrees of freedom will be a central theme in lowering the overhead of FTQC.

\newpage

\section{Dynamical codes}\label{sec:dynamical}

As we saw in~\cref{sec:subsystem}, we do not necessarily need commuting observables to define a codespace and to extract error syndromes. In a subsystem code, its codespace $Q$ is stabilized by the phaseless Pauli operators in the center of the gauge group $\mathcal{G}$ generated by the set of non-commuting check measurements.
% such that the codespace stays invariant up to some the same as we measure the gauge checks.
However, in extracting error syndromes we do actually measure these non-commuting checks which effectively evolve the codestate $\ket{\overline{\psi}}$ that we are measuring, and therefore the effective codespace at a given time.

\begin{figure}[!t]
    \centering
    \includegraphics[width=0.5\linewidth]{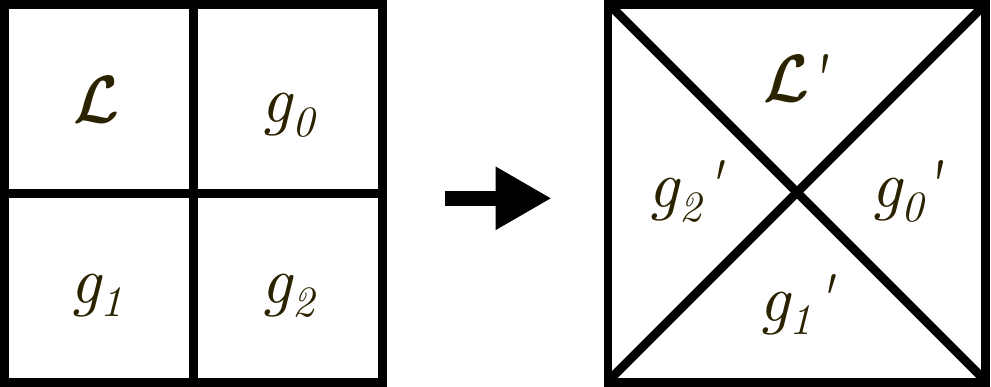}
    \caption{A change of logical subspace and syndrome subspaces.
    A subspace stabilizer code $Q$ (left-hand side) consists of logical subspace $\mathcal{L}$ which is the joint $+1$ eigenspace of three stabilizer generators $g_0,g_1,g_2$, and subspaces corresponding to $-1$ for each stabilizer generator $g_0,g_1,g_2$.
    Any operation performed on a code state $\ket{\psi}$ of $Q$, such as measurements or gates, may change the code $Q$ to a new subspace stabilizer code $Q'$ (right-hand side) with different subspace $\mathcal{L}'$ and three stabilizer generators $g_0',g_1',g_2'$.}
    \label{fig:dynamic_subspaces}
\end{figure}

\eczoo[\emph{Dynamical codes}]{da} need not have a fixed codespace, as long as logical information remains preserved throughout the syndrome extraction process. For a codespace $Q(\mathcal{S}_t)$ stabilized by a stabilizer group $\mathcal{S}_t$ at time $t$, measuring an observable $s'$ that anticommutes with some stabilizers in $\mathcal{S}_t$ on codestate $\ket{\overline{\psi}_t} \in Q(\mathcal{S}_t)$ evolves it to a state $\ket{\overline{\psi}_{t+1}}$ in codespace $Q(\mathcal{S}_{t+1})$, which is distinct from $Q(\mathcal{S}_t)$ but retains the encoded information. For an illustration, see~\cref{fig:dynamic_subspaces}. As we will see, dynamical codes are \emph{distinct} from subsystem codes. For a sequence of non-commuting check measurements $C_1,C_2,\ldots$ that defines a dynamical code that encodes a $K$-dimensional codespace, a subsystem code with gauge group generated by the same check measurements, that is, $\mathcal{G}=\<C_1\cup C_2\cup \dots\>$, may not allow any logical information to be encoded in it.

\subsection{``Hidden'' dynamical logical qubit}\label{subsec:dynamical-hidden}
\begin{figure}[!b]
    \centering
    \includegraphics[width=0.5\linewidth]{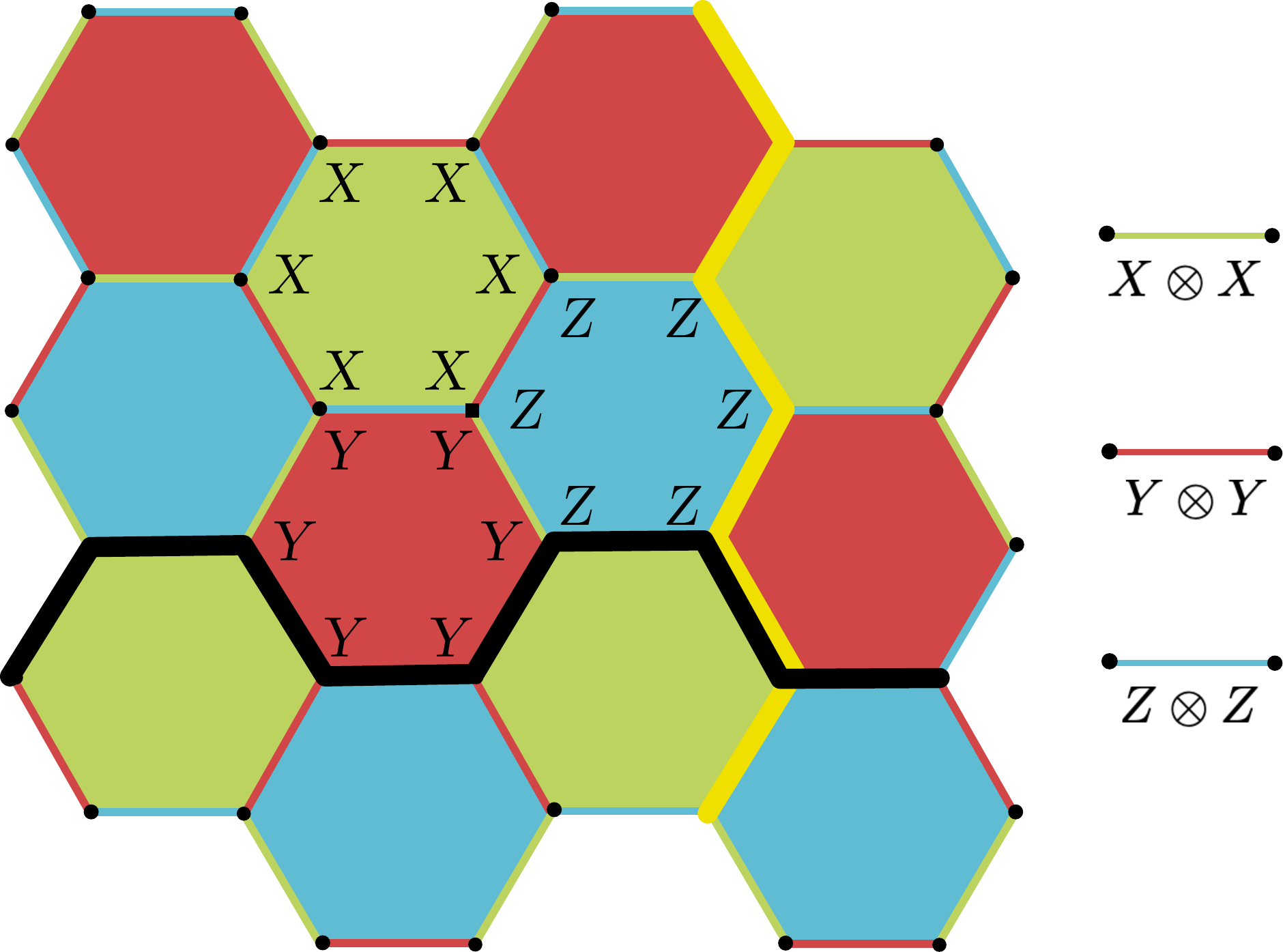}
    \caption{A $36$-qubit Kitaev's honeycomb code on a hexagonal ``honeycomb'' lattice with degree 3 on the surface of a torus.
    The lattice has a periodic boundary condition: The bottom and top boundaries represent the same set of edges, similarly for its left and right boundaries.
    Colors of hexagonal plaquettes and edges determine 6-qubit stabilizers and 2-qubit check measurements, respectively.
    In addition to the 6-qubit plaquette stabilizers, there are two stabilizers that loop around the torus along the vertical (yellow line) and horizontal (black line) directions.
    }
    \label{fig:honeycomb_lattice}
\end{figure}

As discussed in~\cref{sec:subsystem}, an $\llbracket n,k,g,d \rrbracket$ stabilizer subsystem code has $n$ physical $q$-dimensional systems, a $q^k$-dimensional logical subspace, a $q^g$-dimensional gauge subspace, and a $q^r$-dimensional error syndrome subspace. For some choice of gauge group $\mathcal{G}$, we could end up with gauges and stabilizers that take up ``too much'' space such that $g+r=n$, leaving no room to encode any logical information. This is the case with Kitaev's honeycomb code~\cite{kitaev2006anyons}, which is a qubit subsystem stabilizer code defined on a hexagonal lattice on the surface of a torus where qubits are placed on the vertices (see~\cref{fig:honeycomb_lattice}). The lattice edges and faces are colored with green, red, or blue such that any three edges sharing a vertex must have different colors and two hexagonal faces at the end of each edge have the same color as that edge.

This code has a gauge group $\mathcal{G}$ generated by two-qubit Pauli measurements on the edges of the lattice where the edge color determines its corresponding Pauli observable, while the plaquette colors determine the type of 6-qubit stabilizer on that plaquette\footnote{The original Kitaev code's check measurements are determined by the orientation of the edges such that each plaquette stabilizer is of the same type. Here we adopt the three-colored honeycomb code convention used in Ref.~\cite{gidney2021fault-tolerant}, which does not change the point we are trying to make here.}. The stabilizer group $\mathcal{S}$ is generated by 6-qubit Pauli operators on the hexagonal faces and two Pauli operators along two loops around the horizontal and vertical directions on the torus (black and yellow lines in~\cref{fig:honeycomb_lattice}).

If there are $n_p$ hexagonal faces on the lattice, then there are $3n_p$ edges and $n=2n_p$ qubits (vertices). Taking the product of all two-qubit Pauli operators $P_iP_j$ that generate the gauge group $\mathcal{G}$ of Kitaev's honeycomb code, gives us the identity, that is, $\prod_{i,j} P_iP_j = I$. Thus, the stabilizer code has $3n_p-1$ independent generators. On the other hand, the product of the 6-qubit Pauli operators on the faces is also equal to the identity, and so the independent generators of $\mathcal{S}$ consist of $n_p - 1$ 6-qubit Pauli operators on the faces and two loops wrapping around the torus in perpendicular directions. This gives us $n_p + 1$ independent generators of $\mathcal{S}$. This in turn gives us a space of $2^r=2^{n_p+1}$ dimensions that are allocated for the syndrome stabilizers, whereas a space of $2^g = 2^{n_p-1}$ dimensions are allocated for the gauge qubits (as $\mathcal{S}\subseteq\mathcal{G}$, $g=(3n_p-1 - n_p-1)/2$). Since there are $n=2n_p$ qubits on the lattice, the number of qubits taken up by the gauge qubits and the stabilizers is $n_p - 1 + n_p + 1 = 2n_p = n$, which shows that there is no encoded logical information.

Later, in~\cref{subsec:dynamical_hastings_haah}, we will see how by shifting our perspective from a \emph{fixed} codespace to a \emph{dynamical} codespace, one can \emph{dynamically} encode two logical qubits in Kitaev's code by looking at how the stabilizer group of the code evolves as a sequence of non-commuting check measurements are performed. This idea is basically what gave rise to the first dynamical code: the \eczoo[Hastings-Haah honeycomb Floquet code]{floquet}~\cite{hastings2021dynamically}. To gain a better understanding of how a codespace changes as we perform non-commuting check measurements while preserving logical information, we will first discuss in~\cref{sec:4_qubit_dynamical_code} a simpler example of a dynamical code consisting only of four data qubits. In this example, we will also see the phenomenon where one dynamical logical qubit is present in this 4-qubit code when viewed as a dynamical code, although there are no logical qubits present when the check measurements are taken to be the checks for a subsystem code.

\subsection{4-qubit dynamical code}\label{sec:4_qubit_dynamical_code}
\begin{figure}[tb]
    \centering
    \includegraphics[width=0.4\linewidth]{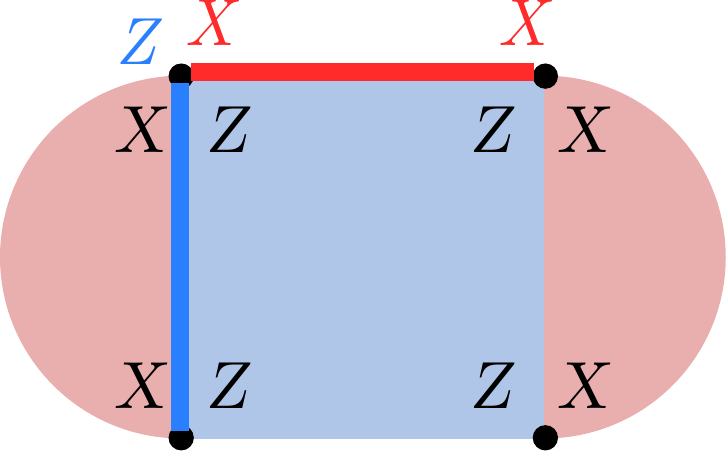}
    \caption{Four-qubit rotated surface code.
    We label the top-left qubit as qubit 1, top-right qubit as qubit 2, bottom-left qubit 3, and bottom-right qubit 4. 
    The red colored half-circles represent the $X_1X_3$ (left) and $X_2X_4$ (right) stabilizer generator of the code, whereas the blue square represents the $Z_1Z_2Z_3Z_4$ stabilizer generator.
    The bright red line and bright blue line represent the $X$-type logical operator $\overline{X}=X_1X_2$ and the $Z$-type logical operator $\overline{Z}=Z_1Z_3$, respectively.}
    \label{fig:d2_surface_code}
\end{figure}

In this section, we discuss a simple dynamical code that uses only four data qubits. We start by considering the 4-qubit rotated surface code, which is a stabilizer error-detecting code with parameters $\stabcode{4}{1}{2}$. Its stabilizer group $\mathcal{S}_0$ is generated by three stabilizer generators:
\begin{equation}
    X_1X_3\,, \quad X_2X_4\,, \quad Z_1Z_2Z_3Z_4\,,
\end{equation}
and logical operator representatives $\overline{X}=X_1X_2$ and $\overline{Z}=Z_1Z_3$, where we denote $P_j$ as a 4-qubit Pauli operator with Pauli operator $P$ on the $j\numth$ qubit and identity everywhere else. The 4-qubit rotated surface code can therefore be illustrated as in~\cref{fig:d2_surface_code}.

To better visualize the Pauli operators over the code, matching the illustration in~\cref{fig:d2_surface_code}, we denote a four-qubit Pauli operator $P=P_1P_2P_3P_4$ over these four data qubits as
\begin{equation}
    P =
    \begin{matrix}
        P_1 P_2 \\
        P_3 P_4
    \end{matrix}\,,
\end{equation}
so that the stabilizer generators and logical operators of $\mathcal{S}_0$ can be written as
\begin{equation}
    g_0 = \begin{matrix}
        ZZ\\
        ZZ
    \end{matrix}
    \quad,\quad
    g_1 = \begin{matrix}
        XI\\
        XI
    \end{matrix}
    \quad,\quad
    g_2 = \begin{matrix}
        IX\\
        IX
    \end{matrix}
    \quad,\quad
    \overline{X} = \begin{matrix}
        XX\\
        II
    \end{matrix}
    \quad,\quad
    \overline{Z} = \begin{matrix}
        ZI\\
        ZI
    \end{matrix}\,.
\end{equation}
We will adopt this notation in analyzing how the codespace and stabilizer group changes after we measure an observable $s'$ that anticommutes with some of the stabilizers of the four-qubit code.

\subsubsection{Updating stabilizer group}\label{subsubsec:updating-stabilizer-group}
We now denote $Q(\mathcal{S}_0)$ as the codespace of the 4-qubit rotated surface code. Then, consider the following measurements and how a codestate in $Q(\mathcal{S}_0)$ evolves upon these measurements:
\begin{enumerate}
    \item Measure $s_0'=\begin{matrix}
            IZ\\
            ZI
        \end{matrix}$ and $s_1'=\begin{matrix}
            ZI\\
            IZ
        \end{matrix}$ on arbitrary codestate $\ket{\overline{\psi}} \in Q(\mathcal{S}_0)$.
        
    \item $s_0'$ anticommutes with generators $g_1=\begin{matrix}
            IX\\
            IX
        \end{matrix}$,
        $g_2=\begin{matrix}
            XI\\
            XI
        \end{matrix}$.
        So, outcome $o_0=\pm 1$ of $s_0'$ is random and we obtain an updated state $\ket{\overline{\varphi}} = \Pi_o^{s_0'}\ket{\overline{\psi}}$, for $\Pi_o^{s_0'} = \frac{I+os_0'}{2}$ being the projector onto the $o$-eigenspace of Pauli observable $s_0'$. Note that the updated state $\ket{\overline{\varphi}}$ is stabilized by $os_0'$.

    \item Since $g_1,g_2$ anticommute with $s_0'$, then $g_1\Pi_o^{s_0'} = \Pi_{-o}^{s_0'} g_1$, the updated state $\ket{\overline{\varphi}}$ is not stabilized by $g_1,g_2$.
        
    \item $s_0'$ commutes with stabilizer
        $s_3 = g_1g_2 = \begin{matrix}
            XX\\
            XX
        \end{matrix}$. 
        So by using $\Pi_o^{s_0'}s_3 = s_3\Pi_o^{s_0'}$ we have
        \begin{equation}
            s_3\ket{\overline{\varphi}} = s_3\Pi_o^{s_0'}\ket{\overline{\psi}} = s_3\Pi_o^{s_0'}s_3\ket{\overline{\psi}} = s_3s_3(\Pi_o^{s_0'}\ket{\overline{\psi}}) = \ket{\overline{\varphi}} \;,
        \end{equation}
        which shows that $\ket{\overline{\varphi}}$ is stabilized by $s_3$.

    \item Since $g_0 = \begin{matrix}
            ZZ\\
            ZZ
        \end{matrix}$ commutes with $s_0'$, then $|\overline{\varphi}\>$ is stabilized by $g_0$.
\end{enumerate}
To summarize, the updated codestate $|\overline{\varphi}\>$ is now stabilized by Paulis
\begin{equation}
\begin{gathered}
    os_0' = o\begin{matrix}
            IZ\\
            ZI
        \end{matrix} \;,\quad
    s_3 = g_1g_2 = \begin{matrix}
            XX\\
            XX
        \end{matrix} \;,\quad
    g_0 = \begin{matrix}
            ZZ\\
            ZZ
        \end{matrix} \;.
\end{gathered}
\end{equation}
Note that it is also stabilized by 
\begin{equation}
    os_0' g_0 = o\begin{matrix}
        ZI\\
        IZ
    \end{matrix} =
    os_1'\,,
\end{equation}
which means that the outcomes of measurements $s_0'$ and $s_1'$ on initial codestate $\ket{\overline{\psi}}$ must have a parity of $0$, that is, their measurement outcomes are equal.
More generally in the absence of errors, the product of measurements $s_0'$ and $s_1'$ must be equal to the eigenvalue of the eigenspace of $g_0$ that the initial codestate $\ket{\overline{\psi}}$ is in.

\begin{figure}
    \centering
    \includegraphics[width=0.65\linewidth]{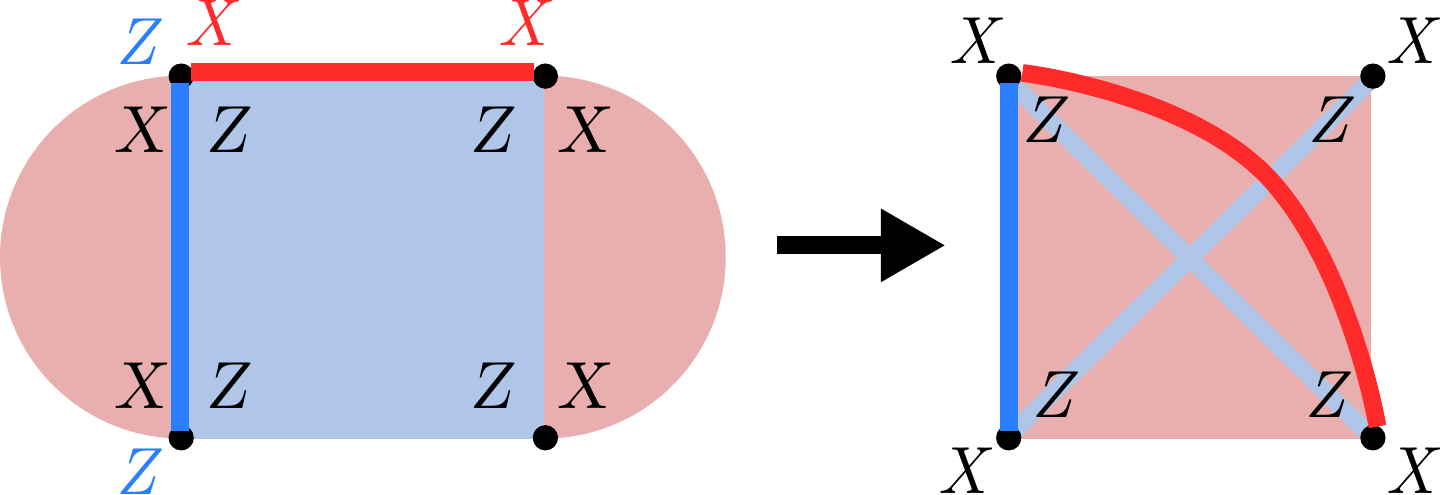}
    \caption{Stabilizer group and logical operator update from the 4-qubit rotated surface code (left) to another 4-qubit code (right).
    Vertices represent qubits, while red (resp. blue) faces represent $X$ (resp. $Z$) stabilizer generators.
    }
    \label{fig:stabilizer_update_4_qubit_code}
\end{figure}

Thus after we measure $s_0',s_1'$ on a codestate $\ket{\overline{\psi}}$ in $Q(\mathcal{S}_0)$, we remove all elements of stabilizer group $\mathcal{S}_0$ that anticommute with either $s_0'$ or $s_1'$ and take the rest to define a new stabilizer group $\mathcal{S}_1$ consisting of: (1) $\pm s_0'$ and $\pm s_1'$ (where the signs depend on the measurement outcomes), and (2) all elements of $\mathcal{S}_0$ that commute with both $s_0'$ and $s_1'$. The new stabilizer group $\mathcal{S}_1$ stabilizes the new codestate $|\overline{\varphi}\>$ obtained after measuring observables $s_0',s_1'$ on initial codestate $\ket{\overline{\psi}}$. We denote this stabilizer group update as
% Stabilizer group update upon measuring $s_0',s_1'$ (remove noncommuting stabilizers, keep commuting ones, add the measurements):
\begin{equation}
    \mathcal{S}_0 = 
    \left\<
    \begin{matrix}
        ZZ\\
        ZZ
    \end{matrix} \;,\;
    \begin{matrix}
        IX\\
        IX
    \end{matrix} \;,\;
    \begin{matrix}
        XI\\
        XI
    \end{matrix}
    \right\>
    \;
    \xrightarrow[\begin{matrix}
        ZI\\
        IZ
    \end{matrix}]{\begin{matrix}
        IZ\\
        ZI
    \end{matrix}}
    \;
    \mathcal{S}_1 = 
    \left\<
    \begin{matrix}
        XX\\
        XX
    \end{matrix} \;,\;
    \begin{matrix}
        ZI\\
        IZ
    \end{matrix} \;,\;
    \begin{matrix}
        IZ\\
        ZI
    \end{matrix}
    \right\> \;.
\end{equation}
This stabilizer group update is illustrated in~\cref{fig:stabilizer_update_4_qubit_code}.

We now consider the logical operator representatives of the initial code $Q(\mathcal{S}_0)$:
\begin{equation}
    \overline{X} = \begin{matrix}
        XX\\
        II
    \end{matrix} 
    \quad\text{and}\quad
    \overline{Z} = \begin{matrix}
        ZI\\
        ZI
    \end{matrix} \;.
\end{equation}
The logical eigenstates $|\overline{0}\>,|\overline{1}\>$ of $\overline{Z}$ after measuring $s_0'$ with outcome $o$ become
\begin{equation}
    |\overline{i}'\> = \Pi_o^{s_0'}|\overline{i}\> \;,
\end{equation}
for $i\in\{0,1\}$.
Since $\overline{Z}$ commutes with $s_0'$, we have
\begin{equation}
    \overline{Z}|\overline{i}'\> = \overline{Z}\Pi_o^{s_0'}|\overline{i}\> = \Pi_o^{s_0'} \overline{Z}|\overline{i}\> = (-1)^i|\overline{i}'\> \;.
\end{equation}
Thus, we can see that the updated eigenstates of $\overline{Z}$ are still eigenstates of $\overline{Z}$. On the other hand, the eigenstates of logical $\overline{X}$, given by
\begin{equation}
    |\overline{+}\> = \frac{|\overline{0}\>+|\overline{1}\>}{\sqrt{2}}
    \quad\text{and}\quad
    |\overline{-}\> = \frac{|\overline{0}\>-|\overline{1}\>}{\sqrt{2}} \;,
\end{equation}
are updated to $|\overline{\pm}'\> = \Pi_o^{s_0'}|\overline{\pm}\>$. Note that $\overline{X}$ anticommutes with $s_0'$, so it cannot be a logical operator of the new stabilizer code $Q(\mathcal{S}_1)$. However, a different logical operator representative
\begin{equation}
    \overline{X}' = g_1\overline{X} = \begin{matrix}
        XI\\
        IX
    \end{matrix}
\end{equation}
does commute with $s_0'$ (and $s_1'$). The action on this logical operator representative on the updated codestate is
\begin{equation}
    \overline{X}'|\overline{\pm}'\> = \overline{X}' \Pi_o^{s_0'}|\overline{\pm}\> = \Pi_o^{s_0'} \overline{X}' |\overline{\pm}\> = \pm |\overline{\pm}'\>\,,
\end{equation}
which indicates that the updated eigenstates are the eigenstates of logical operator representative $\overline{X}'$. Note that logical operator $\overline{X}'$ remains a logical operator after measuring $s_0'$ because we define it as a product between the logical operator representative $\overline{X}$ that we first pick and a stabilizer $g_1$ of the initial stabilizer group $\mathcal{S}_0$. Both $\overline{X}$ and $g_1$ individually anticommute with $s_0'$ so that their product $\overline{X}'=\overline{X}g_1$ commutes with $s_0'$. The logical operator updates are illustrated in ~\cref{fig:stabilizer_update_4_qubit_code}.

\paragraph{General stabilizer group and logical operator update rules.}
As we have seen in the stabilizer group update and logical operator update in the 4-qubit code example above, measuring Pauli observables $s_0',s_1'$ that anticommute with some of the original stabilizers in $\mathcal{S}_0$ evolves a codestate $\ket{\overline{\psi}}\in Q(\mathcal{S}_0)$ to a new codestate $|\overline{\varphi}\>$ stabilized by an updated stabilizer group $\mathcal{S}_1$ and a new set of logical operators. In fact, one can update any initial stabilizer code $\mathcal{S}$ and its logical operators after a Pauli measurement $s'$ is performed, according to whether $s$ commutes with the elements of $\mathcal{S}$ and the logical operators. We note that update rules for stabilizer group and logical operators have been known and used for simulating stabilizer circuits (see e.g.~\cite[Section 7]{gottesman1998heisenberg}).

\begin{lembox}[label={lem:stabilizer_update}]{{Stabilizer group update rules}}
    Update rules for an initial stabilizer group $\mathcal{S}=\<G\>$ with generators $G=\{g_1,\dots,g_r\}$ to stabilizer group $\mathcal{S}'=\<G'\>$ after Pauli measurement $P$ with outcome $o=\pm 1$ :
    \begin{enumerate}
        \item If $\pm P\in\mathcal{S}$, then $\mathcal{S}' = \mathcal{S}$, that is, the stabilizer stays the same.
        In this case, outcome $o$ is deterministic.
        
        \item If $\pm P$ is not in $\mathcal{S}$ but $P$ commutes with all $s\in\mathcal{S}$ (so $P\in\mathcal{N}(\mathcal{S})\setminus\mathcal{S}$ is a logical operator), then $\mathcal{S}' = \<oP,g_1,\dots,g_r\>$. That is, add $oP$ to $G$ to get $G'=\{oP\}\cup G$.
        
        \item If $P$ does not commute with elements in a subset $W = \{g_1,\dots,g_l\} \subseteq G$: 
        \begin{enumerate}
            \item Let $W' = \{g_1g_2,g_1g_3,\dots,g_1g_l\}$ ($|W'|=l-1$).
            All $g'\in W'$ commute with $P$.
            (Note that $\mathcal{S}=\<g_1,W',G\setminus W\>$, so now $P$ anticommutes only with $g_1$.)
            
            \item Update $G$ to $G'=\{oP\}\cup W'\cup (G\setminus W)$ and get $\mathcal{S}'=\<G'\>$.
        \end{enumerate}
        Here we replace generator $g^*$ with $oP$, replace $W\setminus g_1$ with $W'$, and keep the generators that commute with $P$.
        % Generators in $W'$ could be constructed by
        % and $b_1,\dots,b_m\in[D]\backslash\{0\}$ such that $g_jP = \omega^{b_j} Pg_j$.
        % Then, we fix an arbitrary Pauli $g_1 \in W$ and define set $W' = \{g_j g_1^{c_j} : g_j\in W\backslash\{g_1\} \;\textup{and}\; c_j\in[D] \,,\, c_jb_1=-b_j\}$.
    \end{enumerate}
\end{lembox}

A proof of this lemma can be found in Ref.~\cite{tanggara2025simpleconstructionquditfloquet} for general qudit stabilizer group update (which thus includes qubit stabilizer update as a special case). We give some remarks on each of the rules:
\begin{enumerate}
    \item Rule 1 is essentially applied when we measure one of the stabilizers in the initial stabilizer group $\mathcal{S}$.
    \item Rule 2 is applied when we measure a logical operator and thus changes the logical state (which one should not do when extracting error syndromes). Note that if we start with an $\llbracket n,k \rrbracket$ stabilizer code, then we increase the syndrome dimension to $n-k+1$ since we are adding a stabilizer, so that we are left with $k-1$ logical qubits. The updated stabilizer code has parameters $\llbracket n,k-1 \rrbracket$.
    \item Rule 3 is used when we measure a $s'$ that anticommutes with some elements of $\mathcal{S}$. We used this rule in the 4-qubit code example above, where we removed stabilizer generator $g_1$ after measuring $s_0'$ because they anticommute, updated $g_1$ (which also anticommute with $s_0'$) to $g_0g_1$, and kept the generator $g_3$.
\end{enumerate}

Similarly, we can use a set of rules to update logical operator representatives of initial stabilizer group $\mathcal{S}$ after measuring some Pauli observable $s'$.

\begin{lembox}[label={lem:logical_operator_update}]{Logical operator update rules}
    Update rules for logical operator (representative) $L$ of an initial stabilizer group $\mathcal{S}=\<G\>$ with generators $G=\{g_1,\ldots,g_r\}$ to logical $L'$ of stabilizer group $\mathcal{S}'=\<G'\>$ after Pauli measurement $P$ with outcome $o=\pm 1$:
    \begin{enumerate}
        \item If $P=L$ then we are measuring logical operator $L$. $L$ is now a stabilizer in $\mathcal{S}'$.
        
        \item If $PL = LP$, then $L$ remains a logical operator of $\mathcal{S}'$.
        
        \item If $P$ does not commute with $L$ but commutes with all $s\in\mathcal{S}$, then $L$ is updated to $L' = oP$, which is now a stabilizer in $\mathcal{S}'$.
    
        \item If $P$ does not commute with $L$ and some stabilizers in $\mathcal{S}$, then update $L$ to $L' = g_1 L$, where $g_1$ is the stabilizer removed in stabilizer update rule 4.
    \end{enumerate}
\end{lembox}

We give some remarks on each of the logical operator update rules above:
\begin{enumerate}
    \item Rule 1: we measure logical operator $L$. This is basically rule 2 of the stabilizer update rules above where we measure a logical operator so that it is in the updated stabilizer group.
    
    \item Rule 2: measure $P$ that commutes with $L$.
    If $|\overline{j}\>$ is the $+1$ eigenstate of $L$ updated to $|\overline{j}'\> = \Pi_o^P|\overline{j}\>$, then
    \begin{equation}
        L|\overline{j}'\> = L\Pi_o^P|\overline{j}\> = \Pi_o^PL|\overline{j}\> = \Pi_o^PL|\overline{j}\> = |\overline{j}'\> \;.
    \end{equation}
    
    \item Rule 3: $L$ is updated to $L' = oP$. When $P$ does not commute with $L$ but commutes with all $s\in\mathcal{S}$, then $P\in\mathcal{N}(\mathcal{S})\setminus\mathcal{S}$ (since all $s\in\mathcal{S}$ commute with $L$). So $P$ is a representative of a logical operator of $\mathcal{S}$ anticommuting with $L$, thus measuring $s'$ on $+1$ eigenstate $|\overline{j}\>$ of $L$ gives
    \begin{equation}
        \Pi_+^P|\overline{j}\>
        \quad\text{or}\quad
        \Pi_-^P|\overline{j}\>
    \end{equation}
    at random, stabilized by $+P$ or $-P$, respectively (measuring $P$ means doing a logical measurement that anticommutes with $L$, we are adding 1 stabilizer $oP$.)

    \item Rule 4: $L$ is updated to $L' = g_1L$. $P$ does not commute with $L$, but $g_1L$ is another representative of the same logical operator. Since both $g_1$ and $L$ anticommute with $oP$, if $|\overline{j}\>$ is the $+1$ eigenstate of $L$ updated to $|\overline{j}'\> = \Pi_o^P|\overline{j}\>$, then
    \begin{equation}
        L'|\overline{j}'\> = g_1L\Pi_o^P|\overline{j}\> = g_1\Pi_{-o}^PL|\overline{j}\> = \Pi_o^Pg_1L|\overline{j}\> = \Pi_o^P|\overline{j}\> = |\overline{j}'\>\,.
    \end{equation}
    That is, $|\overline{j}'\>$ is the $+1$ eigenstate of $L'$ (we used $L|\overline{j}\>=|\overline{j}\>$ and $g_1|\overline{j}\>=|\overline{j}\>$).
\end{enumerate}

\subsubsection{Four-qubit Floquet code}\label{subsubsec:four-qubit-floquet-code}
\begin{figure}
    \centering
    \includegraphics[width=0.75\linewidth]{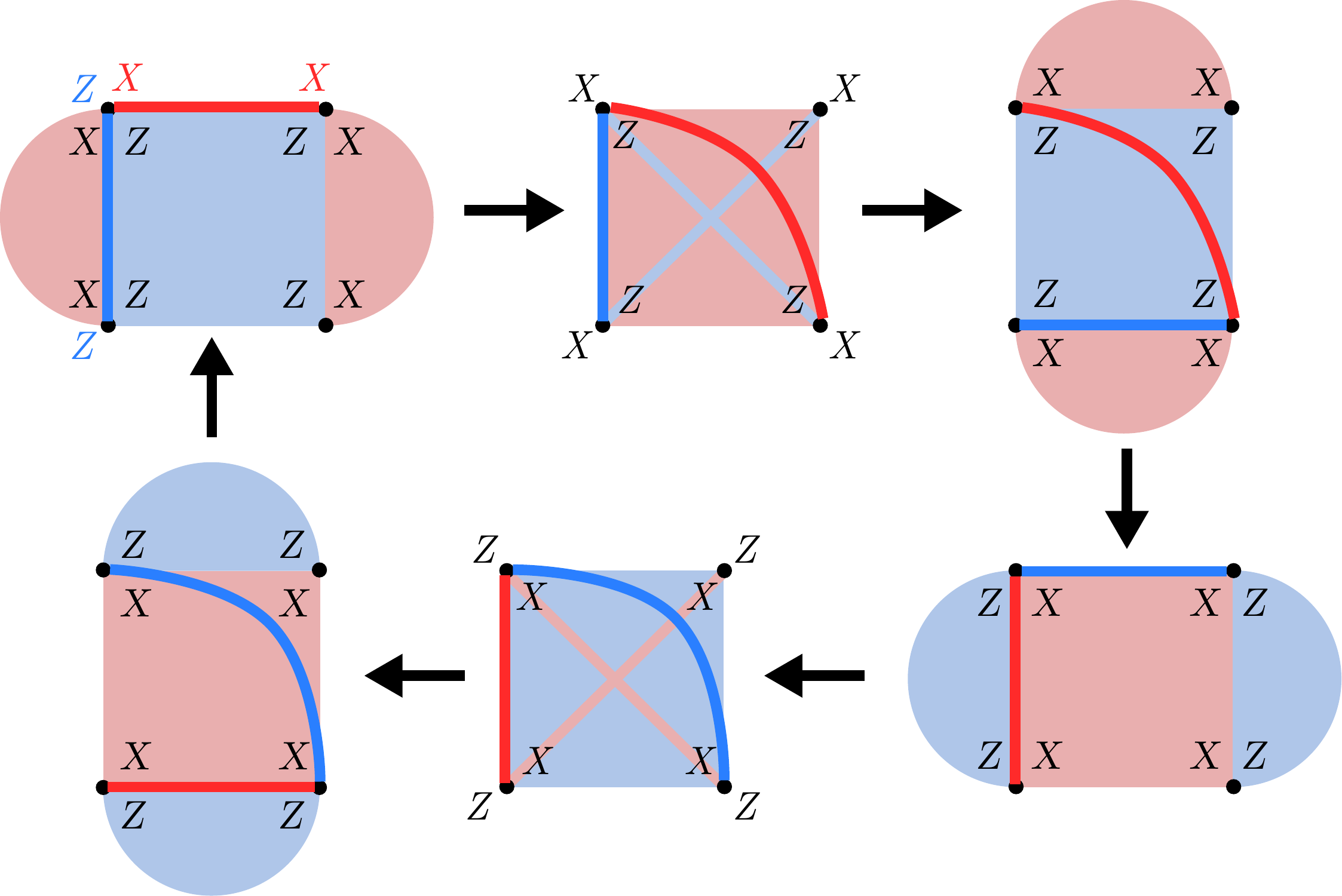}
    \caption{Four-qubit dynamical code with period-6 measurement schedule given in~\cref{eqn:dynamical_surface_code_schedule_ISG}. 
    Red (resp. blue) faces represent $Z$-type (resp. $X$-type) stabilizers for each ISG.
    Bright red (resp. bright blue) lines represent $X$-type (resp. $Z$-type) logical operator representative for each ISG.}
    \label{fig:four_qubit_floquet_code}
\end{figure}

By rule 3 of the stabilizer update rules in~\cref{lem:stabilizer_update}, the size of the updated stabilizer group $\mathcal{S}'$ is preserved whenever we measure a $s'$ that anticommutes with some elements of initial stabilizer group $\mathcal{S}$. Thus, logical information is always preserved by measuring Pauli observable $s'$ such that $s's = -ss'$ for some stabilizer $s\in\mathcal{S}$. By doing this periodically, we can construct a period-6 measurement schedule on 4-qubits so that we have stabilizer groups evolving as
\begin{equation}
    \mathcal{S}_0 \mapsto \dots \mapsto \mathcal{S}_5 \mapsto \mathcal{S}_0 \mapsto \dots
\end{equation}
Each of these six stabilizer groups are called an \emph{instantaneous stabilizer group} (ISG). Since the original 4-qubit rotated surface code encodes one logical qubit, all ISGs also encode one logical qubit. The measurement sequence of this 4-qubit dynamical code can be illustrated as follows:
\begin{equation}
\begin{aligned}
    &\mathcal{S}_0 
    \xrightarrow[\begin{matrix}
        ZI\\
        IZ
    \end{matrix}]{\begin{matrix}
        IZ\\
        ZI
    \end{matrix}}
    \mathcal{S}_1 
    \xrightarrow[\begin{matrix}
        II\\
        XX
    \end{matrix}]{\begin{matrix}
        XX\\
        II
    \end{matrix}}
    \mathcal{S}_2 
    \xrightarrow[\begin{matrix}
        IZ\\
        IZ
    \end{matrix}]{\begin{matrix}
        ZI\\
        ZI
    \end{matrix}}
    \mathcal{S}_3 
    \xrightarrow[\begin{matrix}
        IX\\
        XI
    \end{matrix}]{\begin{matrix}
        XI\\
        IX
    \end{matrix}}
    \mathcal{S}_4 
    \xrightarrow[\begin{matrix}
        II\\
        ZZ
    \end{matrix}]{\begin{matrix}
        ZZ\\
        II
    \end{matrix}}
    \mathcal{S}_5
    \xrightarrow[\begin{matrix}
        IX\\
        IX
    \end{matrix}]{\begin{matrix}
        XI\\
        XI
    \end{matrix}}
    \mathcal{S}_0 
    \rightarrow\dots
\end{aligned}
\end{equation}
This period-6 measurement schedule gives us a period-6 evolution of ISGs:
\begin{equation*}
    \begin{tikzpicture}[
      >=Latex,
      node/.style={circle,draw,thick,minimum size=7mm,inner sep=0pt,font=\small},
      edge/.style={->,thick}
    ]
        % Nodes (circles with text labels inside)
        \node[node] (A) at (90:2cm)  {$\mathcal{S}_0$};
        \node[node] (B) at (150:2cm) {$\mathcal{S}_1$};
        \node[node] (C) at (210:2cm) {$\mathcal{S}_2$};
        \node[node] (D) at (270:2cm)  {$\mathcal{S}_3$};
        \node[node] (E) at (330:2cm) {$\mathcal{S}_4$};
        \node[node] (F) at (30:2cm) {$\mathcal{S}_5$};
        % \node[node] (A) {$\mathcal{S}_0$};
        % \node[node, below=1cm of A, right=2.2cm of A] (B) {$\mathcal{S}_1$};
        % \node[node, below=1.7cm of A] (C) {$\mathcal{S}_2$};
        % Directed edges
        \draw[edge] (A) -- (B);
        \draw[edge] (B) -- (C);
        \draw[edge] (C) -- (D);
        \draw[edge] (D) -- (E);
        \draw[edge] (E) -- (F);
        \draw[edge] (F) -- (A);
        % Optional: curved edge and self-loop
        % \draw[edge,bend left=20] (C) to (B);
        % \draw[edge,loop above] (A) to (A);
    \end{tikzpicture}
\end{equation*}
The explicit form can be obtained by applying the update rules in~\cref{lem:stabilizer_update}:
\begin{equation}\label{eqn:dynamical_surface_code_schedule_ISG}
\begin{aligned}
    &\mathcal{S}_0 = \left\<
    \begin{matrix}
        XI\\
        XI
    \end{matrix} \,,\,
    \begin{matrix}
        IX\\
        IX
    \end{matrix} \,,\,
    \begin{matrix}
        ZZ\\
        ZZ
    \end{matrix} 
    \right\> 
    \xrightarrow[\begin{matrix}
        ZI\\
        IZ
    \end{matrix}]{\begin{matrix}
        IZ\\
        ZI
    \end{matrix}}
    \mathcal{S}_1 = \left\<
    \begin{matrix}
        XX\\
        XX
    \end{matrix} \,,\,
    \begin{matrix}
        ZI\\
        IZ
    \end{matrix} \,,\,
    \begin{matrix}
        IZ\\
        ZI
    \end{matrix} 
    \right\> 
    \\
    &\xrightarrow[\begin{matrix}
        II\\
        XX
    \end{matrix}]{\begin{matrix}
        XX\\
        II
    \end{matrix}}
    \mathcal{S}_2 = \left\<
    \begin{matrix}
        XX\\
        II
    \end{matrix} \,,\,
    \begin{matrix}
        II\\
        XX
    \end{matrix} \,,\,
    \begin{matrix}
        ZZ\\
        ZZ
    \end{matrix} 
    \right\> 
    \xrightarrow[\begin{matrix}
        IZ\\
        IZ
    \end{matrix}]{\begin{matrix}
        ZI\\
        ZI
    \end{matrix}}
    \mathcal{S}_3 = \left\<
    \begin{matrix}
        XX\\
        XX
    \end{matrix} \,,\,
    \begin{matrix}
        ZI\\
        ZI
    \end{matrix} \,,\,
    \begin{matrix}
        IZ\\
        IZ
    \end{matrix} 
    \right\>
    \\
    &\xrightarrow[\begin{matrix}
        IX\\
        XI
    \end{matrix}]{\begin{matrix}
        XI\\
        IX
    \end{matrix}}
    \mathcal{S}_4 = \left\<
    \begin{matrix}
        XI\\
        IX
    \end{matrix} \,,\,
    \begin{matrix}
        IX\\
        XI
    \end{matrix} \,,\,
    \begin{matrix}
        ZZ\\
        ZZ
    \end{matrix} 
    \right\>
    \xrightarrow[\begin{matrix}
        II\\
        ZZ
    \end{matrix}]{\begin{matrix}
        ZZ\\
        II
    \end{matrix}}
    \mathcal{S}_5 = \left\<
    \begin{matrix}
        XX\\
        XX
    \end{matrix} \,,\,
    \begin{matrix}
        ZZ\\
        II
    \end{matrix} \,,\,
    \begin{matrix}
        II\\
        ZZ
    \end{matrix} 
    \right\>
    \xrightarrow[\begin{matrix}
        IX\\
        IX
    \end{matrix}]{\begin{matrix}
        XI\\
        XI
    \end{matrix}}
    \mathcal{S}_0 \,.
\end{aligned}
\end{equation}
So as to not clutter the notation, we assume that all check measurements have an outcome of $+1$. In practice, these checks have random values, as we have seen in the update from ISG $\mathcal{S}_0$ to $\mathcal{S}_1$, so that the two-qubit generators in each ISG would have a $\pm 1$ sign, although their parity must be $+1$ (since the 4-qubit $X$ and $Z$ stabilizers must be in all of the ISGs). The stabilizers and logical operators of each ISG in this dynamical code are illustrated in~\cref{fig:four_qubit_floquet_code}.

By measuring two disjoint weight-2 $Z$-type Paulis in evolving from ISG $\mathcal{S}_0$ to $\mathcal{S}_1$, we are measuring the syndrome of
\begin{equation}
    \begin{matrix}
        ZZ\\
        ZZ
    \end{matrix}\,,
\end{equation}
which is a stabilizer in all ISGs. Similarly, by measuring two disjoint weight-2 $X$-type Paulis in evolving from ISG $\mathcal{S}_1$ to $\mathcal{S}_2$, we are measuring the syndrome of
\begin{equation}
    \begin{matrix}
        XX\\
        XX
    \end{matrix}\,,
\end{equation}
which is a stabilizer in all ISGs. So, we alternately measure these two weight-4 stabilizers over all six rounds. Thus, every two rounds we obtain the syndrome of the weight-4 $X$ stabilizer (resp., weight-4 $Z$ stabilizer), which allows us to detect a single qubit $Z$ error (resp., single-qubit $X$ error) between these two rounds.

\paragraph{Four-qubit dynamical code as a subsystem code.}
As we have mentioned, this 4-qubit dynamical code does not encode any logical qubit when viewed as a subsystem code. To see this, we first construct a gauge group $\mathcal{G}$ generated by all check measurements:
\begin{equation}
    % \scalebox{0.9}{
    \mathcal{G} = 
    \left\<
        \begin{matrix}
            XX\\
            XX
        \end{matrix} \;,\;
        \begin{matrix}
            ZZ\\
            ZZ
        \end{matrix} \;,\;
        \begin{matrix}
            IZ\\
            ZI
        \end{matrix} \;,\;
        \begin{matrix}
            XX\\
            II
        \end{matrix}\;,\;
        \begin{matrix}
            ZI\\
            ZI
        \end{matrix} \;,\;
        \begin{matrix}
            XI\\
            IX
        \end{matrix} \;,\;
        \begin{matrix}
            II\\
            ZZ
        \end{matrix} \;,\;
        \begin{matrix}
            XI\\
            XI
        \end{matrix}
        \right\> \;,
    % }
\end{equation}
where we pick just one out of two check measurements in each round and add the 4-qubit $X$ and $Z$ Paulis, since the product of two check measurements in each rounds is either a 4-qubit $X$ or 4-qubit $Z$ Pauli. Note that some of these gauge group generators are still redundant. The four weight-2 $Z$ checks can be generated by only three of them, and similarly with the four weight-2 $X$ checks. So we can generate the gauge group only with 6 generators as:
\begin{equation}
    \mathcal{G} =
        \left\<
        \begin{matrix}
            XX\\
            XX
        \end{matrix} \;,\;
        \begin{matrix}
            ZZ\\
            ZZ
        \end{matrix}\;,\; 
        \begin{matrix}
            XX\\
            II
        \end{matrix} \;,\;
        \begin{matrix}
            IZ\\
            ZI
        \end{matrix} \;,\;
        \begin{matrix}
            XI\\
            IX
        \end{matrix} \;,\;
        \begin{matrix}
            II\\
            ZZ
        \end{matrix}
    \right\> \,.
\end{equation}
Note that each of the first two generators
\begin{equation}
    g_1'=
    \begin{matrix}
        XX\\
        XX
    \end{matrix}\,,\quad
    g_2'=
    \begin{matrix}
        ZZ\\
        ZZ
    \end{matrix}
\end{equation}
commute with all the other generators, so that they are in the center of $\mathcal{G}$, that is, $Z(\mathcal{G}) = \<\pm i, g_1',g_2'\>$. Thus, they generate the stabilizer group of this subsystem code: $\mathcal{S}=\<g_0',g_1'\>$. On the other hand,
\begin{equation}
    g_3'=
    \begin{matrix}
        XX\\
        II
    \end{matrix}\,,\quad
    g_4'=
    \begin{matrix}
        IZ\\
        ZI
    \end{matrix}
\end{equation}
anticommute with each other but commute with all the other generators. Therefore, they are an $X,Z$ operator pair of a gauge qubit. Similarly, 
\begin{equation}
    g_5'=
    \begin{matrix}
        XI\\
        IX
    \end{matrix}\,,\quad
    g_6'=
    \begin{matrix}
        II\\
        ZZ
    \end{matrix}
\end{equation}
is also an $X,Z$ operator pair of a gauge qubit. Therefore, we have $r=2$ stabilizer generators and $g=2$ gauge qubits:
\begin{equation}
    \mathcal{G} =
        \left\<
        \underbrace{
        \begin{matrix}
            XX\\
            XX
        \end{matrix} \;,\;
        \begin{matrix}
            ZZ\\
            ZZ
        \end{matrix}}_{\text{2 stabilizers}}\;,\; 
        \underbrace{
        \begin{matrix}
            XX\\
            II
        \end{matrix} \;,\;
        \begin{matrix}
            IZ\\
            ZI
        \end{matrix}}_{\text{gauge-q 1}} \;,\;
        \underbrace{
        \begin{matrix}
            XI\\
            IX
        \end{matrix} \;,\;
        \begin{matrix}
            II\\
            ZZ
        \end{matrix} }_{\text{gauge-q 2}}
    \right\> \,,
\end{equation}
which means that there is $k = n-r-g = 4-2-2=0$ logical qubits, as claimed.

\subsection{Hastings-Haah honeycomb Floquet code}\label{subsec:dynamical_hastings_haah}
Now let us return to the Kitaev's honeycomb code~\cite{kitaev2006anyons}, discussed in~\cref{subsec:dynamical-hidden} which is a subsystem code with zero logical qubits. We will show that by picking an appropriate measurement schedule, each consisting of a set of 2-qubit measurements, we can actually encode two logical qubits \emph{dynamically}. This was discovered by Hastings and Haah in their seminal work~\cite{hastings2021dynamically} in 2021, 15 years after Kitaev introduced the subsystem honeycomb code.

\begin{figure}
    \centering
    \includegraphics[width=0.55\linewidth]{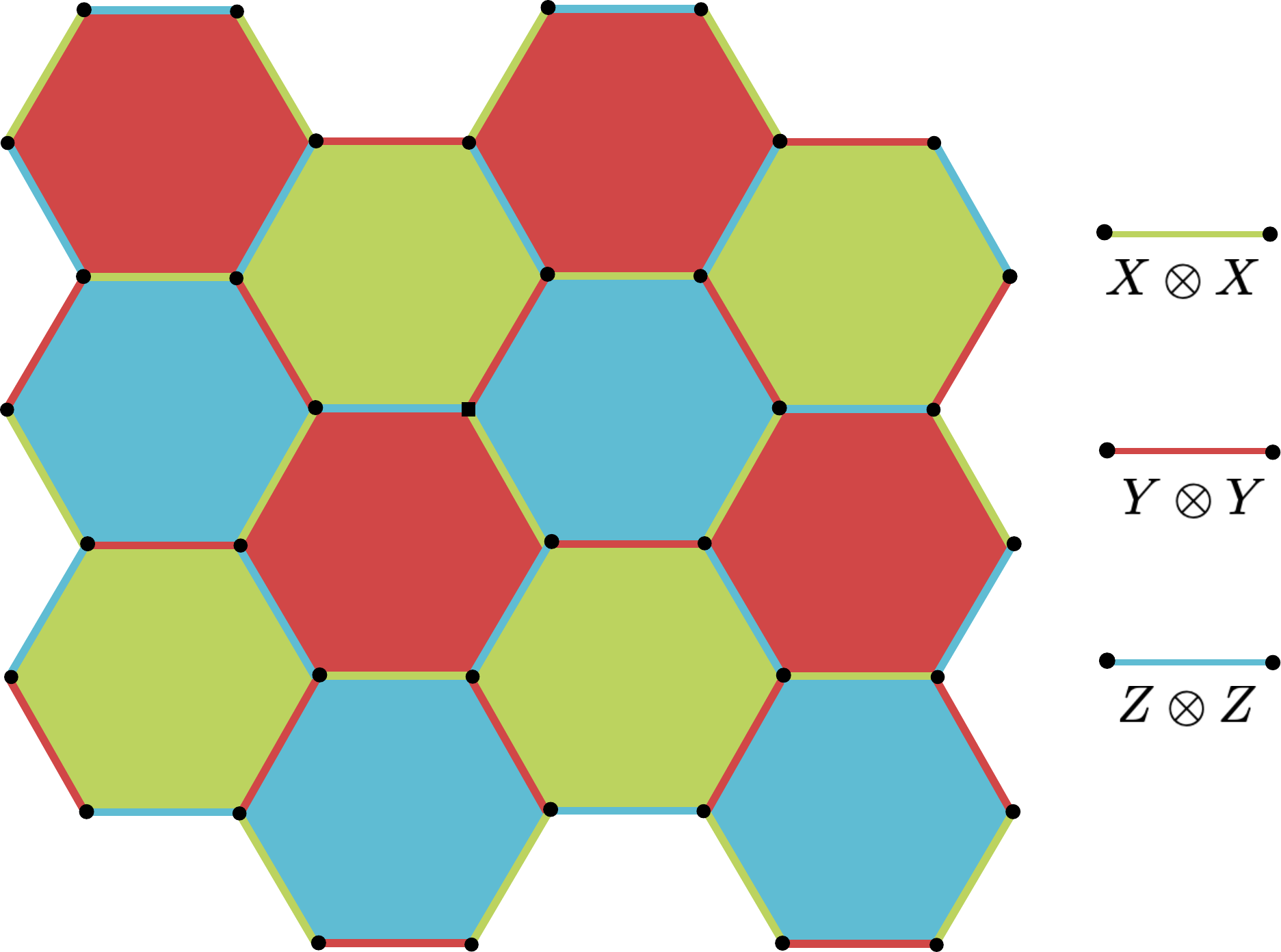}
    \caption{Honeycomb lattice for a $36$ qubit honeycomb code.
    Vertices in the lattice represent qubits, edges represent two-qubit check measurements, and hexagonal plaquettes represent stabilizers of the code.
    Edge color indicates its check measurement basis, where 2-qubit $X$ measurements are performed on green edges, 2-qubit $Y$ measurements on red edges, and 2-qubit $Z$ measurements on blue edges.
    Similarly, green plaquettes are 6-qubit $X$ stabilizers, red plaquettes are 6-qubit $Y$ stabilizers, and blue plaquettes are 6-qubit $Z$ stabilizers.
    }
    \label{fig:honeycomb_lattice36}
\end{figure}

Recall the construction of Kitaev's code by tessellating the surface of a torus with hexagons, resulting in a regular hexagonal lattice, and then coloring the edges and faces with three colors (see~\cref{fig:honeycomb_lattice36} and caption therein). Since all checks of the same color commute (in fact, they act on disjoint qubits), we can measure all of them in the same round, simultaneously. So, let us consider a period-3 measurement schedule cycling between the green, red, and blue checks:
\begin{equation}
    \text{\textcolor{ForestGreen}{green} $\rightarrow$ \textcolor{red}{red} $\rightarrow$ \textcolor{blue}{blue} $\rightarrow$ \textcolor{ForestGreen}{green} $\rightarrow$ \textcolor{red}{red} $\rightarrow\dots$} \;.
\end{equation}
Let us denote the set of green, red, and blue checks as $C_g,C_r,C_b$, respectively. Also denote the set of green, red, and blue plaquette stabilizers as $A_g,A_r,A_b$, respectively. Let us first focus on a red check measurement around a hexagonal plaquette and assume that we have just measured all green checks so that we start with a stabilizer group $\mathcal{S}=\<C_g\>$. Note that a blue plaquette is surrounded by three red $Y$ checks and three green $X$ checks. Then, we measure the three red checks one by one, starting from the pre-existing three green $X\otimes X$ checks, as indicated in the following figure:
\begin{center}
\includegraphics[width=0.75\linewidth]{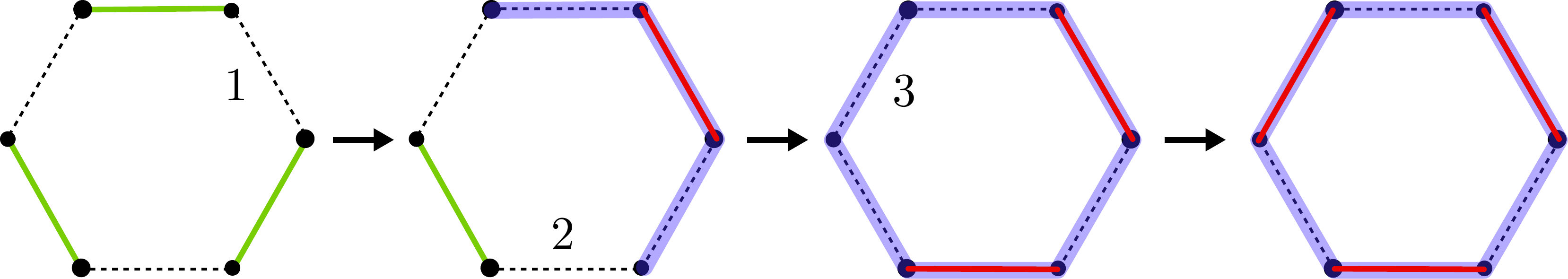}
\end{center}
Here, we label the three edges where the red $Y\otimes Y$ checks are measured as 1, 2, and 3. Let us describe what happens to the stabilizer generators on this plaquette as we update them using the stabilizer update rules in~\cref{lem:stabilizer_update} as illustrated in the diagram above from left to right:
\begin{enumerate}
    \item We start with stabilizer generators $g_1,g_2,g_3$ (three green $X$ checks) on this plaquette.
    
    \item Measure red check 1 $r_1$:\\
    (Rule 3) Remove one noncommuting green check $g_1$, update the other weight-2 green check $g_2$ to $g_1g_2$ (resulting in weight-4 $X$ stabilizer highlighted in blue), put red check 1 $r_1$ in the stabilizer group.
    The updated stabilizers on this plaquette are generated by: $r_1,g_1g_2,g_3$.

    \item Measure red check 2 $r_2$:\\
    (Rule 3) Remove the one remaining green check $g_3$ since it anticommutes with $r_2$.
    Update the weight-4 $X$ check $g_1g_2$ to $g_1g_2g_3$. 
    Put red check 2 $r_2$ in the stabilizer group.
    The updated stabilizers on this plaquette is generated by: $r_1,r_2,g_1g_2g_3$.

    \item Measure red check 3 $r_3$:\\
    (Rule 2) $r_3$ commutes with $r_1,r_2,g_3g_1g_2$, so we put red check 3 $r_3$ in the stabilizer group.
    The updated stabilizers on this plaquette is generated by: $r_1,r_2,r_3,g_1g_2g_3$.
\end{enumerate}

Note that by the end of the red check round, there is one weight-6 $X$ stabilizer generator and three weight-2 $Y$ stabilizer generators. Also, note that the product of all of these four stabilizers is a 6-qubit Pauli $Z$ operator, which is also a stabilizer, that is, $g_3g_1g_2 r_1r_2r_3=ZZZZZZ$. This is a blue $Z$ plaquette stabilizer in $A_b$ whose syndrome we have just obtained by measuring the three green checks $g_1,g_2,g_3$ (in the assumed previous round) and the three red checks $r_1,r_2,r_3$ around it.

In the next round, we measure all blue $Z$ checks and perform the same update steps on each green plaquettes. Each green plaquette is surrounded by three red $Y$ checks $r_1,r_2,r_3$, which will be updated to their product $r_1r_2r_3$ with three blue $Z$ checks $b_1,b_2,b_3$ surrounding this plaquette. Then, by taking their product, we obtain a 6-qubit green $X$ plaquette stabilizer $r_1r_2r_3b_1b_2b_3 = XXXXXX$ in $A_g$. Then, the remaining stabilizers on this plaquette are the three blue $Z$ checks $b_1,b_2,b_3$ and the 6-qubit stabilizer $r_1r_2r_3b_1b_2b_3 = XXXXXX$ in $A_g$. A similar analysis on each red plaquette in the next round where we measure green $X$ checks will give us three green $X$ checks $g_1,g_2,g_3$ and a 6-qubit stabilizer $g_1g_2g_3b_1b_2b_3 = YYYYYY \in A_r$ as the stabilizer generators on each red plaquette.

In summary, the above procedure initializes the honeycomb Floquet code as follows:
\begin{enumerate}
    \item At round $r=-4$ we start with ISG $\mathcal{S}_{-4}$ consisting only of the identity $I^{\otimes n}$.

    \item At round $r=-3$ the ISG is $\mathcal{S}_{-3} = \<C_g\>$.

    \item At round $r=-2$ the ISG is $\mathcal{S}_{-2} = \<A_b,C_r\>$.

    \item At round $r=-1$ the ISG is $\mathcal{S}_{-1} = \<A_g,A_b,C_b\>$.

    \item At round $r=0$ the ISG is $\mathcal{S}_0 = \<A_r,A_g,A_b,C_g\>$.
    
    \item At round $r>0$ the ISG is $\mathcal{S}_r = \<A_r,A_g,A_b,C_l\>$ where $l=g$ if $r=0\mod{3}$ and $l=r$ if $r=1\mod{3}$ and $l=b$ if $r=2\mod{3}$.
\end{enumerate}
So, we have obtained a period-3 measurement sequence that evolves the ISGs in period-3:
\begin{equation*}
    \begin{tikzpicture}[
      >=Latex,
      node/.style={circle,draw,thick,minimum size=5mm,inner sep=0pt,font=\small},
      edge/.style={->,thick}
    ]
        % Nodes (circles with text labels inside)
        \node[node] (A) at (90:0.8cm)  {$\mathcal{S}_0$};
        \node[node] (B) at (210:0.8cm) {$\mathcal{S}_1$};
        \node[node] (C) at (330:0.8cm) {$\mathcal{S}_2$};
        % \node[node] (A) {$\mathcal{S}_0$};
        % \node[node, below=1cm of A, right=2.2cm of A] (B) {$\mathcal{S}_1$};
        % \node[node, below=1.7cm of A] (C) {$\mathcal{S}_2$};
        % Directed edges
        \draw[edge] (A) --node[midway,left]{$\textcolor{red}{C_r}$} (B);
        \draw[edge] (B) --node[midway,below]{$\textcolor{blue}{C_b}$} (C);
        \draw[edge] (C) --node[midway,right]{$\textcolor{ForestGreen}{C_g}$} (A);
        % Optional: curved edge and self-loop
        % \draw[edge,bend left=20] (C) to (B);
        % \draw[edge,loop above] (A) to (A);
    \end{tikzpicture}
\end{equation*}
The checks in each round, the three ISGs and the plaquette stabilizer syndromes obtained in each round can be summarized as follows:
\begin{center}
\begin{tabular}{|c|ccc|}
    \hline
    round $i$ (mod 3) & 0 & 1 & 2 \\
    checks & \textcolor{ForestGreen}{green} $X$ & \textcolor{red}{red} $Y$ & \textcolor{blue}{blue} $Z$ \\
    ISG $\mathcal{S}_i$ & $\mathcal{S}_0=\<A_r,A_g,A_b,C_g\>$ & $\mathcal{S}_1=\<A_r,A_g,A_b,C_r\>$ & $\mathcal{S}_2=\<A_r,A_g,A_b,C_b\>$ \\
    syndromes & $\textcolor{red}{A_r}(Y)$ & $\textcolor{blue}{A_b}(Z)$ & $\textcolor{ForestGreen}{A_g}(X)$ \\
    \hline
\end{tabular} 
\end{center}

% \includegraphics[width=0.6\linewidth]{img/existing codes/ISG dynamics.png}

% \begin{tabular}{p{5cm} c}
%     \includegraphics[width=0.7\linewidth]{img/existing codes/ISG dynamics.png}
    
%     \includegraphics[width=0.9\linewidth]{img/toric code/torus.png} & \includegraphics[width=0.5\linewidth]{img/honeycomb lattice.pdf}
% \end{tabular}

\paragraph{Honeycomb Floquet code as a subsystem code.}
When we take the check measurements of the honeycomb Floquet code to generate a gauge group $\mathcal{G}$, we will recover Kitaev's subsystem honeycomb code with this gauge group which does not allow any logical information to be encoded. The honeycomb Floquet code however does encode two logical qubits, dynamically. We can see this by counting the number of independent stabilizer generators for each ISG $\mathcal{S}_i$.

First, note that the honeycomb lattice has $n$ qubits, $n/2$ plaquettes, and $3n/2$ edges. Let us first focus on ISG $\mathcal{S}_0=\<A_r,A_g,A_b,C_g\>$ and consider the following facts:
\begin{enumerate}
    \item There are $\frac{n}{2}$ plaquette stabilizers $A=A_r\cup A_g\cup A_b$.

    \item Green checks $X\otimes X$ are $\frac{1}{3}$ of all edges: $\frac{1}{3}\frac{3n}{2} = \frac{n}{2}$ edge stabilizers (all green checks).

    \item Plaquette stabilizers are not independent (product of all plaquettes is $I$).
    So, there are $\frac{n}{2}-1$ independent plaquette stabilizers.
    (e.g., take all plaquettes, except for a green plaquette stabilizer to form an independent generator.)

    \item Green checks $C_g$ and the remaining $\frac{n}{2}-1$ plaquette stabilizers in $A$ are not independent since their product is $I$.
    (i.e., take the product between all green $X$ checks and all green $X$ plaquette stabilizers.)
    So there are $\frac{n}{2}-1$ independent green checks.
\end{enumerate}
From point 3 and 4 above, $\mathcal{S}_0$ has $\frac{n}{2}-1$ 2-qubit green $X$ checks and $\frac{n}{2}-1$ 6-qubit plaquette stabilizers as a set of independent generators. Thus, there are $r = \frac{n}{2}-1 + \frac{n}{2}-1 = n-2$ independent stabilizer generators in total and 
\begin{equation}
    k=n - r = n-(n-2)=2
\end{equation}
logical qubits. By doing a similar analysis to $\mathcal{S}_1$ and $\mathcal{S}_2$, we will reach to the same conclusion.

We remark that the difference between ISG $\mathcal{S}_i$ and the stabilizer group of Kitaev's honeycomb code is that the non-trivial loops are not in any of the ISGs $\mathcal{S}_i$ since it is never measured in the honeycomb Floquet code's measurement schedule. Namely, in Kitaev's code, the value of non-trivial loop operators $\ell_1,\ell_2$ are fixed to $+1$, that is, measuring them will give $+1$ outcome.
On the other hand in ISG $\mathcal{S}_i$ of the honeycomb Floquet code, they are logical operators for the two logical qubits, that is, there exists another pair of non-trivial loop operators $\ell_1',\ell_2'$ such that $\ell_1\ell_1' = -\ell_1'\ell_1$ and $\ell_2\ell_2' = -\ell_2'\ell_2$ but $\ell_1'$ commutes with both $\ell_2,\ell_2'$ and $\ell_2'$ commutes with both $\ell_1,\ell_1'$.

\subsection{Conservation of logical information in dynamical codes}\label{subsec:dynamical-conservation}
As measurements in general change the state of a quantum system, it is essential to ensure that check measurements do not destroy any encoded logical information while also extracting error syndromes. This property, which we call the \emph{conservation of logical information}, is necessary for fault-tolerant quantum memory implemented through a dynamical code. In particular, one must ensure that the check measurements in round $r+1$ of the measurement schedule induces a mapping from any code state $\ket{\psi^{(r)}}$ of the previous round $r$ to a code state $\ket{\psi^{(r+1)}}$ of round $r+1$ such that the code state amplitudes are conserved. More formally, round $r+1$ check measurements $\mathcal{M}_{r+1}$ induce a mapping of the form
\begin{equation}\label{eqn:codestate_logical_conservation}
    \ket{\psi^{(r)}} = \sum_{j}{\alpha_j \ket{j^{(r)}} \mapsto \ket{\psi^{(r+1)}}} = \sum_{j}{\alpha_j\ket{j^{(r+1)}}}
\end{equation}
for any amplitude $\{\alpha_j\}_j$, where $\{j^{(r+1)}\}_j$ is an arbitrary basis of the logical codespace in round $r$.

The conservation of logical information has been noted since the conception of dynamical codes, where Hastings and Haah explicated the electric-magnetic automorphism of the honeycomb Floquet code's dynamical logical operator in consecutive rounds~\cite[Section 1.4]{hastings2021dynamically} (such a logical operator is referred to as the \emph{outer} logical operator by Hastings and Haah). Namely, for a representative of a dynamical logical operator $\overline{L}^{(r)}$ of ISG $\mathcal{S}^{(r)}$ in round $r$ and logical operator $\overline{L}^{(r+1)}$ of ISG $\mathcal{S}^{(r+1)}$ in round $r+1$, their product $\overline{L}^{(r+1)} \overline{L}^{(r)}$ is equal to a round-$r$ stabilizer $s_r\in\mathcal{S}^{(r)}$. Since the product between $\overline{L}^{(r+1)}$ and a stabilizer $s^{(r+1)}\in\mathcal{S}^{(r+1)}$ gives a different representative of the same logical operation, this is equivalent to
\begin{equation}\label{eqn:logical_conservation}
    s^{(r)}\overline{L}^{(r)} = s^{(r+1)}\overline{L}^{(r+1)}\,.
\end{equation}
Alam and Rieffel show~\cite{alam2024dynamical} that this equality is a sufficient condition for the conservation of logical information in~\cref{eqn:codestate_logical_conservation} to hold (see~\cite[Eq. 2 and Appendix A]{alam2024dynamical}).

\newpage

\section{Bosonic codes}\label{sec:bosonic}

In this chapter, we take a brief detour into continuous-variable (CV) encodings, which differ in a fundamental way from the finite-dimensional setting emphasized so far. Previous chapters focused on codes built from many finite-dimensional subsystems whose tensor-product structure enables entangled codewords with protection against environmental noise. In the CV setting, the physical carrier is intrinsically infinite dimensional, which enables \eczoo[monolithic encodings]{single_subsystem}: codespaces embedded in the Hilbert space of a single physical system rather than distributed across multiple subsystems. Monolithic encodings can also be realized in single high-dimensional qudits~\cite{mezzadri2024fault,lim2023fault}, but continuous-variable platforms make such single-system embeddings especially natural. Prominent instances include bosonic encodings in oscillator-like degrees of freedom~\cite{albert2025bosonic}, as well as related constructions using atomic and molecular structure~\cite{albert2020robustc,jain2024codes}. With this perspective in place, we now begin our study of bosonic codes.

\subsection{Bases of a simple harmonic oscillator}\label{subsec:sho-bases}
Bosonic quantum error correction refers to encoding quantum information in physical systems that can be modeled as simple harmonic oscillators (SHOs). The oscillator modes could be electromagnetic (e.g., microwave cavities, optical cavities, optical fibers, or free space) or mechanical (e.g., vibrations of atomic or molecular nuclei, acoustic resonators, nanomechanical resonators, or trapped-ion crystals). The appeal of bosonic encodings is that a single oscillator mode already has a large, well-structured Hilbert space. Rather than assembling protection solely from many two-level systems, one can try to use this internal structure directly. This leads to code families whose protection mechanisms are naturally expressed in Fock space, phase space, or the coherent-state picture, which we will properly introduce below. These code families provide a vast playground to explore for quantum error correction.

SHOs allow monolithic encodings because their Hilbert spaces are infinite dimensional and admit continuous-variable representations. In particular, an oscillator state can be described in the position or momentum basis, whose generalized basis states \(\ket{x}\) and \(\ket{p}\) are labeled by real numbers \(x,p\in\R\).\footnote{Since eigenvalues and operators are often denoted by the same symbols in continuous-variable systems, we adopt the $\hat{\phantom{x}}$ notation for operators in this chapter. Ket notation is also somewhat overloaded: the same symbol $\ket{\cdot}$ will be used for position ($x$), momentum ($p$), Fock ($n$), and coherent states ($\alpha$). Usually, the intended basis is clear from context; when needed, we add subscripts such as $\ket{\cdot}_x$ and $\ket{\cdot}_p$. As a rule of thumb, integer, complex-number, and real-number labels typically denote Fock states, coherent states, and position or momentum states, respectively.} These generalized basis states are not normalizable physical states, but they make the continuous-variable structure of an oscillator explicit. The canonical position and momentum bases are summarized in~\cref{box:canonical-basis}.

\begin{explbox}[label={box:canonical-basis}]{Position and momentum bases}
    \centering
    \begin{tblr}{width=\linewidth,colspec={X[l]X[c]X[c]}}
        & \textbf{Position basis} & \textbf{Momentum basis} \\
        States & $\{\,\ket{x} \, | \,x\in \R\}$ & $\{\,\ket{p} \, | \,p\in \R\}$ \\
        Eigenoperators & $\hat{x}\ket{x}=x\ket{x}$ & $\hat{p}\ket{p}=p\ket{p}$ \\
        Orthonormality & $\braket{x}{y} = \delta(x-y)$ & $\braket{p}{q} = \delta(p-q)$ \\
        Completeness & $\int_\R dx \ket{x}\bra{x}=\hat{I}$ & $\int_\R dp \ket{p}\bra{p}=\hat{I}$\\
        Conversion & $\braket{p}{x}=\frac{1}{\sqrt{2\pi}}e^{-ipx}$ & $\braket{x}{p}=\frac{1}{\sqrt{2\pi}}e^{ipx}$
    \end{tblr}
\end{explbox}

Throughout this chapter, we use dimensionless quadratures and set \(\hbar=1\), so that \([\hat{x},\hat{p}]=i\). The energy spectrum of an SHO is governed by the Hamiltonian
\begin{equation}\label{eq:bosonic-hamiltonian}
    \hat{H} = \frac{\omega}{2}\left(\hat{p}^2 + \hat{x}^2\right)\,,
\end{equation}
where \(\omega\) depends on the particular system at hand. A natural basis of states for any quantum system is the set of energy eigenvectors. We find them by solving the Schr\"{o}dinger equation
\begin{equation}\label{eq:schrodinger-sho}
    \hat{H}\ket{n} = E_n\ket{n}, 
\end{equation} 
uncovering the energy eigenstates
\begin{equation}\label{eq:position-basis-sho-energy-states}
    \ket{n} = \int_\R dx \frac{1}{\sqrt{2^n n!}} \frac{1}{\pi^{1/4}}
    e^{-\frac{ x^2}{2}}
    H_n(x)\ket{x}
\end{equation}
with energies 
\begin{equation}\label{eq:sho-energy}
    E_n = \omega \left(n+\frac{1}{2}\right)\,.
\end{equation}
Here, $H_n$ denotes the $n\numth$ Hermite polynomial~\cite{arfken2005mathematical}. These states look cumbersome and are practically impossible to work with. A simple change of basis eliminates this complexity and makes the SHO one of the simplest and most widely studied quantum systems. For this, we introduce the \emph{annihilation} ($\hat{a}$), \emph{creation} ($\hat{a}^\dagger$) and \emph{number} ($\hat{n}$) operators:
\begin{align}
    \hat{a} =& \frac{1}{\sqrt{2}}\left(\hat{x}+i\hat{p}\right)\,,\\
     \hat{a}^\dagger =& \frac{1}{\sqrt{2}}\left(\hat{x}-i\hat{p}\right)\,,\\
     \hat{n} =& \hat{a}^\dagger \hat{a}\,.
\end{align}
Using these definitions, we can rewrite the Hamiltonian as
\begin{equation}\label{eq:bosonic-hamiltonian-fock}
    \hat{H} = \omega \left(\hat{n}+\frac{1}{2}\right)\,.
\end{equation}
This is already reminiscent of~\cref{eq:sho-energy} because every energy eigenstate $\ket{n}$ is exactly the eigenstate of $\hat{n}$ with eigenvalue $n$. Thus, we introduce the \emph{Fock basis} for the SHO system below.

\begin{explbox}[label={box:fock-basis}]{Fock basis}
    \centering
    \begin{tblr}{width=\linewidth,colspec={X[l] c}}    
        States & $\{\,\ket{n} \, | \,n\in \Z_{\geq 0}\}$\\
        Eigenoperator & $\hat{n}\ket{n}=n\ket{n}$\\
        Orthonormality & $\braket{n}{m} = \delta_{nm}$\\
        Completeness & $\sum_{n=0}^\infty \ket{n}\bra{n}=\hat{I}$\\
        Conversion & $\braket{x}{n}=\frac{1}{\sqrt{2^n n!}} \frac{1}{\pi^{1/4}}
        e^{-\frac{ x^2}{2}}
        H_n(x)$ 
    \end{tblr}
\end{explbox}

The action of the annihilation and creation operators on the Fock states can be simply described as
\begin{align}
    \hat{a}\ket{n}&=\sqrt{n}\ket{n-1},\\
    \hat{a}^\dagger\ket{n}&=\sqrt{n+1}\ket{n+1}\,.
\end{align}
One can imagine the annihilation (creation) operator ``annihilating'' (``creating'') one quantum of energy in the system and therefore reducing (increasing) the total energy by one unit.

Finally, we introduce \emph{coherent states} for the SHO below in~\cref{box:coherent-basis}.

\begin{explbox}[label={box:coherent-basis}]{Coherent-state basis}
    \centering
    \begin{tblr}{width=\linewidth,colspec={X[l] X[c]}}
        States & $\{\,\ket{\alpha} \, | \,\alpha\in \C\}$\\
        Eigenoperator & $\hat{a}\ket{\alpha}=\alpha\ket{\alpha}$\\
        Overlap & $\braket{\alpha}{\beta} = e^{-\frac{|\alpha|^2}{2} - \frac{|\beta|^2}{2} +\alpha^\star \beta}$\\
        Overcompleteness & $\int \frac{d^2\alpha}{\pi} \ket{\alpha}\bra{\alpha}=\hat{I}$\\
        Conversion & $\braket{n}{\alpha}=e^{-\frac{|\alpha|^2}{2}}\frac{\alpha^n}{\sqrt{n!}}$ 
    \end{tblr}
\end{explbox}

Coherent states are not exactly orthogonal, but the magnitude of their overlap $|\langle \alpha | \beta \rangle |=e^{-|\alpha-\beta|^2/2}$ becomes negligible for large separations $|\alpha -\beta|$.

\subsection{Error channels}\label{subsec:error-channels}
Possessing a differently structured Hilbert space, bosonic systems are naturally prone to different kinds of noise channels than qubits. We describe some of the most prominent bosonic error channels in this subsection.

\begin{defbox}[label={box:lindblad-noise}]{Lindblad description of quantum channels}
    A \emph{quantum channel} is a map that takes an input density matrix $\rho(0)$ to an output density matrix $\rho(t)$. One common way to describe such a channel is through a set of \emph{Kraus operators} $\{\hat{E}_k\}$:
    \begin{equation}
        \rho(t)=\sum_k \hat{E}_k \rho(0) \hat{E}_k^\dagger,
    \end{equation}
    with $\sum_k \hat{E}_k^\dagger \hat{E}_k=\hat{I}$. Each Kraus operator can be viewed as one possible error process contributing to the overall noisy evolution.\vspace{2.5mm}
    
    When the noise acts continuously in time and memory effects can be neglected, it is often convenient to describe the dynamics instead by a \emph{Lindblad equation}
    \begin{equation}
        \frac{d\rho}{dt}=-i[\hat{H},\rho]+\sum_j \kappa_j \left(\hat{L}_j \rho \hat{L}_j^\dagger-\frac{1}{2}\left\{\hat{L}_j^\dagger \hat{L}_j,\rho\right\}\right),
    \end{equation}
    where the Hamiltonian $\hat{H}$ describes coherent evolution and the operators $\hat{L}_j$ describe dissipative processes such as photon loss or dephasing. Solving this differential equation for a finite time interval produces a quantum channel, which can then be written in Kraus form. Thus, the Lindblad equation describes the infinitesimal time evolution, while the Kraus operators describe the resulting finite-time channel.
\end{defbox}

\subsubsection{Photon loss}\label{subsubsec:photon-loss}
\emph{Photon loss}~\cite{albert2018performance} (a.k.a. the pure-loss or bosonic amplitude-damping channel) is the most common source of incoherent errors in optical and microwave cavities, which are among the most developed platforms for realizing bosonic quantum computing. In the continuous-time description of~\cref{box:lindblad-noise}, evolution under this channel is described by the Lindblad equation
\begin{equation}\label{eq:lindlblad-photon-loss}
    \frac{d\rho}{dt}=\kappa\left(\hat{a}\rho\hat{a}^\dagger-\frac{1}{2}\{\hat{n},\rho\}\right),
\end{equation}
where $\rho$ is the density matrix of the state and $\kappa$ is the excitation-loss rate determined by the physical setup. This channel is equivalently described in the Kraus representation as
\begin{align}
    \label{eq:kraus-evolution}
    \rho(t)= &\sum_{\ell=0}^\infty \hat{E}_\ell \rho(0) \hat{E}_\ell^\dagger,\\
       \label{eq:kraus-photon-loss} \hat{E}_\ell = & \left(\frac{\gamma}{1-\gamma}\right)^{\ell/2} \frac{\hat{a}^\ell}{\sqrt{\ell!}}(1-\gamma)^{\hat{n}/2}\quad\forall \ell\geq 0,
\end{align}
where $\gamma = 1-e^{-\kappa t}$. The $\ell\numth$ Kraus operator induces the loss of $\ell$ photons. The relative weight of this process decreases exponentially with $\ell$ at a fixed time. Thus, multi-photon losses are exponentially suppressed. This is analogous to high-weight Pauli errors being exponentially suppressed in qubit systems. 

At an intuitive level, photon loss causes the oscillator to drift toward lower-energy states by randomly removing some number of excitations. This makes codes with well-separated Fock-state support especially natural for detecting and correcting such errors. Each photon loss is also accompanied by a \emph{back-action} term $(1-\gamma)^{\hat{n}/2}$. Notably, even for the zero-loss operator $\hat{E}_0$, the system experiences some redistribution of probabilities due to this back-action. This process favors lower-energy states by damping higher Fock states. Consequently, only the vacuum state $\ket{0}$ survives in the infinite-time limit since 
\begin{equation}
    \lim_{t\rightarrow \infty}(1-\gamma)^{n/2}=\delta_{n0}\,.
\end{equation}

\subsubsection{Dephasing}\label{subsubsec:dephasing}
The second most common incoherent error in electromagnetic cavities is \emph{dephasing}, caused by fluctuations in the cavity frequency. A general description of the dephasing noise channel $\mathcal{N}_\text{dep}$~\cite{leviant2022quantum} acting on a density matrix $\rho$ is
\begin{align}\label{eq:dephasing-bosonic}
    \mathcal{N}_\text{dep}(\rho)=&\int_{-\infty}^\infty d\phi p(\phi) e^{-i\hat{n}\phi}\rho e^{i\hat{n}\phi} \\
    =& \sum_{m,n\in \Z_{\geq 0}} \rho_{mn}\left[\int_{-\infty}^\infty d\phi p(\phi)e^{-i(m-n)\phi}\right]\ket{m}\bra{n}
\end{align}
where
\begin{equation}
    p(\phi)=\frac{1}{\sqrt{2\pi\gamma_\text{dep}}}e^{-\phi^2/{2\gamma_\text{dep}}}
\end{equation}
is a probability distribution and $\rho_{mn}$ are the matrix elements of $\rho$ written in the Fock basis. This noise channel induces Fock-number-dependent phase shifts $\phi$ to the state governed by some probability distribution. The probability distribution $p(\phi)$ is peaked at $\phi=0$ and decays with increasing $|\phi|$, indicating that larger dephasing events are less likely. The Lindblad equation for this channel is 
\begin{equation}\label{eq:lindlblad-dephasing}
    \frac{d\rho}{dt}=\kappa_\text{dep}\left(\hat{n}\rho\hat{n}^\dagger-\frac{1}{2}\{\hat{n}^2,\rho\}\right),
\end{equation}
and the Kraus operators are
\begin{equation}\label{eq:kraus-dephasing}
    \hat{E}_k = \sqrt{\frac{\gamma_\text{dep}^k}{k!}}e^{-\frac{\gamma_\text{dep}}{2}\hat{n}^2}\hat{n}^k\quad \forall k\geq 0
\end{equation}
where $\gamma_\text{dep}=\kappa_\text{dep}t$. Just as in the photon-loss channel, these Kraus operators contain a back-action term, here \(e^{-\gamma_\text{dep}\hat{n}^2/2}\). For example, the \(k=0\) Kraus operator, corresponding to the no-jump dephasing branch, still redistributes the population in favor of the lower energy Fock states. However, it is important to keep in mind that the full dephasing channel does not redistribute Fock-state populations; when the measurement record is ignored, only the coherences between different Fock states decay.

For the purposes of code construction, one typically simplifies the Kraus operators to be proportional to $\hat{a}^\ell$ and $\hat{n}^k$ as the loss and dephasing error operators. The back-action terms are justifiably ignored due to their relatively negligible contributions at small time scales when $\hat{a}^\ell$ and $\hat{n}^k$ dominate. In simple terms, dephasing preserves photon number but scrambles the relative phases between different Fock components. Codes that organize information by symmetry in phase space or by carefully chosen parity structure can therefore be made robust against such noise.

\subsubsection{Additive white Gaussian noise (AWGN)}\label{subsubsec:AWGN}
The Gaussian displacement noise~\cite{conrad2024fabulous} evolves a state $\rho$ as 
\begin{equation}\label{eq:noise-agwn}
    \int_{-\infty}^{\infty}d^2\alpha \, p_{\sigma}(\alpha)\hat{D}(\alpha)\rho \hat{D}(\alpha)^\dagger\,,
\end{equation}
where
\begin{equation}
    p_\sigma(\alpha)=\frac{1}{\sigma^2 2\pi}e^{-|\alpha|^2/2\sigma^2}
\end{equation}
is the probability of the occurrence of a displacement by $\alpha$. The displacement operator
\begin{equation}\label{eq:displacement-operator}
    \hat{D}(\alpha)=\exp\left(\alpha\hat{a}^\dagger - \alpha^\star\hat{a}\right)
\end{equation}
displaces a coherent state $\ket{\beta}$ by an amount $\alpha$, up to a phase:
\begin{equation}\label{eq:displacement-action}
    \hat{D}(\alpha)\ket{\beta} = e^{(\alpha \beta^\star-\beta\alpha^\star)/2}\ket{\beta+\alpha}\,.
\end{equation}
Its action on the position- and momentum-basis states is
\begin{align}
    \hat{D}(\alpha)\ket{x} &= e^{i\sqrt{2}\alpha_I x+ i \alpha_R\alpha_I}\ket{x+\sqrt{2}\alpha_R},\\
    \hat{D}(\alpha)\ket{p} &= e^{-i\sqrt{2}\alpha_R p- i \alpha_R\alpha_I}\ket{p+\sqrt{2}\alpha_I}\,,
\end{align}
where $\alpha = \alpha_R + i \alpha_I$ for $\alpha_R,\alpha_I\in \mathbb{R}$. Thus, it displaces the quadratures of the state, up to a phase. In the special case where the displacement parameter is purely real or purely imaginary, it displaces only one quadrature. These special cases correspond to the operators $e^{-i\alpha \hat{p}}$ and $e^{i\alpha \hat{x}}$, which displace $x$ and $p$ by an amount $\alpha\in\R$, respectively.

\cref{eq:noise-agwn} can be viewed as Kraus evolution with a continuum of Kraus operators $\hat{E}_{\alpha} = \sqrt{p_\sigma (\alpha)}\hat{D}(\alpha)$. Thus, the action of the channel is to displace the state's quadratures by random amounts governed by a Gaussian probability distribution peaked at zero. 

\begin{exerbox}[label={exer:position-momentum-displacements}]{Position and momentum displacement operators}
    Prove that the operators $e^{-i\alpha \hat{p}}$ and $e^{i\alpha \hat{x}}$ act as position and momentum displacements, respectively.
\end{exerbox}

While we described some of the most prominent noise channels here, our list is not exhaustive. The actual noise model for a system can vary depending on the implementation and the architecture used. For example, superconducting cavities can be afflicted by higher-order nonlinearities inherited from the Josephson junctions. It is common to use a transmon (two-level-system) qubit for control and measurements of bosonic states, which makes the system prone to leakage and additional errors. For the purposes of this work, we will describe codes only in the context of photon loss, dephasing, and AWGN channels but encourage the reader to look at Refs.~\cite{notarnicola2024jointa,puri2017engineering,valadares2024ondemand,ofek2016demonstrating,zhang2022driveinduced,pietikainen2024strategies,boissonneault2008nonlinear,tsunoda2023errordetectable} for more details on other kinds of noise channels.

\subsection{Code examples}\label{subsec:code-examples}
The following sections describe several well-studied codes. These codes are not only among the most popular choices in practice, but also provide a useful baseline for thinking about various aspects of bosonic error correction. Very roughly, the main families below protect information in different geometric ways: dual-rail and binomial codes use separation in Fock space, GKP codes use a lattice in phase space, and cat codes use separated constellations of coherent states. Keeping these three pictures in mind makes it easier to see why different bosonic codes are naturally matched to their respective noise models. We begin with codes based on the Fock-state description of the Hilbert space. 

\subsubsection{Fock-state codes}\label{subsubsec:fock-state-codes}
The {\eczoo[dual-rail code]{dual_rail}} is arguably the simplest bosonic code, with logical information encoded in joint states of two bosonic modes (or ``rails''). These two modes can be realized either as two orthogonal modes of the same resonator or as two separate oscillators, depending on the physical platform. It detects up to one photon-loss error with the code states
\begin{equation}
    \begin{aligned}
        \ket{0}_L =& \ket{0,1}\,,\\
        \ket{1}_L =& \ket{1,0}\,.
    \end{aligned}
\end{equation}
Here, the two numbers in $\ket{n_1,n_2}$ represent the Fock-state occupation numbers of the two individual modes. Photon loss can be detected simply by measuring the total photon number, which is $1$ for the codespace and $0$ for the erroneous state $\ket{0,0}$. Since the code cannot correct any errors, it relies on post-selection: after detection, one either keeps or resets the state depending on the outcome. Despite not correcting any errors, it is a popular choice due to ease of implementation and has enjoyed many experimental demonstrations~\cite{koottandavida2023erasure,lu2023highfidelity,levine2024demonstrating,huang2025logical,chou2023demonstrating,mehta2025biaspreserving,alexander2024manufacturable}.

\eczoo[Binomial codes]{binomial}~\cite{michael2016new} are single-mode, Fock-state-based encodings that can be engineered to protect against any desired number of losses and any desired dephasing order. The canonical binomial code, which corrects up to one photon-loss error, that is, the error set $\{\mathbb{I},\hat{a}\}$, admits the basis
\begin{equation}\label{eq:canonical-binomial}
    \begin{aligned}
        \ket{0}_L &= \frac{1}{\sqrt{2}}\left(\ket{0}+\ket{4}\right)\,,\\
        \ket{1}_L &= \ket{2}\,.
    \end{aligned}
\end{equation}
The syndrome for this code can be measured by determining the parity of the total photon number, that is, by measuring the operator $e^{i\pi\hat{n}}$, with the following outcomes:
\begin{itemize}
    \item $+1$ implies no error,
    \item $-1$ implies a photon-loss error.
\end{itemize}
In general, binomial codes are a family of $q$-dimensional codes described by parameters $(N,S)$, with code states $\{\ket{i}_L\mid i\in\mathbb{Z}_q\}$ admitting a basis
\begin{equation}
    \ket{i}_L
    = \frac{1}{\sqrt{q^N}}
    \sum_{\substack{p=0 \\ p \equiv i \,(\mathrm{mod}\; q)}}^{(q-1)(N+1)}
    \sqrt{\binom{N+1}{p}_{q}}\, \ket{p(S+1)}\,.
\end{equation}
Here, $\binom{N}{p}_{q}$ are the extended binomial coefficients, defined as the coefficient of $x^p$ in $(1+x+x^2+\cdots+x^{q-1})^N$. They can be computed recursively using
\begin{equation}
    \begin{aligned}
        \binom{N}{p}_1 &= 1\,,\\
        \binom{N}{p}_q &= \sum_{k=0}^N \binom{N}{k}\binom{k}{p-k}_{q-1}\,.
    \end{aligned}
\end{equation}
An $(N,S)$ code protects against $L$ photon losses, $G$ photon gains, and dephasing up to order $D$, where $S=L+G$ and $N=\max(L,G,2D)$. For intuition, consider a code that protects against up to $L$ photon losses and no photon gains. This code has $S=L$, meaning that each codeword is supported on Fock states spaced by at least $L+1$ from the support of every other codeword. Hence, any photon loss $\hat{a}^\ell$ with $\ell\leq S$ cannot mix two codewords. This structure ensures that the off-diagonal Knill-Laflamme conditions of the form $\bra{\overline{i}}\hat{a}^{\dagger m_1}\hat{a}^{m_2}\ket{\overline{j}}$ are satisfied for all $m_1,m_2\leq L$ and $i\neq j$. The remaining on-diagonal conditions are satisfied by the choice of binomial coefficients; see Ref.~\cite[Appendix C]{michael2016new}. The syndrome measurement for a general member of the family is more sophisticated, and we refer the interested reader to Ref.~\cite[Sec. VI.D]{albert2018performance} for details.

\begin{exerbox}[label={exer:binomial-code-parameters}]{Binomial code parameters}
    Determine the parameters $q,N$ and $S$ for the canonical binomial code in~\cref{eq:canonical-binomial}.
\end{exerbox}

\subsubsection{Phase space codes}\label{subsubsec:phase-space-codes}
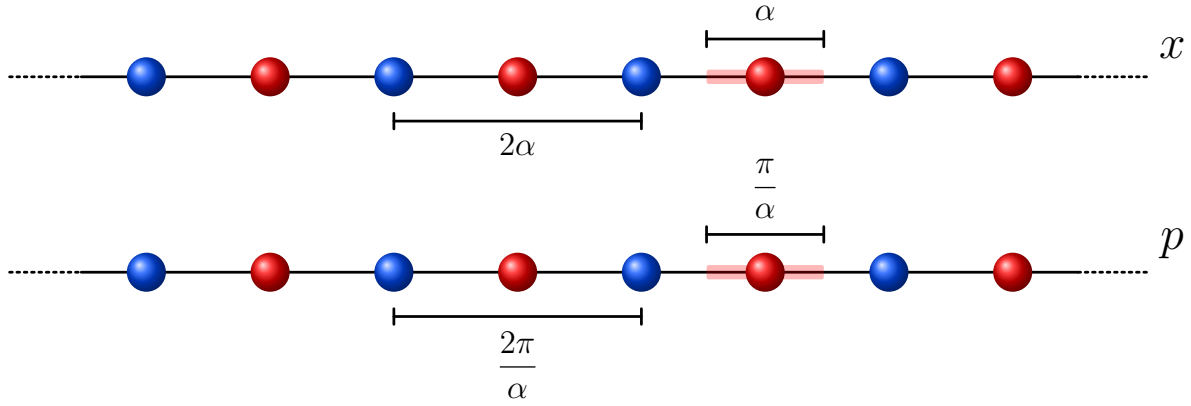
\begin{figure}[tb]
    \centering
    \resizebox{0.95\linewidth}{!}{%
    \begin{tikzpicture}[line cap=round,line join=round]
        
        % ---------- Styles ----------
        \tikzset{
          rail/.style={black, line width=1.2pt},
          cont/.style={black, dotted, line width=1.2pt},
          blueball/.style={circle, shade, ball color=blue!70!cyan, minimum size=6mm, inner sep=0pt},
          redball/.style={circle, shade, ball color=red, minimum size=6mm, inner sep=0pt},
          hilite/.style={draw=none, fill=red, fill opacity=0.28, rounded corners=1pt}
        }
        
        % ---------- Parameters ----------
        \def\sep{1.9}
        \def\ytop{3.0}
        \def\ybot{0.0}
        
        % ---------- Rails ----------
        \draw[cont] (-2.1,\ytop) -- (-1.0,\ytop);
        \draw[rail] (-1.0,\ytop) -- ({7*\sep+1.0},\ytop);
        \draw[cont] ({7*\sep+1.0},\ytop) -- ({7*\sep+2.1},\ytop);
        
        \draw[cont] (-2.1,\ybot) -- (-1.0,\ybot);
        \draw[rail] (-1.0,\ybot) -- ({7*\sep+1.0},\ybot);
        \draw[cont] ({7*\sep+1.0},\ybot) -- ({7*\sep+2.1},\ybot);
        
        % ---------- Highlight regions ----------
        \fill[hilite] ({5*\sep-0.90}, {\ytop-0.11}) rectangle ({5*\sep+0.90}, {\ytop+0.11});
        \fill[hilite] ({5*\sep-0.90}, {\ybot-0.11}) rectangle ({5*\sep+0.90}, {\ybot+0.11});
        
        % ---------- Lattice points ----------
        \foreach \i in {0,1,2,3,4,5,6,7}{
            \pgfmathsetmacro{\x}{\i*\sep}
            \ifodd\i
                \node[redball]  (T\i) at (\x,\ytop) {};
                \node[redball]  (B\i) at (\x,\ybot) {};
            \else
                \node[blueball] (T\i) at (\x,\ytop) {};
                \node[blueball] (B\i) at (\x,\ybot) {};
            \fi
        }
        
        % ---------- Top: alpha ----------
        \draw[rail] ($(T5)+(-0.9,0.58)$) -- ($(T5)+(0.9,0.58)$);
        \draw[rail] ($(T5)+(-0.9,0.46)$) -- ($(T5)+(-0.9,0.70)$);
        \draw[rail] ($(T5)+(0.9,0.46)$) -- ($(T5)+(0.9,0.70)$);
        \node[font=\Large] at ($(T5)+(0,0.98)$) {$\alpha$};
        
        % ---------- Top: 2 alpha ----------
        \draw[rail] ($(T2)+(0,-0.68)$) -- ($(T4)+(0,-0.68)$);
        \draw[rail] ($(T2)+(0,-0.56)$) -- ($(T2)+(0,-0.80)$);
        \draw[rail] ($(T4)+(0,-0.56)$) -- ($(T4)+(0,-0.80)$);
        \node[font=\Large] at ($(T2)!0.5!(T4)+(0,-1.03)$) {$2\alpha$};
        
        % ---------- Bottom: pi/alpha ----------
        \draw[rail] ($(B5)+(-0.9,0.58)$) -- ($(B5)+(0.9,0.58)$);
        \draw[rail] ($(B5)+(-0.9,0.46)$) -- ($(B5)+(-0.9,0.70)$);
        \draw[rail] ($(B5)+(0.9,0.46)$) -- ($(B5)+(0.9,0.70)$);
        \node[font=\Large] at ($(B5)+(0,1.28)$) {$\dfrac{\pi}{\alpha}$};
        
        % ---------- Bottom: 2pi/alpha ----------
        \draw[rail] ($(B2)+(0,-0.68)$) -- ($(B4)+(0,-0.68)$);
        \draw[rail] ($(B2)+(0,-0.56)$) -- ($(B2)+(0,-0.80)$);
        \draw[rail] ($(B4)+(0,-0.56)$) -- ($(B4)+(0,-0.80)$);
        \node[font=\Large] at ($(B2)!0.5!(B4)+(0,-1.42)$) {$\dfrac{2\pi}{\alpha}$};
        
        % ---------- Axis labels ----------
        \node[font=\huge] at ({7*\sep+2.45}, {\ytop+0.45}) {$x$};
        \node[font=\huge] at ({7*\sep+2.45}, {\ybot+0.45}) {$p$};
        
    \end{tikzpicture}%
    }
    \caption{Code states for the $d = 2$ GKP code are depicted by red and blue spheres. The red region shows the shifts for which the errors are correctable if the initial state was the red state. 
    }
    \label{fig:gkp-qubit}
\end{figure}

Phase space codes are most conveniently defined using the position and momentum bases. Arguably the most studied bosonic family, the \eczoo[Gottesman-Kitaev-Preskill (GKP) codes]{multimodegkp} are examples of phase-space codes. They are designed to protect predominantly against the AWGN (or displacement) noise model, but can still provide exceptional practical performance against photon loss~\cite{albert2018performance}. They first appeared in the year 2000 in the seminal paper~\cite{gottesman2001encodingb} by its namesakes: Daniel Gottesman, Alexei Kitaev, and John Preskill. For a canonical GKP code (see~\cref{fig:gkp-qubit}), the code states are 
\begin{equation}\label{eq:GKP-codestates}
    \ket{i}_{L} = \sum_{s = -\infty}^{\infty}\ket{\alpha(i + ds)}_x\,.
\end{equation}
We label the position and momentum bases by the subscripts $x,p$, respectively. The central idea is to encode information into a periodic grid in phase space so that sufficiently small shifts can be rounded back to the nearest lattice point. In this sense, GKP codes convert small continuous displacement errors into correctable syndromes.

Position displacement errors $e^{-i\Delta_x\hat{p}}$ for $|\Delta_x|\leq \alpha/2$ can be corrected by projecting the position-displaced state onto its nearest codeword. An equivalent description is given in terms of the stabilizers
\begin{align}\label{eq:GKPstab}
    \hat{S}_{Z} &= e^{2 \pi i \hat{x} / \alpha} \\
    \hat{S}_{X} &= e^{-i d \hat{p} \alpha}\,.
\end{align}
Since $\hat{S}_X$ is a displacement operator that shifts position states by an amount $d\alpha$, the code states are equal superpositions of position states spaced by $d\alpha$. The $\hat{S}_Z$ stabilizer suggests an equivalent description via equally spaced momentum states, and indeed the code states (up to a change of logical basis) can be written in the momentum basis as
\begin{equation}
    \ket{j}_L=\sum_{s=-\infty}^{\infty}\ket{\frac{2\pi}{d\alpha}(j+ds)}_p\,.
\end{equation}
They analogously correct displacement errors $e^{i\Delta_p \hat{x}}$ for $|\Delta_p|\leq \frac{\pi}{d\alpha}$.

Strictly speaking, the states in~\cref{eq:GKP-codestates} are idealized and have infinite energy. Physical implementations use finite-energy approximate GKP states, which retain the same geometric intuition while only approximately realizing the perfect lattice picture.

In general, GKP-code stabilizers are isomorphic to symplectic integral lattices~\cite{conrad2022gottesmankitaevpreskill}. This correspondence can be used to define other kinds of single-mode and multi-mode GKP codes with interesting properties. We refer the interested reader to Jonathan Conrad's work for further exploration in this direction~\cite{conrad2022gottesmankitaevpreskill,conrad2024fabulous,conrad2024latticesa,conrad2025continuousvariable}. Several broader reviews~\cite{terhal2020scalable,grimsmo2021quantum,brady2024advances,wolf2025no} may also be helpful to the reader in their exploration of the rich world of GKP codes.

\subsubsection{Coherent state codes}\label{subsubsec:coherent-state-codes}
Just like Fock- and phase-space-based encodings, one can think of codes defined natively in the coherent-state basis. Like Fock states, coherent states are closely tied to energy, but unlike the Fock basis, there is a continuum of coherent states with the same expected photon number. For fixed $|\alpha|$, this family of states $\ket{e^{i\theta}\alpha}$ is parameterized by $\theta$ and traces a circle in the complex plane:
\begin{equation}
    \bra{e^{i\theta}\alpha}\hat{n}\ket{e^{i\theta}\alpha} = |\alpha|^2 \qquad \forall\, 0\leq \theta<2\pi\,.
\end{equation}
Coherent-state-based encodings typically use this manifold of equal-energy states to encode logical information. Although more challenging in practice, recent works have begun to explore other kinds of coherent-state-based encodings as well~\cite{yang2025quantum}.

\begin{figure}[tb]
    \centering
    \resizebox{0.95\linewidth}{!}{%
    \begin{tikzpicture}[line cap=round,line join=round]
        
        % -------------------------------------------------
        % Styles
        % -------------------------------------------------
        \tikzset{
          blueball/.style={circle, shade, ball color=blue!70!cyan, minimum size=6mm, inner sep=0pt},
          redball/.style={circle, shade, ball color=red, minimum size=6mm, inner sep=0pt},
          purpleball/.style={circle, shade, ball color=magenta!85!blue, minimum size=6mm, inner sep=0pt},
          orangeball/.style={circle, shade, ball color=orange!95!red, minimum size=6mm, inner sep=0pt},
          hilite/.style={draw=none, fill=red, fill opacity=0.35},
          ring/.style={black, line width=1.1pt},
          gridline/.style={black, line width=1.0pt}
        }
        
        % -------------------------------------------------
        % Left panel: ring with highlighted arc
        % -------------------------------------------------
        \begin{scope}[shift={(1,0)}]
          \def\R{2.25}
          \def\rin{1.92}
          \def\rout{2.52}
          \def\ang{45}
        
          % Main circle
          \draw[ring] (0,0) circle (\R);
        
          % Highlighted arc sector (narrower)
          \path[hilite]
            (\ang:\rin) arc[start angle=\ang,end angle=-\ang,radius=\rin]
            -- (-\ang:\rout)
            arc[start angle=-\ang,end angle=\ang,radius=\rout]
            -- cycle;
        
          % Outer angle marker arc
          \draw[ring] (\ang:2.92) arc[start angle=\ang,end angle=-\ang,radius=2.92];
        
          % End ticks as vertical bars
          \draw[ring] ($( \ang:2.92)+(45:-0.2)$) -- ($( \ang:2.92)+(45:0.2)$);
          \draw[ring] ($(-\ang:2.92)+(135:-0.2)$) -- ($(-\ang:2.92)+(135:0.2)$);
        
          % Ring nodes
          \node[blueball] at ( 90:\R) {};
          \node[redball]  at (  0:\R) {};
          \node[blueball] at (-90:\R) {};
          \node[redball]  at (180:\R) {};
        
          % pi/2 label
          \node[font=\LARGE] at (3.5,0) {$\dfrac{\pi}{2}$};
        \end{scope}
        
        % -------------------------------------------------
        % Middle legend: vertically centered to left panel
        % -------------------------------------------------
        \begin{scope}[shift={(7.10,0.00)}]
          \node[redball] at (-1,0.55) {};
          \node[font=\LARGE, anchor=west] at (-0.38,0.55) {: $\ket{0}_L$};
        
          \node[blueball] at (-1,-0.55) {};
          \node[font=\LARGE, anchor=west] at (-0.38,-0.55) {: $\ket{1}_L$};
        \end{scope}
        
        % -------------------------------------------------
        % Right panel: ladder
        % -------------------------------------------------
        \begin{scope}[shift={(11.75,-2.88)}]
          \def\w{0.78}
          \def\h{0.72}
        
          % Dotted continuation
          \draw[dotted, line width=1.2pt] (\w/2,8*\h) -- (\w/2,8*\h+0.95);
        
          % Outer rectangle
          \draw[gridline] (0,0) rectangle (\w,8*\h);
        
          % Horizontal dividers
          \foreach \k in {1,2,3,4,5,6,7}{
            \draw[gridline] (0,\k*\h) -- (\w,\k*\h);
          }
        
          % Occupied cells
          \node[purpleball] at (\w/2,0.5*\h) {};
          \node[orangeball] at (\w/2,2.5*\h) {};
          \node[purpleball] at (\w/2,4.5*\h) {};
          \node[orangeball] at (\w/2,6.5*\h) {};
        
          % Numbers
          \foreach \k in {0,1,2,3,4,5,6,7}{
            \node[font=\Large, anchor=west] at (\w+0.28,\k*\h+0.5*\h) {\k};
          }
        \end{scope}
        
        % -------------------------------------------------
        % Right legend: vertically centered to ladder
        % -------------------------------------------------
        
        \begin{scope}[shift={(14.05,0.00)}]
          \node[purpleball] at (0.5,0.55) {};
          \node[font=\LARGE, anchor=west] at (1.12,0.55) {: $\ket{+}_L$};
        
          \node[orangeball] at (0.5,-0.55) {};
          \node[font=\LARGE, anchor=west] at (1.12,-0.55) {: $\ket{-}_L$};
        \end{scope}
        
    \end{tikzpicture}%
    }
    \caption{(Left) Computational basis states for the cat code are visualized in the coherent state basis. The red region shows the dephasing shifts for which the errors are correctable if the initial state was the red state. (Right) Hadamard basis states for the cat code are visualized in the Fock basis.}
    \label{fig:four-legged-cat}
\end{figure}
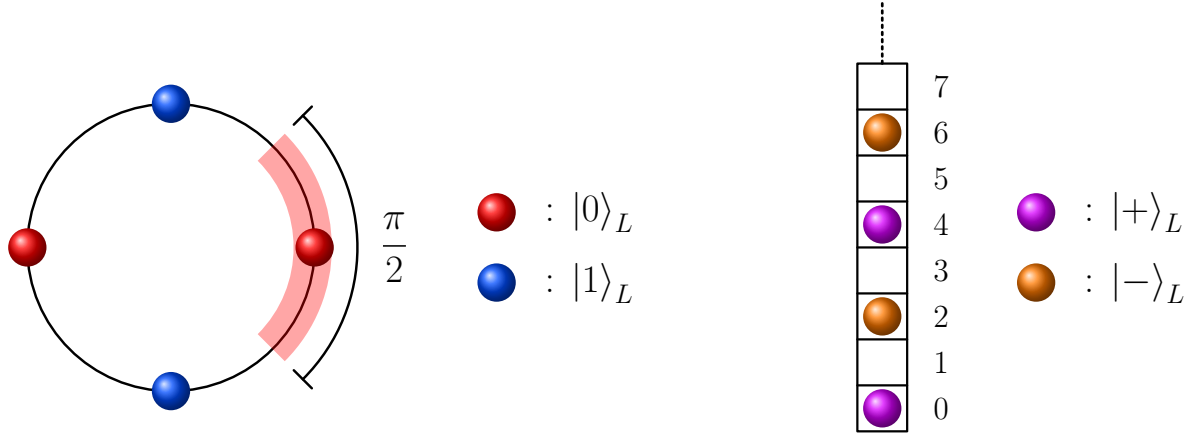

Equal-energy coherent states of the form $\ket{e^{i\theta}\alpha}$ are in one-to-one correspondence with the points on a circle. \eczoo[Cat code]{cat} states are equal superpositions of equiangular points on that circle; see~\cref{fig:four-legged-cat}. One such cat code has code states that are superpositions of the antipodal pairs $\{\alpha,-\alpha\}$ and $\{i\alpha,-i\alpha\}$ in the complex plane:
\begin{equation}\label{eq:four-legged-cat}
    \begin{aligned}
        \ket{0}_L &= \frac{1}{\mathcal{N}}\left(\ket{\alpha}+\ket{-\alpha}\right)\,,\\
        \ket{1}_L &= \frac{1}{\mathcal{N}}\left(\ket{i\alpha}+\ket{-i\alpha}\right)\,,
    \end{aligned}
\end{equation}
where $\mathcal{N}$ is a normalization factor. This code can correct up to one photon loss and protect against dephasing errors of the form $e^{i\phi\hat{n}}$ for $|\phi|<\pi/4$. It is easy to see from~\cref{fig:four-legged-cat} that the code states can be unambiguously distinguished if the rotation angle $\phi$ is less than $\pi/4$. To gain intuition for loss correction, it is useful to view the code states in the Fock basis:
\begin{equation}\label{eq:four-legged-cat-X-basis}
    \begin{aligned}
        \ket{+}_L &= \frac{1}{\sqrt{2}}\left(\ket{0}_L+\ket{1}_L\right) \propto \sum_{n=0\textrm{ mod }4} \frac{\alpha^n}{\sqrt{n!}}\ket{n}\,,\\
        \ket{-}_L &= \frac{1}{\sqrt{2}}\left(\ket{0}_L-\ket{1}_L\right) \propto \sum_{n=2\textrm{ mod } 4} \frac{\alpha^n}{\sqrt{n!}}\ket{n}\,.
    \end{aligned}
\end{equation}
Since the two code states are supported on Fock numbers differing by at least $2$, any state resulting from a single photon loss can be mapped unambiguously to its original code state by measuring the photon number modulo $4$. This Fock-state separation leads to satisfaction of the off-diagonal Knill--Laflamme conditions, similar to the case of binomial codes. The remaining conditions are satisfied by the choice of antipodal points for each codeword.

More generally, one can define a cat code in which each code state corresponds to an equal superposition of the vertices of a $p$-gon inscribed in a circle. By taking symmetrically rotated copies of the polygon, one obtains different codewords and hence a logical space of arbitrary dimension $d$. Such a $p$-gon code corrects up to $p-1$ photon-loss errors and dephasing errors for angles of magnitude up to $\pi/(pd)$. The code states can be written in the computational basis as
\begin{equation}\label{eq:general-cat-code}
    \ket{j}_L = \sum_{s=0}^{p-1}\ket{\exp\left[\frac{2\pi i}{p}\left(\frac{j}{d}+s\right)\right]\alpha}.
\end{equation}

Cat states can be generalized to the multi-mode setting in various nontrivial ways. \eczoo[Quantum spherical codes]{qsc}~\cite{jain2023quantuma} and \eczoo[tiger codes]{tiger}~\cite{xu2025letting} take superpositions of discrete and continuous constellations in higher-dimensional complex spheres, while quantum cubature codes~\cite{yang2025quantum} consider constellations spread across multiple spherical shells of different radii, that is, different energies. Many of them boast superior performance compared to trivial multi-mode generalizations based on independent copies of single-mode codes or simple concatenation schemes, and have opened up fascinating new directions in bosonic coding theory.

\begin{exerbox}[label={exer:normalization-factor}]{Coherent state normalization}
    Why is the normalization factor $\mathcal{N}$ in~\cref{eq:four-legged-cat} different from $\frac{1}{\sqrt{2}}$? Calculate $\mathcal{N}$ as a function of $\alpha$. [Hint: consider the overlap between different coherent states.]
\end{exerbox}

Taken together, these examples illustrate the main themes of bosonic coding theory. Dual-rail codes emphasize simplicity and error detection, binomial codes use carefully engineered Fock-space structure, GKP codes are especially well adapted to small displacement noise, and cat codes exploit geometric separation among coherent states. Which family is most attractive in practice often depends on the dominant noise, the available control operations, and the experimental platform. Overall, which code is the ultimate winner is still an open question.

\subsection{Further reading}\label{subsec:bosonic-further-reading}
This chapter should serve as a starting point for anyone interested in bosonic quantum error correction. While we have tried to make it reasonably self-contained, many interesting aspects of bosonic codes lie beyond the present discussion. As with Clifford gates in qubit systems, bosonic systems admit an ``easy'' set of gates known as \emph{Gaussian gates}. There is a theorem analogous to the Eastin-Knill no-go result~\cite{eastin2009restrictions}, which says that this set alone cannot suffice for universal computation. Hence, computation with bosonic codes typically requires more sophisticated protocols to achieve universality~\cite{lloyd1999quantum,weedbrook2012gaussian}. Another interesting direction is concatenation with qubit-based encodings~\cite{lee2024faulttolerant,noh2020faulttolerant,vuillot2019quantum,zhang2023concatenation}. Such codes use an outer qubit-code structure in which each individual qubit is itself a protected bosonic system. These proposals can reduce the effective physical noise afflicting the constituent qubits and hence lead to better overall performance. Control of bosonic qubits is another important topic and differs substantially from conventional qubit systems. In fact, ordinary qubits are often used as ancillas to control and measure bosonic states. These interactions can introduce additional noise into the bosonic system of interest. Engineering fault-tolerant control protocols for bosonic states remains an important area of research~\cite{grimsmo2021quantum,brady2024advances,ma2021quantum,ma2025scalable,you2024crosstalkrobust,xu2024faulttolerant}.

\newpage

\section{Decoders}\label{sec:decoders}

To protect quantum information from noise, we encode logical qubits into a quantum error-correcting code. Noise then perturbs the encoded state. Although the physical state of the code has changed, the logical information may still be preserved. To continue the computation, one must apply a recovery operation that returns the system to the codespace without changing the encoded quantum information. Because direct measurement of the error would generally destroy the logical state, quantum error correction proceeds indirectly. One measures \emph{syndromes}, that is, observables that reveal information about the error while commuting with the logical degrees of freedom. A \emph{decoder} is the classical algorithm that takes this syndrome data and outputs a suitable recovery operation~\cite{terhal2015quantum}. Decoders therefore form the interface between noisy quantum hardware and fault-tolerant logical computation.

A simple back-of-the-envelope estimate already reveals the scale of the challenge: the number of possible syndrome outcomes, and hence the number of candidate error patterns, typically grows exponentially with the number of physical qubits. At the same time, the decoder must run concurrently with the quantum processor and provide recovery operations on the fly, often on microsecond to millisecond time scales depending on the hardware platform. Thus, a useful decoder must be not only accurate, but also fast and scalable. In practice, decoder design is therefore governed by a three-way trade-off between accuracy, latency, and resource overhead~\cite{terhal2015quantum,skoric2023parallel}.

\subsection{The decoding problem}\label{subsec:decoders-decoding-problem}
Let us turn towards an explicit formulation of decoding, starting with a short recap of concepts introduced in earlier sections.
Suppose a QEC code is subject to a set of $n_\text{e}$ possible errors $\{E_i\}_{i=1}^{n_\text{e}}$, where error $E_i$ occurs with probability $p_i$. The goal of decoding is to infer from syndrome information which error has occurred and the corresponding recovery operation that should be applied.

Because measuring the errors directly would typically destroy the encoded state, we instead measure a set of $n_\text{s}$ syndrome observables $\{S_j\}_{j=1}^{n_\text{s}}$ that commute with the logical operators. For qubit stabilizer codes, these syndrome observables are the stabilizers of the code, namely multi-qubit Pauli operators of the form
\begin{equation}
    S_j=\bigotimes_{i=1}^n P_i\,,
\end{equation}
with single-qubit Pauli operators $P_i\in\{I,X,Y,Z\}$. Measuring $S_j$ yields $s_j\in\{0,1\}$, where we map Pauli eigenvalue $+1\rightarrow s_j=0$, and eigenvalue $-1\rightarrow s_j=1$. In the absence of errors, we have $s_j=0, \forall j$, while $s_j=1$ indicates that the $j$th stabilizer has flipped due to errors.

The relation between possible errors and syndrome outcomes is encoded by the parity-check matrix $H\in\{0,1\}^{n_\text{s}\times n_\text{e}}$. %, where $H_{ji}=1$ if error $E_i$ flips syndrome measurement $S_j$, and $H_{ji}=0$ otherwise. %Now, we measure the syndromes to infer which errors have occurred. Each error can either flip a particular syndrome measurement, or leave it unchanged. 
%This is encapsulated in the so-called \emph{parity-check matrix} $H\in \{0,1\}^{n_\text{s}\times n_\text{e}}$. %. 
Its rows correspond to the $n_\text{s}$ syndrome measurements, and its columns correspond to the $n_\text{e}$ possible errors. We have that $H_{ij}=1$ when the $j\numth$ error flips the $i\numth$ syndrome measurement; otherwise it is zero.  Thus, given an error configuration $\mathbf{e} \in \{0,1\}^{n_\text{e}}$, the syndrome is given by $\mathbf{s}=H\mathbf{e}$, where matrix-vector multiplication is performed modulo 2. 
Thus, one can think of the syndrome measurements as providing circumstantial information about the error.

%Then, the decoder receives as input the measured syndrome bit string $\mathbf{s}\in\{0,1\}^{n_\text{s}}$, the error model $\{p_i\}_i$, and the parity-check matrix $H$.

The decoder must then identify a recovery operation that restores the code to the codespace while acting trivially on the logical information. Importantly, successful recovery does not require reconstructing the exact microscopic error. Rather, it is enough to find a correction whose combined action with the physical error is trivial on the logical subspace. For stabilizer codes under Pauli noise, this often reduces to applying an appropriate Pauli correction. We illustrate this process in detail for the Steane code in~\cref{ex:Decoding1,ex:Decoding2,ex:Decoding3}.

\begin{example}[label={ex:Decoding1}]{How to decode Steane code (part 1)}
    We now present a hands-on example of decoding, showing how to decode the \eczoo[Steane code]{steane}~\cite{steane1996multiple-particle}, which we introduced in~\cref{ex:css-steane-code}. As a short recap, the Steane code is a CSS code with parameters $\stabcode{7}{1}{3}$. Here, 7 physical qubits encode 1 logical qubit and the code distance is 3, so it can correct any \emph{single-qubit} Pauli error.\vspace{2.5mm}
    
    After the code is subject to an error $E$ and we have measured the stabilizer syndromes, we infer a Pauli recovery operator $F$ (typically the one with the lowest Pauli weight) such that $FE$ is a stabilizer and therefore acts trivially on the logical state. The Steane code is a CSS code, so we can decode $X$-type errors and $Z$-type errors \emph{independently}, using the same machinery. It is built from the classical $[7,4,3]$ Hamming code. One convenient choice of parity-check matrix is
    \begin{equation}
        H \;=\;
        \begin{pmatrix}
        1&1&1&0&1&0&0\\
        1&1&0&1&0&1&0\\
        1&0&1&1&0&0&1
        \end{pmatrix}\,.
    \end{equation}
    A CSS code is specified by two binary parity-check matrices $H_X$ for $X$-type checks and $H_Z$ for $Z$-type checks. For the Steane code, we take
    \begin{equation}
        H_X = H_Z = H\,.
    \end{equation}
    We now label the physical qubits $1,2,\dots,7$. Each row of $H$ defines:
    \begin{itemize}
        \item $X$-type stabilizer generator: put an $X$ on each qubit where the row has a $1$;
        \item $Z$-type stabilizer generator: put a $Z$ on each qubit where the row has a $1$.
    \end{itemize}
    Concretely, if the first row is $(1,1,1,0,1,0,0)$, then the corresponding generators are
    \begin{equation}
        S^{(1)}_X = X_1 X_2 X_3 X_5\,,\qquad S^{(1)}_Z = Z_1 Z_2 Z_3 Z_5\,,
    \end{equation}
    and similarly for rows 2 and 3. Altogether, there are 3 $X$-type and 3 $Z$-type stabilizers, that is, 6 independent generators, leaving $7-6=1$ logical qubit.
\end{example}

\begin{example}[label={ex:Decoding2}]{How to decode Steane code (part 2)}
    Notably, for CSS codes such as the Steane code, we can decode $X$ and $Z$ errors separately. Suppose the physical error is a Pauli operator
    \begin{equation}
        E = X^{\mathbf{e}_X} Z^{\mathbf{e}_Z},
    \end{equation}
    where $\mathbf{e}_X,\mathbf{e}_Z \in \{0,1\}^7$ indicate on which qubits an $X$ or $Z$ component occurs (and $Y$ corresponds to both bits being $1$ on the same qubit).
    \begin{itemize}
        \item Measuring \emph{$Z$-type} stabilizers detects \emph{$X$-type} errors: $X$ anticommutes with $Z$.
        \item Measuring \emph{$X$-type} stabilizers detects \emph{$Z$-type} errors: $Z$ anticommutes with $X$.
    \end{itemize}
    Thus, we get two syndrome vectors:
    \begin{equation}
        \mathbf{s}_X \in \{0,1\}^3 \quad \text{(from $Z$-type checks, diagnosing $X$ errors)}\,,
    \end{equation}
    \begin{equation}
        \mathbf{s}_Z \in \{0,1\}^3 \quad \text{(from $X$-type checks, diagnosing $Z$ errors)}\,.
    \end{equation}
    For the Steane code with $H_X = H_Z = H$, the syndrome equations are simply
    \begin{equation}
        \mathbf{s}_X = H\,\mathbf{e}_X \pmod 2\,,
        \qquad
        \mathbf{s}_Z = H\,\mathbf{e}_Z \pmod 2\,.
    \end{equation}
    Because the Steane code has distance 3, it corrects any single-qubit Pauli error. The most likely error is the one with the highest probability, which corresponds to the error of lowest weight. If we assume that at most one $X$ error occurred, then $\mathbf{e}_X$ is either the all-zero vector or a vector with a single $1$.
\end{example}

\subsection{Optimization formulation of decoding}\label{subsec:decoders-decodingproblem}
%
%We now sharpen the previous discussion into an explicit optimization problem for qubit decoding.

%Now, we measure the syndromes to infer which errors have occurred. Each error can either flip a particular syndrome measurement, or leave it unchanged. 
%This is encapsulated in the so-called \emph{parity-check matrix} $H\in \{0,1\}^{n_\text{s}\times n_\text{e}}$. %. Its rows correspond to the $n_\text{s}$ syndrome measurements, and its columns correspond to the $n_\text{e}$ possible errors. We have that $H_{ij}=1$ when the $i$th syndrome measurement is flipped by the $j$th error; otherwise it is zero. 

%Thus, given an error configuration $\mathbf{e}$, the syndrome is given by $\mathbf{s}=H\mathbf{e}$, where matrix-vector multiplication is performed modulo $2$. Thus, one can think of the syndrome measurements as providing circumstantial information about the error.
We now sharpen the previous discussion by explicitly showing how to decode.
While the syndromes give us information about what errors may have occurred, the syndromes alone do not point to a unique error.
The main hurdle is that there are many possible errors that fit the syndrome constraint. 
In fact, the kernel of the parity check matrix $H$ is the
set of undetectable error configurations; modulo stabilizers, nontrivial elements are logical operators.
%In particular, we can always add elements of the kernel of $H$, that is, stabilizers and logical operators, to $\mathbf{e}$ without changing $\mathbf{s}$. 
Thus, a natural approach to decoding is to find the error $\mathbf{e}^\ast$ with the highest probability that is consistent with the syndrome. This is called \emph{most likely error} (MLE) decoding.
%A natural first target is therefore to find the most probable error configuration $\mathbf{e}^\ast$ consistent with the measured syndrome. 

Formally, we assume that our code can be subject to a set of $n_\text{e}$ errors $\{E_i\}_i$, with the $i\numth$ error occurring with probability $p_i$. 
The total probability of observing a particular error configuration $\mathbf{e}\in\{0,1\}^{n_\text{e}}$ is given by
\begin{equation}\label{eq:decoder-probs}
    P(\mathbf{e})=\prod_i (1-p_i)^{1-e_i} p_i^{e_i}=\prod_i (1-p_i)\prod_i \left(\frac{p_i}{1-p_i}\right)^{e_i}\,.
\end{equation}
Then, error configuration $\mathbf{e}$ is achieved by the following maximization problem:
\begin{align}\label{eq:decoder-MLE}
    \text{max}_{\mathbf{e}} \ln(P(\mathbf{e}))&=C-\sum_i w_i e_i\\
    \text{s.t. }\mathbf{s}&=H\mathbf{e}\,,
\end{align} 
where we have applied logarithms to~\cref{eq:decoder-probs} for convenience, defined weights $w_i=\ln((1-p_i)/p_i)$ and an irrelevant constant $C=\sum_i \ln(1-p_i)$. The weights $w_i$ represent the Bayesian prior encoding our knowledge of the error probabilities. Assuming all error probabilities are the same, that is, $p_i=p$, then the optimal solution is the error with the lowest weight.

We note that performing MLE decoding requires accurate prior information about both the possible errors that can occur, as well as the probabilities with which each error occurs. Thus, a necessary precondition for QEC is that the noise of the physical platform is well-characterized. 

\begin{example}[label={ex:Decoding3}]{How to decode Steane code: Part 3}
    We conclude the Steane code example with the simplest concrete decoder: a lookup rule. Because the Steane code is small, one can decode single-qubit errors by matching the observed syndrome to a precomputed table of likely corrections. Assuming that each qubit $j$ is subject to the same error probability $p_j=p$, we now find the error of lowest weight that matches the syndrome. For a $d=3$ code, this corresponds to at most a single $X$ error and a single $Z$ error.\vspace{2.5mm}
    
    Now, for decoding, let $H_{(:,j)}$ denote the $j\numth$ column of $H$ as a 3-bit vector. Then,
    \begin{quote}
        \emph{Assuming the most likely error is a single $X$ error on qubit $j$, we have}
        \begin{equation}
            \mathbf{s}_X = H_{(:,j)}\,.
        \end{equation}
    \end{quote}
    To decode, we compare $\mathbf{s}_X$ to the columns of $H$. If it matches column $j$, infer an $X$ error on qubit $j$. The same holds for $Z$ errors using $\mathbf{s}_Z$.\vspace{2.5mm}
    
    However, what about $Y$ errors?
    A single $Y$ error on qubit $j$ can be decomposed as $Y_j = i X_j Z_j$, so it flips \emph{both} syndrome types. Therefore, independent decoding of $X$ and $Z$ will produce corrections $X_j$ and $Z_j$, whose product is $Y_j$ (up to phase), correcting the error.\vspace{2.5mm}
    
    Now, we give the decoding algorithm step-by-step. As input, we are given
    \begin{itemize}
        \item Parity check matrix $H_X$ and $H_Z$,
        \item measured syndrome bits for the $Z$-type stabilizers, $\mathbf{s}_X \in \{0,1\}^3$, and for the $X$-type stabilizers, $\mathbf{s}_Z \in \{0,1\}^3$.
        %\item measured syndrome bits for the three $X$-type stabilizers: $\mathbf{s}_Z \in \{0,1\}^3$.
    \end{itemize}
    We assume that we measured
    \begin{equation}
        \mathbf{s}_X = (1,0,1), \qquad \mathbf{s}_Z = (0,0,0)\,.
    \end{equation}
    \begin{itemize}
        \item $\mathbf{s}_Z=0$ suggests no $Z$ error.
        \item $\mathbf{s}_X=(1,0,1)$: find the column of $H$ equal to $(1,0,1)$, which corresponds to column $3$ and thus to an $X_3$ error; then apply the correction $E = X_3$.
    \end{itemize}
    %Let us also give a more formal pseuod-code for lookup-decoding of the Steane Code:
    
    %\begin{verbatim}
    %function decode_steane(sX, sZ):
    %    # Decode X part from Z-check syndrome sX
    %    if sX == (0,0,0):
    %        eX_hat = no X corrections
    %    else:
    %        find j in {1..7} such that col[j] == sX
    %        eX_hat = apply X on qubit j
    %
    %    # Decode Z part from X-check syndrome sZ
    %    if sZ == (0,0,0):
    %        eZ_hat = no Z corrections
    %    else:
    %        find j in {1..7} such that col[j] == sZ
    %        eZ_hat = apply Z on qubit j
    %
    %    return E_hat = (product of X corrections) * (product of Z corrections)
    %\end{verbatim}
    
    The above lookup decoding is perfect for \emph{at most one error}. If two physical errors occur, the syndrome is the XOR (mod 2) of two columns, which may coincide with a single column. Then, the minimum-weight decoder might choose the wrong hypothesis (a weight-1 correction) even though the true error had weight 2. This is unavoidable with codes of distance $d=3$: the code is only guaranteed to correct errors of weight $\lfloor (d-1)/2\rfloor \le 1$. For more general codes with more qubits, lookup decoders are not practical as the number of possible correction operators usually scales exponentially in code size. 
\end{example}

\subsection{Degeneracy}\label{subsec:degeneracy}
The optimization problem above identifies the most likely physical error, but in QEC this is not always the quantity of operational
interest. The reason is that quantum codes possess an intrinsically quantum feature that has no direct classical analogue: \emph{degeneracy}~\cite{panteleev2021degenerate}. 

In a classical error-correcting code, for a fixed transmitted codeword $c$,
distinct error patterns $e \neq e'$ produce distinct received words,
$c+e \neq c+e'$. Thus, the received word uniquely determines the error pattern
relative to $c$, and identifying the most likely error pattern is equivalent to
identifying the corresponding most likely corrupted word.

%Classical error-correcting codes are \emph{non-degenerate}: different physical errors necessarily produce different corrupted codewords. 
%Thus, identifying the most likely error is equivalent to identifying the most likely corrupted codeword.

Quantum codes behave differently. They may be \emph{degenerate}, meaning that
distinct physical errors can have the same action on the codespace, up to an
irrelevant global phase~\cite{demarti2024decoding}. More precisely, two errors
$E$ and $E'$ are degenerate if, for every code state $\ket{\psi}$,
\begin{equation}
    E\ket{\psi} = e^{i\theta} E'\ket{\psi}
\end{equation}
for some global phase $\theta$. In this case, the two errors are indistinguishable from
the perspective of the encoded logical information~\cite{iyer2015hardness}.

%Quantum codes behave differently. They are generally \emph{degenerate}, meaning that multiple distinct physical errors can act identically on the code~\cite{demarti2024decoding}. 
%Two errors $E$ and $E'$ are called degenerate if they have the same effect on all code states $\ket{\psi}$, that is,
%\begin{equation}
%    E\ket{\psi} = E'\ket{\psi}\,.
%\end{equation}
%In this case, the two errors are indistinguishable from the perspective of logical information~\cite{iyer2015hardness}.

For stabilizer codes, degeneracy arises from the stabilizer group. A stabilizer
operator $S$ acts trivially on every code state,
\begin{equation}
    S\ket{\psi} = \ket{\psi}.
\end{equation}
Consequently, if an error $E$ occurs, then any error of the form
\begin{equation}
    E' = ES
\end{equation}
has the same action on the codespace, since
\begin{equation}
    E'\ket{\psi} = ES\ket{\psi} = E\ket{\psi}.
\end{equation}
Thus, $E$ and $ES$ represent the same physical effect on the encoded state.
They also produce the same syndrome, because multiplication by a stabilizer does
not change the commutation relations with the stabilizer generators.

% The origin of degeneracy lies in the stabilizers of the code. 
% Stabilizers $S$ leave every code state invariant,
% \begin{equation}
%     S\ket{\psi} = \ket{\psi}\,.
% \end{equation}
% Consequently, if an error $E$ occurs, then any error of the form
% \begin{equation}
%     E' = ES
% \end{equation}
% acts identically on the encoded state. 
% In particular, $E$ and $E'$ produce the same syndrome and also have the same effect on the logical space. 
% All such errors therefore belong to the same \emph{error class}.
% Mathematically, these classes correspond to cosets of the stabilizer group. 
% For a given syndrome, it then remains to find out how each coset affects the logical information. 
% %In particular, for a code encoding $k$ logical qubits, the logical Pauli group contains
% %\begin{equation}
% %$4^k$
% %\end{equation}
% %operators, corresponding to all combinations of the logical operators $I$, $\bar{X}$, $\bar{Y}$, and $\bar{Z}$ on each logical qubit. 
% %Then, each 

Therefore, in a degenerate quantum code, decoding should not necessarily be
viewed as selecting the single most likely physical error. Rather, one should
identify the most likely equivalence class of errors, where errors differing by
stabilizers are identified. Mathematically, these equivalence classes are cosets
of the stabilizer group. For a fixed syndrome, the decoder must determine which
stabilizer coset, or equivalently which induced logical operation, is most
likely.

Formally, let $\mathcal{P}_n$ be the $n$-qubit Pauli group, $\mathcal{S} \subset \mathcal{P}_n$ the stabilizer group, and $\mathcal{N}(\mathcal{S})$ its normalizer. 
The logical Pauli operators are given by the quotient group
\begin{equation}
    \mathcal{L} \cong \mathcal{N}(\mathcal{S}) / \mathcal{S}\,.
\end{equation}
Any error $E \in \mathcal{P}_n$ can be decomposed (up to an overall phase) as
\begin{equation}
    E = T(s)\, S\, L\,,
\end{equation}
where $T(s) \in \mathcal{P}_n$ is a \emph{pure error} (or \emph{syndrome representative}) that produces the syndrome $s$, $S \in \mathcal{S}$ is a stabilizer, and $L \in \mathcal{L}$ is a logical Pauli operator.
The roles of these components are as follows. The operator $T(s)$ determines the measured syndrome, while the stabilizer $S$ does not affect either the syndrome or the logical state, that is, $S\ket{\psi} = \ket{\psi}$, while the logical operator $L$ fully captures the effect of $E$ on the encoded information.
%\begin{align}
%\text{(syndrome)} \quad & \mathrm{syn}(E) = \mathrm{syn}(T(s)), \\
%\text{(stabilizer)} \quad & S\ket{\psi} = \ket{\psi}, \\
%\text{(logical action)} \quad & E\ket{\psi} = L\ket{\psi}.
%\end{align}

%\begin{itemize}
%    \item The operator $T(s)$ determines the measured syndrome.
%    \item The stabilizer $S$ does not affect either the syndrome or the logical state.
%    \item The logical operator $L$ fully captures the effect of $E$ on the encoded information.
%\end{itemize}

%\paragraph{Interpretation}

This decomposition shows that the Pauli group can be partitioned as
\begin{equation}
    \mathcal{P}_n = \bigsqcup_{s} \; T(s)\, \mathcal{S}\, \mathcal{L}\,,
\end{equation}
that is, errors are grouped first by syndrome $s$, and within each syndrome class further by their logical action. From this perspective, decoding consists of the following steps
\begin{enumerate}
    \item Using the syndrome to identify the coset $T(s)\mathcal{S}$,
    \item Inferring the most likely logical operator $L$ within that coset.
\end{enumerate}
A physical error $E$, after projection back onto the codespace via $T(s)$, induces one of these logical operations $L$ on the encoded state. 
Errors that share the same syndrome and induce the same logical operator are said to belong to the same \emph{logical class}. 
Intuitively, the syndrome determines the stabilizer part of the error, while the logical class determines how the logical information is affected.

The decoder's goal is therefore not necessarily to identify the exact physical error, but rather to determine the most probable logical class $[L]$ consistent with the observed syndrome $s$. 
Thus, it is sufficient for the decoder to infer the most likely logical class of the error and then apply the corresponding logical correction operator $C$. If after correction a nontrivial logical operator remains, then a \emph{logical error} has occurred: the syndrome has been corrected, but the encoded information has been altered.

%Decoding is successful if the combined operation $CE$ acts trivially on the logical state. 
%Equivalently, the residual error must be a stabilizer element, meaning the system returns to the codespace with no net logical operation.

%\paragraph{Example: degeneracy in the surface code}

%Degeneracy can be understood particularly easily in the surface code. 
%Consider a pair of neighboring $Z$ errors acting on adjacent qubits along an edge of the lattice. 
%Such a pair of errors produces a trivial syndrome because it is equivalent to applying a $Z$-type stabilizer (a plaquette operator). 
%As a result, many different configurations of physical errors correspond to the same stabilizer operation and therefore act identically on the encoded state.

%More generally, an error chain and the same chain multiplied by a plaquette stabilizer correspond to two different physical error patterns that produce the same syndrome and have the same logical effect. 
%From the decoder's perspective these errors are indistinguishable and belong to the same degenerate error class.

%This degeneracy is crucial in practical decoding. 
%For example, in the surface code there are often many distinct error chains with the same endpoints (i.e. the same syndrome). 
%Although each individual chain may have small probability, their \emph{combined probability} can exceed that of another chain with slightly smaller weight. 
%An optimal decoder must therefore consider the \emph{total probability of all degenerate errors} that correspond to the same logical class.

%\paragraph{Degenerate maximum-likelihood decoding}

The optimal decoding strategy is therefore \emph{degenerate maximum-likelihood decoding} (DMLD).
Instead of identifying the single most probable error, DMLD sums the probabilities of all errors belonging to the same logical class and chooses the logical class with the highest total probability given the syndrome. 
Unfortunately, computing this quantity is usually extremely difficult, making optimal decoding a computationally challenging problem~\cite{iyer2015hardness}.

A common relaxation is MLE, which we introduced above, which simply searches for the single most likely physical error while ignoring degeneracy.\footnote{MLE is also called quantum maximum likelihood decoding, following the notation of Ref.~\cite{iyer2015hardness}. Note that both DMLD and MLE are maximum-likelihood problems, just over different probability distributions.} 
Because it neglects the probability contributions of other degenerate errors, MLE may in general return a suboptimal correction. 
However, its computational complexity is lower: MLE is only an \NP-hard problem. While this implies that MLE takes in the worst case exponential time to solve, it is still relatively easier to compute than the QMLD which is a more difficult $\#\P$-complete problem~\cite{hsieh2011np,iyer2015hardness}.\footnote{To be more formal, \NP is the class of decision problems for which a proposed solution can be checked efficiently, but finding the solution may be difficult. An \NP-hard problem is at least as hard as every problem in \NP, so solving such problems is expected to take exponential time in the worst case. A $\#\P$-complete problem asks not just whether solutions exist, but also how many, which is substantially more difficult.}
Further, MLE is amenable to straightforward relaxations, which has led to many efficient heuristic decoders that can solve the decoding problem quite well in practice.

\subsection{How to measure syndromes}\label{subsec:measuresyndrome}
\begin{figure}[tb]
    \centering
    \includegraphics[width=0.4\textwidth]{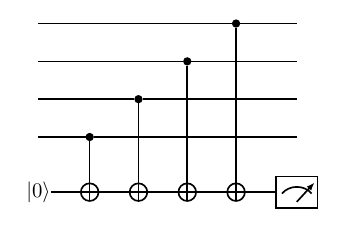}
    \includegraphics[width=0.4\textwidth]{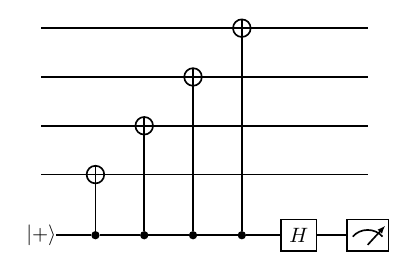}
    \caption{Measurement protocol for $Z$-type (left) and $X$-type (right) stabilizers, here shown for weight $4$ stabilizers.}
    \label{fig:stabilizer_meas}
\end{figure}

Let us turn now towards the more practical aspects of decoding. %Given a particular code, how do we extract the syndromes from the code?
%We begin with the physical origin of the data that a decoder receives: the syndrome measurements. 
For example, given a QEC code, how do we measure syndromes? For stabilizer codes, this corresponds to measuring the Pauli operators of the stabilizers. In particular, for CSS codes, whose stabilizer checks are either purely $Z$-type or purely $X$-type, one can measure syndromes using the circuit shown in~\cref{fig:stabilizer_meas}.

Here, one initializes an ancillary qubit in the $\ket{0}$ state for $Z$-type stabilizers, while $\ket{+}=\frac{1}{\sqrt{2}}(\ket{0}+\ket{1})$ for $X$-type stabilizers. Then, CNOT gates are implemented for each data qubit that the stabilizer has support on. For example, for a code with weight 4-stabilizers such as the surface or Steane code, this measurement involves 4 data qubits. The syndrome outcome is recorded by measuring the auxiliary qubit, where for an $X$-type stabilizer we have to measure in the eigenbasis of $\ket{+}$, which requires an additional Hadamard gate $H$.

\subsection{Noise models for QEC simulation}\label{subsec:decoders-noise-models}
%To evaluate the performance of a QEC code and a decoder, we can simulate its operation under a given noise model. 
To model the decoding process, we need an accurate description of the errors that can occur and their corresponding probabilities.
On the one hand, this is essential when decoding actual experiments, where our noise model should reflect the experimental reality. On the other hand, this is also essential for classical simulation of the codes and to understand their error correction capabilities.
This raises the question of how to model errors in the first place. In fact, the choice of noise model represents a trade-off between physical realism and computational tractability. This has led to a hierarchy of standard simulation models in the literature:
\begin{enumerate}
    \item \textbf{Code-capacity model}: This is the most idealized model. It assumes that the syndrome extraction process (including all ancilla operations and measurements) is perfectly error-free. Noise, typically i.i.d. depolarizing noise, is applied only to the data qubits once per QEC cycle. This model is not realistic but is invaluable for theoretical studies, as it allows for the calculation of a code's maximum possible error tolerance (its ``code capacity'') and provides an upper bound on its performance. 
    \item \textbf{Phenomenological model}~\cite{dennis2002topological,fowler2012surface}: This model takes a step towards realism by acknowledging that syndrome measurements are also faulty. It introduces a probability of measurement error, which effectively flips the classical syndrome bits after they are extracted. This also encapsulates a simplified simulation of errors associated with qubit initialization (state preparation) and reset. However, it still abstracts away the details of the gate-level errors that occur during the syndrome extraction circuit.
    \item \textbf{Circuit-level model}: This is the most comprehensive and realistic noise model by explicitly simulating the QEC circuit. It accounts for all potential sources of error within the quantum circuit used for a QEC cycle. A noise channel is applied after every physical operation, such as single-qubit gates, two-qubit gates, qubit idling (decoherence), state preparation, and measurement readout.
\end{enumerate}

Luckily, these noise models can be simulated efficiently for one of the most important cases: stabilizer codes under Clifford noise models, which includes Pauli noise models. Here, all operations are Clifford and can be efficiently simulated using a tableau formalism in $O(n^3)$ time~\cite{gottesman1998heisenberg}. 
For more general noise models, such as coherent noise, simulation methods for specific settings have also been found~\cite{bravyi2018correcting}.

\subsection{Detector model}\label{subsec:detector-model}
So far, we described decoding in terms of syndrome vectors and parity-check matrices. That language is natural for introducing decoding algebraically, especially for code-capacity and phenomenological models. However, for repeated rounds of syndrome extraction and for circuit-level noise, it is often more convenient to work with the \emph{detector model}. In this representation one tracks changes in measurement outcomes over time, which leads to a particularly clean graphical picture used by many modern circuit-level decoders.

In practical QEC, especially for large codes, it is often inconvenient to work directly with Pauli operators and stabilizer generators. Instead, one can reformulate the problem in terms of \emph{detectors} and \emph{detection events}, which leads to a very clean graphical decoding picture~\cite{gidney2021stim,bravyi2024high}. The key idea is to track \emph{changes} in stabilizer measurement outcomes rather than the outcomes themselves.

Consider repeated rounds of stabilizer measurements. 
Each stabilizer measurement produces a binary outcome (e.g., $\pm 1$). 
A \emph{detector} is defined as a function of one or more measurement outcomes such that, in the absence of errors, it always evaluates to $+1$ (or equivalently, $0$ in binary form). A \emph{detection event} occurs when a detector flips from its expected value. 
Equivalently, a detector outputs $-1$.
For example, in a circuit with repeated measurements, a common detector is the \emph{parity} of a stabilizer measurement between two consecutive rounds:
\begin{equation}
    d_i(t) = m_i(t) \cdot m_i(t-1)\,,
\end{equation}
where $m_i(t)$ is the measurement outcome of stabilizer $i$ at time $t$. 
If no error occurs, the measurement is stable and $d_i(t)=+1$. 
A flip indicates that an error has occurred.

Each physical error (e.g., a Pauli error on a qubit or a measurement error) produces a characteristic pattern of detection events.
We can represent this as a mapping
\begin{equation}
    E \;\longmapsto\; \mathbf{d}(E) \in \mathbb{F}_2^N\,,
\end{equation}
where $\mathbf{d}(E)$ is a binary vector indicating which detectors click.

A crucial property is that most physical errors trigger a \emph{small number} of detectors. As such, the pattern of detection events depends only on the error up to stabilizers.
Thus, the detector model naturally incorporates degeneracy as different physical errors may produce the same detection pattern.
This framework is widely used because it separates the \emph{physics} (error processes) from the \emph{inference problem} (decoding), enabling efficient and scalable decoders. 

\subsection{Decoder landscape}\label{subsec:decoders-landscape}
In this section, we give a brief overview of the landscape of decoders for QEC. No single decoder is uniformly best: the preferred choice depends on the code family, the noise model, and the available latency budget. Broadly speaking, different decoder families are favored in different regimes. For topological codes such as the surface code, minimum-weight perfect matching (MWPM) and Union-Find are among the most important practical baselines, with MWPM emphasizing accuracy and Union-Find emphasizing speed. For qLDPC codes, belief-propagation-based decoders and their post-processed variants play a particularly central role, while tensor-network decoders are among the strongest high-accuracy approaches for structured local codes. Finally, for more general detector-based circuit-level decoding beyond graph-like settings, newer search-based approaches such as decision-tree and Tesseract decoders provide a flexible alternative.
For quick reference, we summarize the main trade-offs between the most common decoder families discussed in this chapter in~\cref{tab:decoder-comparison}.

\begin{table}[!ht]
    \centering
    \scriptsize
    \begin{tblr}{
        width=\linewidth,
        colspec={X[l] X[l] Q[c,m] X[l] X[l]},
        row{1}={font=\bfseries},
        cell{2-Z}{3}={halign=c},
        hlines
    }
        Decoder & Best for & Speed & Main advantage & Main drawback  \\
        \hline
        Lookup table (\cref{ex:Decoding3}) & small toy codes; pedagogy & very fast & easiest to teach and implement & does not scale \\
        MIP / exact MLE (\cref{subsec:decoders-mip}) & small generic stabilizer codes; benchmarking & slow & clean near-optimal baseline & exponential cost in practice \\
        LP / SOS relaxations (\cref{subsec:decoders-mip})& approximate optimization-based decoding & moderate--slow & systematic relaxations of MLE & can yield weak or costly relaxations \\
        MWPM (\cref{subsec:decoders-mwpm})& surface code; graph-like detector models & moderate--fast & strong practical baseline for topological codes & not naturally suited to arbitrary codes \\
        Union-Find (\cref{subsec:decoders-union-find})& low-latency surface-code decoding & very fast & near-linear and hardware-friendly & usually less accurate than MWPM \\
        Belief propagation (BP) (\cref{subsec:decoders-bp})& qLDPC and sparse-check codes & fast & scales well; strong hybrid building block & short loops and degeneracy hurt plain BP \\
        Tensor networks (\cref{subsec:decoders-tensor-network})& high-accuracy decoding for structured local codes & slow & excellent accuracy; captures degeneracy & computationally expensive \\
        Machine learning (\cref{subsec:decoders-machine-learning})& correlated noise; soft information & fast inference & flexible once trained & training is costly \\
        Decision-tree / Tesseract (\cref{subsec:decoders-tesseract})& general detector-model decoding & moderate--slow & very flexible, including circuit-level settings & runtime can grow quickly \\
        Max-SAT / QUBO (\cref{subsec:decoders-maxsat})& optimization-based exact or near-exact decoding & slow & leverages mature optimizers and special hardware & usually too slow for real-time use \\
        Renormalization group (\cref{subsec:decoders-renormalization})& fast multiscale topological decoding & fast & parallelizable and low-latency & may neglect global error correlations \\
    \end{tblr}
    \caption{Comparison of commonly used quantum decoders. The speed labels are qualitative and meant only as a rough guide.}
    \label{tab:decoder-comparison}
\end{table}

\subsection{Mixed-integer programming decoders and relaxations}\label{subsec:decoders-mip}
In general, the MLE problem in~\cref{eq:decoder-MLE} can be formulated as a mixed-integer program (MIP), or for specific choices of error probabilities $p_i$ as an integer program~\cite{landahl2011fault}. In general, these constitute \NP-hard problems, so such MIP decoders likely require exponential time to find the optimal solution.

However, if one is satisfied with potentially non-optimal solutions, one can relax the MLE problem so that decoding can be achieved in polynomial time.
An example of such a relaxation is given by linear program (LP) decoders~\cite{li2018lp,fawzi2021linear}. LP decoders relax the original binary constraint $e_i\in\{0,1\}$ into a continuous optimization problem with $e_i'\in[0,1]$. Then, the LP can be solved in polynomial time using standard solvers. After optimization, the real-valued $e_i'$ is rounded to the nearest binary number to give a valid error configuration. However, LP decoders often struggle with quantum codes, yielding nonphysical fractional solutions that result in logical errors. This issue can be improved by adding post-processing steps~\cite{gu2025powerlimitationslinearprogramming}.

A recent work proposed a systematic relaxation of MLE using sum-of-squares (SOS) or Lasserre hierarchies~\cite{basak2026hierarchical}. 
Here, one approximates the constraint that $e_i$ is binary by using polynomials of sum of squares. The resulting relaxed problem can be solved as a semidefinite program (SDP). The maximal allowed degree $\ell$ of the sum of squares polynomials allows one to systematically trade-off between accuracy and runtime. While for small $\ell$, we see an improvement in runtime, this also results in lower accuracy, whereas a higher value for $\ell$ yields better decoding solutions but requires more time.
Indeed, one finds that with increasing $\ell$, the problem converges to the optimal solution. In fact, there is a certificate of optimality via the rank-loop condition: above a certain $\ell$, the SOS decoding problem is guaranteed to be optimal. 
Numerically, one finds that the lowest level $\ell=1$ has limited accuracy, while even $\ell \geq 2$ shows good performance close to the optimal MLE. 

%First, one observes that one can equivalently replace the binary constraint $e_i\in\{0,1\}$ with a continuous real number $e_i'\in[0,1]$ with the additional constraint $g_i=e_i'^2-e_i'=0$. 
%Now, approximates $g_i$ using $\ell+1$ Here, one first observes that the 

\subsection{Minimum weight perfect matching}\label{subsec:decoders-mwpm}
For topological codes, the most important accuracy-oriented baseline is minimum weight perfect matching (MWPM). The MWPM algorithm is the canonical and most widely studied decoder for the surface code. It maps the decoding problem onto a graph problem, which can then be solved using the blossom algorithm in polynomial time~\cite{edmonds1965paths}. In the worst case, time $O(n^3)$ is required, though in practice much better decoding times have been observed.
However, MWPM in its standard formulation requires that the decoding problem has a graph structure, that is, each error can flip at most two syndromes~\cite{higgott2025sparse}. This condition is satisfied for surface codes, while for more general codes one has to consider relaxations of the decoding problem. 
The core reason that MWPM works so well for the surface code is that a single local data-qubit error typically creates either two nearby detection events or one event together with a boundary. Decoding can therefore be reformulated as the problem of pairing up defects in the most likely way. The decoding process of vanilla MWPM proceeds as follows:
\begin{enumerate}
    \item \textbf{Graph construction}: A graph is constructed where the vertices represent the locations of non-trivial syndrome bits (often called ``defects'' or ``anyons''). An additional set of ``boundary'' vertices is included to handle error chains that terminate at the edges of the code.
    \item \textbf{Edge weighting}: An edge is drawn between every pair of vertices. The weight of an edge is calculated as the negative logarithm of the probability of the most likely physical error chain that could create that pair of defects. For simple i.i.d. noise, this weight is often approximated by the Manhattan distance between the vertices on the qubit lattice. 
    \item \textbf{Matching}: The algorithm then finds a \emph{perfect matching} on this graph (a set of edges that pairs up all vertices with no vertex left over) such that the sum of the weights of the edges in the matching is minimized. This is a standard problem in computer science that can be solved efficiently with algorithms like Edmonds' blossom algorithm~\cite{edmonds1965paths}.
    \item \textbf{Correction}: The minimum-weight matching corresponds to the most probable set of error chains that explains the observed syndrome. A correction operator is then constructed by applying Pauli operators along the paths indicated by the matching.
\end{enumerate}
MWPM is highly accurate for surface codes under independent, depolarizing noise, achieving thresholds near the optimal limit. As a simple example, if two nearby detection events appear in the lattice, MWPM will tend to connect them by the shortest and therefore most probable error chain; if a defect lies close to a boundary, it may instead be cheaper to match it to that boundary. In this way, the matching problem directly encodes the competition between different plausible recovery paths.

Note that there have been many extensions of MWPM to address its restriction to graph-like decoding problems. For example, the color code, an important code class, does not na\"{i}vely support MWPM. However, modifications of MWPM exist that allow extensions to color codes~\cite{sahay2022decoder,gidney2023new}.
Further, recent work proposed decoding via the Minimum-Weight Parity Factor (MWPF) problem, which generalizes MWPM to hypergraphs~\cite{wu2025minimum}. MWPF decoding works on general decoding problems, thus making it far more versatile for generic codes. We note that this is an ongoing and popular research topic, with many other extensions existing.

\subsection{Union-Find}\label{subsec:decoders-union-find}
If one is willing to trade some accuracy for substantially lower latency, a natural alternative is the Union-Find (UF) decoder. The UF decoder is a faster, nearly linear-time alternative to MWPM, making it a compelling candidate for real-time decoding on large-scale systems~\cite{delfosse2021almost}. The UF algorithm operates by growing clusters of defects:
\begin{enumerate}
    \item Each defect initially forms its own cluster.
    \item The algorithm iterates through all possible single-qubit errors in increasing order of size.
    \item For each potential error, it checks if it connects two existing clusters or a cluster to a boundary. If so, the clusters are merged (the ``union'' operation).
    \item This process continues until only clusters with even numbers of defects remain, which are then corrected internally.
\end{enumerate}
The UF decoder's primary advantage is its speed, with a complexity close to $O(n \alpha(n))$, where $\alpha(n)$ is the extremely slow-growing inverse Ackermann function. This makes it significantly faster than MWPM for large codes. This speed comes at the cost of slightly lower accuracy; the threshold achieved by UF decoders is typically a bit lower than that of MWPM under standard noise models. However, its hardware-friendly structure has inspired highly optimized implementations like Riverlane's Collision Clustering decoder~\cite{barber2025real}, designed for deployment on FPGAs.

\subsection{Belief propagation}\label{subsec:decoders-bp}
For sparse qLDPC codes, the graphical structure of the decoding problem can be quite different from that of topological matching-based decoders. In this regime, belief propagation (BP) has been established as a central algorithmic idea. BP is an iterative message-passing algorithm~\cite{kschischang2001factor,mackay2003information}. It is widely used in wireless communication~\cite{richardson2008modern}, and has also found widespread application in QEC decoding~\cite{poulin2008iterative}. 
To decode, BP operates on the code's Tanner graph. This is a bipartite graph consisting of variable nodes that represent errors, and check nodes that represent stabilizer constraints.
%\begin{itemize}
%    \item \emph{variable nodes} (representing qubits or error variables), and
%    \item \emph{check nodes} (representing stabilizer constraints).
%\end{itemize}
An edge connects a variable node to a check node if the corresponding qubit participates in that stabilizer.

Given a measured syndrome $s$, the goal of BP is to estimate for each qubit $i$ the \emph{marginal probability}
\begin{equation}
    P(e_i = 1 \mid s)\,,
\end{equation}
that is, the probability that an error occurred on that qubit conditioned on the observed syndrome.
Instead of solving this global inference problem directly (which is computationally hard), BP approximates it by passing \emph{local messages} along the edges of the Tanner graph. Each edge in the graph carries two types of messages:
\begin{itemize}
    \item $m_{i \to a}$: from variable node $i$ to check node $a$,
    \item $m_{a \to i}$: from check node $a$ to variable node $i$.
\end{itemize}
Intuitively, a variable node tells a check, ``I believe I am in error with probability $p$,'' while a check node tells a variable, ``Given the syndrome constraint and what other qubits are doing, you should (or should not) have an error.'' Messages are initialized using the physical error model. 
For example, for independent bit-flip noise with error rate $p$, one initializes
\begin{equation}
m_{i \to a}^{(0)} = \log\frac{1-p}{p}\,,
\end{equation}
often expressed in terms of \emph{log-likelihood ratios} (LLRs) for numerical stability.

Each check node enforces a parity constraint, that is, the sum modulo 2 of the neighboring error variables must match the syndrome.
The message from check node $a$ to variable node $i$ is computed from all incoming messages except $i$:
\begin{equation}
    m_{a \to i} = 2 \tanh^{-1} \left( 
    \prod_{j \in N(a) \setminus i} \tanh\left(\frac{m_{j \to a}}{2}\right)
    \right)\,,
\end{equation}
where $N(a)$ denotes the neighbors of check $a$.

Each variable node combines its prior belief (from the noise model), and incoming messages from all neighboring checks except $a$.
The update rule is:
\begin{equation}
    m_{i \to a} = \ell_i + \sum_{b \in N(i) \setminus a} m_{b \to i},,
\end{equation}
where $\ell_i$ is the initial log-likelihood ratio. These updates are repeated iteratively. After a fixed number of iterations (or upon convergence), one computes the \emph{belief} at each variable node as
\begin{equation}
    \Lambda_i = \ell_i + \sum_{a \in N(i)} m_{a \to i}\,.
\end{equation}
At the end, a hard decision is then made to decode the error, for example, via
\begin{equation}
    e_i =
    \begin{cases}
        1 & \text{if } \Lambda_i < 0\,, \\
        0 & \text{otherwise.}
    \end{cases}
    \end{equation}
%The resulting error estimate $\hat{E}$ is checked against the syndrome. If it matches, decoding succeeds.

BP works very well on tree graphs (graphs without loops), where it computes the true marginal probabilities. 
These updates are very efficient, with each iteration requiring only local message updates along edges of the Tanner graph, leading to a computational complexity that scales linearly with the number of qubits per iteration. However, Tanner graphs of quantum codes typically contain many \emph{short loops} as stabilizers often overlap heavily. 
These loops cause correlations that BP does not properly account for, leading to overconfident or inconsistent beliefs, or convergence to incorrect solutions.
Further, degeneracy means that multiple errors correspond to the same syndrome.
%\paragraph{BP for quantum codes}
%Applying BP directly to quantum codes introduces additional challenges. Stabilizers overlap heavily, creating many short cycles. Further, degeneracy means that multiple errors correspond to the same syndrome.
As a result, standard BP often performs poorly for qLDPC codes. 
However, BP can be integrated within other methods, making it one of the most widely used building blocks for modern quantum decoders.

As such, several methods have been proposed to improve BP decoding in the quantum setting, of which we mention a few examples here. For example, using automorphisms to randomly permute the Tanner graph can help mitigate trapping in loops~\cite{koutsioumpas2025automorphism}. Further, one can run BP multiple times with different initializations~\cite{muller2025improved} or iterate over the least reliable bits~\cite{ye2025beam}.
Also, one can enhance BP with classical decoding techniques such as ordered statistics decoding (OSD)~\cite{roffe2020decoding} or localized statistics decoding (LSD)~\cite{hillmann2024localized}.
These approaches significantly improve performance. This has led BP-based decoders being one of the most popular choices for qLDPC codes, driving their recent popularity and applications~\cite{bravyi2024high}.

\subsection{Tensor network-based decoders}\label{subsec:decoders-tensor-network}
Tensor network (TN) decoders are among the most powerful known decoding methods, achieving very high accuracy by directly approximating DMLD~\cite{ferris2014tensor,bravyi2014efficient,chubb2021general}. 
Their strength comes from reformulating decoding as a problem in statistical physics and evaluating it using tensor network techniques.

Recall that DMLD requires computing, for each logical class $L$, the total probability
\begin{equation}
    P(L \mid s) = \sum_{E \in \mathcal{E}(s,L)} P(E)\,,
\end{equation}
that is, summing the error probabilities $P(E)$ over all physical errors $E\in\mathcal{E}(s,L)$ consistent with the syndrome $s$ and logical operator $L$.
This sum can be interpreted as a \emph{partition function}:
\begin{equation}
    Z_L = \sum_{E \in \mathcal{E}(s,L)} e^{-\beta H(E)}\,,
\end{equation}
where $H(E)$ is an effective energy (typically proportional to the weight of $E$) and $\beta$ is an inverse temperature determined by the noise model. Thus, decoding becomes equivalent to evaluating a classical statistical mechanics model, that is, \emph{computing the partition function over all error configurations consistent with the syndrome.}

The key insight is that this partition function can be written as a \emph{tensor network}.
Each component of the code is represented as a tensor: variable tensors encode the probability of error on each qubit, constraint tensors enforce stabilizer (parity) constraints, and boundary tensors enforce the observed syndrome and logical class.
The full probability $Z_L$ is obtained by \emph{contracting} the network.
%\begin{equation}
%Z_L = \mathrm{tTr} \left( \bigotimes_i T_i \right),
%\end{equation}
%where $\mathrm{tTr}$ denotes contraction over all shared indices.
%Graphically, this corresponds to:
%\begin{itemize}
%    \item nodes = tensors,
%    \item edges = shared indices (summed over),
%    \item contraction = summing over all error configurations.
%\end{itemize}

%\paragraph{Example: surface code}

%For the surface code, the tensor network lives on a 2D lattice, where
%\begin{itemize}
%    each qubit corresponds to a tensor with indices representing its error state, each stabilizer corresponds to a tensor enforcing parity constraints, and the syndrome modifies the local tensors.
%Error configurations correspond to paths (error chains), and the tensor contraction sums over all such paths consistent with the syndrome.

%\paragraph{Decoding procedure}

The decoding algorithm proceeds as follows. First, we construct the tensor network corresponding to the code, noise model, and observed syndrome. Then, for each logical class $L$, we modify boundary conditions to enforce the corresponding logical operator, and contract the tensor network to compute the partition function $Z_L$. Finally, we select the most likely logical class
\begin{equation}
    L^* = \arg\max_L Z_L\,.
\end{equation}

%\paragraph{Why tensor networks work}

%Tensor networks efficiently exploit the \emph{local structure} of quantum codes:
%\begin{itemize}
%    \item Stabilizers only involve a small number of qubits,
%    \item The resulting graphical model has low connectivity,
%    \item Many correlations can be captured compactly.
%\end{itemize}

Crucially, TN decoders \emph{sum over all degenerate errors}, rather than selecting a single configuration. 
This allows them to approximate true DMLD much more accurately than methods such as MWPM or BP.
Exact contraction of a tensor network generally scales exponentially with the tree width of the graph. 
For 2D codes such as the surface code, this leads to a complexity scaling roughly as
\begin{equation}
    O(\exp(c \sqrt{n}))\,,
\end{equation}
which is significantly better than brute-force enumeration but still super-polynomial.

In practice, approximate contraction methods are used, such as matrix product state (MPS) contraction, where one contracts the network row by row with controlled bond dimension. A powerful hybrid approach combines BP with tensor networks~\cite{alkabetz2021tensor}. Here, BP is used to approximate marginal distributions, and these marginals guide or simplify the tensor contraction. This significantly reduces computational cost while maintaining high accuracy.

%\paragraph{Summary}

%Tensor network decoders reformulate decoding as a partition function evaluation:
%\begin{itemize}
%    \item Errors $\rightarrow$ configurations of a statistical model,
%    \item Probabilities $\rightarrow$ Boltzmann weights,
%    \item Decoding $\rightarrow$ tensor contraction.
%\end{itemize}

%By summing over all degenerate errors, they provide near-optimal decoding performance, at the cost of higher computational complexity.

%Tensor Network (TN) decoders are the state-of-the-art in achieving high decoding accuracy. They achieve near-optimal performance by directly tackling the full DMLD through the powerful mathematical framework of tensor networks~\cite{ferris2014tensor,bravyi2014efficient,chubb2021general}.
%They map the DMLD problem onto a classical spin problem, where they compute its partition function at a specific temperature via tensor contractions. 
%While the computational complexity scales exponentially with problem size, efficient relaxations based on BP-aided tensor network contraction exist~\cite{alkabetz2021tensor}.

\subsection{Machine learning-based decoders}\label{subsec:decoders-machine-learning}
Unlike the previous model-based decoders, machine learning (ML) decoders aim to learn the map from syndrome to correction directly from data. ML decoders therefore represent a paradigm shift from the model-based approaches described above. Instead of relying on explicit, predefined noise models, ML decoders learn the mapping from syndrome to correction by training on data drawn from simulations or actual experiments~\cite{varsamopoulos2017decoding}.

To implement ML decoders, neural networks have been a popular choice, and they can also handle highly correlated noise, even beyond circuit-level noise descriptions~\cite{torlai2017neural,varbanov2025neural}.
Recent studies have trained ML decoders on simulated data, which can often be generated efficiently. Further, the decoder's performance can be refined using experimental data~\cite{bausch2024learning}. 
Because ML models learn directly from data, they appear to perform well on correlated errors and can also incorporate soft information such as analogue read-out signals~\cite{varbanov2025neural}.
A potential downside of ML decoders is the need for large quantities of training data and the corresponding computational time needed to train them, which may scale unfavorably with code distance.

\subsection{Tesseract and decision-tree decoders}\label{subsec:decoders-tesseract} 
For more general detector-model decoding beyond graph-like settings, recent works have proposed the decision-tree decoder~\cite{ott2025decision} and the closely related Tesseract decoder for QEC codes~\cite{beni2025tesseract}. Both decoders frame decoding as a graph-search problem in a high-dimensional space and are able to decode circuit-level noise models. %It was introduced to efficiently decode circuit-level noise—where errors occur not only on data qubits but also during gates, measurement, and idle periods.

These decoders take as input the detector error model and the resulting syndrome bits. They then construct a search tree whose nodes correspond to partial error configurations. The decoders use A*-style informed search with a carefully designed heuristic to explore only the most promising branches~\cite{hart1968formal}. 
The goal is to find the lowest-cost set of physical errors consistent with the observed syndrome.
By using the established A* search with an admissible heuristic, these decoders achieve high accuracy while pruning most of the search space~\cite{grbic2026accelerating}. The Tesseract and decision-tree decoders use different heuristics to improve decoding performance. 
Further, the decision-tree decoder can be combined with other methods such as BP.

Notably, both algorithms can be applied to any stabilizer code including code-capacity and circuit-level noise models, and are not tied to regular lattice structures or graph-like errors. Also, by tuning the search parameters, the decoders can be made equivalent to the optimal maximum likelihood decoding problem, at the cost of vastly increased runtime.

%Overall, the Tesseract decoder is a general, exact, and flexible decoding algorithm that complements fast approximate decoders like matching, especially in regimes where circuit-level correlations matter.

\subsection{Combinatorial decoders}\label{subsec:decoders-maxsat}
A different route is to translate decoding into a standard combinatorial optimization problem and then exploit mature solver technology. Recent works have proposed decoders that map the decoding problem onto combinatorial problems and then solve the resulting combinatorial instance.

In particular, the Max-SAT decoder maps MLE to the Max-SAT problem: Max-SAT is an optimization problem in which one aims to find an assignment of binary variables that satisfies the maximum number of clauses~\cite{berent2023decoding,noormandipour2024}. 
For example, a Max-SAT instance with two Boolean variables $e_1$ and $e_2$ may consist of the three clauses
\begin{equation}
    (e_1\lor e_2)\land(e_1\lor\lnot e_2)\land(\lnot e_1\lor e_2)\,,
\end{equation}
where $\lnot$ is the negation of the literal, $\lor$ indicates a logical OR, and $\land$ indicates a logical AND. In the above example, the choice $e_1=e_2=1$ satisfies all three clauses and is therefore the solution of the problem. 
One can rewrite the syndrome constraints of the decoding problem into a set of such Max-SAT clauses~\cite{noormandipour2024}. 

In Ref.~\cite{berent2023decoding}, this has been framed as the so-called lights-out problem, where one has to find a configuration of switches (corresponding to errors) that is able to turn off all lights (corresponding to syndromes). In fact, this lights-out problem can then be seen as a Max-SAT problem. 
%Then, for ease of numerical optimization, the problem is mapped onto clauses of three variables only, which can be achieved using auxiliary variables~\cite{berent2023decoding,noormandipour2024}.
The Max-SAT decoder demonstrates high accuracy, even for the color code, which is usually difficult to decode. Straightforward extensions to the phenomenological noise model and even to circuit-level noise exist. However, the Max-SAT decoder is relatively slow and is therefore more focused on high accuracy.

Recent works have also proposed decoding using the coherent Ising machine. 
In particular, one first maps MLE to a quadratic unconstrained binary optimization (QUBO) problem~\cite{fujisaki2022practical,fujisaki2023quantum,yugo2024ising}. This represents MLE as an energy-optimization problem over binary variables. Notably, the syndrome constraints require the implementation of additional constraints in the QUBO energy problem. This can be done via energy penalty terms, which penalize violations of the syndrome constraints by assigning them a high energy in the QUBO. 
Then, the QUBO is solved using numerical implementations of the coherent Ising machine. 

\subsection{Renormalization group decoders}\label{subsec:decoders-renormalization}
Returning to fast multiscale methods, renormalization-group (RG) decoders are a family of hierarchical decoding algorithms for topological quantum error-correcting codes, with their most notable application being the surface code~\cite{duclos2010fast}. 
Their central idea is to solve the decoding problem coarsely at large scales first, and then refine the solution by recursively zooming in on smaller scales, mimicking the coarse-graining procedures of the renormalization group in statistical physics.

As a basic idea, the renormalization-group decoder partitions the code into small blocks~\cite{duclos2010fast}. Within each block, it computes either a local decoding decision or an effective logical error that represents the block's behavior at a coarse scale.
This block is then replaced by an effective ``super-qubit,'' and the syndrome information is passed to the next, coarser level. The procedure then repeats from finer to coarser scales, combining small regions into larger ones while summarizing their error behavior.
At the top level, the entire lattice is reduced to a very small effective code, which can be decoded exactly with a brute-force decoder.
Then, we need to propagate the correction decision to the original lattice. For this, we reverse the information flow, moving from the coarse to the fine scale, applying the correction operations.
This creates a multiscale description of the error, efficiently capturing correlations.

Especially for topological codes, the renormalization-group decoder exploits two key structural properties. First, there is locality of error propagation: errors spread only to nearby stabilizers, so small regions can often be decoded nearly independently.
Second, topological codes are usually self-similar, and the decoding problem retains the same structure at larger scales.
Thus, instead of dealing with an exponentially large global configuration space in the full decoding problem, the renormalization-group decoder breaks the problem into manageable chunks.
We also note that by including soft information from a belief-propagation decoder, the performance of renormalization-group decoding can be markedly enhanced~\cite{poulin2008iterative}. In particular, performance beyond MWPM has been reported. 
Renormalization-group decoders offer near-linear decoding time and are directly parallelizable, which is essential in practical applications~\cite{duclos2010fast,duclos2013fault}.
As such, they are especially useful when one needs low-latency decoding, for example in real-time error correction for a quantum computer.

%Several advancements over the original proposal~\cite{poulin2008iterative} have been put forward, such as soft-decision renormalization group decoders where belief propagation or soft syndrome information is included. This improves accuracy by avoiding hard commitment to early decoding decisions. These better capture correlated noise and avoid artificial biases from coarse-graining.

%5. Performance and Limitations Advantages

%Fast: near-linear runtime, excellent for hardware-level decoders
%Parallelizable: local operations map to hardware accelerators (FPGA/ASIC)
%Flexible: works beyond Abelian codes and beyond Pauli noise
%Still somewhat suboptimal compared to specialized decoders such as MWPM or tensor-network decoders.

%Hard to tune block rules for correlated noise or biased noise.
%Coarse-graining can introduce approximations that accumulate and slightly depress the threshold.

%Despite these challenges, RG decoders remain a core idea in the landscape of fast, scalable quantum decoders.

%A useful way to summarize the decoder landscape is the following. 
In this tutorial, we can only cover a small subset of the vast landscape of decoders. The large number of decoding approaches originates from the complex tradeoff between accuracy, speed and latency.
Exact or near-exact formulations best capture the underlying maximum-likelihood problem but are usually too slow for real-time use. Matching-based, Union-Find, and renormalization-group decoders are especially attractive for topological codes and low-latency settings, while BP-based methods are central for qLDPC codes. Tensor-network and search-based decoders offer higher accuracy in structured settings, and machine-learning approaches may become particularly appealing when realistic soft information or strongly correlated noise is available. Overall, practical decoding is less about finding a universally best algorithm, and instead about adapting the decoder to the code, the noise, and the quantum hardware constraints.

\newpage

\section{Fundamental bounds in quantum error correction}\label{sec:bounds}

The development of quantum error correction (QEC) is fundamentally a study of trade-offs. To protect fragile quantum information, one must encode it redundantly, using a larger number of physical systems (e.g., qubits) to represent a smaller amount of logical information. This process is characterized by a set of key parameters, most notably the number of physical qubits $n$, the number of encoded logical qubits $k$, and the code distance $d$. The ratio $R = k/n$ is called the \emph{encoding rate} and is a measure of efficiency, while the distance $d$ quantifies the code's error-correcting capability. Before considering the practical constraints imposed by physical hardware, it is essential to understand the absolute theoretical limits that govern these parameters. The bounds discussed in this section apply to any quantum code, regardless of how it might be implemented, and thus represent the ideal performance ceiling.

\subsection{The quantum Hamming bound (sphere-packing limit)}\label{subsec:bounds-quantum-hamming-bound}
The \emph{quantum Hamming bound}, also known as the sphere-packing bound, provides a powerful upper bound on the achievable encoding rate of a QEC code by relating the code's parameters to the geometry of the underlying Hilbert space~\cite{gottesman1996class,ashikhmin1997remarks}. It is a direct quantum analogue of the classical Hamming bound, which constrains classical error-correcting codes through a similar volume-based argument~\cite{hamming1950error}. The central idea behind the quantum Hamming bound is that for a code to be able to unambiguously correct a certain set of errors, the quantum states resulting from those errors must be distinguishable. Specifically, for a \emph{non-degenerate} code, each correctable error operator, when applied to the code's subspace, must map it to a new subspace that is orthogonal to the subspaces generated by all other correctable errors. Here, non-degeneracy means that distinct correctable errors have distinct effects on the codespace, or equivalently produce distinct error syndromes.

Consider the total $2^n$-dimensional Hilbert space of $n$ physical qubits. A QEC code defines a smaller $2^k$-dimensional subspace within the full $2^n$-dimensional Hilbert space, known as the \emph{codespace}, where logical information is stored. An error-correcting code with distance $d$ can correct any error affecting up to $t = \lfloor(d-1)/2\rfloor$ qubits. When one of these correctable errors occurs, the state is perturbed and moved out of the original codespace into a new, error-affected subspace. For the error to be correctable, this new subspace must be uniquely identifiable. The quantum Hamming bound formalizes the constraint that the total ``volume'' of all of these distinct, orthogonal error subspaces cannot exceed the total available volume of the Hilbert space.

For a non-degenerate $\stabcode{n}{k}{d}$ quantum code over qubits, which can correct any error of weight up to $t = \lfloor(d-1)/2\rfloor$, the quantum Hamming bound is expressed as:
\begin{equation}
    2^k \sum_{j=0}^{t} \binom{n}{j} 3^j \le 2^n\,.\label{eqn:quantum-hamming-bound}
\end{equation}
Here, the term $\sum_{j=0}^{t} \binom{n}{j} 3^j$ represents the total number of distinct correctable Pauli errors of weight up to $t$. The binomial coefficient $\binom{n}{j}$ counts the number of ways to choose $j$ qubits out of $n$ for an error to occur. The factor of $3^j$ accounts for the three possible types of non-trivial Pauli errors ($X$, $Y$, or $Z$) that can occur at each of those $j$ locations. The inequality states that the dimension of the codespace, $2^k$, multiplied by the number of distinct orthogonal subspaces it can be mapped to by correctable errors, must be less than or equal to the total dimension of the ambient Hilbert space, $2^n$. 

The proof of the quantum Hamming bound for non-degenerate codes is a direct consequence of the sphere-packing argument. Let $\mathcal{C}$ be the $2^k$-dimensional codespace and let $\{E_a\}$ be the set of correctable Pauli error operators with weight at most $t$. The error correction conditions for a non-degenerate code require that for any two distinct correctable errors $E_a$ and $E_b$, the resulting error spaces $E_a\mathcal{C}$ and $E_b\mathcal{C}$ are mutually orthogonal. Each of these error spaces has the same dimension as the original codespace, namely $2^k$. The total number of such correctable errors (including the identity for the no-error case) is $N_E = \sum_{j=0}^{t} \binom{n}{j} 3^j$. Since all $N_E$ of these $2^k$-dimensional subspaces must fit within the total $2^n$-dimensional Hilbert space as mutually orthogonal subspaces, the sum of their dimensions cannot exceed the total dimension. This directly yields the inequality $N_E \cdot 2^k \le 2^n$, which is the quantum Hamming bound. If a QEC code saturates this bound, it is said to be a \emph{perfect} quantum code. Such codes are exceptionally rare as they represent the most efficient possible packing of error spheres within the Hilbert space. We saw one example of this in~\cref{subsec:stabilizer-five-qubit-perfect-code}, which we highlight again in the following example:

\begin{example}[label={ex:5-1-3-code}]{The $\stabcode{5}{1}{3}$ code}
    The $\stabcode{5}{1}{3}$ code is the canonical example of a perfect quantum code. It encodes $k=1$ logical qubit into $n=5$ physical qubits and has a distance of $d=3$, meaning it can correct any single-qubit error ($t=1$). Plugging these parameters into the quantum Hamming bound in~\cref{eqn:quantum-hamming-bound} gives
    \begin{equation}
        2^1 \left( \binom{5}{0}3^0 + \binom{5}{1}3^1 \right) = 2 \left( 1 + 5 \cdot 3 \right) = 2(16) = 32\,.
    \end{equation}
    Since the dimension of the ambient Hilbert space is $2^5 = 32$, the bound is saturated ($32 \le 32$), confirming that the $\stabcode{5}{1}{3}$ code is perfect. 
\end{example}

On the other hand, the Steane code is \emph{not} a perfect code:

\begin{example}[label={ex:steane-code}]{The $\stabcode{7}{1}{3}$ Steane code}
    The $\stabcode{7}{1}{3}$ code, or Steane code, is another well-known code with $n=7$, $k=1$, and $d=3$ ($t=1$). For this code, the left-hand side of the inequality is
    \begin{equation}
        2^1 \left( \binom{7}{0}3^0 + \binom{7}{1}3^1 \right) = 2 \left( 1 + 7 \cdot 3 \right) = 2(22) = 44\,.
    \end{equation}
    The total dimension of the Hilbert space is $2^7 = 128$. Since $44 < 128$, the Steane code does not saturate the bound and is therefore not a perfect code. This illustrates that there is ``wasted space'' in the Hilbert space that is not utilized for correcting single-qubit errors.
\end{example}

The standard proof of the quantum Hamming bound relies critically on the assumption of non-degeneracy, where each correctable error corresponds to a unique error syndrome and maps the codespace to a distinct orthogonal subspace. A degenerate code is one where multiple distinct physical errors can produce the same error syndrome. This opens up the possibility of a more efficient packing of information, as different error subspaces are no longer required to be orthogonal. This leads to one of the most significant and long-standing open questions in quantum coding theory:
\begin{open}[label={open:hamming-bound}]{}
    Can a degenerate quantum code violate the quantum Hamming bound?
\end{open}
For decades, it was conjectured that all quantum codes must obey this bound. A significant body of work has shown this to be true for many large and important classes of codes. For instance, it has been proven that single and double error-correcting binary stabilizer codes, as well as many types of CSS codes, cannot beat the bound. For example, CSS codes over a prime power alphabet $q \geq 5$ cannot beat the quantum Hamming bound~\cite{sarvepalli2010degenerate}. However, a fully general proof that covers all possible degenerate codes remains elusive. The discovery of a code that provably violates the bound would be a major theoretical breakthrough, suggesting that our understanding of the ultimate limits of information packing in quantum systems is incomplete. While some numerical studies have suggested the possibility of small violations, no definitive counterexample has been found. 

\subsection{The quantum Singleton bound (information-theoretic limit)}\label{subsec:bounds-singleton}
The \emph{quantum Singleton bound} establishes a different kind of limit on the parameters of a quantum error-correcting code~\cite{knill1996threshold,calderbank1997quantum,ashikhmin1997remarks}. Instead of relying on a geometric sphere-packing argument, its origins are deeply information-theoretic, stemming from fundamental principles of quantum mechanics such as the no-cloning theorem. The bound provides a direct trade-off between the amount of redundancy used ($n-k$) and the error-correcting capability ($d$). The core idea is that if a code can correct the erasure of any $d-1$ qubits, it means the full logical information must be recoverable from the remaining $n-(d-1)$ qubits.

The quantum Singleton bound arises from considering the implications of this recoverability on the distribution of quantum information across the physical qubits. For any $\stabcode{n}{k}{d}$ qubit error-correcting code, the quantum Singleton bound is given by the inequality:
\begin{equation}
    n - k \ge 2(d - 1)\,.\label{eqn:quantum-singleton-bound}
\end{equation}
This is stricter than its classical counterpart, which is given by
\begin{equation}
    n_{\mathrm{cl}} - k_{\mathrm{cl}} \ge d_{\mathrm{cl}} - 1\,.\label{eqn:classical-singleton-bound}
\end{equation}
The factor of 2 in the quantum version reflects the need to protect against both bit-flip ($X$) and phase-flip ($Z$) errors.

A valuable insight into the proof of the quantum Singleton bound comes from a no-cloning argument. Consider a code with distance $d$. This implies that if any set of $d-1$ qubits are lost (an erasure error), their state can be perfectly reconstructed by performing a recovery operation on the remaining $n-(d-1)$ qubits. We partition the $n$ physical qubits into three sets: $A$, $B$, and $C$, where $\abs{A} = d-1$ and $\abs{B} = d-1$. The remaining qubits are in set $C$, so $\abs{C} = n - 2(d-1)$. Because the code has distance $d$, the logical information of the $k$ encoded qubits must be fully recoverable from the qubits in sets $B$ and $C$ alone (since the $d-1$ qubits in $A$ could have been erased). Similarly, the logical information must also be fully recoverable from the qubits in sets $A$ and $C$ alone (since the $d-1$ qubits in $B$ could have been erased).

If $k > \abs{C}$, or equivalently $k > n - 2(d-1)$, the subsystem $C$ alone is too small to contain the full $k$-qubit logical state. Thus, the recovery from $B \cup C$ must use information contained in $B$, while the recovery from $A \cup C$ must use information contained in $A$. This would imply that the $k$ qubits of logical information are simultaneously present in the system $B \cup C$ and the system $A \cup C$. Since $A$ and $B$ are disjoint, this would mean that the information has been effectively cloned onto two different physical locations, which is forbidden by the no-cloning theorem. Therefore, to avoid this contradiction, we must have $k \le n - 2(d-1)$, which we can rearrange to give the quantum Singleton bound in~\cref{eqn:quantum-singleton-bound}.

A \emph{quantum maximal distance separable (MDS) code} is a quantum error-correcting code that saturates the quantum Singleton bound, that is, one that satisfies the relation $n - k = 2(d - 1)$. These codes are optimal in the sense that they achieve the maximum possible distance for a given number of physical and logical qubits.

\begin{example}[label={ex:5-1-3-mds-code-4-2-2-mds-code}]{The $\stabcode{5}{1}{3}$ and $\stabcode{4}{2}{2}$ codes}
   The $\stabcode{5}{1}{3}$ code is an MDS code, since $5 - 1 = 4 = 2(3-1)$. The $\stabcode{4}{2}{2}$ code is also MDS, as $4 - 2 = 2 = 2(2-1)$. 
\end{example}

The search for and classification of quantum MDS codes is an active area of research in quantum error correction.

\subsection{The quantum Gilbert-Varshamov bound (existence guarantee)}\label{subsec:quantum-gv-bound}
While the quantum Hamming and Singleton bounds define the upper limits of what is possible in quantum error correction, they do not tell us whether codes approaching these limits actually exist. The quantum Gilbert-Varshamov (GV) bound addresses this by providing a lower bound on the performance of the best possible codes, thereby guaranteeing the existence of ``good'' quantum codes~\cite{calderbank1997quantum,jin2011quantum}. The GV bound is fundamentally a non-constructive existence proof. The core idea is to show, through a counting argument, that the space of all possible quantum codes is too large to be composed entirely of ``bad'' codes (i.e., codes with a distance less than some target value $d$). By demonstrating that the number of potential codes exceeds the number of ways a code can fail to have distance $d$, one proves that at least one code with the desired distance must exist.  

The quantum GV bound for the existence of an $\stabcode{n}{k}{d}$ stabilizer code is often stated as an $\stabcode{n}{k}{d}$ stabilizer code existing if the inequality
\begin{equation}
    \sum_{j=0}^{d-1} \binom{n}{j} 3^j \leq 2^{n-k}
\end{equation}
is satisfied. The proof typically proceeds by a probabilistic method or a direct counting argument within the space of stabilizer codes. One can consider the set of all possible stabilizer groups that define $\llbracket n,k \rrbracket$ codes. Then, one calculates an upper bound on the number of these codes that have a ``bad'' logical operator, that is, a non-trivial logical operator with weight less than $d$. If this number is smaller than the total number of $\llbracket n,k \rrbracket$ codes, it follows that at least one code must exist that has no such low-weight logical operators, and therefore has a distance of at least $d$. This greedy selection of stabilizers is a common way to approach the proof. 

This result is profound because it proves the existence of \emph{asymptotically good quantum codes.} This is via the GV bound in the asymptotic limit, where the number of physical qubits $n$ becomes very large. In this regime, the bound guarantees the existence of code families for which both the rate $R = k/n$ and the relative distance $\delta = d/n$ are non-zero constants. The asymptotic form of the quantum GV bound is often expressed as:
\begin{equation}
    R \ge 1 - H\left(\delta\right) - \delta\log_2{3}\,,
\end{equation}
where $H(p) = -p \log_2{p} - (1-p)\log_2{(1-p)}$ is the binary entropy function. The asymptotically good quantum codes are the theoretical ideal for scalable, fault-tolerant quantum computation, as they offer protection against errors with a constant resource overhead. The GV bound assures us that such powerful codes are not mathematically impossible.

The Hamming and Singleton bounds act as a ``ceiling,'' establishing the absolute best-case performance that any QEC code can hope to achieve. They delineate the impossible, setting firm limits on the trade-off between information density (rate) and robustness (distance). In contrast, the GV bound acts as a ``floor.'' It does not describe the best possible code but rather guarantees a minimum level of performance that is achievable. It assures us that codes with a certain combination of rate and distance must exist. The space between the GV ``floor'' and the Hamming/Singleton ``ceiling'' represents the vast, largely unexplored territory of possible code parameters where optimal codes may lie, and it is where much of the effort of quantum error correction research is directed.

\begin{open}[label={gv}]{}
    Note that the GV bound's proof is non-constructive; it is a powerful statement of existence that offers no explicit recipe for building the codes it promises. This gives rise to one of the foremost challenges in quantum coding theory: the search for explicit and efficient constructions of code families that can achieve the performance promised by the GV bound. 
\end{open}

The ultimate goal is to find constructive families of asymptotically good quantum codes. This quest has been a primary motivator for decades of research and leads directly to the development of sophisticated code families, such as the qLDPC codes that we will discuss in~\cref{sec:qldpc}, which represent the most promising attempts to turn the GV bound's abstract promise into a physical reality.

\subsection{The constraint of geometric locality}\label{subsec:geometric-locality-constraint}
The bounds discussed so far describe a theoretical ideal, a world where any qubit can interact with any other qubit to perform the complex multi-body measurements required for error correction. However, the physical reality of most leading quantum computing platforms is one of severe constraints on connectivity. Interactions are often limited to nearest-neighbor or next-nearest-neighbor qubits on a physical lattice. This constraint of geometric locality fundamentally alters the rules of the game, imposing harsh new trade-offs on code parameters and giving rise to a distinct class of bounds that have profoundly shaped the direction of experimental quantum error correction.

\begin{defbox}[label={def:geometrically-local-code}]{Geometrically local code}
    A quantum code is said to be \emph{geometrically local} in $D$ spatial dimensions if its constituent physical qubits can be arranged on a $D$-dimensional lattice (e.g., a two-dimensional square grid) such that all checks (the operators measured to detect errors) act only on qubits that are physically close to one another. More formally, the qubits involved in any single check measurement must be contained within a ball of a fixed radius, where this radius is a constant that does not scale with the total number of qubits, $n$. 
\end{defbox}

This is a very strong constraint that directly reflects the physical limitations of architectures like superconducting circuits or quantum dots, where fabricating long-range connections is technologically challenging. 

\subsubsection{The Bravyi-Poulin-Terhal bound}\label{subsubsec:bpt-bound}
The Bravyi-Poulin-Terhal (BPT) bound is a landmark result that quantifies the severe limitations imposed by geometric locality in two dimensions~\cite{bravyi2010tradeoffs}. It establishes a trade-off between the number of encoded logical qubits ($k$), the code distance ($d$), and the total number of physical qubits ($n$) for any 2D geometrically local code. For any quantum code defined by geometrically local commuting projectors on a two-dimensional lattice, the BPT bound states:
\begin{equation}
    k d^2 \le cn\,,
\end{equation}
where $c$ is a constant that depends on the specifics of the lattice and the interaction range but is independent of $n$, $k$, or $d$. The implications of the BPT bound are profound:
\begin{enumerate}
    \item If one desires a code with a finite rate ($R = k/n$ is a constant, so $k = \Theta(n)$), then the bound implies $d^2 = \bigOh{1}$, meaning the distance must be a constant. Such a code cannot offer scalable protection against errors as the system size grows.
    \item Conversely, if one desires a scalable distance (e.g., $d$ grows as a power of $n$, like $d = \Theta(\sqrt{n})$), then the bound implies $kn = \bigOh{n}$, which means $k$ must be a constant, $k = \bigOh{1}$. Such a code has a rate that vanishes asymptotically in $n$ ($R \to 0$) and suffers from enormous overhead.
\end{enumerate}
Thus, the BPT bound implies that no two-dimensional geometrically-local code can be asymptotically good. The intuition behind the BPT bound involves ideas from geometry and quantum information theory. One approach to proving it involves a clever partitioning of the two-dimensional lattice. Imagine an $L \times L$ square lattice of physical qubits, so $n=L^2$. The lattice can be divided into $d$ vertical strips, each of width approximately $L/d$. A logical operator must span the lattice to change the topological information, and a distance-$d$ code must have logical operators of weight at least $d$. One can construct a ``smooth'' logical operator that acts on qubits primarily within one of these thin, vertical strips. Using fundamental properties of entanglement entropy, specifically that local quantum operations can only generate an amount of entanglement proportional to the size of the boundary they act across, one can bound the information content ($k$) that can be encoded. The argument essentially shows that the ability to correct errors across a region of size $d$ (the distance) limits the density of logical information ($k/n$) that can be stored, leading directly to the $kd^2 \le \bigOh{n}$ trade-off.

\begin{example}[label={ex:surface-code-bpt-bound}]{The surface code saturates the BPT bound}
     On an $L \times L$ grid, a standard planar surface code uses $n \approx 2L^2$ physical qubits to encode $k=1$ logical qubit with a distance of $d=L$. Plugging these parameters into the BPT bound:
     \begin{equation}
         kd^2 = (1) \cdot L^2 = \bigOh{n}
     \end{equation}
     This shows that the surface code perfectly saturates the BPT bound. It is not an under-performing code; rather, it is an \emph{optimal} code given the stringent constraints of two-dimensional geometric locality.
\end{example}

The BPT bound can be generalized for codes that are geometrically local in $D$ spatial dimensions:
\begin{equation}
    k d^{2/(D-1)} \le \bigOh{n}\,.
\end{equation}
This result, derived in subsequent work, shows that increasing the spatial dimension of the hardware can alleviate the trade-off. For any fixed dimension $D$, it remains impossible to construct an asymptotically good code with only local interactions. For example, in three dimensions ($D=3$), the bound becomes $kd \le \bigOh{n}$. This allows for a linear distance ($d = \Theta(n)$) with a constant number of logical qubits ($k = \bigOh{1}$), or a logarithmic distance with a polylogarithmic number of logical qubits, but not linear scaling for both.

\subsubsection{The Baspin-Krishna bound}\label{subsubsec:baspin-krishna-bound}
The BPT bound applies to codes with strict geometric locality. However, modern hardware and recent code constructions often explore relaxing this locality, such as by allowing a limited number of long-range connections. Baspin and Krishna extended the study of geometric constraints by analyzing the \emph{connectivity graph} associated with a quantum code~\cite{baspin2022connectivity}. Rather than assuming a rigid embedding in a $D$-dimensional Euclidean space, they established trade-offs based on the \emph{graph separator} of the code's connectivity graph, where vertices are qubits and edges denote participation in the same parity check. A separator is a subset of vertices whose removal disconnects the graph into disjoint, roughly equal-sized components. Graphs that can be embedded in low dimensions have small separators, whereas highly connected graphs, like expanders, lack small separators.

The Baspin-Krishna bound demonstrates that the size of the graph separator places a hard upper limit on the achievable dimension $k$ and distance $d$. An interesting conclusion of this work is the necessity of expansion: unless the connectivity graph of a qLDPC code contains an \emph{expander graph}, the code parameters are severely limited. Therefore, an expander graph structure is a necessary condition to construct asymptotically ``good'' qLDPC codes (where $k = \Theta(n)$ and $d = \Theta(n)$). Furthermore, they quantified the ``cost'' of outperforming geometrically local codes~\cite{baspin2022quantifying}. For instance, to build a qLDPC code on a two-dimensional layout that achieves a distance $d \propto n^{1/2 + \epsilon}$ (beating the $O(\sqrt{n})$ BPT limit), the architecture must include $\Omega(n^{1/2 + \epsilon})$ nonlocal interactions of length at least $\Tilde{\Omega}(n^\epsilon)$. This reveals that mildly relaxing locality is insufficient to achieve optimal code parameters; significant, highly nonlocal wiring is fundamentally required.

\subsection{Bounds from optimization theory}\label{subsec:lp-sdp-bounds}
While analytical bounds like the quantum Hamming and Singleton bounds provide fundamental limits on code parameters, they often leave a significant gap where it is unknown if codes exist. Tighter bounds, especially for specific, finite block lengths, can be obtained numerically using optimization theory, specifically linear programming (LP) and semidefinite programming (SDP). These bounds treat the search for a quantum code as an optimization problem.

\subsubsection{Weight enumerators and the MacWilliams identity}\label{subsubsec:weight-enumerators}
The foundation of the LP bound is the \emph{quantum weight enumerator}, which generalizes concepts from classical coding theory~\cite{shor1997quantum,rains1998quantum}. For a quantum error-correcting code with code projector $\Pi$ encoding $K = 2^k$ logical states in $n$ physical qubits, we define the \emph{weight enumerator} coefficients $A_j$ and the \emph{dual weight enumerator} coefficients $B_j$ as:
\begin{align}
    A_j &= \frac{1}{K^2} \sum_{P \in \hat{\mathcal{P}}_n \mid \text{wt}(P)=j} |\text{Tr}(P\Pi)|^2\,, \\
    B_j &= \frac{1}{K} \sum_{P \in \hat{\mathcal{P}}_n \mid \text{wt}(P)=j} \text{Tr}(P\Pi P^\dagger \Pi)\,,
\end{align}
where $\hat{\mathcal{P}}_n$ is the $n$-qubit Pauli group (ignoring overall phases) and $\text{wt}(P)$ is the weight of Pauli operator $P$ (i.e., the number of non-identity terms). For a stabilizer code with stabilizer $\mathcal{S}$, $A_j$ simply counts the number of elements in $\mathcal{S}$ of weight $j$, and $B_j$ counts the number of normalizer (logical) operators $N(\mathcal{S})$ of weight $j$.

These coefficients are encapsulated in the polynomials $A(x) = \sum_{j=0}^n A_j x^j$ and $B(x) = \sum_{j=0}^n B_j x^j$. Shor and Laflamme showed~\cite{shor1997quantum} that they are intimately related by the \emph{quantum MacWilliams identity}:
\begin{equation}
    B(x) = \frac{K}{2^n}(1+3x)^n A\left(\frac{1-x}{1+3x}\right)\,.
\end{equation}
Additionally, Rains introduced the \emph{quantum shadow enumerator}~\cite{rains1999quantum} $Sh(x) = \sum_{j=0}^n Sh_j x^j$, where the coefficients $Sh_j$ are defined as:
\begin{equation}
    Sh_j = \frac{1}{K} \sum_{P \in \hat{\mathcal{P}}_n \mid \text{wt}(P)=j} \text{Tr}(P\Pi P^\dagger Y^{\otimes n} \Pi^* Y^{\otimes n})\,,
\end{equation}
with $\Pi^*$ being the complex conjugate of the projector $\Pi$. The shadow enumerator relates to $A(x)$ via a similar identity:
\begin{equation}
    Sh(x) = \frac{K}{2^n}(1+3x)^n A\left(\frac{x-1}{1+3x}\right)\,.
\end{equation}

\subsubsection{The linear programming bound}\label{subsubsec:linear-programming-bound}
The existence of a quantum code imposes strict inequalities on these enumerators, giving the quantum linear programming (LP) bound~\cite{rains1998quantum,rains2002monotonicity}. Specifically, any valid $((n,K,d))$ QEC code must satisfy the following:
\begin{enumerate}
    \item $A_0 = B_0 = 1$, that is, there is exactly one Pauli of weight 0: the identity.
    \item $A_j \ge 0$, $B_j \ge 0$, and $Sh_j \ge 0$ for all $j \in \{0, \dots, n\}$.
    \item $B_j - A_j \ge 0$ for all $j$, that is, undetectable errors are elements of the normalizer that are not in the stabilizer.
    \item $A_j = B_j$ for $1 \le j < d$, or the error correction condition: a code of distance $d$ has no undetectable errors of weight less than $d$.
\end{enumerate}
To find an upper bound on $K$ for a given $n$ and $d$, we treat the coefficients $A_j$, $B_j$, and $Sh_j$ as unknown variables in an LP. We seek to check if there is any feasible assignment of $A_j$ that satisfies all the linear equality and inequality constraints simultaneously, paired with the MacWilliams and shadow identity mappings. If the LP is infeasible for a target set of parameters $((n, K, d))$, then no such quantum code can possibly exist. For instance, LP constraints elegantly prove the non-existence of a $((3, 2, 2))$ quantum code. Writing down the MacWilliams relations for $n=3$ and setting $A_0=B_0=1$ and $A_1=B_1$ inevitably forces $Sh_0 < 0$, which strictly violates the $Sh_j \ge 0$ constraint.

\subsubsection{Semidefinite programming bounds}\label{subsubsec:sdp-bounds}
While LP bounds effectively rule out many parameter regimes by enforcing linear constraints on weight enumerators, they provide only necessary conditions. Semidefinite programming (SDP) bounds~\cite{munne2024sdp} generalize the LP approach by enforcing the global positive semidefiniteness of underlying (moment) matrices, yielding strictly tighter limits. By turning the search for quantum codes into a hierarchy of feasibility SDPs, the SDP framework has successfully provided infeasibility certificates proving that codes with parameters such as $((8, 9, 3))_2$ and $((10, 5, 4))_2$ do not exist, cases that traditional LP bounds enhanced with shadow inequalities could not resolve~\cite{munne2024sdp}. 

\newpage

\section{Quantum low-density parity-check codes}\label{sec:qldpc}

We now introduce \emph{low-density parity-check} (LDPC) codes, a promising class of quantum error-correcting codes. The development of quantum low-density parity-check (qLDPC) codes is deeply rooted in the principles of classical error correction. In his seminal 1960 doctoral thesis~\cite{gallager1960low}, Robert Gallager introduced a class of codes that, while initially overlooked, would eventually revolutionize digital communication. These were the LDPC codes. Gallager showed that randomly generated LDPC codes could, with high probability, achieve the Gilbert–Varshamov bound, indicating excellent error-correction capabilities. Gallager's code construction did not receive much attention until 1982, when Michael Tanner introduced~\cite{tanner1981recursive} the use of bipartite graphs (now called \emph{Tanner graphs}) as a useful tool for analyzing LDPC codes. Then, over the next decade or so, LDPC codes underwent further development, led by researchers such as David MacKay and others. During this time period, several LDPC code constructions were introduced, such as the high-performing turbo codes, introduced in 1993~\cite{berrou1993near}. Today, classical LDPC codes play a vital role in technology all around us, such as Wi-Fi for wireless Internet connections, DVB-T for television broadcasting, and 5G for cellular communication.

\subsection{The LDPC code construction}\label{subsec:ldpc-construction}
We start by defining \emph{classical} LDPC codes, as LDPC codes started out in the classical domain before being extended to quantum codes. A classical LDPC code is a binary\footnote{We assume in this chapter that we are working with bits and qubits, though there are constructions for $q$-ary classical and qLDPC codes.} linear code $\mathcal{C} \subseteq \FF_2^n$ that admits a sparse \emph{parity-check matrix}, which we denote by $H \in \FF_2^{m \times n}$. This means that for a family of LDPC codes $\{\mathcal{C}_n\}$, the number of 1's in any given row or column of $H$ is small and, crucially, bounded by a constant that does not grow with the code's block length $n$. If we denote by $x$ a codeword of the code, then $x$ lives in the kernel of $H$. That is, the code is given by
\begin{equation}
    \mathcal{C} = \{x \in \FF_2^n : Hx = 0\}\,.
\end{equation}
Another way to think of classical LDPC codes is that the number of bits involved in each check and the number of checks acting on a given bit are both bound by a constant for all members of the code family.

With this in mind, we introduce qLDPC codes. Much of the early work on qLDPC codes was done by Daniel Gottesman~\cite{gottesman2014faulttolerantquantumcomputationconstant} and later expanded upon in a seminal work~\cite{breuckmann2021quantum} by Breuckmann and Eberhardt in 2021. A quantum error-correcting code is a qLDPC code if it satisfies the following two conditions:
\begin{enumerate}
    \item The weight of each check (i.e., the number of qubits it acts on) is bounded by a constant, $w$.
    \item The degree of each qubit (i.e., the number of checks it participates in) is bounded by a constant, $l$.
\end{enumerate}
Since $w$ and $l$ are constants, they do not grow as the total code size $n$ increases. More formally, we have the following:

\begin{defbox}[label={def:quantum-ldpc-code}]{Quantum low-density parity-check (qLDPC) code}
    An $\llbracket n, k \rrbracket$ stabilizer code is an $(r,c)$-LDPC code if there exists a choice of generators $\{M_1, \ldots, M_{n-k}\}$ for the stabilizer $Q$ of the code such that $\wt{M_i} \leq r$ for all $i$ and the number of generators that act nontrivially (i.e., as $X$, $Y$, or $Z$) on qubit $j$ is at most $c$ for all $j \in [n]$.
\end{defbox}

Based on this definition, every $\llbracket n, k \rrbracket$ stabilizer code is automatically an $(n, n-k)$-LDPC code, but we are usually interested in small values of $r$ and $c$. Furthermore, we are often interested in the so-called "good LDPC codes," which meet the following criteria:
\begin{itemize}
    \item \textbf{Constant rate}: $\frac{k}{n} \to R > 0$ as $n \to \infty$,
    \item \textbf{Linear distance}: $\frac{d}{n} \to \delta > 0$ as $n \to \infty$.
\end{itemize}
Thus, we want $k = Rn$ and $d = \delta n$, but recall from the Bravyi-Poulin-Terhal bound that we discussed in~\cref{subsubsec:bpt-bound} that, for local commuting projector codes (which includes stabilizer codes and thus qLDPC codes), on a $D$-dimensional lattice, we have
\begin{equation}
    kd^{2/(D-1)} = \bigOh{n}\,.
\end{equation}
If we consider a $D=2$-dimensional lattice and take $k = Rn$ and $d = \delta n$, then plugging these values in, we find
\begin{equation}
    kd^2 = \bigOh{n^3} \neq \bigOh{n}\,,
\end{equation}
thus violating the bound. Thus, we cannot take $k$ and $d$ scaling linearly with $n$. In the classical setting, a random parity-check matrix immediately gives us a good LDPC code, but this is not necessarily the case for qLDPC codes. Thus, much of current QEC research into qLDPC codes focuses on finding code constructions that give good LDPC codes, using tools from various fields of mathematics and computer science. 

However, note that the BPT bound only applies when we restrict our system to local interactions; if we allow for nonlocal interactions, we may be able to do better than the BPT bound. The natural question is then to ask how much nonlocality is needed to allow for codes that do better than the BPT bound allows in local settings. Dai and Li answered this question in 2024~\cite{dai2025locality}, where they showed that $\bigOmega{\max\{k,d\}}$ interactions of length $\bigOmega{\max\{d/\sqrt{n}, (kd^2/n)^{1/4}\}}$ are needed to beat the BPT bound in a 2D embedding.

For practical implementation on physical hardware, it is often desired to arrange the qubits in a physical lattice, where qubits can only interact with their nearest neighbor on the lattice. However, in qLDPC codes, the checks are often allowed to connect qubits which are far apart, as long as the overall connection graph remains sparse. All geometrically local codes are, by definition, qLDPC codes, but the converse is not true. As such, this is a central theme of modern QEC research: while geometric locality enforces the severe constraints from the BPT bound that we saw above, the weaker constraint of nonlocal sparsity may still allow for the construction of highly performant codes. In what follows, we will discuss qLDPC codes that may potentially violate the BPT bound.

\subsection{The hypergraph product construction}\label{subsec:ldpc-hgp}
We first consider the \emph{hypergraph product construction}, which yields \emph{hypergraph product} (HGP) codes, initially introduced by Tillich and Z\'{e}mor~\cite{tillich2014quantum}. HGP codes have a fixed nonzero rate and distance that grows with the square root of the number of physical qubits $n$. The hypergraph product takes the tensor product of two classical code chain complexes to build a new, valid quantum chain complex. Given two classical "seed" codes $C_1 = \ker{H_1}$ and $C_2 = \ker{H_2}$ for parity-check matrices $H_1$ (of size $r_1 \times n_1$) and $H_2$ (of size $r_2 \times n_2$), respectively, we obtain a quantum CSS code with the following $X$- and $Z$-type parity-check matrices:
\begin{align}
    H_X &=
    \begin{pmatrix}
        H_1 \otimes I_{n_2} & I_{r_1} \otimes H_2^\intercal
    \end{pmatrix}\,,\\
    H_Z &=
    \begin{pmatrix}
        I_{n_1} \otimes H_2 & H_1^\intercal \otimes I_{r_2}
    \end{pmatrix}\,,
\end{align}
where $I_m$ is the $m \times m$ identity matrix. The construction guarantees that the CSS condition is satisfied:
\begin{align}
    H_X H_Z^\intercal &= (H_1 \otimes I_{n_2})(I_{n_1} \otimes H_2)^\intercal + (I_{r_1} \otimes H_2^\intercal)(H_1^\intercal \otimes I_{r_2})^\intercal\\
    &= H_1 \otimes H_2^\intercal + H_1 \otimes H_2^\intercal\\
    &= 0 \pmod{2}\,.
\end{align}
If the input matrices $H_1$ and $H_2$ are sparse, the resulting $H_X$ and $H_Z$ are also sparse, thus producing a qLDPC code.

The parameters of the resulting $\stabcode{n}{k}{d}$ quantum code are determined by the parameters of the classical seed codes $C_1$ and $C_2$ used to generate the HGP code, $[n_1, k_1, d_1]$ and $[n_2, k_2, d_2]$. The number of physical qubits in the HGP code is
\begin{equation}
    n = n_1n_2 + r_1r_2\,,
\end{equation}
the number of logical qubits is
\begin{equation}
    k = k_1k_2 + k_1^\intercal k_2^\intercal\,,
\end{equation}
where $k_i^\intercal$ is the dimension of the transposed code $C_i^\intercal$ of $C_i$, and the distance is lower bounded by
\begin{equation}
    d \geq \min\{d_1, d_2, d_1^\intercal, d_2^\intercal\}\,,
\end{equation}
where $d_i^\intercal$ is the distance of the transposed code $C_i^\intercal$ of $C_i$ (where $d_i^\intercal = \infty$ if $k_i^\intercal = 0$). We now look at the toric code through this construction.

\begin{example}[label={ex:toric-code}]{Toric code}
    Let $C_1 = C_2 = C_{\mathrm{rep}}$ be the $[L, 1, L]$ classical repetition code. This code has codewords $\{00 \ldots 0, 11 \ldots 1\}$ and its parity-check matrix, assuming periodic boundary conditions, is an $L \times L$ circulant matrix:
    \begin{equation}
        H_{\mathrm{rep}} =
        \begin{pmatrix}
            1 & 1 & 0 & \dots & 0\\
            0 & 1 & 1 & \dots & 0\\
            \vdots & & \ddots & & \vdots\\
            0 & \dots & 0 & 1 & 1\\
            1 & 0 & \dots & 0 & 1
        \end{pmatrix}\,.
    \end{equation}
    For this code, $n_1 = n_2 = r_1 = r_2 = L$, $k_1 = k_2 = 1$, and $d_1 = d_2 = L$. The transposed code is also a repetition code, so $k_1^\intercal = k_2^\intercal = 1$ and $d_1^\intercal = d_2^\intercal = L$. Applying the hypergraph product formulae from above yield a quantum code with parameters:
    \begin{enumerate}
        \item \textbf{Physical qubits}: $n = L \cdot L + L \cdot L = 2L^2$\,,
        \item \textbf{Logical qubits}: $k = 1 \cdot 1 + 1 \cdot 1 = 2$\,,
        \item \textbf{Distance}:  $d = \min\{L, L, L, L\} = L$\,.
    \end{enumerate}
    Thus, as expected, we see that the toric code, when analyzed using the HGP construction, is a $\stabcode{2L^2}{2}{L}$ code.
\end{example}

The HGP construction elegantly solves the central challenge of ensuring the CSS commutativity condition, $H_X H_Z^\intercal = 0$, by leveraging the mathematical structure of homological algebra, which we introduced in~\cref{subsubsec:topological-homology}. A CSS code can be viewed as a chain complex, which is a sequence of vector spaces connected by linear maps (boundary operators, $\partial$) such that the composition of any two consecutive maps is zero: $\partial_i \circ \partial_{i+1} = 0$. Take the following chain complex:
\begin{equation}
    \cdots \xrightarrow{\partial_{j+1}} \mathcal{A}_j \xrightarrow{\partial_j} \mathcal{A}_{j-1} \xrightarrow{\partial_{j-1}} \cdots \xrightarrow{\partial_1} \mathcal{A}_0 \xrightarrow{\partial_0=0} 0\,,
\end{equation}
where each $\mathcal{A}_j$ is a vector space over $\FF_q$ and $\partial_j : \mathcal{A}_j \to \mathcal{A}_{j-1}$ is a boundary operator. Recall also the fundamental lemma of homology, which states that $\partial_j \circ \partial_{j+1} = 0$, that is, the boundary of a boundary is zero. Then, the homology group is
\begin{equation}
    H_j(\mathcal{A}_j) = \ker{\partial_j} / \im{\partial_{j+1}}\,.
\end{equation}
Using this, we can take a tensor product of chain complexes. That is, given chain complexes $(\mathcal{A}, \partial^{\mathcal{A}})$ and $(\mathcal{B}, \partial^{\mathcal{B}})$, if we define the tensor product
\begin{equation}
    \mathcal{C} = \mathcal{A} \otimes \mathcal{B}\,,
\end{equation}
then
\begin{equation}
    \mathcal{C}_k = \bigoplus_{i+j=k}{\mathcal{A}_i \otimes_{\FF_q} \mathcal{B}_j}
\end{equation}
and
\begin{equation}
    \partial_k^{\mathcal{C}} : \mathcal{C}_{k} \to \mathcal{C}_{k-1}\,.
\end{equation}
We can define the boundary operators as follows. Let $a \in \mathcal{A}_j$ and $b \in \mathcal{B}_j$. Then,
\begin{equation}
    \partial_k^{\mathcal{C}}(a \otimes b) = (\partial^{\mathcal{A}}a) \otimes b + (-1)^{i}a \otimes \partial^{\mathcal{B}}b\,,
\end{equation}
where we get the negative sign from the \emph{Koszul sign rule}. Thus, we take a classical code to get a length-2 chain complex $C : C_1 \xrightarrow{\partial_C} C_0$, where $\partial_C$ represents the parity-check matrix $H$ of the code. If we take two such codes over $\FF_q$, $C : C_1 \xrightarrow{\partial_C} C_0$ and $D_1 : \xrightarrow{\partial_D} D_0$, then we can build the following chain complex:
\begin{align}
    (C \otimes D)_2 &= C_1 \otimes D_1\,,\\
    (C \otimes D)_1 &= C_1 \otimes D_0 \oplus C_0 \otimes D_1\,,\\
    (C \otimes D)_0 &= C_0 \otimes D_0\,,
\end{align}
where we can use the Koszul sign rule to construct the boundary operators as we did above. We will find two such boundary operators, one of which represents the $X$-type parity-check matrix $H_X$ and one that represents the $Z$-type parity-check matrix $H_Z$.

\subsection{LDPC codes from algebraic constructions}\label{sec:ldpc-algebraic}
While we analyzed the HGP codes from the perspective of homological algebra, we can also find codes using polynomials. In this section, we consider one promising and popular code construction, the \emph{bivariate bicycle codes}, to see how we can construct codes from polynomials. While these codes do not yield parameters that are as good as other constructions asymptotically, they are mathematically easier to work with and give good, practical, and experimentally realizable codes for realistic values of $n$.

Bivariate bicycle codes were popularized in a recent paper from IBM~\cite{bravyi2024high}, though they are a specific instance of a more general class of codes, namely the two-block group algebra codes~\cite{lin2023quantum}. To define a bivariate bicycle code, we define two polynomials,
\begin{equation}
    \begin{aligned}
        A &= A_1 + A_2 + A_3\,,\\
        B &= B_1 + B_2 + B_3\,,
    \end{aligned}
\end{equation}
where $A_i$ and $B_i$ are powers of the commuting matrices $x = S_\ell \otimes I_m$ and $y = I_\ell \otimes S_m$ for some integers $\ell$ and $m$. $I_m$ is, as usual, the $m \times m$ identity matrix and $S_\ell$ is the $\ell \times \ell$ \emph{cyclic shift matrix}, which is such that the $i\numth$ row has one element equal to 1 in column $(i+1) \pmod{\ell}$. This is easiest to understand by looking at the first two cyclic shift matrices:
\begin{equation}
    S_2 =
    \begin{pmatrix}
        0 & 1\\
        1 & 0
    \end{pmatrix}
    \quad \text{and} \quad
    S_3 =
    \begin{pmatrix}
        0 & 1 & 0\\
        0 & 0 & 1\\
        1 & 0 & 0
    \end{pmatrix}\,,
\end{equation}
and so on. In the polynomials $A$ and $B$, we choose $A_i$ and $B_i$ such that each term is unique. That is, $A_i \neq A_j$ and $B_i \neq B_j$ for $i \neq j$ and $A_i \neq B_j$ for all $i$ and $j$. There are a few useful properties of the objects we have introduced here to define the code. First, note that $x$ and $y$ commute:
\begin{align}
    xy &= (S_\ell \otimes I_m)(I_\ell \otimes S_m) = S_\ell \otimes S_m\,,\\
    yx &= (I_\ell \otimes S_m)(S_\ell \otimes S_m) = S_\ell \otimes S_m\,,\\
    \implies xy &= yx\,.
\end{align}
Also, note that $x^\ell = y^m = I_{\ell m}$ and, by extension, $A$ and $B$ commute.

\begin{exerbox}[label={exer:bb-code-x-y-}]{Cyclic matrix properties}
    \begin{enumerate}[label=(\alph*)]
        \item Check by explicit calculation that $x^\ell = y^m = I_{\ell m}$.
        \item Show that $A$ and $B$ commute.
    \end{enumerate}
\end{exerbox}

With $A$ and $B$ in hand, we can define the parity-check matrices of the bivariate bicycle code. As these codes are a type of two-block group algebra code, we see that structure reflected in the parity-check matrices:
\begin{equation}
    H_X = [A \mid B]\quad \text{and} \quad H_Z = [B^\intercal \mid A^\intercal]\,.
\end{equation}
We can check that the CSS condition holds, as we would expect:
\begin{align}
    H_X H_Z^\intercal &= AB + BA\\
    &= AB + AB\\
    &= 0\,,
\end{align}
where the arithmetic is done modulo 2. Given all of this, we find the code parameters to be the following:

\begin{thmbox}[label={thm:bb-code-parameters}]{Bivariate bicycle code parameters}
    The matrices $A, B \in \FF_2^{\ell m \times \ell m}$ defined above give a bivariate bicycle code with parameters $\stabcode{n}{k}{d}$, where
    \begin{align}
        n &= 2\ell m\,,\\
        k &= 2\dim{(\ker{A} \cap \ker{B})}\,,\\
        d &= \min\{d_X, d_Z\}\,,
    \end{align}
    where the $X$ and $Z$ distances are given as
    \begin{align}
        d_X &= \min\{\abs{v} : v \in \ker{H_Z} \backslash \mathrm{rs}(H_X)\}\,,\\
        d_Z &= \min\{\abs{v} : v \in \ker{H_X} \backslash \mathrm{rs}(H_Z)\}\,,
    \end{align}
    where $\mathrm{rs}(\cdot)$ is the row space and where $\abs{v}$ is the Hamming weight of the vector $v \in \FF_2^n$. In particular, $d_X = d_Z$, so $d = d_X = d_Z$.
\end{thmbox}

We do not prove the code parameters here, though interested readers can find the proof in Ref.~\cite{bravyi2024high}. We highlight that, even though these codes do not beat the square-root barrier, they are easy to work with numerically to search for new codes and well-designed for near-term hardware~\cite{voss2025multivariate}. As such, bivariate bicycle codes may be one error-correcting code that bring the field closer to large-scale, error corrected quantum devices.

\subsection{\texorpdfstring{Breaking the $\sqrt{n}$ barrier: advanced constructions}{Breaking the square root barrier: advanced constructions}}
For many years, qLDPC codes could not admit distances that scaled better than $O(\sqrt{n})$. Then, researchers introduced several new code constructions that managed to finally break through this barrier. We discuss two such codes in this section: fiber bundle codes and lifted product codes.

\begin{explbox}[label={box:quantum-nlts-conjecture}]{Quantum NLTS conjecture}
    The question of going beyond the square-root barrier has implications beyond quantum error correction. For example, the quantum \emph{no low energy trivial state} (NLTS) conjecture, now a theorem, was proven using properties of good LDPC codes. The quantum NLTS conjecture is related to the \emph{quantum PCP conjecture}, which is one of the biggest open questions in quantum complexity theory. It states that if there is a family $\{H_n, \alpha(n), \beta(n)\}$ of Hamiltonians such that $\beta(n) - \alpha(n) = \bigOh{1}$, where $H_n$ is a local Hamiltonian with $\poly(n)$ terms, then deciding if the ground state energy of $H_n$ is less than or equal to $\alpha(n)$ or greater than or equal to $\beta(n)$ is \QMA-hard.\vspace{2.5mm}

    This statement implies the NLTS theorem, which says that there exists a fixed constant $\epsilon > 0$ and an explicit family of $\ell$-local Hamiltonians $\vb{H}$ for an infinite set of integer values $n$ in which $\vb{H}$ acts on $n$ particles, consists of $m = \bigTheta{n}$ local terms, such that for any family of states $\psi$ satisfying
    \begin{equation}
        \Tr[\vb{H}\psi] \leq \epsilon m + \lambda_{\min}(\vb{H})\,,
    \end{equation}
    the minimum depth of any quantum circuit generating $\psi$ grows faster than any constant. Quantum PCP implies NLTS, so NLTS is a necessary condition for the quantum PCP conjecture to be true. In particular, if the NLTS theorem is false and the quantum PCP conjecture is true, then this implies that $\NP = \QMA$, which is believed unlikely to be true.\vspace{2.5mm}
    
    However, when good qLDPC codes were introduced by Panteleev and Kalachev~\cite{panteleev2022asymptotically,panteleev2022quantum}, the quantum NLTS theorem was able to be proven by Anshu, Breuckmann, and Nirkhe~\cite{anshu2023nlts}. The proof for this theorem is beyond the scope of this tutorial, but very briefly, the existence of good LDPC codes provided families of codes that have enough nonlocal structure to construct an NLTS Hamiltonian. This then allowed Anshu, Breuckmann, and Nirkhe to prove the theorem. Proving the quantum NLTS theorem is a necessary, but not sufficient, condition for the quantum PCP conjecture to be true, so while progress has been made towards proving this, there still remains some work to be done.
\end{explbox}

\subsubsection{Fiber bundle codes}\label{subsubsec:qldpc-fiber-bundle}
The first qLDPC code that was introduced to break through the square root distance barrier was introduced by Hastings, Haah, and O'Donnell~\cite{hastings2021fiber} and is based on \emph{fiber bundles}. \eczoo[Fiber bundle codes]{fiber_bundle} achieve a distance of $\Omega\lp \frac{n^{3/5}}{\polylog(n)} \rp$ and a code dimension of $\Tilde{\Theta}(n^{3/5})$ on $n$ physical qubits. We will see below how these code parameters are derived, but we begin by taking a brief but fun mathematical detour to understand fiber bundles. Then, we use what we learn to develop the formalism of fiber bundle codes.

\begin{figure}[t]
    \centering
    \subfloat[]{\includegraphics[scale=0.5]{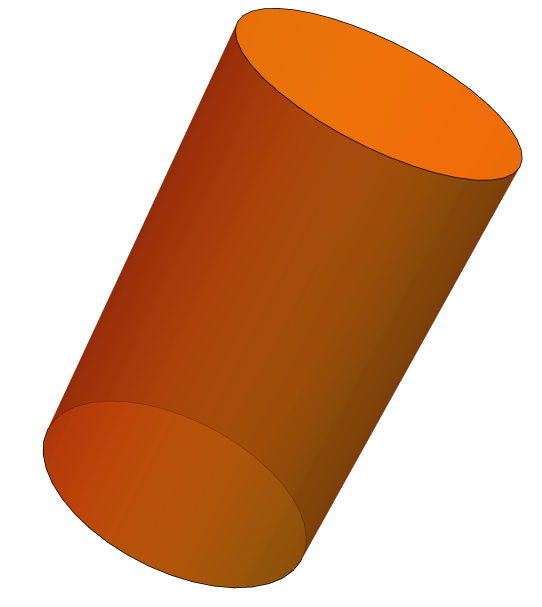}\label{subfig:cylinder}}
    \hspace{15mm}
    \subfloat[]{\includegraphics[scale=0.5]{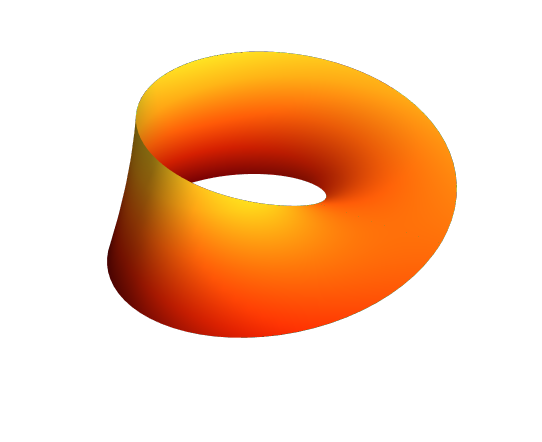}}
    \caption{(a) A cylinder (a trivial fiber bundle) and (b) a M\"{o}bius strip (a simple but nontrivial fiber bundle).}
    \label{fig:mobius-strip-cylinder}
\end{figure}

Fiber bundle theory is a branch of topology that is of great interest beyond just error correction. Informally, a fiber bundle is a space that \emph{locally} looks like a product space, but \emph{globally} is not necessarily a product space, that is, it can have twists. The classic comparison is between a cylinder and a M\"{o}bius strip, where they both locally look the same (an interval $\times$ a line segment) but the M\"{o}bius strip, of course, has a twist; see~\cref{fig:mobius-strip-cylinder}.

We now define things more carefully. We start by defining fiber bundles themselves:

\begin{defbox}[label={def:fiber-bundle}]{Fiber bundle}
    A \emph{fiber bundle} is a set of four objects $(E,B,\pi,F)$, usually written as
    \begin{equation}
        F \hookrightarrow E \xrightarrow{\pi} B\,,
    \end{equation}
    where
    \begin{itemize}
        \item $B$ is the \emph{base space}
        \item $E$ is the \emph{total space}
        \item $\pi : E \to B$ is a \emph{projection}
        \item $F$ is a \emph{fiber}.
    \end{itemize}
\end{defbox}

This is rather abstract, so we walk through this somewhat intuitively, using a cylinder as a guiding trivial fiber bundle to help us visualize what's going on. Here, the base space $B$ is the underlying space on which we are moving. For a cylinder, the base space is taken to be a circle, which we denote by $S^1$ (i.e., a 1D sphere). Then, the fiber $F$ is something that we attach to each point in the base space. This can be a set, vector space, group, or some other mathematical object. In our cylinder mental model, the fiber is taken to be a line segment or, often, the real line $\RR$. Think of us dropping a line from above down to a point on the circle; these lines are the fibers (since they are lines in this case, they really are like fibers!). Then, the total space is the \emph{product} of the base space and the fibers, that is, the combined set of points on the circle and the corresponding fibers. Mathematically, we write this as $S = B \times F$, but we took $B = S^1$ and $F = \RR$, so a cylinder can, in fiber bundle language, be written as $S_{\text{cylinder}} = S^1 \times \RR$. Finally, the map $\pi : E \to B$ tells us which point in the base space a given fiber sits over. Equivalently, given a base point $b \in B$, the map $\pi^{-1}(b)$ gives us a fiber that looks like $F$.

We now add a ``twist'' to go from a trivial fiber bundle to something nontrivial, like a M\"{o}bius strip. First, note that if we zoom into a small region of the base space, call it $U \subseteq B$, then the bundle looks like
\begin{equation}
    \pi^{-1}(U) \cong U \times F\,,
\end{equation}
that is, a local product space. Now for the twist. Imagine we cover the base space with many overlapping patches $U, V, \ldots$. Then, for an overlap region $U \cap V$, we may have two different local coordinate systems for the fiber. We relate these via a \emph{gluing map}:
\begin{equation}
    f \mapsto g_{UV}(x)f\,,
\end{equation}
where $x \in U \cap V$ and $g_{UV}(x)$ is some symmetry of the fiber. For example, this could be a vector space or a group. For the trivial bundle, the map is chosen to do nothing. For a nontrivial bundle, such as the M\"{o}bius strip, though, the map adds a twist, such that if we go around a loop in the base space, we accumulate a nontrivial transformation. Thus, continuing with this M\"{o}bius strip example for a moment, the base space is $B = S^1$ and the fiber $F$ is a line segment, so that locally everything looks like a cylinder. However, globally, when we go once around the circle, the fiber flips, that is, $f \mapsto -f$. This flip is exactly the ``twist'' we have been talking about. Thus, the M\"{o}bius strip is fundamentally a different mathematical object than a cylinder, even though it is described locally in the same way. Other examples of fiber bundles include the tangent bundle and the Hopf fibration, each of which is important in areas of mathematics and physics. We won't describe those here, but there are abundant resources for interested readers to learn more about these and other fiber bundles.

Now that we have some intuition for what fiber bundles are and how they connect to familiar objects like cylinders and M\"{o}bius strips, we dive into some specifics that will help us develop fiber bundle error-correcting codes. We start by connecting the base space, fiber, and bundle to chain complexes, which we introduced back in~\cref{subsubsec:topological-homology}. We mostly follow the presentation in Ref.~\cite{hastings2021fiber}. Let $B$ be a base space and $F$ be a fiber, as above. Then, we can construct the bundle $E$ by taking tensor products of the component chain vector spaces. Note that these vector spaces are taken over $\FF_2$ since we will be working with qubits; analogously defined vector spaces over $\FF_q$ for $q > 2$ exist for qudit fiber bundle codes~\cite{spencer2025qudit}. An example of these chain vector spaces like the following:
\begin{equation}
    \begin{gathered}
    \xymatrix{
    \cdots \ar[d] &\ar[l] \cdots \ar[d]& B_\mu \otimes F_\nu \ar[d]^{I \otimes \partial_\nu}\ar[l]_{\partial_\mu \otimes I}\\
    B_0 \otimes F_1 \ar[d]^{I \otimes \partial_1} &\ar[l]_{\partial_1 \otimes I} B_1 \otimes F_1 \ar[d]^{I \otimes \partial_1}& \vdots \ar[l] \ar[d]\\
    B_0 \otimes F_0 &\ar[l]_{\partial_1 \otimes I} B_1 \otimes F_0 & \vdots \ar[l]
    }
    \end{gathered}
\end{equation}
Here, each $B_\mu$ and $F_\nu$ is a space and we can connect different products of spaces using the boundary operators $\partial_\mu \otimes I$ and $I \otimes \partial_\nu$. We define the chain space $E_\rho$ of the bundle as the following direct sum:
\begin{equation}
    E_\rho = \bigoplus_{\mu + \nu = \rho}{E_{\mu,\nu}}\,,
\end{equation}
where $E_{\mu,\nu} = B_\mu \otimes F_\nu$ for $\rho \geq 0$. Then, we define the boundary map for each $E_{\mu,\nu}$ as
\begin{equation}
    \partial_\rho^E\eval_{(\mu,\nu)} = I \otimes \partial_\nu^F + \partial_\mu^B \otimes I\,.
\end{equation}
All of this was for non-twisted bundles, that is, the trivial bundles that we considered above. To extend this to twisted bundles, we first assume that the fiber has an automorphism group $G$, which is a collection of permutation actions on a set of $\nu$-cells such that the boundary operator commutes with the permutation. That is, for each $g \in G$ and $f^\nu \in F_\nu$, we have
\begin{equation}
    g\partial f^\nu = \partial gf^\nu\,.\label{eqn:fiber-automorphism}
\end{equation}
Then, we can define a \emph{twisted boundary map} as follows:

\begin{defbox}[label={def:twisted-boundary-map}]{Twisted boundary map}
    Given a fiber automorphism $G$ as in~\cref{eqn:fiber-automorphism}, a \emph{connection} $\varphi$ of a bundle is an arbitrary assignment of an automorphism group element that is called a \emph{twist}, where for each pair of a base cell and one of its boundary cells, we have
    \begin{equation}
        \{(b,a) : b, a \text{ are cells such that } a \in \partial b\} \xrightarrow{\varphi} G\,.
    \end{equation}
    Then, a \emph{twisted boundary map} $\partial^E$ is defined by $\varphi$ as
    \begin{align}
        \partial^E_{(0,\nu)}(b^0 \otimes f) &= b^0 \otimes \partial f\,,\\
        \partial^E_{(1,\nu)}(b^1 \otimes f) &= b^1 \otimes \partial f + \sum_{a^0 \in \partial b^1}{a^0 \otimes \varphi(b^1,a^0)f}\,.
    \end{align}
\end{defbox}
Here, we only consider up to 1-complexes, but this can be generalized to high-dimensional complexes as well, as long as we retain the constraint that $\partial^E \partial^E = 0$, that is, the boundary of a boundary is 0. With the complexes defined, we can now consider the (co)homology groups. The first homology and cohomology of the bundle, $H_1(E)$ and $H^1(E)$, respectively, are isomorphic to the homology and cohomology groups of the base space, $H_1(B)$ and $H^1(B)$. This implies that $b_1(E) = b_1(B)$. Then, we can define the \emph{bundle projection}:

\begin{defbox}[label={def:bundle-projection}]{Bundle projection}
    The \emph{bundle projection} $\Pi_\rho : E_\rho \to B_\rho$ is defined as
    \begin{align}
        b^\rho \otimes f^0 &\mapsto b^\rho\,,\\
        b^{\rho - j} \otimes f^j &\mapsto 0 \text{ for } j > 0
    \end{align}
    for all $\rho$-cells $b^\rho$ and $(\rho-j)$-cells $b^{\rho-j}$ of the base and 0-cells $f^0$ and $j$-cells $f^j$ of the fiber. The bundle projection also forms a chain map.
\end{defbox}

With this in mind, we have the following result:

\begin{thmbox}[label={thm:bundle-projection-properties}]{Bundle projection properties}
    The bundle projection induces a vector space isomorphism $\Pi_* : H_1(E) \to H_1(B)$ and $\Pi^* : H^1(B) \to H^1(E)$ if all of the following conditions are true:
    \begin{itemize}
        \item $B$ is a 1-complex
        \item The boundary $\partial f^1$ of any fiber 1-cell $f^1$ has even weight
        \item Every fiber 0-chain of even weight is a boundary
        \item $H_0(B) = 0$, that is, the zeroth Betti number vanishes
        \item Every fiber automorphism acts trivially on $H_1(F)$.
    \end{itemize}
\end{thmbox}

We refer the reader to~\cite[Sec. 2.4]{hastings2021fiber} for a proof of this result. This forms the last piece of background on fiber bundles that we need to define fiber bundle codes, so without further ado, we present the following definition of fiber bundle codes:

\begin{defbox}[label={def:fiber-bundle-code}]{Fiber bundle code}
    A \emph{fiber bundle code} is a CSS code whose logical operators are associated with one-dimensional homology and cohomology groups of the twisted bundle complex $E_2 \to E_1 \to E_0$, constructed from the circle fiber $F_1 \to F_0$ and a base $B_1 \to B_0$.
\end{defbox}

To determine the code parameters, we first define $n_F = m_F = \ell^2$ for an odd integer $\ell$. We then choose a random classical code for the base $B$ using $n_B$ bits and $m_B$ parity-checks, which we can represent using a Tanner graph that has $n_B$ right-vertices and $m_B$ left-vertices. In particular, we take $m_B = \frac{3}{4}n_B$ and we take the vertices to have degree $\Theta(\log^2{n_B})$. Then, the minimum distance is $\Omega(n_B)$ and the parity checks are linearly independent with high probability. That is, the bundle $E$ has $n = n_B \cdot m_F + m_B \cdot n_F$ 1-cells, which correspond to physical qubits, and the total number of cells is $(n_B + m_B) \cdot (n_F + m_F)$. This gives us $\Theta(n_B)$ logical qubits with $X$- and $Z$-distances being given by $d_X = \Omega(m_F/\log^2{n_B})$ and $d_Z = \Omega(n_B \cdot m_F^{1/2}/\log^2{n_B})$, respectively, for $n_B \geq m_F$. If we take $n_B \sim m_F$, then we have $n = \Theta(n_B^2)$ physical qubits, $k = \Theta(n^{1/2})$ logical qubits, and $X$- and $Z$-distances $d_X = \Omega(n^{1/2}/\log^2{n})$ and $d_Z = \Omega(n^{3/4}/\log^2{n})$, respectively. We can perform a procedure called \emph{distance balancing} to improve these parameters further, yielding fiber bundle codes with parameters $\stabcode{n}{\Theta(\sqrt{n})}{\Omega(n^{3/5}/\polylog(n))}$, which beats the square-root distance barrier that plagued qLDPC codes until the introduction of these codes.

\subsubsection{Lifted product codes}\label{subsubsec:lifted-product-codes}
Around the same time that the fiber bundle codes were introduced, the \emph{lifted product codes} were also proposed. This code construction was introduced by Panteleev and Kalachev~\cite{panteleev2022quantum} originally under the name of \emph{generalized hypergraph product codes}~\cite{panteleev2021degenerate}, but later renamed to lifted product codes. The lifted product code is a Galois-qudit code (see~\cref{subsec:qudit-stabilizers}), which means it is defined over the field $\FF_q$ for prime power $q$, but we can consider $q = 2$ for the purposes of this tutorial.

Lifted product codes are CSS LDPC codes that generalize the hypergraph product codes that we explored in~\cref{subsec:ldpc-hgp}. Like the hypergraph product codes, lifted product codes take as their starting point a sparse parity-check matrix of a classical binary code and result in a constant-rate code. However, unlike hypergraph product codes, lifted product codes are able to achieved a \emph{linear distance}, while hypergraph product codes suffer from the square-root distance barrier that we discussed above. As such, lifted product codes are also \emph{good} qLDPC codes.

Recall, hypergraph product codes work with binary matrices over the field $\FF_2$. Lifted product codes extend this to work over a group algebra $\FF_2[G]$. For instance, the simplest version of this starts from sparse matrices over the ring
\begin{equation}
    R_\ell = \FF_2[x]/(x^\ell - 1)\,,
\end{equation}
where each element $x^s$ from the ring can be represented as an $\ell \times \ell$ cyclic shift matrix $P^s$. Here, $P$ is the permutation matrix that shifts the standard basis by one position. For example, the $3 \times 3$ permutation matrix that does this is
\begin{equation}
    P =
    \begin{pmatrix}
        0 & 0 & 1\\
        1 & 0 & 0\\
        0 & 1 & 0
    \end{pmatrix}\,.
\end{equation}
A matrix defined over $R_\ell$ can be expanded into a larger binary matrix by taking $0 \mapsto 0$ and $x^s \mapsto P^s$. Here, $0$ is the $\ell \times \ell$ all-zero matrix. This process is called a \emph{lift}, hence the name of the codes. Thus, lifted product codes follow the same hypergraph product code tensor-product structure, but this is done over $R_\ell$ (or, more generally, over $\FF_2[G]$) before expanding into binary blocks.

\begin{example}[label={example:small-lp-code}]{Small lifted product code}
    Let $\ell = 3$ such that
    \begin{equation}
        P = 
        \begin{pmatrix}
            0 & 0 & 1\\
            1 & 0 & 0\\
            0 & 1 & 0
        \end{pmatrix}
    \end{equation}
    and take
    \begin{equation}
        C = 
        \begin{pmatrix}
            1 & x & 0\\
            0 & 1 & x^2
        \end{pmatrix}\,.
    \end{equation}
    Using the mapping $x^s \mapsto P^s$, we expand $C$ into blocks as
    \begin{equation}
        C' = 
        \begin{pmatrix}
            I_3 & P & 0\\
            0 & I_3 & P^2
        \end{pmatrix}\,,
    \end{equation}
    which is a $6 \times 9$ binary matrix, larger than the original $2 \times 3$ matrix, but preserving the weight of each column and row.
\end{example}

Lifted product codes follow the same chain complex construction that we used to describe hypergraph product codes, except now the chain complexes are taken over the group algebra. Thus, we have
\begin{equation}
    C_2 \xrightarrow{\partial_2} C_1 \xrightarrow{\partial_1} C_0\,,
\end{equation}
where the fundamental lemma of homology is still satisfied:
\begin{equation}
    \partial_1\partial_2 = 0\,.
\end{equation}
Taking the parity-check matrices to be
\begin{equation}
    H_X = \partial_1\,,\quad H_Z = \partial_2^\intercal\,,
\end{equation}
the CSS condition is automatically satisfied:
\begin{equation}
    H_XH_Z^\intercal = \partial_1(\partial_2^\intercal)^\intercal = \partial_1\partial_2 = 0\,.
\end{equation}
Equivalently, given the parity-check matrices $H_1$ and $H_2$ for the classical seed codes $C_1$ and $C_2$, respectively, we can construct the $X$- and $Z$-type parity-check matrices for the lifted product code as
\begin{equation}
    H_X = [H_1 \otimes I \mid I \otimes H_2^\intercal]\,,\quad H_Z = [I \otimes H_2 \mid H_1^\intercal \otimes I]\,,
\end{equation}
except we replace the 1s with permutation matrices, as discussed above. This lift introduces a controlled global structure that takes the code beyond a local tensor-product construction, which enables lifted product codes to escape the square-root distance barrier.

\begin{exerbox}[label={exer:hgp-from-lp-codes}]{HGP codes from LP codes}
    Show that the lifted product code construction reduces to the hypergraph product code construction when taking $\ell = 1$ in the ring $R_\ell = \FF_2[x]/(x^\ell - 1)$ and simplifying the resulting $H_X$ and $H_Z$.
\end{exerbox}

There are two ways to introduce the lift: in an abelian way and in a non-abelian way. The abelian lifts are built from $R_\ell = \FF_2[x]/(x^\ell - 1)$ or $\FF_2[G]$ for some abelian group $G$. The resulting codes are often quasi-cyclic and easier to describe and implement explicitly. This  construction is what was initially proposed by Panteleev and Kalachev~\cite{panteleev2022quantum}, the resulting codes use $n$ physical qubits to encode
\begin{equation}
    k = \bigTheta{\log{n}}
\end{equation}
logical qubits with a distance scaling as
\begin{equation}
    d = \bigTheta{\frac{n}{\log{n}}}\,.
\end{equation}
As the number of logical qubits scales as $\bigTheta{\log{n}}$, this code is not constant-rate and thus not asymptotically good.

\begin{exerbox}[label={exer:abelian-lp-code-not-good}]{Abelian LP codes are not good LDPC codes}
    Explain why the code dimension scaling as
    \begin{equation}
        k = \bigTheta{\log{n}}
    \end{equation}
    makes the abelian lifted product codes not asymptotically good.
\end{exerbox}

The non-abelian construction~\cite{panteleev2022asymptotically}, on the other hand, allows for the group algebra to be taken over a non-abelian group. The resulting codes have constant-rate and improved code parameters
\begin{equation}
    k = \bigTheta{n}\,,\quad d = \bigTheta{n}\,,
\end{equation}
enabling asymptotically ``good'' codes. Thus, lifted product codes are one of the most promising code constructions, yielding great code parameters. However, \emph{decoding}~\cite{golowich2024decoding,bhattacharyya2025decoding,raveendran2025minimum,pradhan2025linear} remains a challenge for lifted product codes and other types of quasi-cyclic codes, delaying their deployment in practical, real-world instances.

\newpage

\section{Qudit codes}\label{sec:qudit}

Much of quantum information science has been built upon the qubit. However, this two-level system, while powerful, represents only the simplest instance of a unit of quantum information. A more general and potentially more potent framework emerges from the study of qudits, which are the higher-dimensional generalizations of qubits and allow us to work with multi-level systems. In this chapter, we give an overview of qudit error-correcting codes.

\subsection{Motivation for and basic background on qudits}\label{subsec:qudit-motivation}
Before we explore the motivation for qudit codes, we need to understand the motivation for qudits to begin with; after all, shouldn't two-level systems (i.e., qubits) be enough? It turns out, though, that we can gain a lot by considering higher dimensional Hilbert spaces.

A qudit is simply the $q$-dimensional generalization of the qubit, for $q \geq 2$. This larger state space allows for the storage of more information per computational unit and reduced circuit depth. For example, for qubits, a controlled-$U$ gate only allows one to apply a unitary gate $U$ if the control qubit is in the state $\ket{1}$ or do nothing if the control qubit is in the state $\ket{0}$. However, if we were working with qutrits, where we have $q=3$ basis states $\ket{0}$, $\ket{1}$, and $\ket{2}$, then we think of more complex control operations: For example,  do nothing if the control qutrit is in the state $\ket{0}$, apply the gate $U_1$ if the qutrit is in the state $\ket{1}$, and apply the gate $U_2$ if the qutrit is in the state $\ket{2}$. However, if we restrict ourselves to qubit systems, such conditional operations require two control qubits where, for example, we do nothing if the two control qubits are in the state $\ket{00}$, apply $U_1$ if they are in state $\ket{01}$ or $\ket{10}$, and apply $U_2$ if they are in the state $\ket{11}$. Implementing this qubit conditional unitaries may also require more gates (e.g., to check for the parity or to do binary arithmetic on the two control qubits).

More formally, a $q$-dimensional qudit system can be represented by a $q$-dimensional Hilbert space $\mathcal{H}_q$. The Hilbert space $\mathcal{H}_q$ is spanned by $q$ orthonormal vectors $\{\ket{0}, \ket{1}, \ldots, \ket{q-1}\}$ such that any single-qudit state can be written as
\begin{equation}
    \ket{\psi} = \alpha_0\ket{0} + \alpha_1\ket{1} + \ldots + \alpha_{q-1}\ket{q-1}\,,
\end{equation}
where, as usual, $\alpha_i \in \CC$ and
\begin{equation}
    \abs{\alpha_0}^2 + \abs{\alpha_1}^2 + \ldots + \abs{\alpha_{q-1}}^2 = 1\,.
\end{equation}
Just as for qubits, if we have $n$ qudits, then the whole state is a vector in the Hilbert space $\mathcal{H}_q^{\otimes n}$, spanned by $q^n$ orthonormal basis vectors.

To perform quantum computation, we need a set of gates, but these are no longer the qubit gates we have been working with throughout this tutorial. For qudits, though, we still require gates to be unitary operators, where any unitary quantum gate $U$ can be arbitrarily approximated by a universal gate set $\{U_k\}_k \subseteq U(q^n)$, where $U(q^n)$ is the set of $q^n \times q^n$ unitary matrices; of course, for qubits, we had $q=2$. Also just like for the case of qubits, there is more than one such universal gate set and the choice of gate set can be informed by factors like hardware constraints.

To take two specific examples of $q$-dimensional generalizations of common gates, consider the qudit generalizations of the Pauli-$X$ and $Z$ gates, which we denote as $X_q$ and $Z_q$, respectively. These gates operate on basis states $\ket{j}$ as
\begin{align}
    X_q\ket{j} &= \ket{j+1\pmod{q}}\,,\\
    Z_q\ket{j} &= \omega^j\ket{j}\,,
\end{align}
where $\omega$ is the $q\numth$ root of unity:
\begin{equation}
    \omega = e^{2\pi i/q}\,.
\end{equation}
In the basis of the eigenstates of $Z_q$, these gates have the form
\begin{equation}
    X_q =
    \begin{pmatrix}
        0 & 0 & \cdots & 0 & 1\\
        1 & 0 & \cdots & 0 & 0\\
        0 & 1 & \cdots & 0 & 0\\
        \vdots & \vdots & \ddots & \vdots & \vdots\\
        0 & 0 & \cdots & 1 & 0
    \end{pmatrix}
\end{equation}
and 
\begin{equation}
    Z_q =
    \begin{pmatrix}
        1 & 0 & 0 & \cdots & 0\\
        0 & \omega & 0 & \cdots & 0\\
        0 & 0 & \omega^2 & \cdots & 0\\
        \vdots & \vdots & \vdots & \ddots & 0\\
        0 & 0 & 0 & \cdots & \omega^{q-1}
    \end{pmatrix}\,,
\end{equation}
With these in mind, we can then generalize the Pauli group. To do so, we now consider in more detail the specific dimension $q$, which, until now, we have been assuming to be any integer. We start by assuming that $q=p$ for some prime integer $p$. For much of the rest of this subsection, we follow the presentation in Daniel Gottesman's lecture notes~\cite{gottesman2024surviving}.

\begin{defbox}[label={def:single-qudit-prime-pauli}]{Single-qudit prime Pauli group}
    Let $p > 2$ be a prime integer. Then, the \emph{single-qudit Pauli group} $\mathcal{P}_1(p)$ is defined as the set $\{\omega^a X_p^b Z_p^c\}$, where $a,b,c \in \{0, 1, \ldots, p-1\}$ and where $\omega$ is the $p\numth$ root of unity and $X_p$ and $Z_p$ are as defined above for $q=p$.  
\end{defbox}

For $n$ qudits, the Pauli group is, of course, just $\mathcal{P}_n(p)^{\otimes n}$ and we denote by $\hat{\mathcal{P}}_n(p) = \mathcal{P}_n(p)/\{\omega^a I\}$ the Pauli group without phases, just as we did for the qubit Pauli group. The commutation relation between the $X_p$ and $Z_p$ operators is then
\begin{equation}
    Z_pX_p = \omega X_pZ_p\,.
\end{equation}
One can also represent the qudit Pauli group using the symplectic form, but we refer the reader to~\cite[Sec. 8.1.1]{gottesman2024surviving} for more details on this.

If, instead, we consider a \emph{prime power}, such as $q=p^m$ for prime $p$ and $m > 1$, it is easiest to consider the Pauli group as $\mathcal{P}_n(q) = \mathcal{P}_m(p)^{\otimes n}$, that is, as $m$ copies of the $p$-dimensional Pauli group. We now bring in some of the mathematical machinery from the next section on finite field theory, where we label each basis state of a $q$-dimensional qudit by an element of the finite field $\FF_q$. With this in mind, we define the Pauli group for prime power dimensions as follows:

\begin{defbox}[label={def:prime-power-pauli-group}]{Prime power Pauli group}
    Let $\beta \in \FF_q$ and let $X_q^\beta$ and $Z_q^\beta$ be the usual generalized Pauli-$X$ and $Z$ operators, respectively, such that
    \begin{align}
        X^\beta\ket{a} &= \ket{a+\beta}\,,\\
        Z^\beta\ket{a} &= \omega^{\mathrm{tr}(\beta a)}\ket{a}\,,
    \end{align}
    where $a \in \FF_q$ and $\mathrm{tr} : \FF_q \to \FF_p$ is the trace function that operates as $\mathrm{tr}(a) = \defsum{j=0}{m-1}{a^{p^j}}$. Then, the Pauli group $\mathcal{P}_n(q)$ is comprised of elements of the form
    \begin{equation}
        \eta\omega^b\bigotimes_{j=1}^{n}{X^{\beta_j}Z^{a_j}}\,,
    \end{equation}
    where $b \in \FF_p$ and $\beta_j, a_j \in \FF_q$.
    Here, $\eta=1$ if $q$ is odd, and $\eta\in\{1, i\}$ if $q$ is even. 
    We denote $\hat{\mathcal{P}}_n(q)$ as the set of Pauli representatives without phases, that is,
    \begin{equation}
        \bigotimes_{j=1}^{n}{X^{\beta_j}Z^{a_j}}\,.
    \end{equation}
\end{defbox}

Now, the commutation relation between the $X$ and $Z$ operators is
\begin{equation}
    Z^aX^\beta = \omega^{\tr(a\beta)}X^\beta Z^a\,.
\end{equation}
We again refer the reader to~\cite[Sec. 8.1.2.]{gottesman2024surviving} for further details on this.

Finally, we consider what happens when we have a qudit of dimension $q$ that is not a prime power. The most straightforward way to deal with this is to factor $q$ into its prime factors and then treat each prime factor separately with its own Pauli group (e.g., $\mathcal{P}_1(2) \otimes \mathcal{P}_1(3)$ for $q=6$). Alternatively, we could make use of the \emph{Weyl-Heisenberg group}. We do not often consider qudits of such a non-prime-power composite (as opposed to prime) order, so this is all we will say about this for now.

\subsection{Finite field theory}\label{subsec:qudit-finite}
While the mathematical formalism for qubits is technically based on the finite field $\FF_2$, we do not often need to worry about the details of fields beyond some basic notions like the addition and multiplication rules (e.g., $1+1=0\pmod{2}$). However, when we expand the local Hilbert space, we need to consider more general fields and mathematical abstractions. In this section, we introduce \emph{finite field theory} (also called \emph{Galois} field theory, after its short-lived founder, \'{E}variste Galois (1811--1832)), as well as some other important notions from algebra that we will find useful.

We assume that the reader is familiar with the group theoretic concepts from~\cref{subsec:stabilizer-group-theory-primer}; if not, we encourage the reader to review that section now. Before we get to finite fields, we need to introduce \emph{rings}:

\begin{defbox}[label={def:ring}]{Ring}
    A \emph{ring} $R$ is a set that is closed under two binary operations: (1) addition: $x+y\in R$ for any $x,y\in R$ and (2) multiplication: $xy\in R$ for any $x,y\in R$. Addition and multiplication satisfy the following properties for all $x,y,z\in R$: (1) associativity: $(x+y)+z=x+(y+z)$ and $(xy)z=x(yz)$, (2) addition commutativity $x+y=y+x$, and (3) distributivity $x(y+z) = xy+xz$. Also, there exists a multiplicative identity element $1$ and an additive identity element $0$. For each element $x\in R$ there exists a unique additive inverse $y\in R$, i.e. a unique $y\in R$ such that $x+y=0$.
\end{defbox}

If the multiplication in ring $R$ is commutative, that is, $xy=yx$ for all $x,y \in R$, then we say that the ring is a \emph{commutative ring}. An important concept from ring theory is that of ring \emph{ideals}:

\begin{defbox}[label={def:ideal}]{Ring ideal}
    Let $R$ be a ring. A subset $I \subseteq R$ is called an \emph{ideal} if $I$ is an additive subgroup (meaning that if $x, y \in I$, then so too is $x + y$) and for any element $y \in R$ and $x \in I$, then $yx \in I$.
\end{defbox}

A \emph{generated ideal} is constructed as follows. Given a subset $S \subseteq R$ of a ring $R$, the ideal generated by $S$, which we denote by $(S)$ (though some references also write this as $\langle S \rangle$), is the set of elements of the form $\sum_{i}{r_i a_i}$, where the $a_i \in S$ and the $r_i \in R$. The generator elements $\{a_i\}$ are not unique. For example, in the ring of integers, we have $(4, 10) = (2)$. An ideal $I \subseteq R$ of a ring $R$ is further said to be \emph{maximal} if the only other ideal to fully contain $I$ (i.e., contain all of the elements in $I$) is $R$ itself. A \emph{quotient ring} $R/I$ of an ideal $I \subseteq R$ is a ring itself whose elements are the cosets of $I$ in $R$ with induced operations (addition and multiplication). Combining the notions of a quotient ring and a maximal ideal, it is a fact that $R/I$ is a field if and only if $I$ is a maximal ideal; we will see this below more explicitly.

With this background on ring theory, we can now define a finite field. We start by defining a field in general:

\begin{defbox}[label={def:field}]{Field}
    A \emph{field} is a commutative ring in which every nonzero element has a multiplicative inverse.
\end{defbox}

Then, more specifically, we have the simple following definition:

\begin{defbox}[label={def:finite-field}]{Finite field}
    A \emph{finite field} is a field with a finite number of elements.
\end{defbox}

We denote a field with $q \in \NN$ elements by $\FF_q$, though the notation $GF(q)$ (for \emph{G}alois \emph{f}ield is also common). If $q$ is prime, then $\FF_q$ is isomorphic to $\ZZ/q\ZZ$, that is, the integers modulo $q$. We write this as $\FF_q \cong \ZZ/q\ZZ$. If $q = p^m$ for some prime $p$ and integer $m \geq 2$ (i.e., $q$ is a \emph{prime power}), then $\FF_q$ is \emph{not} isomorphic to $\ZZ/q\ZZ$. We call $p$ the \emph{characteristic} of the field. The number of elements in a finite field, which we call the \emph{order} of the field (i.e., $q$), is always a prime or a prime power; there are no finite fields of order that is a \emph{composite} number, such as 6. We can formalize this in the following theorem:

\begin{thmbox}[label={def:finite-field-existence-uniqueness}]{Existence and uniqueness theorem for finite fields}
    For every $q = p^m$ for prime $p$ and integer $m \geq 1$, there exists a field with $q$ elements. Furthermore, any two fields with $q$ elements are isomorphic. Thus, the field $\FF_q$ is unique.
\end{thmbox}

This is a well-known result and the proof is beyond the scope of this tutorial, but we refer the reader to any good set of lecture notes or textbooks~\cite{dummit2003abstract} on finite field theory. We can construct a finite field using polynomials, specifically \emph{polynomial quotients}. To see this, we need one additional concept, that of \emph{irreducibility}:

\begin{defbox}[label={def:polynomial-irreducibility}]{Polynomial irreducibility}
    A polynomial $f(x) \in \FF_q[x]$ is \emph{irreducible} if it cannot be factored into two polynomials $g(x), h(x) \in \FF_q[x]$ such that the degrees of both $g(x)$ and $h(x)$ are less than the degree of $f(x)$.
\end{defbox}

With this in mind, we can construct a finite field $\FF_q$ with $q = p^m$ elements, for prime $p$ and $m \geq 1$ as follows. Let $f(x) \in \FF_p[x]$ be an irreducible polynomial, as defined in~\cref{def:polynomial-irreducibility}. Then,
\begin{equation}
    \FF_q \cong \FF_p[x]/(f(x))\,.
\end{equation}
Note the parentheses around $f(x)$, denoting this as a ring ideal, as defined in~\cref{def:ideal}. Thus, we define higher-order finite fields in terms of their characteristic (i.e., $p$) field. One of the simplest examples is the construction of the field $\FF_4$:

\begin{example}[label={ex:f4}]{Finite field with four elements}
    Let $p=2$ and $f(x) = x^2 + x + 1$, which is irreducible over $\FF_2$. We can construct $\FF_4$ as
    \begin{equation}
        \FF_4 = \FF_2[x]/(x^2+x+1)\,,
    \end{equation}
    where the four elements of $\FF_4$ are $\{0,1,x,x+1\}$. Arithmetic is then done modulo $x^2+x+1$. $\FF_4$ is the smallest non-prime field and is thus important for understanding higher-order finite fields.
\end{example}

Notably, every function $f: \FF_q \to \FF_q$ can be represented as a polynomial. We can also consider finite fields from a vector space perspective, as we hinted at in~\cref{subsec:qudit-motivation}. That is, the field $\FF_{p^m}$ can be worked with as an $m$-dimensional vector space over $\FF_p$. This allows us to use the vast toolbox of linear algebra where we keep in mind that arithmetic operations are done over $\ZZ/p\ZZ \cong \FF_p$.

To conclude, we highlight some interesting properties of finite fields. First, we note that the nonzero elements of a finite field $\FF_q$, which we denote by $\FF_q^\times$, form a \emph{cyclic group} under multiplication:
\begin{equation}
    \FF_q^\times \cong C_{q-1}\,.
\end{equation}
A cyclic group $C_{q-1}$ is simply a group that is generated by a single element under multiplication. To connect this concept to finite fields, we note that there is some element $\alpha \in \FF_q^\times$ that we call \emph{primitive} that gives us the whole of $\FF_q^\times$:
\begin{equation}
    \FF_q^\times = \{\alpha^k : 0 \leq k < q-1\}\,.
\end{equation}
We can also denote this as $\FF_q^\times = \langle\alpha\rangle$. Furthermore, there are exactly $\varphi(q-1)$ primitive elements in any multiplicative group, where $\varphi(n)$ is a function called \emph{Euler's totient function}, which counts the number of positive integers up to $n$ that are coprime to $n$. Finally, we note that a finite field $\FF_{p^n}$ has a subfield of size $\FF_{p^m}$ if and only if $m|n$, that is, $m$ divides $n$. Each such subfield is unique.

\subsection{Major families of qudit stabilizer codes}\label{subsec:qudit-stabilizers}
The main class of qudit codes that we focus on in this chapter is that of \emph{qudit stabilizer codes}. There are two axes along which we can define a qudit stabilizer code: the algebra upon which the code is built and the construction style. By the ``algebra,'' we mean the group or field that all of our subsequent operations are done over, which is, of course, determined by the local dimension of the qudit. Thus, for a qudit code with local qudit dimension $q$, we could have codes defined over the integers modulo $q$, $\ZZ_q$, or a finite field $\FF_q$, depending on what $q$ is, to name two examples. The ``construction'' is the other ingredient that gives us a qudit code, and this determines the structure of the code. Examples include CSS, non-CSS, topological, homological, LDPC, and graph-based codes.

\subsubsection{Algebra}\label{subsubsec:qudit-algebra}
We start by looking at different qudit stabilizer codes based on their algebra, that is, what group or field the local qudit dimension is defined over.

\paragraph{Modular-qudit stabilizer codes.}
We start by considering \eczoo[modular-qudit stabilizer codes]{qudit_stabilizer}. These codes are the most straightforward generalization of qubit stabilizer codes and were introduced concurrently with the qubit formalism in Daniel Gottesman's PhD thesis~\cite{gottesman1997stabilizer}. As was the case for qubits, the fundamental group that we work with to build up our stabilizer codes is the Pauli group, except we now consider the \emph{modular-qudit Pauli group}~\cite{appleby2005symmetric}, which we define as follows:

\begin{defbox}[label={def:modular-qudit-pauli-group}]{Modular qudit Pauli group}
    Let $\mathcal{H}_q$ be the Hilbert space of one modular qudit with computational basis
    \begin{equation}
        \{\ket{k} : k \in \ZZ_q\}\,.
    \end{equation}
    Define the single-qudit Pauli operators
    \begin{equation}
        X\ket{k} = \ket{k+1\pmod{q}}\,, \quad Z\ket{k} = e^{2\pi ik/q}\ket{k}\,.
    \end{equation}
    For $n$ qudits and vectors $a,b \in \ZZ_q^n$, write
    \begin{equation}
        X(a) = X^{a_1} \otimes \cdots \otimes X^{a_n}\,,\quad Z(b) = Z^{b_1} \otimes \cdots \otimes Z^{b_n}\,.
    \end{equation}
    Then, the \emph{modular qudit Pauli group} on $n$ qudits is the subgroup of unitary operators that is generated by all $X(a), Z(b)$, and a global phase $\tau = e^{2\pi i/q}$ that is a $q\numth$ root of unity. In other words,
    \begin{equation}
        \mathcal{P}_{n,q}^{\mathrm{mod}} = \{\tau^c X(a)Z(b) : c \in \ZZ_{2q}, a,b \in \ZZ_q^n\}\,.
    \end{equation}
\end{defbox}

Now, we represent Pauli strings as vectors in $\ZZ_q^{2n}$, where $n$ is the number of qudits and $q$ is the local qudit dimension. Here, $q$ can be \emph{any} integer; we are not restricted to, for example, primes or prime powers. The object $\ZZ_q^{2n}$ is now a \emph{module}, not necessarily just a vector space; see~\cref{box:modules} for a brief introduction to modules. A stabilizer is then a commuting set $\mathcal{S}$ of these generalized Pauli strings and the code is taken to be the joint +1 eigenspace of $\mathcal{S}$. Any single-qudit Pauli string $X_a Z_b$ for $a,b \in \ZZ_q$ can be written using a \emph{modular symplectic representation} $(a|b) \in \ZZ_q^2$, analogous to the binary symplectic representation that we used for qubits. This, of course, extends naturally to the case for $n$-qudit systems, where we can write $(\vb{a}|\vb{b}) \in \ZZ_q^{2n}$.

We denote a modular-qudit stabilizer code encoding into a codespace of dimension $K = q^k$ into $n$ physical qudits with distance $d$ by $\stabcode{n}{k}{d}_{\ZZ_q}$, where $q$ is the local qudit dimension. If $q$ is prime, then $K$ is a power of $q$, but if $q$ is composite (e.g., $q=4=2^2$ or $q = 6 = 2 \times 3$), then $K = q^n/\abs{\mathcal{S}}$ (see e.g.~\cite[Theorem 1]{gheorghiu2014standard}), where $\abs{\mathcal{S}}$ now does not need to be a power of $q$ and so $k$ is not necessarily an integer. For example, consider $q=4$ and fourth root of unity $\omega = e^{2\pi i/4} = i$ so that the generalized Paulis are
\begin{equation}
    X = \sum_{j\in\ZZ_4}\ket{j+1}\bra{j}
    \quad\text{and}\quad
    Z = \ketbra{0} + i \ketbra{1} - \ketbra{2} - i\ketbra{3}\,.
\end{equation}
For a stabilizer group $\mathcal{S}=\<Z^2\> = \{I,Z^2\}$, the corresponding codespace is a two-dimensional subspace spanned by $\{\ket{0}, \ket{2}\}$ since $Z^2 = \ketbra{0} - \ketbra{1} + \ketbra{2} - \ketbra{3}$. The dimension of the codespace can be clarified by $K=q/|\mathcal{S}| = 4/2 = 2$, since the stabilizer group is $|\mathcal{S}|=2$. Thus, the dimension is not any power of $4$, that is, $K=2 \neq 4^k$ for any positive integer $k$.

A $\stabcode{n}{k}{d}_{\ZZ_q}$ modular-qudit code can detect errors on up to $d-1$ qudits and can correct erasure errors on up to $d-1$ qudits. The code can correct arbitrary errors on at most $\lfloor \frac{d-1}{2} \rfloor$ qudits. Finally, we note that we can also describe the modular-qudit stabilizer code using its parity-check matrix, $H = (A|B)$, where each row $(a|b)\in\ZZ_q^{2n}$ of $H$ represents a stabilizer generator $X_aZ_b$.

\begin{explbox}[label={box:modules}]{Modules}
    A \emph{module} can be thought of as a generalization of a vector space, where the field of scalars is replaced by a ring. The ring can be either commutative or non-commutative. Modules have important applications in commutative algebra, homological algebra, algebraic geometry, and algebraic topology. For example, in commutative algebra, ideals and quotient rings (which we defined above in~\cref{subsec:qudit-finite}) are modules. Much of the work with modules has to do with extending the properties of vector spaces to well-behaved rings, but there are several complications that arise when doing this. For example, not all modules have a basis, which is, of course, untrue for vector spaces.\vspace{2.5mm}

    Let $R$ be a ring (see~\cref{def:ring}). A \emph{left $R$-module} is a set $M$ with
    \begin{enumerate}
        \item a binary operation $+$ on $M$ in which $M$ is an abelian group and
        \item an action of $R$ on $M$ (i.e., a map $R \times M \to M$) denoted by $rm$ for all $r \in R$ and $m \in M$ satisfying:
        \begin{enumerate}
            \item $(r+s)m = rm + sm$ for all $r,s \in R$, $m \in M$,
            \item $(rs)m = r(sm)$ for all $r,s \in R$, $m \in M$, and
            \item $r(m+n) = rm + rn$ for all $r \in R$, $m,n \in M$.
        \end{enumerate}
        If $R$ has an identity element 1, then we further have a \emph{unital} condition, that is,
        \begin{enumerate}
            \item[(d)] $1m = m$ for all $m \in M$.
        \end{enumerate}
        In this work, we assume rings are unital.
    \end{enumerate}
    \emph{Right} $R$-modules are defined analogously and if $R$ is commutative, then we can turn a left $R$-module $M$ into a right $R$-module by taking $rm = mr$ for $r \in R$ and $m \in M$.\vspace{2.5mm}

    A common example of a module is a $\ZZ$-module, where $\ZZ$ is, of course, the set of integers. Let $R = \ZZ$ and $A$ be any abelian group with binary operation $+$. To make $A$ into a $\ZZ$-module, we define for any $n \in \ZZ$ and $a \in A$ the quantity $na$, where
    \begin{equation}
        na =
        \begin{cases}
            a + a + \ldots + a \text{ ($n$ times)} & \text{if } n > 0\\
            0 & \text{if } n = 0\\
            -a - a - \ldots - a \text{ ($n$ times)} & \text{if n < 0}\,,
        \end{cases}
    \end{equation}
    where 0 is the additive identity in $A$. This is the only possible action of $\ZZ$ on $A$ making it a unital $\ZZ$-module. Since $A$ was taken to be \emph{any} abelian group, we can conclude that \emph{every} abelian group is a $\ZZ$-module.

    If $A$ is an abelian group with element $x$ of finite order $n$, then $nx = 0$. That is, for $\ZZ$-modules, there exist nonzero elements $x\in A$ such that $nx = 0$ for $n \neq 0$, which is not true for vector spaces. If $A$ has order $m$, then $mx = 0$ for all $x \in A$ and we say that $A$ is a module over $\ZZ/m\ZZ$. Furthermore, if $p$ is prime and $A$ is abelian such that $px = 0$ for all $x \in A$, then $A$ is a $\ZZ/p\ZZ$-module. However, we know that $\ZZ/p\ZZ \cong \FF_p$, so $A$ is also an $\FF_p$-module, which is just a vector space over $\FF_p$.
\end{explbox}

\paragraph{Galois-qudit stabilizer codes.}
While for modular-qudit stabilizer codes we allowed $q$ to be \emph{any} positive integer, if we restrict to qudits that have a local dimension that is a prime power, then we get \eczoo[\emph{Galois-qudit stabilizer codes}]{galois_stabilizer}~\cite{wills2026review}. That is, we take $q = p^m$ for some prime $p$ and $m \geq 1$. Thus, the algebra is now based on the \emph{finite field} $\FF_q$. The basis states are therefore labeled by elements from $\FF_q$. Thus, a single-qudit Pauli operator $X_aZ_b$ for $a,b \in \FF_q$ is written as $(a|b) \in \FF_q^2$. The notation for $n$ Galois qudits follows analogously. Furthermore, the Pauli operators are defined such that their commutation relations are determined by the \emph{trace-symplectic inner product}. That is, given two Galois-qudit stabilizers on $n$ qudits described by $(\vb{a}|\vb{b}), (\vb{a}'|\vb{b}') \in \FF_q^{2n}$, they commute if and only if the following quantity, the trace-symplectic inner product, is zero:
\begin{equation}
    \Tr[\vb{a}\cdot\vb{b}' - \vb{a}'\vb{b}] = \sum_{i=1}^{n}{\Tr[a_ib_i' - a_i'b_i]} = 0\,.
\end{equation}
Here, the trace is the \emph{field trace} (see~\cref{def:prime-power-pauli-group}). One nice property of Galois-qudit stabilizer codes is that they can be mapped to \emph{self-orthogonal additive codes} over $\FF_{q^2}$. That is, given an $n$-qudit Pauli stabilizer, we can represent it as a length-$n$ vector over the field $\FF_{q^2}$. Then, given a basis $(\beta, \beta^q)$ for $\FF_{q^2}$ over $\FF_q$, a vector $(a|b) \in \FF_q^2$ can be mapped to an element $a\beta + b\beta^q \in \FF_{q^2}$.

Just as we did for modular-qudit stabilizer codes, we denote a Galois-qudit stabilizer code using $n$ Galois-qudits to encode $k$ logical qudits with codespace dimension $K = q^k$ and distance $d$ using the notation $\stabcode{n}{k}{d}_q$. Note, though, the subscript $q$ rather than $\ZZ_q$ to denote that these codes are defined over a finite field rather than a ring. Furthermore, just as was the case for modular-qudit stabilizer codes, the logical codespace dimension does not need to be an integer, as we can write $K = q^{n-r/m}$, where $r$ is the number of generators of the stabilizer group, and the quantity $r/m$ is not necessarily an integer. The distance means the same thing as it did for modular-qudit stabilizer codes as well, namely, the minimum weight of a Galois-qudit Pauli string yielding a nontrivial logical operator. Thus, the code $\stabcode{n}{k}{d}_q$ detects and corrects erasure errors on $d-1$ qudits and corrects errors on at most $\lfloor \frac{d-1}{2} \rfloor$ qudits.

\paragraph{True Galois-qudit stabilizer codes.}
Finally, we can consider a subclass of Galois-qudit stabilizer codes whose stabilizers' symplectic representation forms a linear subspace. Such codes are called \eczoo[\emph{true Galois-qudit stabilizer codes}]{galois_true_stabilizer}. Here, the vectors comprising the stabilizer group form a subspace in which addition and multiplication are closed by elements of $\FF_q$. Note that this definition adds \emph{multiplication}, where for the Galois-qudit stabilizer codes in the previous section, we only had closure under addition. The number of generators $r$ is thus an integer multiple of $m$, so the quotient $r/m$ is an integer. Thus, the code has a logical codespace dimension $K = q^{n-r/m}$, which is now always an integer, which is nice. Everything else about these codes (e.g., the parity-check matrices, the code parameters, etc.) is the same as for the Galois-qudit stabilizer codes.

\subsubsection{Construction style}\label{subsubsec:qudit-construction}
While the algebraic structure told us how to represent the stabilizers and the commutation relations between Pauli operators, the \emph{code construction style} tells us how to build the commuting check operators. We consider four construction styles: CSS, non-CSS, topological/homological, and LDPC.

\paragraph{CSS-type.}
Consider the true Galois-qudit stabilizer codes from above. If we have such a code and we have a set of stabilizer generators that are either $X$- or $Z$-type Pauli strings, then we have a \emph{Galois-qudit CSS code}. As such, these codes can be defined using chain complexes over $\FF_q$ by extending the techniques that map CSS codes to homology groups for qubits to qudits. The stabilizer generator matrix is now nicely given as
\begin{equation}
    H = 
    \begin{pmatrix}
        0 & H_X\\
        H_Z & 0
    \end{pmatrix}\,,
\end{equation}
where we must have the CSS condition satisfied:
\begin{equation}
    H_X H_Z^\intercal = 0\,.
\end{equation}
Of course, the nice thing about CSS qudit codes is that they can be derived from $q$-ary linear codes, where we take one code to be an $[n, k_X, d_X]_q$ code $C_X$ and the other to be an $[n,k_Z,d_Z]_q$ code $C_Z$ such that $C_X^\perp \subseteq C_Z$. The quantum code parameters are then $\stabcode{n}{k}{d}_q$, where $k = k_X + k_Z - n$ and $d \geq \min\{d_X, d_Z\}$. In particular,
\begin{equation}
    d_X = \min\{\mathrm{wt}(c) : c \in C_Z \setminus C_X^\perp\}\,,\quad d_Z = \min\{\mathrm{wt}(c) : c \in C_X \setminus C_Z^\perp\}\,.
\end{equation}
Furthermore, we associate $H_X$ with the parity-check matrix of the classical code $C_X$ and $H_Z$ as the parity-check matrix of the classical code $C_Z$.

\paragraph{Non-CSS-type.} For non-CSS-type qudit codes, we do not have the nice separation between $X$- and $Z$-type stabilizers. These codes generally refer to self-orthogonal codes under a given symplectic form and they relate to the self-orthogonal additive codes over $\FF_{q^2}$ that we mentioned above. This allows for the development of quantum Hamming codes, quantum BCH codes, and other quantum analogues of popular classical $q$-ary codes over $\FF_q$.

\paragraph{Topological and homological.} As we alluded to above in our discussion of CSS-type qudit codes, we can define qudit stabilizer codes from \emph{chain complexes} and take advantage of the tools from \emph{homological algebra}, which we considered earlier in this tutorial (see, for example,~\cref{subsubsec:topological-homology}). Basic examples of such codes include qudit versions of the surface code and the toric code. For example, for qudit surface codes, stabilizer generators are of $X$- and $Z$-type and they act on only a few qudits. We associate these with the stars and plaquettes of a tessellation of a two-dimensional surface. Other examples include the qudit color codes~\cite{watson2015qudit}, which are just qudit generalizations of the qubit color codes.

\paragraph{LDPC.} Finally, there exist qudit generalizations of several major classes of LDPC codes, where the stabilizers have the benefit of being sparse. Examples of such qudit LDPC codes~\cite{andriyanova2012new,spencer2025qudit} include qudit generalizations of the currently-popular bivariate bicycle codes, qudit hypergraph product codes (also very popular), subsystem hypergraph (SHYPS) codes, high-dimensional expander codes, and fiber bundle codes. Qudit LDPC codes inherit many of the favorable qualities of qubit LDPC codes, such as reduced physical qudit overhead, high encoding rates, and sparsity in the check operators, requiring fewer check measurements.

\subsection{Advances in code bounds and asymptotic performance}\label{subsec:qudit-advances}
Recall the quantum Singleton bound for qubits from~\cref{subsec:bounds-singleton}, which gives us the relation
\begin{equation}
    n - k \geq 2(d-1)
\end{equation}
for an $\stabcode{n}{k}{d}$ code. It turns out that this is the same for qudits~\cite{rains1997nonbinary}, except now we have a different size for our logical subspace, $K = q^k$, instead of $K = 2^k$. Thus, there is no difference in the tradeoff between $n$, $k$, and $d$ when going from qubits to qudits, but we do gain a larger Hilbert space. It is this larger Hilbert space that allows for a richer landscape of MDS codes (those that saturate the quantum Singleton bound), though. Thus, it is easier to find codes over $\FF_q$ for $q > 2$ that saturate this bound compared to restricting ourselves to $\FF_2$.

We do, however, get a different quantum Hamming bound, which we discussed in~\cref{subsec:bounds-quantum-hamming-bound}, when we take $q > 2$. Recall for qubits we had the following relation, where we define $t = \lfloor \frac{d-1}{2} \rfloor$:
\begin{equation}
    2^k\sum_{j=0}^{t}{\binom{n}{j}3^j} \leq 2^n\,.
\end{equation}
For qudits of local dimension $q$, we have
\begin{equation}
    q^k\sum_{j=0}^{t}{\binom{n}{j}(q^2-1)^j} \leq q^n\,,
\end{equation}
which shows that each qudit has $q^2-1$ nontrivial generalized Pauli errors~\cite{aly2007note}. Note that this only holds for nondegenerate codes; finding the Hamming bound for degenerate codes remains an open problem.

Now we discuss some asymptotics. We define the \emph{asymptotic rate} as
\begin{equation}
    R = \lim_{n \to \infty}{\frac{\log_q{K}}{n}} = \lim_{n \to \infty}{\frac{k}{n}}
\end{equation}
and the \emph{relative distance} as
\begin{equation}
    \delta = \lim_{n \to \infty}{\frac{d}{n}}\,.
\end{equation}
The $q$-ary Gilbert-Varshamov bound lower bound is given by
\begin{equation}
    R \geq 1 - \delta\log_q(q+1) - H_q(\delta)\,,
\end{equation}
where
\begin{equation}
    H_q(x) = x\log_q(q-1) - x\log_q{x} - (1-x)\log_q(1-x)
\end{equation}
is the $q$-ary entropy function~\cite{jin2011quantum}. As $q$ grows, the right hand side also grows, giving a larger lower bound and thus a better rate, and therefore a better code. There exist codes such that both $R$ and $\delta$ are nonzero for a fixed $q$ and we call these codes \emph{good codes}.

\subsection{Experimental realizations of qudit codes}\label{subsec:qudit-experiment}
There have been several examples of experimental realizations of qudit-based systems using various platforms, such as superconducting~\cite{neeley2009emulation,blok2021quantum,morvan2021qutrit,roy2023two,goss2022high,fedorov2012implementation}, photonic~\cite{lu2020quantum,wang2018multidimensional,chi2022programmable,reimer2019high}, trapped-ion~\cite{low2020practical,hrmo2023native,leupold2018sustained,ringbauer2022universal}, and circuit quantum electrodynamics (QED)~\cite{brock2024quantum} devices. However, these are just experimentally implementing qudit systems; there has been less, but still notable, work on implementing qudit-based \emph{error correction} on real devices. We highlight two examples in this section. The first~\cite{iqbal2025qutrit} is an experimental demonstration of a qutrit toric code, which demonstrated the preparation and measurement of a $\ZZ_3$ toric code ground state on 24 qutrits. The second~\cite{brock2024quantum}, which received quite a bit of attention in the community and the press, demonstrated encoding a logical qutrit and a logical ququart using a single-mode square-lattice GKP qudit. While we did not discuss GKP codes in the context of qudits, we refer the reader to~\cref{sec:bosonic} for more on bosonic codes. Both of these early demonstrations show that implementing error-corrected qudit systems is experimentally viable and pave the way to future, larger-scale experiments.

\newpage

\section{Fault tolerance I}\label{sec:fault-tolerance-pt1}

A quantum error-correcting code protects \emph{states}. In contrast, a fault-tolerant protocol protects \emph{processes}~\cite{shor1996fault-tolerant,knill1996threshold,aharonov1996fault,gottesman2024surviving}. The core difficulty is not only detecting and correcting errors, but ensuring that the mechanism of correction does not create more errors than it removes. A \emph{location} is an elementary circuit component: a one- or two-qubit gate, state preparation, measurement, or an idle step. A \emph{fault} is a location at which the physical operation deviates from the ideal one. An ideal circuit contains abstract logical locations: perfect preparations, perfect gates, perfect waits, and perfect measurements. A fault-tolerant implementation replaces each logical location by a \emph{gadget} operating on encoded data. Examples include an encoded CNOT gadget, an encoded measurement gadget, and an error-correction gadget. In this chapter, we make the first steps in understanding \emph{fault-tolerant quantum computation} (FTQC), the pinnacle of the quantum error correction enterprise.

\subsection{Error propagation}\label{subsec:error-propagation}
Suppose we are given a circuit with faulty components. If each location of this circuit fails with probability at most $p$, then the failure probability associated with a fixed output location scales linearly as $O(p)$. For a computation with $N$ gates, a union bound gives an overall failure probability scaling as $O(Np)$. Error correction becomes useful only when encoded gadgets suppress the effective logical error rate below the underlying physical error rate. Merely performing error correction periodically is not sufficient to prevent the buildup of errors, even if it is applied after every encoded gate. Two reasons for this include:
\begin{enumerate}
    \item \textbf{Error propagation through gates}: Encoded gates can cause errors to propagate. For example, let $U$ be the CNOT gate. The propagation of a Pauli-$X$ error on the control qubit is given by:
    \begin{equation}
        U(X \otimes I)U^\dagger = X \otimes X
    \end{equation}
    Similarly, a $Z$ error on the target qubit propagates backwards:
    \begin{equation}
        U(I \otimes Z)U^\dagger = Z \otimes Z
    \end{equation}
    A single fault in a two-qubit gate can thus introduce two errors in the code block, as seen in~\cref{fig:cnot-propagation}.
    \item \textbf{Errors introduced by correction}: Error correction itself can introduce errors on the encoded qubits if the ancilla qubits or the gates used to measure the syndrome are faulty.
\end{enumerate}

\begin{figure}[b]
  \centering
  \begin{minipage}{0.48\textwidth}
    \centering
    \[
    \begin{quantikz}[column sep=0.25cm]
      \lstick{$c$} & \gate{X} & \ctrl{1} & \qw \\
      \lstick{$t$} & \qw      & \targ{}  & \qw
    \end{quantikz}
    \quad\Longrightarrow\quad
    \begin{quantikz}[column sep=0.25cm]
      \lstick{$c$} & \ctrl{1} & \gate{X} & \qw \\
      \lstick{$t$} & \targ{}  & \gate{X} & \qw
    \end{quantikz}
    \]

    \smallskip
    \small $X$ on the control propagates forward to the target.
  \end{minipage}\hfill
  \begin{minipage}{0.48\textwidth}
    \centering
    \[
    \begin{quantikz}[column sep=0.25cm]
      \lstick{$c$} & \qw      & \ctrl{1} & \qw \\
      \lstick{$t$} & \gate{Z} & \targ{}  & \qw
    \end{quantikz}
    \quad\Longrightarrow\quad
    \begin{quantikz}[column sep=0.25cm]
      \lstick{$c$} & \ctrl{1} & \gate{Z} & \qw \\
      \lstick{$t$} & \targ{}  & \gate{Z} & \qw
    \end{quantikz}
    \]

    \smallskip
    \small $Z$ on the target propagates backward to the control.
  \end{minipage}
  \caption{Pauli error propagation through CNOT. The naive use of two-qubit gates can turn one fault into multiple data errors.}
  \label{fig:cnot-propagation}
\end{figure}

With this in mind, we define fault tolerance as follows:

\begin{defbox}[label={def:fault-tolerance}]{Fault tolerance}
    For a code correcting up to $t$ errors per block, a procedure is \emph{fault-tolerant} if any pattern consisting of at most $s \le t$ faults during the procedure, together with at most $r \le t-s$ pre-existing errors on each input block, produces at most a correctable number of errors on each output block, that is, at most $t$ errors per output block, and does not alter the encoded logical action.  
\end{defbox}

To satisfy the fault tolerance condition, we design encoded operations such that a failure anywhere during the implementation can only propagate to a small number of qubits on each block. The most robust way to achieve this is via \emph{transversality}.

\begin{defbox}[label={def:Transversal Gate}]{Transversal gate}
    Let a code block consist of $n$ physical qubits. An encoded gate acting on one or more blocks is \emph{transversal} if it is implemented as a tensor product of gates that couple only corresponding qubits across blocks and never couple two qubits within the same block. For a one-block gate,
    \begin{equation}
       U_{\mathrm{trans}}=\bigotimes_{i=1}^n U_i. 
    \end{equation}
    For a two-block transversal gate, the $i\numth$ qubit of one block may interact only with the $i\numth$ qubit of the other block.
\end{defbox}

We demonstrate these ideas with the following example using the Steane code:

\begin{example}[label={example:steane-code}]{The Steane code}
    The $7$-qubit Steane code is a $\llbracket7,1,3\rrbracket$ CSS code. A convenient choice of stabilizer generators is
    \begin{align}
        g_1^{(X)} &= X_1X_2X_3X_4, & g_1^{(Z)} &= Z_1Z_2Z_3Z_4\,,\\
        g_2^{(X)} &= X_1X_2X_5X_6, & g_2^{(Z)} &= Z_1Z_2Z_5Z_6\,,\\
        g_3^{(X)} &= X_1X_3X_5X_7, & g_3^{(Z)} &= Z_1Z_3Z_5Z_7\,.
    \end{align}
    The codespace is the simultaneous $+1$ eigenspace of these six commuting operators. The logical operators are $\overline{Z} \equiv Z_1 Z_2 Z_3 Z_4 Z_5 Z_6 Z_7$ and $\overline{X} \equiv X_1 X_2 X_3 X_4 X_5 X_6 X_7$.\vspace{2.5mm}
    
    The Steane code is self-dual as a CSS code, which implies that transversal Hadamard acts logically:
    \begin{equation}
        \barH = H^{\otimes 7},
        \qquad
        H^{\otimes 7} \barX H^{\otimes 7} = \barZ,
        \qquad
        H^{\otimes 7} \barZ H^{\otimes 7}= \barX. 
    \end{equation}
    Likewise, transversal CNOT between two Steane blocks implements the logical CNOT,
    \begin{equation}
       \overline{\mathrm{CNOT}} = \mathrm{CNOT}^{\otimes 7}. 
    \end{equation}
    A phase gate is also transversal up to the usual convention for the logical phase. With the logical operators above, one finds
    \begin{equation}
     S^{\otimes 7}\,\barX\,(S^{\otimes 7})^{\dagger} = -i\barX\barZ \qquad\text{and}
    \quad
    S^{\otimes 7}\,\barZ\,(S^{\otimes 7})^{\dagger} = \barZ \,,   
    \end{equation}
    so $S^{\otimes 7}$ implements $\barS^{\dagger}$ and $(S^{\dagger})^{\otimes 7}$ implements $\barS$.
\end{example}

\subsection{Syndrome extraction schemes}\label{subsec:syndrome-extraction-schemes}
Consider an \(n\)-qubit stabilizer code with stabilizer group \(S \subset \mathcal{P}_n\) and generators
\begin{equation}
    S = \langle S_1,\dots,S_r \rangle\,,\qquad S_j \in \{I,X,Y,Z\}^{\otimes n}\,.
\end{equation}
For a CSS code, we can choose generators such that
\begin{equation}
    S = \langle S^{(X)}_1,\dots,S^{(X)}_{r_X}, S^{(Z)}_1,\dots,S^{(Z)}_{r_Z} \rangle\,,
\end{equation}
where each \(S^{(X)}_a\) contains only \(X\) and \(I\), and each \(S^{(Z)}_b\) contains only \(Z\) and \(I\). The goal of \emph{syndrome extraction} is to measure the eigenvalues \(s_j \in \{\pm 1\}\) of the generators \(S_j\) fault-tolerantly, that is, such that a single fault does not propagate to multiple data qubits. We will discuss three different fault-tolerant schemes for syndrome extraction, in the sense formalized later in \cref{def:Fault-tolerant error-correction gadget}.

\subsubsection{Shor scheme}\label{sec:shor_EC}
Peter Shor introduced~\cite{shor1996fault-tolerant} the first method for fault-tolerant syndrome extraction. Shor's method extracts syndromes stabilizer-by-stabilizer, utilizing cat states or GHZ states. For a weight-$w$ stabilizer, Shor's scheme prepares an ancilla in the $w$-qubit state:
\begin{equation}
    \ket{\text{cat}_w} = \frac{1}{\sqrt{2}} \left( \ket{0}^{\otimes w} + \ket{1}^{\otimes w} \right)\,.
\end{equation}
The protocol proceeds in three steps:
\begin{enumerate}
    \item \textbf{Preparation and verification}: The cat state is prepared using a cascade of CNOT gates. However, a single fault during this preparation circuit could easily create a weight-2 error in the cat state. To prevent this from corrupting the data qubits, the cat state must be rigorously verified. Note that a cat state is a repetition code, one uses additional ancilla qubits to measure the parity of adjacent qubits in the cat state (checking that it is indeed in the repetition space). If any verification check fails, the corrupted cat state is instantly discarded, and a new one is prepared.

     \item \textbf{Transversal CNOT gates}: Once accepted by the verification checks, the cat state interacts with the data block via transversal CNOT gates. Crucially, the $i\numth$ qubit of the cat state interacts exclusively with the $i\numth$ data qubit involved in the stabilizer. Because the interaction is transversal, an error on the cat state can only ever propagate to a single data qubit.

     \item \textbf{Measurement and majority voting}: After the coupling, the ancilla is measured and the parity of its outcomes yields the stabilizer eigenvalue. Because ancilla preparation and measurement are noisy, a single extraction of the same stabilizer is not usually trusted. Instead, the measurement is repeated an odd number of times and a majority vote is taken. The data are corrected only after a consistent syndrome is obtained or after a decoding rule is applied to the repeated outcomes.
\end{enumerate}
Shor extraction is conceptually clean and applies to general stabilizer codes, not only CSS codes. Its main cost is ancilla overhead and time: a separate cat state is needed for each stabilizer measurement, and repeated measurements are typically required.

\subsubsection{Steane scheme}\label{subsubsec:steane-scheme}
Steane extraction is tailored to CSS codes and exploits the fact that the $X$-type and $Z$-type stabilizers arise from classical linear parity checks~\cite{steane1996error}. Instead of measuring each stabilizer individually, one uses an entire encoded ancilla block. To extract the syndrome of $Z$-type stabilizers (equivalently, to diagnose $X$ errors on the data), prepare an ancilla in the encoded state $\ket{\overline 0}$. Apply a transversal CNOT from the data block to the ancilla block. Then measure the ancilla transversally in the computational basis and classically decode the outcome using the parity-check matrix of the underlying classical code. The measured bit string contains the same syndrome information that would be produced by measuring the $Z$-type stabilizers directly.

Similarly, to extract the syndrome of $X$-type stabilizers (equivalently, to diagnose $Z$ errors), prepare $\ket{\overline{+}}=\barH\ket{\overline{0}}$, apply a transversal CNOT in the opposite orientation, measure the ancilla transversally in the $X$ basis, and classically decode the resulting outcomes. The fault tolerance mechanism is twofold:
\begin{enumerate}
    \item The data-ancilla interaction is transversal, so one gate fault creates at most one error in each block.
    \item The ancilla is itself an encoded state, so the syndrome is extracted in a highly redundant form that can be decoded classically.
\end{enumerate}
The subtle point is ancilla preparation. A fault during the preparation of $\ket{\overline 0}$ or $\ket{\overline +}$ may create a correlated, high-weight ancilla error. If the ancilla is used without further checks, that correlated error can propagate through the transversal CNOT gates to several data qubits. Therefore, the ancilla must either be \emph{verified} before use or \emph{decoded} afterward so that dangerous correlated faults can be identified and compensated in the Pauli frame. Both approaches appear in the literature; ancilla decoding is especially important when post-selection is expensive. Compared with Shor extraction, Steane extraction uses larger ancillas but fewer rounds of interaction and measurement. On CSS codes such as the Steane code itself, it is usually far more efficient.

\subsubsection{Knill scheme}\label{subsubsec:knill-scheme}
Knill's method performs error correction by \emph{teleporting} the logical state through a verified encoded Bell pair~\cite{knill2004scalablequantumcomputationpresence}.  Prepare two encoded blocks in the logical Bell state
\begin{equation}
    \ket{\overline{\Phi^+}}=\frac{1}{\sqrt2}\bigl(\ket{\overline 0\overline 0}+\ket{\overline 1\overline 1}\bigr)\,.
\end{equation}
One half of this Bell pair is then Bell-measured against the incoming data block, using only transversal gates and transversal measurements. As in ordinary quantum teleportation, the unknown logical state is transferred to the second half of the Bell pair, up to a Pauli correction determined by the Bell-measurement outcomes. The key observation is that the Bell measurement simultaneously reveals the error syndrome. In effect, teleportation and error correction are fused into a single step. Any correction can be tracked in software as an update of the \emph{Pauli frame}, so physical recovery operations need not be applied immediately. The Knill scheme has two important practical features:
\begin{enumerate}
    \item Ancilla verification can be done aggressively \emph{offline}. Bad Bell pairs are discarded before they ever interact with the data.
    \item Because the data block is consumed by a teleportation measurement, many faults are converted into known Pauli-frame updates rather than lingering as unknown physical errors.
\end{enumerate}
This often leads to excellent threshold estimates in architectures that can afford abundant ancilla factories. The tradeoff is higher preparation overhead and more demanding ancilla logistics.

\begin{table}[tb]
    \centering
    \begin{tblr}{
        width=\linewidth,
        colspec={X[c,m]X[c,m]X[c,m]X[c,m]},
        row{1}={font=\bfseries},
        cell{2-Z}{3}={halign=c},
        hlines}
        Scheme & Ancilla resource & Fault-tolerance mechanism & Typical tradeoff\\
        \hline
        Shor & Verified $w$-qubit cat state for a weight-$w$ stabilizer & One-to-one ancilla-data coupling; repeated measurement and majority vote & Broad applicability, but many ancillas and repeated rounds\\
        Steane & Verified encoded $|\overline{0}\rangle$ and $|\overline{+}\rangle$ blocks & Transversal block-wise syndrome transfer plus classical decoding & Very efficient for CSS codes, but ancilla preparation is nontrivial\\
        Knill & Verified encoded Bell pairs & Teleportation through a clean ancilla; corrections absorbed into the Pauli frame & High thresholds and flexible scheduling, but large ancilla overhead
    \end{tblr}
    \caption{Tradeoffs between various syndrome extraction schemes.}
    \label{tab:syndrome-extraction-schemes}
\end{table}

\subsection{Formalism for fault tolerance}\label{subsec:ft-formalism}
To make the threshold theorem precise, one must state exactly what a fault-tolerant gadget is required to do~\cite{aliferis2005quantum,gottesman2024surviving}. Let $\code$ be an $\llbracket n,k,2t+1 \rrbracket$ code, so that up to $t$ physical errors are correctable on each block.

\subsubsection{Error sets and filters}\label{subsubsec:error-sets-filters}
For a subset $R\subseteq \{1,\dots,n\}$ of the $n$ qubit indices, let $\mathcal{E}(R)$ denote the linear span of Pauli operators supported on $R$. The subspace of states with at most $r$ errors relative to the codespace is
\begin{equation}
    \mathcal V_r = \sum_{|R|\le r} \mathcal{E}(R)\code.
\end{equation}
An ideal \emph{$r$-filter} is the projector onto $\mathcal V_r$. Informally, an $n$-qubit block state passes an $r$-filter exactly when it differs from some codeword by errors on at most $r$ qubits. This language is convenient because fault tolerance is not merely a statement about the final decoded state, it is also a statement about the intermediate encoded state remaining inside the correctable sector.

\begin{defbox}[label={def:Fault-tolerant error-correction gadget}]{Fault-tolerant error-correction gadget}\label{def:FT_EC_gadget}
    An error-correction gadget for an $\llbracket n,k,2t+1 \rrbracket$ code is \emph{$t$-fault-tolerant} if for every $r,s\ge 0$ with $r+s\le t$ the following hold:
    \begin{enumerate}
        \item \textbf{Recovery property.} If the gadget contains at most $s$ faults, then its output passes an $s$-filter.
        \item \textbf{Correctness property.} If the input passes an $r$-filter and the gadget contains at most $s$ faults, then ideal decoding of the output gives the same state as ideal decoding of the input.
    \end{enumerate}
\end{defbox}

The first property says that faults occurring during error correction leave behind at most a correctable number of residual errors. The second property says that if the total number of pre-existing input errors and faults inside the gadget is at most $t$, then the output is correctly decodable to the same logical state. For distance-3 codes ($t=1$), the definition reduces to two memorable conditions:
\begin{enumerate}
  \item If one fault occurs during error correction, the output block has at most one error;
  \item If the input block has at most one error and the gadget is otherwise fault-free, the output is logically correct.
\end{enumerate}

\subsubsection{Fault-tolerant gate gadgets}\label{subsec:fault-tolerant-gate-gadgets}
A gate gadget must also satisfy two requirements. First, it must control how pre-existing errors and internal faults propagate to the outputs. Second, after decoding, it must implement the intended logical gate.

\begin{defbox}[label={def:Fault-tolerant gate gadget}]{Fault-tolerant gate gadget}
    Consider a logical gate acting on one or more encoded blocks of an $\llbracket n,k,2t+1 \rrbracket$ code. A gate gadget is \emph{$t$-fault-tolerant} if for every $r,s\ge 0$ with $r+s\le t$:
    \begin{enumerate}
        \item \textbf{Error-propagation property.} If each input block passes an $r$-filter and the gadget contains at most $s$ faults, then each output block passes an $(r+s)$-filter.
        \item \textbf{Logical correctness.} If each input block passes an $r$-filter and the gadget contains at most $s$ faults, then ideal decoding of the output is the same as applying the ideal logical gate to the ideally decoded input. Equivalently, if $D$ denotes ideal decoding and $\mathcal{G}^{U}_{s}$ denotes any implementation of the logical gate $U$ with at most $s$ faults, then logical correctness requires
        \begin{equation}
            D\!\left(\mathcal{G}^{U}_{s}(\rho)\right) = U D(\rho) U^\dagger
        \end{equation}
        for every admissible input state $\rho$ satisfying the filter condition above.
    \end{enumerate}
\end{defbox}

For transversal gadgets, the propagation property is usually immediate: a single physical fault affects only the corresponding output qubits, so the number of errors per block grows additively rather than catastrophically.

\subsubsection{Rectangles and extended rectangles}\label{subsec:rectangles}
Threshold proofs are organized around the replacement of each ideal logical location by a level-1 gadget. The gadget implementing a logical gate together with the error-correction gadgets that follow it on every output block is called a \emph{rectangle} (Rec). If one includes the leading error-correction gadgets on the input blocks as well, one obtains an \emph{extended rectangle} (exRec).

The exRec is the natural unit of analysis because faults in the leading error correction can combine with faults inside the gate gadget. A distance-3 exRec is called \emph{good} if it contains at most one fault. The usual exRec-correctness lemma states that every good exRec simulates the ideal logical location correctly. This lemma is the bridge from local gadget properties to the global threshold theorem.

\subsubsection{Strong and weak fault-tolerant error correction}\label{subsec:strong-weak-ft}
In the literature, one encounters both \emph{strong} and \emph{weak} notions of fault-tolerant error correction~\cite{tansuwannont2023adaptive,gottesman2024surviving}:
\begin{itemize}
  \item \emph{Strong fault-tolerant EC} requires the recovery property unconditionally: if the EC gadget contains at most $s$ faults, then regardless of the detailed input state the output lies within the $s$-error sector of the code. This notion is highly composable and is especially convenient in recursive proofs.
  \item \emph{Weak fault-tolerant EC} imposes correctness only for the class of inputs that can arise during a fault-tolerant computation, for example inputs already known to pass an $r$-filter for suitable $r$. In other words, the gadget is guaranteed to work on \emph{admissible} states even if it need not behave well on arbitrary adversarial inputs.
\end{itemize}
Weak fault tolerance is often sufficient in practice because encoded gadgets are never fed completely arbitrary states; they are fed outputs of previous gadgets that already satisfy filter conditions. Strong fault tolerance is conceptually cleaner and easier to compose, but it can be more expensive to realize. The distinction matters when optimizing ancilla verification, post-selection, and decoder scheduling.

\subsection{Noise models}\label{subsec:noise-models}
Threshold theorems do not apply to arbitrary noise without assumptions. The meaning of the physical error rate depends on the class of allowed correlations in space and time. Three broad models appear repeatedly.

\subsubsection{Stochastic noise}\label{subsubsec:stochastic-noise}
The simplest model assigns a fault probability to each location. In an \emph{independent stochastic} model, each location fails independently with probability $p$. More general and more useful is \emph{local stochastic noise}: for every set $R$ of locations~\cite{aliferis2005quantum},
\begin{equation}
    \Pr[\text{all locations in }R\text{ are faulty}] \le p^{\abs{R}}\,.
\end{equation}
This bound allows arbitrary correlations among faults, provided large correlated sets are exponentially suppressed. The local stochastic model is particularly convenient for combinatorial threshold proofs because the probability of a malignant set of $m$ faulty locations is at most $p^m$. A common special case is stochastic Pauli noise, in which each faulty location is followed by a Pauli channel,
\begin{equation}
    \mathcal N(\rho)=\sum_{P\in\Pauli_n} q_P P\rho P\,, \qquad \sum_P q_P=1\,.
\end{equation}
Because stabilizer propagation is easy to analyze under Pauli channels, this model is ubiquitous in simulations and decoder benchmarking.

\subsubsection{Non-Markovian noise}\label{subsubsec:non-markovian-noise}
A Markovian noise model describes each time step by a completely positive trace-preserving map that depends only on the current system state. In contrast, \emph{non-Markovian} noise allows the environment to retain memory. A standard Hamiltonian model writes
\begin{equation}
    H(t)=H_S(t)+H_B(t)+H_{SB}(t)\,,
\end{equation}
where $H_{SB}(t)$ couples the computer to its environment. Errors need not be stochastic events with well-defined probabilities; they may interfere coherently across time.

Threshold theorems nevertheless survive under suitable locality assumptions~\cite{aharonov1996fault}. Roughly speaking, if the system-bath coupling is local and sufficiently weak in operator norm, then the sum of all bad fault paths remains bounded and concatenation still suppresses the effective logical noise. The proof is more delicate because amplitudes, not probabilities, are combined, and coherent cancellation or reinforcement must be controlled.

\subsubsection{Adversarial noise}\label{subsubsec:adversarial-noise}
The strongest model allows the faulty operation at each bad location to be chosen adversarially, possibly depending on the entire fault pattern, subject only to a locality or norm bound. This worst-case viewpoint is often formulated in terms of the diamond norm distance between the implemented and ideal operations.

Adversarial noise is mathematically severe: one no longer assumes that errors are random Pauli flips or even stochastic events. The strength of fault tolerance in this regime comes from geometry and locality rather than from any favorable probabilistic structure. Rigorous thresholds exist for several adversarial or coherent models, but the proven threshold constants are generally smaller than the thresholds observed under phenomenological stochastic noise.

\begin{table}[tb]
    \centering
    \begin{tblr}{
        width=\linewidth,
        colspec={X[c,m]X[c,m]X[c,m]},
        row{1}={font=\bfseries},
        cell{2-Z}{3}={halign=c},
        hlines}
        Noise model & Typical assumption & What threshold proofs track\\ \hline
        Independent & Fault sets of size $m$ have probability at most $p^m$ & Counting malignant fault sets inside gadgets\\
        Non-Markovian coherent noise & Local system-bath coupling with small operator norm & Bounding the norm of the sum of bad fault-path amplitudes\\
        Adversarial local noise & Arbitrary faulty maps subject to locality and strength bounds & Worst-case logical deviation, often in diamond norm
    \end{tblr}
    \caption{Comparison of various noise models.}
    \label{tab:noise-models}
\end{table}

\subsection{Threshold theorem}\label{subsec:threshold-thm}
The discovery of QEC was a monumental step, but it introduced a seemingly fatal paradox: the very circuits used to perform error correction (encoding, syndrome measurement, and recovery) are themselves composed of noisy physical gates. This raises a critical question: does the process of error correction introduce more errors than it fixes? The answer, which provides the theoretical foundation for all of scalable quantum computing, is given by the \emph{threshold theorem}. Informally, it states: 

\begin{quote}
    If the error rate $p$ of each physical component in a quantum computer (qubit preparation, gates, measurements, memory) is below a certain constant value, the accuracy threshold $p_{\mathrm{th}}$, then it is possible to perform an arbitrarily long quantum computation with an arbitrarily low logical error rate $\epsilon$.
\end{quote}

The overhead required to achieve this fault tolerance, in terms of the number of physical qubits and gates, scales efficiently, typically as a polynomial in the logarithm of the circuit size and $1/\epsilon$. The theorem is best understood through a quantitative recursion. Let $p_L^{(\ell)}$ denote the failure probability of a level-$\ell$ exRec, and let $A$ be an upper bound on the number of malignant sets of size $t+1$ inside the largest level-$1$ exRec. Then for sufficiently small $p$,
\begin{equation}
    p_L^{(1)} \le A p^{t+1} + O(p^{t+2})\,.\label{eq:level1bound}
\end{equation}
At higher concatenation levels the same reasoning applies recursively,
\begin{equation}
    p_L^{(\ell+1)} \le A\bigl(p_L^{(\ell)}\bigr)^{t+1}\,.\label{eq:recursion}
\end{equation}
The logic is simple: a level-$(\ell+1)$ exRec can fail only if enough level-$\ell$ sub-gadgets fail in a malignant pattern.

\subsubsection{Distance-3 case}\label{subsubsec:distance-3-threshold}
For the most common pedagogical case $t=1$,~\cref{eq:recursion} becomes
\begin{equation}
    p_L^{(\ell+1)}\le A\bigl(p_L^{(\ell)}\bigr)^2\,.
\end{equation}
Defining $\lambda = Ap$, one obtains by induction
\begin{equation}
    p_L^{(\ell)} \le \frac{1}{A}\lambda^{2^{\ell}}\,.\label{eq:doubleexp}
\end{equation}
Hence, if $\lambda < 1$, or equivalently if
\begin{equation}
    p < p_{\mathrm{th}} \approx \frac{1}{A}\,,
\end{equation}
then the logical error rate decreases doubly exponentially with the concatenation level. This is the hallmark of a threshold.

\subsubsection{General $t$}\label{subsubsec:general-t}
For a code correcting $t$ errors, rescale
\begin{equation}
    y_{\ell} = A^{1/t}p_L^{(\ell)}\,.
\end{equation}
Then,~\cref{eq:recursion} implies $y_{\ell+1}\le y_{\ell}^{t+1}$, so
\begin{equation}
p_L^{(\ell)} \le A^{-1/t}\bigl(A^{1/t}p\bigr)^{(t+1)^{\ell}}.
\label{eq:generalrecursion}
\end{equation}
Thus, the threshold estimate is
\begin{equation}
    p_{\mathrm{th}} \approx A^{-1/t},
\end{equation}
up to higher-order corrections and model-dependent constants.

\subsubsection{Why the overhead is efficient}\label{subsubsec:why-overhead-is-efficient}
Suppose an ideal computation contains $N$ logical locations. If each level-$\ell$ location fails with probability at most $p_L^{(\ell)}$, then the total failure probability is at most $Np_L^{(\ell)}$ by the union bound. Requiring $Np_L^{(\ell)}\le \epsilon$ and using~\cref{eq:generalrecursion} shows that it suffices to choose
\begin{equation}
    (t+1)^{\ell}=O\!\left(\log\frac{N}{\epsilon}\right)\,,\qquad \ell = O\!\left(\log\log\frac{N}{\epsilon}\right).
\end{equation}
If one level of encoding uses $n$ physical qubits per logical qubit, then $\ell$ levels use $n^{\ell}$ physical qubits, so the space overhead scales as
\begin{equation}
    n^{\ell}=\left(\log\frac{N}{\epsilon}\right)^{\log_{t+1} n}\,,
\end{equation}
which is polynomial in $\log(N/\epsilon)$. The gate overhead obeys a similar polylogarithmic bound. We note that the constant $p_{\mathrm{th}}$ is not universal. It depends on the code, the gadget construction, the decoder, the connectivity constraints, and the noise model. Rigorous threshold values derived from malignant-set counting are typically more conservative than thresholds estimated numerically for specific architectures. Concatenation is the recursive architecture that turns the local gadget properties of the previous sections into asymptotically reliable computation~\cite{knill1996threshold,aliferis2005quantum,gottesman2024surviving}.

\subsection{Recursive encoding}\label{subsec:recursive-encoding}
Let $\mathcal C$ be an $\llbracket n,k,d \rrbracket$ code. In one level of encoding, each logical block is mapped to $n$ physical qubits. In two levels of concatenation, each of those $n$ qubits is itself re-encoded using the same code. Continuing recursively, a level-$\ell$ encoded block uses $n^{\ell}$ physical qubits. For self-concatenation of an $\llbracket n,1,d \rrbracket$ code, the effective parameters are $\llbracket n^{\ell},1,d^{\ell} \rrbracket$. The multiplicative distance law follows because an error must cause a logical fault at the outer level and, within each affected outer-level qubit, must also cause a logical fault at the next inner level, and so on recursively. As an example, concatenating the Steane $\llbracket 7,1,3 \rrbracket$ code twice yields a $\llbracket 49,1,9 \rrbracket$ code, and concatenating by $\ell$ levels gives a $\llbracket 7^{\ell},1,3^{\ell} \rrbracket$ code. 

\subsection{Recursive gadgets}\label{subsec:recursive-gadgets}
Concatenation is not merely recursive state encoding. Every logical location in the ideal circuit is replaced by a level-1 gadget. Then every physical location inside that gadget is replaced by its own level-$1$ gadget to create a level-2 gadget, and so on. Thus a level-$\ell$ encoded gate is a gadget built from level-$(\ell-1)$ encoded gates, measurements, preparations, and error-correction steps. This hierarchical construction ensures self-similarity: once one proves that good level-1 exRecs are correct, the same argument applies at every higher level because a level-$\ell$ exRec behaves as a single effective location for the level-$(\ell+1)$ simulation.

\subsection{Why concatenation works}\label{subsec:why-concatenation}
Suppose a code corrects one error and the level-$1$ gadgets are designed so that any single fault inside an exRec is harmless. Then a logical failure at level $1$ requires at least two faults in a malignant arrangement. At level $2$, a logical failure requires at least two level-$1$ exRec failures, each of which already requires two physical faults, so the dominant contribution is fourth order. Repeating this recursively yields the double-exponential suppression in~\cref{eq:doubleexp}. This recursive suppression is the conceptual heart of concatenated fault tolerance: \emph{the effective error model becomes progressively more benign at higher levels of encoding, provided the physical noise starts below threshold}.

\subsection{Why transversality is powerful}\label{subsec:why-transversality}
Suppose a code corrects one error per block. If a single $X$ error enters a transversal CNOT on the control block, the error may spread to the corresponding qubit in the target block, but each block still contains only one error. The gate is therefore compatible with distance-3 correction. This is exactly the right notion of safe error spread: faults may duplicate across blocks, but they may not proliferate \emph{within} a block. Many CSS codes admit useful transversal gates. For example:
\begin{itemize}
    \item transversal CNOT is logical CNOT for any CSS code built from nested classical codes;
    \item self-dual CSS codes often admit transversal Hadamard;
    \item doubly-even CSS codes can admit a transversal phase gate, sometimes up to adjoint convention, as in the Steane code.
\end{itemize}
Despite their usefulness, transversal gates alone can never provide a universal gate set on a finite-dimensional quantum code. In fact, the \emph{Eastin-Knill theorem} says that no quantum error-correcting code that detects arbitrary single-subsystem errors can realize a universal set of logical gates using only transversal unitary gates~\cite{eastin2009restrictions}. It does not say that transversal gates are rare or unhelpful; rather, it says that \emph{any} finite code with nontrivial error-detecting power must sacrifice universality if one insists on transversality alone. In practice, this means that a fault-tolerant architecture needs an additional ingredient beyond transversal Clifford operations. Common remedies include magic-state distillation and injection, code switching, gauge fixing, lattice surgery, and pieceable fault tolerance.

Transversality is a special case of a broader geometric concept: \emph{locality-preserving logical gates}. These are logical operations that map local errors to local errors and can therefore be implemented without long-range spread of damage. For topological stabilizer codes, such gates are severely constrained. The Bravyi-K\"{o}nig theorem says that for a $D$-dimensional topological stabilizer code, every locality-preserving logical gate belongs to the $D\numth$ level of the Clifford hierarchy~\cite{bravyi2013classification}. The consequence is especially striking in two dimensions. In a two-dimensional topological stabilizer code, every locality-preserving logical gate is Clifford. Therefore, no two-dimensional topological stabilizer code can realize a universal fault-tolerant gate set by transversal or any other locality-preserving mechanism alone. In three dimensions one may obtain logical gates from the third level of the Clifford hierarchy, such as CCZ in suitable codes, but even that is still not universal by itself.

Transversal gates are indispensable because they are simple, local, and naturally fault-tolerant. But the Eastin-Knill and Bravyi-K\"{o}nig theorems show that fault-tolerant universality cannot come from transversality alone. Modern architectures therefore combine a ``cheap'' fault-tolerant gate set, usually Clifford operations, with a more expensive resource-generation procedure that supplies the missing non-Clifford ingredients.

\newpage

\section{Fault tolerance II}\label{sec:fault-tolerance-pt2}

In this second part on fault tolerance, we will cover three topics: (1) the seminal work on a constant-overhead FTQC scheme that uses a family of qLDPC codes in~\cref{sec:constant_overhead_ftqc,sec:qldpc_noiseless_decoding_threshold,sec:qldpc_noisy_decoding_threshold,sec:constant_overhead_ftqc_threshold}, (2) a magic state distillation scheme that distills 10 noisy magic states to two magic states with suppressed error in~\cref{sec:magic_state_distillation}, (3) code-switching protocols that allows for an implementation of a universal transversal gate set by using two codes.

We will first discuss the proof of the existence of a decoding threshold for exponential error suppression in the code distance for qLDPC codes in the case of ideal syndrome extraction (\cref{sec:qldpc_noiseless_decoding_threshold}) and noisy syndrome extraction (\cref{sec:qldpc_noisy_decoding_threshold}).
Then, in~\cref{sec:constant_overhead_ftqc_threshold}, we will discuss the overall FTQC scheme and derive a threshold for the entire scheme in detail.
Next, in~\cref{sec:magic_state_distillation} we will discuss a simple magic state distillation scheme based on a 4-qubit error-detecting code and show how it distills 10 copies of noisy magic states to two magic states with quadratic suppression of their noise rate.
Lastly, in~\cref{subsec:code_switching}, we will discuss two code-switching protocols as an alternative protocol to magic state distillation to enable a universal FTQC.

\subsection{Constant-overhead FTQC scheme using qLDPC codes}\label{sec:constant_overhead_ftqc}
Here we discuss the FTQC scheme with a family of qLDPC codes with non-vanishing asymptotic encoding and distance rate proposed in~\cite{gottesman2014faulttolerantquantumcomputationconstant}.
Namely, we consider a family of qLDPC codes $Q_1,Q_2,\dots$ with parameters $\llbracket n_i,k_i,d_i\rrbracket$ such that each code has a non-vanishing asymptotic encoding rate $R=\lim_{i\rightarrow\infty} \frac{k_i}{n_i} > 0$ and a non-constant distance $\lim_{i\rightarrow\infty} d_i =\infty$.
We will first discuss two important results showing that there exists a decoder for such a family of codes such that errors are suppressed exponentially in the code distance $d_i$ whenever physical error rates are below some threshold values.
First, in~\cref{sec:qldpc_noiseless_decoding_threshold} we will show a decoding threshold when the syndrome extraction circuit is assumed to be ideal.
Then, in~\cref{sec:qldpc_noisy_decoding_threshold} we will show that such decoding threshold still exists even when the syndrome extraction circuit is noisy, particularly by doing multiple cycles of syndrome extraction.
Lastly in~\cref{sec:constant_overhead_ftqc_threshold}, we will give a detailed proof on the existence of a threshold for this constant-overhead FTQC scheme.

Here we will consider a \emph{local stochastic noise} model where the probability of faults of a certain size occurring decays exponentially with the fault size.

\begin{defbox}[label=def:local_stochastic_noise]{Local stochastic noise}
    For a quantum circuit $\mathcal{C}$ with locations $\loc(\mathcal{C})$, a local stochastic noise model with rate $p$ implies that the probability of faults on all locations in a subset $S\subseteq\loc(\mathcal{C})$ is given by:
    \begin{equation}
        \Pr[\text{All locations in $S$ are faulty}] \leq p^{|S|} \,.
    \end{equation}
    For an $n$-qubit system, a local stochastic noise with rate $p$ implies that the probability of faults on all qubits in a set $S\subseteq[n]$ is given by
    \begin{equation}
        \Pr[\text{All qubits in $S$ are faulty}] \leq p^{|S|} \,.
    \end{equation}
\end{defbox}

\subsection{Decoding threshold for ideal syndrome extraction}\label{sec:qldpc_noiseless_decoding_threshold}
Theorem 3 of Ref.~\cite{gottesman2014faulttolerantquantumcomputationconstant} shows that for a family $\{Q_i\}_i$ of $(r,c)$-qLDPC codes with parameters $\llbracket n_i,k_i,d_i \rrbracket$ with $n_i\rightarrow\infty$, $d_i\rightarrow\infty$, local stochastic noise with rate $p$, there exists a decoding algorithm and a threshold $p_0>0$ such that the logical error rate is $p_L = O(n_i(p/p_0)^{d_i/2})$ whenever $p<p_0$.
This result generalizes the result by Kovalev and Pryadko~\cite{Kovalev_2013} on the existence of a threshold for such qLDPC codes under independent noise model to local stochastic adversarial noise model.
We go through the proof of this result.

To obtain this threshold, we first consider a decoding algorithm that receives a syndrome string $s(F)\in\Z_2^l$ (corresponding to generators $g_1,\dots,g_l$ of stabilizer group $\mathcal{S}_i$) from Pauli error $F\in\hat{\mathcal{P}}_{n_i}$ and outputs a correction Pauli operator $E\in\hat{\mathcal{P}}_{n_i}$ with lowest weight that is consistent with syndrome $s(F)$ (choosing one at random if there are multiple minimum-weight consistent Paulis).
This decoder is successful whenever $EF\in\mathcal{S}_i$ and fails whenever $EF\in\mathcal{N}(\mathcal{S}_i)\setminus\mathcal{S}_i$, since $EF$ is a logical operation and has a trivial syndrome as $s(EF)=s(E)+s(F)=0^l$ because $s(E)=s(F)$.
When the latter happens, the decoding procedure results in a logical error.
Now we introduce a graph we call the \emph{code adjacency graph} that captures how the qubits are connected by the stabilizers.

\begin{defbox}[label=def:code_adjacency_graph]{Code adjacency graph}
    For an $n$-qubit stabilizer code $Q$ with stabilizer generators $g_1,\dots,g_l$, a simple graph $G$ is the \emph{code adjacency graph} of $Q$ if its vertices $V(G)$ and edges $E(G)$ are given by
    \begin{equation}
    \begin{gathered}
        V(G) = [1,n] 
        \quad\text{and}\\
        E(G) = \{(u,v) : \exists i\in[1,l],\; u,v\in\supp(g_i)\} \,.
    \end{gathered}
    \end{equation}
    Namely, the vertices of $G$ are labels for the $n$ qubits and two vertices are connected by an edge iff the two corresponding qubits are in the support of a stabilizer generator.
\end{defbox}

For an $(r,c)$-qLDPC code, each qubit is involved in at most $c$ generators and each generator has support on at most $r$ qubits.
Therefore, the degree of each vertex $u$ in the adjacency graph $G$ of an $(r,c)$-qLDPC as defined in~\cref{def:code_adjacency_graph} is bounded as
\begin{equation}\label{eqn:qldpc_adjacency_graph_bound}
    \deg(u) \leq (r-1)c \,.
\end{equation}
Note also that a generator $g_i$ corresponds to a clique in $G$, since qubits involved in $g_i$ are connected to each other.

Now, let us call a subset of vertices $W\subseteq V$ a \emph{cluster} if the induced subgraph $G(W)$ is connected.
For a Pauli $P\in\hat{\mathcal{P}}_{n_i}$, we say that a cluster $W\subseteq \supp(P)$ is a  \emph{disjoint cluster} of $P$ whenever there are no qubits in $\supp(P)$ that are in the border of $W$, that is, $\forall v\in\supp(P)\setminus W, \forall u\in W, (u,v)\notin E(G)$.
A disjoint cluster $W$ of $P$ is \emph{minimal} if there is no $N\subsetneq W$ such that $N$ is a  disjoint cluster of $P$.
Let us denote the set of all minimal disjoint cluster of $P$ as $\mathrm{MDC}(P)$. Now we will show the following lemma that connects failure of the decoding scheme above to the structure of errors across the $n_i$ qubits.

\begin{lembox}[label={lemma:decoder-fails}]{Decoder failure}
    If the decoder fails, then there exists a set of connected qubits $W\subseteq V(G)$ of size $|W| \geq d_i$ such that at least $|W|/2$ of qubits in it are faulty.
\end{lembox}

\begin{proof}
    Suppose that the decoding fails, that is, it chooses a correction Pauli $E$ such that $EF\in\mathcal{N}(\mathcal{S})\setminus\mathcal{S}$.
    Then, consider a cluster $M\in\mathrm{MDC}(EF)$ and $EF|_M$, the restriction of $EF$ to qubits in cluster $M$.
    Note that $EF|_M$ has the following properties:
    \begin{enumerate}
        \item $\supp(EF|_M) = M$, since if there is a qubit $q\in M$ that is not in $\supp(EF|_M)$ then we can always find a subset $N\subseteq M\setminus\{q\}$ that is a disjoint cluster of $EF$, contradicting that $M$ is minimal.
        \item $s(EF|_M)=0^l$ (it has a trivial syndrome). 
        Restriction $EF|_M$ corresponds to $l'\leq l$ syndromes (say, $g_1,\dots,g_{l'}$) while $EF|_{V\setminus M}$ corresponds to the remaining $l-l'$ (say, $g_{l'+1},\dots,g_l$). 
        This is because $\supp(EF|_M)$ is disconnected from $\supp(EF|_{V\setminus M})$, that is, $(u,v)\notin E$ for all $u\in\supp(EF|_M)$ and $v\in\supp(EF|_{V\setminus M})$ (by the definition of a disjoint cluster).
        So, $s(EF|_M)$ is a length-$l'$ sub-string of $s(EF)=0^l$.
        \item $EF|_M\in\mathcal{N}(\mathcal{S}_i)$.
        The only generators with support on $M$ are $g_1,\dots,g_{l'}$, each giving syndrome $0$.
        So $EF|_M$ commutes with each of these $l'$ generators, thus commute with all generators.
        \item There exists $M\in\mdc(EF)$ where $EF|_M \notin\mathcal{S}_i$ (it is not a stabilizer).
        If there is no such $M$, then all $M\in\mdc(EF)$ is a stabilizer, which means that $EF$ is also a stabilizer, contradicting the assumption that the decoding has failed.
    \end{enumerate}
    
    Consider $W\in\mdc(EF)$ such that $EF|_W \notin\mathcal{S}_i$ (c.f., property 4 above).
    By property 3, it also holds that $EF|_W\in\mathcal{N}(\mathcal{S}_i)$.
    Thus, for any such $W\in\mdc(EF)$, it holds that $|W|\geq d_i$ since $EF|_W \in\mathcal{N}(\mathcal{S}_i)\setminus\mathcal{S}_i$, that is, $EF|_W$ is a non-trivial logical operator of $\mathcal{S}_i$ so that its weight must be at least the code distance $d_i$ since otherwise it is detectable. Now we show that there are at least $|W|/2$ faulty qubits in $W$, that is, $\wt(F|_W) \geq |W|/2$.
    First, we show that $\wt(E|_W)\leq \wt(F|_W)$.
    The decoder always chooses a minimum weight Pauli $E$ satisfying $s(E)=s(F)$ so that 
    \begin{equation}
        \wt(E|_W)+\wt(E|_{V\setminus W}) = \wt(E)\leq\wt(F) = \wt(F|_W)+\wt(F|_{V\setminus W}) \,.
    \end{equation}
    If $\wt(E|_W)> \wt(F|_W)$ then we can choose $E'$ such that $E'|_W=F|_W$ and $E'|_{V\setminus W}=E|_{V\setminus W}$ so that $s(E')=s(F|_W)+s(E|_{V\setminus W}) = s(F|_W)+s(F|_{V\setminus W}) =s(F)$ and $\wt(E')<\wt(E)$ ($s(E|_{V\setminus W})=s(F|_{V\setminus W})$ follows from property 2 above that $s(EF|_{V\setminus W})=0$).
    This contradicts that $E$ is a minimum weight error with syndrome equal to $s(F)$, therefore $\wt(E|_W)\leq \wt(F|_W)$ must be true.
    So,
    \begin{equation}
        |W| = \wt(EF|_W) \leq \wt(E|_W)+\wt(F|_W) \leq 2\wt(F|_W)
    \end{equation}
    which implies that $\wt(F|_W) \geq |W|/2$, completing the proof
\end{proof}

We have shown that a decoding failure implies that there exists a set of connected qubits $W\subseteq V(G)$ of size $|W| \geq d_i$ such that at least $|W|/2$ of the qubits in it are faulty.
Thus, the probability of a logical error resulting from decoding failure for a Pauli error $F$ and (incorrect) correction Pauli $E$ can be bounded as
\begin{equation}\label{eqn:qldpc_decoding_logical_error_bound1}
\begin{aligned}
    \Pr[\text{Logical error}] &= \Pr[EF\in\mathcal{N}(\mathcal{S}_i)\setminus\mathcal{S}_i] \\
    &\leq \Pr{\exists W\subseteq V(G) : W\text{ connected}\,,\; |W|\geq d_i \,,\; \wt(F|_W)\geq\lceil|W|/2\rceil} \\
    &\leq \sum_{\substack{ W\subseteq V(G) :\\ |W|\geq d_i \,,\\ W\text{ connected} }} \Pr[\wt(F|_W)\geq\lceil|W|/2\rceil] \,,
\end{aligned}
\end{equation}
by using a union bound to obtain the second inequality.
To obtain an upper bound for the logical error in terms of physical error rate $p$ and the code parameters, we need to put a bound on the number of the connected sets in the sum of~\cref{eqn:qldpc_decoding_logical_error_bound1} and the probability of more than half erroneous qubits in a set.

For a qubit $v\in V(G)$, the number of subsets of connected qubits $W\subseteq V(G)$ of size $|W|=q$ containing $v$ can be upper bounded using Lemma 2 of Ref.~\cite{gottesman2014faulttolerantquantumcomputationconstant} as
\begin{equation}\label{eqn:number_of_clusters_upper_bound}
    |\{ W\subseteq V(G) : |W|=q \,,\; W\text{ connected} \,,\; v\in W \}| \leq \deg(v)e^{q-1} \,.
\end{equation}
Since there are $n_i$ qubits in total and the degree of each qubit is bounded by $(r-1)c$ (c.f.~\cref{eqn:qldpc_adjacency_graph_bound}), therefore the total number of connected subsets of size $q$ in $G$ is bounded as
\begin{equation}
    |\{W\subseteq V(G) : |W|=q \,,\; \text{$W$ connected}\}| \leq n_i ((r-1)c e)^{q-1} \,.
\end{equation}
On the other hand, for a subset $W\subseteq V(G)$, the probability that there are $\lceil |W|/2\rceil$ erroneous qubits in $W$ is bounded by
\begin{equation}
\begin{aligned}
    \Pr[\text{$\lceil q/2\rceil$ errors in $W$}] &\leq \sum_{T\subseteq W : |T|=\lceil |W|/2\rceil} \Pr[\text{all qubits in $T$ are faulty}] \\
    &\leq \binom{q}{\lceil q/2\rceil} p^{\lceil q/2\rceil} 
    \leq 2^q p^{\lceil q/2\rceil}
\end{aligned}
\end{equation}
by the union bound and the local stochastic noise assumption (c.f.~\cref{def:local_stochastic_noise}).

Thus, we can further bound the inequality in  ~\cref{eqn:qldpc_decoding_logical_error_bound1} as
\begin{equation}
    \begin{aligned}
        \Pr[\text{Logical error}] &\leq \sum_{\substack{ W\subseteq V(G) :\\ |W|\geq d_i \,,\\ W\text{ connected} }} \Pr[\wt(F|_W)\geq\lceil|W|/2\rceil] \\
        &\leq \sum_{q=d_i}^\infty |\{ W\subseteq V(G) : |W|=q \,,\; W\text{ connected} \,,\; v\in W \}| \\
            &\quad \times \Pr[\text{$\lceil q/2\rceil$ errors in $W$}] \\
        &\leq \sum_{q=1}^\infty n_i((r-1)c e)^{q-1} 2^q p^{\lceil q/2\rceil} \\
        &\leq \frac{n_i}{(r-1)ce} \sum_{q=1}^\infty (2(r-1)ce\sqrt{p})^q \,. 
    \end{aligned}
\end{equation}
The series $\sum_{q=1}^\infty (2(r-1)ce\sqrt{p})^q$ is a geometric series and therefore converges to $\frac{(2(r-1)ce\sqrt{p})^{d_i}}{1-(2(r-1)ce\sqrt{p})}$ whenever $2(r-1)ce\sqrt{p} < 1$.
So we can define the threshold as $p_0 = (2(r-1)ce)^{-2}$ so that whenever $p<p_0$ we have $(2(r-1)ce\sqrt{p} < 1$ and by denoting $\alpha=(r-1)ce$ we have
\begin{equation}
\begin{aligned}
    \Pr[\text{Logical error}] &\leq \frac{n_i}{\alpha} \frac{(2\alpha \sqrt{p})^{d_i}}{1-(2\alpha\sqrt{p})} 
    \leq \frac{n_i}{\alpha(1-(2\alpha\sqrt{p}))} \Big(\frac{p}{p_0}\Big)^{d_i/2} \,,
\end{aligned}
\end{equation}
since $2\alpha \sqrt{p} = \sqrt{p/p_0}$.
So, the logical error probability $p_L$ is bounded as $p_L \leq O(n_i (p/p_0)^{d_i/2})$ as claimed (the $\sqrt{p}$ in the denominator is absorbed into the $O$ and $\alpha=(r-1)ce$ is a constant independent of the code parameter).

\subsection{Decoding threshold for noisy syndrome extraction}\label{sec:qldpc_noisy_decoding_threshold}
In the last section, we showed that a family of qLDPC codes $\llbracket n_i,k_i,d_i\rrbracket$ with growing physical qubits and distance ($n_i\rightarrow\infty$ and $d_i\rightarrow\infty$ as $i\rightarrow\infty$) can be decoded with exponential logical error suppression whenever the physical error rate is below a certain threshold, under local stochastic noise model.
However, this assumes that the syndrome extraction process itself is noiseless, that is, the syndrome extraction (SE) circuit implementing stabilizer measurements is ideal.
In a realistic scenario, components inside the SE circuit can themselves be faulty.
To model a realistic scenario more faithfully, besides qubit errors on the input state, we also need to consider a circuit-level noise model in the QEC procedure, which consists of both qubit errors and measurement errors in the SE circuit.

Theorem 4 of~\cite{gottesman2014faulttolerantquantumcomputationconstant} shows that even if SE circuits are noisy, a qLDPC code family whose parameters satisfy the aforementioned properties can also be decoded with exponential logical error suppression given that initial error rate per qubit $\Tilde{p}$, error rate per qubit at each SE cycle $p$, and bit-flip error rate per syndrome bit $q$ are below certain thresholds, under the circuit-level local stochastic noise model.
Such error suppression is obtained when we repeat the SE circuit $T=d_i$ many times where syndromes between subsequent SE cycles are compared to infer syndromes reliably.
In this section, we discuss the proof of this result in detail.

\subsubsection{Errors in the syndrome extraction circuit}\label{subsubsec:errors-syndrome-extraction}
As mentioned above, we first need to consider input errors, syndrome bit errors (resulting from faulty measurements), and gate errors.
For this, we consider the following errors at each step $t\in[1,T]$ in the SE cycle:
\begin{enumerate}
    \item $F_t\in\mathcal{P}_n$ the actual $n$-qubit physical Pauli errors at timestep $t$ and $E_t\in\mathcal{P}_n$ physical Pauli errors deduced by the decoder (i.e., which physical errors the decoder thinks happened at round $t$).
    \item $B_t\in\Z_2^l$ the set of actual syndrome bit errors at timestep $t$ and $C_t\in\Z_2^l$ are the sets of syndrome bit errors deduced by the decoder (i.e., which bit-flip errors the decoder thinks happened in round $t$).
    \item $F_1$ and $E_1$ are actual and deduced input errors, respectively.
\end{enumerate}
Physical errors in $F_t$ are those that can trigger a detection in round $t$ SE.
We can express physical error $F_t$ and syndrome bit error $B_t$ as $F_t=F_{t,1}\otimes\dots\otimes F_{t,n}$ and $B_t = (B_{t,1},\dots,B_{t,l})$ where $F_{t,q}$ are Pauli errors on qubit $q$ at timestep $t$ and $B_{t,j}$ are bit-flip errors on syndrome measurement for stabilizer generator $g_j$ at round $t$.
To illustrate this, consider the following example.

\begin{example}[label={ex:repetition_SE_circuit}]{Repetition code syndrome extraction circuit}
    A 3-round SE circuit for the 4-qubit bit-flip repetition code is given by
    \begin{equation}\label{eqn:4q_repetition_SE_circuit}
    \begin{gathered}
        \resizebox{0.85\linewidth}{!}{
        \begin{quantikz}
            & \zmeas{2}{M_{ZZ}} \gategroup[4,steps=3,style={dashed,rounded corners,fill=blue!05, inner xsep=2pt},background,label style={label position=above,anchor=north,yshift=0.4cm}]{{SE cycle 1}} \wire[d][4][""{above,pos=-0.3}]{c} & & & 
                & \zmeas{2}{M_{ZZ}} \gategroup[4,steps=3,style={dashed,rounded corners,fill=blue!05, inner xsep=2pt},background,label style={label position=above,anchor=north,yshift=0.4cm}]{{SE cycle 2}} \wire[d][4][""{above,pos=-0.3}]{c} & & & 
                & \zmeas{2}{M_{ZZ}} \gategroup[4,steps=3,style={dashed,rounded corners,fill=blue!05, inner xsep=2pt},background,label style={label position=above,anchor=north,yshift=0.4cm}]{{SE cycle 3}} \wire[d][4][""{above,pos=-0.3}]{c} & & & \\
            & & \zmeas{2}{M_{ZZ}} \wire[d][3][""{above,pos=-0.3}]{c} & &
                 & & \zmeas{2}{M_{ZZ}} \wire[d][3][""{above,pos=-0.3}]{c} & & 
                 & & \zmeas{2}{M_{ZZ}} \wire[d][3][""{above,pos=-0.3}]{c} & & \\
            & & & \zmeas{2}{M_{ZZ}} \wire[d][2][""{above,pos=-0.3}]{c} & 
                 &&& \zmeas{2}{M_{ZZ}} \wire[d][2][""{above,pos=-0.3}]{c} & 
                 &&& \zmeas{2}{M_{ZZ}} \wire[d][2][""{above,pos=-0.3}]{c} & \\
            & & & &
                & & & &
                & & & & \\
            \setwiretype{n} & \invgate{s_{1,1}} & \invgate{s_{1,2}} & \invgate{s_{1,3}} &
                & \invgate{s_{2,1}} & \invgate{s_{2,2}} & \invgate{s_{2,3}} &
                & \invgate{s_{3,1}} & \invgate{s_{3,2}} & \invgate{s_{3,3}} &
        \end{quantikz}
        }
    \end{gathered}
    \end{equation}
    A noisy implementation of this 3-round SE circuit is therefore given by
    \begin{equation}
    \begin{gathered}
        \resizebox{0.9\linewidth}{!}{
        \begin{quantikz}
            & \qnoise{F_{1,1}} & \zmeas{2}{M_{ZZ}} \gategroup[4,steps=3,style={dashed,rounded corners,fill=blue!05, inner xsep=2pt},background,label style={label position=above,anchor=north,yshift=0.4cm}]{{SE cycle 1 w/ error $F_1$}} \wire[d][5][""{above,pos=-0.3}]{c} & & & \qnoise{F_{2,1}} & 
                & \zmeas{2}{M_{ZZ}} \gategroup[4,steps=3,style={dashed,rounded corners,fill=blue!05, inner xsep=2pt},background,label style={label position=above,anchor=north,yshift=0.4cm}]{{SE cycle 2 w/ error $F_2$}} \wire[d][5][""{above,pos=-0.3}]{c} & & &  \qnoise{F_{3,1}} &
                & \zmeas{2}{M_{ZZ}} \gategroup[4,steps=3,style={dashed,rounded corners,fill=blue!05, inner xsep=2pt},background,label style={label position=above,anchor=north,yshift=0.4cm}]{{SE cycle 2 w/ error $F_3$}} \wire[d][5][""{above,pos=-0.3}]{c} & & & \\
            & \qnoise{F_{1,2}} & & \zmeas{2}{M_{ZZ}} \wire[d][4][""{above,pos=-0.3}]{c} & & \qnoise{F_{2,2}} &
                 & & \zmeas{2}{M_{ZZ}} \wire[d][4][""{above,pos=-0.3}]{c} & &  \qnoise{F_{3,2}} &
                 & & \zmeas{2}{M_{ZZ}} \wire[d][4][""{above,pos=-0.3}]{c} & & \\
            & \qnoise{F_{1,3}} & & & \zmeas{2}{M_{ZZ}} \wire[d][3][""{above,pos=-0.3}]{c} & \qnoise{F_{2,3}} &
                 &&& \zmeas{2}{M_{ZZ}} \wire[d][3][""{above,pos=-0.3}]{c} & \qnoise{F_{3,3}} &
                 &&& \zmeas{2}{M_{ZZ}} \wire[d][3][""{above,pos=-0.3}]{c} & \\
            & \qnoise{F_{1,4}} & & & & \qnoise{F_{2,4}} &
                & & & & \qnoise{F_{3,4}} &
                & & & & \\
            \setwiretype{n} & & \qnoise{B_{1,1}} & \qnoise{B_{1,2}} & \qnoise{B_{1,3}} & &
                & \qnoise{B_{2,1}} & \qnoise{B_{2,2}} & \qnoise{B_{2,3}} & &
                & \qnoise{B_{3,1}} & \qnoise{B_{3,2}} & \qnoise{B_{3,3}} & \\
            \setwiretype{n} & & \invgate{s_{1,1}} & \invgate{s_{1,2}} & \invgate{s_{1,3}} & &
                & \invgate{s_{2,1}} & \invgate{s_{2,2}} & \invgate{s_{2,3}} & &
                & \invgate{s_{3,1}} & \invgate{s_{3,2}} & \invgate{s_{3,3}} &
        \end{quantikz}
        }
    \end{gathered}
    \end{equation}
\end{example}

Now, let $\sigma(F_t)\in\Z_2^l$ and $\sigma(E_t)\in\Z_2^l$ be the syndromes for Pauli errors $F_t$ and $E_t$, respectively, before accounting for syndrome bit errors.
$\sigma$ maps Pauli errors to bit strings so that the syndrome of the product of two errors is the sum of their individual syndromes, that is, $\sigma(FF') = \sigma(F) +\sigma(F')$. In other words, $\sigma$ is a homomorphism from Pauli errors to bit strings.
Note that the syndrome $s_t$ that we observe from timestep $t$ SE takes into account the cumulative errors $D_{t-1} = F_{t-1}\dots F_1$ from timesteps $1$ to $t-1$ as well as timestep-$t$ physical errors $F_t$ and measurement errors $B_t$.
So the syndrome with perfect measurement from time-$t$ SE is given by $\sigma(F_t D_{t-1}) = \sigma(F_t) + \sigma(D_{t-1})$ and thus
\begin{equation}
    s_t = \sigma(F_t D_{t-1}) + B_t = \sigma(F_t) + \sigma(D_{t-1}) + B_t \,.
\end{equation}
We call $s_t$ the \emph{observed} syndrome from SE at timestep $t$ (note that $s_t$ depends on $F_1,\dots,F_t$ and $B_t$, but not on $B_1,\dots,B_{t-1}$ since these are errors on syndrome measurements from timestep $1$ to $t$, which do not affect the measurement outcomes at timestep $t$).
To further clarify this, consider the following example.

\begin{example}[label={ex:syndromes}]{Observed syndromes and cumulative errors}
    Consider a syndrome extraction for a 4-qubit repetition code illustrated in~\cref{ex:repetition_SE_circuit} stabilizer generators $g_1,g_2,g_3$.
    Note that the cumulative errors at the end of timestep $1$ is $D_1 = F_1$, and suppose that it flips $g_1$ and $g_2$ so that $\sigma(F_1) = (1,0,1)$ where there is no measurement errors, that is, $B_1=(0,0,0)$.
    Thus, the syndromes observed at $t=1$ SE is
    \begin{equation}
        s_1 = \sigma(F_1) + B_1 = \sigma(F_1) = (1,0,1) \,.
    \end{equation}
    If there are no errors between timesteps $1$ and $2$, that is, $F_2=I$ and $\sigma(F_2)=(0,0,0)$, and for measurement errors $B_2=(0,0,1)$ in the second round of SE, then the observed syndrome at $t=2$ is
    \begin{equation}
        s_2 = \sigma(F_2 D_1) + B_2 = \sigma(F_1) + B_2 = (1,0,0) \,.
    \end{equation}
    Namely, the observed syndrome $s_2$ from $t=2$ SE is the observed syndrome from SE at $t=1$ with its third bit flipped due to measurement error $B_2=(0,0,1)$.
\end{example}

The observed syndrome from timestep-$t$ SE cycle is given by $\sigma(F_t) + B_t$, where syndrome bit errors $B_t$ are taken into account.
Also, we define the change between observed syndromes at timestep $t-1$ and $t$ as
\begin{equation}
\begin{aligned}
    \Delta_t &= 
    \begin{cases}
        s_{t-1}+s_t \quad&,\quad t\geq2\\
        s_1 \quad&,\quad t=1\,.
    \end{cases}
\end{aligned}
\end{equation}
Since $D_t=F_tD_{t-1}$ therefore we can write the observed syndrome change $\Delta_t$ as
\begin{equation}
\begin{aligned}
    \Delta_t &=
    \begin{cases}
        \sigma(F_t) + B_t + B_{t-1} \quad&,\quad t\geq2\\
        \sigma(F_t) + B_t \quad&,\quad t=1\,,
    \end{cases}
\end{aligned}
\end{equation}
where the syndrome at timestep $t=1$ is compared to the all-zero syndrome, that is, the syndrome corresponding to an ideal initial codestate $|\overline{\psi}\>$.
Syndrome change $\Delta_t = \sigma(F_t) + B_t + B_{t-1}$ is basically the difference between observed syndromes between $t-1$ and $t$ SE rounds.

\begin{example}[label={ex:syndrome_change}]{Observed syndrome change}
    Continuing~\cref{ex:syndromes}, recall that the syndromes due to physical errors $F_1$ and $F_2$ are given by $\sigma(F_1) = (1,0,1)$ and $\sigma(F_2)=(0,0,0)$, respectively.
    Note that syndrome change $\Delta_2$ takes into account \emph{new} errors $F_2$ between timesteps $1$ and $2$, as well as measurement errors $B_1=(0,0,0)$ and $B_2=(0,0,1)$.
    So, it is given by
    \begin{equation}
        \Delta_2 = s_1+s_2 = ( \sigma(F_1) + B_1 ) + (\sigma(F_2D_1) + B_2) = \sigma(F_2) + B_1 + B_2 = (0,0,1) \,,
    \end{equation}
    indicating that the syndrome $g_3$ has been flipped (in this case due to measurement error $B_2$).

    Now suppose that at $t=3$ SE round, there are physical error $F_3$ such that $\sigma(F_3)=(0,1,0)$ and measurement error $B_3=(0,1,0)$ so that we observe syndrome $s_3=(0,0,0)$.
    Recall that $s_2=\sigma(F_2D_1)+B_2 = (1,0,0)$, which means that the first syndrome bit is flipped from $t=2$ to $t=3$, that is, $s_2+s_3=(1,0,0)$.
    
    Let us now verify this using the syndrome change formula $\Delta_3 = \sigma(F_3)+B_3+B_2$ which takes into account \emph{new} error $F_3$ between timestep $2$ and $3$, as well as measurement errors $B_2$ and $B_3$.
    Since $D_2 = F_2D_1$, we have
    \begin{equation}
    \begin{aligned}
        (1,0,0) = s_2+s_3 =
        \Delta_3 
        &= \sigma(F_3) + B_2 + B_3 = (1,0,0) \,,
    \end{aligned}
    \end{equation}
    which indicates that we observe that the first syndrome bit is flipped.
\end{example}

Syndrome change $\Delta_t$ provides information to the decoder about any errors between timestep $t$ and $t-1$.
Up to the point after SE at timestep $t$ is completed, the decoder have deduced errors $E_1,E_2,\dots,E_t$ consistent with observed syndromes $s_1,\dots,s_t$, namely 
\begin{equation}
\begin{gathered}
    \sigma(F_1) +B_1 = s_1 = \sigma(E_1) +C_1 \,,\\
    \sigma(F_2F_1) +B_2+B_1 = s_2 = \sigma(E_2E_1) +C_2+C_1 \,,\\
    \vdots\\
    \sigma(F_t\dots F_1) +B_t+B_{t-1} = s_t = \sigma(E_t\dots E_1) +C_t+C_{t-1} \,.
\end{gathered}
\end{equation}
Since $\sigma(F_t\dots F_1) = \sigma(F_t)+\sigma(F_{t-1}\dots F_1)$ and $\sigma(E_t\dots E_1) = \sigma(E_t)+\sigma(E_{t-1}\dots E_1)$, we obtain a consistency relation between actual errors and deduced errors:
\begin{equation}\label{eqn:syndrome_change_consistency}
\begin{gathered}
    \sigma(F_t) + B_t + B_{t-1} = \Delta_t = \sigma(E_t) + C_t + C_{t-1}  \quad,\;\forall t\geq2 \\
    \text{and}\\
    \sigma(F_1) + B_1 = \Delta_1 = \sigma(E_1) + C_1 \,.
\end{gathered}
\end{equation}
Errors $E_1,\dots,E_T$ deduced by the decoders over the $T$ SE cycles are the effective correction performed by the decoder.
So over the $T$ SE cycles, the overall effect of both errors and corrections are given by $\prod_{t=1}^T E_tF_t$, where $E_tF_t$ is the discrepancy between what the decoder thinks happened at round $t$, that is, $E_t$, and the actual error $F_t$.
Thus $\prod_{t=1}^T E_tF_t$ is the overall discrepancy in the entire $T$-cycle SE circuit.
So, if the decoder deduced all errors accurately, that is, $E_t=F_t$, then there would be no discrepancy at the end of the SE cycles, that is, $\prod_{t=1}^T E_tF_t = \prod_{t=1}^T F_tF_t = I$.

Now note that by consistency condition in~\cref{eqn:syndrome_change_consistency}, we have
\begin{equation}\label{eqn:syndrome_consistency}
\begin{gathered}
    \sigma(F_tE_t) = \sigma(F_t) + \sigma(E_t) = B_t + B_{t-1} + C_t + C_{t-1} \quad,\; \forall t\geq2 \\
    \text{and}\\
    \sigma(F_1E_1) = \sigma(F_1) + \sigma(E_1) = B_1 + C_1 \,.
\end{gathered}
\end{equation}
Suppose that we do a perfect syndrome measurement after cycle $T$ (no measurement errors and no physical errors after cycle $T$).
This gives us syndrome
\begin{equation}
\begin{aligned}
    \sigma\Big(\prod_{t=1}^T E_tF_t\Big) 
    &= \sum_{t=1}^T \sigma(E_t) + \sigma(F_t) \\
    &= B_1+C_1 + \sum_{t=2}^T B_t + B_{t-1} + C_t + C_{t-1} \\
    &= B_T+C_T
\end{aligned}
\end{equation}
since there are pairs of $B_t$ and $C_t$ for all $t$ in the summation which cancels out, except for $B_T,C_T$.
With this syndrome the decoder will deduce a Pauli error $G$ which is consistent with syndrome $\sigma(\prod_{t=1}^T E_tF_t) = B_T+C_T$ and has minimum weight, that is,
\begin{equation}
    G \in \arg\min\bigg\{\wt(G) : \sigma(G)=\sigma\Big( \prod_{t=1}^T E_tF_t \Big) \bigg\} \,.
\end{equation}

\subsubsection{Syndrome adjacency graph and error clusters}\label{subsubsec:syndrome-adjacency-graph}
Now that we have laid the groundwork for syndromes and syndrome errors, before proceeding to the main body of the proof for the error correction threshold of qLDPC codes with syndrome measurement errors, we need to define one more object to help us with relating qubit errors, syndrome errors, and the stabilizer checks.
Similar to the proof for a threshold for qLDPC codes with no measurement errors in~\cref{sec:qldpc_noiseless_decoding_threshold}, we will represent qubit errors and measurement bit-flip errors using a graph that take into consideration both data qubits and syndrome bits over $T$ SE cycles.

\begin{defbox}[label=def:syndrome_adjacency_graph]{Syndrome adjacency graph}
    For an $n$-qubit stabilizer code $Q$ with stabilizer generators $g_1,\dots,g_l$ and a positive integer $T$, a simple graph $\Gamma$ is the \emph{syndrome adjacency graph} of $Q$ over $T$ SE cycles if its vertices $V(\Gamma) = X(\Gamma) \cup B(\Gamma)$ and edges $E(\Gamma) = K(\Gamma)\cup L(\Gamma)$ are given by
    \begin{equation}
    \begin{gathered}
        X(\Gamma) = \{(x,t): x\in[1,n], t\in[1,T+1]\}
        \quad,\\
        B(\Gamma) = \{(b,t): b\in[1,l], t\in[1,T]\}
        \quad,\\
        K(\Gamma) = \big\{ ((x,t),(y,t))\in X(\Gamma)\times X(\Gamma) : \exists i\in[1,l],\; x,y\in\supp(g_i) \big\} 
        \quad,\\
        L(\Gamma) = \big\{ ((b,t),(x,t'))\in B(\Gamma)\times X(\Gamma) : x\in\supp(g_b) ,\; t'\in\{t,t+1\} \big\} \,.
    \end{gathered}
    \end{equation}
    Namely, qubit vertices $X(\Gamma)$ are labels for the $n$ qubits at before and after each $T$ SE cycles, syndrome vertices $B(\Gamma)$ are labels of syndrome bits in each of the $T$ SE cycles, edges $K(\Gamma)$ connects any two qubits at timestep $t$ involved in a stabilizer generator, and edges $L(\Gamma)$ connects syndrome vertices at timestep $t$ to qubit vertices at timestep $t$ and $t+1$.
\end{defbox}

For an $(r,c)$-qLDPC code family $Q_1,Q_2,\dots$ with parameters $\llbracket n_1,k_i,d_i\rrbracket$, each qubit vertex $(x,t)$ is connected to at most other $r-1$ qubit vertices at timestep $t$, connected to at most $c$ syndrome vertices at timestep $t$, and to at most $c$ syndrome vertices at timestep $t+1$.
So the degree of $(x,t)$ is at most $2c + r-1$.
On the other hand, each syndrome vertex $(b,t)$ is connected to at most $r$ qubit vertices at timestep $t-1$ and at most $r$ qubit vertices at timestep $t$.
Thus, any syndrome vertex $(b,t)$ has degree of at most $2r$.
So, a syndrome adjacency graph $\Gamma$ corresponding to any code $Q_i$ from an $(r,c)$-qLDPC code family $Q_1,Q_2,\dots$ must have degree of at most $\max\{2c + r-1, 2r\}$, which is a constant in $n_i$.

Now consider actual errors $F_1,\dots,F_T$ and $B_1,\dots,B_T$, deduced errors $E_1,\dots,E_T$ and $C_1,\dots,C_T$, and final correction $G$.
Then consider the following sets of ``marked'' vertices:
\begin{equation}
\begin{gathered}
    X_\mrk = \{ (x,t)\in X(\Gamma) : t\in[1,T] \,,\; (x,t)\in\supp(E_tF_t) \} \,,\\
    B_\mrk = \{ (b,t)\in B(\Gamma) : (B_t+C_t)_b = 1 \} \,,\\
    X_\mrk^{T+1} = \{ (x,T+1)\in X(\Gamma) :  (x,T+1)\in\supp(G) \} \,.
\end{gathered}
\end{equation}
Namely, we mark qubit vertices $(x,t)\in X(\Gamma)$ if qubit $x$ at time $t\in\{1,\dots,T\}$ is in the support of Pauli error $E_tF_t$, qubit vertices $(x,T+1)$ at time $T+1$ that are in the support of final correction Pauli $G$, and syndrome bit vertices $(b,t)$ at time $t\in\{1,\dots,T\}$ that are in the support of $B_t+C_t$.
For the rest of the proof, we will look into largest subsets of marked vertices from $V_\mrk = X_\mrk\cup B_\mrk\cup X_\mrk^{t+1}$ that induces a connected subgraph which we call a \emph{cluster}.
More precisely, $K\subseteq V_\mrk$ is a cluster if (1) for all pairs of vertices $u,v\in K$ there exist a sequence of edges $(u,w_1),(w_1,w_2),\dots,(w_l,v)$ such that $w_1,\dots,w_l\in K$ and (2) for all vertices $v\in K$, all marked neighbors of $v$ are $K$.
As an illustration, consider the following example.

\begin{example}[label={ex:syndrome_adj_graph_clusters}]{Syndrome adjacency graph and clusters}
    Consider the 3-round SE circuit for the 4-qubit bit-flip repetition code in~\cref{ex:repetition_SE_circuit} with errors $F$ and final correction $G$: 
    \begin{equation}
    \begin{gathered}
        \resizebox{0.9\linewidth}{!}{
        \begin{quantikz}
            \lstick[1]{$q_1$} & \qnoise{I} & \zmeas{2}{M_{ZZ}} \gategroup[4,steps=3,style={dashed,rounded corners,fill=blue!05, inner xsep=2pt},background,label style={label position=above,anchor=north,yshift=0.4cm}]{{SE cycle 1 w/ error $F_1$}} \wire[d][5][""{above,pos=-0.3}]{c} & & & \qnoise{I} & 
                & \zmeas{2}{M_{ZZ}} \gategroup[4,steps=3,style={dashed,rounded corners,fill=blue!05, inner xsep=2pt},background,label style={label position=above,anchor=north,yshift=0.4cm}]{{SE cycle 2 w/ error $F_2$}} \wire[d][5][""{above,pos=-0.3}]{c} & & &  \qnoise{I} &
                & \zmeas{2}{M_{ZZ}} \gategroup[4,steps=3,style={dashed,rounded corners,fill=blue!05, inner xsep=2pt},background,label style={label position=above,anchor=north,yshift=0.4cm}]{{SE cycle 2 w/ error $F_3$}} \wire[d][5][""{above,pos=-0.3}]{c} & & & \qnoise{I} & \\
            \lstick[1]{$q_2$} & \qnoise{I} & & \zmeas{2}{M_{ZZ}} \wire[d][4][""{above,pos=-0.3}]{c} & & \qnoise{F_{2,2}} &
                 & & \zmeas{2}{M_{ZZ}} \wire[d][4][""{above,pos=-0.3}]{c} & &  \qnoise{F_{3,2}} &
                 & & \zmeas{2}{M_{ZZ}} \wire[d][4][""{above,pos=-0.3}]{c} & & \qnoise{I} & \\
            \lstick[1]{$q_3$} & \qnoise{I} & & & \zmeas{2}{M_{ZZ}} \wire[d][3][""{above,pos=-0.3}]{c} & \qnoise{F_{2,3}} &
                 &&& \zmeas{2}{M_{ZZ}} \wire[d][3][""{above,pos=-0.3}]{c} & \qnoise{I} &
                 &&& \zmeas{2}{M_{ZZ}} \wire[d][3][""{above,pos=-0.3}]{c} & \qnoise{G_3} & \\
            \lstick[1]{$q_4$} & \qnoise{I} & & & & \qnoise{I} &
                & & & & \qnoise{F_{3,4}} &
                & & & & \qnoise{G_4} & \\
            \setwiretype{n} & & \qnoise{0} & \qnoise{0} & \qnoise{0} & &
                & \qnoise{1} & \qnoise{1} & \qnoise{0} & &
                & \qnoise{0} & \qnoise{0} & \qnoise{1} & \\
            \setwiretype{n} & & \invgate{s_{1,1}} & \invgate{s_{1,2}} & \invgate{s_{1,3}} & &
                & \invgate{s_{2,1}} & \invgate{s_{2,2}} & \invgate{s_{2,3}} & &
                & \invgate{s_{3,1}} & \invgate{s_{3,2}} & \invgate{s_{3,3}} &
        \end{quantikz}
        }
    \end{gathered}
    \end{equation}
    Its syndrome adjacency graph is given by:
    
    \resizebox{0.75\linewidth}{!}{
    \begin{tikzpicture}[
        >=Latex,
        qubit/.style={circle, draw, thick, minimum size=8mm, inner sep=0pt},
        check/.style={rectangle, draw, thick, rounded corners=2pt, minimum width=8mm, minimum height=6mm, inner sep=1pt},
        clusteredge/.style={red, very thick},
        clusternode/.style={draw=red, very thick, fill=red!12},
        every node/.style={font=\small},
        xscale=1.6,
        yscale=1.4
    ]
        %---------------------------------
        % Time labels
        %---------------------------------
        \node at (0,3)   {$t=1$};
        \node at (0,1)   {$t=2$};
        \node at (0,-1)  {$t=3$};
        \node at (0,-3)  {$t=4$};
        
        %---------------------------------
        % Qubit nodes at each time slice
        %---------------------------------
        % t = 1
        \node[qubit] (q11) at (1,3) {$q_1$};
        \node[qubit] (q21) at (2,3) {$q_2$};
        \node[qubit] (q31) at (3,3) {$q_3$};
        \node[qubit] (q41) at (4,3) {$q_4$};
        
        % t = 2
        \node[qubit] (q12) at (1,1) {$q_1$};
        \node[qubit] (q22) at (2,1) {$q_2$};
        \node[qubit] (q32) at (3,1) {$q_3$};
        \node[qubit] (q42) at (4,1) {$q_4$};
        
        % t = 3
        \node[qubit] (q13) at (1,-1) {$q_1$};
        \node[qubit] (q23) at (2,-1) {$q_2$};
        \node[qubit] (q33) at (3,-1) {$q_3$};
        \node[qubit] (q43) at (4,-1) {$q_4$};
        
        % t = 4
        \node[qubit] (q14) at (1,-3) {$q_1$};
        \node[qubit] (q24) at (2,-3) {$q_2$};
        \node[qubit] (q34) at (3,-3) {$q_3$};
        \node[qubit] (q44) at (4,-3) {$q_4$};
        
        %---------------------------------
        % Check nodes (syndrome bits) between time slices
        % Checks: Z1Z2, Z2Z3, Z3Z4
        %---------------------------------
        % between t=1 and t=2
        \node[check] (s12a) at (1.5,2) {$s_{1,1}$};
        \node[check] (s23a) at (2.5,2) {$s_{1,2}$};
        \node[check] (s34a) at (3.5,2) {$s_{1,3}$};
        
        % between t=2 and t=3
        \node[check] (s12b) at (1.5,0) {$s_{2,1}$};
        \node[check] (s23b) at (2.5,0) {$s_{2,2}$};
        \node[check] (s34b) at (3.5,0) {$s_{2,3}$};
        
        % between t=3 and t=4
        \node[check] (s12c) at (1.5,-2) {$s_{3,1}$};
        \node[check] (s23c) at (2.5,-2) {$s_{3,2}$};
        \node[check] (s34c) at (3.5,-2) {$s_{3,3}$};
        
        %---------------------------------
        % Horizontal code-adjacency edges within each time slice
        % (q1-q2-q3-q4 chain)
        %---------------------------------
        \draw[thick] (q11) -- (q21);
        \draw[thick] (q21) -- (q31);
        \draw[thick] (q31) -- (q41);
        
        \draw[thick] (q12) -- (q22);
        \draw[thick] (q22) -- (q32);
        \draw[thick] (q32) -- (q42);
        
        \draw[thick] (q13) -- (q23);
        \draw[thick] (q23) -- (q33);
        \draw[thick] (q33) -- (q43);
        
        \draw[thick] (q14) -- (q24);
        \draw[thick] (q24) -- (q34);
        \draw[thick] (q34) -- (q44);
        
        %---------------------------------
        % Syndrome adjacency edges
        %---------------------------------
        
        % s12 between t=1 and t=2
        \draw[thick] (s12a) -- (q11);
        \draw[thick] (s12a) -- (q21);
        \draw[thick] (s12a) -- (q12);
        \draw[thick] (s12a) -- (q22);
        
        % s23 between t=1 and t=2
        \draw[thick] (s23a) -- (q21);
        \draw[thick] (s23a) -- (q31);
        \draw[thick] (s23a) -- (q22);
        \draw[thick] (s23a) -- (q32);
        
        % s34 between t=1 and t=2
        \draw[thick] (s34a) -- (q31);
        \draw[thick] (s34a) -- (q41);
        \draw[thick] (s34a) -- (q32);
        \draw[thick] (s34a) -- (q42);
        
        % s12 between t=2 and t=3
        \draw[thick] (s12b) -- (q12);
        \draw[thick] (s12b) -- (q22);
        \draw[thick] (s12b) -- (q13);
        \draw[thick] (s12b) -- (q23);
        
        % s23 between t=2 and t=3
        \draw[thick] (s23b) -- (q22);
        \draw[thick] (s23b) -- (q32);
        \draw[thick] (s23b) -- (q23);
        \draw[thick] (s23b) -- (q33);
        
        % s34 between t=2 and t=3
        \draw[thick] (s34b) -- (q32);
        \draw[thick] (s34b) -- (q42);
        \draw[thick] (s34b) -- (q33);
        \draw[thick] (s34b) -- (q43);
        
        % s12 between t=3 and t=4
        \draw[thick] (s12c) -- (q13);
        \draw[thick] (s12c) -- (q23);
        \draw[thick] (s12c) -- (q14);
        \draw[thick] (s12c) -- (q24);
        
        % s23 between t=3 and t=4
        \draw[thick] (s23c) -- (q23);
        \draw[thick] (s23c) -- (q33);
        \draw[thick] (s23c) -- (q24);
        \draw[thick] (s23c) -- (q34);
        
        % s34 between t=3 and t=4
        \draw[thick] (s34c) -- (q33);
        \draw[thick] (s34c) -- (q43);
        \draw[thick] (s34c) -- (q34);
        \draw[thick] (s34c) -- (q44);
        
        %---------------------------------
        % Highlighted error clusters in red
        %---------------------------------
        
        % Cluster 1: a small spacetime cluster around q2,q3 and s23 between t=2 and t=3
        \node[qubit, clusternode] (cq22) at (2,1) {$q_2$};
        \node[qubit, clusternode] (cq32) at (3,1) {$q_3$};
        \node[check, clusternode] (cs12b) at (1.5,0) {$s_{2,1}$};
        \node[check, clusternode] (cs23b) at (2.5,0) {$s_{2,2}$};
        \node[qubit, clusternode] (cq23) at (2,-1) {$q_2$};
        
        \draw[clusteredge] (cs12b) -- (cq22);
        \draw[clusteredge] (cs12b) -- (cq23);
        \draw[clusteredge] (cq22) -- (cq32);
        \draw[clusteredge] (cs23b) -- (cq22);
        \draw[clusteredge] (cs23b) -- (cq23);
        \draw[clusteredge] (cs23b) -- (cq32);
        
        % \node[
        %     draw=red, very thick, rounded corners,
        %     inner sep=4pt,
        %     fit=(cq22)(cq32)(cs23b)(cq23)
        % ] {};
        
        % Cluster 2: another cluster near the bottom right
        \node[qubit, clusternode] (cq43) at (4,-1) {$q_4$};
        \node[check, clusternode] (cs34c) at (3.5,-2) {$s_{3,3}$};
        \node[qubit, clusternode] (cq44) at (4,-3) {$q_4$};
        \node[qubit, clusternode] (cq34) at (3,-3) {$q_3$};
        
        \draw[clusteredge] (cs34c) -- (cq43);
        \draw[clusteredge] (cq44) -- (cq34);
        \draw[clusteredge] (cs34c) -- (cq34);
        \draw[clusteredge] (cs34c) -- (cq44);
        
        %---------------------------------
        % Optional legend
        %---------------------------------
        \node[qubit, scale=0.8, label=right:{data qubit $(q_i,t)$}] at (5.3,1.8) {};
        \node[check, scale=0.8, label=right:{syndrome bit $(s_{t,j},t)$}] at (5.3,0.9) {};
        \node[draw=red, very thick, rounded corners, minimum width=7mm, minimum height=5mm, fill=red!12,
          label=right:{marked vertices}] at (5.3,0.0) {};
    \end{tikzpicture}
    }
    
    There are exactly two clusters in the graph: one involving data qubits and syndrome bits at timesteps 2 and 3, and the other involving a syndrome bit at timestep 3 and data qubits at timesteps 3 and 4. 
\end{example}

\subsubsection{Properties of error clusters}\label{subsubsec:properties-error-clusters}
As we discussed above, a cluster is a connected induced graph of the syndrome adjacency graph $\Gamma$ consisting of marked nodes determined by qubit errors $F=F_1,\dots,F_T$, syndrome errors $B=B_1,\dots,B_T$, and the errors $(E=E_1,\dots,E_T,\; C=C_1,\dots,C_T,\; G)$ deduced by the decoder throughout the circuit.
Clusters in $\Gamma$ capture how all errors affect the final residual error $P=G\prod_{t=1}^T E_tF_t$ at the end of the circuit, which determines whether the decoding is successful.
In the following, we will show properties of clusters in a series of lemmas that we use later to characterize logical errors and bound logical error rates to obtain thresholds.
First, we show that the residual error $P_K$ caused by errors in cluster $K$ must have a trivial syndrome.

\begin{lembox}[label={lem:connected_cluster_trivial_syndrome}]{Connected cluster trivial syndrome}
    For a connected cluster $K$, let $P_K = G|_K\prod_{t=1}^T E_t|_K F_t|_K$ where $F$ is the actual errors, $E$ the deduced errors, $G$ the final correction, and $E|_K$ is the restriction of Pauli $E$ to $K$. 
    Then $\sigma(P_K)=0$, that is, it has a trivial syndrome.
\end{lembox}

\begin{proof}
    Recall that for all bits $b\in\{1,\dots,l\}$ we have the consistency condition
    \begin{equation}
    \begin{gathered}
        \sigma(E_tF_t)_b = (B_{t-1}+B_t+C_{t-1}+C_t)_b \,,\;\forall t\in[2,T] \\
        \text{and} \\
        \sigma(E_1F_1)_b = (B_1+C_1)_b \,,
    \end{gathered}
    \end{equation}
    by~\cref{eqn:syndrome_consistency}.
    Let us write $E_t = E_t|_K E_t|_{K'}$ and $F_t = F_t|_K F_t|_{K'}$ where $K'$ is the complement of cluster $K$, so that the consistency conditions become
    \begin{equation}\label{eqn:consistency_cluster}
    \begin{gathered}
        \sigma(E_t|_K F_t|_K)_b + \sigma(E_t|_{K'}F_t|_{K'})_b = (B_{t-1}+C_{t-1})_b + (B_t+C_t)_b \,,\;\forall t\in[2,T] \\
        \text{and}\\
        \sigma(E_1|_K F_1|_K)_b + \sigma(E_1|_{K'} F_1|_{K'})_b = (B_1+C_1)_b \,.
    \end{gathered}
    \end{equation}
    By the definition of a cluster, all marked qubit vertices $(x,t)$ and $(x,t+1)$ in $K'$ for any qubit $x\in[n_i]$ are not adjacent to syndrome bit vertex $(b,t)\in K$ since all marked qubit vertices adjacent to $(b,t)$ must be in $K$.
    Namely, errors $E_t|_{K'}F_t|_{K'}$ and $E_{t+1}|_{K'} F_{t+1}|_{K'}$ are not in the support of stabilizer generator $g_b$ and therefore must have trivial $b\numth$ syndrome: $\sigma(E_t|_{K'}F_t|_{K'})_b = \sigma(E_{t+1}|_{K'} F_{t+1}|_{K'})_b = 0$.
    Similarly, for $(b,t)\in K'$ errors $E_t|_KF_t|_K$ and $E_{t+1}|_K F_{t+1}|_K$ are not in the support of stabilizer generator $g_{b}$, thus its $b\numth$ syndrome is trivial.
    Thus, for all $b\in[l]$,
    \begin{equation}
    \begin{gathered}
        (b,t)\in K \Rightarrow \sigma(E_t|_{K'}F_t|_{K'})_b = \sigma(E_{t+1}|_{K'}F_{t+1}|_{K'})_b = 0 \\
        \quad\text{and}\quad\\
        (b,t)\in K' \Rightarrow \sigma(E_t|_K F_t|_K)_b = \sigma(E_{t+1}|_K F_{t+1}|_K)_b = 0 \,.
    \end{gathered}
    \end{equation}
    Therefore, for $(b,t)\in K$, consistency conditions in~\cref{eqn:consistency_cluster} become
    \begin{equation}
    \begin{gathered}
        \sigma(E_t|_K F_t|_K)_b =
            \begin{cases}
                (B_{t-1}+C_{t-1})_b + (B_t+C_t)_b \quad&,\;\text{if }(b,t),(b,t-1)\in K \\
                0 \quad&,\;\text{otherwise}
            \end{cases} 
            \,,\;\forall t\in[2,T] \\
        \text{and}\\
        \sigma(E_1|_K F_1|_K)_b =
            \begin{cases}
                (B_1+C_1)_b \quad&,\;\text{if }(b,1)\in K \\
                0 \quad&,\;\text{otherwise.}
            \end{cases}
    \end{gathered}
    \end{equation}
    Thus, the syndrome of the total physical errors from timestep $1$ to $T$ is 
    \begin{equation}
        \sigma\Big(\prod_{t=1}^T E_t|_K F_t|_K\Big)_b =
        \begin{cases}
            (B_T+C_T)_b \quad&,\;\text{if }(b,T)\in K \\
            0 \quad&,\;\text{otherwise.}
        \end{cases}
    \end{equation}
    Similarly, for timestep $T+1$, errors $G|_{K'}$ are not in the support of stabilizer $g_b$ for all $(b,T)\in K$ and errors $G|_K$ are not in the support of stabilizer $g_{b'}$ for all $(b',T)\in K'$ so that
    \begin{equation}
    \begin{gathered}
        (b,T)\in K \Rightarrow \sigma(G|_{K'})_b = 0
        \quad\text{and}\quad
        (b',T)\in K' \Rightarrow \sigma(G|_{K})_{b'} = 0\,,
    \end{gathered}
    \end{equation}
    which implies that 
    \begin{equation}
        \sigma(G|_K)_b =
        \begin{cases}
            (B_T+C_T)_b \;&,\quad\text{if }(b,T)\in K \\
            0 \;&,\quad\text{otherwise.}
        \end{cases}
    \end{equation}
    So, it holds that
    \begin{equation}
    \begin{aligned}
        \sigma(P_K) = \sigma\Big(G|_K\prod_{t=1}^T E_t|_K F_t|_K\Big) 
        &= \sigma(G|_K) + \sigma\Big(\prod_{t=1}^T E_t|_K F_t|_K\Big)
        = 0 \,,
    \end{aligned}
    \end{equation}
    which completes the proof.
\end{proof}

The lemma above shows that errors $P_K = G|_K\prod_{t=1}^T E_t|_K F_t|_K$ in a cluster $K$ do not trigger a syndrome flip.
In other words, if we can decompose $P$ into clusters $K_1,\dots,K_m$, then $\sigma(K_i)=0$ for all $i\in[m]$, and from the proof of~\cref{lem:connected_cluster_trivial_syndrome}, this is due to any two nodes $(\xi,\tau)\in K_i$ and $(\zeta,\tau')\in K_j$ for $j\neq i$ must not share an edge and therefore cannot trigger the same syndrome check $(b,t)$.

The syndrome triviality of $P_K$ could mean two things: (1) $P_K$ is a non-trivial logical error or (2) $P_K$ is a stabilizer.
The following lemma shows that if the residual error $P=G\prod_{t=1}^T E_t F_t$ at the end of the $T$ cycles is a non-trivial logical operator (thus resulting in a logical error), then there is a cluster $K$ which contains errors $P_K$ which itself is a logical operator and therefore has weight of at least the distance $d$ of the code.

\begin{lembox}[label={lem:error_weight_lower_bound_decoding_failure}]{Error weight lower bound}
    If $P = G\prod_{t=1}^T E_tF_t$ is a non-trivial logical operator of qLDPC code $Q$ with stabilizer group $\mathcal{S}$ and parameter $\llbracket n,k,d\rrbracket$ (i.e., $P\in\mathcal{N}(\mathcal{S})\setminus\mathcal{S}$), then there exists a cluster $K$ such that $\wt(P_K) \geq d$ for $P_K = G|_K \prod_{t=1}^T E_t|_K F_t|_K$.
\end{lembox}

\begin{proof}
    Let $H = \prod_{t=1}^T E_tF_t$ so that $P=GH$ and assume that it is a non-trivial logical operator, that is, $GH\in\mathcal{N}(\mathcal{S})\setminus\mathcal{S}$.
    We will then show that $P_K$ is in $\mathcal{N}(\mathcal{S})\setminus\mathcal{S}$ for at least one cluster $K$ of $P$, that is, it is a logical operator.

    Consider clusters $K_1,\dots,K_m$ of $P$ so that $P=\prod_{i=1}^m P_{K_i}$.
    By~\cref{lem:connected_cluster_trivial_syndrome}, for each $i\in[m]$ we have $\sigma(P_{K_i})=0$ so that $P_{K_i}\in\mathcal{N}(\mathcal{S})$.
    Now, assume that $P_{K_i}\in\mathcal{S}$ for all $i\in[m]$.
    Then this means that $P=\prod_{i=1}^m P_{K_i} \in\mathcal{S}$, which contradicts the assumption that $GH\in\mathcal{N}(\mathcal{S})\setminus\mathcal{S}$.
    So, $P_K\in\mathcal{N}(\mathcal{S})\setminus\mathcal{S}$ must hold.
    All non-trivial logical operator of $Q$ must have weight of at least $d$.
    So, $\wt(P_K)\geq d$.
\end{proof}

By how we set our decoder, it chooses minimum weight Paulis that are consistent with the observed syndrome (c.f. consistency conditions in~\cref{eqn:syndrome_change_consistency}).
Now we show that the minimum weight property of deduced errors restricted to any cluster $K$ also holds.

\begin{lembox}[label={lem:total_error_weight_inequality}]{Total error weight}
    Consider cluster $K$ and minimum weight deduced errors
    \begin{equation}
        \{E_1,C_1,\dots,E_T,C_T\} \in \arg\min \wt(E,C)\,,
    \end{equation}
    where $\wt(E,C) = \sum_{t=1}^T \wt(E_t) + \wt(C_t)$ and the minimization is over errors $\{E_t,C_t : t\in[T]\}$ satisfying consistency constraints in~\cref{eqn:syndrome_change_consistency}.
    Then, it holds that
    \begin{equation}\label{eqn:cluster_minimum_weight}
    \begin{aligned}
        \sum_{t=1}^T \wt(E_t|_K) + \wt(C_t|_K) \leq \sum_{t=1}^T \wt(F_t|_K) + \wt(B_t|_K) \,,
    \end{aligned}
    \end{equation}
    where $E_t|_K$ is the errors at timestep $t$ restricted to cluster $K$.
\end{lembox}

\begin{proof}
    Consider a cluster $K$ and its complement $K'$ and (not necessarily minimum weight) deduced errors $E_1',C_1',\dots,E_T',C_T'$ such that
    \begin{equation}
    \begin{gathered}
        E_t'|_K = F_t|_K \,,\quad
        C_t'|_K = B_t|_K \,,\quad
        E_t'|_{K'} = E_t|_K \,,\;\text{and}\quad
        C_t'|_{K'} = C_t|_K \,.
    \end{gathered}
    \end{equation}
    Namely, $(E',C')$ are equal to actual errors $(F,B)$ in cluster $K$, but equal to minimum weight deduced errors $(E,C)$ in $K'$.
    Note that $E_1',C_1',\dots,E_T',C_T'$ satisfy the constraints in~\cref{eqn:syndrome_change_consistency} since for $t\geq2$
    \begin{equation}
    \begin{gathered}
        \sigma(E_t'|_K) + C_t'|_K + C_{t-1}'|_K = \sigma(F_t|_K) + B_t|_K + B_{t-1}|_K \quad, \\
        \sigma(E_t'|_{K'}) + C_t'|_{K'} + C_{t-1}'|_{K'} = \sigma(E_t|_{K'}) + C_t|_{K'} + C_{t-1}|_{K'} = \sigma(F_t|_{K'}) + B_t|_{K'} + B_{t-1}|_{K'} \,, \\
        \text{and}\\
        \sigma(E_1'|_K) + C_1'|_K = \sigma(F_1|_K) + B_1|_K \quad,\\
        \sigma(E_1'|_{K'}) + C_1'|_{K'} = \sigma(E_1|_K) + C_1|_K = \sigma(F_1|_{K'}) + B_1|_{K'} \,.
    \end{gathered}
    \end{equation}
    (See the proof of~\cref{lem:connected_cluster_trivial_syndrome}.)
    Since $\wt(E_t'|_{K'})+\wt(C_t'|_{K'}) = \wt(E_t|_{K'})+\wt(C_t|_{K'})$ and $\wt(E_t'|_{K'})+\wt(C_t'|_{K'}) = \wt(F_t|_{K'})+\wt(B_t|_{K'})$, then we have
    \begin{equation}
    \begin{gathered}
        \wt(E',C') - \wt(E,C) = \bigg( \sum_{t=1}^T \wt(F_t|_K)+\wt(B_t|_K) \bigg) - \bigg( \sum_{t=1}^T \wt(E_t|_K)+\wt(C_t|_K) \bigg) \,.
    \end{gathered}
    \end{equation}
    Since $E,C$ are chosen as consistent errors with minimum weight, then $\wt(E',C') \geq \wt(E,C)$ must hold.
    Thus~\cref{eqn:cluster_minimum_weight} holds.
\end{proof}

Recall that clusters are defined by both the actual (qubit and syndrome) errors $F,B$ as well as the errors $E,C,G$ deduced by the decoders.
As such, it is not clear how many of the actual errors occurred in a cluster.
In the following lemma, we show that the total number of actual errors $F,B$ in each cluster $K$ must be either at least half of the size of $K$, or at least a quarter od the size of $K$.

\begin{lembox}[label={lem:cluster_faults_lower_bound}]{Cluster faults lower bound}
    Consider a cluster $K$ with respect to qubit errors $E_1F_1,\dots,E_TF_T,G$ and syndrome bit errors $B_1,C_1,\dots,B_T,C_T$.
    If $K$ contains no vertices of the form $(x,T+1)$ (i.e., $G|_K = I$), then the number of faults $A_K$ in cluster $K$ satisfies
    \begin{equation}
        A_K = \sum_{t=1}^T \wt(F_t|_K) + \wt(B_t|_K)  \geq \frac{|K|}{2} \,.
    \end{equation}
    Also, if $K$ contains at least one vertex of the form $(x,T+1)$ (i.e., $G|_K \neq I$), then
    \begin{equation}
        A_K = \sum_{t=1}^T \wt(F_t|_K) + \wt(B_t|_K)  \geq \frac{|K|}{4} \,.
    \end{equation}
\end{lembox}

\begin{proof}
    Consider the total weight of all deduced errors $D_K$, all qubit errors $U_K$, and all syndrome errors $V_K$ in cluster $K$:
    \begin{equation}
    \begin{gathered}
        D_K = \sum_{t=1}^T \wt(E_t|_K) + \wt(C_t|_K) \,,\quad
        U_K = \sum_{t=1}^T \wt(F_t|_K) + \wt(E_t|_K) \,,\\
        \text{and}\quad
        V_K = \sum_{t=1}^T \wt(B_t|_K) + \wt(C_t|_K)
        \,.
    \end{gathered}
    \end{equation}
    Now recall that each qubit node $(q,t)$ for $1\leq t\leq T$ is marked if qubit $q$ is in the support of $E_tF_t$, each syndrome bit node $(b,t)$ is marked if the $b\numth$ bit of $B_t+C_t$ is 1, and each qubit node $(q,T+1)$ is marked if $G|_q\neq I$.
    Denote the number of marked qubits in $K$ from timestep $1$ to $T$ by $s_q$, the number of marked syndrome bits in $K$ by $s_c$, and the number of marked qubits in $T+1$ by $s_g=\wt(G|_K)$, so that $|K|=s_c+s_q+s_g$. 

    First we will show that $A_K\geq |K|/2$ assuming that $G|_K=I$ (so, $s_g=0$).
    Note that $U_K\geq s_q$ since for each qubit node $(q,t)\in K$ either $E_t|_q\neq I$ or $F_t|_q\neq I$.
    Similarly, we have $V_K\geq s_q$ since for each syndrome bit node $(b,t)\in K$ either $C_t|_q= 1$ or $B_t|_q=1$.
    Then, we have
    \begin{equation}
        A_K+D_K = U_K+V_K \geq s_c+s_q = |K| \,.
    \end{equation}
    By~\cref{lem:total_error_weight_inequality}, it holds that $D_K\leq A_K$.
    So, we obtain
    \begin{equation}
        |K| \leq A_K+D_K \leq 2A_K \,.
    \end{equation}
    which implies that $A_K\geq |K|/2$ as claimed.

    Now we will show that $A_K\geq |K|/4$ if $G|_K\neq I$.
    We will first show an upper bound to the number of vertices in $G|_K$ as $s_g=\wt(G|_K)\leq U_K$.
    First note that $\sigma(G) = B_T+C_T = \sigma(\prod_{t=1}^T E_tF_t)$ by the definition of correction $G$.
    This is true for all the errors and correction restricted to $K$, that is, $\sigma(G|_K) = B_T|_K+C_T|_K = \sigma(\prod_{t=1}^T E_t|_K F_t|_K)$ since $\sigma(G|_K)_b = (B_T|_K+C_T|_K)_b$ whenever $(b,T)\in K$ and $\sigma(G|_K)_b = 0$ otherwise as all marked qubit nodes $(q,T+1)$ adjacent to a $(b,T)\in K$ must be in $K$.
    By a similar argument, $\sigma(\prod_{t=1}^T E_t|_K F_t|_K)_b = (B_T|_K+C_T|_K)_b$ whenever $(b,T)\in K$ and $\sigma(\prod_{t=1}^T E_t|_K F_t|_K)_b = 0$ otherwise.
    (This is the same line of argument given in the proof of~\cref{lem:connected_cluster_trivial_syndrome}).
    Since $G$ is chosen to be the smallest weight operator such that  $\sigma(G) = \sigma(\prod_{t=1}^T E_tF_t)$, we have
    \begin{equation}
        s_g = \wt(G|_K) \leq \wt\Big(\prod_{t=1}^T E_t|_K F_t|_K\Big) \leq \sum_{t=1}^T \wt(E_t|_K F_t|_K) \leq U_K \,.
    \end{equation}
    By the argument for the $G|_K=I$ case above, we have $A_K+D_K \geq s_c+s_q$.
    Then, since $|K| = s_q+s_c+s_g$ and $U_K\leq A_K+D_K$ we obtain
    \begin{equation}
        |K| \leq A_K+D_K + U_K \leq 2(A_K+D_K) \,.
    \end{equation}
    Since $D_K\leq A_K$.
    By~\cref{lem:total_error_weight_inequality}, we have
    \begin{equation}
        |K| \leq 4 A_K\,,
    \end{equation}
    which implies that $A_K\geq |K|/4$, as claimed.
\end{proof}

So far we have not considered the probability of a certain number of actual errors in a cluster.
We now show an upper bound to the probability of at least $m$ actual errors (or, ``faults'') in a cluster $K$ (in fact we show this for any subset of vertices that are not necessarily connected).

\begin{lembox}[label={lem:faults_in_cluster_probability}]{Probability of faults in cluster}
    For a subset of vertices $U\subseteq V(\Gamma)$ and local-stochastic noise model with error parameter $p_*$, it holds that
    \begin{equation}
        \Pr\Big[\text{There are at least $m$ faults in $U$}\Big] \leq p_*^m 2^{|U|} \,.
    \end{equation}
\end{lembox}

\begin{proof}
    Note that by a union bound and by the definition of the local-stochastic noise model, we have
    \begin{equation}
    \begin{aligned}
        \Pr\Big[\text{There are at least $m$ faults in $U$}\Big] 
        &\leq \sum_{V\subseteq U : |V|\geq m} \Pr[\text{All qubits in $V$ are faulty}] \\
        &\leq \sum_{V\subseteq U : |V|\geq m} p_*^m \,.
    \end{aligned}
    \end{equation}
    There are $\binom{|U|}{k}$ subsets of $U$ with size $k$.
    Thus,
    \begin{equation}
    \begin{aligned}
        \Pr\Big[\text{There are at least $m$ faults in $U$}\Big] 
        &\leq \sum_{k=m}^{|U|} \binom{|U|}{k} p_*^k \\
        &\leq p_*^m \sum_{k=m}^{|U|} \binom{|U|}{k} 
        \leq p_*^m 2^{|U|}\,,
    \end{aligned}
    \end{equation}
    since $\sum_{k=0}^s \binom{s}{k} = 2^s$.
\end{proof}

Now we show the last property of clusters in a graph.
This is the same property we used in the proof of decoding threshold without syndrome errors (c.f.~\cref{eqn:number_of_clusters_upper_bound} of~\cref{sec:qldpc_noiseless_decoding_threshold}), which uses Lemma 2 of~\cite{gottesman2014faulttolerantquantumcomputationconstant}.

\begin{lembox}[label={lem:number_of_clusters_upper_bound}]{{Upper bound on number of clusters~\cite[Lemma 2]{gottesman2014faulttolerantquantumcomputationconstant}}}
    For any graph $\Gamma$ with maximum degree $z'$, the number of connected clusters $K$ of size $|K|=s$ containing vertex $v\in V(\Gamma)$ is at most $(z'e)^{s-1}$ and the total number of connected clusters of size $s$ is at most $|V(\Gamma)| (z'e)^{s-1}$.
\end{lembox}

\paragraph{Bounding logical error probability}
Now we will bound the logical error rate that grows either as $O(nT(p'/p_i)^d)$, $O(n(p'/p_f)^d)$, $O(n(p''/p_i)^d)$, or $O(n(p''/p_f)^d)$, according to different types of clusters causing the logical error, for some $p_i, p_f$ and for $p'=\max\{p,q\}$ and $p''=\max\{\Tilde{p},p,q\}$.
First, let us consider errors and corrections $G,E_1,B_1,\dots,E_T,B_T,F_1,C_1,\dots,F_T,C_T$ so that a decoding failure results in a logical error $P=G \prod_{t=1}^T E_tF_t \in\mathcal{N}(\mathcal{S})\setminus\mathcal{S}$, which restriction to a cluster $K$ yields $P_K = G|_K \prod_{t=1}^T (E_tF_t)|_K$, with actual number of faults $A_K = \sum_{t=1}^T \wt(F_t|_K)+\wt(B_t|_K)$.
By~\cref{lem:error_weight_lower_bound_decoding_failure}, a logical error implies the existence of cluster $K$ such that $\wt(P_K) \geq d$, so that the size $|K|$ of such ``bad'' cluster must satisfy $|K|\geq \wt(P_K)\geq d$.
So, the probability of a logical error can be bounded as
\begin{equation}
\begin{aligned}
    \Pr[\text{Logical error}] &\leq \Pr[\text{There is a cluster $K$ s.t. $\wt(P_K) \geq d$}] \\
    &\leq \Pr[\text{There is a cluster $K$ s.t. $|K|\geq d$}] \\
    &\leq \sum_{s\geq d} \Pr[\text{There is a cluster $K$ s.t. $|K|=s$}] \,.
\end{aligned}
\end{equation}

Now, let us consider four types of logical errors according to the bad cluster $K$ causing it:
\begin{enumerate}
    \item $K$ does not contain both initial errors and final errors (i.e., $(E_1F_1)|_K=I$ and $G|_K=I$).
    \item $K$ does not contain any final errors but contains some initial errors (i.e., $(E_1F_1)|_K= I$ and $G|_K\neq I$).
    \item $K$ does not contain any initial errors but contains some final errors (i.e., $(E_1F_1)|_K\neq I$ and $G|_K= I$).
    \item $K$ contains both initial errors and final errors (i.e., $(E_1F_1)|_K\neq I$ and $G|_K\neq I$).
\end{enumerate}
By~\cref{lem:cluster_faults_lower_bound}, the number of faults $A_K = \sum_{t=1}^T \wt(F_t|_K)+\wt(B_t|_K)$ in cluster $K$ must satisfy
\begin{equation}
\begin{gathered}
    A_K \geq |K|/2 \quad\text{if }G|_K= I \quad\text{and}\quad
    A_K \geq |K|/4 \quad\text{if }G|_K\neq I \,,
\end{gathered}
\end{equation}
Then for a cluster $K$, we can use~\cref{lem:faults_in_cluster_probability} to bound the probability of number of faults in $K$ for each of these four cases as
\begin{equation}
\begin{aligned}
    \Pr[\text{At least $|K|/2$ faults in $K$}] \leq 
        (p')^{|K|/2}2^{|K|} \quad&,\quad\text{if }(E_1F_1)|_K=I \,,\; G|_K=I \,,\\
    \Pr[\text{At least $|K|/4$ faults in $K$}] \leq 
        (p')^{|K|/4}2^{|K|} \quad&,\quad\text{if }(E_1F_1)|_K=I \,,\; G|_K\neq I \,,\\
    \Pr[\text{At least $|K|/2$ faults in $K$}] \leq 
        (p'')^{|K|/2}2^{|K|} \quad&,\quad\text{if }(E_1F_1)|_K\neq I \,,\; G|_K=I \,,\\
    \Pr[\text{At least $|K|/4$ faults in $K$}] \leq 
        (p'')^{|K|/4}2^{|K|} \quad&,\quad\text{if }(E_1F_1)|_K\neq I \,,\; G|_K\neq I \,,
\end{aligned}
\end{equation}
for $p'=\max\{p,q\}$ and $p''=\max\{\Tilde{p},p,q\}$.

For case 1, we only need to consider vertices contained from timestep $2$ to $T$ since the cluster $K$ is contained in this interval.
Thus there are $nT$ vertices that could be contained in such $K$, and by~\cref{lem:number_of_clusters_upper_bound} there are at most $(nT)(z'e)^{s-1}$ such clusters of size $s$.
Hence,
\begin{equation}
\begin{aligned}
    \Pr[\text{Logical error type 1}] 
    &\leq \sum_{s\geq d} \Pr[\exists\text{ a type 1 cluster $K$ s.t. $|K|=s$}] \\
    &\leq \sum_{K:|K|=s\geq d} (nT)(z'e)^{s-1} \Pr[\text{At least $|K|/2$ faults in $K$}] \\
    &\leq \frac{nT}{z'e}\sum_{s\geq d} (2z'e\sqrt{p'})^s \,.
\end{aligned}
\end{equation}
Since $\sum_{a\geq d} c^a = a^d/(1-a)$ for all $a<1$, if $2z'e\sqrt{p'}<1$ we have
\begin{equation}
    \Pr[\text{Logical error type 1}] 
    \leq \frac{nT}{z'e} \frac{(2z'e\sqrt{p'})^d}{1-2z'e\sqrt{p'}} \,.
\end{equation}
Defining $p_i=(2z'e)^{-2}$, we have
\begin{equation}
    \Pr[\text{Logical error type 1}] 
    \leq \frac{nT}{z'e (1-2z'e\sqrt{p'})} \Big(\frac{p'}{p_i}\Big)^{\frac{d}{2}} \,.
\end{equation}

For case 2, the cluster $K$ must contain at least one of the $n$ qubits at timestep $T+1$.
So, by~\cref{lem:number_of_clusters_upper_bound} there are at most $n(z'e)^{s-1}$ such clusters of size $s$.
Hence,
\begin{equation}
\begin{aligned}
    \Pr[\text{Logical error type 2}] 
    &\leq \sum_{s\geq d} \Pr[\exists\text{ a type 2 cluster $K$ s.t. $|K|=s$}] \\
    &\leq \sum_{K:|K|=s\geq d} n(z'e)^{s-1} \Pr[\text{At least $|K|/4$ faults in $K$}] \\
    &\leq \frac{n}{z'e}\sum_{s\geq d} (2z'e(p')^{\frac{1}{4}})^s \,.
\end{aligned}
\end{equation}
If $2z'e(p')^{\frac{1}{4}}<1$, then we have
\begin{equation}
    \Pr[\text{Logical error type 2}] 
    \leq \frac{n}{z'e} \frac{(2z'e(p')^{\frac{1}{4}})^d}{1-2z'e(p')^{\frac{1}{4}}} \,.
\end{equation}
Defining $p_f=(2z'e)^{-4}$, we have
\begin{equation}
    \Pr[\text{Logical error type 2}] 
    \leq \frac{n}{z'e (1-2z'e(p')^{\frac{1}{4}})} \Big(\frac{p'}{p_f}\Big)^{\frac{d}{4}} \,.
\end{equation}

For case 3, the cluster $K$ must contain at least one of the $n$ qubits at timestep $1$.
So, by~\cref{lem:number_of_clusters_upper_bound} there are at most $n(z'e)^{s-1}$ such clusters of size $s$.
Hence,
\begin{equation}
\begin{aligned}
    \Pr[\text{Logical error type 3}] 
    &\leq \sum_{s\geq d} \Pr[\exists\text{ a type 3 cluster $K$ s.t. $|K|=s$}] \\
    &\leq \sum_{K:|K|=s\geq d} n(z'e)^{s-1} \Pr[\text{At least $|K|/2$ faults in $K$}] \\
    &\leq \frac{n}{z'e}\sum_{s\geq d} (2z'e\sqrt{p''})^s \,.
\end{aligned}
\end{equation}
If $2z'e\sqrt{p''}<1$, then we have
\begin{equation}
    \Pr[\text{Logical error type 3}] 
    \leq \frac{n}{z'e} \frac{(2z'e\sqrt{p''})^d}{1-2z'e\sqrt{p''}} \,.
\end{equation}
Thus, we can again use $p_i=(2z'e)^{-2}$ to obtain
\begin{equation}
    \Pr[\text{Logical error type 3}] 
    \leq \frac{n}{z'e (1-2z'e\sqrt{p''})} \Big(\frac{p''}{p_i}\Big)^{\frac{d}{2}} \,.
\end{equation}

For case 4, cluster $K$ must contain at least $2T+1$ vertices since it spans all qubits from timestep $1$ to $T+1$ as well as at least $T$ syndrome bit vertices in between those $T+1$ layers of qubit vertices, that is, $|K|\geq 2T+1$.
So, by~\cref{lem:number_of_clusters_upper_bound} there are at most $(2T+1)(z'e)^{s-1}$ such clusters of size $s$.
Hence,
\begin{equation}
\begin{aligned}
    \Pr[\text{Logical error type 4}] 
    &\leq \sum_{s\geq 2T+1} \Pr[\exists\text{ a type 4 cluster $K$ s.t. $|K|=s$}] \\
    &\leq \sum_{K:|K|=s\geq 2T+1} n(z'e)^{s-1} \Pr[\text{At least $|K|/4$ faults in $K$}] \\
    &\leq \frac{n}{z'e} \sum_{s\geq 2T+1} (2z'e(p')^{\frac{1}{4}})^s \,.
\end{aligned}
\end{equation}
If $2z'e(p')^{\frac{1}{4}}<1$, then we have
\begin{equation}
    \Pr[\text{Logical error type 4}] 
    \leq \frac{n}{z'e} \frac{(2z'e(p'')^{\frac{1}{4}})^{2T+1}}{1-2z'e(p'')^{\frac{1}{4}}} \,.
\end{equation}
Thus, for $p_f=(2z'e)^{-4}$, we have
\begin{equation}
    \Pr[\text{Logical error type 4}] 
    \leq \frac{n}{z'e (1-2z'e(p'')^{\frac{1}{4}})} \Big(\frac{p''}{p_f}\Big)^{\frac{1}{4}} \Big(\frac{p''}{p_f}\Big)^{\frac{T}{2}} \,.
\end{equation}

To summarize, the probability of each type of logical errors is bounded as:
\begin{equation}\label{eqn:logical_error_types_proba}
\begin{gathered}
    \Pr[\text{Logical error type 1}] 
    \leq \frac{nT}{z'e (1-2z'e\sqrt{p'})} \Big(\frac{p'}{p_i}\Big)^{\frac{d}{2}} \,,\\
    \Pr[\text{Logical error type 2}] 
    \leq \frac{n}{z'e (1-2z'e(p')^{\frac{1}{4}})} \Big(\frac{p'}{p_f}\Big)^{\frac{d}{4}} \,,\\
    \Pr[\text{Logical error type 3}] 
    \leq \frac{n}{z'e (1-2z'e\sqrt{p''})} \Big(\frac{p''}{p_i}\Big)^{\frac{d}{2}} \,, \\
    \Pr[\text{Logical error type 4}] 
    \leq \frac{n}{z'e (1-2z'e(p'')^{\frac{1}{4}})} \Big(\frac{p''}{p_f}\Big)^{\frac{1}{4}} \Big(\frac{p''}{p_f}\Big)^{\frac{d}{2}}\,,
\end{gathered}
\end{equation}
where we set the total number of timesteps as $T=d$ for $p_f = (2z'e)^{-4}$ and $p_i = (2z'e)^{-2}$.

\subsubsection{Residual error rate}\label{subsubsec:residual-error-rate}
We now show that the remaining residual errors after $T$ rounds of syndrome extraction are local stochastic when the decoding is successful.
Note that a successful decoding means that $G\prod_{t=1}^T E_tF_t$ is in the stabilizer group $\mathcal{S}$, whereas decoding failure means that it is in $\mathcal{N}(\mathcal{S})\setminus\mathcal{S}$.
For an initial codestate $\rho$, the actual state after $T$ syndrome extraction rounds with faults $F=F_T\dots F_1$ and inferred errors $E=E_T\dots E_1$ is 
\begin{equation}
    F\rho F^\dag = \Big(E\prod_{t=1}^T E_tF_t\Big)\rho \Big(E\prod_{t=1}^T E_tF_t\Big)^\dag  
    = \Big(EG^\dag G\prod_{t=1}^T E_tF_t \Big)\rho \Big(EG^\dag G\prod_{t=1}^T E_tF_t \Big)^\dag\,,
\end{equation}
since $F = E\prod_{t=1}^T E_tF_t$ (up to some phase).
If $G\prod_{t=1}^T E_tF_t \in\mathcal{S}$, then it acts trivially to codestate $\rho$ and we obtain
\begin{equation}
    F\rho F^\dag = EG \rho GE \,.
\end{equation}
Since $E$ is the errors that the decoder has inferred throughout the $T$ rounds, then the actual remaining residual error is $G$\footnote{Here, $E$ can be removed because it is the frame that is applied in software (not a gate that is actually performed) to the codestate once it is inferred by the decoder.}. Since $G$ is the residual error that will act as an initial error in the next SE cycle, we need to determine its error rate.

If the decoding results in no logical error and there is no cluster spanning from timestep $1$ to $T+1$, then the correction operator $G$ is local stochastic (c.f.~\cref{def:local_stochastic_noise}) over the output $n$ qubits with error rate $p_0/3$ for some $p_0\in(0,1)$ if qubit error rate $p$ and syndrome bit error rate $q$ are small enough.
Namely, if $G \prod_{t=1}^T E_tF_t \notin\mathcal{N}(\mathcal{S})\setminus\mathcal{S}$ and $G|_K= I$ or $E|_1F|_1= I$, then the probability of errors on output qubits $S$ is
\begin{equation}\label{eqn:local_stochastic_output_successful_decoding}
    \Pr\Big[\text{errors } S\subseteq \supp(G)\Big] \leq (p_0/3)^{|S|} \,,
\end{equation}
whenever $p_0 > 3(4z'e^2\sqrt{p'})$ and $e(1-2z'e(p')^{1/4})\geq1$, where $p'=\max\{p,q\}$.

Consider a subset $S$ of the $n$ qubits at timestep $T+1$ of size $|S|=a$.
Then consider all clusters $K_1,K_2,\dots,K_m$ intersecting $S$ from $E_1F_1,\dots,E_TF_T,G$, that is, all clusters $K_j$ such that $K_j\cap S \neq \emptyset$ and $(\bigcup_j K_j)|_{T+1} = S \subseteq \supp(G)$.
Now let $s_{j,q} = \sum_{t=1}^T \wt(E_t|_{K_j} F_t|_{K_j})$ be the number of marked qubit vertices from timestep $1$ to $T$ in cluster $K_j$ and $s_{j,c} = \sum_{t=1}^T \wt(B_t|_{K_j} C_t|_{K_j})$ the marked syndrome vertices and $s_{j,g} = \wt(G|_{K_j})$ the marked qubit vertices at timestep $T+1$.
So, the size of each such cluster is $|K_j|=s_{j,q}+s_{j,c}+s_{j,g}$ and the total size of all clusters $K_1,\dots,K_m$ intersecting $S$ is 
\begin{equation}
    s = \sum_{j=1}^m s_{j,q}+s_{j,c}+s_{j,g} \,,
\end{equation}
since all clusters are disjoint.

Now note that the total size of output qubits in clusters intersecting $S$ is $\sum_j s_{j,g} = a$.
Since $G|_K$ is a minimum-weight error with syndrome $\sigma(G|_K) = B_T|_K + C_T|_K = \sigma(\prod_{t=1}^T E_t|_K F_t|_K)$, we have 
\begin{equation}
    \wt(G|_K) \leq \wt\Big(\prod_{t=1}^T E_t|_K F_t|_K\Big) \leq \sum_{t=1}^K \wt(E_t|_K F_t|_K)
\end{equation}
by the same argument as given in the proof of~\cref{lem:total_error_weight_inequality}.
Thus for the clusters intersecting $S$ we have $s_{j,g} \leq s_{j,q}$, and therefore the total size of the union of clusters $K_1,\dots,K_m$ is lower bounded as
\begin{equation}\label{eqn:total_size_clusters_containing_final_qubits}
\begin{aligned}
    s = \sum_{j=1}^m s_{j,q}+s_{j,c}+s_{j,g} \geq \sum_{j=1}^m 2s_{j,g} \geq 2a \,.
\end{aligned}
\end{equation}
Now, since all clusters $K_1,\dots,K_m$ intersecting $S$ has some vertices in timestep $T+1$ in it, by~\cref{lem:cluster_faults_lower_bound} the total number of actual faults $A_j$ in each cluster $K_j$ is lower bounded as
\begin{equation}
    A_j = \sum_{t=1}^T \wt(F_t|_{K_j})+\wt(B_t|_{K_j}) \geq \frac{|K_j|}{4} \,.
\end{equation}
Thus the total number of actual faults in the union of clusters $K_1,\dots,K_m$ is lower bounded as
\begin{equation}\label{eqn:actual_faults_in_clusters_containing_final_qubits}
    \sum_{j=1}^m A_j \geq \frac{1}{4}\sum_{j=1}^m |K_j| = \frac{s}{4} \,.
\end{equation}
So, by~\cref{lem:faults_in_cluster_probability} the probability of at least $s/4$ faults in $K_1,\dots,K_m$ is
\begin{equation}
    \Pr[s/4 \text{ faults in }K_1,\dots,K_m] \leq p^{s/4} 2^s \,.
\end{equation}
By Lemma 2 of~\cite{gottesman2014faulttolerantquantumcomputationconstant}, the number of sets $U$ of size $s$ that are a union of connected clusters in a graph $\Gamma$ with maximum degree $z'$ such that $U$ contains a subset of vertices $S$ of size $|S|=a$ is upper-bounded by $e^{a-1} (z'e)^{s-a}$.
Thus, the probability of clusters $K_1,\dots,K_m$ such that sets $(\bigcup_j K_j)|_{T+1} = S$ such that $|S|=a$, given that the decoding succeeded, is upper-bounded as
\begin{equation}
\begin{aligned}
    \Pr[\Big( \bigcup_j K_j \Big)|_{T+1} = S \,,\; S\subseteq \supp(G) \,,\; |S|=a] 
        &\leq \sum_{s\geq 2a} e^{a-1} (z'e)^{s-a} (p')^{s/4} 2^s \\
    &= \frac{1}{e(z')^a} \sum_{s\geq 2a} (2z'e(p')^{\frac{1}{4}})^s \,,
\end{aligned}
\end{equation}
since the total size $s=\sum_j|K_j|$ of clusters $K_1,K_2,\dots$ must be at least $2a$ by~\cref{eqn:total_size_clusters_containing_final_qubits} and the total number of actual faults across them must be at least $s/4$ by~\cref{eqn:actual_faults_in_clusters_containing_final_qubits}.
Note that we use $p'=\max\{p,q\}$, which is the maximum between qubit error rate $p$ and syndrome error rate $q$, since we assume that clusters with support from timestep 1 to $T+1$ is considered as a decoding failure.
Since we only consider all clusters containing some output qubit vertices, these clusters cannot contain initial qubits if we assume that the decoding succeeded.
Lastly, if $2z'e(p')^{\frac{1}{4}}<1$ then we can use the geometric sum identity $\sum_{s\geq t} c^s = \frac{c^t}{1-c}$ for $c<1$ to obtain
\begin{equation}\label{eqn:output_error_proba_bound}
\begin{aligned}
    \Pr[\Big( \bigcup_j K_j \Big)|_{T+1} = S\subseteq \supp(G)]
    &\leq \frac{1}{e(z')^{|S|}} \frac{(2z'e(p')^{1/4})^{2|S|}}{(1-2z'e(p')^{1/4})} \\
    &= \frac{(4z'e^2\sqrt{p'})^{|S|}}{e(1-2z'e(p')^{1/4})} \,.
\end{aligned}
\end{equation}
Thus if $p_0 > 3(4z'e^2\sqrt{p'})$ and $e(1-2z'e(p')^{1/4})\geq1$, then we have
\begin{equation}
    \Pr[S\subseteq \supp(G)] \leq \Big(\frac{p_0}{3}\Big)^{|S|} \,,
\end{equation}
which means that output error $G$ is local stochastic with rate $p_0/3$.

\subsubsection{Bounding logical error rates}\label{subsubsec:bounding-logical-error-rates}
Lastly let us pick the thresholds $p_0,p_1,p_2$ for the initial errors $\Tilde{p}$, qubit SE errors $p$, and syndrome bit errors $q$, respectively, so that the local-stochastic output error in~\cref{eqn:local_stochastic_output_successful_decoding} is satisfied and the logical error rates in~\cref{eqn:logical_error_types_proba} are bounded by functions exponential in code distance $d$. 

First, recall the logical error probability bounds for the four different types of errors given in~\cref{eqn:logical_error_types_proba}:
\begin{equation}
\begin{gathered}
    \Pr[\text{Logical error type 1}] 
    \leq \frac{nT}{z'e (1-2z'e\sqrt{p'})} \Big(\frac{p'}{p_f}\Big)^{\frac{d}{2}} \,,\\
    \Pr[\text{Logical error type 2}] 
    \leq \frac{n}{z'e (1-2z'e(p')^{\frac{1}{4}})} \Big(\frac{p'}{p_f}\Big)^{\frac{d}{4}} \,,\\
    \Pr[\text{Logical error type 3}] 
    \leq \frac{n}{z'e (1-2z'e\sqrt{p''})} \Big(\frac{p''}{p_f}\Big)^{\frac{d}{2}} \,, \\
    \Pr[\text{Logical error type 4}] 
    \leq \frac{n}{z'e (1-2z'e(p'')^{\frac{1}{4}})} \Big(\frac{p''}{p_f}\Big)^{\frac{1}{4}} \Big(\frac{p''}{p_f}\Big)^{\frac{d}{2}}
\end{gathered}
\end{equation}
for $p'=\max\{p,q\}$ and $p''=\max\{\Tilde{p},p,q\}$ and $p_f = (2z'e)^{-4}$~\footnote{
    Note that we use $p_f$ for all the bounds, since the bounds in~\cref{eqn:logical_error_types_proba} with $p_i=(2z'e)^{-2}$ is stricter compared to ones with $p_f<p_i$).
}.
Let us set our first threshold $p_0$ as $p_0=p_f=(2z'e)^{-4}$.

In the output error rate analysis above, we need $p'=\max\{p,q\}$ to be small enough such that $p_0 > 3(4z'e^2\sqrt{p'})$ and $e(1-2z'e(p')^{1/4})\geq1$.
So,
\begin{equation}
    p'\leq \min\Big\{ \Big(\frac{p_0}{3(4z'e^2\sqrt{p'})}\Big)^2 , \Big(\frac{e-1}{2z'e}\Big)^4 \Big\} \,.
\end{equation}
Thus, we need to set threshold $p_1$ and $p_2$ so that $p\leq p_1$ and $q\leq p_2$, such that the upper bound on $p'$ holds.
This can be satisfied by setting
\begin{equation}
    p_1=p_2 = \frac{p_f^2}{144(z')^2e^4} = (192(z')^5e^6)^{-2} \,.
\end{equation}
Thus whenever $\Tilde{p}<p_0$, $p<p_1$, and $q<p_2$, the logical error rates are bounded as
\begin{equation}
\begin{gathered}
    \Pr[\text{Logical error type 1}] 
    \leq O\Big(nT \Big(\frac{p'}{p_0}\Big)^{\frac{d}{2}} \Big) \,,\\
    \Pr[\text{Logical error type 2}] 
    \leq O\Big(n \Big(\frac{p'}{p_0}\Big)^{\frac{d}{4}} \Big) \,,\\
    \Pr[\text{Logical error type 3}] 
    \leq O\Big(n \Big(\frac{p''}{p_0}\Big)^{\frac{d}{2}} \Big) \,, \\
    \Pr[\text{Logical error type 4}] 
    \leq O\Big(n \Big(\frac{p''}{p_0}\Big)^{\frac{d}{2}} \Big) \,.
\end{gathered}
\end{equation}

\subsection{Threshold theorem for constant-overhead FTQC scheme}\label{sec:constant_overhead_ftqc_threshold}
\begingroup
\renewcommand{\H}{{{H}}}
\renewcommand{\CH}{{\mathrm{CH}}}
\renewcommand{\S}{{{S}}}
\newcommand{\T}{{{T}}}
\newcommand{\swp}{{\mathrm{SWAP}}}
\renewcommand{\cnot}{{\mathrm{CNOT}}}

Gottesman's FT scheme~\cite{gottesman2014faulttolerantquantumcomputationconstant} simulates logical quantum circuit \(C\) with \(k\) qubits and \(f(k)\) \emph{locations} using $O(k)$ physical qubits at any given time and $O(f(k)\poly k)$ circuit locations.
Namely, this scheme has a constant $\frac{O(k)}{k} = O(1)$ space overhead and polynomial $\frac{O(f(k)\poly k)}{f(k)} = O(\poly k)$ time overhead.
For simplicity, we also assume that the circuit $C$ consists of Clifford gates $\{\H,\cnot,\S\}$ and a non-Clifford gate $\T$.

The code family $\{Q_i\}$ with parameters $\llbracket n_i,k_i,d_i\rrbracket$ used in the scheme is $(r,c)$-LDPC: each stabilizer generator acts on at most $r$ qubits and each qubit involved in at most $c$ generators, for constants $r,c$ independent of $i$.
Moreover, we also demand some additional properties of the code family:
\begin{enumerate}
    \item Encoding rate $R = \lim_{i\rightarrow\infty} k_i / n_i > 0$.
    \item Distance $d_i \to \infty$.
    \item Growth of number of physical qubits: $0<n_i - n_{i-1} < n_{i-1}^\beta$ for $\beta > 0$.
    \item If thresholds $\Tilde{p}<p_0$, $p<p_1$, and $q<p_2$ are satisfied and $T(n_i)$ SE rounds are performed, then the code is decodable with logical error rate $p_L=O(1/g(n_i))$, for $g(n)=\Omega(\poly n)$ and $T(n)=\poly n$.
\end{enumerate}
Other than the qLDPC code family, below we list key ingredients and ideas that go into this FTQC scheme, each of which we discuss in detail:
\begin{enumerate}
    \item Block-encode $k$ logical qubits into $M= \lceil\frac{k}{k_i}\rceil$ blocks of \((r,c)\)-qLDPC codes \(Q_i\) satisfying the conditions above (for an illustration comparing this block encoding scheme to the concatenated code encoding in~\cref{sec:fault-tolerance-pt1}, see~\cref{fig:concat_vs_block}).
    \item \emph{Sequential FT circuit}: one logical operation at a time.
    \item Cycle error correction (EC) across \(M\) blocks of $Q_i$, sharing ancillas.
    \item \emph{Shor-style syndrome extraction (SE)} with cat-state ancillas.
    \item \emph{Knill-style teleportation} for logical gates and measurements with offline ancilla preparation using concatenated codes.
\end{enumerate}
First, though, we discuss the fault tolerance and noise model used for the scheme.

\begin{figure}
    \centering
    \includegraphics[width=0.6\columnwidth]{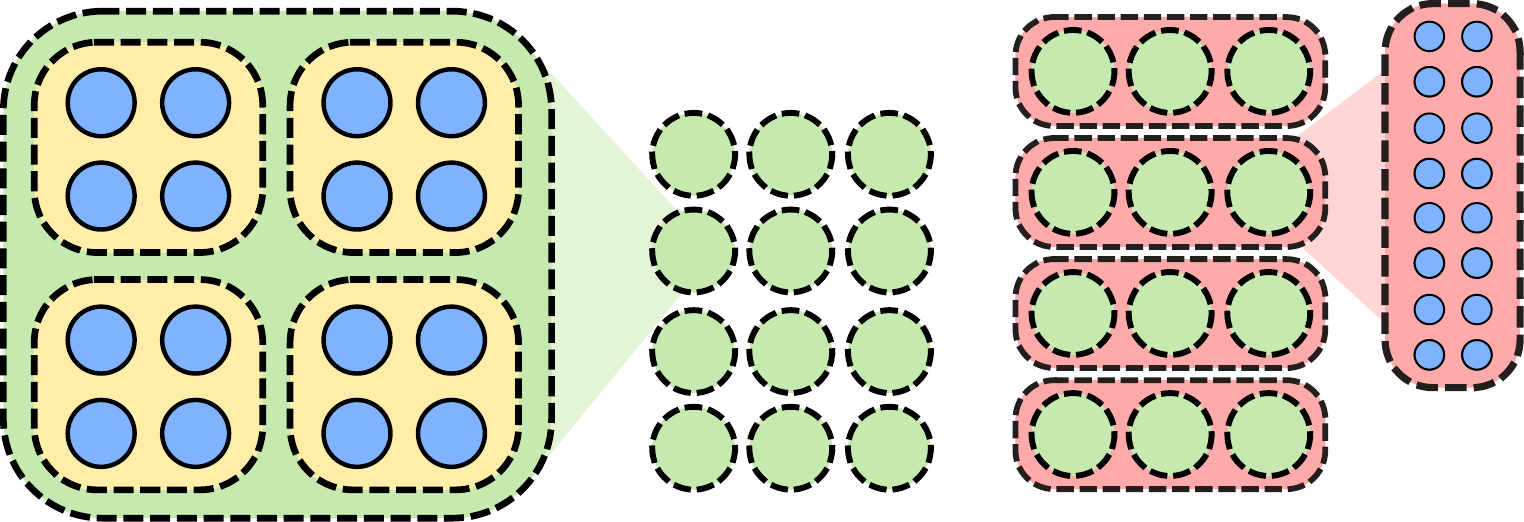}
    \caption{Illustration of encoding by concatenation (left) and block encoding (right).
    In both figures, blue circles represent physical qubits and green circles represent logical qubits.
    On the left-hand side, each green circle represent an outer $\llbracket4,1\rrbracket$ code, where each one of its four qubits (yellow, rounded squares) is itself a $\llbracket4,1\rrbracket$ code (where the four physical qubits are the blue circles inside a yellow rounded square).
    Thus, the 12 logical qubits use $12\times16 = 192$ physical qubits through two levels of concatenation.
    On the right-hand side, each red, rounded rectangle is a $\llbracket16,3\rrbracket$ code, where the 16 physical qubits are represented by blue circles and the 3 logical qubits by green circles.
    Thus, with 4 copies of the $\llbracket16,3\rrbracket$ code, we can encode 12 logical qubits using $4\times 16=64$ physical qubits.}
    \label{fig:concat_vs_block}
\end{figure}

\subsubsection{Basic model and simplified model of fault tolerance}\label{subsubsec:basic-model-fault-tolerance}
Here we adopt the \emph{basic model of fault tolerance}.
In this model, we consider locations of a circuit $C$, denoted by $\loc(C)$, as the set of all potentially faulty components, which includes state preparation, gates, measurements, and wait locations.
For example, consider the following circuit with two state preparation locations, one single-qubit gate location, one two-qubit gate location, one measurement location, and two wait locations:
\begin{equation}
    \begin{quantikz}
        \invgate{|0\>} \gategroup[1,steps=1,style={rounded corners,dashed,fill opacity=0, inner xsep=0.3pt, inner ysep=0.3pt},background,label style={label position=above,anchor=north,yshift=0.3cm}]{{\footnotesize prep}} & 
            \gate[1]{\H} \gategroup[1,steps=1,style={rounded corners,dashed,fill opacity=0, inner xsep=0.3pt, inner ysep=0.3pt},background,label style={label position=above,anchor=north,yshift=0.3cm}]{{\footnotesize 1-q gate}} & 
            \ctrl{1} \gategroup[2,steps=1,style={rounded corners,dashed,fill opacity=0, inner xsep=0.3pt, inner ysep=0.3pt},background,label style={label position=below,anchor=north,yshift=-0.2cm}]{{\footnotesize 2-q gate}} & 
            \meter{} \gategroup[1,steps=1,style={rounded corners,dashed,fill opacity=0, inner xsep=0.3pt, inner ysep=0.3pt},background,label style={label position=above,anchor=north,yshift=0.3cm}]{{\footnotesize meas.}} \\
        \invgate{|0\>} \gategroup[1,steps=1,style={rounded corners,dashed,fill opacity=0, inner xsep=0.3pt, inner ysep=0.3pt},background,label style={label position=below,anchor=north,yshift=-0.3cm}]{{\footnotesize prep}} &
            \gategroup[1,steps=1,style={rounded corners,dashed,fill opacity=0, inner xsep=0.3pt, inner ysep=0.3pt},background,label style={label position=below,anchor=north,yshift=0.3cm}]{{\footnotesize wait}} &
            \targ{} & \gategroup[1,steps=1,style={rounded corners,dashed,fill opacity=0, inner xsep=0.3pt, inner ysep=0.3pt},background,label style={label position=below,anchor=north,yshift=0.3cm}]{{\footnotesize wait}} &
    \end{quantikz}
\end{equation}
We adopt the local stochastic noise model as defined in~\cref{def:local_stochastic_noise} so that for an error rate $p$, the probability of faults in a subset of locations $S\subseteq\loc(C)$ is at most $p^{|S|}$.

Later, for the fault tolerance analysis of the error correction cycles, we will instead use a \emph{simplified model of fault tolerance}.
In the simplified model, we instead consider three different types of faults with local-stochastic noise: initial input errors with rate $\Tilde{p}$, output data qubit errors with rate $p_D$, and syndrome bit-flip errors with rate $q$.
Note that this is the model that we use in the qLDPC code decoding threshold in~\cref{sec:qldpc_noisy_decoding_threshold}.
To illustrate this more concretely and to compare between the basic model and simplified model of fault tolerance in syndrome extraction circuits, consider the following $T$ iterations of a generic syndrome extraction circuit in the basic model:
\begin{equation}
    \begin{quantikz}[row sep=0.5cm, column sep=0.5cm, classical gap=0.05cm]
        \setwiretype{n} \slice{} & \gate[style={red!20,fill=red!20}]{|\text{anc}\>} & \ctrl{1} \setwiretype{b} & \gate[style={fill=red!20}]{U} & \meter[style={fill=red!20}]{} & \setwiretype{c} \\
        \setwiretype{b} & & \gate[style={fill=red!20}]{P} & \gate[style={fill=red!20}]{\text{wait}} & & \slice{$\times T$} &
    \end{quantikz}
\end{equation}
Note that here we have ancilla $|\text{anc}\>$ preparation locations, 2-qubit and 1-qubit gate locations, wait locations, and measurement locations where faults may occur with rate $p$.
In the simplified model, we can abstract the locations inside the circuit to a ``black-box'' QEC round circuit with input errors happening before the entire $T$-round QEC cycle with rate $\Tilde{p}$ and for each of the $T$ QEC rounds, syndrome bit-flip errors with rate $q$ and data qubit errors with rate $p_D$:
\begin{equation}
    \begin{quantikz}[row sep=0.5cm, column sep=0.5cm, classical gap=0.05cm]
        \setwiretype{n} & & & \gate[2]{\text{QEC round}} & \gate[1,style={fill=red!20}]{q} \setwiretype{c} & & \setwiretype{n} \\
        \setwiretype{b} & \gate[style={fill=red!20}]{\Tilde{p}} & \slice{} & & \gate[style={fill=red!20}]{p_D} & \slice{$\times T$} & 
    \end{quantikz} \,.
\end{equation}
We summarize the two models in the following table:

\begin{center}
\begin{tabular}{llll}

    \hline
    \textbf{Aspect} & \textbf{Basic model} & \textbf{Simplified model} \\
    \hline
    \hline
    Scope & Full FT protocol: gates, EC, ancillas & Abstract QEC cycle ($T$ SE rounds) \\
    \hline
    Noise & $k$ locations local stoch. $\leq p^k$ & Local stochastic: rates \(\tilde{p}, p, q\) \\
    \hline
    Errors & Location-based (gate/prep/meas/wait) & Initial $\Tilde{p}$, data $p_D$, syndrome $q$ \\
    \hline
\end{tabular}
\end{center}

With these FT models and using the ingredients for the FTQC scheme we listed above, we can show that FTQC with constant space overhead can be achieved whenever error rates are below some thresholds.
In particular, this FTQC scheme uses a family of qLDPC codes $\{Q_i\}_i$ with parameters $\llbracket n_i,k_i,d_i\rrbracket$ satisfying: (1) a non-zero asymptotic rate $R=\lim_{n_i\rightarrow\infty}\frac{k_i}{n_i}$, (2) a condition on the growth of number of physical qubits $n_i$, and (3) existence of a decoding with logical error rate and residual error rate guarantees.
Using this qLDPC code family, we can show that a $k$-qubit logical circuit $C$ with $f(k)$ locations can be simulated by a circuit $\Tilde{C}$ using $O(k)$ qubits at any given time and $O(f(k)\poly(k))$ total number of physical locations, where the output distribution of $\Tilde{C}$ is $\epsilon$-far from $C$.
This is formalized in the following threshold theorem.

\begin{thmbox}[label={thm:constant_overhead_ftqc_threshold}]{{Constant-overhead FTQC threshold~\cite[Theorem 1]{gottesman2014faulttolerantquantumcomputationconstant}}}
    Let $\{Q_i\}$ be a family of QEC codes with:
    \begin{enumerate}
        \item $Q_i$ is an $\llbracket n_i, k_i \rrbracket$ $(r, c)$-LDPC code, for constants $r,c$.
        \item As $n_i \to \infty$, $\frac{k_i}{n_i} \to R>0$.
        \item $0 < n_i - n_{i-1} < n_{i-1}^\beta$ for some $\beta > 0$.
        \item Decoding guarantees local-stochastic noise: 
        Under local stochastic noise (with initialization error rate $\Tilde{p}$ per qubit, error rate in QEC cycle $p_D$ per data qubit per syndrome extraction (SE), syndrome bit flip rate $q$), $\Tilde{p}<p_0$, $p_D<p_1$, and $q<p_2$ implies logical error $O(1/g(n_i))$ after decoding and residual errors $\leq p_0/3$ given a polynomial $T(n_i)$ syndrome extractions each with depth $h(n_i)$ if there is no logical errors.
        ($T,h,g$ are non-decreasing functions.)
    \end{enumerate}
    Choose $\alpha < 1$ such that $0 < \alpha\beta < 1$. 
    Then, for all $\eta > 1$, $\epsilon > 0$ and for all polynomials $f(n) = o(g(n^\alpha))$, there exists $p_T(\eta)$ and $k_0(\eta,f,\epsilon)$ such that for all sequential circuit $C$ with $k > k_0$ qubits and $f(k)$ locations, there exists a simulation $\tilde{C}$ using at most $\eta k / R$ physical qubits.
    If local-stochastic error rate $p$ satisfies $p < p_T(\eta)$, statistical distance between $\Tilde{C}$ and $C$ is at most $\epsilon>0$.
    Furthermore for $\alpha' = \max(\alpha, \alpha\beta)$, simulation $\tilde{C}$ uses: 
    \begin{enumerate}
        \item $O\big( \frac{f(k) T(2 k^{\alpha'}) k^{\alpha'}}{\eta - 1} + f(k) k^{\alpha'} \mathrm{polylog}(k/\epsilon) \big)$ physical locations.
        \item Classical computation time of $O(h(2 k^{\alpha'}) + \log \log(k/\epsilon))$ per logical step.
        \item If $T(n_i)$ polynomial, then $|\Tilde{C}|=\poly k$; if decoding can be done in polynomial time, then classical computation time complexity is also polynomial.
    \end{enumerate}
\end{thmbox}

Note that the parameters $\beta>0$ and $\alpha<1$ determine the number of locations $\loc(\Tilde{C})$ and the number of classical computation steps.
We will later see how these parameters are chosen.
Also, note that this threshold theorem holds for big enough logical circuits $C$, that is, ones that uses $k>k_0$ qubits, where $k_0$ also depends on these parameters.

\subsubsection{Error correction}\label{subsubsec:threshold-error correction}
We now discuss how error correction is performed in this FTQC protocol, the rate at which errors remain, and how much overhead it incurs.
To perform reliable error correction, we measure each syndrome \(T(n_i)=\poly(n_i)\) times, where each syndrome measurement is done using Shor's fault-tolerant error correction (Shor EC) scheme (c.f.~\cref{sec:shor_EC}).
For each stabilizer generator $g$ with weight \(w\leq r\), Shor EC for measuring $g$ consists of three steps: 
\begin{enumerate}
    \item Prepare an ancilla cat state \(\frac{1}{\sqrt{2}} (|0^w\> + |1^w\>)\) fault-tolerantly (will be discussed below).
    \item Apply controlled-Pauli gates from cat qubits to data qubits.
    \item Measure each ancilla qubit in $X$-basis.
\end{enumerate}
For example, the Shor EC circuit to measure the syndrome of $X_1Z_2$ is given by
\begin{equation}
    \begin{quantikz}[row sep=0.5cm, column sep=0.5cm]
        \lstick[2]{$|\text{cat}\> = \frac{|00\>+|11\>}{\sqrt{2}}$} & & \ctrl{2} & \gate{H} & \meter{} & \rstick[2]{compute parity} \setwiretype{c} \\
        & \ctrl{2} & & \gate{H} & \meter{} & \setwiretype{c} \\
        \lstick[2]{\text{data qubits}} & \qw & \gate{X} & \qw & \qw \\
        & \gate{Z} & \qw & \qw & \qw
    \end{quantikz}
\end{equation}
The cat state $\frac{|0^r\>+|1^r\>}{\sqrt{2}}$ over $r$ qubits can be prepared with low fault-probability using a gadget $\mathcal{G}_\mathrm{cat}^r$ that prepares the state and checks for bit-flip errors using pairwise $Z$ measurement and $r'$ ancillas.
We then discard the output if the error check outputs a $-1$ outcome and accept it otherwise.
For example, a gadget $\mathcal{G}_\mathrm{cat}^4$ that prepares $4$-qubit cat code using $r'=1$ check ancilla is given by
\begin{equation}
    \mathcal{G}_\mathrm{cat}^4 =
    \begin{quantikz}
        \lstick[1]{$|0\>$} & & & \targ{} && \ctrl{4} & \rstick[4]{$\frac{|0000\>+|1111\>}{\sqrt{2}}$} \\
        \lstick[1]{$|0\>$} &\gate[1]{\H}& \ctrl{1} & \ctrl{-1} && & \\
        \lstick[1]{$|0\>$} & & \targ{} & \ctrl{1} && & \\
        \lstick[1]{$|0\>$} & & & \targ{} &\ctrl{1} & & \\
        \setwiretype{n}&&&\lstick[1]{$|0\>$}&\targ{} \setwiretype{q} &\targ{}& \meter{Z}
    \end{quantikz}
\end{equation}
where the two-qubit $Z$ measurement checks the parity of the cat state.

Since each stabilizer generator has weight of at most $r$, a Shor EC needs an additional $r'=O(r)$ ancilla qubits per stabilizer generator (since we also need ancillas to check for errors in the cat state preparation).
Since we have $n_i-k_i$ stabilizers to measure for each of the $M=\lceil k/k_i\rceil$ code-blocks, a single round of Shor EC will need $r'M(n_i-k_i)$ ancilla qubits if we measure all stabilizers in all code-blocks simultaneously.
To prevent this blowup, we split the $M$ blocks of $\llbracket n_i,k_i,d_i\rrbracket$ code to $s$ sets, each containing $M/s$ blocks.
Then, we measure only the syndromes of $M/s$ code-blocks at any given time in the QEC cycle while all the other $(s-1)M/s$ blocks wait until we measure all syndromes of the $M/s$ code-blocks before we move on to the next $M/s$ blocks.
This way, we only need additional $r'M(n_i-k_i)/s$ ancilla qubits for QEC that we reuse to do Shor EC on different blocks in cycle.
Since we need to prepare an average of $\frac{1}{1-Ap}$ cat states (as we dispose those that fail the error check), the total number of ancilla qubits we need for QEC is
\begin{equation}\label{eqn:number_of_qubits_for_QEC}
    \frac{r'M(n_i-k_i)}{s(1-Ap)} \,.
\end{equation}

Now let us translate the error rate $p$ in the basic model of fault tolerance to the physical data qubit error rate $p_P$ and syndrome bit-flip error rate $p_B$ to obtain the local-stochastic data qubit error rate $p_D$ and syndrome bit-flip error rate $q$ the simplified model for a QEC cycle.
Let us consider all fault locations in the SE circuit causing a physical data qubit error:
\begin{itemize}
    \item Cat state faults: cat state preparation error can propagate via controlled gates. 
    Each ancilla has \(A\) fault locations (dependent on \(r\)) and each qubit involved in at most $c$ syndrome measurements. 
    So, the probability of qubit error from bad ancilla up to post-selection (with success probability \(\geq 1 - pA\)) is \( \frac{c A p}{1 - p A}\) 
    
    \item Gate faults: at most \(c\) controlled-Pauli gates from each syndrome measurements the qubit is involved in, each with probability \(p\).
    
    \item Wait faults: idling errors during waits. Single syndrome measurement takes \(l\) time steps and measured every \(s\) cycles.
    So each qubit has to wait at most for \(s l\) steps, each contributing to error probability \(p\).
\end{itemize}

Thus, by putting all three error sources together, the physical data error rate $p_D$ is given by:
\begin{equation}
    p_P = \underbrace{ \frac{c Ap}{1 - p A} }_{\text{ancilla propagation after post-selection}} + \underbrace{cp}_{\text{controlled gate error}} + \underbrace{slp}_{\text{wait time error}}
\end{equation}
Note that the data qubit error rate satisfy local stochastic condition: \(\Pr[a \text{ data errors}] \leq p_P^a\) because Shor SE is fault-tolerant (see~\cref{def:FT_EC_gadget}).
Now let us consider all fault locations in the SE circuit causing a syndrome bit-flip error:
\begin{itemize}
    \item Cat state faults: similar to the physical data qubit faults above, faults in the cat state preparation circuit causes bit-flip error with probability \(\frac{A p}{1 - p A}\).
    \item Gate and measurement faults: there are \(r\) controlled-Pauli, \(r\) Hadamard gates, and \(r\) measurements in an SE circuit, where each location fails with probability $p$. 
    Thus the probability of a bit-flip error from gate and measurement faults is \(3r p\).
\end{itemize}
Putting the two error sources together gives a syndrome bit-flip error rate of 
\begin{equation}
    p_B = \underbrace{\frac{Ap}{1 - p A}}_{\text{Cat-state prep}} + \underbrace{rp}_{\text{controlled-Paulis}} + \underbrace{rp}_{\text{Hadamard gates}} + \underbrace{rp}_{\text{measurements}} \,.
\end{equation}
Note that there are correlations between data qubit errors and syndrome errors, where a physical data error in the basic FT model can cause both syndrome bit-flip errors and data qubit errors in the simplified model.
For example, suppose that a qubit $j$ is involved in stabilizers $g_1,\dots,g_5$, measured in this sequence.
Then, suppose that a physical error $E$ at qubit $j$ after measurement $g_3$ but before $g_4$ in the first round of SE, and that this error anticommutes with all five stabilizers.
Then, this error will flip syndromes $g_4,g_5$ in the first SE round and then flip $g_1,g_2,g_3$ in the second SE round while $g_4,g_5$ remains the same (assuming there have been no other errors), as illustrated in the following:
\begin{equation}
    \resizebox{0.9\linewidth}{!}{
    \begin{quantikz}[row sep=0.5cm, column sep=0.5cm, classical gap=0.04cm]
        \;j\;\setwiretype{q} & \gate[2,style={fill=red!0}]{g_1} & \gate[2,style={fill=red!0}]{g_2} & \gate[2,style={fill=red!0}]{g_3} &\gate[style={fill=red!20}]{E}& \gate[2,style={fill=red!0}]{g_4} & \gate[2,style={fill=red!0}]{g_5} &
            \gate[2,style={fill=red!0}]{g_1} & \gate[2,style={fill=red!0}]{g_2} & \gate[2,style={fill=red!0}]{g_3} & \gate[2,style={fill=red!0}]{g_4} & \gate[2,style={fill=red!0}]{g_5} & \\
        \setwiretype{b} &\wire[d][1]{c}&\wire[d][1]{c}&\wire[d][1]{c}&&\wire[d][1]{c}&\wire[d][1]{c}&
            \wire[d][1]{c}&\wire[d][1]{c}&\wire[d][1]{c}&\wire[d][1]{c}&\wire[d][1]{c}& \\
        \setwiretype{n}&o_1&o_2&o_3&&o_4+1&o_5+1&
            o_1+1&o_2+1&o_3+1&o_4+1&o_5+1& 
    \end{quantikz} 
    }
\end{equation}
In the simplified model, this data qubit error can be placed between the first and second SE rounds as
\begin{equation}
    \begin{quantikz}[row sep=0.5cm, column sep=0.5cm, classical gap=0.05cm]
        \setwiretype{n} & & \gate[2]{\text{SE round 1}} & \setwiretype{c} & \setwiretype{n} &
            \gate[2]{\text{SE round 2}} & \setwiretype{c} & \\
        \setwiretype{b} & \slice{} & & \gate[style={fill=red!20}]{E} & \slice{} &
            && \slice{} &
    \end{quantikz}
\end{equation}
but only $g_1$ and $g_2$ syndromes differ between the first and second round.
Thus, we translate this fault $E$ in the basic FT model to a data qubit error $E$ and syndrome bit errors on $g_4,g_5$ in the first round of SE.

On the other hand, if error $E$ at qubit $j$  happens after measurement $g_1$ but before $g_2$ in the second round of SE, it will flip $g_2,\dots,g_5$:
\begin{equation}
    \resizebox{0.9\linewidth}{!}{
    \begin{quantikz}[row sep=0.5cm, column sep=0.5cm, classical gap=0.04cm]
        \;j\;\setwiretype{q} & \gate[2,style={fill=red!0}]{g_1} & \gate[2,style={fill=red!0}]{g_2} & \gate[2,style={fill=red!0}]{g_3} & \gate[2,style={fill=red!0}]{g_4} & \gate[2,style={fill=red!0}]{g_5} &
            \gate[2,style={fill=red!0}]{g_1} &\gate[style={fill=red!20}]{E}& \gate[2,style={fill=red!0}]{g_2} & \gate[2,style={fill=red!0}]{g_3} & \gate[2,style={fill=red!0}]{g_4} & \gate[2,style={fill=red!0}]{g_5} & \\
        \setwiretype{b} &\wire[d][1]{c}&\wire[d][1]{c}&\wire[d][1]{c}&\wire[d][1]{c}&\wire[d][1]{c}&
            \wire[d][1]{c}&&\wire[d][1]{c}&\wire[d][1]{c}&\wire[d][1]{c}&\wire[d][1]{c}& \\
        \setwiretype{n}&o_1&o_2&o_3&o_4&o_5&
            o_1&&o_2+1&o_3+1&o_4+1&o_5+1& 
    \end{quantikz} 
    }
\end{equation}
In this case, we consider $E$ to be an error between round 1 and 2 in the simplified model, as
\begin{equation}
    \begin{quantikz}[row sep=0.5cm, column sep=0.5cm, classical gap=0.05cm]
        \setwiretype{n} & & \gate[2]{\text{SE round 1}} & \setwiretype{c} & \setwiretype{n} &
            \gate[2]{\text{SE round 2}} & \setwiretype{c} & \\
        \setwiretype{b} & \slice{} & & \gate[style={fill=red!20}]{E} & \slice{} &
            && \slice{} &
    \end{quantikz} 
\end{equation}
and we translate this fault $E$ in the basic FT model to a syndrome bit error on $g_1$ in the second round of SE.
Generalizing this to the $(r,c)$-LDPC codes where each qubit is involved in at most $c$ syndrome measurements, a fault in a SE round that flips at most the first $a\leq\lceil c/2\rceil$ syndromes is translated to syndrome errors on these first $a$ syndromes, whereas a fault in a SE round that flips at most the last $b<\lceil c/2\rceil$ syndromes is translated to syndrome errors on these last $b$ syndromes.
Thus, there are at most $\lceil c/2\rceil$ syndrome bit-flip errors (in the simplified FT model) that can be caused by a physical data qubit error in the basic FT model.
 
Now we need to bound the physical data error rate $p_P$ with $p_D$ and physical bit-flip error rate $p_B$ with $q$ such that the data qubit error rate and the syndrome bit error rate are local-stochastic.
By using the above bit-flip error accounting, a physical data error causes at most \(\lceil c/2 \rceil\) syndrome errors.
Therefore, for set \(S\) with \(b_D\) data errors and \(b_B\) syndrome errors in the simplified FT model, we have
\begin{equation}
    b_B = \text{[syndrome bit-flips]} + \text{[from phys. data errors]} \leq a_B + b_D\lceil c/2\rceil \,,
\end{equation}
for some $a_B$ in the interval $[a_{\min}, b_B]$ for $a_{\min} = b_B - b_D \lceil c/2 \rceil$.
Since $b_B= a_{\min} + b_D\lceil c/2\rceil$ and $\sum_{a=k}^\infty p^a = \frac{p^k}{1-p}$, the probability of errors $S$ is bounded as
\begin{equation}
\begin{aligned}
    \Pr[\text{errors }S] &\leq \sum_{a_B = \max(0,a_{\min})}^{b_B} \binom{b_B}{a_B} p_P^{b_D} p_B^{a_B} \\ 
    &\leq 2^{b_B} p_P^{b_D} \sum_{a_B = \max(0, a_{\min})}^{b_B} p_B^{a_B} \\
    &\leq 2^{a_{\min} + b_D\lceil c/2\rceil} p_P^{b_D} \frac{p_B^{a_{\min}}}{(1 - p_B)} \\
    &\leq (2^{\lceil c/2\rceil} p_P)^{b_D} \frac{(2p_B)^{a_{\min}}}{(1 - p_B)} \,.
\end{aligned}
\end{equation}
We need to set local-stochastic rates $p_D$ and $q$ for $S$ as
\begin{equation}
    \Pr[\text{errors }S] \leq p_D^{b_D} q^{b_B} = p_D^{b_D} q^{a_{\min}+b_D\lceil c/2\rceil} = (p_Dq^{\lceil c/2\rceil})^{b_D} q^{a_{\min}} \,.
\end{equation}
This can be satisfied by setting 
\begin{equation}\label{eqn:data_and_syndrome_error_rates}
    p_D = \frac{\sqrt{p_P}}{1 - p_B}
    \quad\text{and}\quad
    q = \max\Big\{ 2 p_P^{1/(c+1)} \,,\; \frac{2 p_B}{1-p_B} \Big\} \,.
\end{equation}
Then the thresholds for data qubit errors and syndrome errors are given by \(p_D < p_1\) and \(q < p_2\).

\subsubsection{Logical gates}\label{subsubsec:logical-gates}
In this FTQC scheme, we use Knill's gate teleportation scheme proposed in~\cite{knill1996threshold}, which allows for an implementation of a universal fault-tolerant gate set~\cite{gottesman1998fault-tolerant} by preparing a logical ancilla state that encodes a gate $U$ using concatenated codes.
Recall that we assume that the logical circuit $C$ contains gates from a Clifford gate set $\H,\S,\cnot$ and a non-Clifford gate $\T$.
Such a protocol that implements a logical unitary $\overline{U}$ on a $\llbracket n_i,k_i\rrbracket$-qLDPC code block $Q_i$, uses two ancilla code blocks $A_1,A_2$ of the $\llbracket n_i,k_i\rrbracket$-qLDPC codes, and consists of the following three steps:
\begin{enumerate}
    \item Using concatenated codes (see~\cite{aharonov1999faulttolerantquantumcomputationconstant}), prepare $2n_i$-qubit state $|\Psi_{k_i}^{\overline{U}}\> = (I\otimes\overline{U})|\overline{\text{Bell}}_{k_i}\>$ over the ancilla code blocks $A_1\otimes A_2$. 
    The state $|\Psi_{k_i}^{\overline{U}}\>$ encodes the unitary gate $\overline{U}$, which is a logical $U$ for the $\llbracket n_i,k_i\rrbracket$ code.
    Here, $|\overline{\text{Bell}}_{k_i}\> = \big( \frac{|\overline{00}\>+|\overline{11}\>}{\sqrt{2}} \big)^{\otimes k_i}$ is the logical Bell state over ancilla blocks $A_1,A_2$. 
    
    \item Teleport the $\overline{U}$ gate via transversal Bell measurement over $Q_i$ and $A_1$ ($\overline{X}_{Q_i}\overline{X}_{A_1}, \overline{Z}_{Q_i}\overline{Z}_{A_1}$ basis).
    
    \item Correct with logical Pauli $\overline{P}_{a,b} = \overline{X}_1^{a_1}\overline{Z}_1^{b_1} \dots \overline{X}_{k_i}^{a_{k_i}}\overline{Z}_{k_i}^{b_{k_i}}$ conjugated by $\overline{U}$ based on outcomes.
\end{enumerate}
The logical gate teleportation circuit in step 2 above, including correction in step 3 is given by
\begin{equation}
    \begin{quantikz}[classical gap=0.04cm]
        \lstick{$|\overline{\psi}\>_{Q_i}$} \setwiretype{b} & \qwbundle{\llbracket n_i,k_i\rrbracket} & & \ctrl{1} \gategroup[2,steps=3,style={blue!10,rounded corners,fill=blue!10,inner xsep=4pt,inner ysep=10pt},background,label style={label position=above,anchor=north,yshift=0.4cm}]{{Transversal Bell measurement}} \gategroup[2,steps=1,style={dashed,rounded corners,fill=blue!0,inner xsep=2.5pt,inner ysep=4pt},background,label style={label position=above,anchor=north,yshift=-0.2cm}]{\small{$\otimes n_i$}} & \gate{\H^{\otimes n_i}} & \meter{Z^{\otimes n_i}} & \setwiretype{c} \wire[l][1]["{b\in\{0,1\}^{k_i}}"{above,pos=-0.9}]{c} & \ctrl{2} & \\
        \lstick[2]{$|\Psi_{k_i}^{\overline{U}}\>$} \setwiretype{b} & \qwbundle{\llbracket n_i,k_i\rrbracket} & & \targ{} & & \meter{Z^{\otimes n_i}} & \setwiretype{c} \wire[l][1]["{a\in\{0,1\}^{k_i}}"{above,pos=-0.9}]{c} & \ctrl{1} & \\
        \setwiretype{b} & \qwbundle{\llbracket n_i,k_i\rrbracket} & & & & & & \gate[style={fill=green!20}]{\overline{U}\overline{P}_{a,b}\overline{U}^\dag} & \rstick{$\overline{U}|\overline{\psi}\>_{Q_i}$}
    \end{quantikz}
\end{equation}
where a transversal $\cnot$ is performed on the first two code-blocks, followed by a transversal Hadamard on the first block, then all qubits in the first two blocks are measured in the $Z$ basis, whose outcomes determine the logical correction unitary $\overline{U}\overline{P}_{a,b}\overline{U}^\dag$ on the third block.
Note that when $U$ is a 2-qubit unitary, such as a CNOT gate, across two code blocks, we prepare a $4n_i$-qubit state $|\Psi_{k_i}^{\overline{U}}\>$ over four blocks of ancilla $A_1^{(1)},A_1^{(2)},A_2^{(1)},A_2^{(2)}$ and do a transversal Bell measurement over two blocks of data qubits $Q_i^{(1)}, Q_i^{(2)}$ and two blocks of ancilla $A_1^{(1)},A_1^{(2)}$.
Then, correction $\overline{U}\overline{P}_{a,b}\overline{U}^\dag$ is done over two ancilla blocks $A_2^{(1)},A_2^{(2)}$.

To prepare a $2n_i$-qubit state $|\Psi_{k_i}^{\overline{U}}\>$, we use an $r$-level concatenated code $C^{\otimes r}$ ancilla factory\footnote{A proof that any state can be prepared using a concatenated code can be found in~\cite[Section 4.2.2]{nguyen2025quantumfaulttoleranceconstantspace}     or~\cite{christandl2025faulttolerantquantuminputoutput}}.
This ancilla factory encodes the $2n_i$ qubits in concatenated code $C^{\otimes r}$, then prepares the state $|\Psi_{k_i}^{\overline{U}}\>$ using a circuit $\Tilde{\mathcal{D}}$ that encodes an ideal circuit $\mathcal{D}$ with $|\mathcal{D}|$ locations that prepares the state $|\Psi_{k_i}^{U}\> = (I\otimes U)|\mathrm{Bell}_{k_i}\>$.
For an $\llbracket n,k\rrbracket$ code with encoder $\mathtt{Enc}$ and $k=2n_i < n$, an ancilla factory preparing $|\Psi_{k_i}^{\overline{U}}\>$ using a two-level concatenation of $\llbracket n,k\rrbracket$ codes can be illustrated as
\begin{equation}
    \begin{quantikz}
        &\qwbundle[]{k}& \gate[1]{\mathtt{Enc}} & \qwbundle[]{n} & \gate[5]{\mathtt{Enc}} &\setwiretype{n}&
            && \gate[5]{\mathtt{Enc}^{-1}} &\qwbundle[]{n}\setwiretype{q}& \gate[1]{\mathtt{Enc}^{-1}} \setwiretype{q} &\qwbundle[]{k}& \rstick{$|\Psi_{k_i}^{\overline{U}}\>$} \\
        &\qwbundle[]{k}& \gate[1]{\mathtt{Enc}} & \qwbundle[]{n} & &\setwiretype{n}&
            && &\qwbundle[]{n}\setwiretype{q}& \gate[1]{\mathtt{Enc}^{-1}} && \\
        &\qwbundle[]{k}& \gate[1]{\mathtt{Enc}} & \qwbundle[]{n} & &\qwbundle[]{n^2}& \gate[1]{\Tilde{\mathcal{D}}}
            &\qwbundle[]{n^2}& &\qwbundle[]{n}\setwiretype{q}& \gate[1]{\mathtt{Enc}^{-1}} && \\
        &\qwbundle[]{k}& \gate[1]{\mathtt{Enc}} & \qwbundle[]{n} & &\setwiretype{n}&
            && &\qwbundle[]{n}\setwiretype{q}& \gate[1]{\mathtt{Enc}^{-1}} && \\
        &\qwbundle[]{k}& \gate[1]{\mathtt{Enc}} & \qwbundle[]{n} & &\setwiretype{n}&
            && &\qwbundle[]{n}\setwiretype{q}& \gate[1]{\mathtt{Enc}^{-1}} && 
    \end{quantikz}
\end{equation}
For logical error rate $\varepsilon_0$, the ancilla factory uses
\begin{equation}\label{eqn:number_of_qubits_ancilla_factory}
    \text{\# qubits in ancilla factory} = O(n_i\polylog (n_i/\varepsilon_0))
\end{equation}
physical qubits and 
\begin{equation}
    \loc(\text{ancilla factory}) = O(|\mathcal{D}|\polylog (|\mathcal{D}|/\varepsilon_0))
\end{equation}
locations for the ancilla factory.
By setting the number of code blocks as $M=\lceil\frac{k}{k_i}\rceil = \polylog k$ and by picking $k_i = \frac{k}{\polylog k}$, so that $n_i=\Theta(k_i) = \Theta( \frac{k}{\polylog k})$, the number of physical qubits used in an ancilla factory is
\begin{equation}
    \text{\# qubits in ancilla factory} = O(n_i\polylog (n_i/\varepsilon_0)) \approx O(k) \,,
\end{equation}
thus maintaining constant space overhead.

Such an ancilla factory is also used to prepare the logical state $|\overline{0}\>$ at the start of the circuit.
On the other hand, it can be used to do logical measurements for general stabilizer code, where the code blocks are encoded into a concatenated code and measured there.
In both $|\overline{0}\>$ state preparation and logical measurement, we also use an additional $O(n_i\polylog (n_i/\varepsilon_0))$ physical qubits.

Lastly, we note that for Clifford gates $U\in\{\H,\S,\cnot\}$, correction $\overline{U}\overline{P}_{a,b}\overline{U}^\dag$ is a logical Pauli which can be applied transversally (or, we can just keep track of it in the Pauli frame).
However, for the non-Clifford $\T$ gate, $\overline{V} = \overline{\T}\overline{P}_{a,b}\overline{\T}^\dag$ is a logical Clifford for which transversal implementation may not be possible on the qLDPC code.
Therefore, we need to implement $\overline{V}$ using gate teleportation with another ancilla $|\Psi_{k_i}^{\overline{V}}\>$.
In this case, after the transversal Bell measurement is performed, the data qubits need to wait for this additional ancilla to be prepared and teleported for correction $\overline{V}$ to be applied.

\subsubsection{Initial error rate in the simplified FT model}\label{subsubsec:initial-error-rate-simplified-ft}
In this FTQC scheme, a QEC cycle consisting of $T$ QEC rounds is performed between subsequent logical gates.
So, faults from the logical gate implementation right before a QEC round are counted as initial errors for that QEC round with local-stochastic error rate $\Tilde{p}$.
In addition to logical gate faults, the initial errors of a QEC cycle also comes from residual errors from the QEC round right before the logical gate.
As an illustration, consider a logical gate $\overline{U}$ and two QEC cycles before and after it:
\begin{equation}
    \begin{quantikz}[classical gap=0.04cm]
        & \setwiretype{n} & \gate[2]{\text{QEC round}} & \setwiretype{c} 
            & \setwiretype{n} & 
            & & \gate[2]{\text{QEC round}} & \setwiretype{c} & \setwiretype{n} \\
        \setwiretype{b} & \slice{} & & \slice{$T\times$} 
            & \gate[style={fill=red!20}]{p_0/3} \gategroup[1,steps=2,style={dashed,rounded corners,fill=red!10, inner xsep=1pt},background,label style={label position=below,anchor=north,yshift=-0.2cm}]{{$\Tilde{p}$}} & \gate[style={fill=red!20}]{\overline{U}} 
            & \slice{} & & \slice{$T\times$} & 
    \end{quantikz}
\end{equation}
Note that the residual error from the previous QEC cycle is local-stochastic with rate $p_0/3$ as we showed in~\cref{sec:qldpc_noisy_decoding_threshold}, whenever the decoding is successful.

From our previous discussion on how logical gates are implemented, we can now analyze how the logical gate faults cause initial errors at the QEC cycle following every logical gate, along with residual errors from the previous QEC cycle, and determine its local-stochastic error rate $\Tilde{p}$.
A physical data qubit error caused by logical gate faults comes from the following two sources: 
\begin{enumerate}
    \item Ancilla factory: if there are $B$ locations in an ancilla factory, then faults in it cause a physical data qubit error with probability $Bp$.
    \item In a teleportation circuit, a data qubit is involved in a Hadamard, a CNOT, a measurement, and a logical Pauli correction.
    Thus faults on these locations will cause a physical data qubit error with probability $4p$.
\end{enumerate}
Note that a logical Clifford $\overline{U}$ may be a two-qubit gate across two code blocks, such as a CNOT, which will propagate residual errors from the previous QEC cycle across code blocks.
So the physical initial error rate $\Tilde{p}_C$ in the simplified error model for QEC cycles after a logical Clifford $\overline{U}$ is at most
\begin{equation}\label{eqn:initial_error_clifford}
    \Tilde{p}_C = \underbrace{2p_0/3}_{\text{residual if }\overline{U}=\overline{\cnot}} + \underbrace{Bp}_{|\Psi_{k_i}^{\overline{U}}\>\text{ prep.}} + \underbrace{p}_{\text{CNOT}} + \underbrace{p}_{\H} + \underbrace{p}_{\text{meas.}} + \underbrace{p}_{\text{Pauli }\overline{U}\overline{P}_{a,b}\overline{U}^\dag} \,.
\end{equation}
On the other hand, for a logical non-Clifford $\overline{U}$ such as the $\T$ gate, we need an additional ancilla to do Clifford correction $\overline{V}=\overline{U}\overline{P}_{a,b}\overline{U}^\dag$ in the gate teleportation circuit.
Thus, the physical initial error rate $\Tilde{p}_{NC}$ in the simplified error model for QEC cycles after a logical non-Clifford $\overline{U}$ is at most
\begin{equation}\label{eqn:initial_error_nonclifford}
    \Tilde{p}_{NC} = \underbrace{p_0/3}_{\text{residual}} + \underbrace{2}_{\text{two tele.+prep.}} \Big( \underbrace{Bp}_{|\Psi_{k_i}^{\overline{U}}\>\text{ or }|\Psi_{k_i}^{\overline{V}}\>} + \underbrace{p}_{\text{CNOT}} + \underbrace{p}_{\H} + \underbrace{p}_{\text{meas.}} + \underbrace{p}_{\overline{U}\overline{P}_{a,b}\overline{U}^\dag\text{ or }\overline{V}\overline{P}_{a,b}\overline{V}^\dag} \Big) \,.
\end{equation}
We can then choose an initial error rate threshold $p_0$ that bounds initial input error rate $\Tilde{p}$ which is given by the largest quantity between $\Tilde{p}_C$ in~\cref{eqn:initial_error_clifford} and $\Tilde{p}_{NC}$ in~\cref{eqn:initial_error_nonclifford}.

\subsubsection{Threshold and logical error rate}\label{subsubsec:threshold-logical-error-rate}
Now let us collect the local-stochastic data qubit error rate $p_D$ and syndrome error rate $q$ from~\cref{eqn:data_and_syndrome_error_rates} and the initial error rate $\Tilde{p} = \max\{\Tilde{p}_C,\Tilde{p}_{NC}\}$ from~\cref{eqn:initial_error_clifford} and~\cref{eqn:initial_error_nonclifford} in the simplified FT model.
We obtain thresholds $p_0,p_1,p_2$ for these error rates as follows:
\begin{equation}
\begin{gathered}
    \Tilde{p} = \max\{\Tilde{p}_C,\Tilde{p}_{NC}\} = \max\bigg\{ \frac{2p_0}{3} + (B+4)p \,,\; \frac{p_0}{3} + 2(B+4)p \bigg\} < p_0
    \quad,\\
    p_D = \frac{\sqrt{p_P}}{1-p_B} < p_1
    \quad,\quad\text{and}\quad
    q = \max\bigg\{2p_P^{1/(c+1)} \,,\; \frac{2p_B}{1-p_B}\bigg\} < p_2 \,.
\end{gathered}
\end{equation}
Then, we choose a threshold $p_T$ such that all 4 inequalities above are satisfied whenever $p<p_T$. We also need the logical error probability to be at most $\varepsilon$.
Because circuit $C$ has $f(k)$ locations, we need at most $\varepsilon/f(k)$ logical error rate per logical location:
\begin{itemize}
    \item From probability $\varepsilon_0$ of ancilla preparation failure from the ancilla factory, we need $\varepsilon_0\leq \varepsilon/(3f(k))$ (at most 2 ancillas per location , for example, for logical $\T$ gates).
    \item From probability $D/g(n_i)$ of QEC failure, we choose $i$ such that $D/g(n_i) \leq \varepsilon/(3f(k))$ by taking $n_i>k^\alpha$ for large $k$ since $f(k) = o(g(k^\alpha))$ for $\alpha>1$.
\end{itemize}
Thus, by using $\varepsilon_0\leq \varepsilon/(3f(k))$ and $D/g(n_i) \leq \varepsilon/(3f(k))$, the logical error rate per logical location can be bounded as
\begin{equation}
\begin{aligned}
    \Pr[\text{Logical location }\ell\text{ fails}] 
    &\leq \Pr[\text{ancilla prep failure at }\ell] + \Pr[\text{QEC failure at }\ell] \\
    &\leq 2\varepsilon_0 + \frac{D}{g(n_i)} \\
    &\leq 3\frac{\varepsilon}{3f(k)} = \frac{\varepsilon}{f(k)} \,.
\end{aligned}
\end{equation}

Since $\Pr[\text{Logical location }\ell\text{ fails}] \leq \frac{\varepsilon}{f(k)}$, we have
\begin{equation}
\begin{aligned}
    p_\text{fail} &:= \Pr[\text{there is a logical failure}]\\ 
    &\leq \sum_{\ell} \Pr[\text{Logical location }\ell\text{ fails}] \leq f(k)\frac{\varepsilon}{f(k)} 
    = \varepsilon  \,.
\end{aligned}
\end{equation}
Let $\xi$ be an output distribution with logical errors.
For ideal output distribution $p_\text{ideal}$ and noisy output distribution $p_\text{noisy} = (1-p_\text{fail}) p_\text{ideal} + p_\text{fail} \xi$, their total variation distance is therefore bounded as
\begin{equation}
\begin{aligned}
    \text{TVD}(p_\text{ideal},p_\text{noisy}) \leq p_\text{fail} \leq \varepsilon \,.
\end{aligned}
\end{equation}

\subsubsection{Constant space overhead}
We now show that this FTQC scheme incurs a constant space overhead.
In particular, for a logical circuit $C$ using $k$ qubits, we can construct a circuit $\Tilde{C}$ simulating $C$ using $O(1)$ qubits at any given time.
Note that the total number of physical qubits used in $\Tilde{C}$ is given by
\begin{equation}
    \text{Total physical qubits} = N_\text{data} + N_\text{QEC} + N_\text{op} \,,
\end{equation}
where $N_\text{data}$ is the number of physical data qubits, $N_\text{QEC}$ is the total number of physical qubits used for ancillas in QEC at any given time, and $N_\text{op}$ is the total number of physical qubits used for ancillas in logical operations (logical gates, measurements, and state preparation) at any given time.

Let us now consider constants $\omega,\lambda,\mu>0$ to be determined later and also recall the positive encoding rate assumption in~\cref{thm:constant_overhead_ftqc_threshold}, that is, $R=\lim_{i\rightarrow\infty} k_i/n_i >0$.
The number of physical data qubits is given by
\begin{equation}
    N_\text{data} = Mn_i\,,
\end{equation}
since we are using $M=\lceil k/k_i\rceil$ blocks of $\llbracket n_i,k_i,d_i\rrbracket$ qLDPC codes.
We take a large positive integer $i$ such that $k_i/n_i\approx R$, so that the bound
\begin{equation}
    N_\text{data} = Mn_i \leq (1+\omega)\frac{k}{R}
\end{equation}
holds for large $k$.\footnote{Note that this is possible since $Mn_i \approx \frac{n_ik}{k_i}$ so that $\lim_{i\rightarrow\infty}Mn_i \approx \lim_{i\rightarrow\infty} \frac{n_ik}{k_i} = \frac{k}{R}$ by the assumption that the asymptotic encoding rate is positive: $R = \lim_{i\rightarrow\infty} \frac{k_i}{n_i} >0$. If instead we have zero asymptotic encoding rate $R = \lim_{i\rightarrow\infty} \frac{k_i}{n_i} = 0$, then $\lim_{i\rightarrow\infty} \frac{n_i}{k_i}$ blows up hence $\lim_{i\rightarrow\infty}Mn_i$ also blows up and we can no longer upper bound the number of data qubits.}

Now, recall that from~\cref{eqn:number_of_qubits_for_QEC}, that the number of physical qubits used for QEC ancillas is given by
\begin{equation}
    N_\text{QEC} = \frac{M(n_i-k_i)r'}{s(1-Ap)} \,,
\end{equation}
since we do QEC only on $M/s$ blocks at a time, where $r'$ is the total physical qubits used per cat state and $1-Ap$ is the success probability in the cat state post-selection (repeat $1/(1-Ap)$ times on average).
We can take a large $s$ such that 
\begin{equation}
    N_\text{QEC} = \frac{M(n_i-k_i)r'}{s(1-Ap)} \leq \lambda\frac{k}{R} \,.
\end{equation}
By~\cref{eqn:number_of_qubits_ancilla_factory}, the number of physical qubits used for each teleportation ancilla $|\Psi_{k_i}^{\overline{U}}\>$ is
\begin{equation}
    N_\text{op} = O(n_i\polylog(n_i/\varepsilon_0)) = c n_i(\log(n_i/\varepsilon_0))^a \,,
\end{equation}
where $\varepsilon_0$ is the logical error rate of the ancilla factory and constants $c,a>0$.
Since we only do one logical operation at a time in $\Tilde{C}$, we need at most $2 c n_i(\log(n_i/\varepsilon_0))^a$ physical qubits (i.e., since we need to prepare two ancillas to implement a logical non-Clifford $\T$).
We can pick a sufficiently large $n_i$ such that
\begin{equation}
    N_\text{op} = c n_i (\log(n_i/\varepsilon_0))^a \leq \mu\frac{k}{R} \,.
\end{equation}
Now, we can choose $\eta>0$ such that
\begin{equation}
    N_\text{tot} = N_\text{data} + N_\text{QEC} + N_\text{op} \leq (1+\omega)\frac{k}{R} + \lambda\frac{k}{R} + \mu\frac{k}{R} 
    \leq \eta\frac{k}{R} \,,
\end{equation}
which gives a constant overhead of $\eta /R$ since both $\eta,R$ are constants.
Note that we still need to pick $\omega,\lambda,\mu>0$ and $i$ such that: 
\begin{equation}
\begin{gathered}
    N_\text{data} = Mn_i \leq (1+\omega)\frac{k}{R} \\
    N_\text{QEC} = \frac{M(n_i-k_i)r'}{s(1-Ap)} \leq \lambda\frac{k}{R} \\
    N_\text{op} = C n_i (\log(n_i/\varepsilon_0))^a \leq \mu\frac{k}{R} \\
    \frac{D}{g(n_i)} < \frac{\varepsilon}{3f(k)} \,.
\end{gathered}
\end{equation}
We do this by picking $\omega>0$ and $\lambda>1$ and $M \leq \lambda k/k_i$ and large $s$, which depend on $n_i$ chosen below.
If $R/(1+\omega) < k_i/n_i < R(1+\omega)$ then we have
\begin{equation}
    N_\text{data} = Mn_i \leq (1+\omega)\frac{k}{R} \quad\text{and}\quad
    N_\text{QEC} = \frac{M(n_i-k_i)r'}{s(1-Ap)} \leq \lambda\frac{k}{R} \,.
\end{equation}
We want a large $i$ such that (1) the logical error after decoding and (2) the logical operations qubit overhead are small:
\begin{equation}
\begin{gathered}
    \frac{D}{g(n_i)} < \frac{\varepsilon}{3f(k)}
    \quad\text{and}\quad
    c n_i (\log(n_i/\varepsilon_0))^a \leq \mu\frac{k}{R} \,.
\end{gathered}
\end{equation}
So, we can pick $i$ such that $n_i \geq k^\alpha$ (since $f(k) = o(g(k^\alpha))$ for large $k$) and
\begin{equation}
    n_i \leq \frac{k}{R \log(k/(R\varepsilon_0))^{a+1} }  \,.
\end{equation}
Because $n_i \leq k_i/R \leq k/R$ and $\varepsilon_0 \leq \varepsilon/(3f(k))$, for $t>0$ it holds that
\begin{equation}
    c n_i (\log(n_i/\varepsilon_0))^a
    \leq \frac{ck}{R\log(k/(R\varepsilon_0))} \Big( 1+O\Big(\frac{\log\log(k/(R\varepsilon_0))}{\log(k/(R\varepsilon_0))}\Big)\Big) < \frac{ck}{tR} \,.
\end{equation}    
Lastly, we show why such an $i$ exists.
Recall that we have assumed $0< n_i-n_{i-1}<n_i^\beta$ for $\beta<1$ for the qLDPC code family.
Also, $\alpha<1$ such that $0<\alpha\beta<1$ (which determines $f(n_i) = o(g(n_i^\alpha))$).
We pick smallest $i$ such that $n_i > k^\alpha$ (so $n_{i-1} < k^\alpha$), so that
\begin{align}
    k^\alpha &< n_i \leq n_{i-1}+n_{i-1}^\beta \\
    &\leq k^\alpha + k^{\alpha\beta} \\
    &\leq 2k^{\alpha'} \\
    &< \frac{k}{R \log(k/(R\varepsilon_0))^{a+1} } \\
    &\leq \frac{k}{R \log(3kf(k)/(R\varepsilon))^{a+1} }
\end{align}
for large enough $k$ where $\alpha' = \min\{\alpha,\alpha\beta\}<1$.
So, $n_i = O(k/\polylog(k))$.

\subsubsection{Total number of locations}\label{subsubsec:total-number-locations}
We now determine the total number of locations in $\Tilde{C}$.
First, let us consider locations in the QEC cycles.
Note that there are $T(n_i)$ SE repetitions each QEC cycle, $sl$ waits in between, a cat state for each of the $n_i-k_i$ stabilizer generators, each using $\frac{A}{1-Ap}$ gates for prep (on average), and $2r$ gates for measurement.
So, the total number of locations per QEC cycle is
\begin{equation}
    \text{\# locations per QEC cycle} \leq T(n_i) \Big( sln_i + (n_i-k_i)\Big( \frac{A}{1-Ap} + 2r \Big)\Big) \,.
\end{equation}
At each logical timestep there are $M/s$ QEC cycles, so the number of locations in QEC cycles at each logical step is
\begin{equation}
\begin{aligned}
    \text{\# QEC cycle locations per logical step} &\leq MT(n_i) \Big( ln_i + \frac{(n_i-k_i)}{s}\Big( \frac{A}{1-Ap} + 2r \Big)\Big) \\
    &= O(n_i T(n_i)) \,.
\end{aligned}
\end{equation}
For the number of locations in each logical gate implementation, note that there are $O(n_i\polylog(n_i/\varepsilon_0)) = O(n_i\polylog(n_if(k)/\varepsilon))$ locations in concatenated code ancilla factory for $|\Psi^{\overline{U}}\>$ per logical location (since $\varepsilon_0\leq \varepsilon/(3f(k))$).
Therefore, the total number of locations in each logical operation is given by
\begin{equation}
    \text{\# locations per logical operation} \leq O\Big( n_iT(n_i) + n_i\polylog(kf(k)/\varepsilon) \Big)
\end{equation}
Since $n_i<2k^{\alpha'}$ for $\alpha'=\max\{\alpha,\alpha\beta\}$, we have $T(n_i) < T(2k^{\alpha'})$ and $\polylog(n_if(k)/\varepsilon) < \polylog(k/\varepsilon)$.
So, total number of locations in $\Tilde{C}$ is
\begin{equation}
    |\Tilde{C}| \leq O\bigg( f(k) k^{\alpha'} \Big( T(2k^{\alpha'}) + \polylog(kf(k)/\varepsilon) \Big) \bigg) \,.
\end{equation}
Lastly, we also need to consider how many steps of classical computation are needed per logical step in $\Tilde{C}$.
Note that each QEC decoding runs in $O(h(n_i)) \leq O(h(2k^{\alpha'}))$ classical steps since $n_i<3k^{\alpha'}$ (recall the assumption in~\cref{thm:constant_overhead_ftqc_threshold} that decoding runs for $h(n_i)$ timesteps for some non-decreasing function $h$).
We also need to consider $O(\log\log\varepsilon_0) = O(\log\log (f(k)/\varepsilon)) = O(\log\log (k/\varepsilon))$ levels of concatenated code ancilla factory where decoding is performed.
So, the total number of classical step per logical step is
\begin{equation}
    \text{\# classical steps per logical step} = O(h(2k^{\alpha'}) + \log\log(k/\varepsilon)) \,.
\end{equation}
With this, we conclude the proof of~\cref{thm:constant_overhead_ftqc_threshold}.

\subsubsection{Further improvements}\label{subsubsec:further-improvements}
One shortcoming of Gottesman's constant-overhead FTQC result is that it makes no promise whether decoding can be done in polynomial time (in general it is indeed a hard problem).
This was solved first in the work by Fawzi \emph{et al.}~\cite{Fawzi_2018}, where the general Gottesman FTQC scheme is instantiated using \eczoo[quantum expander code]{quantum_expander} (which is a \eczoo[hypergraph product code]{hypergraph_product}~\cite{tillich2014quantum} with classical expander codes as seed codes).
The quantum expander code family has a parameter of $\llbracket \Theta(k_i),k_i,\Theta(\sqrt{k_i})\rrbracket$ and enables a single-shot decoder with $O(\log(k_i))$ runtime (even with noisy syndrome measurements).

More recently, a work by Nguyen and Pattison~\cite{nguyen2025quantumfaulttoleranceconstantspace} uses \eczoo[quantum locally-testable codes (LTC)]{qltc} which allows for single-shot decoding that incurs only a constant-time classical computation per logical quantum timestep.
Moreover, they also showed that an improvement from $O(\polylog(k))$ overhead in concatenated ancilla factory to $(\log k)^{o(1)}$ by using optimized stabilizer state and magic state distillation.
\endgroup

\subsection{Magic state distillation}\label{sec:magic_state_distillation}
\begingroup
\renewcommand{\H}{{{H}}}
\renewcommand{\CH}{{\mathrm{CH}}}
\renewcommand{\S}{{{S}}}
\newcommand{\T}{{{T}}}
\newcommand{\swp}{{\mathrm{SWAP}}}
\renewcommand{\cnot}{{\mathrm{CNOT}}}

\renewcommand{\Hg}{{{H}}}
\renewcommand{\Tg}{{{T}}}
\renewcommand{\cnotg}{{\mathrm{CNOT}}}
\renewcommand{\Sg}{{{S}}}
\renewcommand{\swapg}{{\mathrm{SWAP}}}
\renewcommand{\czg}{{\mathrm{CZ}}}
\renewcommand{\Id}{{\mathrm{Id}}}

Magic state distillation is one of the most, if not the most, widely used subroutines in FTQC schemes to enable a universal fault-tolerant computation based on stabilizer codes.
In general, a magic state distillation protocol starts with multiple copies of noisy \emph{magic states} $\rho^{\otimes m}$ followed by an encoding to logical magic states $\overline{\rho}^{\otimes k}$ in some quantum error-correcting code (for $k<m$), where errors can be detected or corrected to produce less noisy magic states $\Tilde{\rho}^{\otimes k}$.
Then, one can implement non-Clifford gates using gate teleportation on $\Tilde{\rho}^{\otimes k}$.
Thus, while the Eastin-Knill theorem obstructs a transversal implementation of a universal transversal logical gate set on any code (see~\cref{subsec:why-transversality}), one could still perform universal quantum computation using a stabilizer code with transversal logical Clifford gates and magic state distillation.
Here, we discuss a magic state distillation protocol proposed in~\cite{meier2012magicstatedistillationfourqubitcode} using the \eczoo[$\llbracket4,2,2\rrbracket$]{stab_4_2_2} error-detecting stabilizer code.

\subsubsection{The $\llbracket4,2,2\rrbracket$ stabilizer code}\label{subsubsec:422-stabilizer-code}
The $\llbracket4,2,2\rrbracket$ code is defined by its stabilizer group $\mathcal{S}=\<s_1,s_2\>$ with logical operator representatives $\overline{X}_1,\overline{Z}_1,\overline{X}_2,\overline{Z}_2$, given by
\begin{equation}\label{eqn:422_code_stabilizers_logicals}
    s_1 = \begin{matrix}
        XX \\ XX
    \end{matrix}
    \quad,\quad
    s_2 = \begin{matrix}
        ZZ \\ ZZ
    \end{matrix}
    \quad,\quad
    \overline{X}_1 = \begin{matrix}
        XX \\ II
    \end{matrix}
    \quad,\quad
    \overline{Z}_1 = \begin{matrix}
        ZI \\ ZI
    \end{matrix}
    \quad,\quad
    \overline{X}_2 = \begin{matrix}
        XI \\ XI
    \end{matrix}
    \quad,\quad
    \overline{Z}_2 = \begin{matrix}
        ZZ \\ II
    \end{matrix}\,,
\end{equation}
where we denote the four qubits in the square arrangement $\begin{matrix}
    P_1 P_2\\
    P_4 P_3
\end{matrix}$
where $P_i$ is the Pauli operator on the $i\numth$ qubit. Also, we note the that logical Hadamard gate $\overline{\H}_1\overline{\H}_2$ on both logical qubits of the $\llbracket4,2,2\rrbracket$ code can be implemented SWAP-transversally.
Namely, it can be implemented by applying $\H^{\otimes4}$ followed by a logical swap $\overline{\swp}_{1,2} = \swp_{2,4}$ (i.e., a physical swap between qubit 2 and 4) since
\begin{equation}
\begin{gathered}
    \H^{\otimes 4}\overline{X}_1\H^{\otimes 4} = \overline{Z}_2 \,,\quad
    \H^{\otimes 4}\overline{Z}_1\H^{\otimes 4} = \overline{X}_2 \,,\quad
    \H^{\otimes 4}\overline{X}_2\H^{\otimes 4} = \overline{Z}_1 \,,\quad
    \H^{\otimes 4}\overline{Z}_2\H^{\otimes 4} = \overline{X}_1 \\
    \text{and}\\
    \overline{\swp}_{1,2}\overline{X}_1\overline{\swp}_{1,2} = \overline{X}_2 \,,\quad
    \overline{\swp}_{1,2}\overline{Z}_1\overline{\swp}_{1,2} = \overline{Z}_2 \,,\\
    \overline{\swp}_{1,2}\overline{X}_2\overline{\swp}_{1,2} = \overline{X}_1 \,,\quad
    \overline{\swp}_{1,2}\overline{Z}_2\overline{\swp}_{1,2} = \overline{Z}_1 \,.
\end{gathered}
\end{equation}

\subsubsection{The $|\H\>$ magic state}\label{subsubsec:H-magic-state}
Here, we consider a magic state $|\H\>$, which is the $+1$ eigenstate of the Hadamard operator $\H$, given by
\begin{equation}
    |\H\> = R_y(\pi/4)|0\> = \cos\frac{\pi}{8}|0\> + \sin\frac{\pi}{8}|1\> \,,
\end{equation}
for rotation gate $R_y(\theta) = e^{-i\theta Y/2}$ about the Pauli-$Y$ axis. 
First let us consider the Hadamard twirling channel $\mathcal{T}_\H(\cdot) = \frac{1}{2}I\cdot I + \frac{1}{2}\H\cdot\H$ on a state $\rho$, which will give a resource state $\rho_\H$:
\begin{equation}
    \rho_\H = \mathcal{T}_\H(\rho) = (1-p)\ketbra{\H} + p\ketbra{-\H}
\end{equation}
for some $p\in[0,1]$ depending on the input $\rho$ and where $|-\H\>$ is the $-1$ eigenstate of $\H$.
Since $Y|\H\> = |-\H\>$, we can write the action of $\mathcal{T}_\H$ on state $\rho$ as
\begin{equation}
    \rho_\H = (1-p)\ketbra{\H} + pY\ketbra{\H}Y \,,
\end{equation}
namely, a Hadamard twirling results in the magic state $|\H\>$ with some probability $1-p$ and $Y$-error faulty magic state $Y|\H\>$ with probability $p$.

Since $|\H\>$ is the $+1$ eigenstate of $\H$, one can determine whether the noisy resource state $\rho_\H$ is the magic state $|\H\>$ or $|-\H\>$ by measuring observable $\H$ (treating $\H$ as an observable, instead of a gate).
A $+1$ outcome then corresponds to a projection onto the $|\H\>$ eigenspace and a $-1$ outcome a projection onto $|-\H\>=Y|\H\>$.
In particular, since the 4-qubit code encodes 2 logical qubits, we need to consider measurement of two-qubit observable $\H_1\H_2$ which projects a two-qubit state onto its $+1$ subspace $\spn\{|\H\>|\H\>,|-\H\>|-\H\>\}$ or its $-1$ subspace $\spn\{|\H\>|-\H\>,|-\H\>|\H\>\}$.
This measurement can be performed by the circuit
\begin{equation}\label{eqn:two_qubit_hadamard_measurement}
    \begin{quantikz}
        \lstick[1]{$|+\>$} & \ctrl{1} & \ctrl{2}& \meter{X} \\
        &\gate[1]{\H}&&&\\
        &&\gate[1]{\H}&\gate[1]{\H}&
    \end{quantikz}
\end{equation}
The control-$\H$ gate is a non-Clifford gate, which can be decomposed as
\begin{equation}\label{eqn:control_H_decomposition}
\begin{gathered}
    \begin{quantikz}
        & \ctrl{1} & \\
        & \gate[1]{\H} &
    \end{quantikz} =
    \begin{quantikz}
        & \bshade{2}{3}{} & \ctrl{1} & & \\
        & \gate[1]{R_y(-\frac{\pi}{4})} & \ctrl{0} & \gate[1]{R_y(\frac{\pi}{4})} &
    \end{quantikz}
\end{gathered}
\end{equation}
whereas the $Y$-rotation gate $R_y(\pm\frac{\pi}{4})$ can be performed using a $|\H\>$ state ancilla, a controlled Pauli gate, and measurement feed-forward (conditioning a Clifford gate on a measurement outcome) as
\begin{equation}\label{eqn:pi4_y_rotation_gadget}
\begin{gathered}
    \begin{quantikz}
        &\gate[1]{R_y(\pm\frac{\pi}{4})}&
    \end{quantikz} =
    \begin{quantikz}
        \setwiretype{n} & \invgate{|\H\>} \bshade{2}{3}{} & \ctrl{1} \setwiretype{q} & \meter{Y} \wire[d][1][""{right,pos= 0.4}]{c} \\
        && \gate[1]{Y} & \gate[1]{R_y(\pm\frac{\pi}{2})} &
    \end{quantikz}
\end{gathered}
\end{equation}
where a $R_y(\pm\frac{\pi}{2})$ gate is applied when the $Y$ measurement outcome is $+1$ and an identity gate is applied otherwise.

\subsubsection{Distilling two magic states using error-detection}\label{subsubsec:distilling-two-magic-states}
Two logical magic states $|\overline{\H}\>$ encoded in the $\llbracket4,2,2\rrbracket$ code can be obtained probabilistically using 10 resource states $\rho_\H$ as follows:
\begin{enumerate}
    \item Encode two resource states $\rho_\H$ to a $\llbracket4,2,2\rrbracket$ code block.
    
    \item Perform the logical $\overline{\H}_1\overline{\H}_2$ measurement on the code block.
    Here, we use a $\llbracket4,2,2\rrbracket$-code gadget that implements the circuit in~\cref{eqn:two_qubit_hadamard_measurement} by decomposing the control Hadamard gate as in~\cref{eqn:control_H_decomposition} and the $\pi/4$ $Y$-rotation gadget as in~\cref{eqn:pi4_y_rotation_gadget}.
    This gadget consumes eight resource states $\rho_\H$, which will be described shortly.

    \item A $-1$ outcome from measuring $\overline{\H}\otimes\overline{\H}$ indicates that one copy of the encoded $\rho_\H$ state has a $\overline{Y}$ error and a $+1$ outcome indicates that either both encoded resource states are $|\H\>$ or $|-\H\>$.
    
    \item Run syndrome extraction of the code by measuring the two stabilizer generators $s_1,s_2$ in~\cref{eqn:422_code_stabilizers_logicals} to detect single-qubit errors, then unencode the $\llbracket4,2,2\rrbracket$ code block.
    
    \item Accept the two output resource states if: (1) $\overline{\H}_1\overline{\H}_2$ measurement outcome is $+1$ and (2) there is no $-1$ syndrome.
    If at least one of the conditions is not satisfied, discard the output state and start over.
\end{enumerate}
It has been shown~\cite{meier2012magicstatedistillationfourqubitcode} that the $\overline{\H}_1\overline{\H}_2$ measurement can be implemented using the circuit $\mathcal{G}_{\overline{\H}_1\overline{\H}_2}$ given by
\begin{equation}
    \resizebox{0.85\linewidth}{!}{
    \begin{quantikz}
        \setwiretype{n} & \invgate{|+\>} \bshadeth{5}{13}{$\mathcal{G}_{\overline{\H}_1\overline{\H}_2}$} & \ctrl{2} \setwiretype{q} \gshade{3}{1}{} & \ctrl{4} \gshade{5}{1}{} 
            &\gategroup[5,steps=7,style={dashed,rounded corners,fill opacity=0,fill=green!10, inner xsep=0.3pt, inner ysep=0.3pt},background,label style={label position=below,anchor=north,yshift=0.3cm}]{{$U$}}&&& \ctrl{2} &&& \ctrl{4} & 
            \ctrl{2} \gshade{3}{1}{} & \ctrl{4} \gshade{5}{1}{} & \meter{X} & \setwiretype{c} \\
        &&& 
            &&&  &&& &&
            &&& \\
        &&\gate[1]{\H}&
            &\gate[1]{\H}&\gate[1]{\S}&\ctrl{2}&\ctrl{0}&\ctrl{1}&\gate[1]{\H}&&
            \gate[1]{\H}&&& \\
        &&&
            &\gate[1]{\H}&&&&\gate[1]{Y}&\gate[1]{Y}&&
            &&& \\
        &&&\gate[1]{\H}
            &&\gate[1]{\S^\dag}&\ctrl{0}& &&\ctrl{-1}&\ctrl{0}&
            &\gate[1]{\H}&& 
    \end{quantikz}
    }
\end{equation}
where the four controlled-$\H$ highlighted in green are themselves gadgets described in~\cref{eqn:control_H_decomposition} and~\cref{eqn:pi4_y_rotation_gadget}.
Since each controlled-$\H$ gadget uses two $Y$-rotations and each $Y$-rotation is implemented by a gadget that uses a single resource state ancilla $\rho_\H$, then the entire circuit $\mathcal{G}_{\overline{\H}_1\overline{\H}_2}$ uses eight resource states, as mentioned earlier in step 2.

A $Y$-rotation gadget in~\cref{eqn:pi4_y_rotation_gadget} with noisy resource state ancilla $\rho_\H$ instead of $|\H\>$ can be described by a $Y$ error at the ancilla with probability $p$ which is equivalent to a $Y$ error at its output
\begin{equation}
    \begin{quantikz}
        \setwiretype{n} & \invgate{|\H\>} \bshade{2}{4}{$\mathcal{G}_{R_y(\pm\frac{\pi}{4})}$} & \rgate{Y} \setwiretype{q} & \ctrl{1} & \meter{Y} \wire[d][1]["{o=\pm 1}"{right,pos= 0.4}]{c} \\
        &&& \gate[1]{Y} & \gate[1]{R_y(\pm\frac{\pi}{2})} &
    \end{quantikz}
    = \begin{quantikz}
        \setwiretype{n} & \invgate{|\H\>} \bshade{2}{3}{$\mathcal{G}_{R_y(\pm\frac{\pi}{4})}$} & \ctrl{1} \setwiretype{q} & \meter{Y} \wire[d][1]["{o=\pm 1}"{right,pos= 0.4}]{c} \\
        && \gate[1]{Y} & \gate[1]{R_y(\pm\frac{\pi}{2})} & \rgate{Y} &
    \end{quantikz}
\end{equation}
Thus, for the controlled-$\H$ gadget $\mathcal{G}_{\CH}$, a $Y$ error on the resource ancilla of the first $Y$ rotation is equivalent to a $Z\otimes Y$ error at the output of $\mathcal{G}_{\CH}$
\begin{equation}
    \begin{quantikz}
        & \bshade{2}{4}{$\mathcal{G}_{\CH}$} && \ctrl{1} & & \\
        & \gate[1]{\mathcal{G}_{R_y(-\frac{\pi}{4})}} & \rgate{Y} & \ctrl{0} & \gate[1]{\mathcal{G}_{R_y(\frac{\pi}{4})}} &
    \end{quantikz}
    =
    \begin{quantikz}
        & \bshade{2}{3}{$\mathcal{G}_{\CH}$} & \ctrl{1} & & \rgate{Z} & \\
        & \gate[1]{\mathcal{G}_{R_y(-\frac{\pi}{4})}} & \ctrl{0} & \gate[1]{\mathcal{G}_{R_y(\frac{\pi}{4})}} & \rgate{Y} &
    \end{quantikz}
\end{equation}
Now, for the two resource states $\rho_\H$ being encoded in the code block, there is a probability $p$ for each copy that it has a $Y$ error.
So for an encoding circuit $\mathcal{G}_\mathrm{enc}$, there is a probability $p$ of a logical error $\overline{Y}_1$ (resp., $\overline{Y}_2$) on logical qubit 1 (resp., logical qubit 2):
\begin{equation}
    \begin{quantikz}
        \lstick[1]{$|\H\>$} & \rgate{Y} & \gate[4]{\mathcal{G}_\mathrm{enc}} & \\
        \lstick[1]{$|\H\>$} & \rgate{Y} & & \\
        \setwiretype{n}&&&\setwiretype{q}\\
        \setwiretype{n}&&&\setwiretype{q}
    \end{quantikz}
    =\begin{quantikz}
        \lstick[1]{$|\H\>$} & \gate[4]{\mathcal{G}_\mathrm{enc}} & \gate[4,style={red!20,fill=red!10}]{\overline{Y}_1\overline{Y}_2} & \\
        \lstick[1]{$|\H\>$} & && \\
        \setwiretype{n}&&\setwiretype{q}&\\
        \setwiretype{n}&&\setwiretype{q}&
    \end{quantikz}
    = \overline{Y}_1\overline{Y}_2|\overline{\H}_1\overline{\H}_2\>
\end{equation}
The full distillation circuit then starts with encoding circuit $\mathcal{G}_\mathrm{enc}$ followed by Hadamard measurement gadget $\mathcal{G}_{\overline{\H}_1\overline{\H}_2}$ and ends with a decoding gadget $\mathcal{G}_\mathrm{dec}$ that performs syndrome extraction and unencode the code block to two physical qubits:
\begin{equation}
    \resizebox{0.92\linewidth}{!}{
    \begin{quantikz}
    \setwiretype{n} &&
            & \invgate{|+\>} \bshadeth{5}{13}{$\mathcal{G}_{\overline{\H}_1\overline{\H}_2}$} & \ctrl{2} \setwiretype{q} \gshade{3}{1}{} & \ctrl{4} \gshade{5}{1}{} 
            &\gategroup[5,steps=7,style={dashed,rounded corners,fill opacity=0,fill=green!10, inner xsep=0.3pt, inner ysep=0.3pt},background,label style={label position=below,anchor=north,yshift=0.3cm}]{{$U$}}&&& \ctrl{2} &&& \ctrl{4} & 
            \ctrl{2} \gshade{3}{1}{} & \ctrl{4} \gshade{5}{1}{} & \meter{X} & \setwiretype{c} \\
        \lstick[1]{$|\H\>$} & \gate[4]{\mathcal{G}_\mathrm{enc}} &
            &&& 
            &&&  &&& &&
            &&& \gate[4]{\mathcal{G}_\mathrm{dec}} & \\
        \lstick[1]{$|\H\>$} & &
            &&\gate[1]{\H}&
            &\gate[1]{\H}&\gate[1]{\S}&\ctrl{2}&\ctrl{0}&\ctrl{1}&\gate[1]{\H}&&
            \gate[1]{\H}&&&& \\
        \setwiretype{n}&&\setwiretype{q}
            &&&
            &\gate[1]{\H}&&&&\gate[1]{Y}&\gate[1]{Y}&&
            &&&&\setwiretype{n} \\
        \setwiretype{n}&&\setwiretype{q}
            &&&\gate[1]{\H}
            &&\gate[1]{\S^\dag}&\ctrl{0}& &&\ctrl{-1}&\ctrl{0}&
            &\gate[1]{\H}&& &\setwiretype{n}
    \end{quantikz}
    }
\end{equation}
Note that a single logical error $\overline{Y}_j$ triggers a $-1$ outcome from the Hadamard measurement gadget and becomes a physical $Y_j$ error on the output state, whereas two logical errors $\overline{Y}_1\overline{Y}_2$ become physical error $Y_1Y_2$ at the output.
A $Y$ error on one of the eight resource ancillas for one of the first two $\mathcal{G}_\CH$ gadgets will become either a $Z_a Y_i$ or $Y_i$ for $i\in\{2,4\}$ ($P_a$ denotes Pauli $P$ on the top measurement ancilla qubit) error at the end of the gadget.
For example, a $Y$ error on the ancilla resource state for first $R_y$ gadget in the second $\mathcal{G}_\CH$ gadget gives us errors
\begin{equation}\label{eqn:full_distillation_circuit}
    \resizebox{0.92\linewidth}{!}{
    \begin{quantikz}
    \setwiretype{n} &&
            & \invgate{|+\>} \bshadeth{5}{14}{$\mathcal{G}_{\overline{\H}_1\overline{\H}_2}$} & \ctrl{2} \setwiretype{q} \gshade{3}{1}{} & \ctrl{4} \gshade{5}{1}{} & \rgate{Z}
            &\gategroup[5,steps=7,style={dashed,rounded corners,fill opacity=0,fill=green!10, inner xsep=0.3pt, inner ysep=0.3pt},background,label style={label position=below,anchor=north,yshift=0.3cm}]{{$U$}}&&& \ctrl{2} &&& \ctrl{4} & 
            \ctrl{2} \gshade{3}{1}{} & \ctrl{4} \gshade{5}{1}{} & \meter{X} & \setwiretype{c} \\
        \lstick[1]{$|\H\>$} & \gate[4]{\mathcal{G}_\mathrm{enc}} &
            &&& &
            &&&  &&& &&
            &&& \gate[4]{\mathcal{G}_\mathrm{dec}} & \\
        \lstick[1]{$|\H\>$} & &
            &&\gate[1]{\H}&&
            &\gate[1]{\H}&\gate[1]{\S}&\ctrl{2}&\ctrl{0}&\ctrl{1}&\gate[1]{\H}&&
            \gate[1]{\H}&&&& \\
        \setwiretype{n}&&\setwiretype{q}
            &&&&
            &\gate[1]{\H}&&&&\gate[1]{Y}&\gate[1]{Y}&&
            &&&&\setwiretype{n} \\
        \setwiretype{n}&&\setwiretype{q}
            &&&\gate[1]{\H}& \rgate{Y}
            &&\gate[1]{\S^\dag}&\ctrl{0}& &&\ctrl{-1}&\ctrl{0}&
            &\gate[1]{\H}&& &\setwiretype{n}
    \end{quantikz}
    }
\end{equation}
It is shown in~\cite{meier2012magicstatedistillationfourqubitcode} that a single fault on any of the resource state ancillas in the $\mathcal{G}_\CH$ gadgets will trigger a syndrome flip in the decoding stage, since the $\llbracket4,2,2\rrbracket$ can detect any single-qubit error.
However, more than one faulty resource state ancillas in the $\mathcal{G}_\CH$ gadgets may not be detected and cause a logical error.

The probability of $k$ out of ten resource states $\rho_\H = (1-p)\ketbra{\H} + p\ketbra{-\H}$ being faulty is given by
\begin{equation}
\begin{gathered}
    \Pr[\text{$k$ faulty resource states}] = \binom{10}{k} p^k(1-p)^{10-k} \,.
\end{gathered}
\end{equation}
A rejection of the output state resulting from a single fault on the resource state, which is detected by either from the Hadamard check $\mathcal{G}_{\overline{\H}_1\overline{\H}_2}$ or the syndrome measurement by $\mathcal{G}_\mathrm{dec}$.
On the other hand, an output state is accepted either when the resources state is non-faulty or if there are multiple faults (evading detection).
So, we have an acceptance probability of
\begin{equation}
\begin{aligned}
    \Pr[\text{output accepted}] &= \Pr[\text{$0$ or $k\geq2$ faulty resource states}] \\
    &= (1-p)^{10} + \sum_{k=2}^{10} \binom{10}{k} p^k(1-p)^{10-k} \\
    &= 1 - O(p) \,.
\end{aligned}
\end{equation}
On the other hand, 2 or more faulty resource states will not trigger any syndrome flips or Hadamard-check flips, but may result in an output error.
So, the probability of bad outputs being accepted is
\begin{equation}
    \Pr[\text{bad output accepted}] \leq \sum_{k=2}^{10} \binom{10}{k} p^k(1-p)^{10-k} = O(p^2) \,.
\end{equation}
An output error happens when we accept a bad output, that is, where at least one of the output states is not a $|\H\>$ state, but no syndrome flips and no Hadamard-check flip.
Thus the probability of output error is
\begin{equation}
\begin{aligned}
    \Pr[\text{output error}] &= \Pr[\text{bad output}|\text{output accepted}] \\
    &\leq \frac{\Pr[\text{bad output accepted}]}{\Pr[\text{output accepted}]} \\
    &= \frac{O(p^2)}{1-O(p)} = O(p^2) \,.
\end{aligned}
\end{equation}
Thus, we have shown that using ten input resource states with error rate $p$ we can use the distillation circuit in~\cref{eqn:full_distillation_circuit} to obtain two copies of magic state with quadratically suppressed error rate $O(p^2)$.

\subsection{Code switching}\label{subsec:code_switching}
As with magic-state distillation, \emph{code switching} is a method to bypass the Eastin-Knill theorem in which one uses two stabilizer codes $Q_1$ and $Q_2$ such that code $Q_1$ (resp., $Q_2$) allows for a set of transversal logical gates $\mathcal{G}_1$ (resp., $\mathcal{G}_2$), where the two combined logical gate sets form a universal gate set.
Then, if we can find two such codes where we can transform one code into the other, preferably in a manner where errors do not propagate much, we can perform a universal FTQC by switching between these two codes to implement any gates from the universal gate set as needed. 

\begin{figure}
    \centering
    \includegraphics[width=0.7\linewidth]{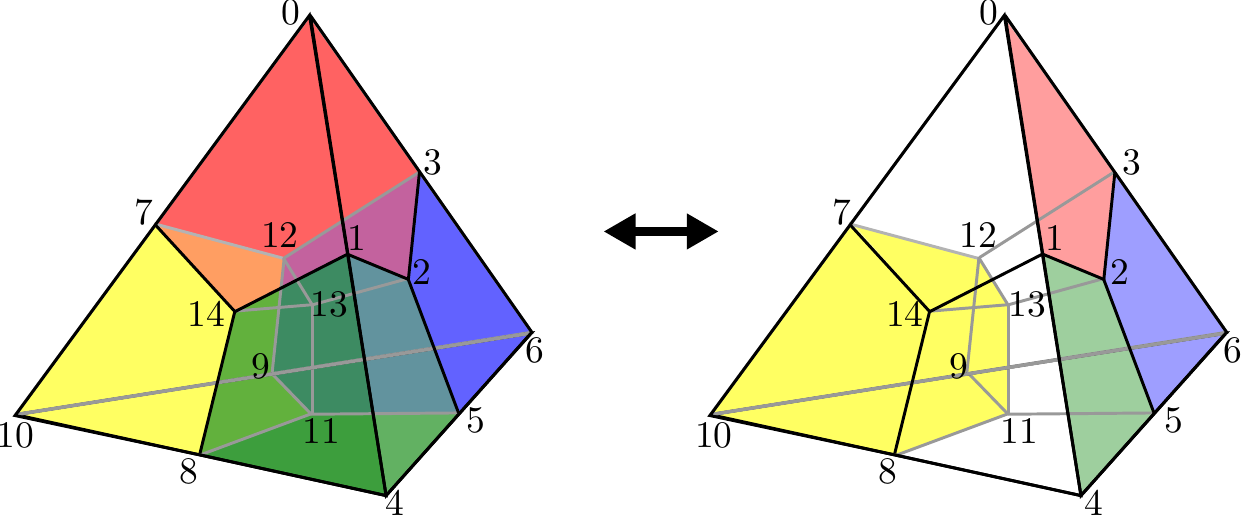}
    \caption{
    Code switching between Steane's $\llbracket7,1,3\rrbracket$ code (right-hand side face of the tetrahedron on the right-hand side) and $\llbracket15,1,3\rrbracket$ tetrahedral 3-D color code (left-hand side).
    (Figure reproduced from~\cite{Khu_2026} with permission from the authors.)
    }
    \label{fig:code_switching}
\end{figure}

One of the most popular pairs of codes to use for code switching is the \eczoo[$\llbracket7,1,3\rrbracket$ Steane code]{steane} and the \eczoo[$\llbracket15,1,3\rrbracket$ tetrahedral 3D color code]{stab_15_1_3} \cite{Paetznick_2013,Anderson_2014} (c.f.~\cref{fig:code_switching}). 
The Steane code allows for transversal logical $\H$ and $\cnot$ gates, while the tetrahedral code allows for transversal logical $\T$ gate, thus forming a universal logical gate set.

As illustrated in~\cref{fig:code_switching}, we need a total of 15 qubits to do this code switching protocol where the Steane code lies only on the 7 qubits (numbered 0 to 6) on one of the triangular faces of the tetrahedron.
When we encode the logical information in the Steane code, the remaining 8 qubits (on the vertices of the yellow hexahedron cell, numbered 7 to 14) do not encode any logical information.
On the other hand, the tetrahedral code has 14 stabilizers: 4 $X$-flavored and 4 $Z$-flavored 4-qubit stabilizers supported on the hexahedra, 3 $Z$-flavored 4-qubit stabilizers on the triangular face numbered 0 to 6, and 3 $Z$-flavored 4-qubit stabilizers on the three interior faces of the yellow hexahedron cell (numbered 7,8,9,11,12,13).

Switching from the tetrahedral code to Steane code can be done by measuring three weight-4 $X$ Paulis on the red, green, and blue faces on the triangular face supporting the Steane code.
Switching from the Steane code to the tetrahedral code can be done by measuring three 4-qubit $Z$ Paulis on the interior faces intersecting the red, blue, and green hexahedra (i.e., faces with vertices numbered 2,5,11,13 and 1,2,13,14 and 2,3,12,13).

A variant of code switching, \emph{transversal code switching}, is a fault-tolerant protocol for universal quantum computation proposed in~\cite{Heusen_2025} which uses a self-dual (CSS) code and a (CSS) \eczoo[tri-orthogonal code]{quantum_triorthogonal}. Since a self-dual code admits a transversal logical Hadamard gate, a tri-orthogonal code admits a transversal logical $\Tg$-gate, and CSS codes admit a transversal logical $\overline{\cnotg}$ gate, we can achieve universal quantum computation.
Note that, as opposed to more standard code-switching protocols that switch between stabilizer groups within a subsystem code, the transversal code-switching protocol requires two separate stabilizer code blocks.
However, fault tolerance of the protocol is guaranteed since all of its operations are transversal.

The protocol involves a transversal implementation of a logical teleportation from one code to another according to which gates one intends to perform.
For a self-dual code, we can choose the Steane code, while for a self-orthogonal code, we can choose the tetrahedral code, so that we use a total of $15+7=22$ qubits.
If we start from the Steane code in logical state $|\overline{\psi}\>_\mathrm{s}$, implementing a transversal logical $\Tg$ gate followed by a transversal logical $\H$ involves: 
\begin{enumerate}
    \item Preparing a logical $|\overline{0}\>_\mathrm{t}$ on the tetrahedral code, 
    
    \item Teleporting the Steane code logical state $|\overline{\psi}\>_\mathrm{s}$ to the tetrahedral code (a transversal $\overline{\cnotg}_{\mathrm{s}\rightarrow\mathrm{t}}$, logical $\overline{X}_\mathrm{s}$ measurement on the Steane code block, and logical Pauli $\overline{Z}_\mathrm{t}$ on the tetrahedral code block conditioned on $-1$ outcome of the $\overline{X}_\mathrm{s}$ measurement), 
    
    \item Applying transversal $\overline{\Tg}_\mathrm{t}$ on the tetrahedral block (resulting in $\overline{\Tg}_\mathrm{t}|\overline{\psi}\>_\mathrm{t}$),
    
    \item Teleporting the tetrahedral code state to the Steane code (a transversal $\overline{\cnotg}_{\mathrm{s}\rightarrow\mathrm{t}}$, logical $\overline{Z}_\mathrm{t}$ measurement on the tetrahedral code block, and logical Pauli $\overline{X}_\mathrm{s}$ on the Steane code block conditioned on $-1$ outcome of the $\overline{Z}_\mathrm{t}$ measurement), and lastly
    
    \item Applying transversal $\overline{\Hg}_\mathrm{s}$ on the Steane code block.
\end{enumerate}
\endgroup

\subsection{Outlook}\label{subsec:outlook}
With \emph{FTQC}, the goal is not only to store quantum information reliably, but also to process it without allowing errors to spread faster than they can be corrected. In practice, this requires not just good code properties, but also a universal logical gate set, control over correlated faults, and an architecture with manageable overhead. A central challenge is the implementation of non-Clifford gates. In many leading architectures, this relies on \emph{magic-state distillation}, a well-developed route to universality, but also one of the most resource-intensive parts of the stack. Much current work therefore aims to reduce its cost through improved protocols, better codes, and hardware-aware optimization. More recently, heuristic alternatives such as \emph{magic-state cultivation} have also been proposed, promising lower-cost preparation of high-fidelity $T$ states~\cite{gidney2024magic}.

At the same time, approaches such as \emph{code switching} and \emph{code deformation} seek to combine the strengths of different encodings, for example by using one code for storage and another for specific logical operations. This reflects a broader shift: fault tolerance is increasingly understood not as a property of a single code, but of an entire stack spanning hardware, control, decoding, and compilation.

Looking further ahead, one of the most exciting directions is \emph{constant-overhead fault tolerance} schemes~\cite{gottesman2014faulttolerantquantumcomputationconstant}. Surface code-based architectures have set the benchmark for locality and thresholds, but typically require substantial overhead as the target logical performance improves. By contrast, qLDPC codes and related constructions raise the prospect of protecting many logical qubits with far fewer physical resources, at least asymptotically. Whether this promise can be realized in practice remains open, since finite-size performance, decoder complexity, connectivity constraints, syndrome-extraction schedules, and realistic noise robustness are all crucial. Still, the possibility of going beyond the overheads of current leading schemes makes this a particularly important frontier.

More broadly, quantum error correction is evolving from the study of individual codes to the design of \emph{fully fault-tolerant systems}. This includes faster and more realistic decoders, codes tailored to specific noise biases and hardware platforms, and tighter integration between correction, calibration, and control. It also requires addressing experimentally important effects such as leakage, correlated and non-Markovian noise, and the latency of real-time classical processing. Progress is therefore likely to come from sustained co-design across theory, experiment, and architecture rather than from any single isolated breakthrough.

At the same time, the goals of the field are becoming clearer. In the near term, a key milestone is the demonstration of \emph{logical qubits that outperform physical qubits in a scalable and reproducible way}, including logical memories, logical gates, and small logical processors. Beyond that lies the transition from proof-of-principle error correction to genuinely useful fault-tolerant computation, where protected logical operations support algorithms of scientific or practical value. Achieving this will require not only lower logical error rates, but also credible resource estimates, robust compilation pipelines, and a better understanding of which applications are most likely to benefit first.

In this sense, quantum error correction is no longer only about suppressing decoherence. It has become the central organizing principle of quantum computing as a whole. The key questions now are not simply which code has the highest threshold or best asymptotic scaling, but which combination of codes, gates, decoders, and hardware can deliver useful logical performance under realistic constraints. The coming years will likely be shaped by this transition: from protecting qubits to engineering scalable fault-tolerant quantum computers.\vspace{5mm}

\subsection*{Acknowledgments}
\addcontentsline{toc}{section}{Acknowledgments}
D.J.S.~is supported by a graduate research fellowship from the Joint Quantum Institute (JQI) at the University of Maryland, College Park and grant ARL (W911NF-24-2-0107).
S.P.J.~is supported by the NSF grant OMA-2120757 (QLCI). A.T.~is supported by a CQT PhD scholarship, a Google PhD fellowship program, and a CQT Young Researcher Career Development Grant. Z.S.~is supported by a CQT PhD scholarship. K.B.~is supported by a Hartree fellowship from the Joint Center for Quantum Information and Computer Science (QuICS) at the University of Maryland, College Park.

\printbibliography
\end{document}